# THÈSE

**En vue de l'obtention du**

## DOCTORAT DE L'UNIVERSITÉ DE TOULOUSE

**Délivré par :**

Université Toulouse 3 Paul Sabatier (UT3 Paul Sabatier)

---

**Présentée et soutenue par :**
**Eric BLANQUIER**
le lundi 30 septembre 2013

Titre :

# The Polyakov, Nambu and Jona-Lasinio model and its applications to describe the sub-nuclear particles

---

**Ecole doctorale et discipline :**
ED GEET : Ingénierie des PLASMAS

**Unité de recherche :**
Laboratoire LAPLACE

**Directeur de Thèse :**
M. Patrice RAYNAUD

**Jury :**

| | |
|---|---|
| Directeur de thèse : | M. RAYNAUD Patrice, Directeur de recherche CNRS, Laboratoire LAPLACE, Toulouse |
| Rapporteurs : | M. CUGNON Joseph, Professeur, Université de Liège, BELGIQUE |
| | M. KNEUR Jean-Loïc, Directeur de recherche CNRS, Laboratoire Coulomb, Montpellier |
| Examinateurs : | M. COSTA Pedro, Chargé de recherche, Université de Coimbra, PORTUGAL |
| | M. HANSEN Hubert, Maître de conférences, Université de Lyon |
| | M. TOUBLANC Dominique, Professeur, Université de Toulouse |

# Acknowledgements

Via this page, I want to quote all the persons who made possible the realization of my thesis project.

Obviously, it concerns my thesis supervisor, Mr. Patrice RAYNAUD, who allowed me to validate my doctorate. Thank you so much Patrice. Thank you also to all the persons who intervened in the framework of the administrative procedures. I want to quote notably Mr. Alain CAZARRE, director of the *Ecole Doctorale Génie électrique, électronique, télécommunication* (ED GEET), Mrs. Marie ESTRUGA, executive assistant in the ED GEET, Mr. Christian LAURENT, director of the LAPLACE laboratory, Mr. François DEMANGEOT, second vice president Research at the University Paul Sabatier (UPS) Toulouse III, Mr. Arnaud LE PADELLEC, vice president of the CEVU in the UPS, Mrs. Monique LEMORT, responsible of the coordination of the doctoral training in the UPS and Mrs. ROSSI and Mrs. IZARIE of the 3$^{rd}$ cycle service in the UPS. Those who know about my thesis know also that the situation was unusual. Explaining why here would lead to quote persons who do not have to be present in this acknowledgements page. So, I will not do it. But, I can mention, because the anecdote is eventually interesting, that I obtained my PhD the 30$^{th}$ of September 2013, i.e. the eve of the 10$^{th}$ anniversary of my first official inscription to a doctorate.

Obviously, in the framework of the acknowledgements, I do not forget my thesis jury. So, I want to thank Mr. Jean Loïc KNEUR and Mr. Joseph CUGNON who accepted to be *rapporteurs* in my jury. I want to mention the work they done in the framework of the evaluation of my thesis. I thank them for our interesting discussions. Congratulation to Mr. KNEUR to have plunged in the depths of the appendix B. Thank you also to Mr. Pedro COSTA and Mr. Hubert HANSEN for their presence in my jury, as *examinateurs*. Their knowledge of the PNJL model was very useful. I want to thank particularly Mr. COSTA, because his help largely exceeded the scope of this thesis. Thank you so much for your support since we know us. Thank you also to Mr. Dominique TOUBLANC for his participation in my jury, as president of this committee. It was a pleasure to see you again after so many years. Thank you for your kindness and for your contribution in the framework of the validation of my diploma.

I want also to thank my family for their support, notably via the computer equipment they allowed me to use, that allowed me to terminate my last simulations. Thank you to the public that participate to the oral. Some words also to thank my contacts in the CSTJF Pau, my *former colleagues*. Thank you for your replies when I told you that I have obtained my PhD. Thank you also to the journals that allowed me to publish my work. Thank you to Rachel for her professionalism. Thank you to JC for his advices concerning my publications. Thank you to my former students. I cannot establish a list here, but I do not forget them. Thank you also to Céline for our collaboration, in the framework of my activity as scientific author.

# Table of contents

# Presentation

The researches devoted to understand the world that surrounds us made that, throughout the centuries, the limits of the human knowledge were successively pushed back farther. In particular, researches were often directed towards a comprehension of the matter on increasingly small scales, with the discovery of the atom, of the atomic nucleus... At the opposite, others preferred to turn the head towards stars, and to try understanding these objects, which appeared eternal to us still little time ago, to develop gradually the vision of an Universe including increasingly massive objects, like the galaxies, the galaxy clusters…

Most disconcerting is that these two ways of research, which seem however going in opposed directions, seem to rejoin themselves. Indeed, to describe the functioning of stars such as the white dwarfs, the pulsars..., the nuclear physics, i.e. the physics of infinitely small, is necessary [1, 2]. Astrophysicists and particles physicists see their work becoming increasingly close, without sometimes realizing it really. Is not there a better object to study nuclear fusion than a star? And, it is without speaking about the study of the first moments of the Universe with the Big Bang model [3, 4]. This bringing together of infinitely small and infinitely large would be similar to the well-known picture of the snake biting his own tail, employed to give the image of a cycle. In that point of view, the vision of a Universe similar to a mathematical fractal appears to guide the thought we have of it, in a more or less conscious way. Rutherford's Planetary Model used to describe the atom could be seen like an illustration of this reasoning, even if probably its author did not have this goal.

A hasty reasoning would then push us to conclude that we finally arrived at the end of the road since, according to what we have just seen, the circle is now complete. The Theory Of Everything, the completion of the ultimate scientific building would be thus for soon... However, this reasoning is of course too hasty, and even *incorrect.* It is incorrect because physics is only one scientific topic among other, describing only one very partial vision of the world. In addition, the history of sciences teaches that when the Man thinks perfectly understanding Nature, this one proves the opposite. The rise of quantum mechanics was thus accompanied by the greatest distress for the persons who thought being able to explain the world with traditional physics. A fractal vision of the Universe is finally a source of interesting analogies to try to understand such or such physical phenomenon [5, 6]. But, an atom is not as a star with planets which turn around it. A model, whatever it is, stays a way of interpreting reality, without *being* reality.

The physics of infinitely small, i.e. the particle physics, still holds surprises. It corresponds to the world of the quarks [7]. These particles currently make figures of elementary particles. They are considered as the "bricks" of the matter: thanks to them, the nucleons can be formed. Starting from the nucleons, the atom nuclei is then constituted. With the atom nuclei and electrons, the atoms are created. With atoms, molecules can be formed, and then the macroscopic matter … Even if it is probable that the existence of particles even smaller may be demonstrated in the future, as the preons [8], the quarks physics is currently a very intense research topic, theoretically and experimentally [9].

Firstly, in experimental physics, with large projects such as the American RHIC or the European LHC, collisions between nuclei with increasingly energies try to highlight the mechanisms that rule the quarks physics. In practice, a difficulty comes to the fact that the quarks cannot be observed in a free state. In normal conditions, they are confined in the traditional matter, i.e. in the nucleons. The goal of the experimenters working on this theme is to try to form a new state of the matter where these quarks are deconfined, during a very short lapse of time, before recombining. This very particular state of the matter is called the quark gluon plasma (QGP) [9–11]. It was probably present in the first moments of the Universe. The QGP is unobservable directly, so it is necessary for experimenters to prove that this state was really formed, via indirect evidences.

The current objective of the theoretical research related to the quark gluon plasma is to understand how this phase cools, and how the quarks/antiquarks combine themselves to form observables particles, i.e. to understand the mechanisms of the quarks/antiquarks hadronization. This transformation is named in the literature as the phase transition between the QGP and the hadronic phase [12, 13], in which the hadronic phase refers to the observable particles: the baryons (including the nucleons) and the mesons. In fact, the quarks physics is correctly described by the Quantum Chromodynamics (QCD) [9, 14]. Even if the QCD equations are known, one does not know how to solve them in the general case. This remark is particularly true in the framework of the involved energies in the QGP/hadronic phase transition. Notably, the quark confinement is still not mastered theoretically.

Thus, to study the hadronization of a quarks/antiquarks system, effective models are frequently used. Among them, we can quote the Nambu and Jona-Lasinio model (NJL) [15, 16]. Even if this model was not designed initially for that purpose, it proved its reliability to describe the quarks physics for a long time, notably thanks to the formulation of the model performed in the 1980s and in the 1990s, e.g. [17]. This approach allowed modeling particles as the quarks and the mesons, to study their behavior at finite temperatures and densities. In addition, the cross-sections associated with reactions between these particles were estimated, notably to investigate the formation of mesons starting from quarks and antiquarks [18]. The NJL model is at the origin of various works. For example, we can quote [19–21], in which the evoked NJL studies were considered again, and it was performed extensions of this topic. It was notably proposed an attempt to model baryons as a bound state of a quark and a diquark [20]. In [21], it was also described an attempt of an NJL simulation describing the cooling of a quarks/antiquarks plasma into mesons. However, the NJL model presents some limitations. In particular, the confinement is absent in this model. So, it was recently proposed a new version of the NJL model, in which a Polyakov loop was added in order to simulate a mechanism of confinement. This version is known as the PNJL model [22]. It was reported in the literature the various advantages of this approach. Among the already performed works, we mention for example the modeling of the quarks and the mesons with the PNJL model [23, 24].

But, to manage to describe the cooling of a quarks/antiquarks plasma into mesons and baryons, already performed works should be completed, in the framework of the NJL and of the PNJL approaches. First, the description of the relevant particles, as quarks, mesons and baryons, was traditionally done according to the temperature, to the density/chemical potential, but more rarely the both in the same time. The phase transition between the QGP and the hadronic phase can be done according to the temperature, to the baryonic density, and the both, thus calculations in the $T, \rho_B$ plane are thus fully interesting. Also, the baryon modeling should be completed in the framework of the NJL model, and then the baryons should be also included in the PNJL description. In the same way, if the reactions using quarks and mesons are well mastered now in the framework of the NJL model, the cross sections of reactions using baryons should be investigated [25]. Notably, the reactions forming the baryons are fully relevant, because they can allow understanding the dynamics of the baryonization of the system. In addition, in the NJL model, the cross-sections were estimated according to the temperature, and more rarely according to the baryonic density. It could be interesting to proceed to an estimation of the cross-sections according to these two parameters, in order to fully understand their influence on the results. Furthermore, PNJL cross-sections are very rare in the literature. Thus, the calculations of these cross-sections should be performed in this framework: the influence of the Polyakov loop on the cross sections is clearly not obvious. About the dynamical models studying the cooling of a quark/antiquark system, the one evoked in [21] does not include baryons. Even if the hadronization of the system is certainly dominated by a mesonization, the description of the baryons' formation in such a model is crucial. Also, in experimental conditions, the matter dominates the antimatter. Thus, a complete hadronization of such systems cannot be done only by the formation of mesons: the baryonization is clearly necessary. Moreover, the role played by the diquarks should also be clarified: what are their contributions in the dynamics? To answer this question, it appears as indispensable to study their formation during the cooling, and then to estimate if they are numerous enough to really intervene. Also, in such a study, it is particularly interesting to see if the used models, NJL and PNJL, could finally allow a complete hadronization of the system. For each evolution proposed here, a systematic comparison of the NJL and PNJL results should be performed, to estimate the concrete modifications induced by the Polyakov loop on the results, at each stage of the work.

In the framework of this thesis, we will consider the points evoked in the previous paragraph. To reach this objective, we propose the following structure: in the chapter 1, we proceed to a rapid overview of some theoretical notions useful in the frameworks of our works. It mainly concerns two topics: the Quantum Chromodynamics, and the group theory. About the QCD, it includes a description of the associated equations and an analysis of the specificities of this model, as the quark confinement. Furthermore, this chapter also evokes notions of group theory. Indeed, an objective of this part is to see how this theory can help us in this work [26]. In the chapter 2, we focus on a description of the NJL and PNJL models. We particularly insist on one side on the approximation to be done in order to obtain the NJL equations. On the other side, we present the inclusion of the Polyakov loop in the NJL model. In this way, we explain the modifications to be done in a pure NJL model to obtain the PNJL one. A first application of these models is presented in this chapter: the calculations of the masses of effective quarks.

In the chapter 3 to 5, we proceed to the modeling of composite particles that intervene in our work. In the chapter 3, the mesons are considered. Even if such particles were also studied in the (P)NJL models, we recover the results available in the literature, and we propose to extend the results according to various aspects. In addition, this chapter is an occasion to present the method to model composite particles, in order to adapt the equations for the other treated particles. These ones are the diquarks and the baryons. Indeed, we will treat the baryons as a bound state formed by a quark and a diquark, so we propose to study the diquarks in the chapter 4. There, we will see the used method, and then we will apply it to model several "families" of diquarks. A comparison between our results and the ones obtained by other approaches is proposed. Then, in the chapter 5, we focus on the baryons modeling. By using some approximations, a method to build the baryons is described. The baryons are then studied, as the other particles, according to the temperature and to the baryonic density. Secondary works are also performed in this chapter, as a modeling of anti-baryons, or a study of the baryons' stability in the $T, \rho_B$ plane.

Afterward, in the chapter 6, we estimate the cross-sections associated with reactions involving the quoted particles. On one side, we consider inelastic reactions. Notably, we study again the mesonization reactions evoked in [18] to recover the results, and then we extend the results following several directions: calculations at non-null baryonic densities, inclusion of the Polyakov loop … Then, several reactions using diquarks and/or baryons are considered, in order to foresee the dominant processes. On the other side, elastic reactions are treated, as [27] and new ones. Finally, in the chapter 7, we focus on the dynamical study of the quarks/antiquarks system. This part of the work constitutes a fascinating challenge, because all the calculations performed in the previous chapter are gathered in the computer code that performs the simulation. More precisely, it concerns the calculations of the masses of the quoted particles (quarks, mesons, diquarks, baryons and their antiparticles) and the estimations of the cross-sections of all the reactions evoked in the previous chapter. In this chapter 7, after explanations about the developed algorithms, some tests are done and commented. For example, the interactions between the particles are studied; a comparison is done between NJL and PNJL results... Then, complete simulations are presented. Thanks to them, the evolution of the system is analyzed, and we will see at this occasion if is possible to obtain a complete hadronization of the system with the (P)NJL models …

# Chapter 1

# Current knowledge about the QGP

## 1. Introduction

The quark physics [1, 2] is a field that concerns various topics. Upon its theoretical aspects, we firstly think about the Quantum Chromodynamics (QCD) [3, 4] used to model the strong interaction inside the nucleons, i.e. in fact the interactions between the quarks. Compared to the well know interactions as the gravitational and the electromagnetic ones, the strong interaction presents some characteristics, as the quark confinement inside the hadrons. The quest of a state in which these quarks could be deconfined, i.e. the quark gluon plasma (QGP), is currently the object of intensive researches. However, nowadays, the QCD equations cannot be solved in the general case, but only in some particular cases [5]. It will motive us in the following chapters to propose an alternative, via effective models. Moreover, the quarks physics implies also powerful mathematical tools, as the group theory [6–11]. In fact, this theory is based on the study of the symmetries of a physical system. This method is applicable to various domain, as crystallography, atomic physics, etc. [8]. In the framework of the particle physics [1], the study of symmetries allows confirming the number of various particles, as the baryons, mesons, … observed experimentally. Furthermore, the existence of symmetries has consequences on the Hamiltonian's writing.

The essential goal of this first chapter is to recall some elements about the relevant aspects in the study of the quarks physics. By the description of the involved theories, another goal is to mention the difficulties we will take into account in our work. In the section 2, some notions of the group theory are presented, and some applications to our work. A finality of this part is to be familiarized with the notations and terminologies used in the continuation of the work. A list of the quarks, mesons and baryons is then established by the use of the group theory. The QCD Lagrangian is recalled in section 3. A description of each term that composes this Lagrangian is proposed. The notion of chiral symmetry is introduced, as the breaking of this symmetry and the associated consequences. Then, in the section 4, we detail the characteristics and difficulties met with the QCD. It concerns the quark confinement, the unsolvability of the QCD equations, and the asymptotic freedom phenomenon. It leads to present the quark gluon plasma and its properties. More precisely, it is mentioned the conditions in which the QGP is expected to be formed, and when/where it was/is expected to exist. At this occasion, a phase diagram is proposed, according to the baryonic and strangeness densities and to the temperature. It permits to present the various objects or phases that exist or are supposed to exist according to the evoked parameters. It allows describing the "landscape" in which we will work in this thesis. In the section 5, we focus on the theoretical study of the QGP, notably via a rapid description of the lattice QCD (LQCD). The limitations of this approach are explained. In the section 6, we present some aspects linked to the experimental study of the QGP. Possible signatures of this phase are presented.

Relevant observables, as the elliptic flow, are also defined. Finally, a rapid overview of the recent experimental results is performed.

# 2. Symmetries

The study of symmetries is a very powerful tool in physics. The idea is to start by studying the existing symmetries of the considered physical system. Thanks to this analysis, it is possible to see that one or several physical quantities are conserved. From there, it is deduced the consequences on the Hamiltonian or Lagrangian describing the system. Indeed, it can be affected by the symmetries. More precisely, the conserved quantities can take part on the writing of our Hamiltonian/Lagrangian.

To illustrate that in a concrete way, a first simple example consists to imagine an isolated particle. This one is not subject to any force. Therefore, its speed is not modified according to the time. So, it implies symmetry by time translation. Thus, there is conservation of its momentum $p$ by the time. As a conclusion, the Hamiltonian or the Lagrangian associated with the particle have not any time-dependent term in their writing.

It is possible to go further. Within the framework of the quantum field theory, the Noether's theorem [10] is a generalization of this approach. This one stipulates that to all continuous symmetry corresponds a conserved current. This one is noted $J^{\mu}$, each $\mu$ is associated with a coordinate: 0 for the temporal coordinate and $1,2,3$ for the space coordinates. This current conservation is mathematically expressed by:

$$\partial_{\mu} J^{\mu} = 0 . \tag{1}$$

Let us consider now a Lagrangian $\mathcal{L}$ of a quantum field $\psi(x)$. The link between the symmetry (expressed indirectly by the preserved current) and the Lagrangian is given in the relation [10, 12]:

$$J^{\mu} = \mathcal{L} \cdot \delta x^{\mu} + \frac{\partial \mathcal{L}}{\partial(\partial_{\mu}\psi)} \cdot \delta\psi . \tag{2}$$

In addition, the scalar $J^{\mu} \cdot J_{\mu}$ can be a Lagrangian term. It is known as an *interaction term.* We will see along our work that this term will be useful, in particular when we will build our effective Lagrangian.

## 2.1 Introduction to group theory

Let us start with a simple example. We consider a macroscopic non-quantum and non-relativist object, and a Euclidean three-dimensional coordinate system. We take an unspecified point $M$ in the object. Its coordinates are:

$$\overrightarrow{OM} = \begin{bmatrix} x \\ y \\ z \end{bmatrix}. \tag{3}$$

Then, a rotation, whose center is the origin $O$, is applied to the object, This rotation is an active one: in other words, the object really rotates, whereas the coordinate system stays fixed. Of course, in a passive rotation, the axes would have undergone the rotation and the object would not have been moved… The rotation is modeled by the way of (4), extracted from [10, 11]. It gives the new coordinates of the point after the rotation:

$$\overrightarrow{OM'} = \begin{bmatrix} x' \\ y' \\ z' \end{bmatrix} = \exp\left(-i\vec{\theta}\cdot\vec{J}\right)\cdot\begin{bmatrix} x \\ y \\ z \end{bmatrix}. \tag{4}$$

The vector $\vec{\theta}$ was introduced in this formula. It specifies the rotation angles according to each axis of the coordinate system. Also, $\vec{J}$ is a vector whose components $J_x, J_y, J_z$ are matrices. A possible definition of these matrices is:

$$J_x = \begin{bmatrix} 0 & 0 & 0 \\ 0 & 0 & -i \\ 0 & i & 0 \end{bmatrix},\ J_y = \begin{bmatrix} 0 & 0 & i \\ 0 & 0 & 0 \\ -i & 0 & 0 \end{bmatrix},\ J_z = \begin{bmatrix} 0 & -i & 0 \\ i & 0 & 0 \\ 0 & 0 & 0 \end{bmatrix}. \tag{5}$$

If the rotation is performed according to the $z$ axis only, the term $\exp\left(-i\vec{\theta}\cdot\vec{J}\right)$ from the equation (4) is then simplified and rewritten as:

$$\exp\left(-i\vec{\theta}\cdot\vec{J}\right) = \exp\left(-i\underbrace{\theta_x}_{=0}\cdot J_x - i\underbrace{\theta_y}_{=0}\cdot J_y - i\theta_z\cdot J_z\right) = \exp\left(\theta_z\cdot\begin{bmatrix} & -1 & \\ 1 & & \\ & & \end{bmatrix}\right)$$

$$= \begin{bmatrix} 1-\frac{\theta_z^2}{2}+\frac{\theta_z^4}{24}+\dots & -\theta_z+\frac{\theta_z^3}{6}+\dots & 0 \\ \theta_z-\frac{\theta_z^3}{6}+\dots & 1-\frac{\theta_z^2}{2}+\frac{\theta_z^4}{24}+\dots & 0 \\ 0 & 0 & 1 \end{bmatrix} = \begin{bmatrix} \cos(\theta_z) & -\sin(\theta_z) & 0 \\ \sin(\theta_z) & \cos(\theta_z) & 0 \\ 0 & 0 & 1 \end{bmatrix}. \tag{6}$$

By simple geometry arguments, it can be checked that (6) corresponds indeed to a rotation of angle $\theta_z$ according to $z$.

To return to the group theory, the matrices $J_x, J_y, J_z$ are the *generators* of the rotations in a three-dimensional space. The objects that undergo rotations *generate* the representation of the symmetry group, which is here $SO(3)$. Indeed, $SO(3)$ is the sets of real matrices like $R(\theta) = \exp\left(-i\vec{\theta}\cdot\vec{J}\right)$, because all of them check the properties *S*, *O*, *3*, where *S* (*special*) means $\det(R)=1$, *O* for orthogonal (${}^T R = R^{-1}$ with ${}^T$ is the transposed operation) and 3 because the *R* matrices are square $3\times 3$ matrices. Here, the objects that undergo the rotations are *vectors*, such as the ones we employed in the example. On the other hand, objects such as the dots do not generate this group. Indeed, they are insensitive to all possible rotations on themselves created via $\exp\left(-i\vec{\theta}\cdot\vec{J}\right)$.

## 2.2 Some useful symmetries

The macroscopic world, described by traditional mechanics, is composed by material objects. These ones can be described by a set of points and vectors. As mentioned above, the points are insensitive to rotations on themselves. They are spin 0 objects, or *scalar* objects. Their symmetry group is $U(1)$. It corresponds to $1\times 1$ matrices (i.e. scalar) of the type $\exp(i\varphi)$, where $\varphi$ is an angle, so a real number. For a vector, it is necessary to make it at least one turn on itself to give it again its aspect, i.e. its initial direction. The vectors are spin 1 object, or *vectorial* objects.

To go further, a spin ½ object requires an even number of rotations, at least two, to give it again its initial aspect. Some of these objects can be particles currently considered as elementary. It concerns the electrons or the quarks, or not elementary particles, as some composite fermions if their spin corresponds to ½. They are *spinor* objects. Their symmetry group is $SU(2)$, i.e. the set of the complex square matrices, with a determinant equal to 1 (*S*) and unitary matrices (*U*), i.e. such as $R^{\dagger}=R^{-1}$. For them, the equivalent of (2) is:

$$u'=\exp\left(-\frac{i}{2}\vec{\theta}\cdot\vec{\tau}\right)\cdot u\,, \tag{7}$$

where $u,u'$ are the objects that could be named "vectors", because they *generate* the representation $SU(2)$. Clearly, in group theory, the concept of *vector* largely exceeds the concept we had in traditional mechanics. In fact, $u,u'$ are called spinors. In (7), $\vec{\tau}$ contains, like $\vec{J}$, three generators. They are the Pauli matrices, appendix B.

In addition, certain particles only require one half turn, at least, to recover their initial position. The most known example of particles checking this property concerns the gravitons. They are the vectors of the gravitational interaction, as well as the photons are those of the electromagnetic interaction. The gravitons are spin 2 particles. They are *tensorial* particles.

Except for these tensorial particles, it is possible to take into account another type of symmetry. We could describe it as "pseudo" symmetry. In a general way, these symmetries use the $\gamma_5$ Dirac matrix. In fact, four relevant symmetries should be underlined in the framework of our work; two of them include this new type of symmetry. The four are gathered in table 1 [13, 14]. The first of the list, i.e. $U_V(1)$, was previously evoked: it concerns scalar objects. If we simply "add" the $\gamma_5$ matrix to it, in its transformation matrix, a new symmetry is obtained: $U_A(1)$, known as *pseudo-scalar* symmetry. In the same way, the inclusion of a $\gamma_5$ matrix into the transformation matrix relating to the vectorial symmetry $SU_V(3)$ gives other symmetry: the *axial* symmetry.

| symmetry name | group theory designation | transformation matrix | conserved currents |
|---|---|---|---|
| scalar | $U_V(1)$ | $\exp(-i\varphi)$ | $J_\mu = \bar{\psi}\gamma_\mu \psi$ |
| pseudo-scalar | $U_A(1)$ | $\exp(-i\varphi\cdot\gamma_5)$ | $J_\mu = \bar{\psi}\gamma_\mu \gamma_5 \psi$ |
| vectorial | $SU_V(3)$ | $\exp\left(-\frac{i}{2}\theta_a\cdot\lambda^a\right)$ | $J_\mu^a = \bar{\psi}\gamma_\mu \lambda^a \psi$ |
| axial | $SU_A(3)$ | $\exp\left(-\frac{i}{2}\theta_a\cdot\lambda^a\gamma_5\right)$ | $J_\mu^a = \bar{\psi}\gamma_\mu \gamma_5 \lambda^a \psi$ |

**Table 1.** Symmetries and characteristics.

In table 1, a column is devoted to the conserved currents. For all the treated symmetries, each of them acts on a $\psi$ field. The column establishes a link with the field theory, and in particular with the concept of conserved current mentioned in the beginning of this chapter. From the table 1, it comes:

$$J_\mu^a = \bar{\psi}\gamma_\mu \Gamma^a \psi, \tag{8}$$

where $\gamma_\mu$ indicates the $\mu^{\text{th}}$ Dirac matrix and $\Gamma^i$ refers to the corresponding symmetry type. For scalar or pseudo-scalar symmetries, we take respectively 1 or $\gamma_5$ (only one current for each $\mu$, the index $a$ is thus useless). On the other hand, for vectorial or axial symmetries, we have respectively $\lambda^a$ or $\gamma_5\lambda^a$. The $\lambda^a$ term makes reference to the $a^{\text{th}}$ propagator of the symmetry group, i.e. we have $a$ currents for one fixed $\mu$.

# 2.3 Application of the group theory to the particles physics

The values of the electric charges or the mass of particles, can allow predicting the existence of symmetry between the particles that come from the same "family" (quarks, mesons...). The application of the group theory formalism made it possible to find theoretically some properties of subatomic particles. Besides, they were close from what is obtained in experiments. In addition, it is possible to anticipate the experimental results, and to fix the number of possible particles for a given type. We propose to see here the main ideas, and then to present the useful particles for our work.

We begin our description with the quarks. They are considered in the standard model as elementary particles [1]. In the framework of this thesis, the three quarks of flavor $u,d,s$ and their corresponding anti-particles $\bar{u},\bar{d},\bar{s}$ are considered. In fact, each one of these triplets generates the same group of symmetry. This one is $SU(3)_f$, the $f$ index referring to flavor. Compared to the group $SU(2)$ met before, the notable difference concerns the size of the matrices. Indeed, it is here $3\times3$ matrices.

A $SU(3)_f$ *vector* have each flavor quark as component, by its wave function $|u\rangle$, $|d\rangle$ or $|s\rangle$. All "rotation" in $SU(3)_f$ is then written:

$$\begin{bmatrix} |u'\rangle \\ |d'\rangle \\ |s'\rangle \end{bmatrix} = \exp\left(-\frac{i}{2}\theta_a \cdot \lambda^a\right) \cdot \begin{bmatrix} |u\rangle \\ |d\rangle \\ |s\rangle \end{bmatrix} , \tag{9}$$

where $\theta_a$ is the $a^{\text{th}}$ component of the "vector $\theta$". This one has 8 components; it is the equivalent of the $\vec{\theta}$ seen in (4). Also, $\lambda^a$ is the $a^{\text{th}}$ $SU(3)_f$ generator. We have 8 different generators. They are $3 \times 3$ matrices and they are clarified in appendix B.

Each component of the vector after the rotation, i.e. in the left hand side of (9), is a linear combination of the $|u\rangle, |d\rangle, |s\rangle$ wave functions. However, for some rotations, each component after rotation can be associated with one distinct quark flavor. In other words, it exists one rotation $\theta$ in which, for example, $|u'\rangle = |d\rangle$, $|d'\rangle = |s\rangle$ and $|s'\rangle = |u\rangle$ … In practice, the application of such a rotation gives good results for the quarks electric charges. But, it is not really the case for the masses, because the $s$ is heavier than the $u$ and $d$ quarks. In this case, $SU(3)_f$ is known as an *approximate* symmetry.

In our example using $SO(3)$, we note the existence of a conserved scalar quantity whatever the applied rotation: the vector norm $\|\overline{OM'}\| = \|\overline{OM}\|$. With $SU(3)_f$, converted scalar quantities also exist. They are frequently noted $Y$ and $I_3$ [1, 15]. The scalar $Y$ is the strong hypercharge, and it is defined by:

$$Y = N_B + S , \tag{10}$$

where $N_B$ is the baryonic number. $N_B = 1/3$ for one quark and $N_B = -1/3$ for one anti-quark. The $S$ indicates the strangeness number: $-1$ for one $s$ quark, 1 for one $\bar{s}$ anti-quark, and 0 for the others. About $I_3$, it is connected to $Y$ and to the particle's electric charge, noted $Q$, by the Gell–Mann and Nishijima relation:

$$I_3 = Q - \frac{Y}{2} . \tag{11}$$

Thanks to these quantities, the quarks and the anti-quarks can be represented in a two-dimension graph, left hand side of figure 1. The quark triplet forms the 3 representation; the anti-quark triplet forms the $\overline{3}$ one. The quarks are the "bricks" of composite particles, i.e. the hadrons. Therefore, by several combinations of quarks and/or anti-quarks, these particles can be created. With the group theory, the process is simple: each quark or anti-quark is treated as a vector in the $I_3, Y$ plane. The mesons are thus obtained by a vectorial summation of a quark and an antiquark vector, right hand side of figure 1. We represented the "family" of the most used mesons, i.e. the pseudo-scalar ones. They have the greatest stability, because they are the lightest, compared to the other mesons.

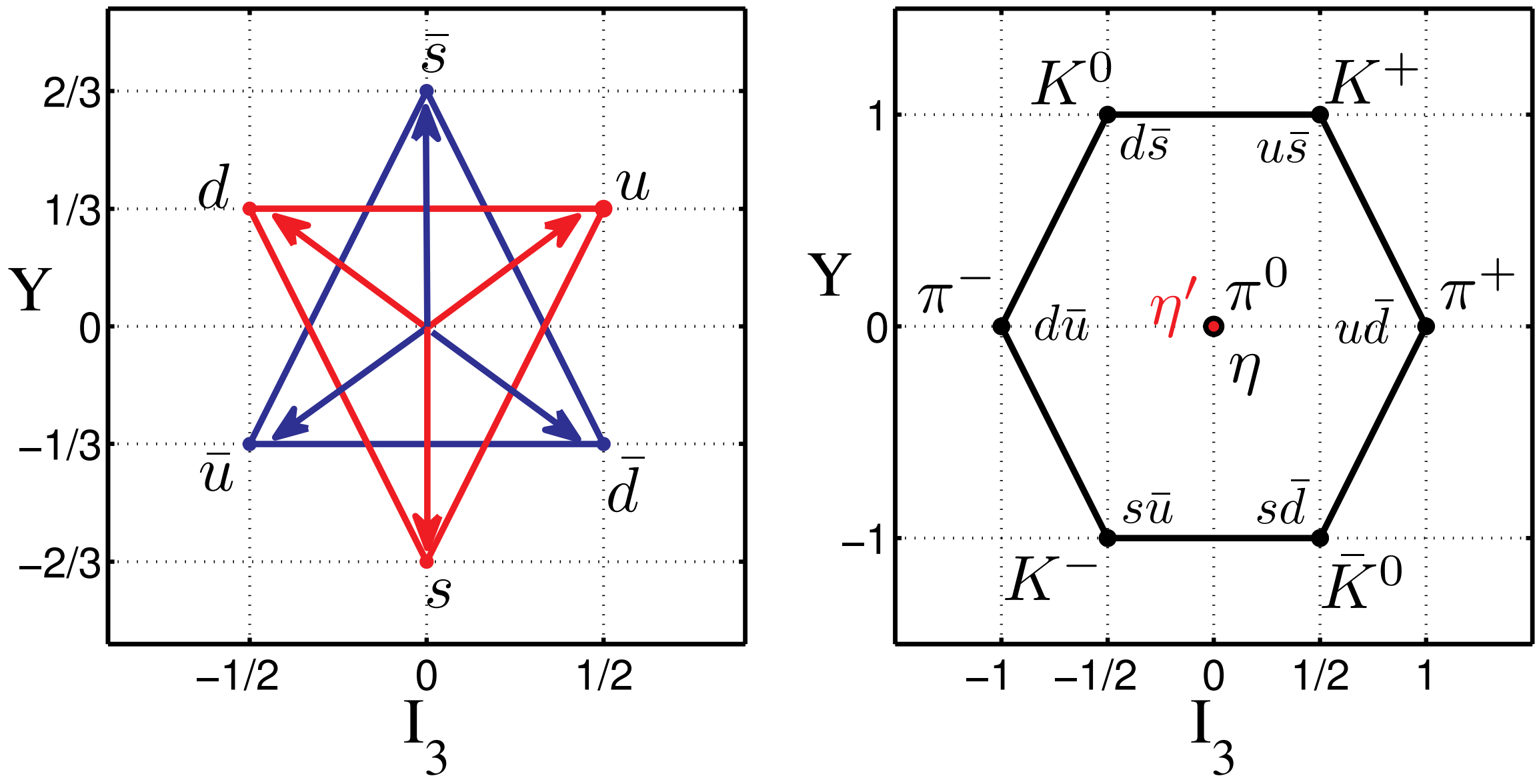

**Figure 1.** Left hand side: $u,d,s$ quarks triplet (representation 3) and anti-quarks $\bar{u},\bar{d},\bar{s}$ triplet (representation $\bar{3}$). Right hand side: Pseudo-scalar mesons, octet and singlet ($\eta'$).

Let us focus on the mesons example. In this case, we saw that a quark/antiquark association is considered. It implies a vectorial space corresponding to the tensor product of the quarks space and the anti-quarks space. It leads to the formal writing of the mesons wave function:

$$|q\bar{q}\rangle \equiv |q\rangle \otimes |\bar{q}\rangle \,. \tag{12}$$

In the group theory formalism, the representation of the mesons' symmetry group is written in the same way. Indeed, by analogy with (12), it is written as a tensor product of the 3 and $\bar{3}$ representations. Thus, we write $3\otimes\bar{3}$. This writing can be modified if we express the representation as the sum of representations that cannot be simplified, known as *irreducible representations*. Finally, it comes [1, 16]:

$$3\otimes\bar{3} = 1\oplus 8 \,. \tag{13}$$

Therefore, the mesons representation is composed by one unit representation, known as trivial representation, and by an 8 representation, i.e. an octet. It explains that we have 9 mesons in the figure 1: 8 octet mesons and the singlet meson $\eta'$. Moreover, in the framework of $SU(3)_f$, the theory indicates the existence of an octet (representation 8) and a decuplet (representation 10) of baryons. Theses baryons are represented in the figure 2, in which we used the same method as the one described for the mesons. Now, we propose to temporarily leave the group theory description to focus on the basic theory that governs the systems we want to model.

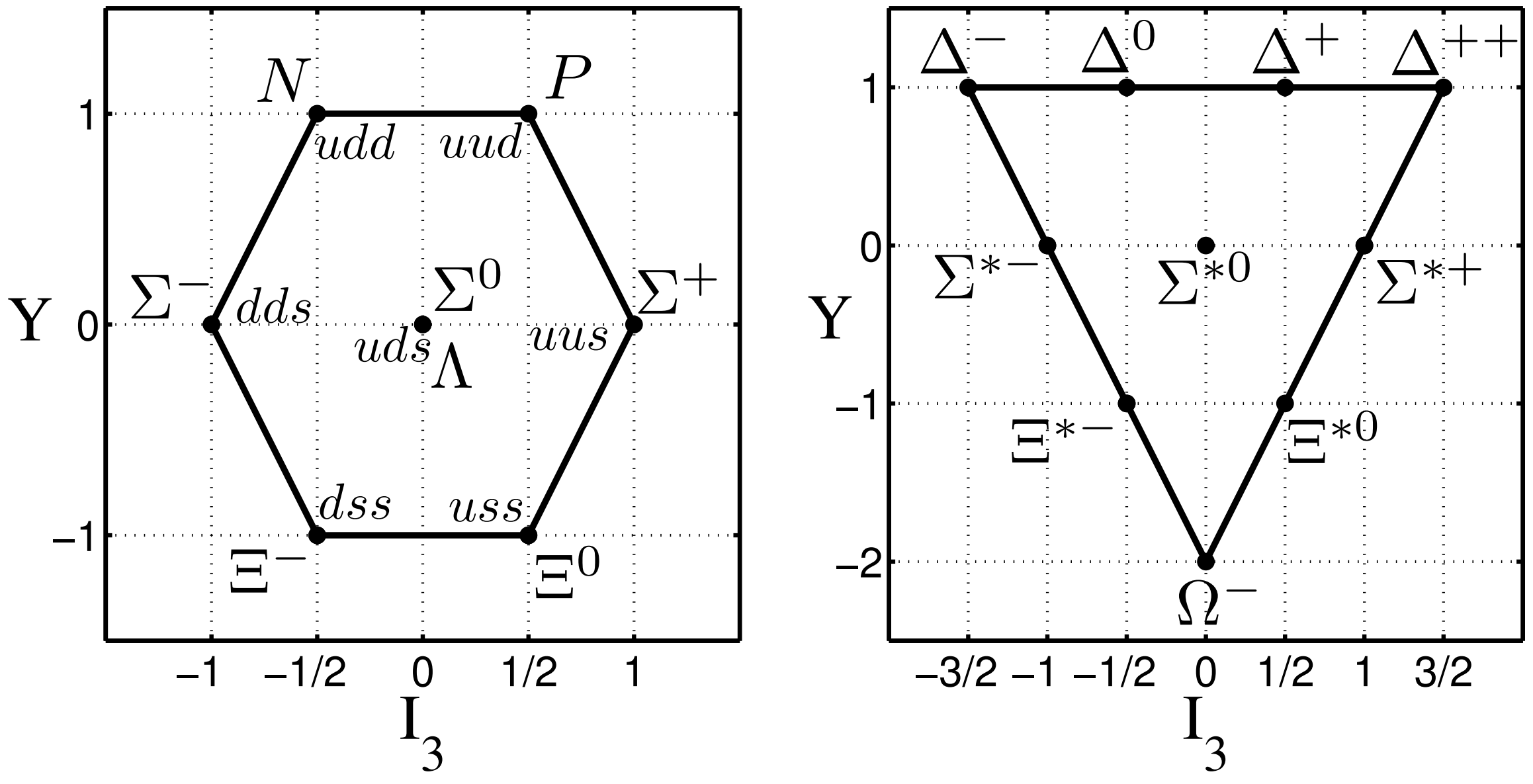

**Figure 2.** The baryons. Left: octet baryons. Right: decuplet baryons.

# 3. The Quantum Chromodynamics

## 3.1 Description of the QCD Lagrangian

The Quantum Chomodynamics (QCD) is the most sophisticated model to describe the quarks physics [2, 4]. Its Lagrangian models the strong interaction, i.e. the interaction undergone by the quarks. The vectors of this interaction are gluons. They are spin 1 massless particles. The QCD Lagrangian is written, in a condensed form [5, 17, 18]:

$$\mathcal{L}_{QCD} = -\frac{1}{4} \cdot G^a_{\mu\nu} \cdot G^{\mu\nu}_a + \sum_f \bar{\psi}_f \left( i\gamma^\mu D_\mu - m_{0f} \right) \psi_f \quad , \tag{14}$$

with:

$$D_\mu = \partial_\mu - i g_s \cdot A_\mu \ , \tag{15}$$

in which we have:

$$A_\mu = \sum_{a=1}^{8} A^a_\mu \cdot \frac{\lambda_a}{2} \text{ writable as } A_\mu = A^a_\mu \cdot \frac{\lambda_a}{2} \text{ (Einstein summation convention).} \tag{16}$$

In (14), $D_\mu$ corresponds to $\partial_\mu$, i.e. the derived operator, in which we apply a gauge transformation, as detailed in (15). $g_s$ is the coupling constant associated with the strong interaction. Also, $G^a_{\mu\nu}$ is the gluon field tensor, expressed as:

$$G^a_{\mu\nu} = F^a_{\mu\nu} + g_s \cdot f^{abc} \cdot A_{b\mu} \cdot A_{c\nu} , \tag{17}$$

in which $F^a_{\mu\nu}$ is defined by:

$$F^a_{\mu\nu} = \partial_\mu A^a_\nu - \partial_\nu A^a_\mu \ . \tag{18}$$

The $A_\mu^a$ refer to the $\mu$ component of the gluon field $A$. The $a$ is a color index. The $\lambda_a$ are the eight $SU(3)$ generators, $f^{abc}$ the associated structure constants, see (9) and appendix B. But now, the symmetry group $SU(3)$ is noted $SU(3)_c$, $c$ like *color*. The $\psi_f$ refer to the quarks fields with flavor *f*. More precisely, the $\psi_f$ are triplets, i.e. 3-component vector. They admit $SU(3)_c$ as symmetry group. They are written, in developed form [4]:

$$\psi_f \equiv \begin{pmatrix} \psi_f^{red} \\ \psi_f^{green} \\ \psi_f^{blue} \end{pmatrix}. \tag{19}$$

The *red, green, blue* subscripts refer to a quantum number specific to Chromodynamics: the color. In the same way, the "anti-values" anti–red, anti–green and anti–blue are associated with anti-quark fields.

The flavor is another quantum number, linked to the considered quark type. We already spoke about it in subsection 2.3. In fact, 6 quark flavors exist in the current model: *u*, *d*, *s*, *c*, *b*, *t*, and so on for the equivalent anti-quarks: anti–*u*, noted $\bar{u}$,… Each of them has a mass known as naked mass, or current mass. This one is noted as $m_{0f}$ in (14). The lightest quark is the $u$, whose mass is about 2 or 3 MeV, whereas the $t$ would have a mass close to 173 GeV, i.e. comparable to the one of a gold nucleus…, see appendix A. Therefore, thermodynamic arguments legitimate the choice to keep only the 3 quarks $u$, $d$ and $s$, as done in the present work. More precisely, the others do not really intervene in the physics we want to describe: they are definitely too heavy to be created in a notable way, even if we will see later the possible role played with the mesons made by *c* and $\bar{c}$ … By analogy with (19), we write the complete quark field as:

$$\psi \equiv \begin{pmatrix} \psi_u \\ \psi_d \\ \psi_s \end{pmatrix}. \tag{20}$$

Each component of the vector presented in (20) can be broken up as in (19), $f = u,d,s$. Also, $\psi$ admits $SU(3)_f$ as symmetry group (9). To finish our analysis, let us imagine that we do not have gluons in the equations. This would correspond to a theory where the quarks would not interact together. So, they could be considered as free particles. The equation (14) would then be written:

$$\mathcal{L}_{free} = \sum_f \bar{\psi}_f \left(i\gamma^\mu \partial_\mu - m_{0f}\right)\psi_f = \sum_f \bar{\psi}_f \left(i\partial\!\!\!/ - m_{0f}\right)\psi_f . \tag{21}$$

Finally, it could be expressed in the condensed form:

$$\mathcal{L}_{free} = \bar{\psi}\left(i\partial\!\!\!/ - m_0\right)\psi, \tag{22}$$

in which $m_0$ corresponds to a matrix defined by:

$$m_0 = \begin{bmatrix} m_{0u} & & \\ & m_{0d} & \\ & & m_{0s} \end{bmatrix}. \qquad (23)$$

Clearly, (22) is the Dirac Lagrangian [19]. It describes the evolution of spin ½ quantum relativistic particles.

# 3.2 The chiral symmetry

An important characteristic related to the QCD Lagrangian is related to the notion of chiral symmetry. More precisely, the quarks can be divided in two categories according to their chirality: right (the spin and the momentum are in the same sense) and left (opposite senses). We note $\psi_R$ a field of "right quarks", and $\psi_L$ a field of "left quarks". A field of quarks can be written as:

$$\psi = \begin{pmatrix} \psi_R \\ \psi_L \end{pmatrix}. \qquad (24)$$

Using the $\gamma^5$ matrix defined in a Weyl (chiral) representation, appendix B,

$$\gamma^5 = \begin{bmatrix} 1_2 & \\ & -1_2 \end{bmatrix}, \qquad (25)$$

we define the projectors upon the left and right states [20]:

$$\psi_L = \frac{1_4 - \gamma^5}{2}\psi, \quad \bar{\psi}_L = \psi_L^+ \gamma_0 = \bar{\psi}\frac{1_4 + \gamma^5}{2}, \quad \psi_R = \frac{1_4 + \gamma^5}{2}\psi, \quad \bar{\psi}_R = \psi_R^+ \gamma_0 = \bar{\psi}\frac{1_4 - \gamma^5}{2}, \qquad (26)$$

where $1_2$ and $1_4$ are respectively the $2\times 2$ and $4\times 4$ identity matrices. If we use these projectors in the QCD Lagrangian (14), and if we do not explicit the flavors, we obtain:

$$\mathcal{L}_{QCD} = -\frac{1}{4}\cdot G^a_{\mu\nu}\cdot G_a^{\mu\nu} + i\bar{\psi}_R\gamma^\mu D_\mu \psi_R + i\bar{\psi}_L\gamma^\mu D_\mu \psi_L - m\cdot\bar{\psi}\psi, \qquad (27)$$

because of the anti-commutation relation $\{\gamma^\mu, \gamma^5\} = 0$, appendix B. Clearly, the terms $i\bar{\psi}_R\gamma^\mu D_\mu \psi_R$ and $i\bar{\psi}_L\gamma^\mu D_\mu \psi_L$ indicates that a left quark can only interact with the other left quarks. In the same way, the right quarks interact with right quarks. This splitting of the left and right quarks corresponds to the chiral symmetry. However, the term $m\cdot\bar{\psi}\psi = m\cdot(\bar{\psi}_R\psi_L + \bar{\psi}_L\psi_R)$ breaks this symmetry, because left and right quarks are "allowed to interact" by this term, in which $m$ is the mass of the quarks. At high temperatures and/or high densities, the quarks' masses are close to the naked masses $m_0$. For the quarks $u$ and $d$, the naked masses are low enough to be able to neglect the $m_0\cdot\bar{\psi}\psi$ term. It wants to say that the chiral symmetry is there an approximate symmetry. One says that the chiral symmetry is *explicitly broken*. However, in some calculations, it is possible to work at the chiral limit, i.e. the quarks naked masses are completely neglected.

At the opposite, at low temperatures and/or densities, quark-antiquark condensates can appear. They are noted traditionally as $\langle\langle\bar{\psi}_f \psi_f\rangle\rangle$, and they are frequently named as *quark condensates* or *chiral condensates*. The antiquark of a condensate can interact with a quark (not from a condensate). By this way, a left quark and a condensate can give a right quark, as illustrated by the figure 3. This coupling breaks the chiral symmetry.

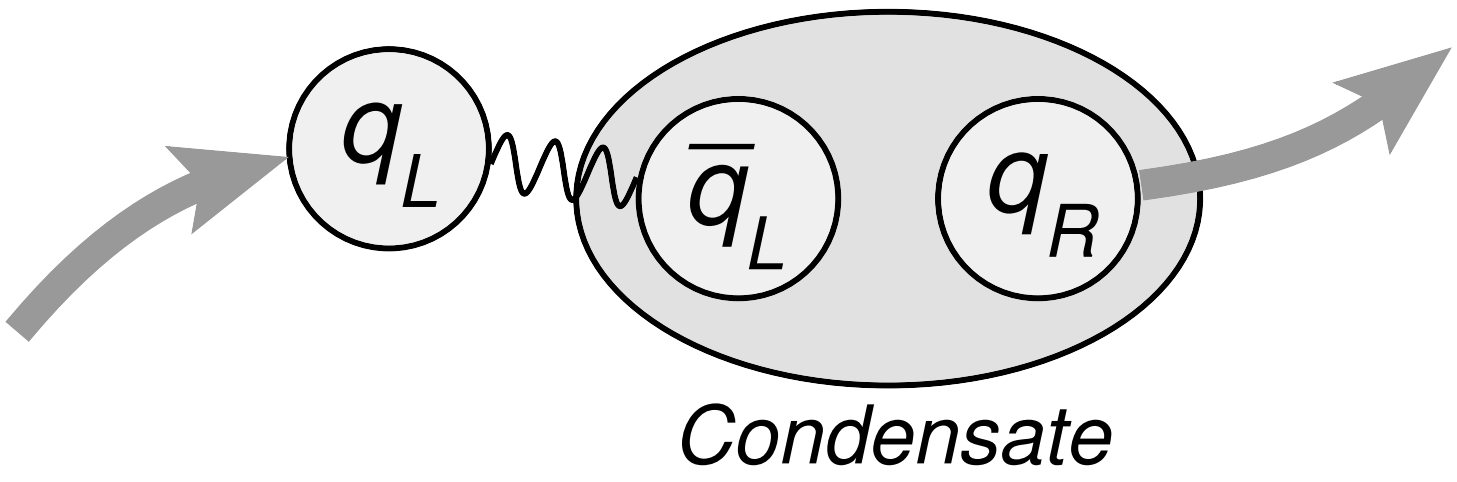

**Figure 3.** Description of the breaking of the chiral symmetry by the condensate. This picture was largely inspired from [5].

In fact, this interaction leads to consider effective quarks masses. Clearly, we consider there a mass *m* largely greater than the naked mass. The term $m \cdot \bar{\psi}\psi$ is responsible of the breaking of the chiral symmetry, thus an increase of *m* allows this term to become non-negligible. In this case, the chiral symmetry is *spontaneous broken*.

# 3.3 The breaking of the chiral symmetry

In physics, the breaking of symmetries presents relevant applications. More precisely, the Goldstone theorem [21, 22] explains that when a continuous symmetry is spontaneously broken, it gives birth to bosons, designated as the Nambu-Goldstone bosons. If the symmetry is exact, these bosons are massless. At the contrary, in the case of approximate symmetries, the (pseudo)-Nambu-Goldstone bosons are massive particles. In the framework of the chiral symmetry with light quarks *u* and *d*, the symmetry is not exact but well verified, thanks to the low naked masses of these quarks. As a consequence, the associate Nambu-Goldstone bosons have got low masses. They correspond to the pions. In fact, the Gell-Mann-Oakes-Renner (GMOR) relation [23] establishes a link between the masse of the pions $m_\pi$, their disintegration constant in the vacuum $f_\pi$, the *u*, *d* quarks naked masses $m_{0u}, m_{0d}$ and the value of the light quark condensate $\langle\langle\bar{\psi}_q \psi_q\rangle\rangle$ [20, 24]:

$$m_\pi^2 \cdot f_\pi^2 = -(m_{0u} + m_{0d}) \cdot \langle\langle\bar{\psi}_q \psi_q\rangle\rangle, \text{ with } \langle\langle\bar{\psi}_q \psi_q\rangle\rangle \equiv \frac{\langle\langle\bar{\psi}_u \psi_u\rangle\rangle + \langle\langle\bar{\psi}_d \psi_d\rangle\rangle}{2}. \quad (28)$$

The strange quark is heavier, so the chiral symmetry involving $SU(3)_f$ quarks is more approximate. It is imagined [25] that the related Nambu-Golstone bosons can be associated with the $\eta$ and kaons in the limit where the *u, d, s* quarks are massless.

Moreover, the Landau theory on phase transition [26] considers the phase transitions for which a symmetry is broken or restored, e.g. according to the temperature. An order parameter is a quantity used in order to study such a phase transition. In the phase in which the symmetry is respected, the order parameter is null. At the opposite, when the symmetry is broken, the order parameter is non-null. The way how the order parameter varies, from null to non-null values (or reversely), teaches us about the order of the phase transition. More precisely, when the order parameter presents a discontinuity between the two phases, we have a first order phase transition. When the order parameter goes brutally to zero, but continuously, it corresponds to a second order phase transition. There, the derivative of the order parameter upon the temperature presents a discontinuity. Also, when the order parameter only converges towards zero, i.e. it has non-null values all the time, we have a crossover. These three configurations are displayed in the figure 4, in which $T_C$ corresponds to the critical temperature of the transition.

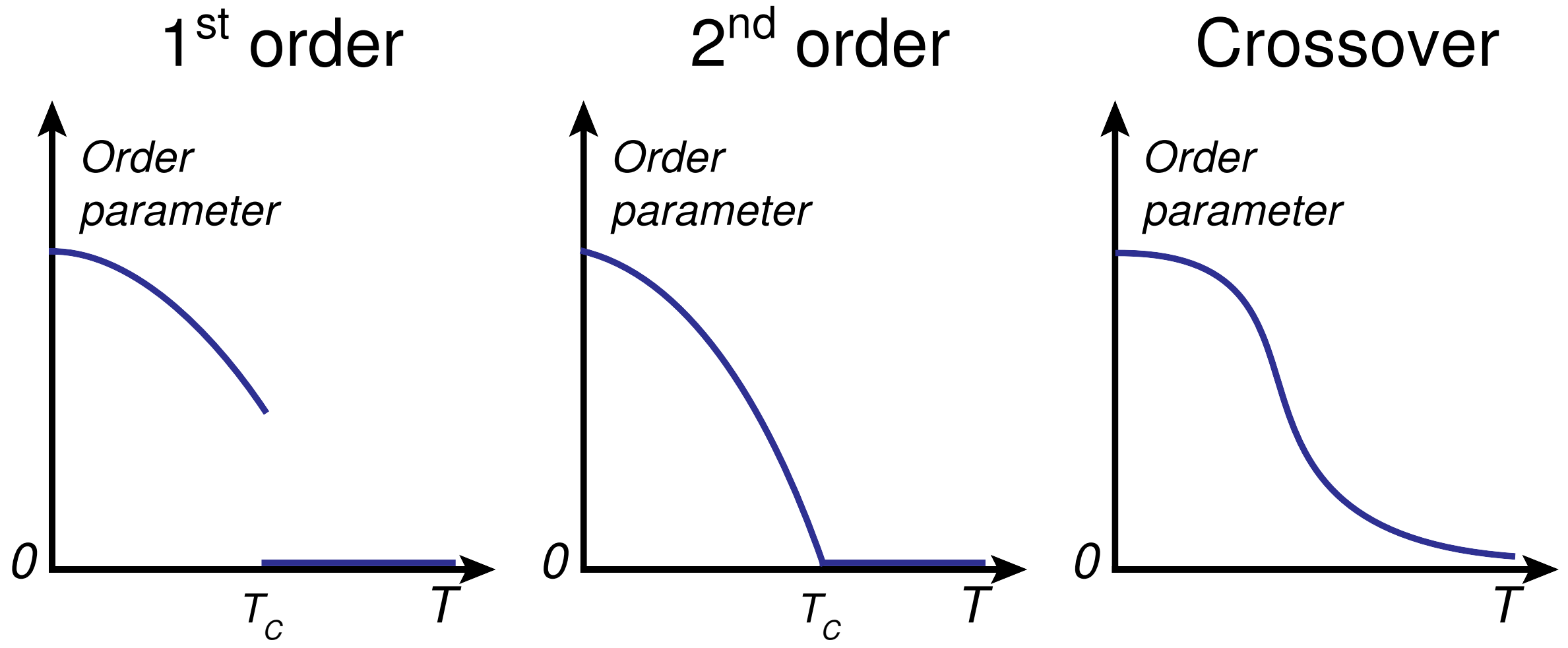

**Figure 4.** Evolutions of the order parameter according to the temperature *T*.

In the case of the chiral symmetry, the value of the quark condensate $\langle\langle \bar{\psi}_q \psi_q \rangle\rangle$ constitutes an order parameter. Clearly, an objective can be to study the restoration of the chiral symmetry at high temperatures, when the value of the condensate tends to zero.

# 4. Characteristics and problems involved in the quarks physics

## 4.1 Quarks confinement inside hadrons

The confinement of the quarks is a notable phenomenon related to the strong interaction [4]. As an example, we consider one quark and one anti-quark. If these two particles are linked together, it corresponds physically to a meson. The strong interaction prohibits the separation of our two particles if no other particle intervenes, figure 6 in subsection 4.3. Therefore, when the distance between these two particles is too large, the strong interaction acts via a springlike force, as an elastic cord. The force prevents the two particles to be found insulated one compared to the other. In fact, it is possible to modify a structure like a meson or a baryon in high energies collisions. However, a quark or an anti-quark will never be found insulated, i.e. in a free state, even if there is creation of quarks/anti-quarks pairs.

Indeed, the standard model stipulates that the only particles being able to exist individually, and by extension being observed, must be objects having a color either null ("black") or a "complete" color ("white" or "anti-white"). This concept of color refers to the color charge carried by the QCD objects, as saw in the subsection 3.1: the color of a quark can be $r, g, b$, and the "anti-colors" of the anti-quarks can be $\bar{r}, \bar{g}, \bar{b}$. For example, a meson built with a *red* quark necessarily has a $\bar{r}$ anti-quark. As a consequence, the mesons have a null color charge: "$r+\bar{r} = black$". Furthermore, the baryons must be composed by three quarks that have the three possible colors. Therefore, the baryons have a "complete" color charge "*white*": "$r+g+b = white$". By extension, we have "anti-white" for the anti-baryons.

In the framework of the group theory, this reasoning can be developed. Clearly, the $SU(3)_f$ (flavor) calculations, as e.g. (13), can be done with $SU(3)_c$ (color). The table 2 proposes such calculations. The representations for which the particles really exist in a free state are only the ones that can be broken up into a sum of representations including the trivial representation 1: they are color invariant, i.e. scalar upon the color [15]. It concerns obviously the mesons and the baryons. Also, similar reasoning could be made for the anti-baryons $\overline{qqq}$.

| quarks/anti-quarks structure | representation and possible simplification | possible structure name | observations |
|---|---|---|---|
| $q$ | $3$ | quark | do not exist in a free state |
| $\bar{q}$ | $\bar{3}$ | anti-quark | do not exist in a free state |
| $q\bar{q}$ | $3 \otimes \bar{3} = 1 \oplus 8$ | meson | checked existence experimentally |
| $qq$ | $3 \otimes 3 = 6 \oplus \bar{3}$ | diquark | cannot exist in a free state, not observed in experiments |
| $qq\bar{q}$ | $3 \otimes 3 \otimes \bar{3} = 3 \oplus 3 \oplus \bar{6} \oplus 15$ | – | do not exist |
| $qqq$ | $3 \otimes 3 \otimes 3 = 1 \oplus 8 \oplus 8 \oplus 10$ | baryon | checked existence experimentally! |
| $qqqq$ | $3 \otimes 3 \otimes 3 \otimes 3 = 3 \oplus 3 \oplus 3 \oplus \bar{6} \oplus \bar{6} \oplus ...$ | – | do not exist |

**Table 2.** Possible structures of quarks/anti-quarks, inspired from [1, 15].

Let us mention now the particular role played by gluons, the vectors of the strong interaction. Concretely, they act by exchanging color with the quarks. So, they can carry a color charge and a different anti-color charge, and are thus non-observable. For example, if a quark carries initially a $r$ color, then it can interact with a gluon carrying $b\bar{r}$, and finally it will carry a $b$ color after interaction. In fact, this description is purely illustrative. In the framework of the quantum mechanics, the gluons' states are mixed, to form linear combinations. For example, it can give $\left(1/\sqrt{2}\right)\cdot\left(r\bar{b}+b\bar{r}\right)$. We have 8 independent linear combinations formed by the color/anti-color possibilities, i.e. in fact 8 gluons. They correspond to the 8 $SU(3)_c$ generators, noted as $\lambda_a$ in section 3. The Gell-Mann matrices, described in appendix B, are a possible representation of these generators, in which $\left(1/\sqrt{2}\right)\cdot\left(r\bar{b}+b\bar{r}\right)$ could be associated with $\lambda_4$. Moreover, it can be underlined the fact that the singlet state $\left(1/\sqrt{3}\right)\cdot\left(r\bar{r}+g\bar{g}+b\bar{b}\right)$, associated with the identity matrix, is not a 9th gluon, because it can be formed by linear combination of the 8 $SU(3)_c$ generators [27].

Also, nothing prohibits the gluons to interact themselves, since their color charges allow it. This gluon self interaction corresponds to the $-\frac{1}{4}\cdot G_{\mu\nu}^{a}\cdot G_{a}^{\mu\nu}$ term of (14). This behavior cannot exist in the electromagnetic interaction. This is because the vectors of this interaction, the photons, do not carry the charge related to the interaction, i.e. the electric charge. About the gluons, their self-interaction suggests the existence of bound states of gluons, the *glueballs* or *gluoniums* [1, 28, 29]. Nevertheless, they were not been observed in experiments. One explanation is they are probably mixed with meson states.

# 4.2 Unsolvability of QCD equations

The interaction between gluons mentioned in subsection 4.1 is an obstacle that makes that one cannot solve the QCD equations in the general case. But, this is not the only reason. We can speak about the coupling constant $\alpha_S$ associated with the strong interaction. This one is connected to $g_s$ present in the equation (15) by the relation $\alpha_s = g_s^{\,2}/4\pi$. In fact, $\alpha_S$ is not really a constant, especially on the energy domain that concerns us [18]. In addition, it has sufficiently strong values to prohibit the use of the perturbative methods, which were used successfully for example in atomic physics or in quantum electrodynamics. Indeed, in these quoted theories, the associated coupling constant corresponds to the famous fine-structure constant $\alpha = 1/137$, which governs the electromagnetic interactions. On the other hand, about $\alpha_S$, the approximate value of 0.1184 is usually found in the literature (for the $Z$ boson's mass). This value is strong enough to makes it possible that a simple interaction between two quarks can involve a great number of gluons, of particles/antiparticles pairs… Therefore, there will be a great quantity of possible events to describe this interaction [1]. Calculating and taking into account all these events is very difficult, notably when we are in the range of energies corresponding to the QCD known as "low energies QCD".

More precisely, this is the range of required energies to describe the hadrons. Moreover, in this case, $\alpha_S$ is much stronger than the value we saw upstream. Thus, one cannot model the quarks inside the hadrons starting from the QCD equations. Since the quarks cannot exist in a free state in normal conditions, and since one cannot model their interaction inside the hadrons, some of their characteristics are not easy to study. In particular, it concerns their mass, designated as their naked mass in the section 3. Values of the quark naked masses are available in the literature, but the uncertainty is rather strong.

# 4.3 Asymptotic freedom phenomenon

In an opposed way to what was seen for low energies, the strong interaction is paradoxically much more "cooperative" at the high energies. Let us imagine a system made by quarks and/or anti-quarks, which are sufficiently close. If the proximity is rather consequent, one obtains the second typical phenomenon of the quarks physics, characterized by the relative weakness of the strong interaction. It corresponds to the asymptotic freedom phenomenon [2, 4]. Moreover, in this case, the methods of perturbative calculations previously evoked are usable [5]. Therefore, this branch of the theory is the *perturbative QCD*. To illustrate that, we consider the figure 5, inspired from [5]. It describes in a very schematic way the evoked phenomenon with baryons. In this figure, the picture (1) corresponds to the ordinary conditions, in a nucleus. If the proximity of several baryons is sufficient, pictures (2) and (3), their quarks can be mixed, since nothing permit to say that such quark belongs to such or such baryon. If we interpret the baryon as a "bag" that confines three quarks (see bag model, evoked later in this chapter), the picture (3) is interpretable as a fusion of the bags. So, the confinement is always present, but, it is exerted on a greater volume and with more quarks than in a single baryon.

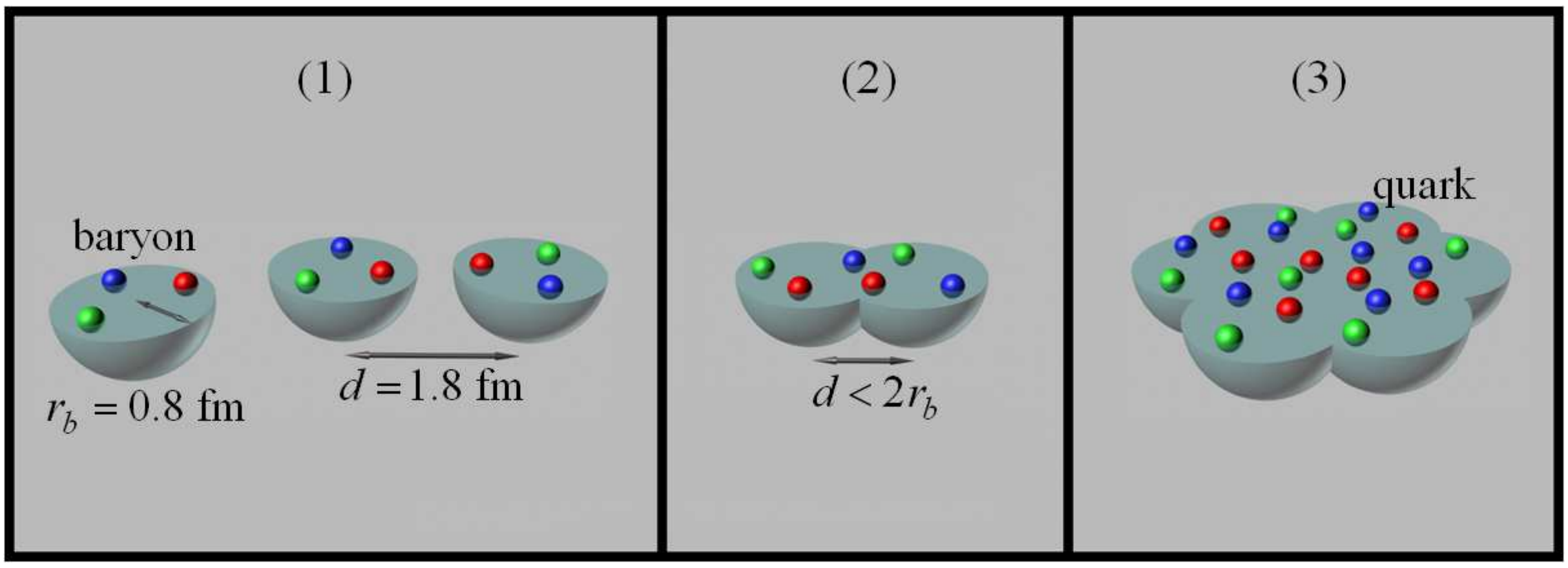

**Figure 5.** Schematization of the "fusion" of several baryons.

As a second example, let us consider one quark and one anti-quark. Firstly, we admit that these two particles do not undergo the effect of other ones, i.e. no influence of the vicinity. In this case, the interaction between these particles is sometimes given by the following formula:

$$V_{q\bar{q}}(r) \equiv -\frac{\alpha}{r} + \kappa \cdot r \,. \tag{29}$$

This potential is represented in the figure 6, directly inspired from [5]. It corresponds to the curve labeled "without screening". We spoke about this in the subsection 4.1. Clearly, the quarks and the anti-quarks cannot be separated. However, such particles are not generally isolated. So, it is necessary to take into account the "screening effect", associated with the other curve in the figure 6. More precisely, if the quark and the anti-quark are remote enough, we could have other quarks/antiquarks between them. It is true especially if the medium is dense enough, i.e. as in the picture (3) in the figure 5. They then could "hide" the interaction between our quark and our antiquark. Finally, the quarks and the antiquarks could be separated. If the screening effect is considered, (29) is then modified to give [14]:

$$V_{q\bar{q}\ \text{screening}}(r) \equiv -\frac{\beta}{r} \cdot \exp(-r/R) \,. \tag{30}$$

The screening effect allows the formation of a phase, within the thermodynamic meaning of this word, where the quarks and gluons can coexist, without apparent manifestation of the confinement phenomenon. This phase is the quark gluon plasma.

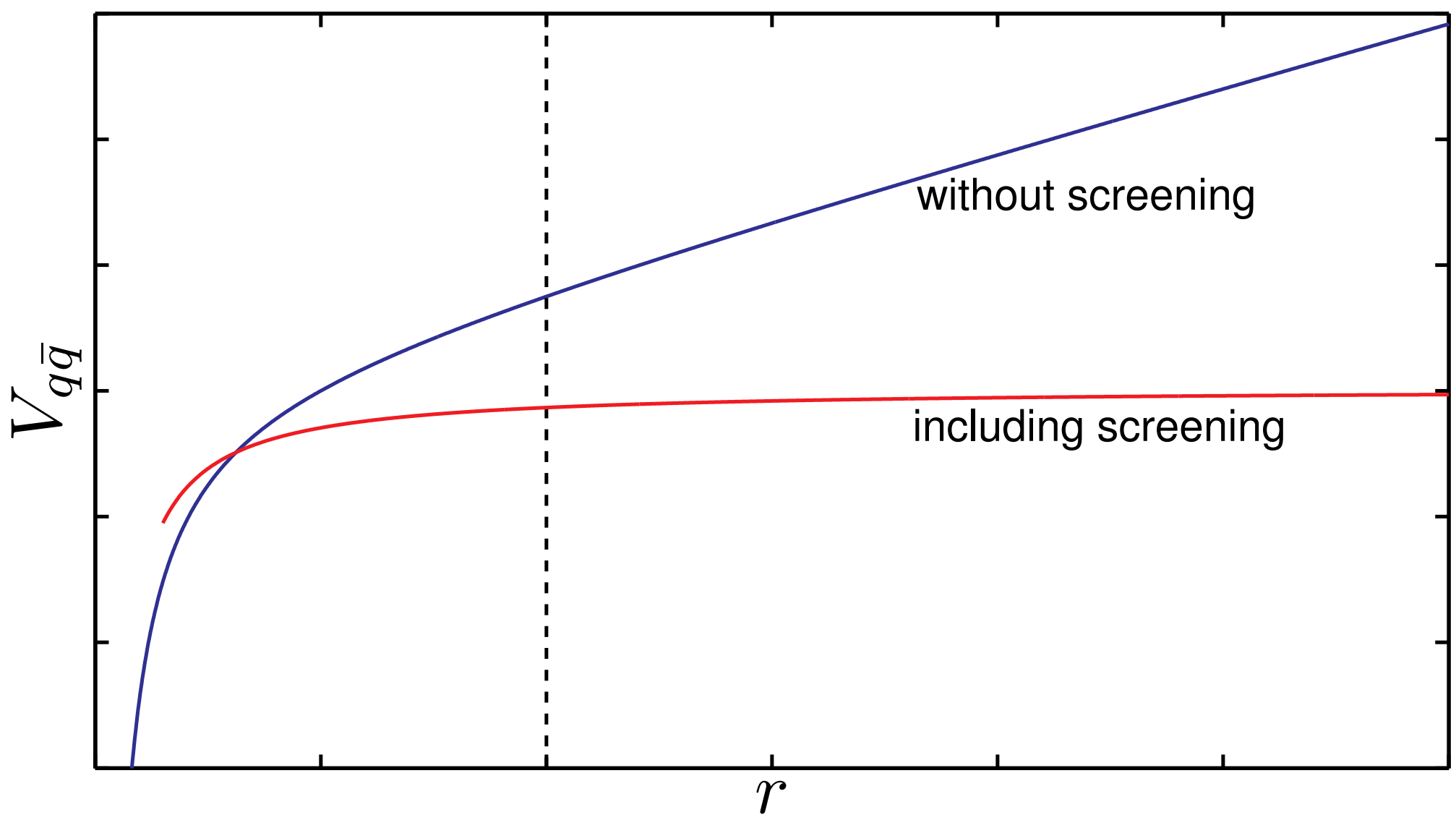

**Figure 6.** Quark-antiquark potential $V_{q\bar{q}}$ function of the distance $r$ between the quark and the antiquark.

# 4.4 The quark gluon plasma and the phase diagram

The quark gluon plasma (QGP) [5, 17] is a very particular state of the matter, since it is the only one where the quarks are deconfined from the hadrons. Nowadays, one knows only two natural physical systems in which such a state is expected to exist or to have existed. The first corresponds to the Universe in its first moments, a few times after the Big Bang. When the temperature was sufficiently low, the initial energy was converted into matter/antimatter. During this phase, the size of the Universe was supposed to be about the current Jupiter orbit. Then, during the cooling, the strong interaction began to act and the quarks became confined inside the hadrons. By a process still badly understood now, as a breaking of the matter/anti-matter symmetry (which would be in favor of the matter), the matter dominated the antimatter. It explains why the matter composes the Physical World such as we know it. The antimatter appears only in some particular cases, as inside the mesons. Clearly, the initial cooling of the Universe constitutes a phase transition of the quark gluon plasma towards the hadronic matter. In this configuration, the temperature is the only parameter that ruled this condensation.

The second physical system is the core of some neutron stars, where the density exceeds several times the ordinary density of an atomic nucleus. The neutron stars are dead stars in the point of view of their thermonuclear activity. So, they are considered as cold objects. In this configuration, a mass comparable to the one of the Sun, i.e. $M_{\odot} \approx 2 \cdot 10^{30}\,\mathrm{kg}$, is contained in a ten-kilometer sphere. This emphasizes another parameter to form QGP: the baryonic density $\rho_B$. In fact, this one is connected to the baryonic number $N_B$ seen in the beginning of this chapter. So, the neutron stars correspond to another possibility of phase transition, but this time the density is the parameter describing the transition. For some massive neutron stars, whose masses largely exceed $M_{\odot}$, it is common to speak about *quark stars*. In the references [30, 31], it is besides explained that these stars cannot be made only by neutrons, but also by quark matter, and even by *strange* matter. So, we speak there about *strange stars* [32].

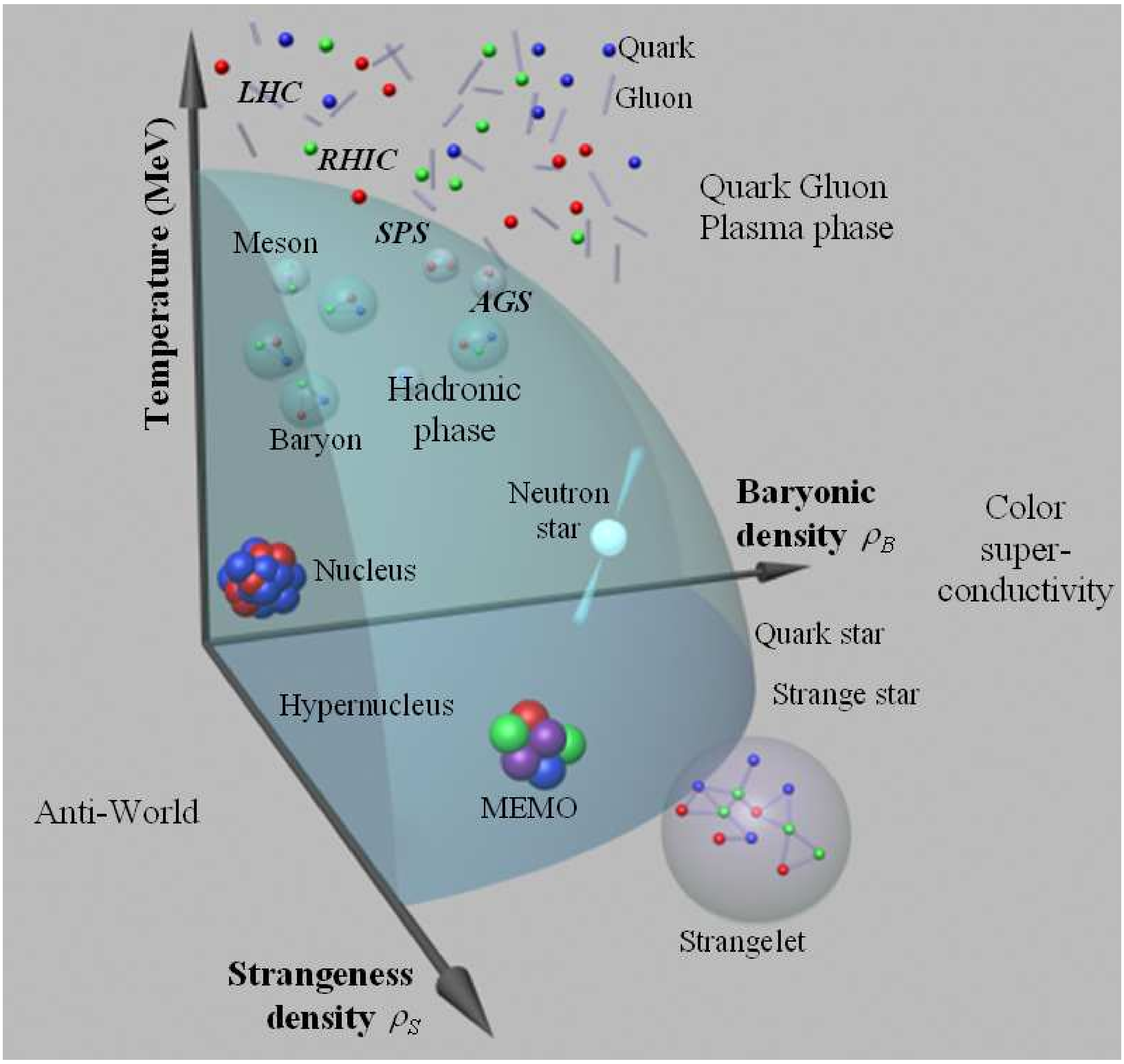

**Figure 7.** Phase diagram of the nuclear and quark matter in the $\rho_B, \rho_S, T$ space.

These two preceding examples are two particular cases. Each time, only one parameter, the temperature or the density, governs the phase transition. In the general case, a graph can be done according to these parameters. This graph is then divided into several zones: one where the quark gluon plasma exists, another one for the hadronic matter… Each zone corresponds to a phase. It forms the *phase diagram* figure 7, inspired by [5] and [30]. In this figure, a new parameter is introduced: the strangeness density $\rho_S$, in reference to the strange quarks and strange matter evoked before. A surface is used to indicate the frontier between the hadronic phase and the other ones. However, in the $\rho_B, \rho_S, T$ space, a mixed phase between the hadronic matter and the QGP phase is expected, especially at high densities and low temperatures. The figure 7 proposes also to summarize the current speculations about some hypothetical exotic objects. They would exist under some temperatures and densities conditions. They are the Metastable Exotic Multihypernuclear Objects (MEMO) [30, 33, 34] or the strangelets [30, 32, 34]. Furthermore, another phase is expected at high densities and low temperatures: the color superconductivity phase, which can be divided into several sub-phases [35–41]. This phase will be evoked again in our work in the next chapter. The AGS, SPS, RHIC and LHC designate the projects devoted to study the QGP phase experimentally. Their positions on the graph indicate the expected zones explored by these experiments. Results related to them are detailed in the section 6.

# 5. Theoretical study: the LQCD

Even if the QCD equations cannot be solved in the general case, some numerical approaches are available to perform calculations. The Lattice QCD (LQCD) is a one these methods. It was proposed in 1974 by Kenneth Wilson [42]. In this method [43–47], the QCD calculations are performed in a 4-dimensional lattice, i.e. three spatial dimensions and the time, in a Euclidian space. It wants to say to the temporal component is written as $x_4 = i \cdot t$. With the LQCD, the average of one observable $A$ can be evaluated with a relation as [46]:

$$\left\langle \hat{A} \right\rangle = \frac{1}{Z} \cdot \int \mathcal{D}U \mathcal{D}\bar{\psi} \mathcal{D}\psi \cdot \hat{A} \cdot \exp(-S), \tag{31}$$

where $\psi$ is a field of quarks. Also, $Z$ is the partition function, defined as:

$$Z = \int \mathcal{D}U \mathcal{D}\bar{\psi} \mathcal{D}\psi \cdot \exp(-S), \tag{32}$$

in which $S$ designates the action. We recall $S = \int \mathrm{d}^4 x \cdot \mathcal{L}$, where $\mathcal{L}$ is the Lagrangian density, also named Lagrangian in practice, as in our work. The action can be spitted in two parts: one associated with the gluons, and the other one for the quarks: $S = S_{gluons} + S_{quarks}$.

## 5.1 The Wilson gauge action and the loops

Firstly, we focus on the gluons. The action of the gluons can be evaluated by the Wilson gauge action, written as [44]:

$$S_{gluons} = \beta \sum_{x,\mu<\nu} \left( 1 - \frac{1}{3} \mathrm{Re}\left( Tr\left( P_{\mu\nu}(x) \right) \right) \right). \tag{33}$$

In this expression, $\beta$ is the inverse of the temperature. Also, $P_{\mu\nu}(x)$ is a plaquette. Following the figure 8, the plaquette is expressed as [48]:

$$P_{\mu\nu}(x) = U_\mu(x) U_\nu(x+\hat{\mu}) U_{-\mu}(x+\hat{\mu}+\hat{\nu}) U_{-\nu}(x+\hat{\nu}), \tag{34}$$

in which $U_\mu(x)$ is a gauge field variable, with $U_\mu(x) \in SU(3)$.

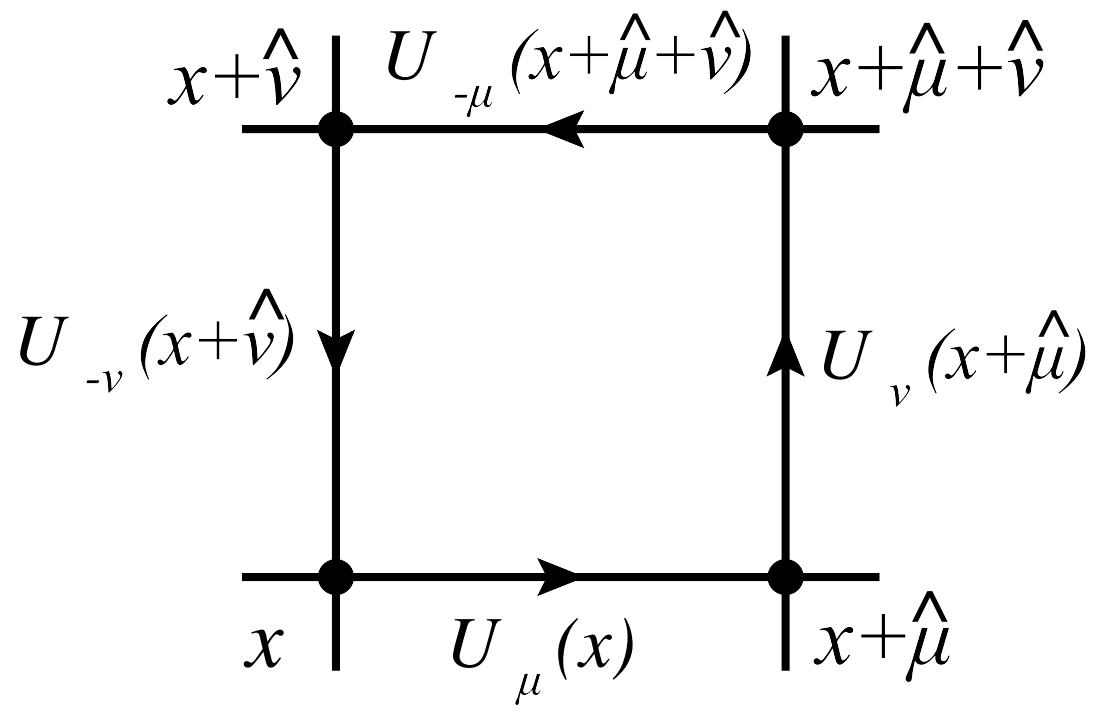

**Figure 8.** A plaquette.

$U_\mu(x)$ is related to the gluons fields $A_\mu$ by $U_\mu(x)=\exp\left(ig\int_x^{x+\mu} d\tilde{x}\cdot A_\mu(\tilde{x})\right)$. However, if we consider that $A_\mu$ is constant during the propagation from $x$ to $x+\hat{\mu}$, (the $A_\mu$ value only changes when we reach the next dot of the lattice), it comes :

$$U_\mu(x)\approx\exp\left(ig\cdot a\cdot A_\mu(x)\right), \tag{35}$$

where $g$ is the coupling constant. Also, $a$ is the parameter of the lattice. More precisely, it corresponds to the distance between two consecutives dots when it is applied to spatial dimensions.

In fact, a plaquette is a particular case of Wilson loops. They are rectangular loops formed in a subspace that include one spatial dimension and the time. Clearly, a plaquette is a $1\times1$ Wilson loop. The Wilson loops are gauge invariant, i.e. under $SU(3)$, as the Polyakov loops. A Polyakov loop is a line according to the time, for which the ends are linked together by boundary conditions. We will see in the chapter 2 that in the framework of the imaginary time formalism, such conditions can be satisfied. A Polyakov line/loop is written as:

$$P(\vec{x})=Tr\left[\prod_{\tau=0}^{N_t-1} U_4(\vec{x},\tau)\right], \text{ with } U_4(\vec{x},\tau)=\exp\left(ig\cdot a\cdot A_4(\vec{x},\tau)\right). \tag{36}$$

In (36), $N_t$ is the number of times considered in the calculation. Physically, as explained i.e. in [44], $P(\vec{x})$ simulates the introduction of a static quark at the position $\vec{x}$. In the same way, one can also define the complex conjugate $P^\dagger(\vec{x})$ of the Polyakov loop (36), in which one replaces $U_4(\vec{x},\tau)$ with $\exp\left(-ig\cdot a\cdot A_4(\vec{x},\tau)\right)$. Clearly, the complex conjugation "reverses the time" of the loop. Physically, one simulates the introduction of a static antiquark. We will see later how the Polyakov loop will be considered in the framework of this thesis.

# 5.2 Action of the quarks and limitations of the LQCD

When the quarks are not included, $S_{quarks}=0$. Such calculations are named pure gauge LQCD calculations. At the opposite, if the quarks are considered, the action of the quarks is written in a Euclidian space as:

$$S_{quarks}=\int d^4x\cdot\overline{\psi}(x)\left(\not{D}+m\right)\psi(x). \tag{37}$$

In the framework of calculations on the lattice, this action can be rewritten for example as [49]:

$$S_{quarks} = \sum_x \bar{\psi}(x) \left( \sum_\mu \gamma^\mu \frac{U_\mu(x)\psi(x+\hat{\mu}) - U_{-\mu}(x)\psi(x-\hat{\mu})}{2a} + m \cdot \psi(x) \right). \quad (38)$$

Anyway, the quarks' action is usually written in a condensed form:

$$S_{quarks}(\bar{\psi},\psi,U) \equiv \sum_x \bar{\psi}(x) M_{xy}[U] \psi(y), \quad (39)$$

in which $M$ is the interaction matrix, also designated as the Dirac operator. In fact, in (32), the integration upon the quarks contribution can be performed [46]; it leads to the expression:

$$Z = \int \mathcal{D}U\mathcal{D}\bar{\psi}\mathcal{D}\psi \cdot \exp(-S) \equiv \int \mathcal{D}U \cdot \left(\det\left(M[U]\right)\right)^{n_f} \cdot \exp\left(-S_{gluons}\right), \quad (40)$$

where $n_f$ is the number of flavor that are considered in the calculations (usually two or three). In the past, an approximation, known as quenched approximation, considered that $\det\left(M[U]\right) = 1$. It leads to neglect local actions, and the fermions loops are suppressed [44, 49, 50]. Nowadays, this approximation tends to be avoided in LQCD calculations. Indeed, this approximation is exact when the masses of the quarks tend towards the infinite. As a consequence, the method could be suitable for heavy quarks, but not for light ones.

In fact, LQCD calculations including quarks know two major limitations. Firstly, compared to pure gauge calculations, the inclusion of the quarks' action leads to increase the calculation time dramatically. In parallel, this time also explodes as soon as the size of the lattice increases. Such calculations require large computer resources, as super-calculators. It thus justified the quenched approximation, at least upon a numerical point of view, that allowed reducing the calculation time by a simplification of the modeling. Without this approximation, another problem appears at non null chemical potential, because in this regime $\det\left(M[U]\right)$ becomes complex. In this case, the Monte-Carlo calculations are invalidated, because in this approach $\exp(-S)$ is interpreted as a density of probability. Furthermore, this determinant term presents oscillations that affect the found values in relations as (31). These numerical problems are designated in the literature as the *fermion sign problem* [48, 51–53].

# 5.3 Some observables and results in the LQCD

The LQCD calculations allow calculating observables, starting from the partition function Z [54]. Notably, the density of free energy is obtained with:

$$f = -\frac{T}{V} \ln Z(T,V), \quad (41)$$

the energy density is:

$$\varepsilon = \frac{T^2}{V} \frac{\partial \ln Z(T,V)}{\partial T}. \quad (42)$$

Also, the pressure can be found with:

$$p = T\frac{\partial \ln Z(T,V)}{\partial V}, \tag{43}$$

or $p = -f$ in the case of large and homogenous systems. With these quantities, one can also estimate the density of entropy $s = \frac{\varepsilon + p}{T}$ or the sound velocity $c_s = \frac{dp}{d\varepsilon}$. In theses relations, the volume is found with $V = (N_x \cdot a)^3$, in which $N_x$ is the size of the lattice. Also, the temperature is obtained with $T = \frac{1}{N_t \cdot a}$, where $N_t$ is the number of times. The link between the time and the temperature will be clarified in the chapter 2, in the framework of the imaginary time formalism.

With the quoted observables, one motivation can be to study the regime for which $\varepsilon = 3p$. As explained for example in [5, 17], this equality is verified when the particles that form the system are massless and without interaction. It corresponds to the Stefan-Boltzmann limit, for which $\frac{\varepsilon}{T^4}$, $\frac{3p}{T^4}$ and $\frac{3s}{4T^3}$ (according to the definition of *s*) converge towards the same value. This behavior is often expected in the literature for QCD when the temperature tends towards the infinite [55].

Moreover, LQCD calculations allow also evaluating the value of the chiral condensate, evoked in the section 3. Clearly, the finality is to study the restoration of the chiral symmetry at high temperatures. More precisely, it was found [56] that the value of the condensate decreases very quickly when the system reaches the critical temperature. After this one, it was observed that the value of the condensate tends towards zero, confirming the expected restoration of the chiral symmetry.

## 5.4 Effectives models

The LQCD results are often considered as references. However, we saw in the subsection 5.2 that LQCD presents limitations. Clearly, LQCD alone cannot allow studying the phase transition between the QGP and the hadronic phase. In fact, some phenomenological models could give some interesting results, and in an easier way than with LQCD. The unsolvability of the QCD equations at low energies is related to some aspects, like confinement. So, some of these models treated this aspect directly, in order to try to mimic the behavior of the confinement. One of them, treated in this thesis, concerns the inclusion of the Polyakov loop in an effective model. Also, as argued previously, the quarks "naked" masses are currently rather badly known. By reversing the problem, since the strong interaction is sufficiently intense to occult strongly this intrinsic characteristic of the quarks, then a model can start from *effective* or *constituent* quarks. The quarks' effective masses do not have anything common with the naked masses, since they include most of the inter-quark interactions. For the *u* quark, the naked mass is about 3 MeV, whereas the effective mass exceeds 300 MeV in the framework of some models, i.e. the mass of a nucleon divided by three. To bind these constituent quarks, a residual interaction can be added in the model. Of course, this interaction is not able to describe the confinement correctly. Therefore, some improvements are imagined to this quark constituent model, to be able to obtain results that are confirmed experimentally. For example, it is sometimes used models like the *M.I.T. bag model,* in order to simulate the confinement, or at least to imitate its aspects [1, 5]. This approach stipulates the existence of an imaginary bag with “impassable” walls. This bag would contain the quarks inside the studied hadron, as in the picture (1) in the figure 5.

# 6. Experimental study of the QGP

We saw in the subsection 4.4 that the Quark Gluon Plasma was/is expected to be present in the first moments of the Universe and in the core of neutron stars. In order to study experimentally this state of the matter “in laboratory”, heavy ion collisions are produced in colliders. If the nucleuses bring enough energy, the QGP phase can be formed. Clearly, it concerns highly relativistic collisions, i.e. the velocity of the nucleuses is very close to the speed of light. The evolution of the QGP is described by a scenario proposed in 1983 by J D Bjorken [57]. In this scenario, figure 9, the description of the system is performed according to the axis along which the nucleuses move, noted as $z$ (figure 10) and according to the time $t$.

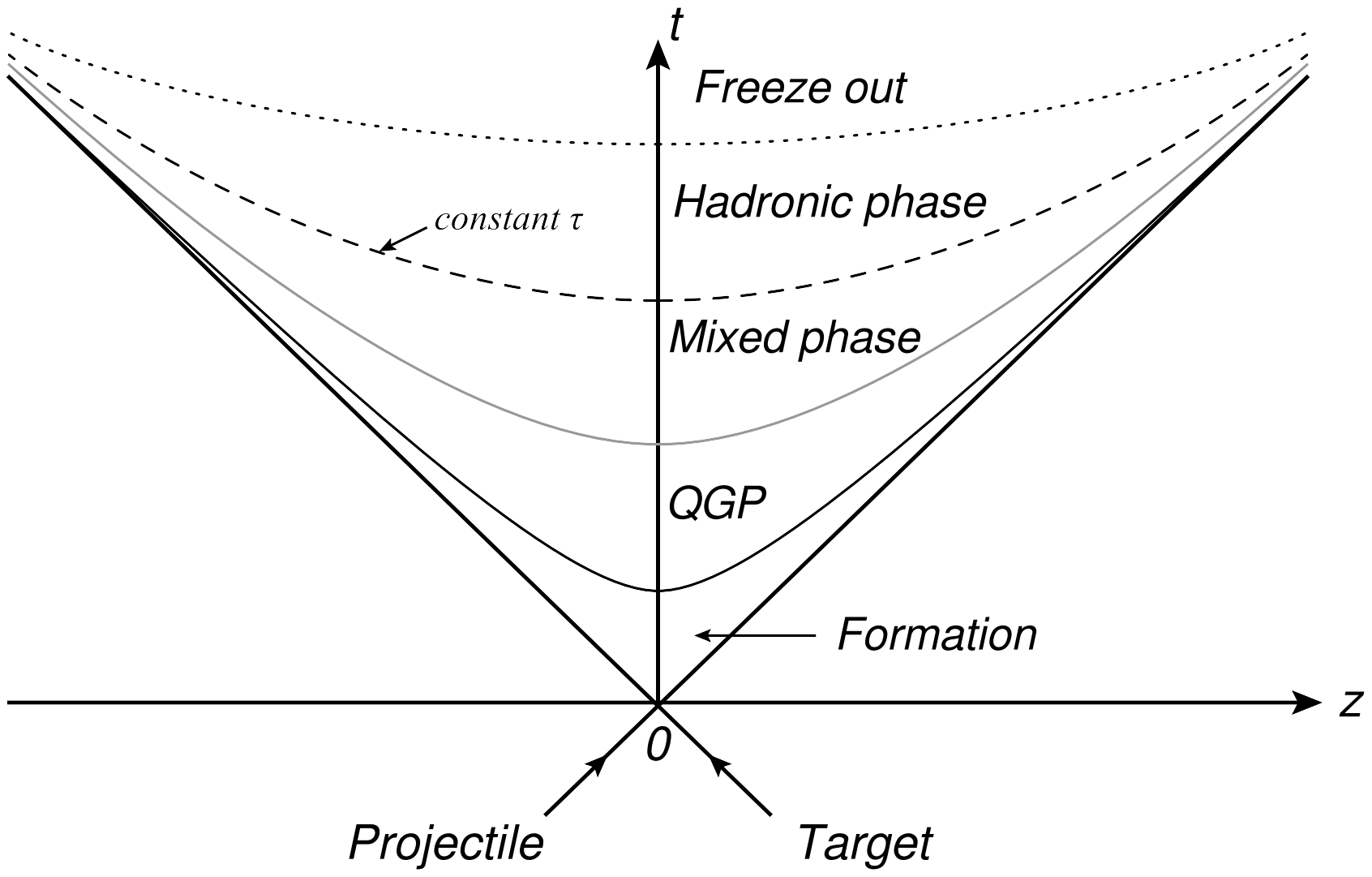

**Figure 9.** Schematic representation of a heavy ion collision forming QGP, according to the Bjorken scenario [5, 57].

In the figure, the hyperbolas correspond to constant proper times $\tau$, related to the time $t$ and the distance $z$ by the relation $\tau = \sqrt{t^2 - z^2}$. In fact, the QGP is expected to be rapidly formed. After that, the quarks, antiquarks and gluons combine themselves to form hadrons. One speaks about hadronization. In some versions of the Bjorken scenario [5], a mixed phase composed by QGP and hadrons is expected. After the hadronization, the system rapidly expands. The freeze out designates to the state for which the outgoing particles no longer interact, because of a too great dilution of the system. The QGP phase is expected to exist only during few $\mathrm{fm/c}$. At the opposite, the produced particles are measured at the level of the detectors largely after the hadronization.

# 6.1 Signatures of the QGP

Because of the quarks' confinement, it is not possible to *observe* the QGP phase directly. As a consequence, the finality is to collect proofs that the QGP was formed during the experiments. Thus, several possible "signatures" of the QGP are studied.

Notably, a scenario designed by Matsui and Satz in 1986 [58] planned a strong decrease of the $J/\psi$ production. A $J/\psi$ is a meson composed by a charmed quark $c$ and an antiquark $\bar{c}$. In heavy ions collisions, $c,\bar{c}$ pairs are formed. According to the scenario, the $c$ and $\bar{c}$ can easily combine themselves in hadronic matter, and so form the $J/\psi$. However, in the QGP phase, the interaction between these quarks and antiquarks is expected to be screened by the other quarks/antiquarks. This effect should prevent them to form $J/\psi$, and should lead to the expected decreasing of their production. However, this possible signature of the QGP was criticized. Indeed, on one side, a regeneration process of the $J/\psi$ was imagined [59, 60], leading to disturb the measurement of the expected decreasing. On the other side, it was also imagined that some $J/\psi$ could also disappear by multiple interactions with the hadronic matter.

Another possible signature of the QGP is the increase of the strangeness, predicted by Rafelski and Müller in the 80s [61, 62]. In fact, reactions as $u+\bar{u}\rightarrow s+\bar{s}$, $d+\bar{d}\rightarrow s+\bar{s}$ and between two gluons $g+g\rightarrow s+\bar{s}$ have strong kinematic threshold, because the strange quarks/antiquarks are heavy particles. Clearly, if the energy of the incoming particle is lower than this threshold, the reaction cannot occur. The high temperatures reached in the QGP phase allow exceeding the threshold. So, a strong production of $s,\bar{s}$ are expected in this phase, leading to a non negligible formation of strange mesons and hyperons as $\Lambda$, $\Xi$, $\Omega$, etc. or anti-hyperons ($\bar{\Omega}$ …).

We can also quote the production of dileptons [63]: $e^+/e^-$ or $\mu^+/\mu^-$. In fact, these particles result from the disintegrations of mesons. The $J/\psi$ disintegrations are particularly known. We can also mention the disintegrations of the vectorial mesons as $\rho,\omega,\phi$. In fact, the temperature of the QGP augments the instability of the mesons, increasing their disintegrations into dileptons. As noted in [5], the study of leptons is relevant because they cannot interact via the strong interaction with the hadronic matter. They are thus expected to allow a reliable description of the QGP. However, a limitation of this approach is to be able to recognize the formation of dileptons resulting from the QGP compared to the ones formed via a hadronic scenario, notably with the Drell-Yan process between two hadrons [64].

Also, other interesting "witnesses" of the QGP are photons that are produced by thermal emission [17]. More precisely, their production obey to a law upon the temperature in $T^4$. As a consequence, a strong photon emission from the QGP is imagined. Furthermore, as the leptons, the photons cannot interact with the QGP particles via strong interaction. However, a limitation of this approach concerns the photons emitted before and after the existence of the QGP phase, that can disturb the analysis. Notably, photons are also emitted by disintegrations reactions, and not only by the expected thermal radiation.

Then, we can also mention the jet quenching phenomenon [65–67]. The partons (quarks and gluons) produced in the first moments of the QGP formation constitute jets of particles. Before their hadronization, these partons jets are expected to cross the formed QGP phase, and there loose energy in a non-negligible way, notably by gluons bremsstrahlung. At the level of the detectors, the hadrons jets resulting from the hadronization of these partons should have strongly reduced momenta: it constitutes the jet quenching.

## 6.2 The elliptic flow $v_2$

Moreover, other observables are also studied, not necessarily to prove the formation of the QGP phase, but to characterize it. Among them, we can particularly quote the elliptic flow $v_2$ [59, 68, 69]. As evoked before, the QGP is experimentally studied by heavy ion collisions, i.e. a meeting between two nucleuses going in opposite senses. They can move along the same axis or along parallel axes. In this case, the minimum distance between the centers of the nucleuses is named impact parameter, and noted as $b$. In this configuration, a fragment of each nucleus does not participate to the collision and continues its moving in straight line. At the opposite, in the figure 10, the ellipsoid corresponds to the merging of parts of these nucleuses.

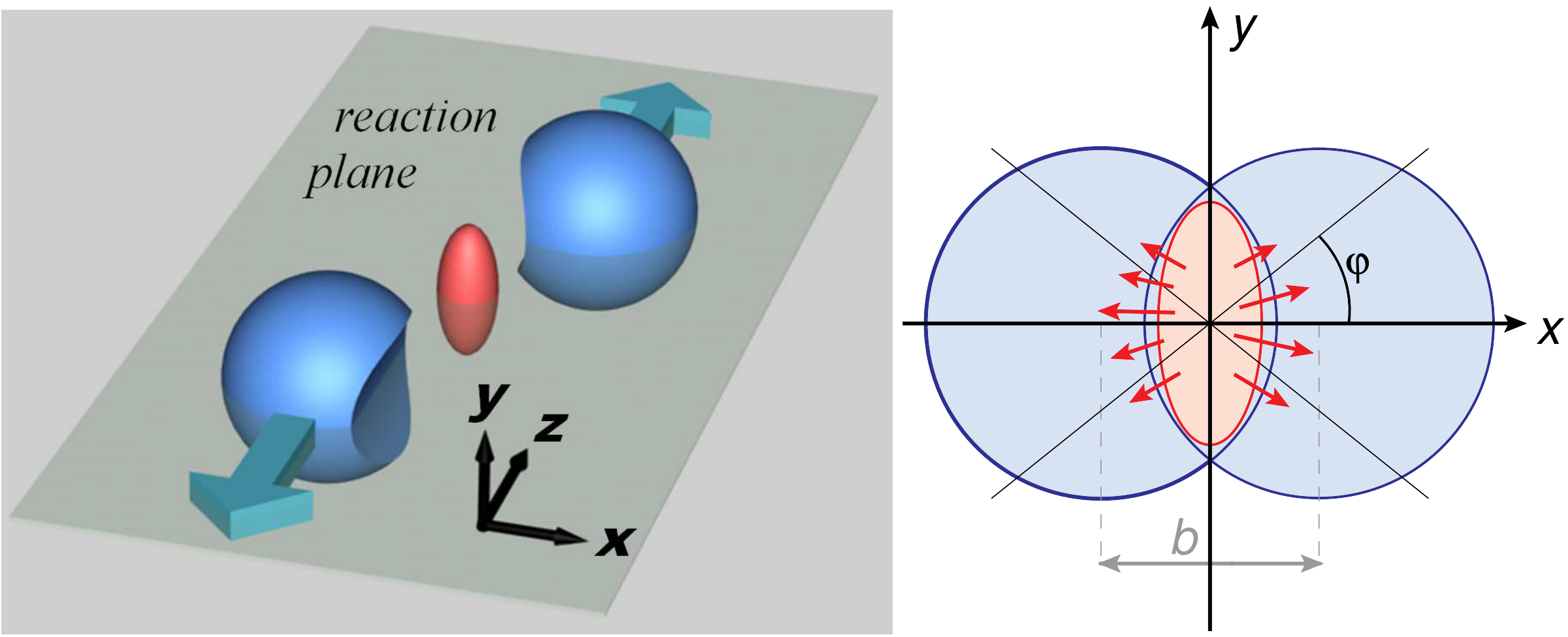

**Figure 10.** Representation of a collision, inspired from [59], and projection upon the *xy* plane.

There, the QGP may be formed. Rapidly after its hadronization, the produced particles move away from this central zone. The distribution of the transverse momenta $p_T$ of these outgoing particles, i.e. along the *x*,*y* plane, constitutes the azimutal distribution. In this plane, one uses the $\varphi$ angle, as in the right hand side of the figure, where $\varphi = \tan^{-1}(p_y / p_x)$ and $p_x, p_y$ are the projections of $p_T$ upon the *x*,*y* axes. The elliptic flow $v_2$ is the second harmonic of the Fourier transform of the azimutal distribution. It is defined as [70]:

$$v_2 = \langle \cos(2\varphi) \rangle , \tag{44}$$

i.e. the average of $\cos(2\varphi)$ upon the outgoing particles. Furthermore, thanks to the mathematical relation:

$$\cos\left(2\cdot\tan^{-1}\left(\frac{p_y}{p_x}\right)\right)=\frac{p_x^2-p_y^2}{p_x^2+p_y^2}\ , \tag{45}$$

we can propose [68]:

$$v_2=\left\langle\frac{p_x^2-p_y^2}{p_x^2+p_y^2}\right\rangle. \tag{46}$$

In fact, $v_2$ gives information about the collective behavior of the particles. More precisely, it allows studying the friction in the matter forming the central zone, thus it gives some properties of the created QGP.

# 6.3 Recent results

Former projects were the Alternating Gradient Synchrotron (AGS) and the Super Proton Synchrotron (SPS). The energy density reached at the AGS was about $0.8\ \mathrm{GeV/fm^3}$, for a temperature close to 150 MeV. In the SPS, it concerned an energy density $2.5\ \mathrm{GeV/fm^3}$, associated with a temperature of 190 MeV [5]. The conditions met at the SPS were expected to be rather close of the ones required to form QGP in the experiments performed in 2000.

Also, one of the most important projects devoted to the study of the QGP is the Relativistic Heavy Ion Collider (RHIC), in the Brookhaven National Laboratory, in the New-York State. In the experiences performed with the RHIC, the energy density is about $5\ \mathrm{GeV/fm^3}$, for a temperature close to 230 MeV [5]. These conditions are favorable to the formation of QGP without ambiguity. More precisely, the creation of this state of the matter in the RHIC was announced in April 2005. In fact, concerning the QGP signatures previously evoked, the jet quenching was observed at the RHIC [71]. Moreover, until the 90s, it was imagined that the QGP was a gas. However, the found results indicated that in the experiments, the GQP phase acts as a liquid [59]. More precisely, the $v_2$ measurements showed the behavior of a perfect fluid, i.e. without friction. Clearly, the vision of free quarks and free gluons, imagined to describe the QGP phase, was not verified by the RHIC results. It leads to consider the sQGP, i.e. a strong interacting QGP [59], and to update the vision of this state of the matter.

The other major project is the Large Hadron Collider (LHC) in the CERN [72]. Energy densities close to $10\ \mathrm{GeV/fm^3}$ and temperatures about 260 MeV were expected [5], i.e. more extremes conditions than in the RHIC. In May 2011, it was announced that QGP was produced with the LHC. The results obtained via the QGP experiments ALICE (A Large Ion Collider Experiment) confirmed the ones found in the RHIC [59]. More precisely, the $v_2$ measurements also go on the sense of a perfect liquid behavior [69]. In the same way, jet quenching was also observed in the LHC [73, 74]. The future experiments planned with the LHC, which will allow increasing the energies and temperatures, are planned for 2015, and then for 2018 [17] …

# 7. Conclusion

In this introducing chapter, we presented some aspects that will directly concern the work we will perform in this thesis. It was underlined two crucial topics strongly related to the quarks physics. Firstly, it concerned the group theory. Starting from rather simple examples, we progressively introduced some notions related to this theory, as the notion of vector, conserved currents, conserved scalar quantities, etc. At this occasion, we presented the quarks, mesons and baryons that will be modeled in the following chapters. Then, we focused on a description of the Quantum Chromodynamics, and the various aspects related to this theory, as the chiral symmetry. Even the QCD Lagrangian writing is finally relatively simple, we cannot hide the complications linked to this theory. Indeed, a correct and complete theoretical modeling of the confinement is still missing today. But, the quarks physics, and by extension the subatomic physics, is also a fantastic topic, because it is linked to various phenomena, as the phase transition between the hadronic matter and the quark gluon plasma, the color superconductivity, etc. Moreover, we presented some experimental aspects. In fact, we saw that recent results are particularly interesting, because they show that the QGP phase acts in the RHIC-LHC experiments as a perfect fluid.

Obviously, this chapter cannot be exhaustive, concerning the description of the group theory, the QCD, the LQCD or the experimental researches. Furthermore, we cannot speak about quarks physics without speaking about quantum field theory or relativity, but these aspects will be developed in the studies performed in the following chapters.

# Chapter 2

# The Nambu Jona-Lasinio model and the Polyakov NJL model

A part of this chapter was published in *J. Phys. G: Nucl. Part. Phys.* **38** 105003

## 1. Introduction

It was seen in the previous chapter that the high energies nuclear physics and the particles physics are correctly modeled by the Quantum Chromodynamics (QCD). However, we also noted that this model cannot be solved in the framework that interested us, i.e. when the quarks are confined inside the hadrons. The lattice QCD (LQCD) is an interesting solution to study quarks physics [1]. In addition, the produced results by this approach are considered as references. But, as noted previously, LQCD requires high computer resources. In the same time, it was reported [2, 3] that LQCD presents some limitations at finite densities. Indeed, the LQCD meets there a difficulty known as the fermion sign problem [4, 5]. As saw in the chapter 1, this numerical problem comes to the contribution of the quarks, via the term $\det(M[U])$ associated with them. More precisely, at finite densities, this determinant becomes complex, invalidating stochastic approaches, as Monte-Carlo. Also, it leads to unwanted fluctuation in the calculations of observables.

Therefore, effective models were developed to try to overcome the difficulties linked to the QCD or its associated numerical methods. Among them, the Nambu and Jona-Lasinio model (NJL) [6, 7] proved its reliability to study the quarks physics for a long time. As seen in this chapter, this model presents interesting qualities. Notably, it allows working at finite temperatures, thanks to the use of the Matsubara formalism [8]. Concretely, as with QCD, it shows the breaking of the chiral symmetry [9], and its restoration at high temperatures. Also, it permits to consider calculations at finite baryonic densities. Furthermore, we will see in the three next chapters that it permits to model mesons, and also baryons by the use of diquarks. Since its creation, the NJL model was progressively improved. The reference [10] proposes a partial chronology of the NJL evolutions. During the 1980s or later, we mention [11, 12] and [9, 13–16]. About the 1990s, we quote [17–19] and [20]. These references proposed an elegant formulation of the NJL approach, making it able to be used in the model quarks physics, using the quarks as degrees of freedom. More recently, we also underline the work performed in [21, 22]. We remark that the model is still the object of ameliorations. Let us note for example the work devoted to the eight quark interaction [23–30].

The main idea of the NJL model is to considered massive gluons, whose interaction become punctual, avoiding the difficulties of QCD. A direct consequence of this simplification is the gluons are absent in the NJL description, at least as dynamical particles. Furthermore, the confinement is not treated by this approach. To try to correct this aspect, it was recently proposed to couple the NJL quarks to a Polyakov loop [31–34], in order to simulate a mechanism of confinement. It formed the Polyakov Nambu Jona-Lasinio model (PNJL) [35–44]. It was reported in the literature that this evolution of the model allows correcting some aspects of the NJL description. It concerns notably the behavior of the PNJL model at low temperatures, compared to a pure NJL one: for example, the suppression of the contribution of colored states in the thermodynamics, as noted in [40]. Also, thanks to a rapid decreasing of the quarks masses (observed hereafter in this chapter), the PNJL approach is more efficient than the NJL one to model the restoration of the chiral symmetry [42]. Furthermore, it was also noted that PNJL results correspond well to data obtained with LQCD [3, 37]. As with the NJL model, several versions of the PNJL model can be found in the recent literature, as the Entangled Polyakov Nambu Jona-Lasino model (EPNJL) [45], proving that the model is actively used and improved. As a whole, in the framework of the (P)NJL models, the masses of the (dressed) quarks are studied according to the temperature [42], and sometimes according to the baryonic chemical potential $\mu_B$ [37]. However, such analyses are relatively less frequent according to the baryonic density $\rho_B$.

As a consequence of these observations, we propose in this chapter to present the NJL model and to detail the modifications that are required to include the Polyakov loop into the model. To reach these objectives, we recall in section 2 the main equations associated with the NJL model. It mainly concerns a presentation of the Matsubara formalism and a study of the NJL Lagrangian. In the section 3, we write the equations to be solved in order to find the masses of the dressed quarks, while considering the temperature and the baryonic density as parameters. Then, in section 4, we propose a description of the PNJL model. We mainly focus on the modifications of the NJL Lagrangian and we consider the new introduced variables, as the one associated with the Polyakov field. Then, we write the additional equations to be considered to find the masses of the quarks. Finally, in section 5, we gather the results associated with the resolution of the equations established in the previous sections. A goal is to recover the results published in the literature, and to present results according to the baryonic density. Such description includes also study in the $T, \rho_B$ plane. We also propose additional results, as a study of the Polyakov field, also in the $T, \rho_B$ plane, and a study of the link between the baryonic chemical potential and the baryonic chemical potential. Furthermore, as in the next chapters, comparisons are performed between the NJL and the PNJL results, in order to conclude on the effect of the inclusion of the Polyakov loop. In addition, we particularity insist on the zones for which the two models give close results, and we propose explanations of these observed similarities.

# 2. Presentation of the Nambu Jona-Lasinio model

## 2.1 The Matsubara formalism

The used formalism, called *imaginary time formalism*, or Matsubara formalism [8], uses the analogy between the thermal factor $\exp(-\beta \cdot E)$ and the factor $\exp(-i \cdot t \cdot H)$, making a correspondence between the inverse of the temperature $T$, noted $\beta$, and $i \cdot t$. It thus explains the denomination of imaginary time. A characteristic of this formalism is the Green functions depend on the imaginary time $i \cdot t$ such as $0 \leq i \cdot t \leq \beta$, and it leads to a periodicity of period $\beta$. Furthermore, the time Fourier transform is replaced by Fourier series. Another consequence is the energies are quantified. They are multiples of $\pi / \beta$ : even multiples for bosons and odd multiples for the fermions. Concretely, when we have to calculate an integral with a four-momentum as an integration variable, we have to proceed to the following transformation:

$$\int \frac{\mathrm{d}^4 p}{(2\pi)^4} \to \frac{i}{\beta} \cdot \sum_{n=-\infty}^{+\infty} \int \frac{\mathrm{d}^3 p}{(2\pi)^3} . \tag{1}$$

The 0 component is the *temporal* component of the four-vector, i.e. energy. This one is not there an *integration* variable but a *summation* variable, because the energy becomes a discrete variable in this formalism. The index *n*, with $n \in \mathbb{Z}$, refers to the $n^{th}$ energy of Matsubara. It is common to speak about Matsubara *frequency*, defined as:

$$\omega_n^{FD} = \frac{(2n+1) \cdot \pi}{\beta} , \tag{2}$$

when the energy is the one of a fermion (FD: Fermi–Dirac), and:

$$\omega_n^{BE} = \frac{2n \cdot \pi}{\beta} , \tag{3}$$

for a boson (BE: Bose–Einstein).

We will see all along this work that this formalism is very present in our model, even if it does not appear in an explicit way. Examples of calculations using imaginary time are available in the end of the chapter 4, devoted to the diquarks. A complete calculation of a fermion propagator is presented there. Also, the appendix D performs calculations in the framework of this formalism. More precisely, the beginning of this appendix presents a general method to perform a summation of a Matsubara frequency.

## 2.2 The basis of the NJL model

The main idea of the NJL model is to consider that the gluons exchanged between quarks/antiquarks have got an effective mass. This mass is supposed to be strong enough compared to the gluons momenta. So, these ones are negligible in the writing of the gluons propagators. As a consequence, the gluon propagator is expressed as a constant and it is reduced to a simple effective factor. Let us study this simplification, step by step. The gluon propagator is initially written, for a gluon with four-momentum $k$ [46]:

$$\mathcal{D}_{\mu\nu}^{a,b}(k)=\delta^{ab}\cdot\frac{1}{k^2}\cdot\left(\eta_{\mu\nu}-(1-\xi_G)\cdot\frac{k_\mu k_\nu}{k^2}\right). \qquad (4)$$

The indices $a$ and $b$ refer to the colors at the two ends of the propagator. The Kronecker symbol $\delta^{ab}$ ensures the conservation of the color charge. $\mu,\nu$ are the indices relating to the four components of a four-vector. $\eta_{\mu\nu}$ is the Minkowski metric, see appendix B. About $\xi_G$, it refers to our gauge choice for our gluonic field. Among the possibilities, the Landau gauge leads to write $\xi_G=0$, or the Feynman gauge $\xi_G=1$. We choose the second one: it permits to simplify (4):

$$\mathcal{D}_{\mu\nu}^{a,b}(k)=\frac{\delta^{ab}\cdot\eta_{\mu\nu}}{k^2}. \qquad (5)$$

In spite of this choice, we cannot be satisfied. Indeed, during the interaction between a quark and an antiquark, the exchanged gluon probably interacts with other gluons and/or with quark/antiquark pairs, as in the left hand side of figure 1. This behavior is explained by the high value of the strong interaction coupling constant $\alpha_S$ [47].

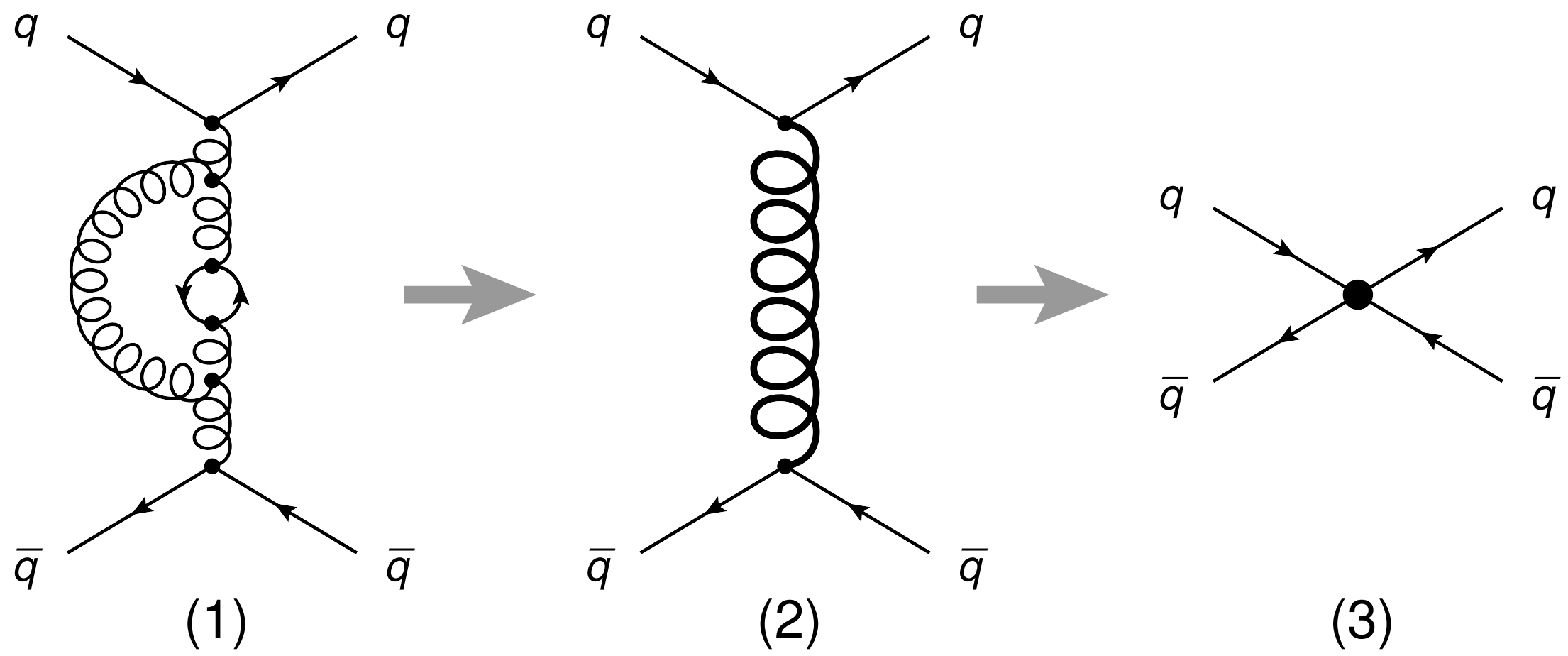

**Figure 1.** Schematization of the NJL approach.

(1) Interaction between a quark and an antiquark in the QCD theory
(2) Acquisition of the effective gluon mass
(3) Gluon in the NJL model

This aspect leads to very difficult calculations. If the gluon has got an effective mass $\Sigma$, case 2 of figure 1, (5) becomes [48]:

$$\mathcal{D}_{\mu\nu}^{a,b}(k)=\frac{\delta^{ab}\cdot\eta_{\mu\nu}}{k^2-\Sigma^2}. \qquad (6)$$

The physical phenomena that interest us within the framework of this work are for example the hadrons formation starting from the quarks/antiquarks. This example corresponds to the “low energies” QCD. So we go further by neglecting $k^2$ compared to $\Sigma^2$. More precisely, we are not on the mass shell. Then, (6) is written [49]:

$$\mathcal{D}_{\mu\nu}^{a,b}(k) = \frac{\delta^{ab} \cdot \eta_{\mu\nu}}{k^2 - \Sigma^2} \simeq -\frac{\delta^{ab} \cdot \eta_{\mu\nu}}{\Sigma^2}, \tag{7}$$

so that the gluon is reduced to a simple vertex, right hand side of figure 1. In fact, the gluons do not exist in the Nambu and Jona-Lasinio model as dynamical degrees of freedom. Of course, this is not without consequences, as the absence of confinement in the pure NJL model. However, the approximation leads to simplifications of the calculations. The associated Lagrangian is also strongly modified compared to the QCD one. Indeed, let us consider this matrix element:

$$\left(\bar{\psi}\gamma^{\mu}\lambda^{a}\psi\right) \cdot \left(\frac{-g^2}{\Sigma^2}\right) \cdot \left(\bar{\psi}\gamma_{\mu}\lambda^{a}\psi\right), \tag{8}$$

describing the interaction between the quark and the antiquark according to the approximations we have just made. $\psi$ is associated with a quark field. The $g$ is a coupling constant between the gluon and the quark/antiquark. The quantity (8) corresponds to the contraction of the conserved current $J_{\mu}^{a}$ associated with $SU(3)_f$, i.e. with one of the interaction terms of the Lagrangian [19, 20, 49]. So, (8) models here the interaction between a quark and an antiquark. The factor $\kappa = -g^2/\Sigma^2$ is finally an effective interaction constant. The corresponding interaction term of the Lagrangian is then expressed as:

$$\mathcal{L}_{\text{int } q\bar{q}} = \kappa \cdot \sum_{a=1}^{8} \left(\bar{\psi}\gamma_{\mu}\lambda^{a}\psi\right)^2 . \tag{9}$$

# 2.3 The NJL Lagrangian

The most complete NJL Lagrangian used in our study is written as [20, 22, 48, 50, 51]:

$$\begin{aligned}
\mathcal{L}_{\text{NJL}} = & \sum_{f=u,d,s} \bar{\psi}_f \left(i\not{\partial} - m_{0f}\right) \psi_f \\
& + G \cdot \sum_{a=0}^{8} \left[ \left(\bar{\psi}\lambda^a\psi\right)^2 + \left(\bar{\psi}i\gamma_5\lambda^a\psi\right)^2 \right] \\
& - G_V \cdot \sum_{a=0}^{8} \left[ \left(\bar{\psi}\gamma_{\mu}\lambda^a\psi\right)^2 + \left(\bar{\psi}\gamma_{\mu}i\gamma_5\lambda^a\psi\right)^2 \right] \\
& - K \cdot \left[ \det\left(\bar{\psi}(1+\gamma_5)\psi\right) + \det\left(\bar{\psi}(1-\gamma_5)\psi\right) \right] \\
& + \sum_{\alpha} G_{DIQ}^{\alpha} \cdot \sum_{i,j} \left(\bar{\psi}_a \gamma_{\mu} \Gamma_{\alpha}^{i} \psi_b^{C}\right)\left(\bar{\psi}_d^{C} \gamma^{\mu} \Gamma_{\alpha}^{j} \psi_e\right) \cdot \varepsilon^{abc} \cdot \varepsilon_c^{\ de}
\end{aligned} \tag{10}$$

This Lagrangian is the sum of several "sub-Lagrangians". Each one of them has a well defined function in the study of the particles treated in this work. The first term $\sum_{f=u,d,s} \bar{\psi}_f \left(i\not{\partial} - m_{0f}\right) \psi_f$ is the Dirac Lagrangian for a spin ½ particle, in which $m_{0f}$ is the naked masses of the $u,d,s$ quarks. The four remaining terms of (10) are completely new compared to the QCD Lagrangian. They use the constants $G, G_V, K, G_{DIQ}$ of the extended NJL model, see table 1. The terms:

$$\begin{aligned} & G\cdot\sum_{a=0}^{8}\left[\left(\bar{\psi}\lambda^a\psi\right)^2+\left(\bar{\psi}i\gamma_5\lambda^a\psi\right)^2\right] \\ & -G_V\cdot\sum_{a=0}^{8}\left[\left(\bar{\psi}\gamma_\mu\lambda^a\psi\right)^2+\left(\bar{\psi}\gamma_\mu i\gamma_5\lambda^a\psi\right)^2\right] \end{aligned}, \tag{11}$$

correspond to the interaction of a quark with an antiquark. They will be used in the next chapter, to model mesons [52, 53]. They constitute the interaction Lagrangian $\mathcal{L}_{\text{int } q\bar{q}}$, equation (9), after a Fierz transformation [19, 20, 48]. This re-writing of (9) reveals the two distinct terms of (11), associated with the constant $G$ for the first and $-G_V$ for the second. Each of them can be divided into two sub-terms: $\left(\bar{\psi}\lambda^a\psi\right)^2$ that models a quark/antiquark scalar interaction, $\left(\bar{\psi}i\gamma_5\lambda^a\psi\right)^2$ a pseudo-scalar interaction, $\left(\bar{\psi}\gamma_\mu\lambda^a\psi\right)^2$ a vectorial interaction and $\left(\bar{\psi}\gamma_\mu i\gamma_5\lambda^a\psi\right)^2$ an axial interaction. The matrix $\lambda^0=\sqrt{2/3}\cdot 1_3$ is added to the summations over the 8 matrices $\lambda^a$ defined appendix B, i.e. the $SU(3)_f$ generators. $1_3$ is the $3\times 3$ matrix identity. $\lambda^0$ comes directly from the applied Fierz transformation [19, 20]. In (10), the term:

$$-K\cdot\left[\det\left(\bar{\psi}(1+\gamma_5)\psi\right)+\det\left(\bar{\psi}(1-\gamma_5)\psi\right)\right], \tag{12}$$

is known as 't Hooft term. It also intervenes to describe the interactions between the quarks and the antiquarks, therefore for some mesons. This term is added to break in an explicit way the $U_A(1)$ pseudo-scalar symmetry. Clearly, as explained in [20], if this symmetry was respected, it would exist a pseudoscalar meson whose mass would be comparable to the one of the pion. But such a meson was not observed…

Finally, the last term of the Lagrangian is:

$$\sum_{\alpha=S,P,V,A} G_{DIQ}^{\alpha}\cdot\sum_{i,j}\left(\bar{\psi}_a\gamma_\mu\Gamma_\alpha^i\psi_b^C\right)\left(\bar{\psi}_d^C\gamma^\mu\Gamma_\alpha^j\psi_e\right)\cdot\varepsilon^{abc}\cdot\varepsilon_c^{\ de}, \tag{13}$$

which is identified with the Lagrangian interaction term $\mathcal{L}_{\text{int } qq}$, i.e. it models the interaction between two quarks. It will be used to build diquarks in the chapter 4, in which we will detail the summation over $\alpha=S,P,V,A$: $S$ for scalar diquarks, $P$ for pseudo scalar ones, $V$ for vectorial ones and $A$ for axial ones …

# 2.4 Employed NJL parameters

The Nambu and Jona-Lasinio model is not renormalizable. As a consequence, a cut-off, noted $\Lambda$, is used in the calculations. It corresponds in practice to the upper bound of the integrals, whose structure follows (1). Also, the model uses the constants $G, G_V, K, G_{DIQ}$ used in (10), which are more or less connected explicitly to physical quantities. Among these physical quantities, we have the coupling constant $g$ associated with the interactions that are described by the QCD, whose uncertainty is very large at low energies. In the literature, these constants and the cut-off are often gathered, to form a parameter set. The table 1 proposes three different parameter sets. The "RK" one is associated with [53], "P1" with [48, 50], and "EB" that we added while modifying "P1" in the case $m_u \neq m_d$, while taking into account the current constrains upon the quarks naked masses [54]. Furthermore, we will see in the following chapters that $G_V$ and $G_{DIQ}$ were updated in order to increase the agreement between our results and the ones found in the literature. The link between the $G_{DIQ}$ values presented in the table 1 and the ones of $G_{DIQ}^{\alpha}\big|_{\alpha=S,P,V,A}$ will be detailed in chapter 4. Also, the constant $G$ is often designated as $G_S$ in some papers.

| | RK | P1 | EB |
|---|---|---|---|
| $m_{0u}$ | 5.50 | 4.75 | 4.00 |
| $m_{0d}$ | 5.50 | 4.75 | 6.00 |
| $m_{0s}$ | 140.7 | 147.0 | 120.0 |
| cut-off $\Lambda$ | 602.3 | 708.0 | 708.0 |
| $G \cdot \Lambda^2$ | 1.835 | 1.922 | 1.922 |
| $G_V$ | — | $0.310\,G$ | $0.295\,G$ |
| $G_{DIQ}$ | — | $0.705\,G$ | $0.705\,G$ |
| $K \cdot \Lambda^5$ | 12.36 | 10.00 | 10.00 |

**Table 1.** Set of used parameters. The masses and the cut-off are in MeV.

The constants $G_i$ are connected to dimensionless constants $g_i$ by the relation [19]:

$$G_i = \left(\frac{g_i}{\Lambda}\right)^2 . \tag{14}$$

The $G_i$ are thus expressed in $\text{MeV}^{-2}$ in the table 1. In addition, Fierz transformations allow fixing relations between constants [19, 39, 55], and we obtain:

$$G_V = \frac{G}{2} \quad \text{and} \quad G_{DIQ} = \frac{3}{4} \cdot G . \tag{15}$$

However, in practice, these relations are not strictly respected, as observed in various NJL papers and in the table 1.

# 3. Masses of constituent quarks

## 3.1 Gap equations

The determination of the masses of the dressed quarks constitutes a first application of the Nambu Jona-Lasinio model. The masses of the other particles masses will depend on it more or less directly. While indicating by $m_{0f}$ the naked mass of the flavor *f* quark and by $m_f$ its corresponding effective mass, we write [19, 53, 56]:

$$\begin{cases} m_u = m_{0u} - \frac{G \cdot N_c}{\pi^2} \cdot m_u \cdot A(m_u, \mu_u) + \frac{K \cdot N_c^2}{8\pi^4} \cdot m_d \cdot m_s \cdot A(m_d, \mu_d) \cdot A(m_s, \mu_s) \\ m_d = m_{0d} - \frac{G \cdot N_c}{\pi^2} \cdot m_d \cdot A(m_d, \mu_d) + \frac{K \cdot N_c^2}{8\pi^4} \cdot m_s \cdot m_u \cdot A(m_s, \mu_s) \cdot A(m_u, \mu_u), \\ m_s = m_{0s} - \frac{G \cdot N_c}{\pi^2} \cdot m_s \cdot A(m_s, \mu_s) + \frac{K \cdot N_c^2}{8\pi^4} \cdot m_u \cdot m_d \cdot A(m_u, \mu_u) \cdot A(m_d, \mu_d) \end{cases} \tag{16}$$

that forms the gap equations. They are obtained by applying the mean-field approximation, also named Hartree approximation, in the Lagrangian (10) [51, 53]. Because of this approximation, the terms from $\mathcal{L}_{\text{int } qq}$ do not intervene in (16). In these equations, $N_c$ indicates the number of possible different colors, i.e. three. $\mu_f$ is the chemical potential of the flavor *f* quark. Furthermore, we use a generic function *A* that corresponds to a one-quark-loop, [53, 56] and the appendix D:

$$A(m_f, \mu_f, \beta, \Lambda) = \frac{16\pi^2}{\beta} \cdot \sum_n \int \frac{\mathrm{d}^3 p}{(2\pi)^3} \frac{1}{(i \cdot \omega_n + \mu_f)^2 - E_f{}^2}, \tag{17}$$

where $E_f = \sqrt{(\vec{p})^2 + m_f{}^2}$ is the energy of the flavor *f* quark, $\beta = 1/T$ and *T* is the temperature. Furthermore, we have the relation:

$$i \cdot Tr\left(S^f(x-x)\right) = -\frac{m_f}{4\pi^2} \cdot A(m_f, \mu_f), \tag{18}$$

in which *Tr* indicates the trace matrix operation. $S^f$ is the quark propagator of flavor *f* in the imaginary time formalism at finite temperature, here expressed in coordinate space:

$$S^f(\vec{x} - \vec{x}', \tau - \tau') = \frac{i}{\beta} \cdot \sum_n e^{i \cdot \omega_n \cdot (\tau - \tau')} \cdot \int \frac{\mathrm{d}^3 p}{(2\pi)^3} \frac{e^{i \cdot \vec{p} \cdot (\vec{x} - \vec{x}')}}{\gamma_0 \cdot (i \cdot \omega_n + \mu_f) - \vec{\gamma} \cdot \vec{p} - m_f}. \tag{19}$$

The $\vec{x}, \vec{x}'$ are positions and the $\tau, \tau'$ are times. The function *A* is also connected to the chiral condensate value $\left\langle\left\langle \bar{\psi}_f \psi_f \right\rangle\right\rangle$ evoked in the previous chapter. More precisely, we have the relation [19, 57]:

$$\left\langle\left\langle \bar{\psi}_f \psi_f \right\rangle\right\rangle = \frac{m_f \cdot N_c}{4\pi^2} \cdot A(m_f, \mu_f, \beta, \Lambda). \tag{20}$$

Schematically, each line of (16) is equivalent to [17, 20–22, 49]:

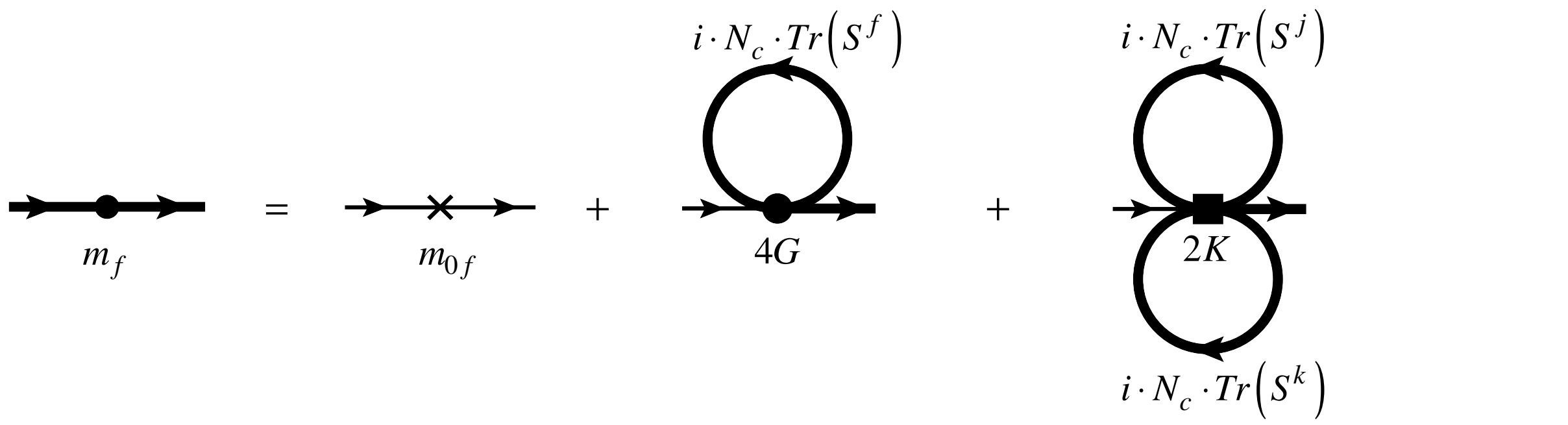

$$m_f \quad = \quad m_{0f} \quad + \quad 4G\cdot\left(i\cdot N_c\cdot Tr\left(S^f\right)\right) \quad + \quad 2K\cdot\left(i\cdot N_c\cdot Tr\left(S^j\right)\right)\cdot\left(i\cdot N_c\cdot Tr\left(S^k\right)\right)$$

or, using the expression of the condensate (20),

$$m_f = m_{0f} - 4G\cdot\left\langle\left\langle\bar{\psi}_f\psi_f\right\rangle\right\rangle + 2K\left\langle\left\langle\bar{\psi}_j\psi_j\right\rangle\right\rangle\left\langle\left\langle\bar{\psi}_k\psi_k\right\rangle\right\rangle\Big|_{\substack{f=u,d,s \\ f\neq j \text{ and } f\neq k}} . \tag{21}$$

Each loop corresponds to a function *A*, which easily enables us to identify the terms of (16). The first term, $m_{0f}$, takes into account the considered quark naked mass. The second term, $4G\cdot\left(i\cdot N_c\cdot Tr\left(S^f\right)\right)$, translates the effect of a flavor *f* loop on the considered quark. The third, $2K\cdot\left(i\cdot N_c\cdot Tr\left(S^j\right)\right)\cdot\left(i\cdot N_c\cdot Tr\left(S^k\right)\right)$, models the effects of the two other quark flavors on the mass of our quark. As in [17, 22], the thick lines are used when the effective masses (dressed quarks) are taken into account, including in the loops, see $m_f$ in (18, 19). The thin lines correspond to the current quarks, associated with the naked masses.

# 3.2 Isospin symmetry

In the literature, and in some parts of our work, the isospin symmetry is considered. It consists to say that the quarks *u* and *d* have the same properties, symbolically « $u = d \equiv q$ ». As a consequence, their masses and their chemical potentials are considered as identical, and they are noted, respectively, $m_q$ and $\mu_q$. Firstly, the finality is to simplify the calculations. This approximation is valid in the majority of the physical systems that are theoretically studied, or in experiments. When the isospin symmetry is satisfied, the set of equations (16) is simplified to give:

$$\begin{cases} m_q = m_{0q} - \dfrac{N_c}{\pi^2}\cdot m_q\cdot A\left(m_q,\mu_q\right)\cdot\left(G - \dfrac{K\cdot N_c}{8\pi^2}\cdot m_s\cdot A\left(m_s,\mu_s\right)\right) \\ m_s = m_{0s} - \dfrac{G\cdot N_c}{\pi^2}\cdot m_s\cdot A\left(m_s,\mu_s\right) + \dfrac{K\cdot N_c^2}{8\pi^4}\cdot\left(m_q\cdot A\left(m_q,\mu_q\right)\right)^2 \end{cases}, \tag{22}$$

where it remains two equations with two unknowns, i.e. the two masses $m_q$ and $m_s$.

# 3.3 Treatment of the densities

For non-null densities calculations, the following formula, resulting from [18, 19], is used:

$$\rho_f = \left\langle\left\langle \psi_f{}^+ \psi_f \right\rangle\right\rangle = \frac{N_c}{\pi^2} \cdot \int dp \cdot p^2 \cdot \left( \frac{1}{1+\exp\left(\beta \cdot \left(E_f - \mu_f\right)\right)} - \frac{1}{1+\exp\left(\beta \cdot \left(E_f + \mu_f\right)\right)} \right). \tag{23}$$

The index *f* indicates the quark flavor (*u*, *d*, *s*). In practice, the integration bounds will be from 0 to $\Lambda$ (cut-off of table 1). In fact, the equation (23) gives the relation between the density $\rho_f$ of the flavor *f* quark and its corresponding chemical potential $\mu_f$ . In the general case, at finite temperature and densities, it is required to solve a set of 6 equations with 6 unknowns. This set of equations is written as:

$$\begin{cases} m_f = m_{0f} - \dfrac{G \cdot N_c}{\pi^2} \cdot m_f \cdot A\left(m_f, \mu_f, T\right) + \dfrac{K \cdot N_c^2}{8\pi^4} \cdot m_j \cdot m_k \cdot A\left(m_j, \mu_j, T\right) \cdot A\left(m_k, \mu_k, T\right) \Bigg|_{\substack{f=u,d,s \\ f \neq j \text{ and } f \neq k}} \\ \rho_f = \dfrac{N_c}{\pi^2} \cdot \displaystyle\int dp \cdot p^2 \cdot \left( \frac{1}{1+\exp\left(\dfrac{E_f - \mu_f}{T}\right)} - \frac{1}{1+\exp\left(\dfrac{E_f + \mu_f}{T}\right)} \right) \Bigg|_{f=u,d,s} \end{cases} \tag{24}$$

The parameters (fixed) are the wanted temperature and densities $\rho_f$. The unknowns are the effective masses $m_f$ and the chemical potentials $\mu_f$. At this occasion, we can mention the existence of a shifting upon the chemical potentials, defined by:

$$\delta\mu_f = G_V \cdot \left\langle\left\langle \psi_f{}^+ \psi_f \right\rangle\right\rangle = G_V \cdot \rho_f , \tag{25}$$

notably evoked in [19, 20, 22, 52], with a factor of two for the two former ones. This shifting translates the effects of the vector interaction, when this sector is taken into account in the Lagrangian, i.e. $G_V \neq 0$. The chemical potential $\mu_f$ appearing in the second line of (24) is a solution of this equation. As a consequence, it corresponds to an effective chemical potential, or "renormalized $\mu$" according to the terminology used in [22]. In other words, it is associated with the chemical potential *after* the shifting [19]. So, it is related to the "real" chemical potential $\mu_{0f}$ (before the shifting) by $\mu_f = \mu_{0f} - \delta\mu_f$. In practice, $\mu_f$ is used in the calculations in which a chemical potential is required, after its estimation via (24). In our work, we consider the temperature and the densities $\rho_f$ as parameters (as input data). Thus, we do not really use $\mu_{0f}$. Anyway, it can be observed that the shifting has a relative poor influence on the results when $G_V < G/3$, as with our parameter sets, table 1.

The equations (24) are reduced to a system with 4 equations and 4 unknowns when we use the isospin symmetry, as in (22). According to subsection 3.2, we write there $\rho_u = \rho_d \equiv \rho_q$. In our work, $\rho_q = N_q/V$ is interpreted as the density of *u* quarks, or the density of *d* quarks, which are necessarily identical to have $m_u = m_d \equiv m_q$ and $\mu_u = \mu_d \equiv \mu_q$. Thus, $\rho_q$ is not the sum of these two densities. In the same way, the quarks number $N_q$ is not be here the *total* number of light quarks, but finally the half. Within the framework of the isospin symmetry, we define the baryonic density as [18]:

$$\rho_B = \frac{2}{3}\rho_q . \tag{26}$$

The equation (26) can be understood by saying that when the isospin symmetry is satisfied, a nucleon is made *on average* by 1.5 quarks *u* and as many quarks *d*, and $\rho_s = 0$. Concretely, in our results, the ratio $\rho_B/\rho_0$ will be used instead of $\rho_B$, where $\rho_0 \approx 0.16\ \mathrm{fm}^{-3}$ corresponds to the ordinary nuclear density.

# 4. The PNJL model

In the NJL model, we saw in the subsection 3.1 that the quarks are coupled to the chiral condensates (21) in the framework of the mean field approximation. As evoked in the introduction of this chapter, the motivation of the PNJL model is to try to correct a major defect of the NJL model, i.e. the absence of confinement, by coupling also the quarks to a Polyakov loop. We will explain in this section how to perform this coupling.

## 4.1 The PNJL Lagrangian

Concretely, we firstly consider the adaptations to be done in the NJL Lagrangian. In the framework of the PNJL model, this one becomes [37, 40, 42]:

$$\mathcal{L}_{PNJL} = \tilde{\mathcal{L}}_{NJL} - \mathcal{U}\left(T,\Phi,\bar{\Phi}\right) + \sum_{f=u,d,s} \mu_f \cdot \bar{\psi}_f \, \gamma^0 \, \psi_f , \tag{27}$$

where $\tilde{\mathcal{L}}_{NJL}$ corresponds to the NJL Lagrangian (10) in which the derivate $\partial\!\!\!/ = \gamma^\mu \partial_\mu$ intervening in the term $\sum_{f=u,d,s} \bar{\psi}_f \left(i\partial\!\!\!/ - m_{0f}\right)\psi_f$ is replaced by $\gamma^\mu D_\mu$, where $D_\mu = \partial_\mu - i \cdot A_\mu$. In this relation, *A* is the Euclidian gauge field associated with gluons. If we do a comparison with the QCD equations of the chapter 1, we note that the strong coupling constant $g_s$ is absorbed by *A*, so here $A_\mu = g_s \cdot A_\mu^a \cdot \lambda_a/2$, e.g. [42]. Also, the term $\sum_{f=u,d,s} \mu_f \cdot \bar{\psi}_f \, \gamma^0 \, \psi_f$ appearing in (27) is added to perform studies at non-null chemical potentials [40], so at non-null densities.

Then, another modification of the NJL Lagrangian concerns the inclusion of an effective potential, noted as $\mathcal{U}\left(T,\Phi,\bar{\Phi}\right)$. This one corresponds to a pure gauge QCD Lagrangian, i.e. without quarks/antiquarks, only gluons. Thus, we can associate it to the $-\frac{1}{4} \cdot G_{\mu\nu}^a \cdot G_a^{\mu\nu}$ term of the QCD Lagrangian, see chapter 1.

The potential $\mathcal{U}(T,\Phi,\bar{\Phi})$ explicitly depends on the temperature *T*, but also on two quantities $\Phi,\bar{\Phi}$. These ones are defined by the relations [37, 40, 42]:

$$\Phi(\vec{x}) = \frac{Tr_c\left\langle\left\langle L(\vec{x})\right\rangle\right\rangle}{N_c} \quad \text{and} \quad \bar{\Phi}(\vec{x}) = \frac{Tr_c\left\langle\left\langle L^\dagger(\vec{x})\right\rangle\right\rangle}{N_c}, \tag{28}$$

where $Tr_c$ is a trace upon the color and $\left\langle\left\langle L(\vec{x})\right\rangle\right\rangle$ is the gauge invariant average of the Polyakov line, noted here as $L(\vec{x})$:

$$L(\vec{x}) = \mathcal{P}\exp\left(i\int_0^\beta A_4(\vec{x},\tau)\cdot d\tau\right). \tag{29}$$

In (29), $\mathcal{P}$ is a path ordering operator, and $\beta = 1/T$. Also, $A_4 = i\cdot A^0$ is the temporal component of the Euclidian gauge field $(\vec{A},A_4)$ evoked above. Physically, $\Phi$ and $\bar{\Phi}$ should correspond, respectively, to the Polyakov loop field and its conjugate. But, as noted in [39], we use their expectation values. In fact, it was also remarked in this reference that $\Phi$ and $\bar{\Phi}$ can be considered as independent variables. Furthermore, the both are real numbers in the framework of the mean field approximation.

In practice, $\Phi$, and by extension $\bar{\Phi}$, is used as an order parameter of the phase transition between a "color confined phase" and a "color deconfined phase"[1]. Furthermore, $\Phi$ is associated with the $\mathbb{Z}_{N_C}$ symmetry, with $N_C = 3$, i.e. to the center of $SU(3)_c$ symmetry of QCD [42]. We recall that the center of a group of symmetry *G* is a set of elements of *G* that commute with all the other elements of *G*. Clearly, $\mathbb{Z}_3$ is formed by the solutions in $\mathbb{C}$ of the equation $x^3 = 1$, i.e. 1, $j = \exp(2i\pi/3)$ and $j^2$. In fact, $\mathbb{Z}_3$ symmetry is broken in the "deconfined regime", for which $\Phi,\bar{\Phi} \to 1$, and restored in the "confined regime" [42], for which $\Phi,\bar{\Phi} \to 0$. This correspondence between $\Phi,\bar{\Phi}$ and these two regimes can be checked if we apply the mean field approximation, which leads to write $A_4(\vec{x},\tau) \equiv A_4$, i.e. a constant field according to the position and the time. There, (28, 29) give, using $A_4 = i\cdot A^0$ [39]:

$$\Phi = \frac{Tr_c\left(\exp(iA_4\cdot\beta)\right)}{N_c} = \frac{Tr_c\left(\exp\left(-A^0/T\right)\right)}{N_c} \propto \exp(-\Delta E/T). \tag{30}$$

Following for example the interpretation of [40], $\Delta E$ is the required energy to add a static quark ("infinite mass") into the system. At finite temperature, in a "confined regime", $\Delta E \to \infty$, so that $\Phi \to 0$. At the opposite, in a "deconfined regime", $\Delta E \to 0$, and thus $\Phi \to 1$. In the same way, the reasoning is similar with $\bar{\Phi}$, associated with the introduction of a static antiquark, according to [4] and the presentation of the Polykov loops performed in the chapter 1.

[1] We use quotation marks to speak about the "confined" and the "deconfined" phases, because they correspond to pure gauge studies, i.e. without quarks, in which the first phase is associated with glueballs, and the second one to deconfined gluons [40]. We can keep the confinement/deconfinement terminology when quarks are added in the modeling, but we cannot say that the Polyakov loop introduces a real color confinement for the quarks…

Now, let us consider the writing of $\mathcal{U}(T,\Phi,\bar{\Phi})$. In fact, in the literature, two possibilities are proposed. For example, in [40, 44], one finds:

$$\frac{\mathcal{U}(T,\Phi,\bar{\Phi})}{T^4} = -\frac{b_2(T)}{2}\cdot\bar{\Phi}\Phi - \frac{b_3}{6}\cdot\left(\Phi^3+\bar{\Phi}^3\right)+\frac{b_4}{4}\cdot\left(\bar{\Phi}\Phi\right)^2, \tag{31}$$

with:

$$b_2(T) = a_0 + a_1\cdot(T_0/T) + a_2\cdot(T_0/T)^2 + a_3\cdot(T_0/T)^3 \tag{32}$$

in which $a_0,a_1,a_2,a_3,b_3,b_4$ are constants, whose values are given in [40]. The other possibility can be found for example in [39, 42], from which we extract the following expression:

$$\frac{\mathcal{U}(T,\Phi,\bar{\Phi})}{T^4} = -\frac{a(T)}{2}\cdot\Phi\bar{\Phi} + b(T)\cdot\ln\left(1-6\Phi\bar{\Phi}+4\left(\Phi^3+\bar{\Phi}^3\right)-3\left(\Phi\bar{\Phi}\right)^2\right), \tag{33}$$

with:

$$a(T) = a_0 + a_1\cdot\left(\frac{T_0}{T}\right) + a_2\cdot\left(\frac{T_0}{T}\right)^2 \quad \text{and} \quad b(T) = b_3\cdot\left(\frac{T_0}{T}\right)^3. \tag{34}$$

The values of the associated constants are reproduced in the table 2.

| $a_0$ | $a_1$ | $a_2$ | $b_3$ | $T_0$ |
|---|---|---|---|---|
| 3.51 | -2.47 | 15.2 | -1.75 | 270 MeV |

**Table 2.** PNJL parameters.

They were chosen by the authors of the quoted papers in order to reproduce pure gauge LQCD data correctly. More precisely, as in [37], these data can concern the energy density $\varepsilon$, the entropy density $s$ and the pressure $p$. Furthermore, as explained in [39], the choice of $a_0 = \frac{16\pi^2}{45} \approx 3{,}51$ was done in order to reach the Stefan-Boltzmann limit at high temperatures, typically $T\to\infty$. Also, as noted in [39], the $b_3 = -0.108\cdot(a_0+a_1+a_2)$ constraint allows obtaining a first order phase transition when $T=T_0$. In fact, the constant $T_0$, also noted $T_D$ in the literature, is the critical deconfinement temperature in a pure gauge theory ($m_{0q}\to\infty$) [35]. This temperature is expected to be higher than the critical temperature of the chiral phase transition in the chiral limit ($m_{0q}\to 0$), whose considered value is $T_c = 170$ MeV e.g. in [34]. In the continuation of the work, we will choose the expression (33) of the potential in our calculations. Indeed, as explained in [39], the potential (33) presents a logarithmic divergence when $\Phi,\bar{\Phi}\to 1$ ("deconfined regime") that limits these two quantities to values lower than 1, in agreement with the expected behavior described before.

Some of theses properties are checked with the figure 2. In the left hand side, we represented $\mathcal{U}/T^4$ (33) as a function of $\Phi$, assuming that $\bar{\Phi}=\Phi$. In the next subsection, we will see that the $\Phi$ values that minimize $\mathcal{U}/T^4$ are relevant in the study. In fact, for $T<T_0$, $\mathcal{U}/T^4$ presents a minimum for $\Phi=0$. When $T=T_0$, $\Phi=0$ is still a value that minimizes the potential, but we also have another minimum associated with a non-null $\Phi$ value. For $T>T_0$, we have only the non-null value. We plotted these $\Phi$ values, noted as $\Phi_{\min}$, according to $T/T_0$, in the right hand side of the figure 2. As explained previously, $\Phi$ is a real number. As a consequence, $\Phi = 0$ is the unique solution for which $\Phi$ satisfies the $\mathbb{Z}_3$ symmetry [33]. Thus, $\Phi$ respects this symmetry when $T < T_0$. When $T$ is higher than $T_0$, $\mathbb{Z}_3$ is spontaneous broken. According to the Landau theory described in the chapter 1, the fact that $\mathcal{U}/T^4$ has got two minimums (for a null and a non-null $\Phi$) when $T = T_0$ allow considering that we have there a first order phase transition between the "confined" and "deconfined" phase. It corresponds to the discontinuity visible in the right hand side of the figure. To conclude, we also verify that $\Phi_{\min} \to 1$ at high temperatures.

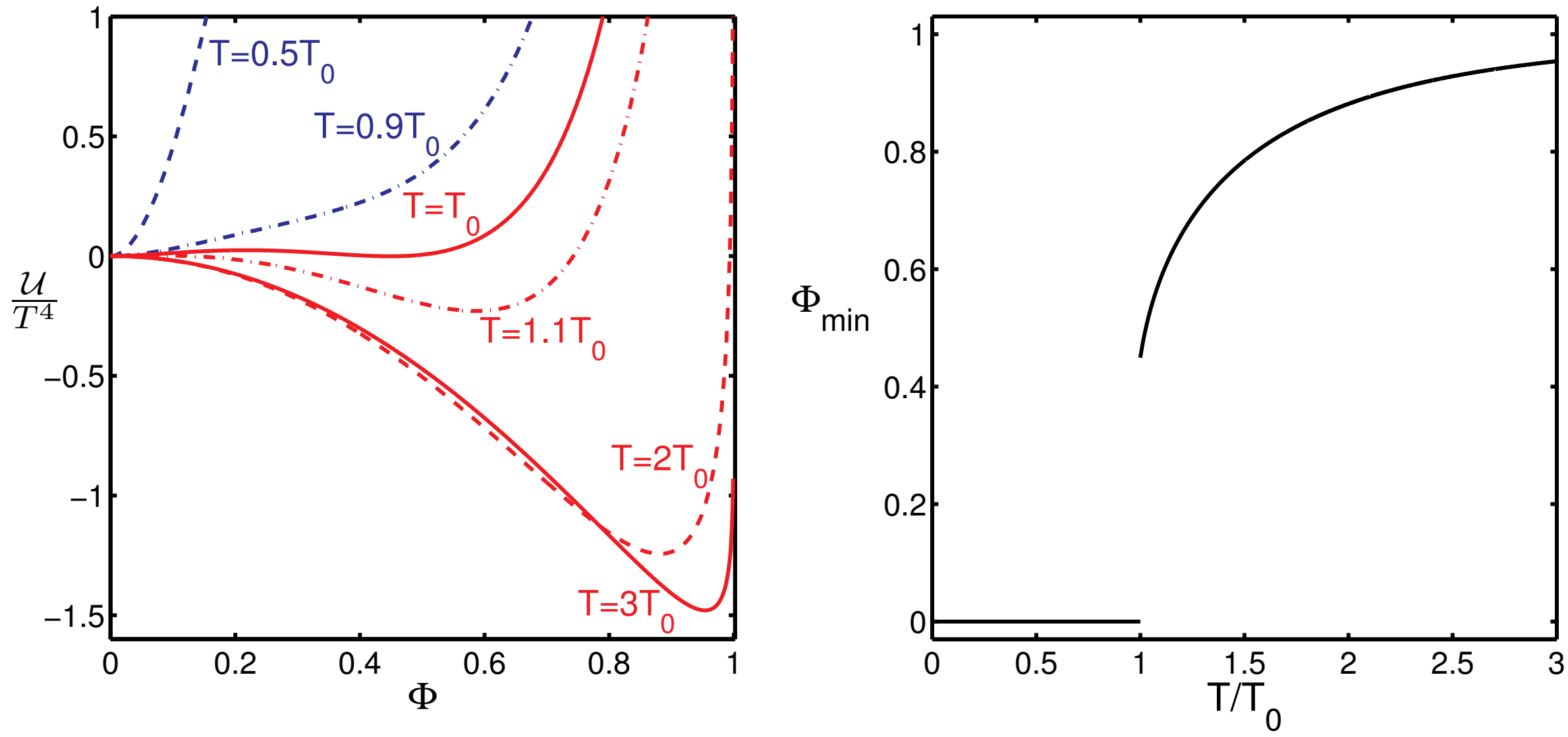

**Figure 2.** Left hand side: behavior of $\mathcal{U}/T^4$ according to $\Phi$, inspired from [39].
Right hand side: evolution of $\Phi_{\min}$ according to $T/T_0$.

## 4.2 The PNJL grand potential

The inclusion of the Polyakov loop also induces modifications of the writing of the grand potential. In the NJL model, the grand potential $\Omega_{\mathrm{NJL}}$ is expressed by the expression [35, 51]:

$$\begin{aligned}\Omega_{\mathrm{NJL}} &= 2G \sum_{f=u,d,s} \left\langle\left\langle \bar{\psi}_f \psi_f \right\rangle\right\rangle^2 - 4K \left\langle\left\langle \bar{\psi}_u \psi_u \right\rangle\right\rangle \left\langle\left\langle \bar{\psi}_d \psi_d \right\rangle\right\rangle \left\langle\left\langle \bar{\psi}_s \psi_s \right\rangle\right\rangle \\ &- 2N_c \sum_{f=u,d,s} \int_0^{\Lambda} \frac{p^2 \cdot dp}{2\pi^2} \cdot \left( E_f + T \cdot \ln\left(Z^+\left(E_f\right)\right) + T \cdot \ln\left(Z^-\left(E_f\right)\right)\right),\end{aligned} \tag{35}$$

in which $E_f = \sqrt{\vec{p}^{\,2} + m_f^{\,2}}$ is the energy of a flavor $f$ quark/antiquark. The expression of the condensates $\left\langle\left\langle \bar{\psi}_f \psi_f \right\rangle\right\rangle$ is given in (20). Furthermore, $Z^+\left(E_f\right)$ and $Z^-\left(E_f\right)$ are, respectively, the partition function of the fermions and anti-fermions, i.e. here the quarks and the antiquarks. They are written as:

$$\begin{cases} Z^+\left(E_f\right) = 1 + \exp\left(-\beta \cdot \left(E_f - \mu_f\right)\right) \\ Z^-\left(E_f\right) = 1 + \exp\left(-\beta \cdot \left(E_f + \mu_f\right)\right) \end{cases}. \tag{36}$$

The Fermi-Dirac distributions for fermions and anti-fermions, noted respectively as $f^+$ and $f^-$ can be obtained by a derivation of the partition functions according to the chemical potential, and it comes:

$$\begin{cases} f^+\left(\beta\cdot\left(E_f-\mu_f\right)\right)=\dfrac{1}{\beta}\cdot\dfrac{\partial\ln\left(Z^+\left(E_f\right)\right)}{\partial\mu_f}=\dfrac{1}{\exp\left(\beta\cdot\left(E_f-\mu_f\right)\right)+1} \\ f^-\left(\beta\cdot\left(E_f+\mu_f\right)\right)=-\dfrac{1}{\beta}\cdot\dfrac{\partial\ln\left(Z^-\left(E_f\right)\right)}{\partial\mu_f}=\dfrac{1}{\exp\left(\beta\cdot\left(E_f+\mu_f\right)\right)+1} \end{cases}. \tag{37}$$

With the inclusion of the Polyakov loop, the grand potential is rewritten as [35, 42]:

$$\begin{aligned} \Omega_{\text{PNJL}} &= \mathcal{U}\left(T,\Phi,\bar{\Phi}\right)+2G\sum_{f=u,d,s}\left\langle\left\langle\bar{\psi}_f\psi_f\right\rangle\right\rangle^2-4K\left\langle\left\langle\bar{\psi}_u\psi_u\right\rangle\right\rangle\left\langle\left\langle\bar{\psi}_d\psi_d\right\rangle\right\rangle\left\langle\left\langle\bar{\psi}_s\psi_s\right\rangle\right\rangle \\ &-2N_c\sum_{f=u,d,s}\int_0^\Lambda\frac{p^2\cdot dp}{2\pi^2}\cdot\left(E_f+\frac{T}{N_c}\cdot Tr_c\ln\left(Z_\Phi^+(E_f)\right)+\frac{T}{N_c}\cdot Tr_c\ln\left(Z_\Phi^-(E_f)\right)\right) \end{aligned}. \tag{38}$$

On one side, a modification is the inclusion of the potential $\mathcal{U}$ previously defined. On the other side, modifications are required at the level of the partition functions. Indeed, they are rewritten as [35, 40, 43, 44]:

$$\begin{cases} Z_\Phi^+\left(E_f\right)=1+L^\dagger\exp\left(-\beta\cdot\left(E_f-\mu_f\right)\right) \\ Z_\Phi^-\left(E_f\right)=1+L\exp\left(-\beta\cdot\left(E_f+\mu_f\right)\right) \end{cases}, \tag{39}$$

where the Polyakov line $L^\dagger$ and $L$ appeared in (28, 29).

To calculate the $Tr_c\ln\left(Z_\Phi^\pm\left(E_f\right)\right)$ terms that appear in (38), in which $Tr_c$ is a trace upon the color, we consider again the mean field approximation evoked in the subsection 4.1. It allows us to rewrite $L$ and $L^\dagger$ as:

$$L(\vec{x})=\mathcal{P}\exp\left(i\int_0^\beta A_4\left(\vec{x},\tau\right)\cdot d\tau\right)\equiv\exp\left(i\beta A_4\right)\text{ and }L^\dagger\left(\vec{x}\right)\equiv\exp\left(-i\beta A_4\right), \tag{40}$$

so that the expression of the Polyakov field $\Phi$ becomes:

$$\Phi=\frac{Tr_c\left(L\right)}{N_c}=\frac{Tr_c\left(\exp\left(i\beta A_4\right)\right)}{N_c}=\frac{1}{3}\cdot\left(\exp\left(i\beta\left(A_4\right)_{11}\right)+\exp\left(i\beta\left(A_4\right)_{22}\right)+\exp\left(i\beta\left(A_4\right)_{33}\right)\right). \tag{41}$$

In the same way, $\bar{\Phi}=\dfrac{Tr_c\left(\exp\left(-i\beta A_4\right)\right)}{N_c}$. Using (40) in the quarks partition function $Z_\Phi^+$ (39), the trace upon the color gives:

$$\begin{aligned} Tr_c\ln\left(Z_\Phi^+\right)&=Tr_c\ln\left(1+L^\dagger\exp\left(-\beta\left(E_f-\mu_f\right)\right)\right)=\ln\left(1+\exp\left(-\beta\left(E_f-\mu_f\right)\right)\cdot\exp\left(-i\beta\left(A_4\right)_{11}\right)\right) \\ &+\ln\left(1+\exp\left(-\beta\left(E_f-\mu_f\right)\right)\cdot\exp\left(-i\beta\left(A_4\right)_{22}\right)\right)+\ln\left(1+\exp\left(-\beta\left(E_f-\mu_f\right)\right)\cdot\exp\left(-i\beta\left(A_4\right)_{33}\right)\right) \end{aligned}, \tag{42}$$

or:

$$Tr_c \ln\left(Z_\Phi^+\right) = \ln\begin{bmatrix} \left(1+\exp\left(-\beta\left(E_f-\mu_f\right)\right)\cdot\exp\left(-i\beta\left(A_4\right)_{11}\right)\right) \\ \times\left(1+\exp\left(-\beta\left(E_f-\mu_f\right)\right)\cdot\exp\left(-i\beta\left(A_4\right)_{22}\right)\right) \\ \times\left(1+\exp\left(-\beta\left(E_f-\mu_f\right)\right)\cdot\exp\left(-i\beta\left(A_4\right)_{33}\right)\right) \end{bmatrix}. \tag{43}$$

If we expand this relation, we obtain:

$$Tr_c \ln\left(Z_\Phi^+\right) = \tag{44}$$

$$\ln\begin{pmatrix} 1+\exp\left(-\beta\left(E_f-\mu_f\right)\right)\cdot\left(\exp\left(-i\beta\left(A_4\right)_{11}\right)+\exp\left(-i\beta\left(A_4\right)_{22}\right)+\exp\left(-i\beta\left(A_4\right)_{33}\right)\right) \\ +\exp\left(-2\beta\left(E_f-\mu_f\right)\right)\cdot\left(\exp\left(-i\beta\left(\left(A_4\right)_{22}+\left(A_4\right)_{33}\right)\right)+\exp\left(-i\beta\left(\left(A_4\right)_{11}+\left(A_4\right)_{33}\right)\right)+\exp\left(-i\beta\left(\left(A_4\right)_{11}+\left(A_4\right)_{22}\right)\right)\right) \\ +\exp\left(-3\beta\left(E_f-\mu_f\right)\right)\cdot\left(\exp\left(-i\beta\left(\left(A_4\right)_{11}+\left(A_4\right)_{22}+\left(A_4\right)_{33}\right)\right)\right) \end{pmatrix}.$$

Immediately, we identify $3\bar{\Phi}$ in the first line of (44). To continue, we recall the possibility evoked in [35] to write $L$ or $L^\dagger$ as diagonal matrices as:

$$L^\dagger \equiv \begin{bmatrix} \exp(-i\beta\phi) & & \\ & \exp(-i\beta\phi') & \\ & & \exp\left(i\beta(\phi+\phi')\right) \end{bmatrix}. \tag{45}$$

This writing enables us to identify each term $\left(A_4\right)_{ii}$, and then we remark that:

$$\begin{cases} \exp\left(-i\beta\left(\left(A_4\right)_{22}+\left(A_4\right)_{33}\right)\right)=\exp\left(i\beta\left(A_4\right)_{11}\right) \\ \exp\left(-i\beta\left(\left(A_4\right)_{11}+\left(A_4\right)_{33}\right)\right)=\exp\left(i\beta\left(A_4\right)_{22}\right), \\ \exp\left(-i\beta\left(\left(A_4\right)_{11}+\left(A_4\right)_{22}\right)\right)=\exp\left(i\beta\left(A_4\right)_{33}\right) \end{cases} \tag{46}$$

and:

$$\exp\left(-i\beta\left(\left(A_4\right)_{11}+\left(A_4\right)_{22}+\left(A_4\right)_{33}\right)\right)=1. \tag{47}$$

As a consequence, (44) can be simplified by the following way:

$$\begin{aligned} Tr_c \ln\left(Z_\Phi^+\right) &= \ln\left(1+\exp\left(-\beta\left(E_f-\mu_f\right)\right)\cdot 3\bar{\Phi}+\exp\left(-2\beta\left(E_f-\mu_f\right)\right)\cdot 3\Phi+\exp\left(-3\beta\left(E_f-\mu_f\right)\right)\cdot 1\right) \\ &= \ln\left(1+3\left(\bar{\Phi}+\Phi\exp\left(-\beta\left(E_f-\mu_f\right)\right)\right)\exp\left(-\beta\left(E_f-\mu_f\right)\right)+\exp\left(-3\beta\left(E_f-\mu_f\right)\right)\right) \end{aligned}, \tag{48}$$

Such a calculation can be remade with $Tr_c \ln\left(Z_\Phi^-\left(E_f\right)\right)$, and we find again the result visible in the PNJL literature [37, 40, 42]:

$$\begin{cases} Tr_c \ln\left(Z_\Phi^+\left(E_f\right)\right) \\ \quad =\ln\left(1+3\left(\bar{\Phi}+\Phi\exp\left(-\beta\left(E_f-\mu_f\right)\right)\right)\exp\left(-\beta\left(E_f-\mu_f\right)\right)+\exp\left(-3\beta\left(E_f-\mu_f\right)\right)\right) \\ Tr_c \ln\left(Z_\Phi^-\left(E_f\right)\right) \\ \quad =\ln\left(1+3\left(\Phi+\bar{\Phi}\exp\left(-\beta\left(E_f+\mu_f\right)\right)\right)\exp\left(-\beta\left(E_f+\mu_f\right)\right)+\exp\left(-3\beta\left(E_f+\mu_f\right)\right)\right) \end{cases}. \tag{49}$$

We recall that the plus sign refers to quarks, and the minus sign to antiquarks. We note that:

$$Tr_c \ln\left(Z_\Phi^-\left(E_f\right)\right) = Tr_c \ln\left(Z_\Phi^+\left(E_f\right)\right)\Big|_{\substack{\Phi\leftrightarrow\bar{\Phi} \\ \mu_f\leftrightarrow-\mu_f}} . \tag{50}$$

These modifications of the partition functions leads to consider a rewriting of the Fermi-Dirac distributions, as explained in [40]. Following the method described by (37), we note these distributions as $f_\Phi^+$ (quarks) and $f_\Phi^-$ (antiquarks), and we write:

$$f_\Phi^\pm\left(\beta\cdot\left(E_f \mp \mu_f\right)\right) = \pm\frac{1}{\beta}\cdot\frac{\partial\left(\frac{1}{N_c}Tr_c \ln\left(Z_\Phi^\pm\right)\right)}{\partial\mu_f} , \tag{51}$$

and so:

$$\begin{cases} f_\Phi^+\left(\beta\cdot\left(E_f-\mu_f\right)\right) \\ = \dfrac{\left(\bar{\Phi}+2\Phi\cdot\exp\left(-\beta\cdot\left(E_f-\mu_f\right)\right)\right)\cdot\exp\left(-\beta\cdot\left(E_f-\mu_f\right)\right)+\exp\left(-3\beta\cdot\left(E_f-\mu_f\right)\right)}{1+3\left(\bar{\Phi}+\Phi\cdot\exp\left(-\beta\cdot\left(E_f-\mu_f\right)\right)\right)\cdot\exp\left(-\beta\cdot\left(E_f-\mu_f\right)\right)+\exp\left(-3\beta\cdot\left(E_f-\mu_f\right)\right)} \\ f_\Phi^-\left(\beta\cdot\left(E_f+\mu_f\right)\right) \\ = \dfrac{\left(\Phi+2\bar{\Phi}\cdot\exp\left(-\beta\cdot\left(E_f+\mu_f\right)\right)\right)\cdot\exp\left(-\beta\cdot\left(E_f+\mu_f\right)\right)+\exp\left(-3\beta\cdot\left(E_f+\mu_f\right)\right)}{1+3\left(\Phi+\bar{\Phi}\cdot\exp\left(-\beta\cdot\left(E_f+\mu_f\right)\right)\right)\cdot\exp\left(-\beta\cdot\left(E_f+\mu_f\right)\right)+\exp\left(-3\beta\cdot\left(E_f+\mu_f\right)\right)} \end{cases} . \tag{52}$$

If $\Phi = \bar{\Phi}$ , we have $f_\Phi{}^\pm(-x) = 1 - f_\Phi{}^\pm(x)$ and, as with the partition function, we check that:

$$f_\Phi^-\left(\beta\cdot\left(E_f+\mu_f\right)\right) = f_\Phi^+\left(\beta\cdot\left(E_f-\mu_f\right)\right)\Big|_{\substack{\Phi\leftrightarrow\bar{\Phi} \\ \mu_f\leftrightarrow-\mu_f}} . \tag{53}$$

As indicated in [40, 42], this update of the Fermi-Dirac statistics is to be applied in all the occasions for which the quarks/antiquarks distributions are required. It notably concerns the modeling of the mesons, diquarks, baryons, and cross-sections calculations.

# 4.3 Gap equations in the PNJL model

Finally, the inclusion of the Polyakov loop has also consequences on the equations to be solved to find the dressed quarks masses. Indeed, the set of equations (24) should be updated in order to take into account the new Fermi-Dirac distributions (52). But, it is also required to minimize the PNJL grand potential according to $\Phi$ and $\bar{\Phi}$. It leads to consider the extra relations to find the values of $\Phi$ and $\bar{\Phi}$ [40, 42]:

$$\frac{\partial\Omega_{\mathrm{PNJL}}}{\partial\Phi} = 0 \qquad \frac{\partial\Omega_{\mathrm{PNJL}}}{\partial\bar{\Phi}} = 0 , \tag{54}$$

with:

$$\frac{\partial\Omega_{\mathrm{PNJL}}}{\partial\Phi}=T^4\cdot\left(-\frac{a(T)}{2}\cdot\bar{\Phi}-6\cdot\frac{b(T)\cdot\left(\bar{\Phi}-2\cdot\Phi^2+\bar{\Phi}^2\cdot\Phi\right)}{1-6\cdot\Phi\cdot\bar{\Phi}+4\cdot\left(\Phi^3+\bar{\Phi}^3\right)-3\cdot\left(\bar{\Phi}\cdot\Phi\right)^2}\right), \qquad (55)$$
$$-6\cdot T\cdot\sum_{f=u,d,s}\int_0^{\Lambda}\frac{p^2\cdot dp}{2\pi^2}\cdot I\left(\Phi,\bar{\Phi},\beta,E_f,\mu_f\right)$$

in which $I$ is defined by:

$$I\left(\Phi,\bar{\Phi},\beta,E_f,\mu_f\right)=$$
$$\frac{\exp\left(-2\beta\left(E_f-\mu_f\right)\right)}{1+3\cdot\left(\bar{\Phi}+\Phi\exp\left(-\beta\left(E_f-\mu_f\right)\right)\right)\exp\left(-\beta\left(E_f-\mu_f\right)\right)+\exp\left(-3\beta\left(E_f-\mu_f\right)\right)}. \qquad (56)$$
$$+\frac{\exp\left(-\beta\left(E_f+\mu_f\right)\right)}{1+3\left(\Phi+\bar{\Phi}\exp\left(-\beta\left(E_f+\mu_f\right)\right)\right)\exp\left(-\beta\left(E_f+\mu_f\right)\right)+\exp\left(-3\beta\left(E_f+\mu_f\right)\right)}$$

Furthermore, we have:

$$\frac{\partial\Omega_{\mathrm{PNJL}}}{\partial\bar{\Phi}}=\left.\frac{\partial\Omega_{\mathrm{PNJL}}}{\partial\Phi}\right|_{\substack{\Phi\leftrightarrow\bar{\Phi}\\ \mu_f\leftrightarrow-\mu_f}}. \qquad (57)$$

In the expression of the PNJL grand potential (38), the potential $\mathcal{U}$ is present. We studied in the subsection 4.1 the behavior of $\mathcal{U}$ as regards its minimums according to $\Phi$. In fact, this behavior will have an influence on the minimization performed in (54). Notably, it wants to say that the properties $0\leq\Phi<1$ and $0\leq\bar{\Phi}<1$ will be verified. Physically, the figure 2 considers only $\mathcal{U}$, i.e. a pure gauge description, whereas the resolution of (54) takes into account the influence of the quarks/antiquarks.

Finally, as argued previously, $\Phi$ and $\bar{\Phi}$ are real numbers in the framework of the mean field approximation. In addition, they can be considered as independent variables. Thus, when the isospin symmetry is not considered and at finite densities, the complete set of equations to be solved has got 8 equations, i.e. (24) and (54), with 8 unknowns. Three of these unknowns are the masses of the *u,d,s* dressed quarks, three are their associated chemical potentials, and two are $\Phi$ and $\bar{\Phi}$. Concretely, such a system is solvable numerically with a root-finding algorithm. Moreover, as evoked in the literature, the PNJL model described here correspond to a minimal coupling of the Polyakov loop with the quarks. It implies as an example that the gap equations (16, 24) do not depend explicitly on the Polyakov loop, except via (52). In a model as the EPNJL one, a difference is $G$ becomes there a function of $\Phi,\bar{\Phi}$ [45].

# 5. Obtained results

## 5.1 Quarks masses

By a resolution of (24), the masses of the NJL dressed quarks, at null temperature and null density, are gathered in the table 3. We considered the three parameter sets defined in the table 1. "RK" and "P1" respect the isospin symmetry, so it explains why the masses of the light quarks $u$ and $d$ are strictly the same. These masses $m_{u,d}$ found with RK are close to $m_{nucleon}/3$, whereas the ones obtained with P1 and EB are greater. However, they are comparable to the masses found in references as [58–60], that we will consider in the chapter devoted to the diquarks. In addition, they stay conform to the range of the possible values established in [22]. The other results presented in this chapter were established with the P1 parameter set.

| | RK | P1 | EB |
|---|---|---|---|
| quark $u$ | 367.65 | 424.23 | 419.10 |
| quark $d$ | 367.65 | 424.23 | 422.31 |
| quark $s$ | 549.48 | 626.49 | 588.17 |

**Table 3.** Effective quarks masses at zero temperature and null density.

The masses of the quarks at finite temperature and null density are represented in the left hand side of the figure 3. Our results permit a comparison between the NJL model and the PNJL one. They are in agreement with the ones published for example in [57] associated with pure NJL data, and [42] that proposed such a comparison with NJL and PNJL approaches. For the two models, we observe that the masses decrease when the temperature is growing. At high temperatures, the masses of the light quarks tend towards their naked values, see table 1. At the opposite, the mass of the strange quarks decrease, but stays rather high, even at $T = 400$ MeV.

At this occasion, it should be recalled that the PNJL model has a limitation according to the temperature, that corresponds to $T \approx 2.5 T_0$ [33, 40, 42]. More precisely, as explained in [33], the effects of the Polyakov loop are optimal around the "deconfinement phase transition", when $T \approx T_0$. But, for higher temperatures, i.e. $T > 2.5 T_0$, these effects come to a saturation, observable in [33] and in the right hand side of the figure 2. In the same time, the contribution on the thermodynamics of the gluons' effective mass becomes important. This contribution is not taken into account in the (P)NJL models. In fact, it is true that a limitation upon the temperature can also be predicted for the pure NJL model. Clearly, NJL is a low energy model, in the sense that it considers "frozen" gluons, subsection 2.2. At high temperatures, this approximation becomes invalid for the same reason: the gluons' degrees of freedom cannot be neglected in this regime. So, in (7), the simplification $k^2 \ll \Sigma^2$ becomes questionable, as the fact that the gluons effective mass $\Sigma$ is considered as a constant in this equation, whereas [33] showed that $\Sigma(T) \propto T$ at high temperatures. A physical limitation of the (P)NJL models upon the density is discussed later in this subsection…

At null temperature, as in [42] for example, the NJL and the PNJL models exactly give the same quarks masses. This behavior will be confirmed in the next chapters with the mesons, diquarks and baryons. Thus, results as the ones seen in the table 3, i.e. calculated at null temperature and null density, will be valid for the two models. An explanation of this behavior comes to the fact that at null density, $\Phi = \bar{\Phi} \to 0$, as confirmed by the figure 7 and [42]. As evoked previously, such a value corresponds to a "confined regime". Using this result in the modified Fermi-Dirac distributions (52), we obtain:

$$f_\Phi^+\left(\beta\left(E_f - \mu_f\right)\right) \xrightarrow{\Phi=\bar{\Phi}\to 0} \frac{\exp\left(-3\beta\left(E_f - \mu_f\right)\right)}{1+\exp\left(-3\beta\left(E_f - \mu_f\right)\right)} = \frac{1}{\exp\left(3\beta\left(E_f - \mu_f\right)\right)+1}. \tag{58}$$

A similar result is found with $f_\Phi^-\left(\beta\cdot\left(E_f + \mu_f\right)\right)$. Except for the factor three inside the exp function in (58), we recover the traditional Fermi-Dirac distributions, used in the framework of the pure NJL model. We are there at low temperatures, thus the factor three is without consequence on our reasoning. In this configuration the NJL and PNJL equations are identical, so the both lead to identical results.

In addition, the NJL results and the PNJL ones tend to be the same at high temperatures, i.e. $T \approx 400$ MeV for the light quarks. The explanations are quite similar. At high temperatures, $\Phi = \bar{\Phi} \to 1$ [40, 42] ("deconfined regime"), so we write:

$$\begin{aligned}
& f_\Phi^+\left(\beta\left(E_f - \mu_f\right)\right) \\
& \xrightarrow{\Phi=\bar{\Phi}\to 1} \frac{\left(1+2\exp\left(-\beta\left(E_f - \mu_f\right)\right)\right)\exp\left(-\beta\left(E_f - \mu_f\right)\right)+\exp\left(-3\beta\left(E_f - \mu_f\right)\right)}{1+3\left(1+\exp\left(-\beta\left(E_f - \mu_f\right)\right)\right)\exp\left(-\beta\left(E_f - \mu_f\right)\right)+\exp\left(-3\beta\left(E_f - \mu_f\right)\right)}, \\
& = \frac{\exp\left(-3\beta\left(E_f - \mu_f\right)\right)\left(1+\exp\left(\beta\left(E_f - \mu_f\right)\right)\right)^2}{\exp\left(-3\beta\left(E_f - \mu_f\right)\right)\left(1+\exp\left(\beta\left(E_f - \mu_f\right)\right)\right)^3} = \frac{1}{\exp\left(\beta\left(E_f - \mu_f\right)\right)+1}
\end{aligned} \tag{59}$$

i.e. also the classical Fermi-Dirac distribution of the quarks. So, at low and high temperatures, the NJL and the PNJL models give comparable results. At the opposite, at moderate temperatures, i.e. $T \approx 200-300$ MeV, differences are observed. Clearly, the PNJL results are shifter towards higher temperatures compared to NJL ones. In addition, as evoked in our introduction, the decreasing of the masses of the PNJL quarks is more brutal compared to NJL quarks. About the PNJL results, we have a zone until $T \approx 250$ MeV for which the masses are rather constant.

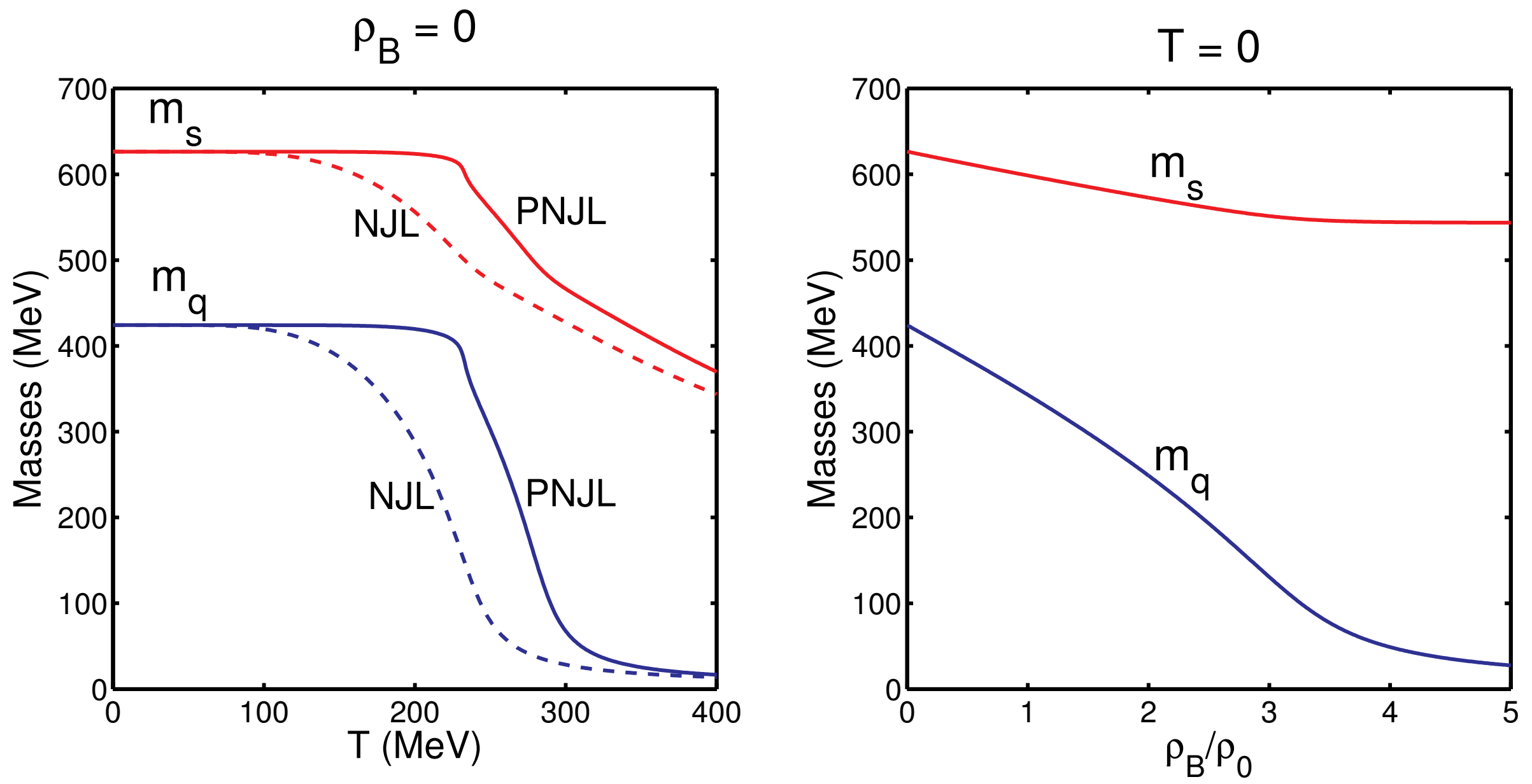

**Figure 3.** Evolution of the mass of the quarks.

The right hand side of the figure 3 indicates the evolution of the quarks masses according to the baryonic density, at null temperature. There, it can be noted that the two models exactly give the same results, whatever be the baryonic density. The explanations associated with (58) stay valid. For the two models, the masses decrease when the baryonic density increase. The strange quark is less affected by a variation of the baryonic density than the light quarks. Indeed, the dependence of the $s$ quarks according to the baryonic density, via the chemical potentials $\mu_u, \mu_d$, only appears in the term of (16) associated with the $\frac{K \cdot N_c^2}{8\pi^4}$ factor, but not in the term linked to $\frac{G \cdot N_c}{\pi^2}$. Such analysis of the quarks masses according to the baryonic density can be found for example in [48], even if the results they found with their "P3" parameter set ($m_{0u} = m_{0d} = 0$, so at the chiral limit) are incorrect: it is indeed possible to obtain quarks masses for $0 < \rho_B < 2.5\rho_0$, whereas this reference indicates the contrary.

Then, we extended the calculations to the $T, \rho_B$ plane. It gives the results presented in the figures 4, 5. The figure 4 shows the evolution of the light quarks masses, and the figure 5 the mass of the strange quark. The left hand side of the figure 4 is in agreement with the equivalent figure published in [19]. Indeed, the aspect of these two graphs is rigorously identical, even if the values are not the same, because we did not take the same parameter set as the one used in this reference. A similar remark can be done with the left hand side of the figure 5 and [61]. Furthermore, these figures allow continuing the comparison between NJL results and PNJL ones. Clearly, the behavior observed in the figure 3 is confirmed, i.e. the NJL and PNJL results coincide at low at high temperatures. Also, at moderate temperatures, the inclusion of the Polyakov loop leads to a shifting of the graphs towards higher temperatures. Clearly, compared to NJL results, the masses values are not modified by the inclusion of the Polyakov loop, but they are simply shifted. In addition, the Polyakov loop has no visible effect on the density, because the shifting is observed only along the temperature axis.

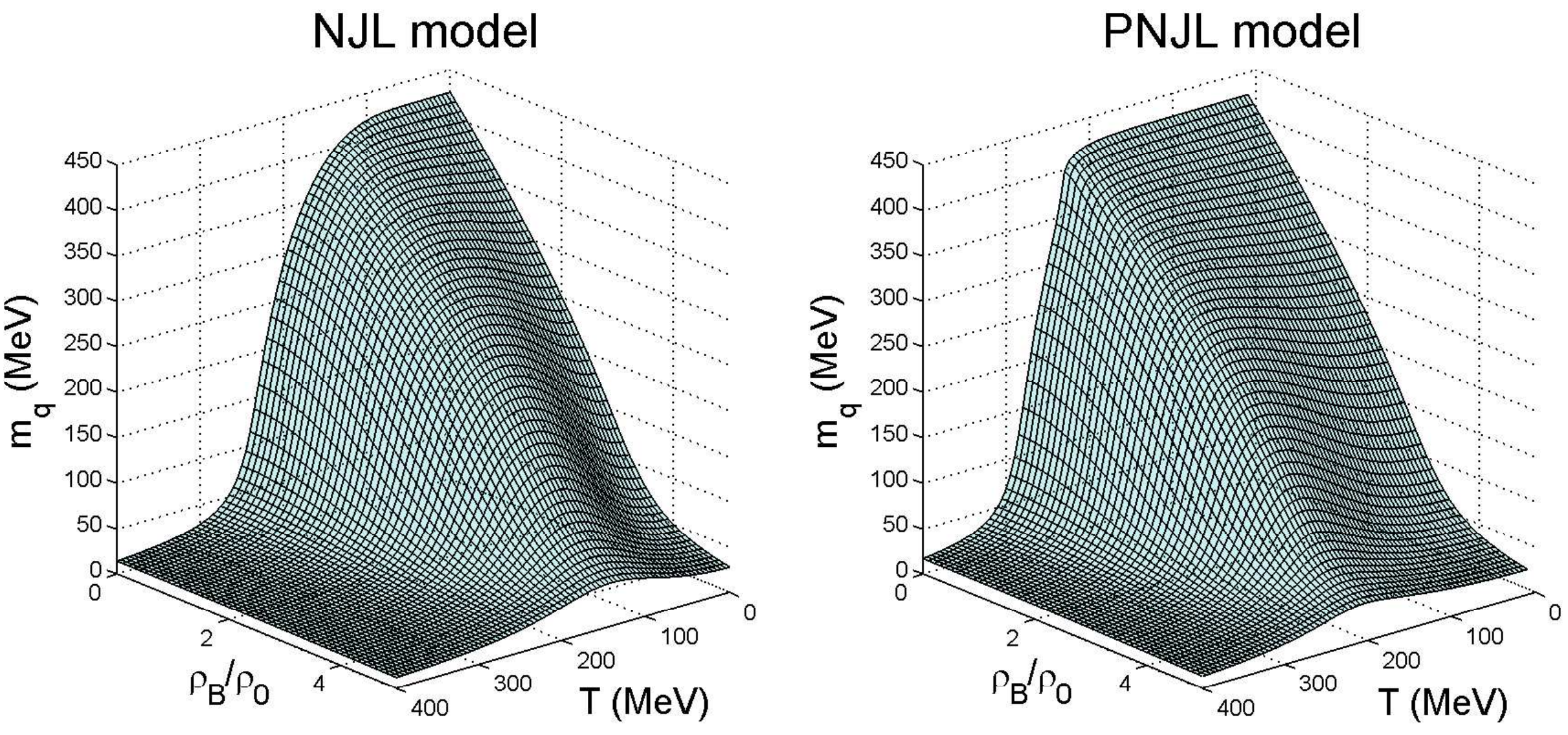

**Figure 4.** Mass of the $q$ quarks according to the temperature $T$ and the baryonic density $\rho_B$.

In fact, we must be prudent about the results performed at finite densities. Indeed, as seen in the chapter 1, some regions, for which the temperature is reduced and the baryonic density is strong enough, are expected to undergo the color superconductivity phenomenon [62–70]. The correct description of this phenomenon requires the use of an adapted formalism, as the Nambu-Gorkov formalism, as reported in the quoted references. However, the frontier between the hadronic, QGP and color superconductivity phases is far for being known perfectly. In [67], it was supposed that the color superconductivity could intervene at $\rho_B \approx 10\rho_0$, i.e. definitively outside of the range of our work ($0 \le \rho_B \le 5\rho_0$). Nevertheless, other papers suggest that the color superconductivity phases could be present before, because they indicate rather reduced baryonic chemical potential values, e.g. [65]. But, it seems to be admitted that the influence of the color superconductivity can be neglected in the zones in which we will work. It concerns the masses of the studied particles, but also the cross-sections of their associated reactions. Furthermore, a consensus seems to consider that the color superconductivity cannot occur for temperatures higher than 100 MeV [65, 67, 69], even if the Polyakov loop is expected to shift this temperature towards higher values. But, clearly, temperatures higher than 150 MeV, i.e. typically above 200 MeV, seem definitely outside the effect of the color superconductivity. The motivations of our work are to investigate the cooling of a quarks/antiquarks system and its hadronization into hadrons. As a consequence, we can be sure that the relevant $T, \rho_B$ zones in which we will work, especially in the simulations performed in the chapter 7, are definitely not concerned by the color superconductivity.

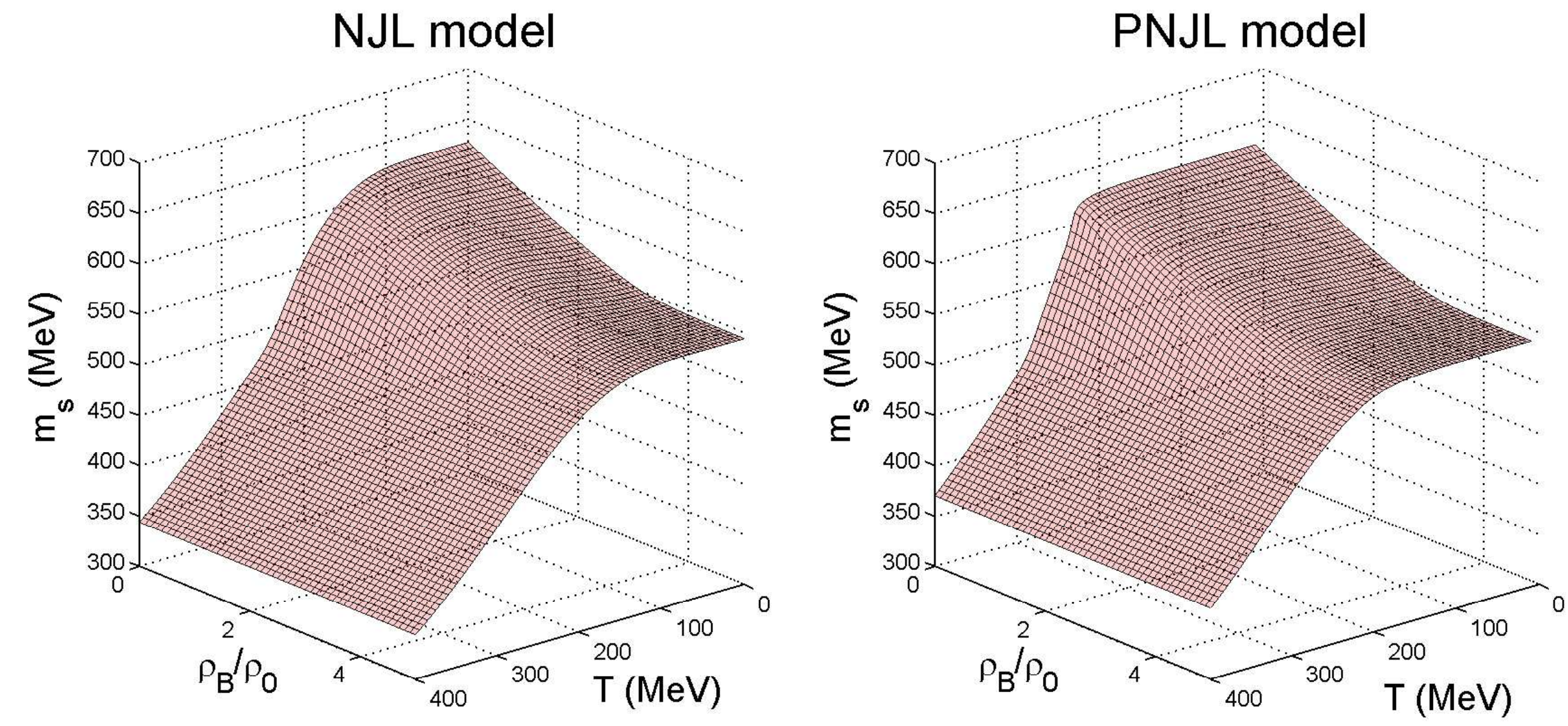

**Figure 5.** Mass of the $s$ quark according to $T$ and $\rho_B$.

# 5.2 Study of the (P)NJL order parameters

We saw that the used gap equations correspond to the coupling of the quarks naked masses with the associated chiral condensates. In addition, with the PNJL model, the quarks are coupled to the Polyakov loop $\Phi$. The chiral condensate is an order parameter related to the chiral symmetry, whereas $\Phi$ is an order parameter of the "confined"/"deconfined" phase transition (rigorously in pure gauge calculations). We represented the evolution of these quantities according to the temperature, at null density, in the left hand side of the figure 6. Immediately, it can be compared to the ones presented in [35, 37]. Qualitatively, the results are very similar.

We plotted the evolution of the chiral condensate $\langle\langle\bar{\psi}_q\psi_q\rangle\rangle$ of the light quarks, normalized by its value at null temperature and null density $\langle\langle\bar{\psi}_q\psi_q\rangle\rangle_0$, in the (P)NJL models. For the both, we observe a decreasing when the temperature is growing up. It goes on the sense of the expected restoration of the chiral symmetry at high temperatures, evoked in the chapter 1. According to the Landau theory concerning the phase transitions, this restoration is performed by the way of a crossover. Indeed, the value of the quark condensate tends towards zero only at high temperatures. Such a result is explainable by the fact that we use the P1 parameter set, in which the naked masses of the light quarks are non-null. At the chiral limit, it is possible to see a second order phase transition [18, 22]: the value of the condensate falls continuously towards zero, when $T = T_c$. In [18], $T_c = 150\text{ MeV}$, versus $T_c = 220\text{ MeV}$ in [22].

If we compare the NJL and PNJL evolutions of the condensate, two differences are mentionable. Firstly, we can study the pseudo-critical temperature[2], noted as $T_c$, for which we suppose that the slope of the curve is maximal, i.e. using the method of [37]. In fact, $T_c$ is stronger in the PNJL model than in the NJL one, respectively 270 MeV against 230 MeV. This result confirms the one found in the literature [35, 37]. However, the PNJL curve of the chiral condensate is less smooth with the potential (33) than with (31) [71]. Thus, our method used to estimate $T_c$ should be improved … Secondly, around $T_c$, the decreasing is faster with the PNJL model. It goes on the sense of the LQCD results [39], see chapter 1. It explains the remark done in [42], reproduced in the introduction of this chapter, that indicates that the description of the restoration of the chiral symmetry is more efficient in the PNJL model than in a pure NJL one. In the right hand side of the figure 6, we extended the study of the light quark condensate in the $T,\rho_B$ plane, for the PNJL model. As previously, the values are normalized by $\langle\langle \bar{\psi}_q \psi_q \rangle\rangle_0$. Obviously, the similarity between this graph and the one displaying the mass of the light quarks (right hand side of figure 4) is striking. As a conclusion, in the gap equation (21) of the light quarks, the $G$ term associated with the coupling to the light quark condensate is dominant. Moreover, using (20), we obtained $\langle\langle \bar{\psi}_q \psi_q \rangle\rangle_0 \approx -(283\ \mathrm{MeV})^3$. This value is comparable, but upper, compared to the "empirical" value $-(250\ \mathrm{MeV})^3$ frequently admitted in the literature [20, 22]. However, this result stays consistent with the GMOR relation evoked in the chapter 1. In addition, the found value is rather close to the $-(287\ \mathrm{MeV})^3$ used in [18] or the $-(283\ \mathrm{MeV})^3$ in [72].

[2] The notion of critical temperature, notably defined with the 1st or 2nd order phase transitions, is frequently extended to crossover transition. In this case, the used term is *pseudo-critical temperature*, as in [35].

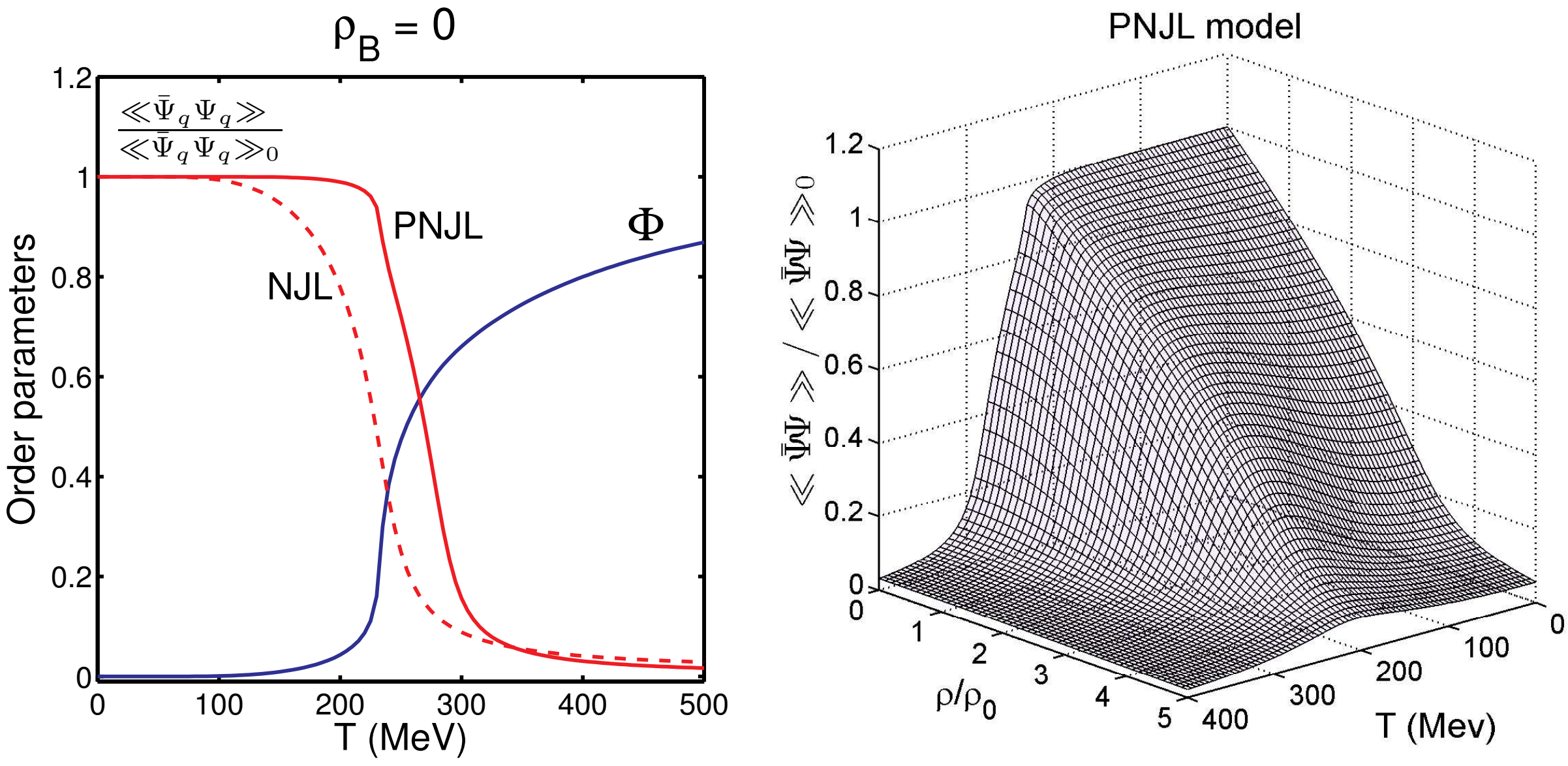

**Figure 6.** Left hand side: evolution of the order parameters according to the temperature. Right hand side: normalized chiral condensate in the $T-\rho_B$ plane, in the PNJL model.

We propose now to study the evolution of $\Phi$. In the right hand side of the figure 2, we observed a first order phase transition when only the effective potential is considered, i.e. no quarks. In the left hand side of the figure 6, the quarks contribution leads to modify this phase transition into a crossover. It confirms the observations done in [35]. Thus, as explained in [40], the symmetry $\mathbb{Z}_3$ is there no exact, but the two regimes "confined"/"deconfined" are still visible in the graph. In fact, in the literature, the Polyakov fields $\Phi$ and $\bar{\Phi}$ were studied according to the temperature, and sometimes for several chemical potentials [37]. So, we propose to complete this analysis by a study in the $T, \rho_B$ plane. In the left hand side of the figure 7, the Polyakov field $\Phi$ is considered, whereas the right hand side of the figure focus on the difference between $\Phi$ and $\bar{\Phi}$. Concerning $\Phi$, we note that it increases when the temperature is growing. The found behavior verifies what was described in the section 4. Indeed, whatever be the baryonic density, $\Phi \to 0$ when $T \to 0$. At high temperatures, $\Phi$ seems to converge toward a finite value. We know that this value is 1, but the convergence in our graph is not strong enough to be able to see it, as in the references using the effective potential (33). In all cases, we can check that $0 \le \Phi < 1$, as mentioned by [39].

Furthermore, the right hand side of the figure 7 confirms some observations found in the literature. First, it was reported that $\Phi = \bar{\Phi}$ at null temperature, but also at null chemical potential [39]. In fact, as confirmed by the figure 8, $\mu_B = 0 \Leftrightarrow \rho_B = 0$. Thus, in the right hand side of the figure 7, the difference between $\Phi$ and $\bar{\Phi}$ is null when $T = 0$ or $\rho_B = 0$. In addition, the convergence of $\Phi$ and $\bar{\Phi}$ at high temperatures (towards 1) induces also a diminution of the difference between them. Our results indicates that this difference is negligible when $T > 300$ MeV. The reference [39] mentions that $\Phi \ne \bar{\Phi}$ at non-null chemical potentials (and non-null temperatures), so at non-null baryonic densities in our graph. Moreover, it was observed in [37], in which the effective potential (31) is used, that $\bar{\Phi} \ge \Phi$ for fixed values of $\mu \ne 0$, whatever be the temperature. In the right hand side of the figure 7, we confirm this affirmation for all the values of the $T, \rho_B$ plane studied in this graph.

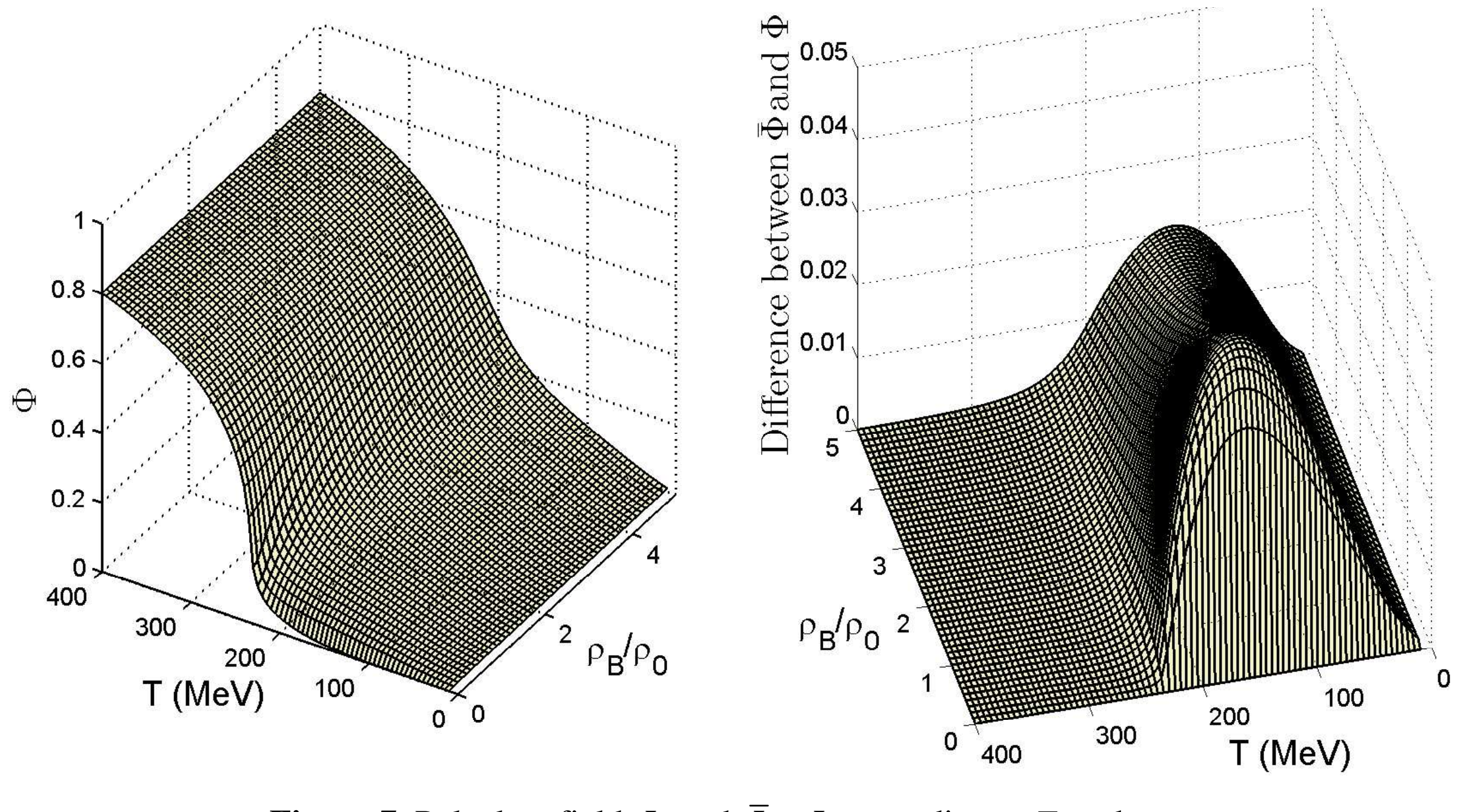

**Figure 7.** Polyakov field $\Phi$ and $\bar{\Phi} - \Phi$ according to $T$ and $\rho_B$.

In fact, these results are in agreement with [4], in which a physical explanation to this behavior was proposed. Clearly, we saw that $\Phi \propto \exp(-\Delta E/T)$, in which $\Delta E$ is the required energy to add a (static) quark in the medium, whereas $\bar{\Phi} \propto \exp(-\Delta \bar{E}/T)$ is linked to the energy $\Delta \bar{E}$ to add a (static) antiquark. At null density, the quarks and the antiquarks present a similar behavior, thus $\bar{\Phi} = \Phi$. When $\mu \ne 0$, this reasoning is no longer valid, so $\Phi \ne \bar{\Phi}$. More precisely, when $\mu > 0$ (positive density), the medium has an excess of quarks over antiquarks.

It leads to $\Delta E > \Delta \bar{E}$, so $\Phi < \bar{\Phi}$, as observed. Obviously, at negative densities, we checked that $\Phi > \bar{\Phi}$. Moreover, at null temperature, we saw that $\Phi = \bar{\Phi} \to 0$. Physically, it corresponds to the "confined phase", so $\Delta E$ and $\Delta \bar{E}$ tend together towards the infinity. In other words, the "confinement" acts in the same way for the quarks and antiquarks. It explains why $\Phi = \bar{\Phi}$ when $T \to 0$. At the opposite, strong temperatures ($T > 300\,\text{MeV}$) correspond to the "deconfined phase". Even if the quarks and antiquarks act differently at finite densities, the "deconfinement" leads to a vanishing of $\Delta E$ and $\Delta \bar{E}$. This is why the difference between $\Phi$ and $\bar{\Phi}$ disappears at high temperatures.

Furthermore, our results leads us to remark that the difference between $\Phi$ and $\bar{\Phi}$ stays always rather modest: it can be considered that $\Phi \approx \bar{\Phi}$, at least in the framework of the part of the $T, \rho_B$ plane explored in our work.

# 5.3 The chemical potential

To conclude this chapter, we propose now to investigate the relation between the baryonic density $\rho_B$ (26) and the light quarks chemical potential $\mu_q$. In the framework of the isospin symmetry, the baryonic density is linked to $\mu_q$ by the relation $\mu_B = 3\mu_q$. Upon numerical aspects, we recall that the density is a parameter (chosen), whereas the chemical potential (an unknown) is found during the numerical solving of the system of equations (24). Also, according to the remark done in the subsection 3.3, we remark that we represent here this chemical potential, i.e. the "effective" one. Anyway, we observe that the shifting $\delta\mu_f$ is lower than 15 MeV when $\mu_{0f} = 500\,\text{MeV}$ and $T \to 0$. It confirms the relatively modest influence of this shifting.

The results are shown in the figure 8, for the NJL and PNJL models. Immediately, we note that the models qualitatively gave similar results. The only difference comes from the shifting of the PNJL graph along the temperature axis, compared to the NJL one. Such a behavior corresponds to what was observed previously with the quarks masses. But, for the two models, the found structure should be commented. First, we focus on the region of the graphs for which the temperature and baryonic density are close to zero. There, a discontinuity is present. Even if this particularity is rather spectacular, it can be explained. Thanks to equations (23, 26), it is easy to check that $\rho_B = 0$ leads to the obvious solution $\mu_q = 0$, whatever be the temperature. On the other hand, at reduced temperature and when $\rho_B \to 0^+$, we observe on the graphs that:

$$\lim_{\substack{\rho_B \to 0 \\ \rho_B > 0}} \mu_q \left(T \simeq 0, \rho_B\right) = m_q > 0\,, \tag{60}$$

which constitutes the discontinuity. This result (60) is explainable by (23). For our reasoning, we consider the NJL version of this equation, i.e. we use the classical Fermi-Dirac distributions (37). At null temperature, these distributions become as Heaviside functions, i.e. "step" functions, so that the term to be integrated in (23, 24):

$$\frac{1}{1+\exp\left(\frac{E_f-\mu_f}{T}\right)}-\frac{1}{1+\exp\left(\frac{E_f+\mu_f}{T}\right)}, \tag{61}$$

looks like a rectangular function upon the $E_f=\sqrt{\vec{p}^2+m_f{}^2}$ variable. As a consequence, when $E_f>\mu_f$, (61) is equal to zero. If the chemical density decreases and tends towards $m_f$, the zone where (61) is null increases and finally occupy all the integration domain of the integral (23). At the limit $\mu_f\to m_f$, the integral, i.e. the calculated density, converges toward $0^+$. With the PNJL model, theses explanations are still valid, especially because we saw equation (58) that the modified Fermi-Dirac distributions looks like the classical ones at reduced temperatures.

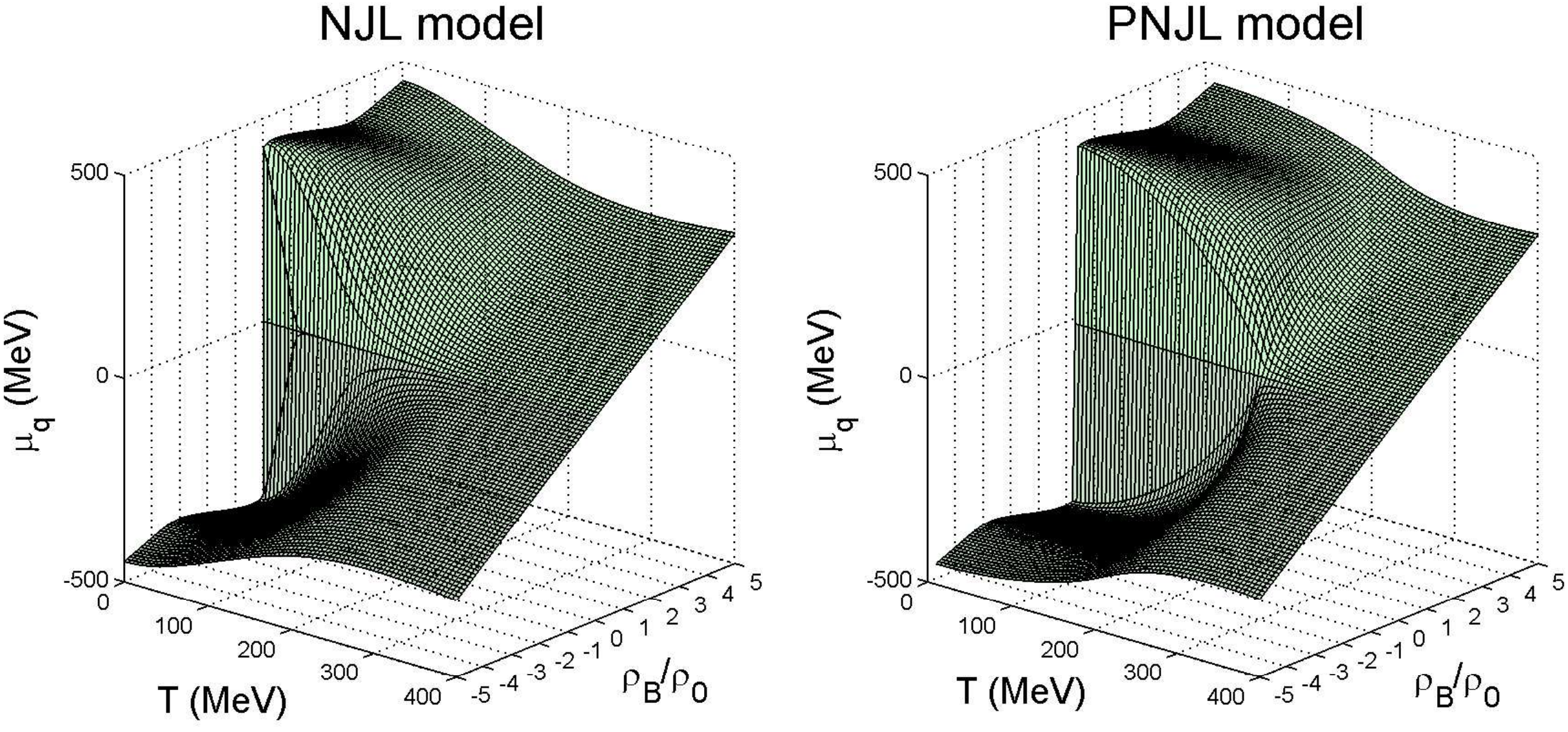

**Figure 8.** Chemical potential $\mu_q$ according to the baryonic density and the temperature.

The results present another particularity. We focus on the zone located at low temperatures, along the zero-temperature axis. For the moment, let us consider only positive baryonic densities. In the NJL approach, during the resolution of (24), a given temperature and baryonic density correspond to one effective light quark mass and one chemical potential $\mu_q$. Nevertheless, the figure 8 indicates that the reciprocal is not true according to $\mu_q$, in the evoked zone. In fact, if $T\approx 0$ and $\rho_B>0$, the function $\mu_q(\rho_B)$ decreases slowly, and then it increases again when $\rho_B$ is growing. As a consequence, $\mu_q(\rho_B)$ is not there a bijection: one chemical potential corresponds in this zone to two different baryonic densities. Furthermore, at reduced temperatures, the chemical potential depends on the baryonic density, but also on the temperature.

On the other hand, with the NJL model, this behavior disappears for a temperature beyond 200 MeV. There, the relation between the baryonic density and the chemical potential tends more and more to a linear behavior. According to figure 8, we find the approximate relation:

$$\mu_q \equiv \mu_u = \mu_d \simeq (62.2 \text{ MeV}) \cdot \frac{\rho_B}{\rho_0} . \tag{62}$$

Of course, this relation is only valid for the parameter set we employed (P1), which respects the isospin symmetry. The relation is also applicable to the PNJL model, but beyond 300 MeV, because of the observed shifting of the PNJL graph towards higher temperatures, leading to an enlargement of the structure described above.

In the figure 8, the graphs were extended to negative densities. We observe that the $\rho_B = 0, \mu_q = 0$ axis corresponds to a symmetry axis, for the two graphs. In the same manner, the discontinuity also exists at reduced temperatures and $\rho_B < 0$:

$$\lim_{\substack{\rho_B \to 0 \\ \rho_B < 0}} \mu_q (T \simeq 0, \rho_B) = -m_q < 0 . \tag{63}$$

Using the thermodynamic definition of the chemical potential, i.e. the required energy to add one particle into the medium, a link between negative densities and antimatter can be made. Indeed, negative densities physically mean that the antimatter dominates the matter. In other words, it was studied there the behavior of quarks plunged in antimatter. In fact, it could be shown that negative chemical potential values do not affect the quarks effective masses. The equation (16) indicates that the masses are calculated while using the chemical potentials only as arguments of the generic function $A$. It was shown equation (13) of the appendix D that this function only considers the chemical potential *absolute value*. In conclusion, for example with figures 4 and 5, we could extend our graphs to negative densities, considering the plane $\rho_B = 0$ as a symmetry plane, for the two figures. The symmetry between matter and antimatter will be used again at several occasions during the modeling of composite particles.

In fact, in the literature, studies at finite $\mu$ are more frequent than at finite densities. As a consequence, it is interesting to see if we can confirm the literature's results from our approach. So, in the left hand side of the figure 9, we plotted the evolution of the chiral condensate and the Polykov loop for several chemical potentials, in the PNJL model. These data appears to be very similar to the ones visible notably in [35, 37]. To establish the link between this graph and the ones performed at finite densities, we represented the used chemical potentials $\mu_q$ in the graph established in the figure 8. We obtained the right hand side of the figure 9. More precisely, for the three $\mu_q$, the associated curve gives the correspondence between the baryonic density and the chemical potential, for each temperature. If this relation is trivial at null chemical potential, it is not the case for the "trajectories" obtained with the three studied $\mu_q$. Then, we used these correspondences in the right hand side of the figure 6 (chiral condensate) and in the left hand side of the figure 7 (Polyakov field). It gives us the figure 10. Firstly, for $\mu_q = 300$ MeV, and by extension for low chemical potentials, the found trajectory is smooth, as the variations of $\langle\langle \bar{\psi}_q \psi_q \rangle\rangle$ and $\Phi$. At this opposite, for $\mu_q = 380$ MeV, the trajectory is different. More precisely, a portion of the curve evolves at constant temperature, close to 150 MeV. It wants to say that at this temperature, several densities correspond to the same chemical potential. With the figure 10, we show that it implies that the order parameters present discontinuities for this temperature, as observable in the left hand side of the figure 9. Similar observations can be made for $\mu_q = 420$ MeV, where the discontinuities are stronger, and occur for a temperature close to 40 MeV.

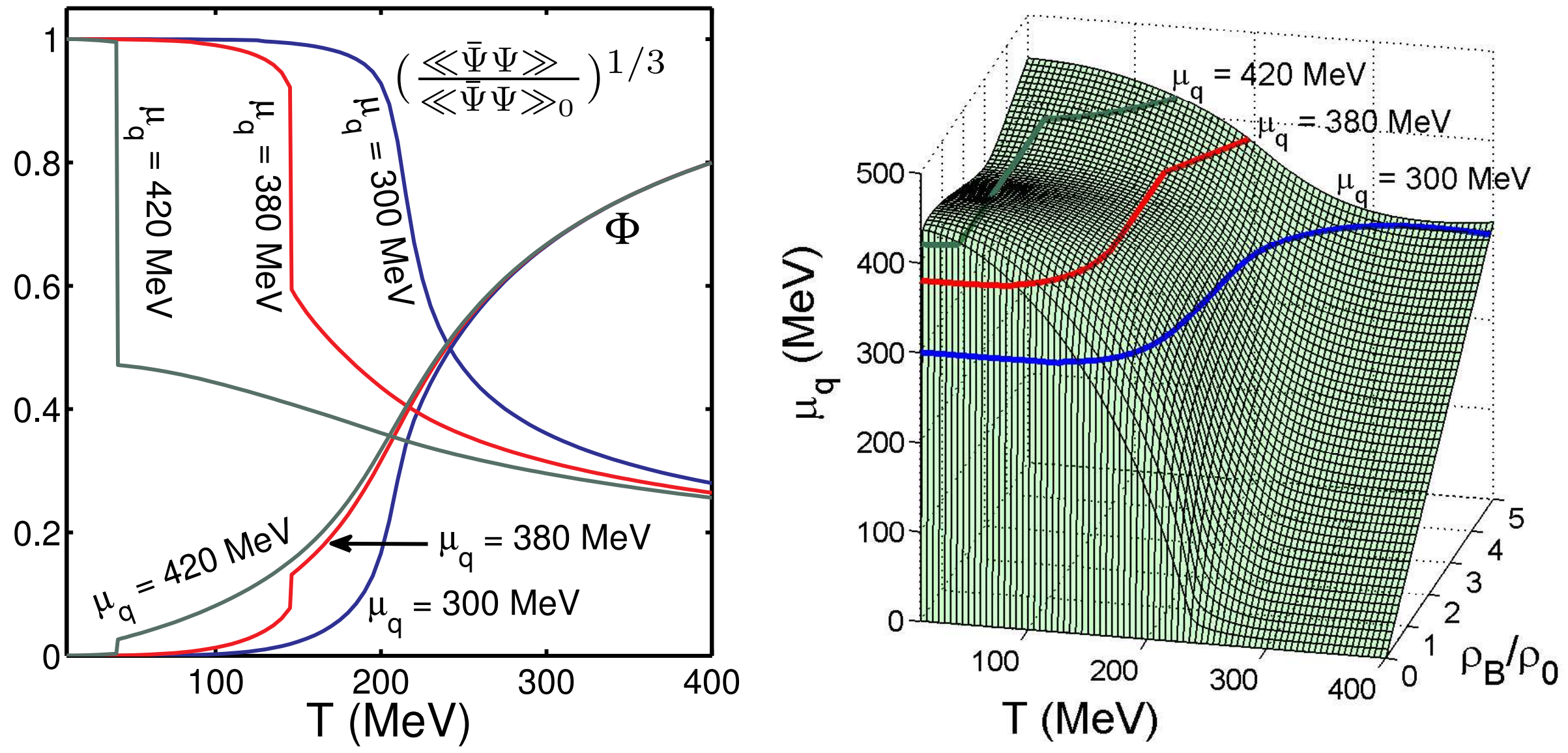

**Figure 9.** Study of the order parameters for several chemical potentials, and associated "trajectories" in the $T-\rho_B$ plane.

Physically, these discontinuities according to the chiral condensate can be interpreted to first order phase transitions [35, 37], neglecting the fact that the curves do not fall to zero after the discontinuity. A possible extension of such a study could be to investigate the critical end point (2$^{nd}$ order) between the crossover and the fist order transition. Calculations at high chemical potentials can also be promising. However, as argued before, these calculations require taking into account the color superconductivity, and an update of the used formalism.

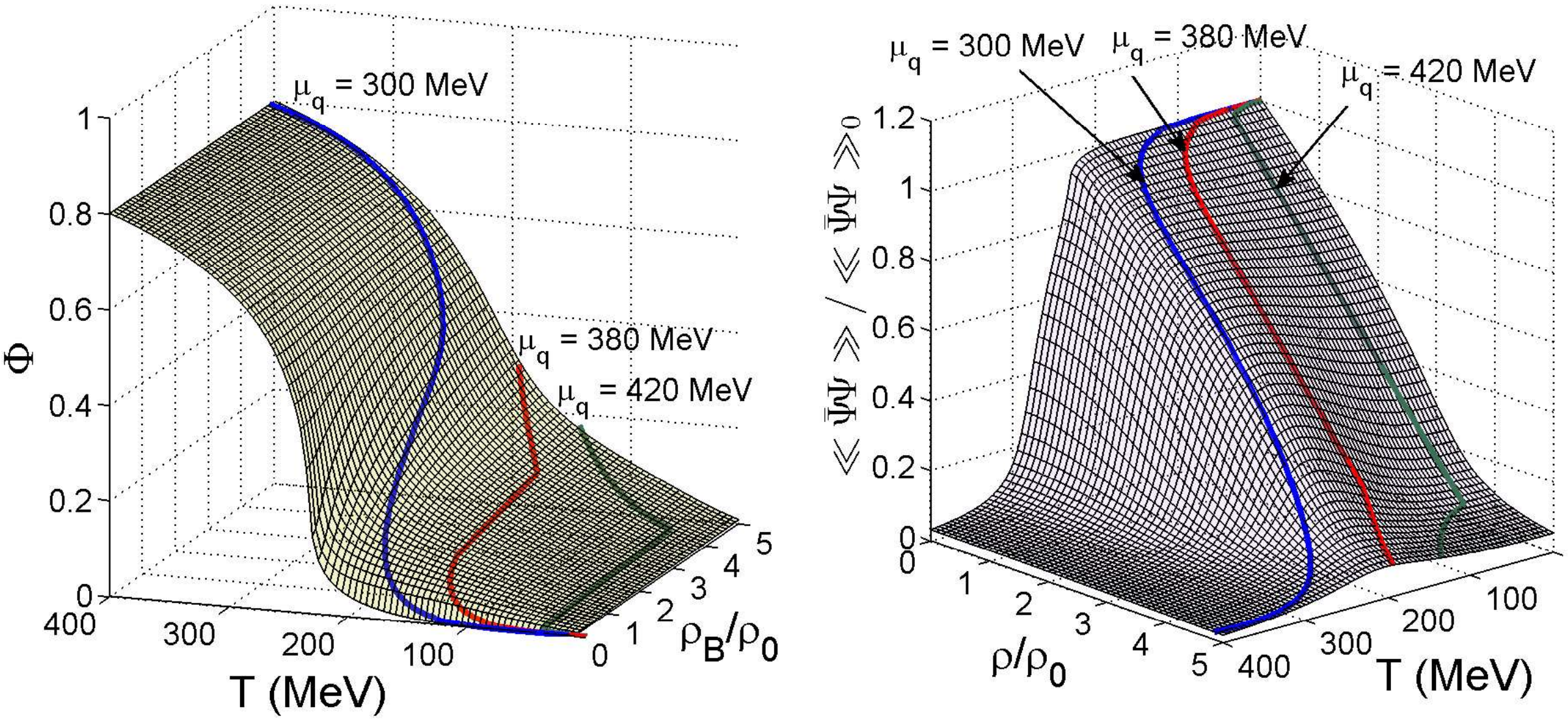

**Figure 10.** Evolution of the order parameters in the $T-\rho_B$ plane, with three fixed chemical potentials.

In the framework of this thesis, we will continue to use the temperature and the densities as study parameters. Indeed, according to a numerical point of view, the observed discontinuities lead to numerical instabilities, unwanted in the framework of a dynamical study. At this opposite, the crossover meets with the density does not present this difficulty.

# 6. Conclusion

In this chapter, we proposed an interesting alternative of the QCD, the NJL model. By the use of “frozen” gluons, this model is usable to study the quarks physics. It was studied the possibility to estimate the masses of effective quarks, by gap equations, and the evolution of these masses according to the temperature and baryonic density. Another interesting aspect of the NJL approach concerns the fact that it shows the restoration of the chiral symmetry, as QCD, at high temperatures and/or high densities. But, because of the absence of confinement, it was studied that the NJL model can be completed, in the framework of the Polyakov Nambu Jona Lasinio model. In this model, the quarks are minimally coupled to a Polyakov loop, whose role is to simulate a mechanism of confinement. The Polyakov loop comes from LQCD pure gauge studies, in which it is used as an order parameter, to describe the phase transition between the “confined” and the “deconfined” phases.

In the numerical results, we showed that the inclusion of the Polyakov loop leads to a shifting of the quarks masses, of the light quark chiral condensate and of the chemical potentials towards higher temperatures. We managed to confirm the results found in the literature, and performed them to the $T, \rho_B$ plane. Such calculations also included the study of the chemical potential of the light quarks in this plane.

# Chapter 3

# Mesons

## 1. Introduction

In the previous chapter, we presented the models that are considered in this thesis, i.e. the NJL and the PNJL models. We saw that these models allow modeling dressed $u,d,s$ quarks, whose masses depend on the temperature and on the baryonic density. In fact, starting from these quarks, one of the first successes of the NJL approach was to model light mesons in a reliable way. Among the NJL references quoted in the previous chapter, we can mention: firstly [1–6], then [7–9], [10] and [11]. These references were followed by other papers that used the performed mesons modeling, as [12–17]. More recently, NJL mesons were considered again, in works as [18–20]. Among the evoked mesons in these references, pseudo-scalar ones are particularly studied. In fact, these mesons are the lightest ones. Moreover, it was experimentally observed a massive production of pseudo-scalar mesons in high energies collisions, notably pions and kaons, e.g. [21]. Also, scalar mesons were also particularly considered in the NJL approach, notably to intervene as propagators in cross-sections calculations [13, 14]. Thanks to the encouraging results encountered with the PNJL model, recent developments of this model included the modeling of PNJL mesons. It mainly concerned pseudo-scalar and scalar mesons [22, 23], even if more exotic mesons were also studied [24]. NJL and PNJL results were compared. As a whole, this work was performed while studying the mesons' mass as a function of the temperature. It was reported [23] that the effect of the inclusion of the Polyakov loop is similar to what was observed for the quarks, i.e. a shift of the mass curve towards higher temperatures, operating a distortion of these curves.

In the framework of the NJL model, the masses of the mesons were estimated according to the temperature, the chemical potential, e.g. [11], or the baryonic density, e.g. [9, 25]. With the PNJL model, the temperature and the chemical potential are also used [22, 23], but not the baryonic density. Moreover, studies in the $T,\rho_B$ plane are rare for the two models, but we can mention [26, 27]. Working in this plane permits to see the complete stability zones of the studied mesons. Furthermore, even if the axial and vectorial mesons were modeled in the first quoted references, the study of these mesons is not frequent in recent works, including in the PNJL model. For these mesons, it could be studied if the agreement with experimental data is correct, as observed for the pseudo-scalar and scalar mesons. Axial mesons are maybe less crucial in the framework of our study, but this remark cannot be true with vectorial ones.

Indeed, the vectorial mesons $\rho$ are important in particle physics. Even if they are heavier than the pions, kaons and $\eta$, they are lighter than the other mesons. Also, they intervene in the pion-pion elastic scattering as propagators [17], and their various decays are particularly studied (decay to a pion pair, to a dilepton, etc.). Also, the works that were previously evoked considers the isospin symmetry. It could be instructive to investigate the modifications if we do not consider this approximation, i.e. if $m_u \neq m_d$. Notably, a question is to see if the agreement with experimental data is better.

In this chapter, these interrogations concerning the mesons are considered, by studying these particles. First, the beginning of the chapter is devoted to present and to explain the equations devoted to the mesons modeling. At this occasion, we indicate the modifications to be done in order to perform the transition from the NJL model towards the PNJL one. These descriptions concern the section 2. The results associated with the pseudo-scalar mesons are presented in section 3. An objective of this section is to recover the results already presented in the literature, i.e. the masses, the widths and the coupling constants of these particles. However, we also consider some aspects less or not treated in the evoked references, as the phase diagram for the *stable* mesons, the study at finite baryonic densities (including negative ones), the $\eta - \eta'$ mixing angle, etc. In section 4, the other mesons are modeled: scalar, vectorial and axial mesons. In these two sections, NJL and PNJL results are compared. In the section 5, we propose to investigate the consequences of the abandon of the isposin symmetry on the obtained results. At this occasion, we insist on the complications induced by such a work for the $\pi_0, \eta, \eta'$ mesons. At the end of this section, a summary table gathers the mesons' masses at null temperature and density. There, a discussion upon the obtained results is proposed.

# 2. Description of the formalism

## 2.1 General method

In the framework of the NJL model, the essential idea is to consider a meson as a quark/antiquark association, forming a loop. This loop is able to reproduces itself ad infinitum. The figure 1 hereafter gives a schematic representation of the applied method [11].

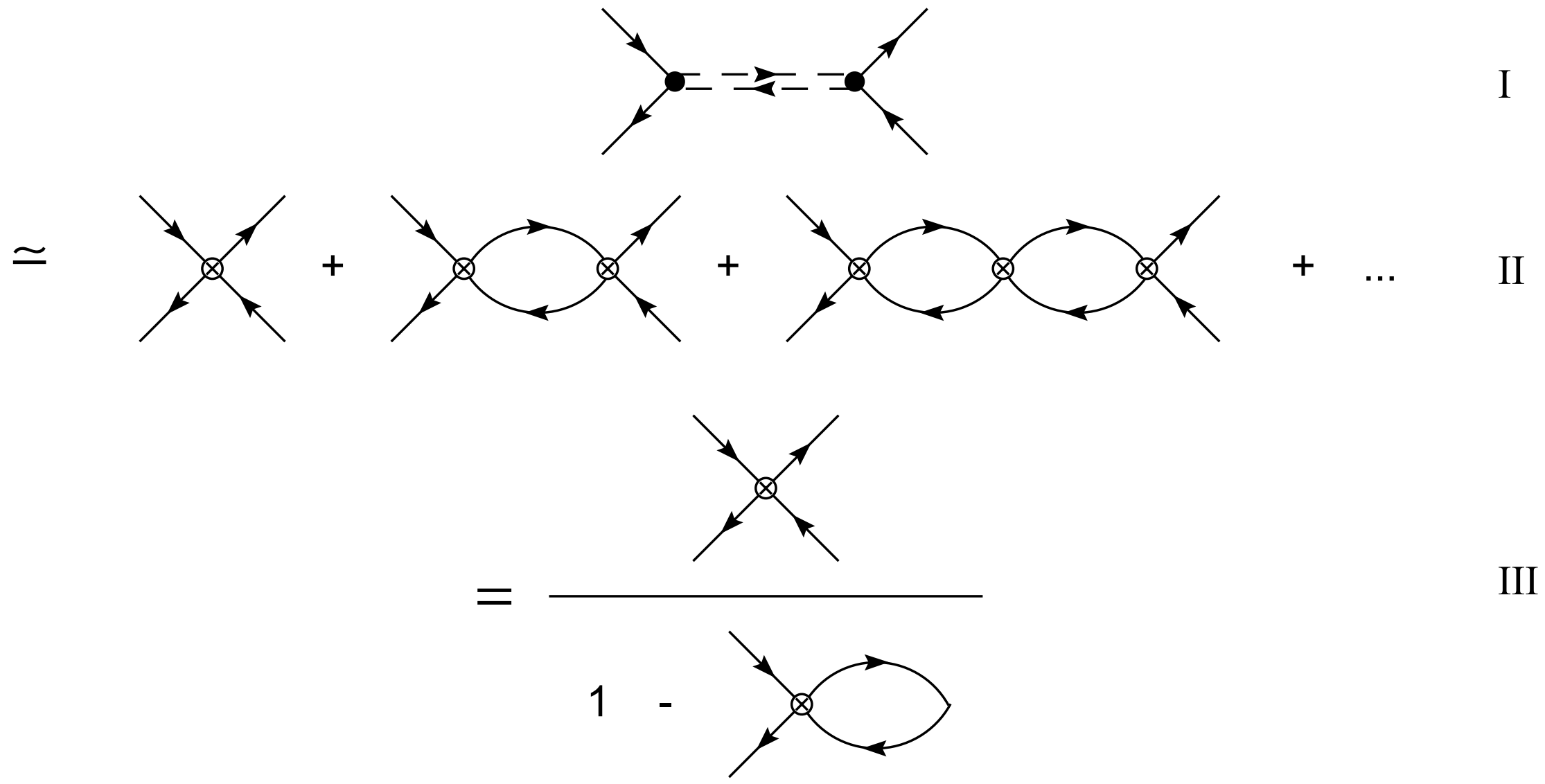

**Figure 1.** Schematization of the approach to treat mesons.

Only the direct term is considered, in the framework of the Random Phase Approximation [32–35], as in [13]. To obtain the mathematical equivalent of figure 1, we take $\otimes \equiv \mathcal{Z}$. It corresponds to an effective coupling at each vertex. For a given quark/antiquark system, this term is a constant. Also, the loop is noted $\Pi$ :

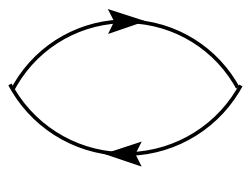

$$\equiv \Pi \quad . \tag{1}$$

This intern loop, made by a quark and by an antiquark, is named irreducible polarization function [11, 13, 15]. The loop is a function of its total four-momentum. In the framework of the Bethe-Salpeter equation, e.g. [8, 9], we identify the interaction in the first line of the figure 1 as the transition matrix T. The Bethe-Salpeter equation gives $\mathrm{T} = \mathcal{Z} + \mathcal{Z} \cdot \Pi \cdot \mathrm{T}$, in which the coupling $\mathcal{Z}$ is associated with a two body interaction kernel [8, 19]. We have then:

$$\mathrm{T} = \mathcal{Z} + \mathcal{Z}\Pi\mathcal{Z} + \mathcal{Z}\Pi\mathcal{Z}\Pi\mathcal{Z} + \mathcal{Z}\Pi\mathcal{Z}\Pi\mathcal{Z}\Pi\mathcal{Z} + \ldots, \tag{2}$$

that models the transition from the first to the second line of the figure. Since $\Pi\mathcal{Z}$ satisfies the property:

$$\lim_{n \to \infty} (\Pi\mathcal{Z})^n = 0, \tag{3}$$

the equation (2) corresponds to a convergent geometrical series, so that the passage from the second line to the third line of the figure 1 is written as:

$$\mathrm{T} = \mathcal{Z} \cdot \left(1 + \Pi\mathcal{Z} + \Pi\mathcal{Z}\Pi\mathcal{Z} + \ldots\right) = \frac{\mathcal{Z}}{1 - \Pi\mathcal{Z}}. \tag{4}$$

The following property is valid for any matrix $A$, if its determinant is not equal to zero,

$$A^{-1} = \frac{1}{\det(A)} \cdot {}^{T}\left(\mathrm{com}(A)\right), \tag{5}$$

where $\mathrm{com}(A)$ indicates the comatrix of $A$ and $^T$ the transposition operation. Applying the relation (5) to (4), it comes:

$$\mathrm{T} = \frac{\mathcal{Z}}{1-\Pi\mathcal{Z}} = \frac{\mathcal{Z}}{\det(1-\Pi\mathcal{Z})} \cdot {}^T\big(\mathrm{com}(1-\Pi\mathcal{Z})\big) \text{ proportional to } \frac{1}{\det(1-\Pi\mathcal{Z})}. \quad (6)$$

If we go back to the first line of the figure 1,

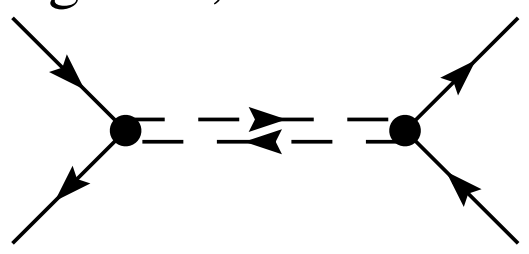

,

we also write, by simple field theory considerations:

$$\mathrm{T} = V^- \cdot \frac{i}{k^2 - m^2} \cdot V^+, \quad (7)$$

where $\frac{1}{k^2 - m^2}$ is the standard propagator of a meson with mass $m$ and four-momentum $k$. Furthermore, $V^-$ and $V^+$ are, respectively, the vertices on the left and right-hand side, materialized by black spots on the figure. These terms will be clarified later, but we just specify that they are non-null and non-infinite numbers. At $k^2 \to m^2$, a divergence is observed in (7). Since (6) and (7) are supposed to be equivalent, this divergence must be found in (6). $\mathcal{Z}$ and $\Pi$ are finite, so the equivalence between the two expressions is satisfied only if:

$$\det(1-\Pi\mathcal{Z}) = 0\Big|_{k^2 \to m^2} . \quad (8)$$

For a meson at rest, its mass is found by solving:

$$\det\left[1-\Pi\left(k_0,\vec{k}\right)\cdot\mathcal{Z}\right] = 0\Big|_{k_0=m,\ \vec{k}=\vec{0}}, \quad (9)$$

whereas for a given momentum $\vec{k}$, the equation is written:

$$\det\left[1-\Pi\left(k_0,\vec{k}\right)\cdot\mathcal{Z}\right] = 0\Big|_{k_0=\sqrt{m^2+\left(\vec{k}\right)^2},\ \vec{k}\text{ fixed}}. \quad (10)$$

Except for the figure 7b, only mesons at rest are used in our numerical calculations. When a meson is considered as stable *by the model* as regards its disintegration into a quark and an antiquark, the equations (9) or (10) to be solved are real, as the obtained mass. Thus, in this regime, the polarization function is a real function. This corresponds to the regime for which the mass of the meson is lower than the sum of the masses of the quark/antiquark that compose it. At the opposite, when this condition is not satisfied, the polarization function is a complex function, and the mass becomes also a complex number, written as:

$$m = m_{\text{physical}} - \frac{i}{2}\cdot\Gamma . \quad (11)$$

$m_{\text{physical}}$ is the real part and is identified to the particle mass, whereas $\Gamma$ forms the imaginary part and is the particle width. A way to understand this behavior consists to say that in a non-relativistic quantum mechanics, the wave function is proportional to:

$$\exp\left(-i\frac{\vec{p}\cdot\vec{x}}{\hbar}\right)=\exp\left(-i\cdot m\cdot\frac{\vec{v}\cdot\vec{x}}{\hbar}\right). \tag{12}$$

If $m$ is a real number, then the exponential keeps a constant modulus, whereas if $m$ is complex, it comes:

$$\begin{aligned}\exp\left(-i\cdot m\cdot\frac{\vec{v}\cdot\vec{x}}{\hbar}\right)&=\exp\left(-i\cdot\left(m_{\text{physical}}-i\frac{\Gamma}{2}\right)\cdot\left(\frac{\vec{v}\cdot\vec{x}}{\hbar}\right)\right)\\&=\exp\left(-i\cdot m_{\text{physical}}\cdot\left(\frac{\vec{v}\cdot\vec{x}}{\hbar}\right)\right)\cdot\exp\left(-\frac{\Gamma}{2}\cdot\left(\frac{\vec{v}\cdot\vec{x}}{\hbar}\right)\right)\end{aligned}. \tag{13}$$

The exponential in the extreme right-hand side of (13) translates the fact that the wave function tends to vanish. It corresponds to the particle instability, indicated by its width $\Gamma$.

## 2.2 Lagrangian associated with the mesons

The relevant part of the NJL Lagrangian that intervenes in the mesons modeling is written as:

$$\begin{aligned}\mathcal{L}=&\sum_{f=u,d,s}\bar{\psi}_f\left(i\partial\!\!\!/-m_{0f}\right)\psi_f\\&+G_S\cdot\sum_{a=0}^{8}\left[\left(\bar{\psi}\lambda^a\psi\right)^2+\left(\bar{\psi}i\gamma_5\lambda^a\psi\right)^2\right]\\&-G_V\cdot\sum_{a=0}^{8}\left[\left(\bar{\psi}\gamma_\mu\lambda^a\psi\right)^2+\left(\bar{\psi}\gamma_\mu i\gamma_5\lambda^a\psi\right)^2\right]\\&-K\cdot\left[\det\left(\bar{\psi}(1+\gamma_5)\psi\right)+\det\left(\bar{\psi}(1-\gamma_5)\psi\right)\right]\end{aligned}. \tag{14}$$

At this stage, a transformation consists to develop the 't Hooft term, i.e. the last term of (14), and to incorporate the obtained sub-terms in the summation $G_S\cdot\sum_{a=0}^{8}\left[\left(\bar{\psi}\lambda^a\psi\right)^2+\left(\bar{\psi}i\gamma_5\lambda^a\psi\right)^2\right]$. In other words, the second and the fourth line of (14) are merged [11, 13], and it comes:

$$
\begin{aligned}
\mathcal{L} = & \sum_{f=u,d,s} \bar{\psi}_f \left( i\partial\!\!\!/ - m_{0f} \right) \psi_f \\
& + \sum_{a=0}^{8} \left[ K_{aa}^{-} \cdot \left( \bar{\psi}\lambda^a \psi \right)^2 + K_{aa}^{+} \cdot \left( \bar{\psi} i\gamma_5 \lambda^a \psi \right)^2 \right] \\
& + K_{30}^{-} \cdot \left( \bar{\psi}\lambda^3 \psi \right) \cdot \left( \bar{\psi}\lambda^0 \psi \right) + K_{30}^{+} \cdot \left( \bar{\psi} i\gamma_5 \lambda^3 \psi \right) \cdot \left( \bar{\psi} i\gamma_5 \lambda^0 \psi \right) \\
& + K_{03}^{-} \cdot \left( \bar{\psi}\lambda^0 \psi \right) \cdot \left( \bar{\psi}\lambda^3 \psi \right) + K_{03}^{+} \cdot \left( \bar{\psi} i\gamma_5 \lambda^0 \psi \right) \cdot \left( \bar{\psi} i\gamma_5 \lambda^3 \psi \right) \\
& + K_{80}^{-} \cdot \left( \bar{\psi}\lambda^8 \psi \right) \cdot \left( \bar{\psi}\lambda^0 \psi \right) + K_{80}^{+} \cdot \left( \bar{\psi} i\gamma_5 \lambda^8 \psi \right) \cdot \left( \bar{\psi} i\gamma_5 \lambda^0 \psi \right) \quad . \\
& + K_{08}^{-} \cdot \left( \bar{\psi}\lambda^0 \psi \right) \cdot \left( \bar{\psi}\lambda^8 \psi \right) + K_{08}^{+} \cdot \left( \bar{\psi} i\gamma_5 \lambda^0 \psi \right) \cdot \left( \bar{\psi} i\gamma_5 \lambda^8 \psi \right) \\
& + K_{83}^{-} \cdot \left( \bar{\psi}\lambda^8 \psi \right) \cdot \left( \bar{\psi}\lambda^3 \psi \right) + K_{83}^{+} \cdot \left( \bar{\psi} i\gamma_5 \lambda^8 \psi \right) \cdot \left( \bar{\psi} i\gamma_5 \lambda^3 \psi \right) \\
& + K_{38}^{-} \cdot \left( \bar{\psi}\lambda^3 \psi \right) \cdot \left( \bar{\psi}\lambda^8 \psi \right) + K_{38}^{+} \cdot \left( \bar{\psi} i\gamma_5 \lambda^3 \psi \right) \cdot \left( \bar{\psi} i\gamma_5 \lambda^8 \psi \right) \\
& - G_V \cdot \sum_{a=0}^{8} \left[ \left( \bar{\psi}\gamma_\mu \lambda^a \psi \right)^2 + \left( \bar{\psi}\gamma_\mu i\gamma_5 \lambda^a \psi \right)^2 \right]
\end{aligned}
\tag{15}
$$

Thus, the term relating to the vectorial channel $-G_V \cdot \sum_{a=0}^{8} \left[ \left( \bar{\psi}\gamma_\mu \lambda^a \psi \right)^2 + \left( \bar{\psi}\gamma_\mu i\gamma_5 \lambda^a \psi \right)^2 \right]$ is not modified. By extension, the vectorial mesons (terms with $\gamma_\mu$) and axial (terms with $\gamma_\mu i\gamma_5$) are not concerned either. The new introduced constants are written as:

$$
\begin{aligned}
& K_{00}^{\pm} = G_S \mp \frac{1}{3} N_C K \cdot \left[ i \cdot Tr\left(S^u\right) + i \cdot Tr\left(S^d\right) + i \cdot Tr\left(S^s\right) \right] \\
& K_{11}^{\pm} = K_{22}^{\pm} = K_{33}^{\pm} = G_S \pm \frac{1}{2} N_C K \cdot \left[ i \cdot Tr\left(S^s\right) \right] \\
& K_{44}^{\pm} = K_{55}^{\pm} = G_S \pm \frac{1}{2} N_C K \cdot \left[ i \cdot Tr\left(S^d\right) \right] \quad , \\
& K_{66}^{\pm} = K_{77}^{\pm} = G_S \pm \frac{1}{2} N_C K \cdot \left[ i \cdot Tr\left(S^u\right) \right] \\
& K_{88}^{\pm} = G_S \pm \frac{1}{6} N_C K \cdot \left[ 2i \cdot Tr\left(S^u\right) + 2i \cdot Tr\left(S^d\right) - i \cdot Tr\left(S^s\right) \right]
\end{aligned}
\tag{16}
$$

and:

$$
\begin{aligned}
& K_{03}^{\pm} = K_{30}^{\pm} = \mp \frac{1}{2\sqrt{6}} N_C K \cdot \left[ i \cdot Tr\left(S^u\right) - i \cdot Tr\left(S^d\right) \right] \\
& K_{08}^{\pm} = K_{80}^{\pm} = \pm \frac{\sqrt{2}}{12} N_C K \cdot \left[ i \cdot Tr\left(S^u\right) + i \cdot Tr\left(S^d\right) - 2i \cdot Tr\left(S^s\right) \right] \quad . \\
& K_{38}^{\pm} = K_{83}^{\pm} = \pm \frac{1}{2\sqrt{3}} N_C K \cdot \left[ i \cdot Tr\left(S^u\right) - i \cdot Tr\left(S^d\right) \right]
\end{aligned}
\tag{17}
$$

The physical quantities used in these equations were defined in the chapter 2. $Tr\left(S_f\right)$ is the trace of the *f* flavor quark propagator $S_f\left(x, x'\right)$ expressed in coordinate space [13–15], with $x' \to x$. This one is related to the quark propagator $S_f\left(p\!\!\!/\right)$ in momentum space, written as:

$$S_{f\ \mathrm{NJL}}\left(\not{p}\right)=\frac{1}{\not{p}+\gamma_0\,\mu_f-m_f}. \tag{18}$$

At this occasion, we indicate that this propagator is rewritten in the PNJL model as [22]:

$$S_{f\ \mathrm{PNJL}}\left(\not{p}\right)=\frac{1}{\not{p}+\gamma_0\left(\mu_f-iA_4\right)-m_f}, \tag{19}$$

that takes into account the dependence on color, via the $A_4$ term.

In the equation (4), according to [8, 11], we have $\mathcal{Z}\equiv 2\cdot K_{ab}^{\pm}$ for the pseudo-scalar mesons (sign +) or scalar mesons (sign –). The choice of the $K_{ab}^{\pm}$ depends on the studied mesons, see table 2 hereafter. Also, $\mathcal{Z}\equiv 2\cdot G_V$ for the axial and vectorial mesons.

## 2.3 Irreducible polarization functions of the mesons

A polarization function $\Pi$, equation (1), can be understood as a two-quark loop. The first has a mass $m_1$, a chemical potential $\mu_1$ and a four-momentum $\left(i\cdot\omega_n,\vec{p}\right)$. About the second quark, its four-momentum is $\left(i\cdot\omega_n-i\cdot\nu_m,\vec{p}-\vec{k}\right)$, see figure 2 [13].

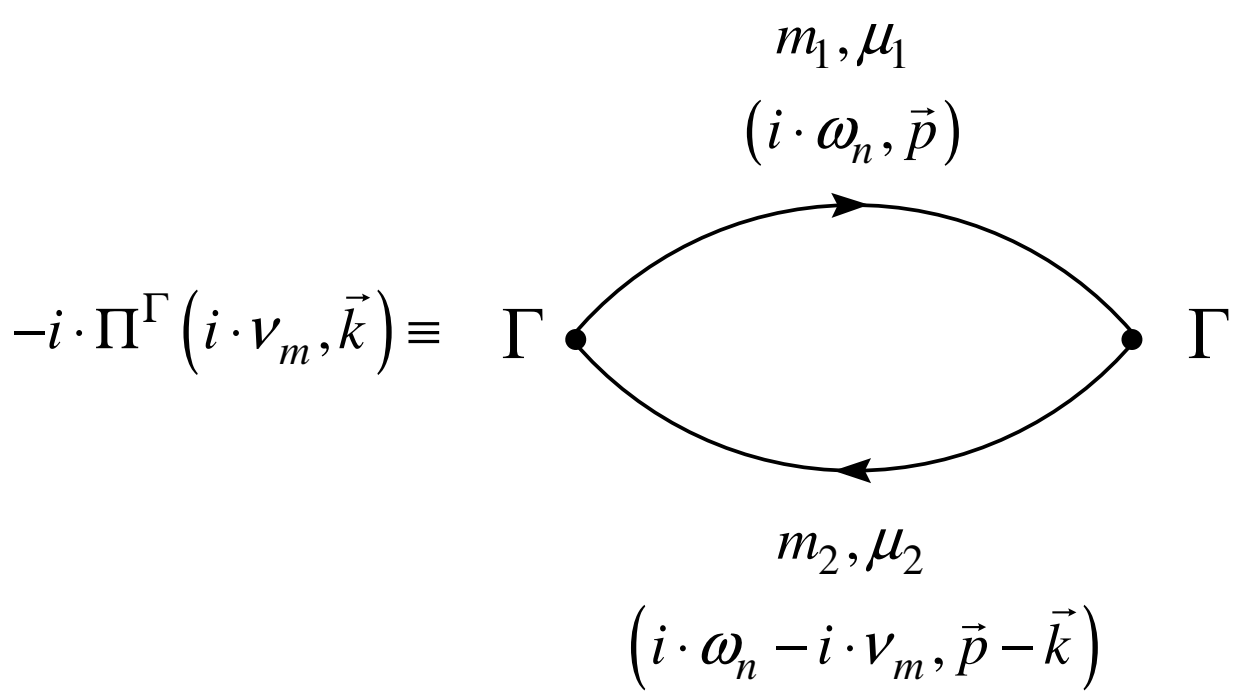

**Figure 2.** Schematic representation of the mesons polarization function.

More precisely, the figure 2 shows that the quark $q_1$ "goes to the right", i.e. towards the future. Therefore, it must be understood as a real quark, whereas the second quark "goes to the left", i.e. towards the past. This one is in fact an antiquark according to the Feynman point of view. Thus, it is noted $\bar{q}_2$, and we write the polarization function as $\Pi_{q_1\bar{q}_2}$. The four-momentum of the loop is $\left(i\cdot\nu_m,\vec{k}\right)$. The $\Gamma$ term in the figure 2 represents the interaction type at each end of the loop, i.e. at each vertex. Four types of interactions are considered. Each of them corresponds to a meson "family", as described in the table 1 [19].

| Considered mesons type | $\Gamma$ value | Corresponding mesons |
|---|---|---|
| pseudo-scalar (P) | $i\gamma^5$ | pion, kaon, $\eta, \eta'$ |
| scalar (S) | 1 | $a_0, K_0^*, f_0, f_0'$ |
| vectorial (V) | $\gamma^\mu$ | $\rho, K^*, \omega, \phi$ |
| axial (A) | $\gamma^\mu \cdot i\gamma^5$ | $a_1, K_1^*, f_1, f_1'$ |

**Table 1.** Values of the $\Gamma$ matrix according to the studied mesons.

The general expression of the meson polarization function is written as [11, 13]:

$$-i \cdot \Pi^{\Gamma}_{q_1\bar{q}_2}\left(i \cdot \nu_m, \vec{k}\right) = -N_C \frac{i}{\beta} \cdot \sum_n \int \frac{d^3 p}{(2\pi)^3} Tr\left(i \cdot S^{f_1}\left(i \cdot \omega_n, \vec{p}\right) \cdot \Gamma \cdot i \cdot S^{f_2}\left(i \cdot \omega_n - i \cdot \nu_m, \vec{p} - \vec{k}\right) \cdot \Gamma\right) , \quad (20)$$

where $S^f\left(\not{p}\right)$ is the flavor *f* quark propagator, defined equations (18, 19) for the NJL and the PNJL models. In the previous chapter, we saw that the inclusion of the Polyakov loop leads to important modifications in the equations to be solved to find the quarks effective masses. However, as argued in [22, 23], the adaptations of the mesons equations consist to replace the classical Fermi-Dirac distributions by the modified ones, equation (41) of chapter 2. In fact, in the numerical calculations, these distributions are used, by the way of generic functions $A$ and $B_0$, in order to estimate (16, 17) and the polarization function (20). More details about these aspects are proposed in the appendix D.

In (20), the Matsubara frequency $i \cdot \omega_n$ of the quark *1* is odd, because the quark is a fermion. This remark is also valid the quark *2*, whose frequency $i \cdot \omega_n - i \cdot \nu_m$ is also odd. The "sum" of these two frequencies gives the Matsubara frequency $i \cdot \nu_m$ of the polarization function. It also represents the frequency of the studied meson. The frequency $i \cdot \nu_m$ is bosonic type: the sum of two odd numbers gives an even number, and of course a meson is a boson.

## 2.4 Clarification of the equations for each meson

For all the mesons quoted in table 2, the equation (4) is written as:

$$M \equiv \mathrm{T} = \frac{2C}{1 - 2C \cdot f \cdot \Pi} . \quad (21)$$

The quantity $M$ is called scattering matrix, as in [9, 13]. It corresponds to the transition matrix T seen previously. Also, it is the meson propagator in the (P)NJL models. The factor *f* added in (21) is a flavor factor. Its value is equal to 2 for all the mesons quoted in table 2. More precisely, we have $\sqrt{2}$ at each vertex, as explained in the appendix C. The choice of $C$ and $\Pi$ depends on the studied meson, as detailed in the table.

| meson | | $C$ | $\Pi$ | meson | | $C$ | $\Pi$ |
|---|---|---|---|---|---|---|---|
| pseudo scalar | $\pi^-$ | ${K_{11}}^+$ | ${\Pi_{d\bar{u}}}^P$ | scalar | $a_0^-$ | ${K_{11}}^-$ | ${\Pi_{d\bar{u}}}^S$ |
| | $\pi^+$ | ${K_{11}}^+$ | ${\Pi_{u\bar{d}}}^P$ | | $a_0^+$ | ${K_{11}}^-$ | ${\Pi_{u\bar{d}}}^S$ |
| | $K^-$ | ${K_{44}}^+$ | ${\Pi_{s\bar{u}}}^P$ | | $K_0^{*-}$ | ${K_{44}}^-$ | ${\Pi_{s\bar{u}}}^S$ |
| | $K^+$ | ${K_{44}}^+$ | ${\Pi_{u\bar{s}}}^P$ | | $K_0^{*+}$ | ${K_{44}}^-$ | ${\Pi_{u\bar{s}}}^S$ |
| | $K^0$ | ${K_{66}}^+$ | ${\Pi_{d\bar{s}}}^P$ | | $K_0^{*0}$ | ${K_{66}}^-$ | ${\Pi_{d\bar{s}}}^S$ |
| | $\bar{K}^0$ | ${K_{66}}^+$ | ${\Pi_{s\bar{d}}}^P$ | | $\bar{K}_0^{*0}$ | ${K_{66}}^-$ | ${\Pi_{s\bar{d}}}^S$ |
| vectorial | $\rho^-$ | $G_V$ | ${\Pi_{d\bar{u}}}^V$ | axial | $a_1^-$ | $G_V$ | ${\Pi_{d\bar{u}}}^A$ |
| | $\rho^+$ | $G_V$ | ${\Pi_{u\bar{d}}}^V$ | | $a_1^+$ | $G_V$ | ${\Pi_{u\bar{d}}}^A$ |
| | $\phi$ | $G_V$ | ${\Pi_{s\bar{s}}}^V$ | | $f_1'$ | $G_V$ | ${\Pi_{s\bar{s}}}^A$ |
| | $K^{*-}$ | $G_V$ | ${\Pi_{s\bar{u}}}^V$ | | $K_1^{*-}$ | $G_V$ | ${\Pi_{s\bar{u}}}^A$ |
| | $K^{*+}$ | $G_V$ | ${\Pi_{u\bar{s}}}^V$ | | $K_1^{*+}$ | $G_V$ | ${\Pi_{u\bar{s}}}^A$ |
| | $K^{*0}$ | $G_V$ | ${\Pi_{d\bar{s}}}^V$ | | $K_1^{*0}$ | $G_V$ | ${\Pi_{d\bar{s}}}^A$ |
| | $\bar{K}^{*0}$ | $G_V$ | ${\Pi_{s\bar{d}}}^V$ | | $\bar{K}_1^{*0}$ | $G_V$ | ${\Pi_{s\bar{d}}}^A$ |

**Table 2.** C and $\Pi$ for each meson.

For these mesons, the equation (8) is rewritten by the form of a scalar expression [13, 15]:

$$1-4\cdot C\cdot\Pi\left(k_0,\vec{k}\right)=0\Big|_{k_0=m,\ \vec{k}=\vec{0}} \quad . \tag{22}$$

The other $SU(3)_f$ mesons not present in the table 2 are the coupled ones. These mesons are $\pi^0,\eta,\eta'$ and their scalar, vectorial and axial equivalents, when they exist. However, if the isospin symmetry is applied, $\pi^0$ is uncoupled from $\eta,\eta'$. In this case, the propagator of $\pi^0$ is identical to those of $\pi^\pm$. But, $\eta,\eta'$ still require a specific treatment. In the following subsection, we propose to reproduce the method explained in [13]. In the end of this chapter, a complementary study extends this method beyond the isospin symmetry.

## 2.5 $\eta-\eta'$ and $f_0-f_0'$ mesons (with the isospin symmetry)

The $\eta,\eta'$ mesons are coupled because of a mixture of the isospin channels $\lambda^0$ and $\lambda^8$. The scattering amplitude $M$ is here a non-diagonal matrix. It is written in the form:

$$M=\begin{bmatrix} M_{00} & M_{08} \\ M_{80} & M_{88} \end{bmatrix}=2K^+\left(1-2\Pi^P K^+\right)^{-1} \quad , \tag{23}$$

with:

$$K^{+}=\begin{bmatrix} K_{00}^{+} & K_{08}^{+} \\ K_{80}^{+} & K_{88}^{+} \end{bmatrix} \quad \text{and} \quad \Pi^{P}=\begin{bmatrix} \frac{2}{3}\cdot\left(2\Pi_{q\bar{q}}^{P}+\Pi_{s\bar{s}}^{P}\right) & \frac{2\sqrt{2}}{3}\cdot\left(\Pi_{q\bar{q}}^{P}-\Pi_{s\bar{s}}^{P}\right) \\ \frac{2\sqrt{2}}{3}\cdot\left(\Pi_{q\bar{q}}^{P}-\Pi_{s\bar{s}}^{P}\right) & \frac{2}{3}\cdot\left(\Pi_{q\bar{q}}^{P}+2\Pi_{s\bar{s}}^{P}\right) \end{bmatrix}. \tag{24}$$

$\Pi_{q\bar{q}}^{P}$ and $\Pi_{s\bar{s}}^{P}$ are polarization functions of pseudo-scalar mesons, respectively for a two-quark $q$ loop and a two-quark $s$ loop. Here, the use of the isospin symmetry enables to have $K_{08}^{+}=K_{80}^{+}$ and $M_{08}=M_{80}$. For the other mesons, we obtained the masses by searching the pole of $M$. Here, we calculate $M^{-1}$, we diagonalize it, and we solve the uncoupled equations system, in which each eigenvalue is posed as equal to 0. It gives:

$$M^{-1}=\frac{1}{2\cdot\det\left(K^{+}\right)}\cdot\begin{bmatrix} \mathcal{A} & \mathcal{B} \\ \mathcal{B} & \mathcal{C} \end{bmatrix}, \tag{25}$$

with:

$$\begin{cases} \mathcal{A}=K_{8}^{+}-\frac{4}{3}\det\left(K^{+}\right)\cdot\left(2\Pi_{q\bar{q}}^{P}+\Pi_{s\bar{s}}^{P}\right) \\ \mathcal{B}=-K_{08}^{+}-\frac{4\sqrt{2}}{3}\det\left(K^{+}\right)\cdot\left(\Pi_{q\bar{q}}^{P}-\Pi_{s\bar{s}}^{P}\right) \\ \mathcal{C}=K_{0}^{+}-\frac{4}{3}\det\left(K^{+}\right)\cdot\left(\Pi_{q\bar{q}}^{P}+2\Pi_{s\bar{s}}^{P}\right) \end{cases}. \tag{26}$$

It comes:

$$M^{-1}\sim\frac{1}{4\cdot\det\left(K^{+}\right)}\cdot\begin{bmatrix} M_{\eta}^{-1} & 0 \\ 0 & M_{\eta'}^{-1} \end{bmatrix}, \quad \text{with} \quad \begin{cases} M_{\eta}^{-1}=\mathcal{A}+\mathcal{C}-\sqrt{\left(\mathcal{A}-\mathcal{C}\right)^{2}+4\mathcal{B}^{2}} \\ M_{\eta'}^{-1}=\mathcal{A}+\mathcal{C}+\sqrt{\left(\mathcal{A}-\mathcal{C}\right)^{2}+4\mathcal{B}^{2}} \end{cases}. \tag{27}$$

We had “uncoupled” $\eta$ and $\eta'$. It remains to solve them: if the particles are at rest, we have:

$$\begin{cases} M_{\eta}^{-1}\left(m_{\eta},\vec{0}\right)=0 \\ M_{\eta'}^{-1}\left(m_{\eta'},\vec{0}\right)=0 \end{cases}. \tag{28}$$

For the scalar mesons $f_0$ and $f_0'$, the method is strictly identical. We only need to replace the pseudo-scalar polarization functions by the scalar ones and the $K^{+}$ by $K^{-}$.

# 3. Results for the pseudo-scalar mesons

## 3.1 Obtained masses

The evolution of the pion, kaon, $\eta$ and $\eta'$ masses according to the temperature is presented in the figure 3. About the NJL curves, the obtained data are in agreement with the ones of the reference [9], as with the ones of [13, 15] and [25–27]. Also, the PNJL data are also in agreement with the corresponding literature [22, 23]. In fact, our PNJL results resemble more to the ones of [23], compared to [22], because of the choice of the effective potential, see chapter 2. As in the other results presented hereafter that use the isospin symmetry, these results were found using the P1 parameter set [19]. It leads to a degeneracy of the $\pi^+$, $\pi^-$, $\pi^0$ mesons on one side (pion curve), and $K^-$, $K^+$, $K^0$, $\bar{K}^0$ mesons on the other side (kaon curve).

Also, as observed in the references, these mesons as found as stable at reduced temperatures, except for the $\eta'$. Indeed, this affirmation is confirmed by the figure 4, in which we represented the widths of the mesons. As explained in (11), the width is associated with the complex part of the mass. When the width is non-null, it reveals that the meson is unstable as regards its disintegration into a quark and an antiquark. When the mass of the meson is equal to the mass of the quark/antiquark pair that constitutes this particle, it corresponds to the temperature known as "Mott temperature", or critical temperature. This temperature marks the frontier between the stability and instability of the meson. As reported in the PNJL literature [22, 23], the inclusion of the Polyakov loop leads to a global increasing of this critical temperature. About $\eta'$, it was found that this meson is unstable, even at null temperature. In addition, we note a brutal halt of the curve, for the two models, that corresponds to a cancelling of the width of this particle. There, the equations associated with this particle become unstable numerically. Some results can be obtained to complete the curve, but the convergence of the equations becomes unreliable. Even if [13] evoked a manifestation of the leak of confinement, it can be observed that the PNJL model does not modify the behavior of this meson.

In the figure 5, the masses of the pseudo-scalar mesons are studied, at null temperature, according to the baryonic density. As explained in the previous chapter, the NJL and the PNJL models coincide at null temperature, whatever the density. It is why we do not specify if the results were found with the NJL model or with the PNJL one. Such studies were performed for example in [19] (left hand side of the figure), even if this reference does not mention the partial removing of degeneracy observed for the kaons. This phenomenon is visible on the left hand side of the figure 5 by the two curves associated with kaons. In fact, ${\Pi_{q\bar{q}}}^P$ is invariant by the exchange of the pair quark/antiquark (for example $u\bar{d} \Leftrightarrow d\bar{u}$), even at non-null baryonic densities. However, ${\Pi_{s\bar{q}}}^P$ (corresponding to $K^-$ and $\bar{K}^0$) is equal to ${\Pi_{q\bar{s}}}^P$ (for $K^+$ and $K^0$) only when $\mu_q = 0$, i.e. when $\rho_B = 0$. We recall that the strangeness density is always fixed to zero in this work. At non-null baryonic densities, ${\Pi_{s\bar{q}}}^P \neq {\Pi_{q\bar{s}}}^P$, and it leads to the observed splitting: the kaons $K^+$ and $K^0$ form the $K^+$ branch, whereas $K^-$ and $\bar{K}^0$ form the $K^-$ branch.

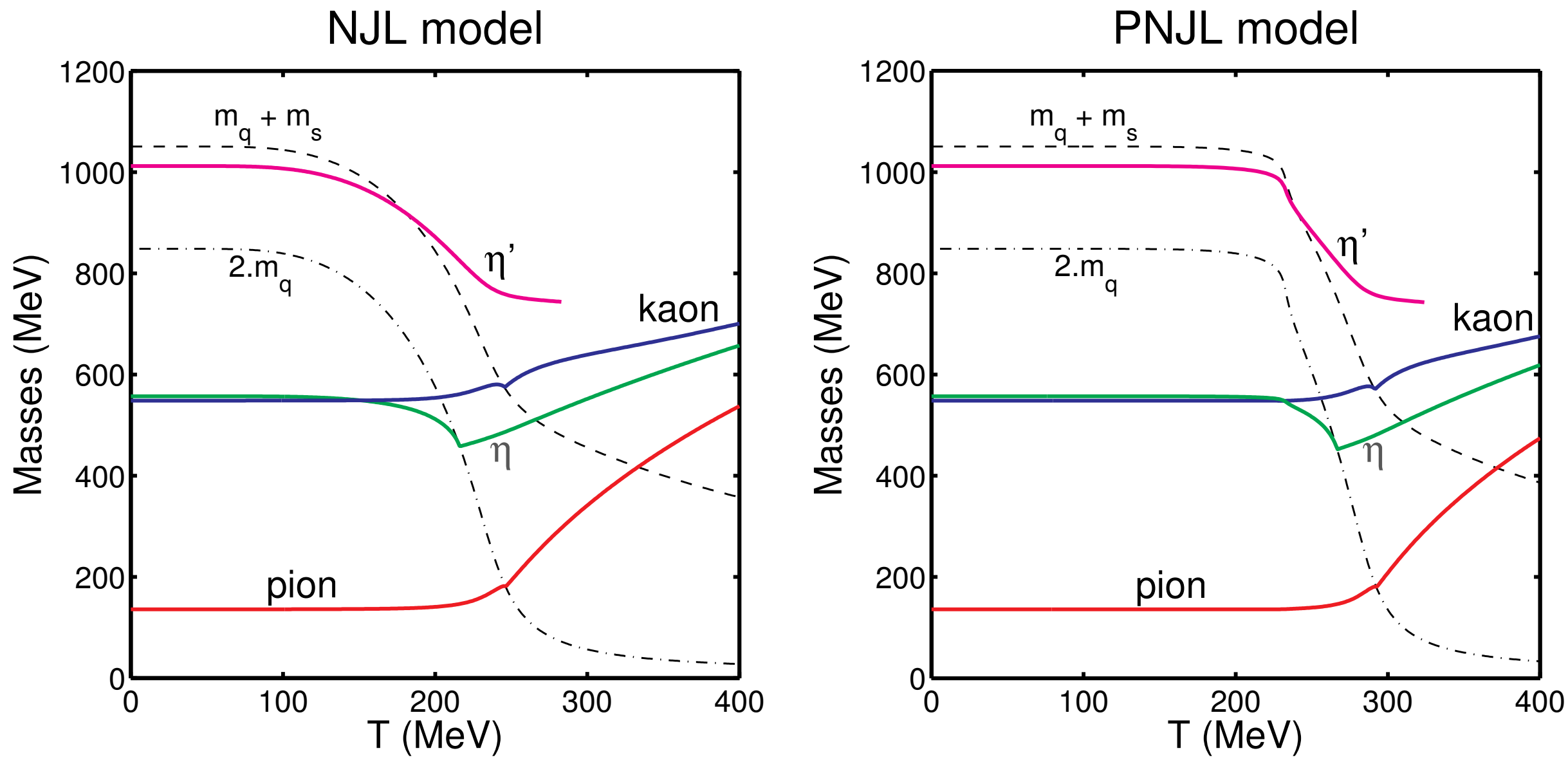

**Figure 3.** Masses of the pseudo-scalar mesons function of the temperature, at null density.

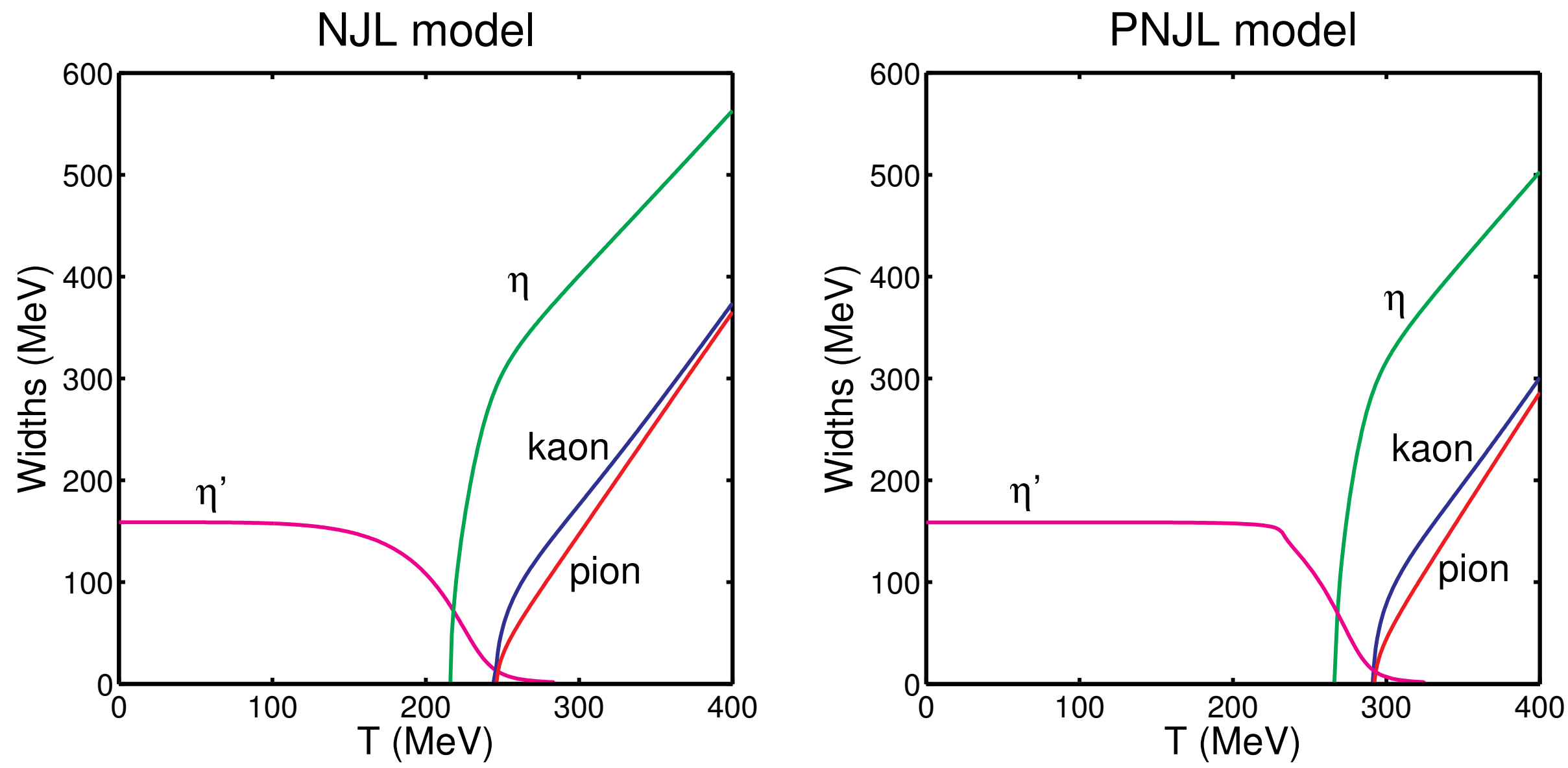

**Figure 4.** Widths of pseudo scalar mesons function of the temperature, at null density.

In addition, in the figure 5, except for the $\eta'$, only the kaons $K^-$ present a transition stable/unstable according to the density. It is materialized in the right hand side of the figure 5 by the fact that the widths of these kaons are null until $\rho_B \approx 3.8\rho_0$. About the pion, kaons $K^+$ and $\eta$, their masses increase with the baryonic density. Then, they become higher than the mass of the quarks/antiquarks that constitute them, but their widths stay null. This apparent strange behavior is in fact explained by the figure 6, in which an "NJL phase diagram" is built for the studied mesons, named there a "diagram of stability/instability". At null baryonic density, along the temperature axis, the critical temperatures found in the left hand side of the figure 3 can be observable, when the curves cross the temperature axis. At the opposite, along the baryonic density axis, only the kaons $K^-$ curve cross this axis. The other curves diverge.

For them, it suggests a soft transition between their stability and instability phases, i.e. a "cross-over transition". About the figure 6, this diagram was performed in the framework of the NJL model, but not in the PNJL one. However, thanks to the results obtained in the figures 4 and 5, we guess that the inclusion of the Polyakov loop should deform the graphs to extend the stability zone of each meson towards higher temperatures.

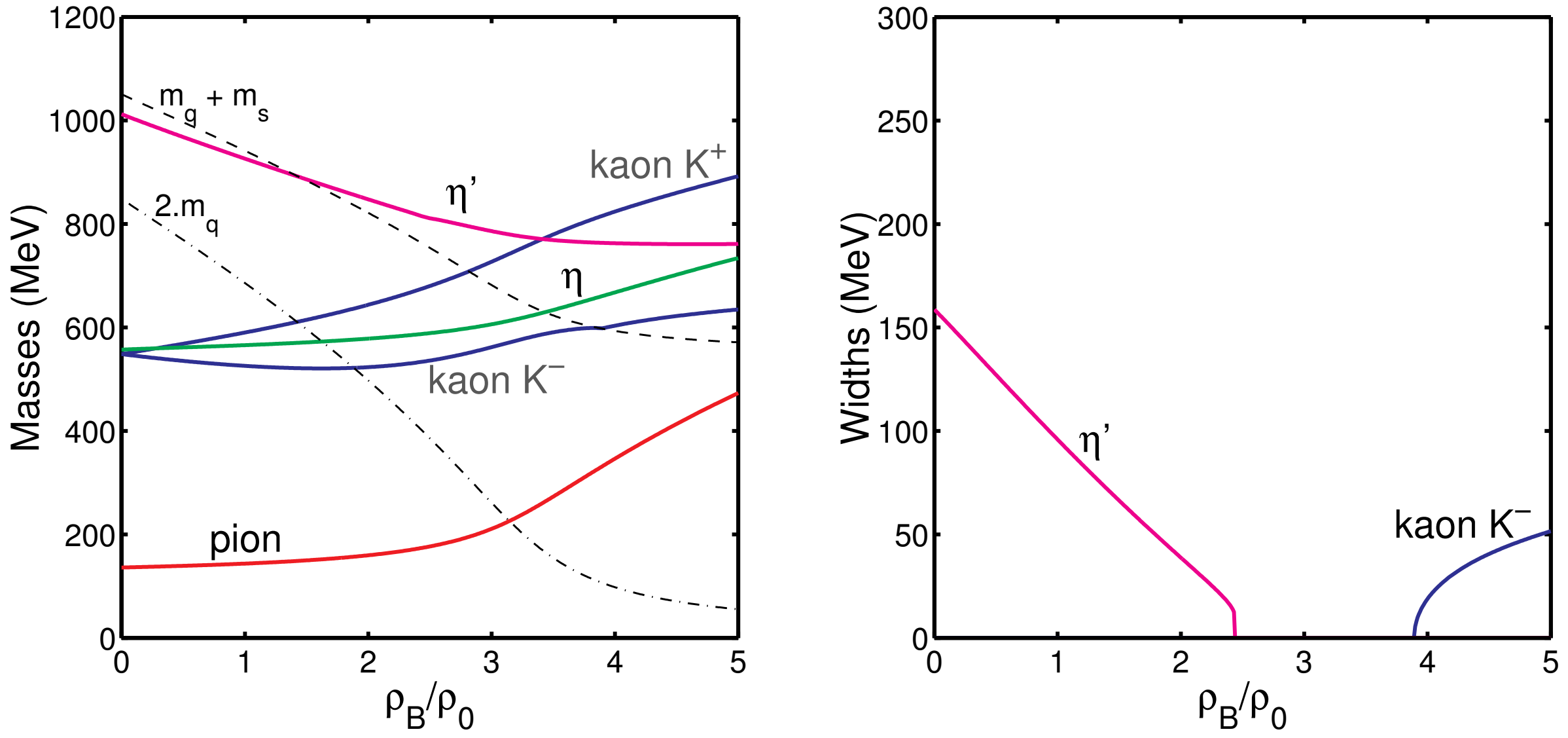

**Figure 5.** Masses and widths of the pseudo-scalar mesons function of the baryonic density, at $T = 0$ .

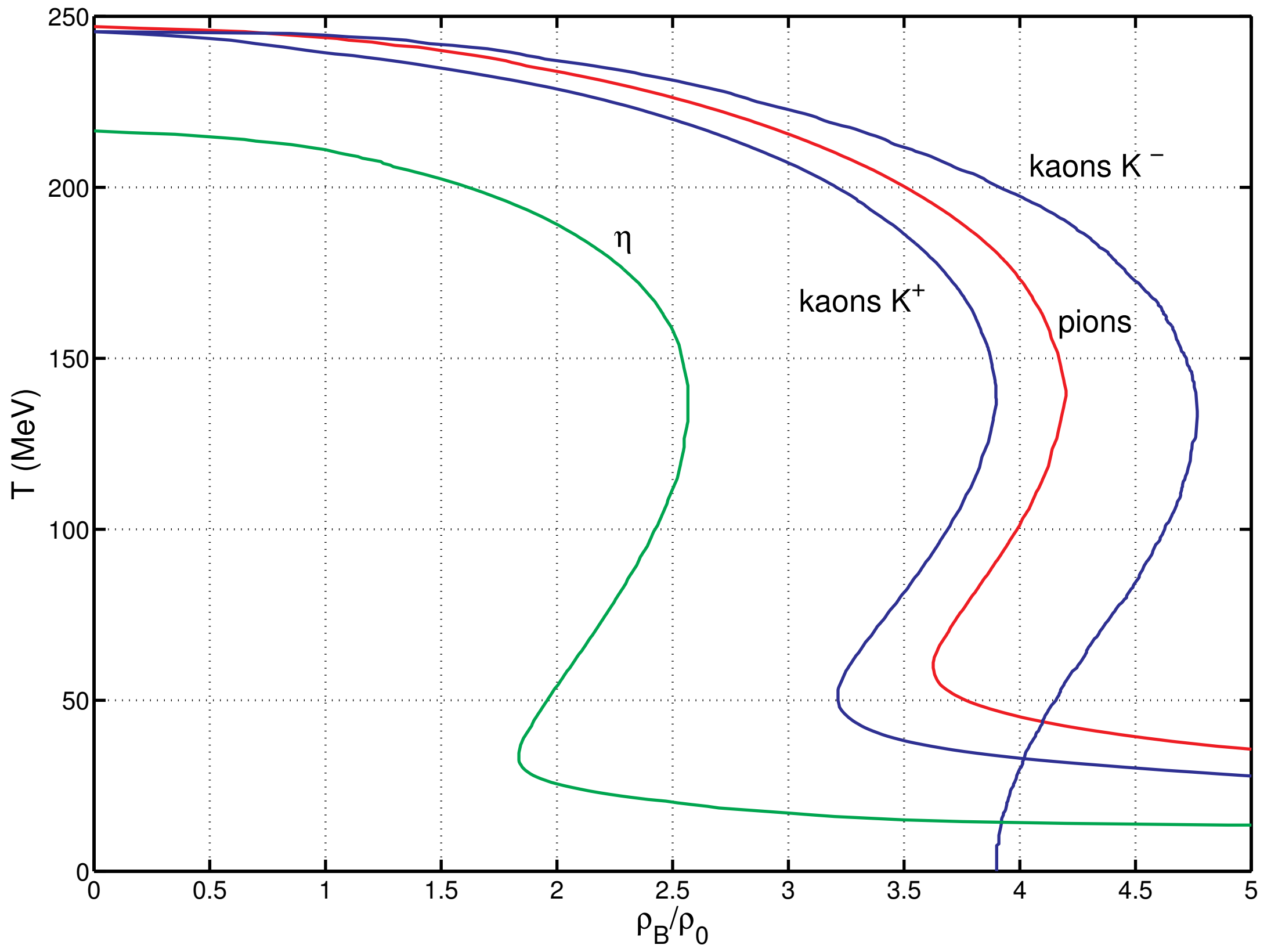

**figure 6.** NJL diagram of stability/instability for the pseudo-scalar mesons; these particles are stables "inside the curve" (in the zone containing the $(0;0)$ point), unstable outside.

The behavior of the mesons was studied at negative baryonic densities. The results were reported in the figure 7a, i.e. in the left hand side of figure 7. As explained in the previous chapter, negative baryonic densities imply that the anti-matter dominates the matter. The obtained results confirm what can be expected, i.e. the $\rho_B = 0$ axis is a symmetry axis of the curves, except for the kaons. For these ones, the passage from positive to negative densities exchanges the $K^+$ and the $K^-$ curves. But, the negative part of the $K^-$ is symmetrical compared to the positive part of the $K^+$, and conversely. A possible application of the results of the figure is to check the validity of the numerical calculations. An asymmetry or a discontinuity of the curves, especially at null density, would be an anomaly.

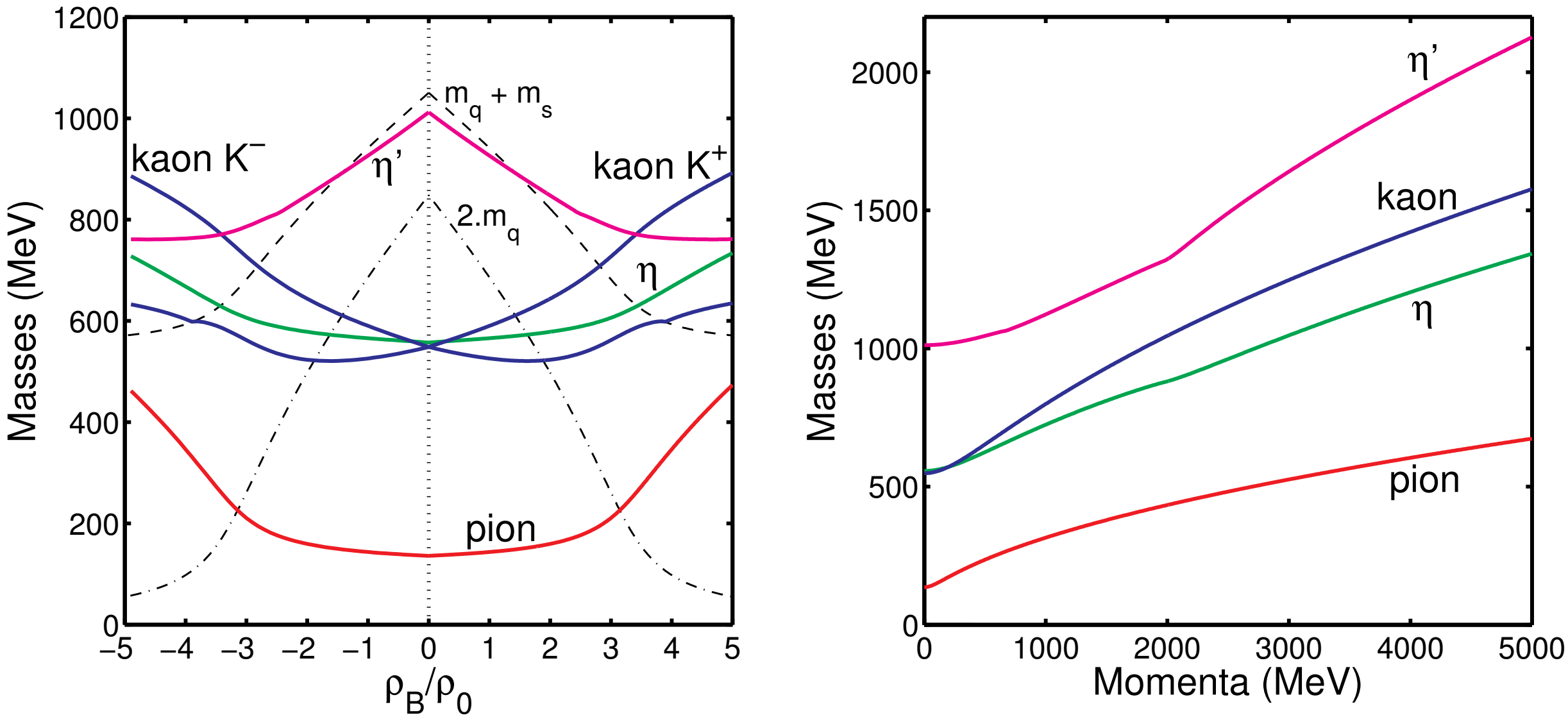

**Figure 7a. (left hand side)** Masses of the pseudo-scalar mesons, according to the baryonic density.
**Figure 7b. (right hand side)** Influence of the momentum on the masses of the pseudo-scalar mesons.

In the figure 7b, the dependence of the mesons' masses concerning their momenta is investigated. More precisely, in this case, the equation (10) is used, instead of (9). It was found that the general tendency is the obtained mass increases when the momentum is growing up. In practice, this dependence is systematically not considered in the (P)NJL works. As observed in the figure, this approximation is justified at reduced momenta.

## 3.2 Coupling constants

A coupling constant is a quantity describing the coupling of a meson with the quark and antiquark that constitute it, at the level of a vertex [11]. This notion is fully usable during cross sections calculations, as in the chapter 6 of this thesis. The $V^-$ and $V^+$ introduced equation (7) are coupling constants. In the literature, this quantity is usually noted $g$. Inspiring us from [13], we take again the expressions of the propagator considered equations (7) and (4, 21). Then, close to their pole, i.e. for $k^2 \to m^2$, we have the relation:

$$\left.\frac{2K}{1-4K\cdot\Pi\left(k_0,\vec{k}\right)}\right|_{k^2=m^2} \approx \left.\frac{-g^2}{k^2-m^2}\right|_{k^2=m^2}, \tag{29}$$

from which we deduce:

$$\left.\frac{1-4K\cdot\Pi\left(k_0,\vec{k}\right)}{2K}\right|_{k^2=m^2} \approx \left.\frac{k^2-m^2}{-g^2}\right|_{k^2=m^2}. \tag{30}$$

By deriving according to $k$, it comes:

$$-2\cdot\left.\frac{\partial\Pi\left(k_0,\vec{k}\right)}{\partial k}\right|_{k^2=m^2} \approx \left.\frac{2k}{-g^2}\right|_{k^2=m^2}. \tag{31}$$

Imposing $k^2=m^2$ is equivalent to write $\begin{cases} k_0 = m \\ \vec{k} = \vec{0} \end{cases}$. So, for a meson at rest, we obtain:

$$g = \sqrt{\frac{m}{\left.\frac{\partial\Pi\left(k_0,\vec{0}\right)}{\partial k_0}\right|_{k_0=m}}}. \tag{32}$$

The equation (32) is still valid in the PNJL model. Only the $\Pi$ loop function should be adapted, at the level of the Fermi-Dirac distributions, as argued before. This formula is applicable to the kaons and to the pions, i.e. to the pseudo-scalar mesons described by the relation (21). For the particles $\eta$ and $\eta'$, the method is more delicate, because of the coupling between these two particles. The process to be used is detailed in [11, 13]. We have previously seen that the mass is *complex* when the particle becomes unstable. This remark is also valid for the polarization function $\Pi$. Clearly, $g$ is a real number when the particle is stable, otherwise $g$ is complex. But, in practice, only the square modulus $|g|^2$ is used in the calculations. As a consequence, the arbitrary choice of the $g$ sign that we implicitly made in (32), in front of the square root, is not important.

The figures 8 and 9 show the behavior of the coupling constants' modulus associated with the pseudo-scalar mesons $\pi, K, \eta$. In the figure 8, their evolution according to the temperature is presented. The results associated with the NJL model are in agreement with references as [13], and the ones associated with the PNJL model qualitatively confirms what was shown in [23]. For the two models, we confirm that the evolution of the coupling constants is reduced for low temperatures. But, a brutal fall of the found values is observed for all the studied curves. In fact, the temperature for which $g \to 0$ corresponds to the critical temperatures of the studied meson. At this critical temperature, the meson has a null binding energy, i.e. the meson's mass is equal to the mass of the quark-antiquark pair that constitutes it. We showed figure 3 that the inclusion of the Polyakov loop leads to an increasing of the critical temperature. So, it explains in figure 8 the observed shifting of the curves between the NJL and the PNJL models. After the critical temperature, the mesons' masses, and by extension the coupling constants, becomes complex, because the mesons are now unstable. Except for $\left|g_{\eta-s\bar{s}}\right|$, it corresponds to an increasing of the found values.

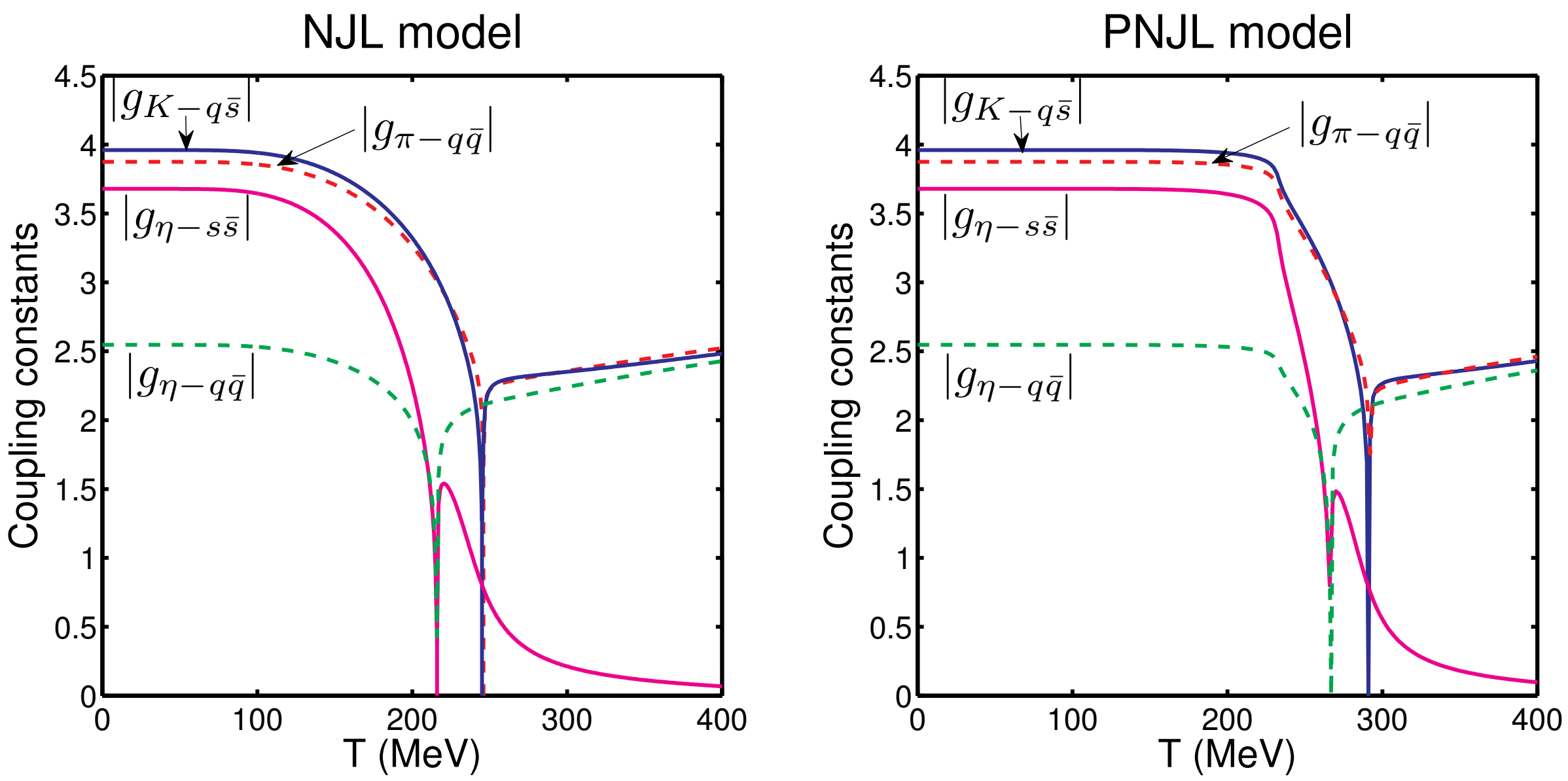

**Figure 8.** Coupling constants of pseudo scalar mesons according to the temperature.

The figure 9 shows the evolution of the coupling constants according to the baryonic density, at null temperature. Such a work was proposed for example in [19, 27], but the first reference does not include results associated with the $\eta$ meson and the $K^-$ curve. In fact, as in figure 5, the $K^-$ curve is the only one that present the same behavior as the one observed according to the temperature, i.e. the brutal decreasing $g \to 0$. About the other curves, the observed variations can suggest a soft transition between their stability and instability regime, as proposed before.

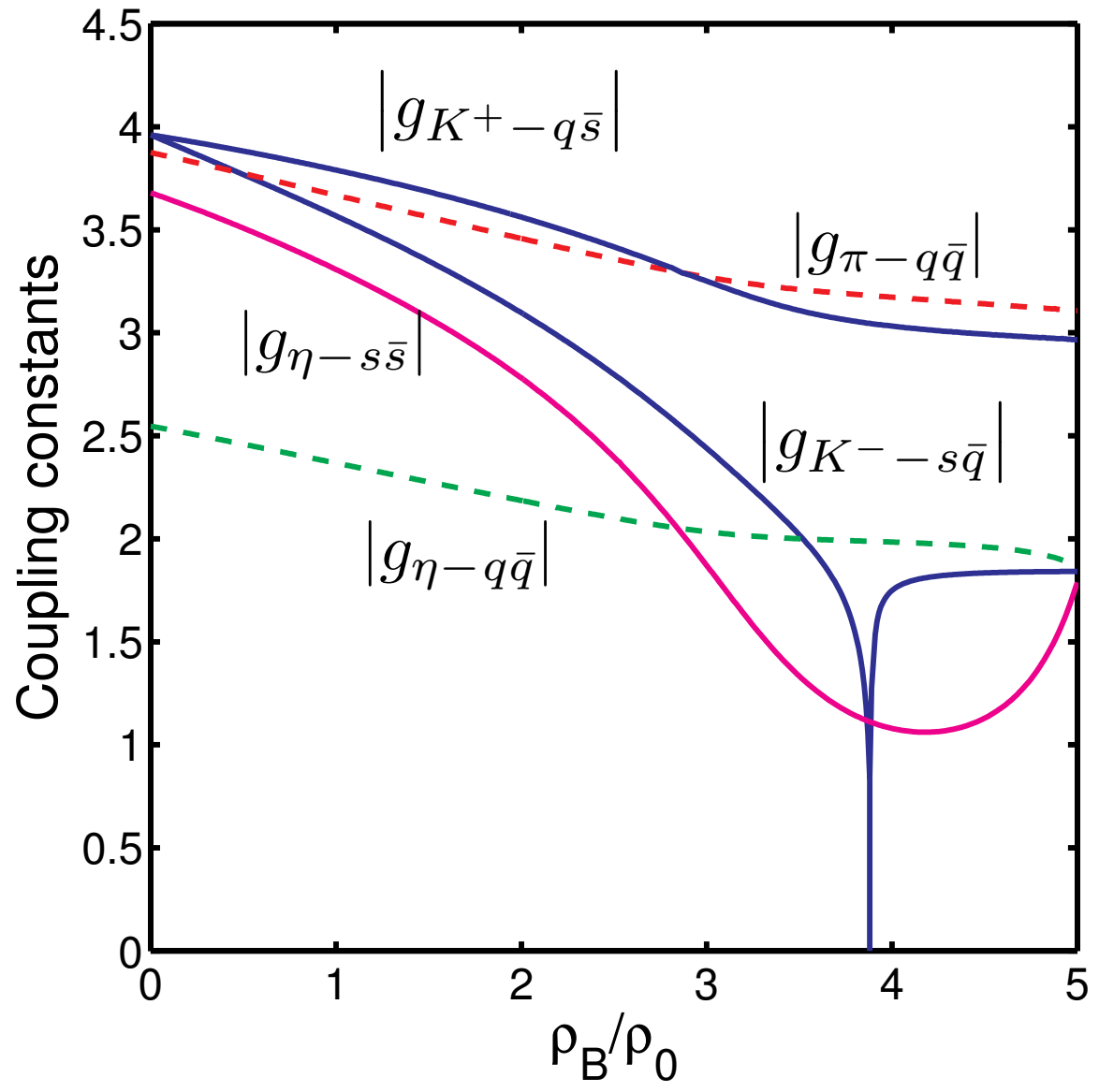

**Figure 9.** Coupling constants of pseudo scalar mesons according to the baryonic density.

# 3.3 $\eta-\eta'$ mixing angle

The Nambu and Jona-Lasinio model, and extension the PNJL one, allows calculating the mixing angle $\eta-\eta'$ [8, 25, 26, 28]. This angle is intrinsic for these models. In other words, it does not require the addition of external data. As explained in the appendix C, this angle is used to estimate the flavor factors involving $\eta$ or $\eta'$. The method usable to estimate the mixing angle $\theta$ is detailed in [11]. Using the equations (23, 24), it consists to consider the quantity:

$$a_\eta = \left.\frac{M_{08}}{M_{00}}\right|_{k^2=m_\eta{}^2} \quad . \tag{33}$$

The angle $\theta$ is connected to $a_\eta$ by the relation:

$$\tan(\theta) = -\frac{1}{a_\eta} \quad . \tag{34}$$

Our numerical results are reported in the figure 10. For all of them, we note that even if $M_{08}$, $M_{00}$ and possibly $m_\eta$ are complex, the mixing angle is always a real number. Furthermore, even if the $\eta'$ mass is not accessible for all the temperatures, figure 3, it does not prevent to find values for $\theta$. This is due to the fact that this mass does not intervene in the determination of $M_{08}$ and $M_{00}$. In addition, in $a_\eta$, we use the $\eta$ mass…

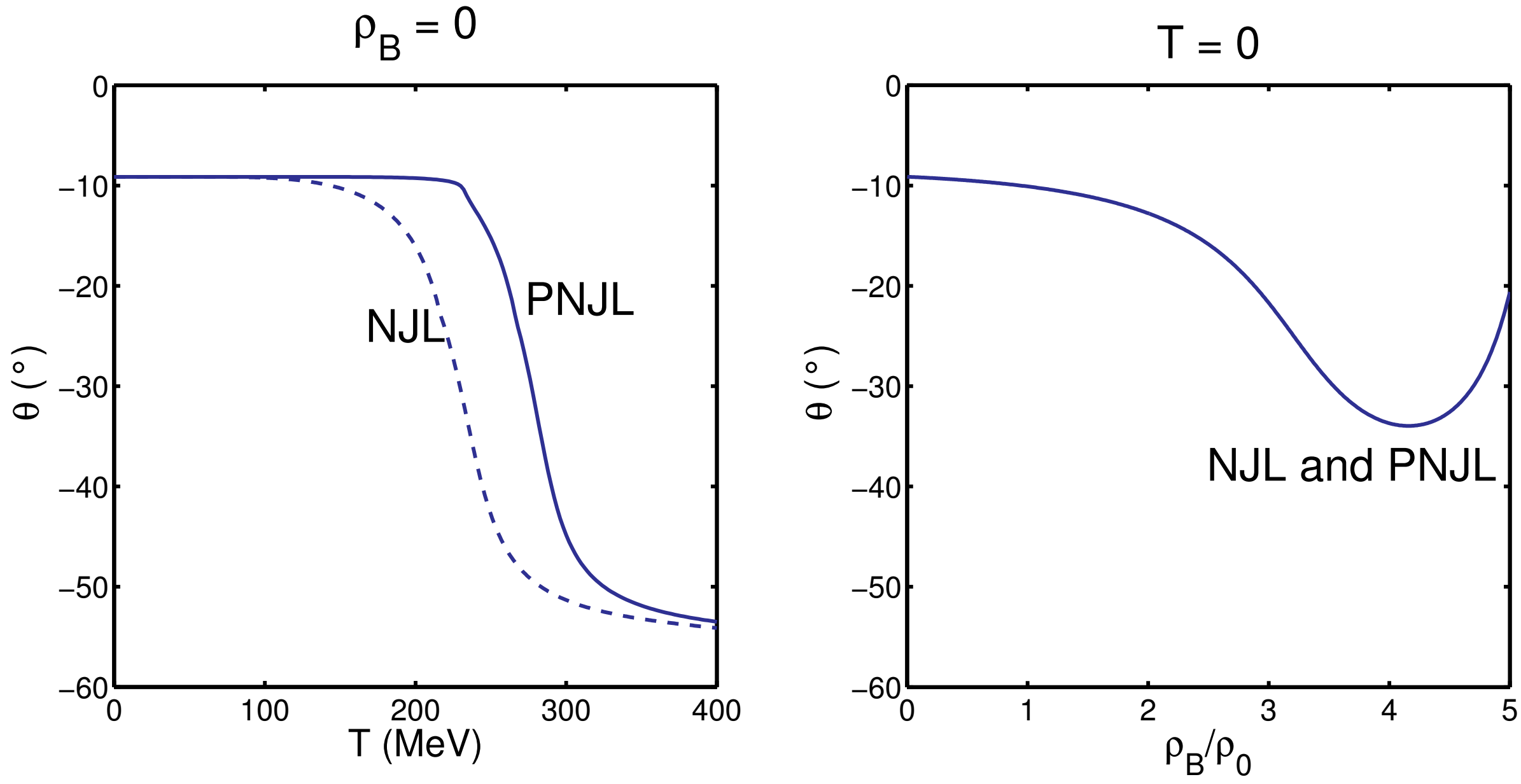

**Figure 10.** Mixing angle function of the temperature, and function of the baryonic density.

The left hand side of the figure studies the evolution of the angle according to the temperature, at null baryonic density. The presented curves are rather close to the ones published in [23]. Compared to this reference, we used the same effective potential $\mathcal{U}$ (for the PNJL curve), but a different parameter set (the P1 one, table 1 of chapter 2). In the figure 3, we remark that the critical temperature of the $\eta$ meson is equal to $T = 216$ MeV with the NJL model, and $T = 267$ MeV with the PNJL one. For each curve of the left hand side of the figure 10, these temperatures correspond to a strong decrease of the found values. In the right hand side of the figure, the evolution of the mixing angle is plotted according to the baryonic density. These calculations were performed at null temperature, so NJL and PNJL values coincide. It is found that the mixing angle decreases until a baryonic density close to $4\rho_0$, and then increases for higher densities. The aspect of this curve, also observed in [26], recall the one found with $\left|g_{\eta-s\bar{s}}\right|$.

# 4. Obtained results for other mesons

## 4.1 Scalar mesons

As the pseudo scalar mesons, the scalar mesons was particularly studied in the (P)NJL literature. At this occasion, it should be noted that the names of these mesons present some differences in the quoted papers. Thus, the meson $f_0$ [9, 30] corresponds to $\sigma$ in [13, 14, 16, 22, 23]. In the same way, $a_0$ [9, 30] is named $\sigma_\pi$ in [13, 14, 16]. $K_0^*$ [30] is equivalent to $\sigma_K$ in [13, 14, 16] or $\kappa$ in [23]. Finally, $f_0'$ [9] corresponds to $\sigma'$ in [13, 14, 16] or $f_0$ in [23, 30], i.e. a resonance of $f_0$ , or $f_0^*$ in [31]. As argued in the section 2, the equations to be solved for pseudo-scalar and scalar mesons are very close [13]. The both use the 't Hooft term $K \cdot \left[ \det\left(\bar{\psi}(1+\gamma_5)\psi\right) + \det\left(\bar{\psi}(1-\gamma_5)\psi\right) \right]$ used in (14).

Our results are shown in the figure 11 to 13. Firstly, in the figure 11, we focus on a study of the scalar mesons' masses according to the temperature, at null density. The data produced with the NJL model can be compared to the ones of [13, 16], and the results associated with the PNJL approach to [22, 23]. As a whole, the aspect of our curves is in agreement with the one of the quoted papers. We also confirm the curve distortion caused by the inclusion of the Polyakov loop. Also, thanks to the figure 12, $f_0$ is the only scalar meson that appears to be stable in our results, at reduced temperatures. Indeed, it width is null until $T \approx 200$ MeV in the NJL model, and until $T \approx 260$ MeV with the PNJL one. The other scalar mesons are unstable, whatever the temperature. In addition, in spite of the behavior of $\eta'$, the meson $f_0'$ presents an evolution that is completely in conformity compared to the other scalar mesons, i.e. no halt of the curve is observed.

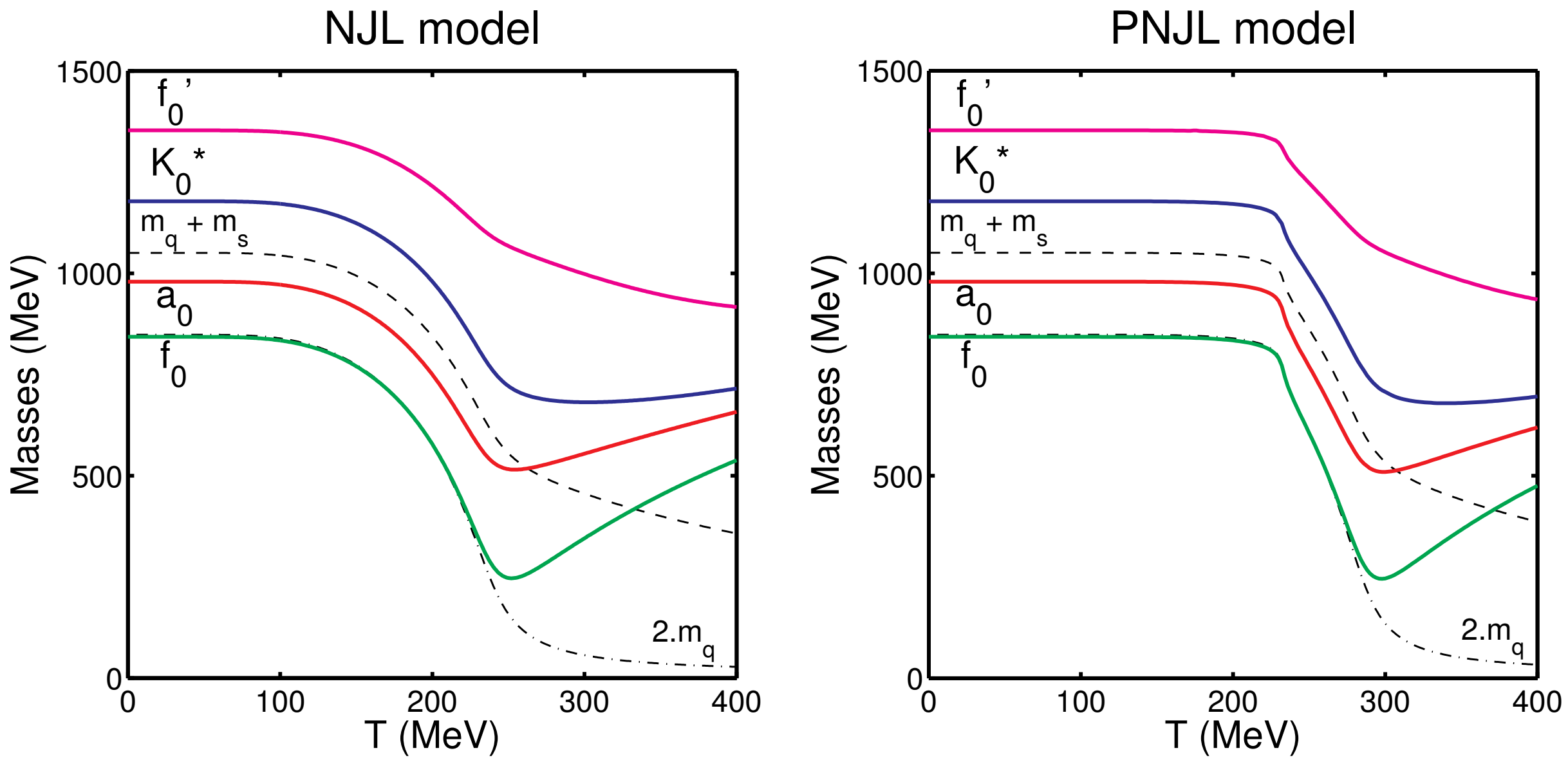

**Figure 11.** Scalar mesons masses function of the temperature.

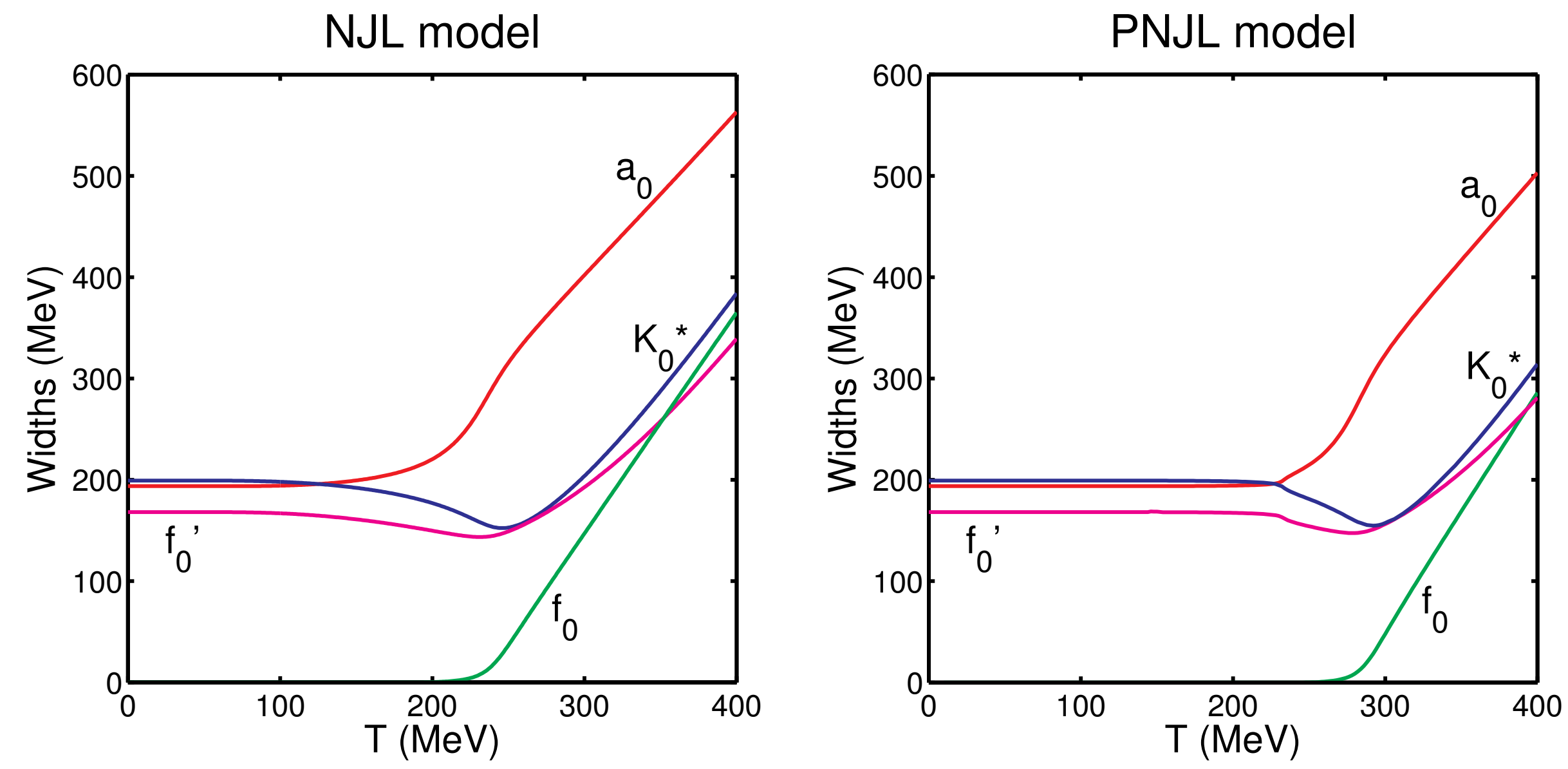

**Figure 12.** Widths of the scalar mesons function of the temperature.

In the figure 13, the mesons are studied according to the baryonic density, at null temperature. The left hand side focuses on the masses. This graph can be compared to the one published in [9]. Even if the used approach and the parameter set are not the same, the results are similar. But, compared to this reference, we added the $K_0^*$ meson. As with its pseudo-scalar partner, the curve splits in two parts, $K_0^{*+}$ and $K_0^{*-}$. The explanations of this behavior are strictly the same as the ones given for the pseudo-scalar mesons. Also, for the $a_0$ and the $K_0^{*+}$ mesons, an angular point is found on the curves, for a density close to $0.5\rho_0$, i.e. when the widths fall to zero. This behavior is an artifact caused by a numerical approximation explained in [15] and in the section 3 of appendix D.

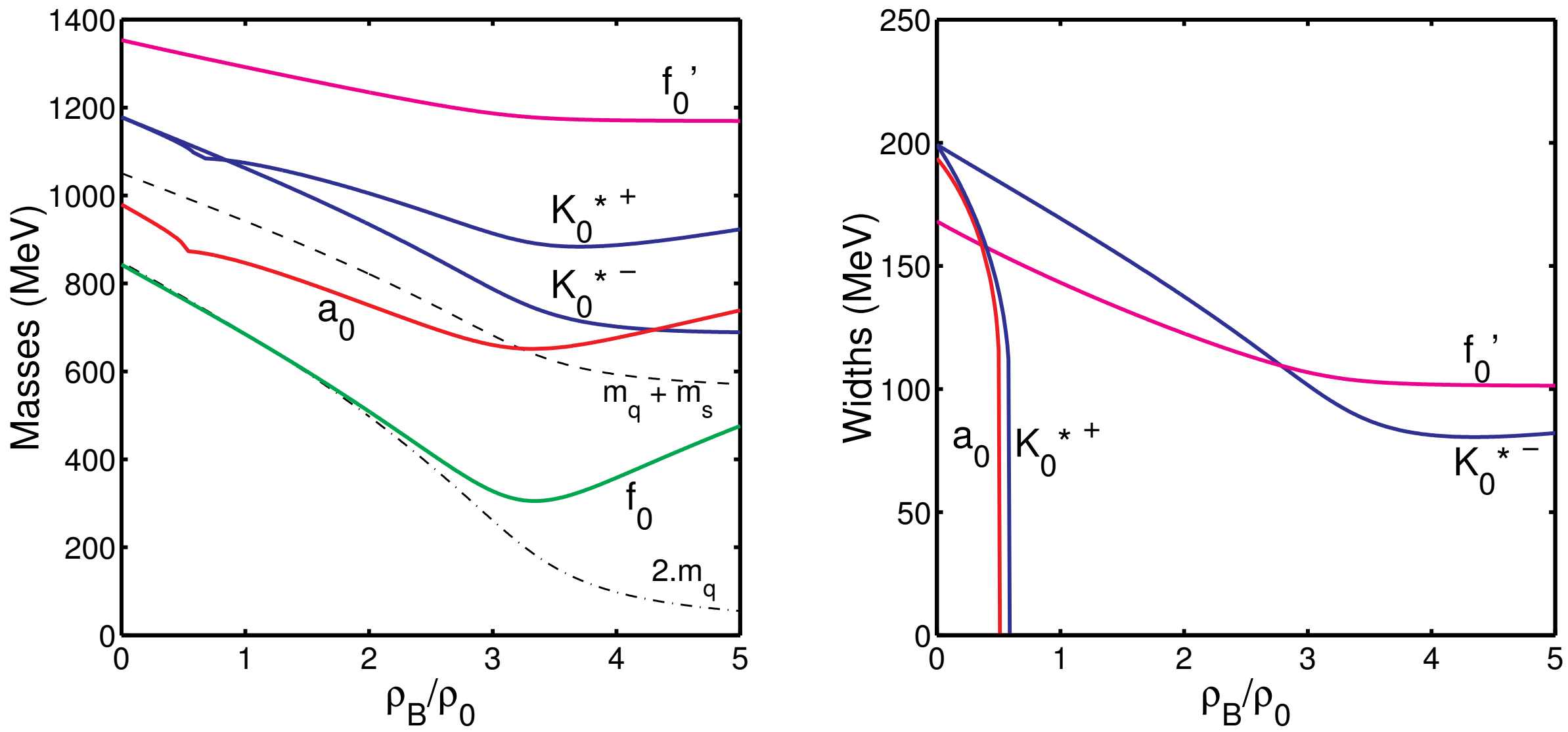

**Figure 13.** Masses and widths of the scalar mesons function of the baryonic density.

Moreover, as in [9], we can remark that the pion and the $f_0$ become degenerate when the baryonic density is strong enough, figures 5 and 13. As mentioned in this reference, this degeneracy occurs when the chiral symmetry is restored for the light quarks. Indeed, when this symmetry is restored, the value of the light quarks condensate is strongly reduced (appearing in (16), (17) via the $i \cdot Tr\left(S^f\right)$ terms), as the light quarks' effective masses that notably appear in the polarization functions. As a consequence, some terms involving these quantities can become negligible in the mesons' equations, and allow this degeneracy. In fact, the phenomenon is also observed according to the temperature, figures 3, 11 and 4, 12. It concerns also other mesons. Indeed, we observe this degeneracy at high temperatures/densities with $a_0 - \eta$, $K^+ – K_0^{*+}$ and $K^- – K_0^{*-}$, even if the convergence is less rapid for the kaons. Obviously, our remarks are valid for NJL model, but also for the PNJL one.

## 4.2 Vectorial mesons

The vectorial mesons were studied in the framework of the NJL model, for example in papers as [3, 7, 8], but the studies of these mesons according to the temperature and the density are relatively rare in this model. This remark is also valid for the PNJL model. Our results are presented in the figures 14 to 16. Without surprise, we conclude with the figure 14 that the inclusion of the Polyakov loop leads to the already observed shifting of the PNJL curves towards higher temperatures. In order to not overload our descriptions, the widths graphs are not included in this document. But, without the widths, we can indicate that the vectorial mesons are stable in the (P)NJL models at reduced temperatures and baryonic densities. This behavior recall the one observed for pseudo-scalar ones, except $\eta'$.

On the other hand, the dependence as regards the temperature is different: the masses of the vectorial mesons decrease continuously until their Mott temperatures. Furthermore, in the framework of the isospin symmetry, the mesons $\rho$ and $\omega$ are degenerate, whatever the temperature and the baryonic density. We recall that these mesons are, respectively, the equivalent of the pseudo scalar mesons $\pi$ and $\eta$. In fact, the 't Hooft term presented in (14) only acts on the scalar and pseudo-scalar mesons, but not on the vectorial ones. As a consequence, the $K_{ab}{}^{+}$ terms used to model the pseudo scalar mesons are replaced here by the constant $G_V$, see table 2. It leads to the observed behavior. Another difference is the vectorial mesons have globally smaller critical temperatures compared to the ones of the pseudo-scalar mesons, figure 14. This remark is also valid for the critical densities, figure 15. As a consequence, in the diagram presented in the figure 16, the areas for which the vectorial mesons are stable are reduced compared to the ones found in the figure 6.

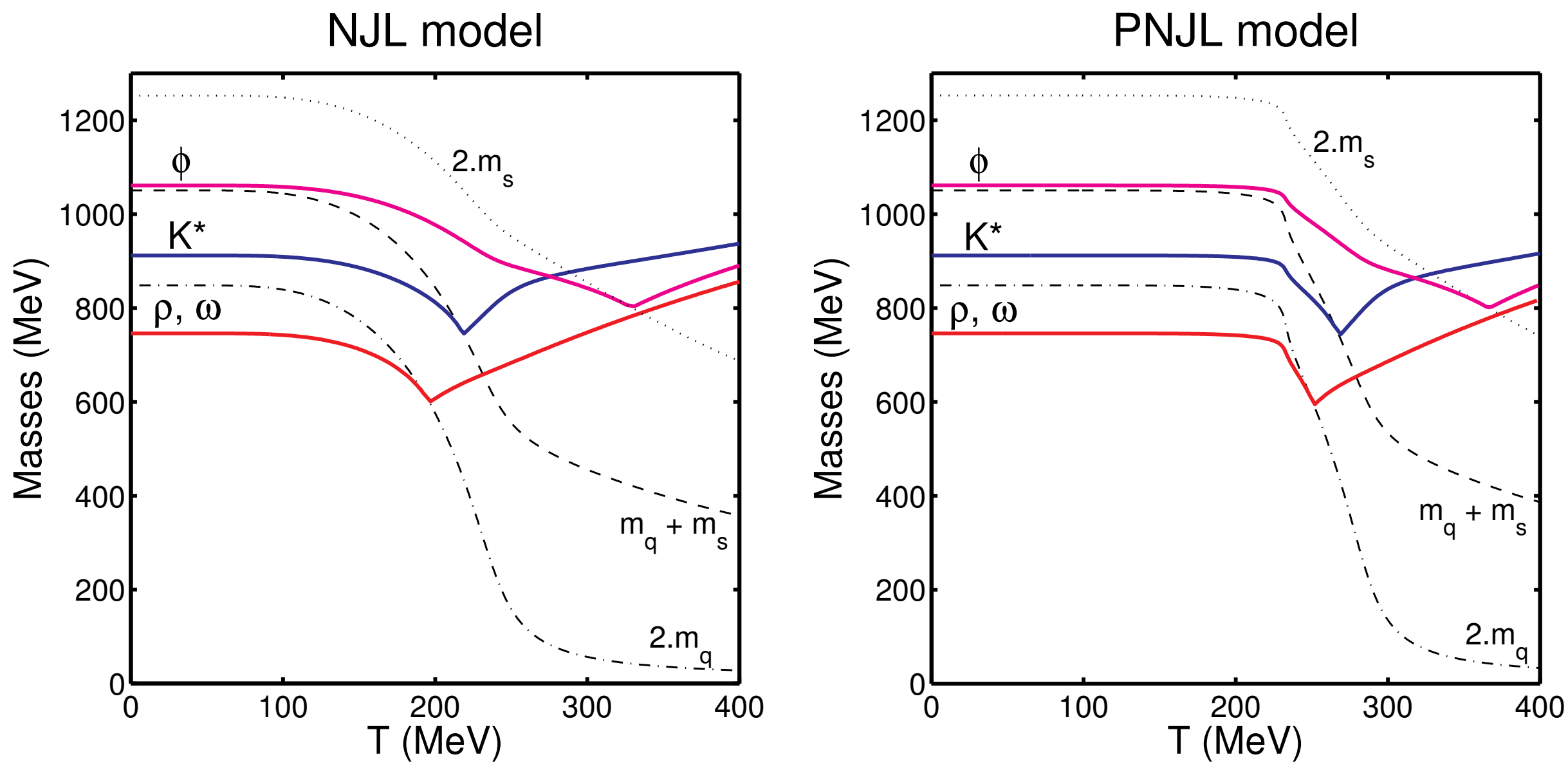

**Figure 14.** Masses of the vectorial mesons function of the temperature.

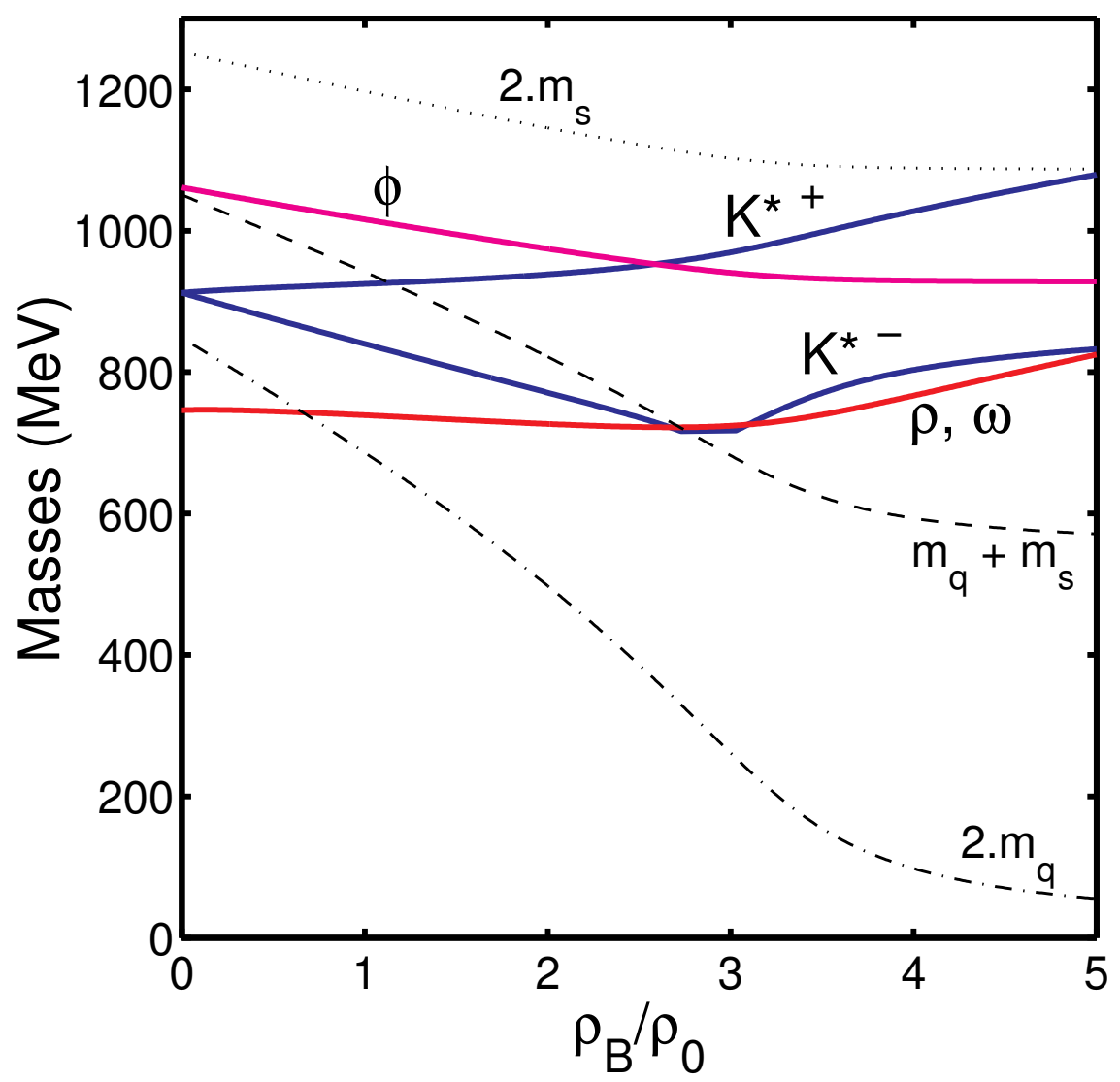

**Figure 15.** Masses of the vectorial mesons function of the baryonic density.

But, similarities can be found between the figures 6 and 16. Notably, the curves associated with the mesons $\rho, \omega, K^{*+}$ diverge towards higher densities at reduced temperatures, without reaching the density axis in the figure 6. This behavior, which we associated with a soft transition, was also observed with the $\pi, \eta, K^{+}$ mesons in the figure 16. On the other hand, the $K^{*-}$ acts as the $K^{-}$, i.e. it seems to present a visible transition according to the density.

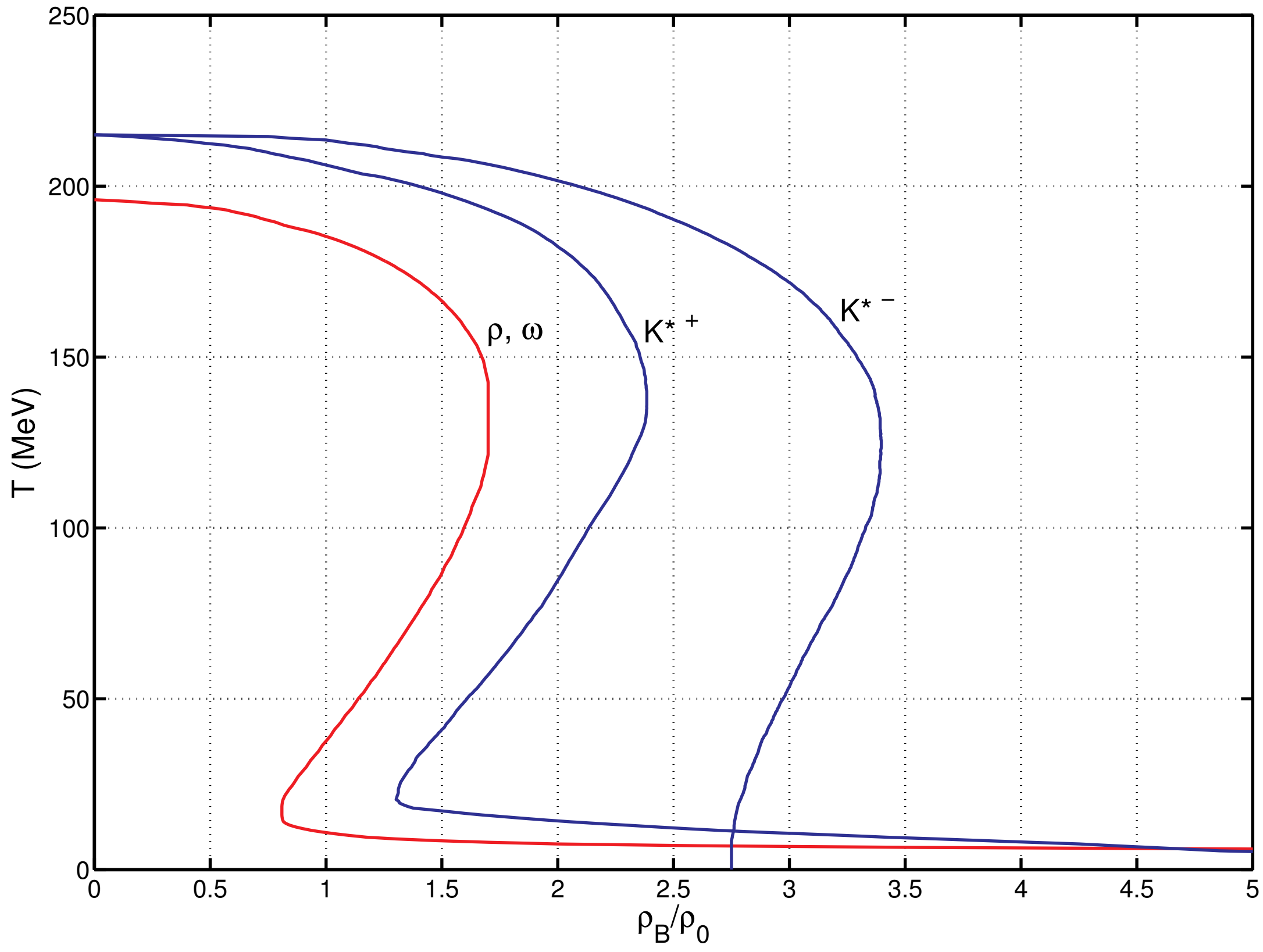

**Figure 16.** NJL diagram of stability/instability for the vectorial mesons.

# 4.3 Axial mesons

Our numerical results, visible in the figures 17 and 18, shows that all the axial mesons studied are unstable at null temperature and null densities, without exception. Their behavior recalls the one of the scalar mesons. However, as with the vectorial mesons, the 't Hooft term is also absent for axial ones. The axial mesons are clearly the heaviest mesons we modeled in this work. Indeed, the $f_1'$, has a mass comparable to the one of a heavy baryon. Also, the artifact observed for the $a_0$ and the $K_0^{*+}$ scalar mesons are reproduced for the $a_1 - f_1$ and $K_1^{+}$ particles. As in the figure 13, this artifact in visible in the figure 18 by angular points, located at $1.8\rho_0$ for $a_1 - f_1$ and $2.4\rho_0$ for $K_1^{+}$.

Moreover, as with the pseudo-scalar and scalar mesons, we also observe that some vectorial and axial mesons become degenerate at high temperatures/densities. Clearly, it concerns $\rho, \omega$ with $a_1, f_1$ on one side, $K^{*+} - K_1^{*+}$ and $K^{*-} - K_1^{*-}$ on the other side, figures 14, 17 and 15, 18. In fact, the only difference between the equations of the vectorial and the axial mesons come from their respective polarization functions, via the $m_1^2 + m_2^2 \pm 4m_1m_2$ terms, appendix D. These ones are thus expected to become negligible at high $T, \rho_B$, in front of the other terms of the polarization functions, if at least one of the two involved quarks is a light quark.

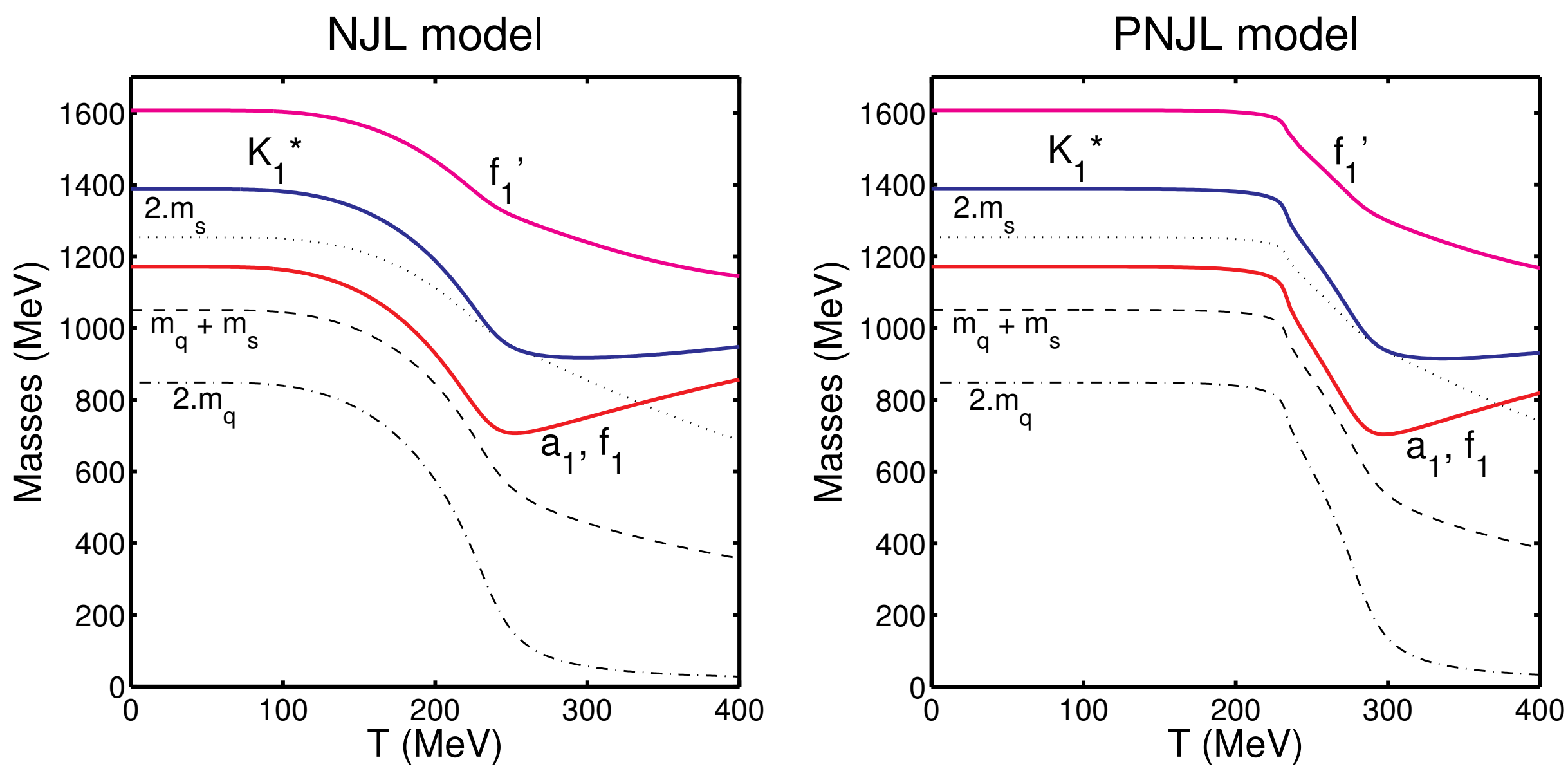

**Figure 17.** Masses of the axial mesons according to the temperature.

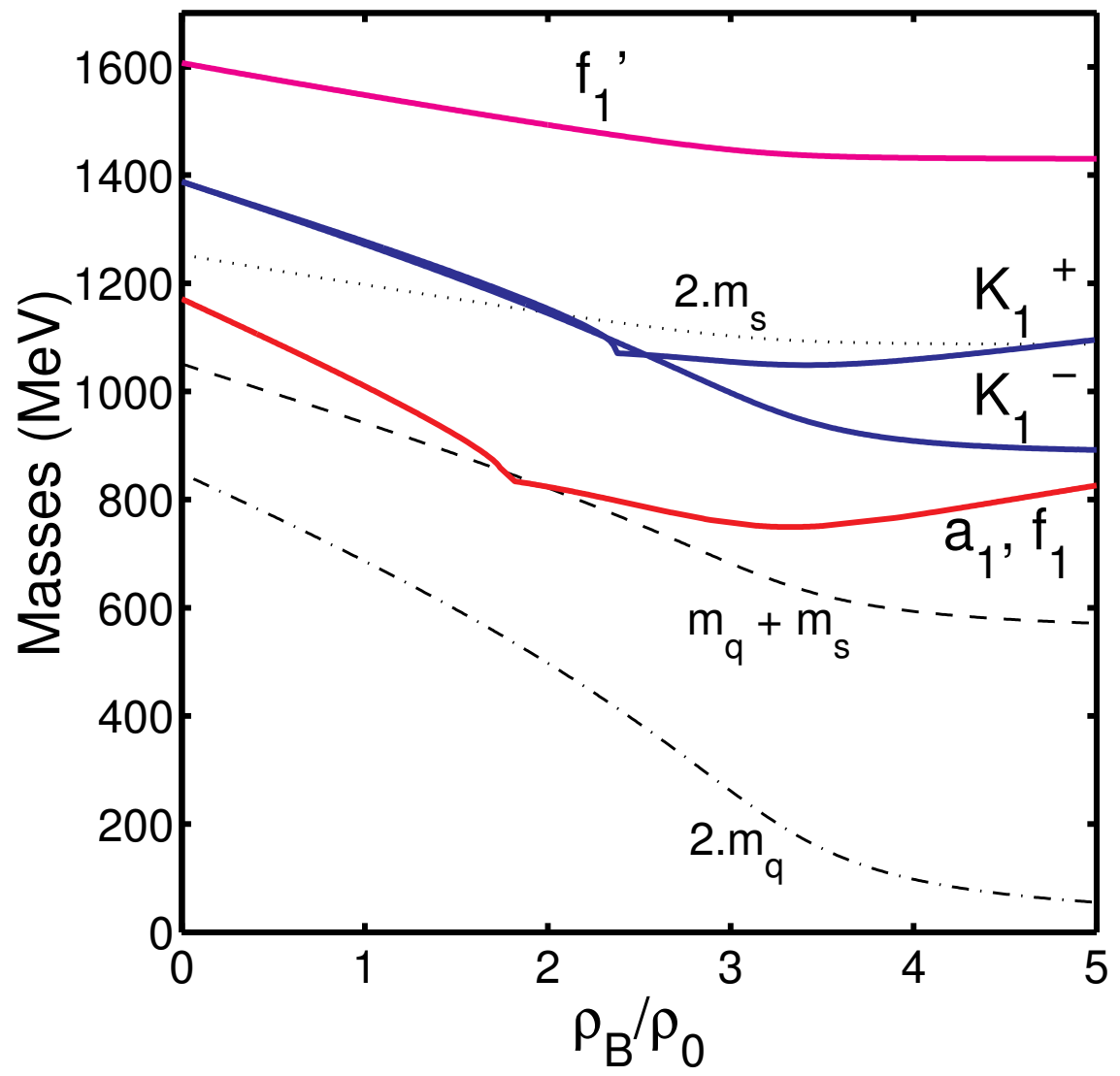

**Figure 18.** Masses of the axial mesons according to the baryonic density.

# 5. Beyond the isospin symmetry

The isospin symmetry is very employed within the framework of this thesis, as in the (P)NJL literature, e.g. [7]. The masses of the *u* and *d* quarks are very close compared to the one of the strange quarks. As a consequence, this approximation is fully justified. As argued in the chapter 2, it leads to simplifications of the calculations. Furthermore, the obtained results are satisfactory. Nevertheless, in some cases, this approximation is not valid, for example when $\mu_u$ (respectively $\rho_u$) is very different of $\mu_d$ (resp. $\rho_d$). This situation is encountered in physical systems like neutron stars or simply with heavy nuclei [19]. These ones are the basic "substrate" for the study of the quark gluon plasma, in heavy ions collisions. In this example, the ratio between neutrons and protons, near to 1.5, creates an asymmetry. In this section, we temporarily do not use the isospin approximation, in order to investigate the consequences on the found values. Of course, the abandon of the isospin symmetry leads to complications of some calculations. Notably, we propose hereafter to consider again the subsection 2.5 and show to the modifications to be performed.

## 5.1 Treatment of the $\pi^0,\eta,\eta'$ mesons when $m_u \neq m_d$

The coupling $\pi^0-\eta-\eta'$ made that the diffusion matrix $M$ associated with these three particles is written on the form of a $3\times3$ matrix:

$$M=\begin{bmatrix} M_{00} & M_{03} & M_{08} \\ M_{30} & M_{33} & M_{38} \\ M_{80} & M_{83} & M_{88} \end{bmatrix}=2K^+\left(1-2\Pi^P K^+\right)^{-1} \quad , \tag{35}$$

with:

$$K^+=\begin{bmatrix} K_{00}^+ & K_{03}^+ & K_{08}^+ \\ K_{30}^+ & K_{33}^+ & K_{38}^+ \\ K_{80}^+ & K_{83}^+ & K_{88}^+ \end{bmatrix} \tag{36}$$

and:

$$\Pi^P=\begin{bmatrix} \Pi_{00}^P & \Pi_{03}^P & \Pi_{08}^P \\ \Pi_{30}^P & \Pi_{33}^P & \Pi_{38}^P \\ \Pi_{80}^P & \Pi_{83}^P & \Pi_{88}^P \end{bmatrix} \quad . \tag{37}$$

If the isospin approximation is not used, the matrix $K^+$ cannot be simplified. In a general way, all its terms are non-null ones. The following property is however still valid:

$$K_{30}^+=K_{03}^+ \,,\; K_{80}^+=K_{08}^+ \text{ and } K_{83}^+=K_{38}^+ \quad . \tag{38}$$

So the matrix $K^+$ is symmetrical. This property is also checked by the matrix $\Pi^P$, which we clarified each term hereafter, i.e. as in [25]:

$$\begin{cases}
\Pi_{00}^P = \frac{2}{3}\cdot\left(\Pi_{u\bar{u}}^P + \Pi_{d\bar{d}}^P + \Pi_{s\bar{s}}^P\right) & \lambda_0\times\lambda_0 = \frac{2}{3}\cdot\begin{bmatrix}1 & & \\ & 1 & \\ & & 1\end{bmatrix} \\
\Pi_{03}^P = \sqrt{\frac{2}{3}}\cdot\left(\Pi_{u\bar{u}}^P - \Pi_{d\bar{d}}^P\right) & \lambda_0\times\lambda_3 = \sqrt{\frac{2}{3}}\cdot\begin{bmatrix}1 & & \\ & -1 & \\ & & \end{bmatrix} \\
\Pi_{08}^P = \frac{\sqrt{2}}{3}\cdot\left(\Pi_{u\bar{u}}^P + \Pi_{d\bar{d}}^P - 2\Pi_{s\bar{s}}^P\right) & \lambda_0\times\lambda_8 = \frac{\sqrt{2}}{3}\cdot\begin{bmatrix}1 & & \\ & 1 & \\ & & -2\end{bmatrix} \\
\Pi_{33}^P = \Pi_{u\bar{u}}^P + \Pi_{d\bar{d}}^P & \lambda_3\times\lambda_3 = \begin{bmatrix}1 & & \\ & 1 & \\ & & \end{bmatrix} \\
\Pi_{38}^P = \frac{\sqrt{3}}{3}\cdot\left(\Pi_{u\bar{u}}^P - \Pi_{d\bar{d}}^P\right) & \lambda_3\times\lambda_8 = \frac{\sqrt{3}}{3}\cdot\begin{bmatrix}1 & & \\ & -1 & \\ & & \end{bmatrix} \\
\Pi_{88}^P = \frac{1}{3}\cdot\left(\Pi_{u\bar{u}}^P + \Pi_{d\bar{d}}^P + 4\Pi_{s\bar{s}}^P\right) & \lambda_8\times\lambda_8 = \frac{1}{3}\cdot\begin{bmatrix}1 & & \\ & 1 & \\ & & 4\end{bmatrix}
\end{cases} \tag{39}$$

The matrices on the right hand side of (39) are used to explain the factors put in front of the $\Pi_{u\bar{u}}^P$, $\Pi_{d\bar{d}}^P$ and $\Pi_{s\bar{s}}^P$. These former terms are the meson pseudo-scalar polarization functions. By analogy with subsection 2.5, the inverse of the matrix $M$ is formally written as:

$$M^{-1} = \frac{1}{2\cdot\det\left(K^+\right)}\cdot\begin{bmatrix}\mathcal{D}_1 & \mathcal{A} & \mathcal{B} \\ \mathcal{A} & \mathcal{D}_2 & \mathcal{C} \\ \mathcal{B} & \mathcal{C} & \mathcal{D}_3\end{bmatrix}, \tag{40}$$

where:

$$\begin{cases}
\mathcal{A} = K_{08}^+\cdot K_{38}^+ - K_{03}^+\cdot K_{88}^+ - 2\cdot\Pi_{03}^P\cdot\det\left(K^+\right) \\
\mathcal{B} = K_{03}^+\cdot K_{38}^+ - K_{08}^+\cdot K_{33}^+ - 2\cdot\Pi_{08}^P\cdot\det\left(K^+\right) \\
\mathcal{C} = K_{08}^+\cdot K_{03}^+ - K_{00}^+\cdot K_{38}^+ - 2\cdot\Pi_{38}^P\cdot\det\left(K^+\right) \\
\mathcal{D}_1 = K_{33}^+\cdot K_{88}^+ - \left(K_{38}^+\right)^2 - 2\cdot\Pi_{00}^P\cdot\det\left(K^+\right) \\
\mathcal{D}_2 = K_{00}^+\cdot K_{88}^+ - \left(K_{08}^+\right)^2 - 2\cdot\Pi_{33}^P\cdot\det\left(K^+\right) \\
\mathcal{D}_3 = K_{00}^+\cdot K_{33}^+ - \left(K_{03}^+\right)^2 - 2\cdot\Pi_{88}^P\cdot\det\left(K^+\right)
\end{cases} \tag{41}$$

and:

$$\det\left(K^+\right) = K_{00}^+\cdot K_{33}^+\cdot K_{88}^+ - K_{00}^+\cdot\left(K_{38}^+\right)^2 - K_{88}^+\cdot\left(K_{03}^+\right)^2 - K_{33}^+\cdot\left(K_{08}^+\right)^2 + 2\cdot K_{03}^+\cdot K_{08}^+\cdot K_{38}^+ . \tag{42}$$

We now need to find the eigenvalues. They correspond to each meson propagator. So, we diagonalize the matrix $M^{-1}$, i.e. we find the roots of its characteristic polynomial. This one is written, on a symbolic way, as:

$$\begin{aligned}&-x^3+\underbrace{\left(\mathcal{D}_1+\mathcal{D}_2+\mathcal{D}_3\right)}_{\alpha}\cdot x^2+\underbrace{\left(\mathcal{A}^2+\mathcal{B}^2+\mathcal{C}^2-\mathcal{D}_1\cdot\mathcal{D}_2-\mathcal{D}_1\cdot\mathcal{D}_3-\mathcal{D}_2\cdot\mathcal{D}_3\right)}_{\beta}\cdot x\\&+\underbrace{\left(2\cdot\mathcal{A}\cdot\mathcal{B}\cdot\mathcal{C}-\mathcal{A}^2\cdot\mathcal{D}_3-\mathcal{B}^2\cdot\mathcal{D}_2-\mathcal{C}^2\cdot\mathcal{D}_1+\mathcal{D}_1\cdot\mathcal{D}_2\cdot\mathcal{D}_3\right)}_{\delta}\end{aligned}\quad. \tag{43}$$

The roots are:

$$\begin{cases}M_{\pi_0}{}^{-1}=\left(-\dfrac{X^{1/3}}{12}+\dfrac{3\cdot\left(-\dfrac{\beta}{3}-\dfrac{\alpha^2}{9}\right)}{X^{1/3}}+\dfrac{\alpha}{3}\right)+\left(\dfrac{X^{1/3}}{6}+\dfrac{6\cdot\left(-\dfrac{\beta}{3}-\dfrac{\alpha^2}{9}\right)}{X^{1/3}}\right)\cdot\dfrac{\sqrt{3}}{2}\cdot i\\[2ex] M_{\eta}{}^{-1}=\left(-\dfrac{X^{1/3}}{12}+\dfrac{3\cdot\left(-\dfrac{\beta}{3}-\dfrac{\alpha^2}{9}\right)}{X^{1/3}}+\dfrac{\alpha}{3}\right)-\left(\dfrac{X^{1/3}}{6}+\dfrac{6\cdot\left(-\dfrac{\beta}{3}-\dfrac{\alpha^2}{9}\right)}{X^{1/3}}\right)\cdot\dfrac{\sqrt{3}}{2}\cdot i\\[2ex] M_{\eta'}{}^{-1}=\dfrac{X^{1/3}}{6}-\dfrac{6\cdot\left(-\dfrac{\beta}{3}-\dfrac{\alpha^2}{9}\right)}{X^{1/3}}+\dfrac{\alpha}{3}\end{cases}\quad, \tag{44}$$

with:

$$\begin{aligned}X=&36\cdot\beta\cdot\alpha+108\cdot\delta+8\cdot\alpha^3\\&+12\cdot\sqrt{-12\cdot\beta^3-3\cdot\beta^2\cdot\alpha^2+54\cdot\beta\cdot\alpha\cdot\delta+81\cdot\delta^2+12\cdot\delta\cdot\alpha^3}\end{aligned}\quad. \tag{45}$$

It is then enough to solve, in the general case:

$$\begin{cases}M_{\pi_0}{}^{-1}\left(\sqrt{m_{\pi_0}{}^2+\left(\vec{k}_{\pi_0}\right)^2},\vec{k}_{\pi_0}\right)=0\\ M_{\eta}{}^{-1}\left(\sqrt{m_{\eta}{}^2+\left(\vec{k}_{\eta}\right)^2},\vec{k}_{\eta}\right)=0\\ M_{\eta'}{}^{-1}\left(\sqrt{m_{\eta'}{}^2+\left(\vec{k}_{\eta'}\right)^2},\vec{k}_{\eta'}\right)=0\end{cases}\quad. \tag{46}$$

The $\vec{k}$ are the momenta of the particles and the $m$ are the wanted masses. Of course, we can choose to study the particle at rest, i.e. $\vec{k}=\vec{0}$, as done previously. To conclude this calculation, it could be relevant to see what we obtain if we consider $m_u=m_d$. In others words, we try to find again the isospin approximation results starting from our formulas. In this case, it is easy to check that:

$$K_{03}^{+}=K_{38}^{+}=0 \quad\text{and}\quad \Pi_{03}^{P}=\Pi_{38}^{P}=0\quad. \tag{47}$$

Taking again the equation (40), we have:

$$M^{-1}=\frac{1}{2\det\left(K^{+}\right)}\cdot\begin{bmatrix}\mathcal{D}_1 & \mathcal{A} & \mathcal{B}\\ \mathcal{A} & \mathcal{D}_2 & \mathcal{C}\\ \mathcal{B} & \mathcal{C} & \mathcal{D}_3\end{bmatrix} \tag{48}$$

$$=\frac{1}{2K_{33}^{+}\cdot\det'}\cdot\begin{bmatrix}K_{33}^{+}\cdot\left(K_{88}^{+}-2\Pi_{00}^{P}\cdot\det'\right) & 0 & K_{33}^{+}\cdot\left(-K_{08}^{+}-2\Pi_{08}^{P}\cdot\det'\right)\\ 0 & K_{33}^{+}\cdot\det'\cdot\left(\frac{1}{K_{33}^{+}}-2\Pi_{33}^{P}\right) & 0\\ K_{33}^{+}\cdot\left(-K_{08}^{+}-2\Pi_{08}^{P}\cdot\det'\right) & 0 & K_{33}^{+}\cdot\left(K_{00}^{+}-2\Pi_{88}^{P}\cdot\det'\right)\end{bmatrix},$$

with:

$$\det\left(K^{+}\right)=K_{33}^{+}\cdot\left(K_{00}^{+}\cdot K_{88}^{+}-\left(K_{08}^{+}\right)^{2}\right)=K_{33}^{+}\cdot\det' \ . \tag{49}$$

This is equivalent to:

$$M^{-1}=\begin{bmatrix}\frac{K_{88}^{+}-2\Pi_{00}^{P}\cdot\det'}{2\det'} & \frac{-K_{08}^{+}-2\Pi_{08}^{P}\cdot\det'}{2\det'} & 0\\ \frac{-K_{08}^{+}-2\Pi_{08}^{P}\cdot\det'}{2\det'} & \frac{K_{00}^{+}-2\Pi_{88}^{P}\cdot\det'}{2\det'} & 0\\ 0 & 0 & \frac{1-2K_{33}^{+}\cdot\Pi_{33}^{P}}{2K_{33}^{+}}\end{bmatrix}\sim\begin{bmatrix}\left[M_{\eta-\eta'}{}^{-1}\right] & 0\\ 0 & \left[M_{\pi_0}{}^{-1}\right]\end{bmatrix} . \tag{50}$$

We find again the equations described in [13], where the isospin approximation was employed. The particle $\pi_0$ becomes uncoupled from $\eta-\eta'$. Its propagator is thus:

$$M_{\pi_0}=\frac{2\cdot K_{33}^{+}}{1-2\cdot K_{33}^{+}\cdot\Pi_{33}^{P}} \ . \tag{51}$$

All this reasoning is of course identical for $a_0{}^0$ and $f_0-f_0{}'$.

# 5.2 Obtained results, discussion

The table 3 gathers the masses and widths of the studied mesons at null temperature and at null density. In this table, the P1 column is associated with the results we found with this parameter set, i.e. associated with the curves described in this chapter. The EB columns concerns the parameter set given in the chapter 2 that does not use the isospin symmetry. The column "experimental values" was constituted thanks of the values given by [29, 30]. However, the $f_0{}'$ and $f_1{}'$ values in this column were estimated also thanks to [31]. In fact, this reference supplies values used in the framework of the QMD/URQMD models. The both consider the isposin symmetry. These data associated with the mesons are reproduced in the appendix A. They constitute another possibility of comparison, especially with our data that use the isospin symmetry. In fact, even if some differences can be mentioned between our P1 data and the ones of the literature (experimental data, [31]), the values stay in the good order of magnitude, notably with the masses. Furthermore, the agreement with (P)NJL papers [7, 13, 16, 19, 22, 23, 25–27] is also correct, aware that the choice of the used parameter set has a great influence on the found values. On the other hand, about the widths, we see dissimilarities with the experimental data. But, we recall that in our work, the notion of width or instability is associated with the meson's disintegration into a quark/antiquark pair, whereas in experimental data, the instability of a meson notably refers to its disintegration into another lighter meson.

| | | | P1 parameter set | | EB parameter set | | Experimental values | |
|---|---|---|---|---|---|---|---|---|
| mesons | | | masses | widths | masses | widths | masses | widths |
| pseudo scalar | $\pi^{\pm}$ | $u\bar{d}, d\bar{u}$ | 135.96 | 0 | 139.46 | 0 | 139.57 | 0 |
| | $\pi^0$ | $mix(u\bar{u}, d\bar{d})$ | 135.96 | 0 | 139.38 | 0 | 134.98 | 0 |
| | $\eta$ | $mix(u\bar{u}, d\bar{d}, s\bar{s})$ | 557.13 | 0 | 517.86 | 0 | 547.85 | 0.00118 |
| | $\eta'$ | $mix(u\bar{u}, d\bar{d}, s\bar{s})$ | 1012.16 | 158.74 | 955.92 | 152.33 | 957.78 | 0.194 |
| | $K^{\pm}$ | $u\bar{s}, s\bar{u}$ | 548.50 | 0 | 493.94 | 0 | 493.677 | 0 |
| | $K^0/\bar{K}^0$ | $d\bar{s}, s\bar{d}$ | 548.50 | 0 | 497.94 | 0 | 497.614 | 0 |
| scalar | $a_0{}^{\pm}$ | $u\bar{d}, d\bar{u}$ | 979.48 | 193.73 | 970.16 | 187.84 | 984.7 | 50–100 |
| | $a_0{}^0$ | $mix(u\bar{u}, d\bar{d})$ | 979.48 | 193.73 | 970.17 | 187.88 | 984.7 | 50–100 |
| | $f_0$ | $mix(u\bar{u}, d\bar{d}, s\bar{s})$ | 843.13 | 0 | 834.07 | 0 | 980 | 40–100 |
| | $f_0{}'$ | $mix(u\bar{u}, d\bar{d}, s\bar{s})$ | 1353.34 | 168.15 | 1274.05 | 161.45 | 1370 | 200 |
| | $K_0^{*\pm}$ | $u\bar{s}, s\bar{u}$ | 1178.14 | 199.28 | 1132.89 | 191.88 | 1429 | 294 |
| | $K_0^{*0}/\bar{K}_0^{*0}$ | $d\bar{s}, s\bar{d}$ | 1178.14 | 199.28 | 1135.99 | 191.94 | 1429 | 294 |
| vectorial | $\rho^{\pm}$ | $u\bar{d}, d\bar{u}$ | 746.09 | 0 | 764.12 | 0 | 775.5 | 146.2 |
| | $\rho^0$ | $mix(u\bar{u}, d\bar{d})$ | 746.09 | 0 | 764.08 | 0 | 775.5 | 146.2 |
| | $\omega$ | $mix(u\bar{u}, d\bar{d}, s\bar{s})$ | 746.09 | 0 | 764.08 | 0 | 782.65 | 8.49 |
| | $\phi$ | $s\bar{s}$ | 1061.18 | 0 | 1025.79 | 0 | 1019.455 | 4.26 |
| | $K^{*\pm}$ | $u\bar{s}, s\bar{u}$ | 912.25 | 0 | 899.96 | 0 | 891.66 | 50.8 |
| | $K^{*0}/\bar{K}^{*0}$ | $d\bar{s}, s\bar{d}$ | 912.25 | 0 | 902.12 | 0 | 896.10 | 50.7 |
| axial | $a_1{}^{\pm}$ | $u\bar{d}, d\bar{u}$ | 1171.06 | 434.01 | 1173.77 | 478.50 | 1230 | 250–600 |
| | $a_1{}^0$ | $mix(u\bar{u}, d\bar{d})$ | 1171.06 | 434.01 | 1173.78 | 478.54 | 1230 | 250–600 |
| | $f_1$ | $mix(u\bar{u}, d\bar{d}, s\bar{s})$ | 1171.06 | 434.01 | 1173.78 | 478.54 | 1281.8 | 24.3 |
| | $f_1{}'$ | $s\bar{s}$ | 1607.45 | 487.39 | 1531.55 | 522.49 | 1512 | 350 |
| | $K_1^{*\pm}$ | $u\bar{s}, s\bar{u}$ | 1387.78 | 464.39 | 1349.69 | 502.58 | 1273 | 90 |
| | $K_1^{*0}/\bar{K}_1^{*0}$ | $d\bar{s}, s\bar{d}$ | 1387.78 | 464.39 | 1353.21 | 503.04 | 1273 | 90 |

**Table 3.** Masses of the mesons at null temperature and at null density.

About the EB data, which motivated the works performed in this section, the agreement with the experimental data is very good. In practice, even if we used the masses of the pions and kaons to calibrate the values of our EB parameter set (notably the quarks naked masses), we can underline the remarkable precision found with the pseudo-scalar mesons. About the $\pi^0,\eta,\eta'$ mesons, we recall that the EB results were found by applying the method described in the subsection 5.1. These results are also correct, even if we expected a greater difference between the mass of the $\pi^0$, compared to the ones of the $\pi^\pm$ pions. Moreover, good results are also observable with vectorial mesons. In fact, generally, the particles found as stables in the model (typically pseudo-scalar and vectorial mesons, except $\eta'$) have masses closer to the experimental values than those that are unstable, i.e. scalar and axial mesons, except $f_0$.

However, concerning the validity of our results, we should have a critical view upon some of our mesons' modeling. About the pseudo-scalar mesons, no real defect could be really noticed, except of course for the $\eta'$. But, on the other hand, we are aware that the heavy masses found for the axial mesons suggest that we may reach there the limit of reliability of the approach. Also, for the vectorial and scalar mesons, stricto sensu the modeling is not complete, simply because some relevant decays are not included in our description. Indeed, it is well known that the $\rho$ vectorial meson has strong chances to decay into two pions instead of disintegrating into one quark and one antiquark. The inclusion of this process in the model can leads to modifications on the results, as the ones of the figure 16. But, we can remark that in our work, or in papers as [13, 14, 17], the scalar mesons and $\rho$ are mainly considered to intervene as propagators in cross-sections calculations.

# 6. Conclusion

In this chapter, devoted to the mesons, we firstly presented the method employed to calculate the mesons masses numerically. We saw it was possible to include a large variety of mesons in our model, for then studying them at finite temperatures and/or densities. We also calculated some derived data, like the coupling constants or the mixing angle between the mesons $\eta$ and $\eta'$. The precision we obtained appeared very promising, in particular when we did not use the isospin symmetry. As a whole, our results are in agreement with the ones found in the associated literature. Notably, we also confirmed the effect of the inclusion of the Polyakov loop on our results, i.e. a shifting of the values towards higher temperatures. It also leads to quasi-constant masses at reduced temperatures. A part of the works presented in this chapter was already published in the quoted references. But, thanks to certain of our graphs, as the "phase diagrams", we also underlined some particular phenomena. It notably concerns the behavior of some pseudo-scalar and vectorial mesons at finite densities and null temperature.

We will see in the two following chapters the construction of the other particles we will use in our work. They will be the diquarks and the baryons. In these chapters, we will remark that the work we will have to perform will be done on the same principle as the one we made here. Beyond mesons, we thus saw a general method to model our composite particles. It justifies that we have devoted in this chapter much time to describe the equations. For the other particles, we will only have to use the method by adapting the equations.

# Chapter 4

# Diquarks

A part of this chapter was published in *J. Phys. G: Nucl. Part. Phys.* **38** 105003

## 1. Introduction

In the previous chapters, we saw the possibilities offered by the (P)NJL models, in order to model quarks, and then mesons. The next step should be now to study baryons with these two models. However, as observed in the chapter 3, the (P)NJL approaches easily treat composite particles as loops. In the example of mesons, we saw that these particles were described as loops of quarks/antiquarks. About the baryons, we have to treat an association of three quarks. The Faddeev equations could allow treating 3-body interaction [1, 2]. They were already been used in the framework of the NJL description [3–5]. In fact, thanks to the representation of the Faddeev equations proposed in [6], a simplification of these equations can be proposed. It consists to consider a two body interaction between two of the quarks, and then to consider a bound state of this two particles group with the third quark. In other words, the three body interactions are neglected. As showed in [6], this simplification is fully justifiable, because three body interactions are expected to be smaller compared to two-body ones. More precisely, in this description, the two body interactions, between two quarks, naturally lead to consider the associated composite particles, i.e. diquarks. As a consequence, this simplification proposes to models the baryons as a bound state of a quark and a diquark, e.g. [7, 8]. It thus justifies the study of diquarks. Furthermore, when we will study reactions allowing the formation of baryons, the diquarks will intervene. It will be there discussed their role at this occasion: propagator in the reactions, intermediate states, the both? Indeed, as observed in the literature, i.e. [6, 9, 10], the diquarks appear to be largely more than a simple phenomenological trick to model the baryons.

But, in the framework of the standard model, we saw in chapter 1 that diquarks are colored objects. As a consequence, they cannot be observed in a free state. It implies that experimental data are not available, notably concerning their masses. Nevertheless, theoretical studies proposed modeling of the diquarks, as [11–14]. Thus, they supply values of the masses of the diquarks. Some approaches were performed with the NJL model, e.g. [15–17], or [18, 19] in the framework of the color superconductivity, but not yet in framework of the PNJL model. It is true that [20] included diquarks contribution terms in the PNJL description, but the finality of this article was not to study the diquarks' masses. As a whole, these references focused on the lightest diquarks, i.e. the scalar diquarks. Some works analyzed the evolution if their mass according to the temperature, or to the baryonic density, e.g. [17].

However, such work is rarer for the other diquarks, i.e. the pseudo-scalar, the axial and vectorial ones. In fact, the axial diquarks are notably expected to allow the modeling of decuplet baryons. In the same time, as remarked with the scalar and axial mesons, some of these diquarks can be considered as “less important” compared to the others. It concerns the pseudo-scalar and vectorial ones. However, we will see in chapter 6 that pseudo-scalar diquarks intervene in baryonization reactions, as propagators. Following this reasoning, the same argument can be applied to vectorial diquarks: in future evolutions of this work, they could be considered as propagators in reactions forming decuplet baryons. Also, other aspect of the diquarks study are still absent in the literature. It concerns the study of the diquarks stability, particularly with the scalar and axial ones, for example in the $T, \rho_B$ plane. It also concerns the treatment of the anti-diquarks, to be able to model anti-baryons …

To treat the points evoked in the previous paragraph, this chapter is organized as follow: firstly, in section 2, we present the equations devoted to the diquarks. It is explained how the work previously performed for mesons can be simply adapted to give us the masses of the diquarks, by using charge conjugate quarks/antiquarks. Obviously, this work includes the adaptation of the NJL equations to obtain a PNJL description. Also, some considerations linked to the group theory are presented. As with the mesons, our theoretical analysis focused also on the establishment of the equations devoted to estimate the diquarks coupling constants. Then, the section 3 presents the results obtained for each quoted diquarks: scalar, pseudo-scalar, axial, vectorial, and also for anti-diquarks. The effect of the Polyakov loop is then underlined for all these results. A summary of the obtained masses is available in the end of the section, allowing us to conclude on the quality of the obtained data. The conclusion is in the section 4. To end this chapter, the detailed calculations that are required to find the propagator of a charge conjugate quark are proposed in section 5.

# 2. Theoretical study

## 2.1 Employed method

To model a diquark, the main idea of the approach consists to take again what was made for mesons in order to adapt it [6, 17]. A meson was treated as a loop of two quarks, one going towards the future and another one going towards the past. This quark is in fact an anti-quark according to the Feynman point of view[1]. For a diquark, it is necessary to replace this anti-quark by a quark. To perform it, a trick consists to replace the particle propagator by its *charge conjugate* propagator. Thus, the “quark” that went towards the past (i.e. the anti-quark) will become virtually a “real quark”, or at least it will act like a real one. The figure 1 hereafter proposes to summarize and to explain the method by the form of diagrams. Rigorously, this approach only “mimics” the behavior of quarks by applying the charge conjugation to anti-quarks.

[1] Feynman R P 1949 The Theory of Positrons *Phys. Rev.* **76** 749–59

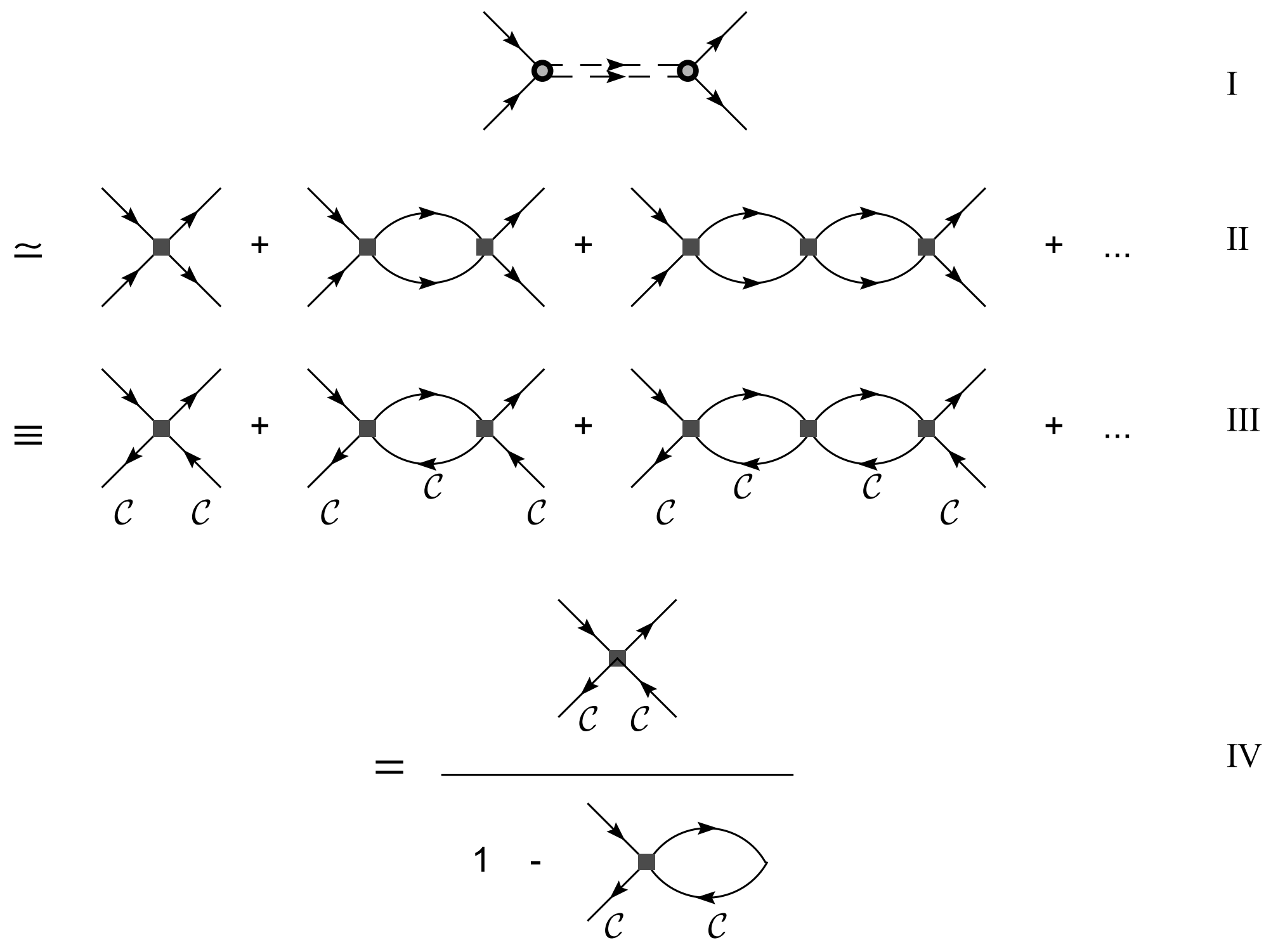

**Figure 1.** Schematization of the method to model the diquarks.

The trick appears in the passage from the second to the third line of the figure 1. Each charge conjugate anti-quark is indicated with the symbol of the charge conjugation $\mathcal{C}$. It concerns the anti-quarks in the loops, but also the external lines. By analogy with the preceding chapter, we write immediately $\blacksquare \equiv \mathcal{Z}$ to describe the effective coupling at each vertex. In the same manner, the loop function is written as:

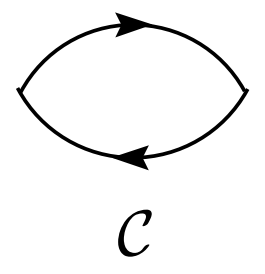

$$\equiv \Pi \quad , \tag{1}$$

so that the transition matrix T is structurally identical to the one seen for the mesons. Starting from the Bethe-Salpeter equation $\mathrm{T} = \mathcal{Z} + \mathcal{Z} \cdot \Pi \cdot \mathrm{T}$, we arrive as in chapter 3 at the relation:

$$\mathrm{T} = \frac{\mathcal{Z}}{1 - \Pi \mathcal{Z}} \ , \tag{2}$$

in which only the $\mathcal{Z}$ and the writing of $\Pi$ differ. In all the cases treated in his chapter, the diquarks are not coupled, as it was the case for some mesons. As a consequence, the equations always uses scalar $\mathcal{Z}$ and $\Pi$. Thus, to obtain the mass $m$ of a diquark with an unspecified momentum $k$, the equation to be solved has the following general form:

$$1 - \Pi\left(k_0, \vec{k}\right) \cdot \mathcal{Z} = 0 \Big|_{k_0 = \sqrt{m^2 + \left(\vec{k}\right)^2},\ \vec{k} \text{ fixed}} . \tag{3}$$

For all the diquarks seen here, the flavor factors are equal to 1 at each vertex implying a diquark and a pair of two quarks, as explained in the appendix C.

## 2.2 Group theory considerations

We begin the analysis by using the work performed in the chapter 1 for the diquarks. Indeed, it was noted that the diquarks can be written as a tensor product of two representations 3 of $SU(3)_f$ associated with the quarks. It leads to two irreducible representations [6],

$$3 \otimes 3 = \overline{3} \oplus 6, \tag{4}$$

so that the diquarks are divided into two groups. The diquarks of the representation 6 can be composed by two identical flavor quarks, whereas this is not possible for those of the representation $\overline{3}$, figure 2.

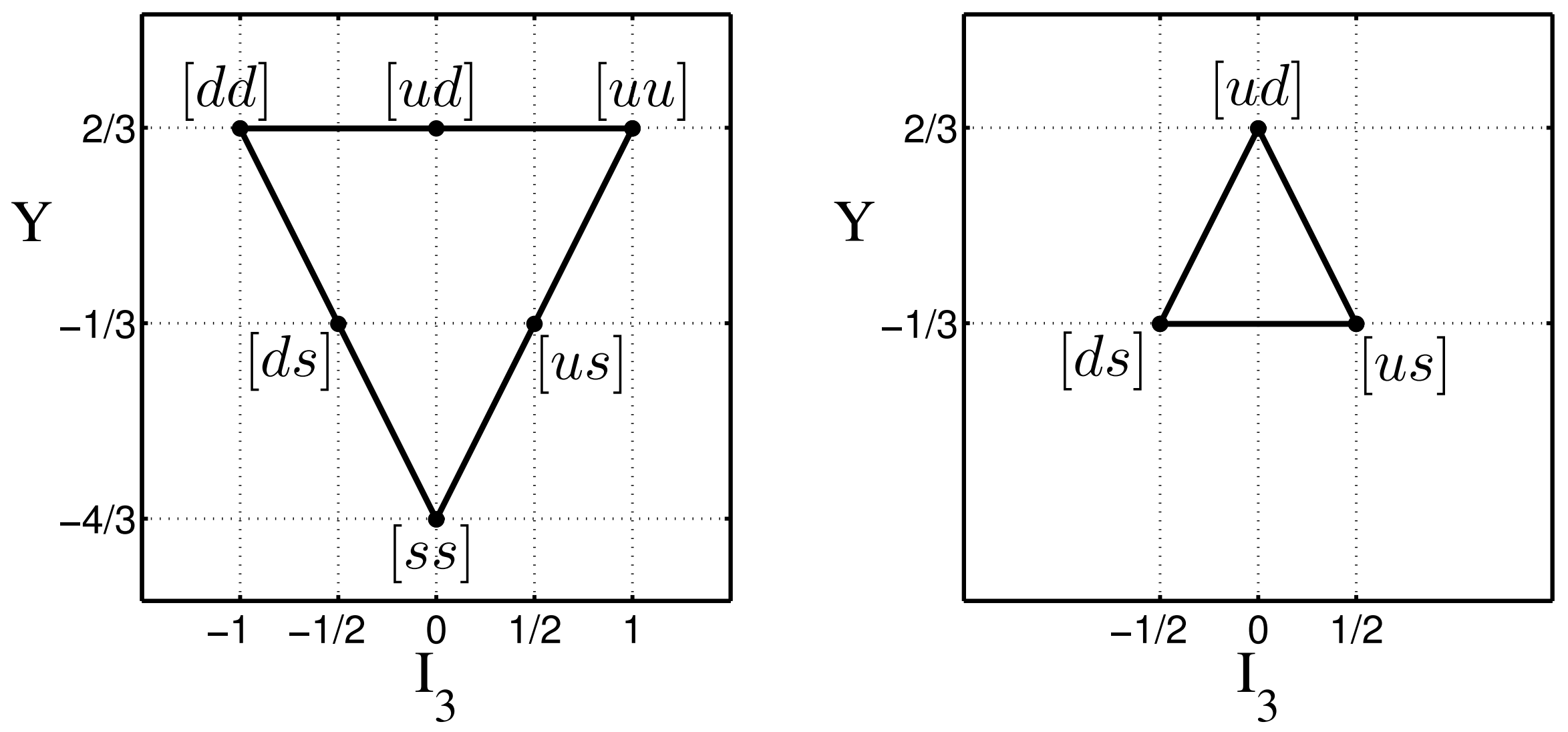

**Figure 2.** Diquarks in the 6 representation (left hand side), and the $\overline{3}$ representation (right hand side).

Indeed, the diquarks of the representation 6 have a flavor symmetrical wave function. For example, for a diquark $[ud]$ from this representation, we have $|[ud]\rangle = \frac{1}{\sqrt{2}}(ud + du)$. On the other hand, the diquarks of the representation $\overline{3}$ have flavor antisymmetric wave functions, so that for a $[ud]$ from this representation $\overline{3}$, the wave function is $|[ud]\rangle = \frac{1}{\sqrt{2}}(ud - du)$ [12].

The same reasoning that we have just made for flavor is feasible in the color space. It thus leads to the same results, i.e. two possible groups. Firstly, it concerns the diquarks that have a color symmetrical wave function, i.e. it corresponds to a 6 representation. The other diquarks have color antisymmetric wave function, with the $\overline{3}$ representation. Nevertheless, as specified in [6], as the finality is to be able to build baryons using diquarks, we badly see how diquarks symmetrical in color could lead to this objective. More precisely, a baryon is a “white” object, therefore composed by quarks with different colors. A diquark made for example by two blue quarks associated with another quark with unspecified color cannot give a baryon. The diquarks modeled in this chapter are thus color antisymmetric. Besides, it allows diquarks as $[uu]$ to respect the Pauli Exclusion Principle.

For mesons, the conserved currents have the form $\bar{\psi}\gamma_\mu\Gamma\psi$. The equivalent conserved currents for the diquarks are obtained while considering the interaction between two quarks. It leads to:

$$\bar{\psi}^C\gamma_\mu\Gamma\psi, \tag{5}$$

where it appears the charge conjugate field $\bar{\psi}^C$. This one is defined according to [6] by:

$$\psi^C = C\gamma_0\psi^* \quad \text{or} \quad \bar{\psi}^C = -\psi^T C^{-1} = \psi^T C. \tag{6}$$

In these equations, * indicates the complex conjugation operator and $^T$ is the transposition operator. The $\Gamma^j$ corresponds to the interaction channel. Also, $C$ is the charge conjugation operator, defined by $C = i\gamma^0\gamma^2$ in the Dirac representation. In addition, like with the mesons, we have several possible types of diquarks. Scalar diquarks and pseudo-scalar ones are spin 0 particle, whereas the spin of the vectorial and axial diquarks is equal to 1. The table 1, inspired from [6], proposes to gather these data. About the conserved currents, we note the introduction of the totally antisymmetric tensor $\varepsilon^{abc}$, in order to respect the fact that the treated diquarks must be color antisymmetric. Each letter *a*, *b*, *c* refers to colors. The color *a* is associated with the charge conjugate field $\psi^T C^{-1}$, and the color *b* to the field $\psi$. As explained, the diquarks can be flavor symmetrical or antisymmetric, which corresponds, respectively, to the representations $6$ and $\bar{3}$. As a consequence, the generators of $SU(3)_f$ and $\lambda^0$, i.e. the 9th $\lambda^j$ that appear in the conserved currents, are divided into two groups. For the flavor symmetrical diquarks, we have $\lambda^S\Big|_{S=0,1,3,4,6,8}$, whereas for the flavor antisymmetric ones, it concerns $\lambda^A\Big|_{A=2,5,7}$.

| Diquark type | Γ value | Conserved currents | Spin | Diquarks representation | Possible diquarks |
|---|---|---|---|---|---|
| Scalar (S) | $i\gamma^5$ | $\psi_a{}^T C^{-1}\gamma_5\lambda^A\psi_b\, i\cdot\varepsilon^{abc}$ | 0 | $\bar{3}$ | $[ud],[us],[ds]$ |
| Pseudo-scalar (P) | 1 | $\psi_a{}^T C^{-1} i\lambda^A\psi_b\, i\cdot\varepsilon^{abc}$ | | | |
| Vectorial (V) | $\gamma^\mu\cdot i\gamma^5$ | $\psi_a{}^T C^{-1} i\gamma_\mu\gamma_5\lambda^A\psi_b\, i\cdot\varepsilon^{abc}$ | 1 | | |
| Axial (A) | $\gamma^\mu$ | $\psi_a{}^T C^{-1} i\gamma_\mu\lambda^S\psi_b\, i\cdot\varepsilon^{abc}$ | | 6 | $[ud],[us],[ds],$ $[uu],[dd],[ss]$ |

**Table 1.** List of the studied diquarks.

Furthermore, a significant point of the table 1 concerns the names of the diquarks types. Indeed, these ones are reversed compared to the one seen for the mesons. For example, the *scalar* diquarks corresponds to a $\gamma_5$ channel. This behavior is due to the $C = i\gamma^0\gamma^2$ operator, as explained in [6, 12].

# 2.3 Diquarks Lagrangian

In the chapter 2, we saw that the NJL Lagrangian part associated with the diquarks, i.e. $\mathcal{L}_{\text{int } qq}$, was written as:

$$\mathcal{L}_{\text{int } qq} = \sum_{\alpha=S,P,V,A} G_{DIQ}^{\alpha} \cdot \sum_{i,j} \left(\bar{\psi}_a \gamma_\mu \Gamma_\alpha^i \psi_b^C\right)\left(\bar{\psi}_d^C \gamma^\mu \Gamma_\alpha^j \psi_e\right) \cdot \varepsilon^{abc} \cdot \varepsilon_c^{\ de} , \tag{7}$$

in which $i, j$ refer to the flavor matrices $\lambda^j$ included in the $\Gamma_\alpha^{i,j}$ term, where $\alpha$ is associated with the four interaction channels: Scalar, Pseudo-scalar, Vectorial and Axial. The equation (7) is now rewritten, in particular to clarify the summation of these four channels. Thanks to a Fierz transformation, it leads to the expression [6, 7, 15, 18]:

$$\begin{aligned}
\mathcal{L}_{\text{int } qq} = G_{DIQ}^{S} \cdot & \sum_{j=2,5,7} \left(\bar{\psi}_a i\gamma_5 \lambda^j C \bar{\psi}_b^T\right)\left(\bar{\psi}_d^T C^{-1} i\gamma_5 \lambda^j \psi_e\right) \cdot \varepsilon^{abc} \cdot \varepsilon_c^{\ de} \\
+ G_{DIQ}^{P} \cdot & \sum_{j=2,5,7} \left(\bar{\psi}_a \lambda^j C \bar{\psi}_b^T\right)\left(\bar{\psi}_d^T C^{-1} \lambda^j \psi_e\right) \cdot \varepsilon^{abc} \cdot \varepsilon_c^{\ de} \\
+ G_{DIQ}^{V} \cdot & \sum_{j=2,5,7} \left(\bar{\psi}_a i\gamma_5 \cdot \gamma_\mu \lambda^j C \bar{\psi}_b^T\right)\left(\bar{\psi}_d^T C^{-1} i\gamma_5 \cdot \gamma^\mu \lambda^j \psi_e\right) \cdot \varepsilon^{abc} \cdot \varepsilon_c^{\ de} \\
+ G_{DIQ}^{A} \cdot & \sum_{j=0,1,3,4,6,8} \left(\bar{\psi}_a \gamma_\mu \lambda^j C \bar{\psi}_b^T\right)\left(\bar{\psi}_d^T C^{-1} \gamma^\mu \lambda^j \psi_e\right) \cdot \varepsilon^{abc} \cdot \varepsilon_c^{\ de}
\end{aligned} . \tag{8}$$

The $a$, $b$, $c$, $d$, $e$ correspond to colors; we note the implicit summation over $c$. In the development performed in (8), each line corresponds to a channel. For each of them, a constant $G_{DIQ}^{\alpha}$ is used. These four constants are related to the constant $G_{DIQ}$ introduced in the chapter 2, during the description of the parameter sets. We have [6]:

$$\begin{cases} G_{DIQ}^{S} = G_{DIQ}^{P} = G_{DIQ} \\ G_{DIQ}^{V} = G_{DIQ}^{A} = \dfrac{G_{DIQ}}{4} \end{cases} . \tag{9}$$

Clearly, the $\mathcal{Z}$ term used in equations (2, 3) is identified with the constants of (9). We recall that the method was similar for the mesons. As a consequence, for a diquark of the type $\alpha$:

$$\mathcal{Z} = 2 \cdot G_{DIQ}^{\alpha} . \tag{10}$$

in which $\alpha$ can be S, P, V or A, see table 1.

# 2.4 Diquark loop function

The polarization function of a diquark $q_1 q_2$ can be written in two following forms, equation (11). In fact, as show in the appendix D, the two-quark-exchange leaves invariant the two forms presented in (11). Clearly, it implies that the two forms are strictly equivalent. The technical details associated with the calculations, which explain the interest of these two forms, were relayed in the appendix E.

$$
\begin{aligned}
&-i\cdot\Pi^{\Gamma}_{q_1q_2}\left(i\cdot\nu_m,\vec{k}\right)=-\frac{i}{\beta}\cdot\sum_n\int\frac{\mathrm{d}^3p}{(2\pi)^3}Tr\left(i\cdot S_{f_1}\left(i\cdot\omega_n-i\cdot\nu_m,\vec{p}-\vec{k}\right)\cdot\Gamma\cdot i\cdot S_{f_2}{}^{C}\left(i\cdot\omega_n,\vec{p}\right)\cdot\Gamma\right)\\
&-i\cdot\Pi^{\Gamma}_{q_1q_2}\left(i\cdot\nu_m,\vec{k}\right)=-\frac{i}{\beta}\cdot\sum_n\int\frac{\mathrm{d}^3p}{(2\pi)^3}Tr\left(i\cdot S_{f_1}{}^{C}\left(i\cdot\omega_n-i\cdot\nu_m,\vec{p}-\vec{k}\right)\cdot\Gamma\cdot i\cdot S_{f_2}\left(i\cdot\omega_n,\vec{p}\right)\cdot\Gamma\right)
\end{aligned}
\quad , \tag{11}
$$

where:

$$
S_f{}^{C}{}_{NJL}\left(\not{p}\right)=\frac{1}{\not{p}-\gamma_0\,\mu_f-m_f}\quad\text{and}\quad S_f{}^{C}{}_{PNJL}\left(\not{p}\right)=\frac{1}{\not{p}-\gamma_0\left(\mu_f-iA_4\right)-m_f} \tag{12}
$$

are the propagators of a charge conjugate quark, in the NJL and in the PNJL models [20, 21]. The complete calculations performed to establish the NJL propagator (12) are gathered at the end of this chapter. About the modifications to be applied to the NJL equations to obtain the PNJL ones, we see that we firstly have to replace $\mu_f$ by $\mu_f - iA_4$ in the quarks/antiquarks propagators (and not elsewhere), as noted in [21]. Furthermore, in the framework of the numerical calculations, the other modifications concern the adaptation of the Fermi-Dirac distributions associated with the quarks/antiquarks. Clearly, this adaptation, suggested in [21, 22] for the mesons, is still applicable for diquarks. We recall that the modified Fermi-Dirac distributions in the PNJL model were established in the chapter 2. In practice, as with the mesons, these modifications concern the $\Pi$ function.

# 2.5 Coupling constants

The method to estimate the coupling constants implying a diquark and a quark-quark pair is the same as the one decribed for the mesons. In fact, only the propagator expression differs:

$$
\left.\frac{4G_{DIQ}}{1-2G_{DIQ}\cdot\Pi\left(k_0,\vec{k}\right)}\right|_{k^2=m^2}\approx\left.\frac{-g^2}{k^2-m^2}\right|_{k^2=m^2}. \tag{13}
$$

It gives:

$$
\left.-\frac{1}{2}\cdot\frac{\partial\Pi\left(k_0,\vec{k}\right)}{\partial k}\right|_{k^2=m^2}\approx\left.\frac{2k}{-g^2}\right|_{k^2=m^2}. \tag{14}
$$

$k^2 = m^2$ is equivalent to write $\begin{cases}k_0=m\\ \vec{k}=\vec{0}\end{cases}$. So, for a diquark at rest, it leads to the formula:

$$
g=\sqrt{\frac{4m}{\left.\frac{\partial\Pi\left(k_0,\vec{0}\right)}{\partial k_0}\right|_{k_0=m}}}. \tag{15}
$$

# 3. Numerical results

In this section, we propose to study each diquark type. As a whole, we use there the P1 parameter set defined in the chapter 2 (using the isospin symmetry). However, in the subsection 3.6, the EB parameter set is also used to estimate the diquarks masses at null temperature and null density.

## 3.1 Scalar diquarks

The results associated with the scalar diquarks' masses and widths are shown in the figures 3 to 5. Firstly, we remark that the obtained NJL curves can be compared to the ones of [15, 17]. Qualitatively, the agreement is correct with these references, notably because we use a rather similar parameter set, i.e. P1. However, differences are quantitatively observed, for example for the mass at null temperature and density. In fact, we take $G_{DIQ}/G = 0.705$ (see chapter 2), whereas [15, 17] seems to considerer different values. As indicated in [15], their choice was fixed in order to obtain good agreement for the nucleons. More precisely, their results can be found again with $G_{DIQ}/G \approx 0.735$, i.e. a value close to the one indicated in [16], but not the $G_{DIQ}/G \approx 0.55$ one announced in [15, 17].

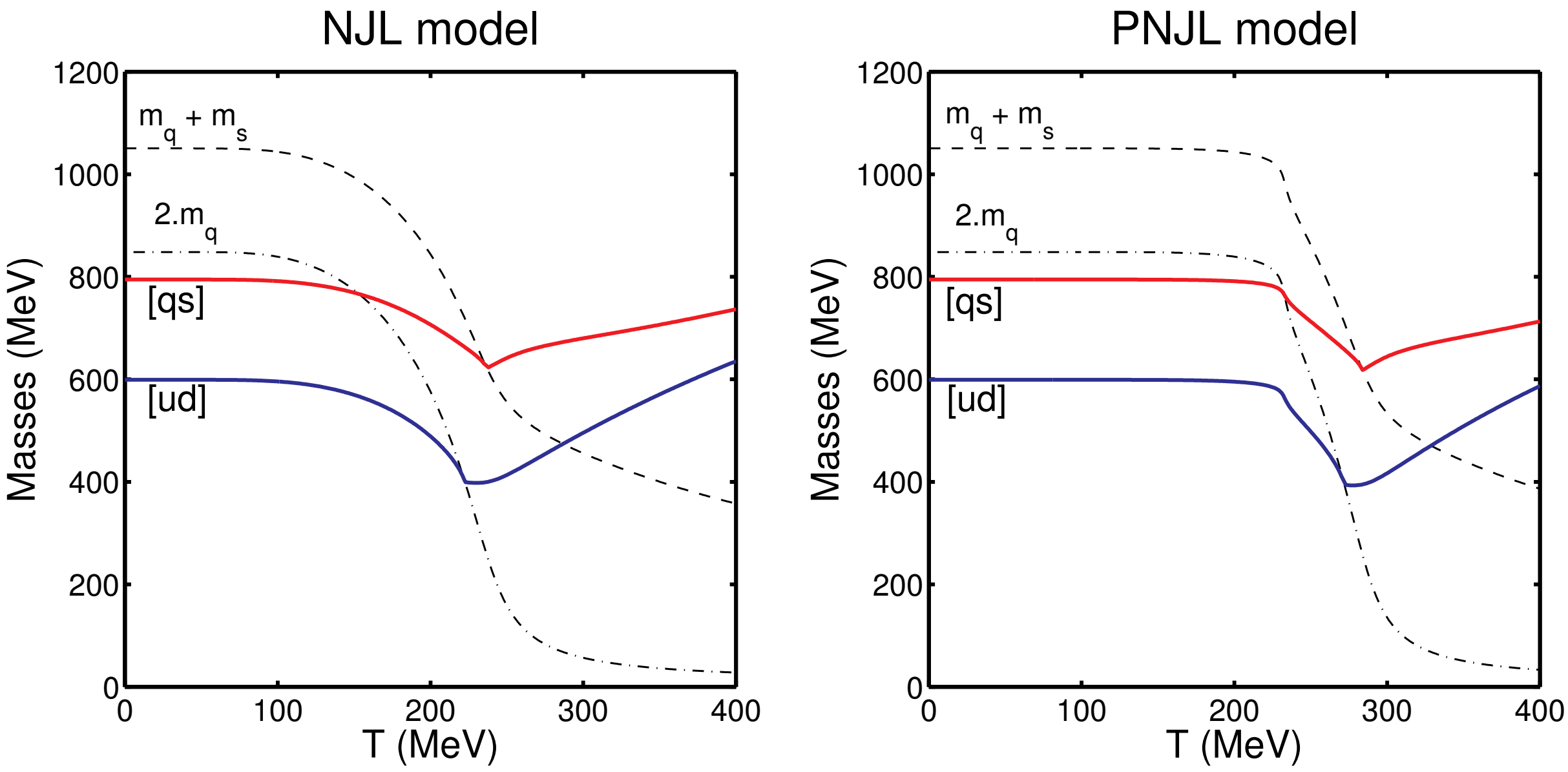

**Figure 3.** Masses of the scalar diquarks according to the temperature.

As confirmed by the study of the widths in the figure 5, the scalar diquarks are stable at reduced temperature and density. Of course, this notion of stability refers to their disintegration into a quark-quark pair. The stability of these diquarks is also observed for the PNJL results. In fact, concerning the differences between the NJL and PNJL descriptions, the figures 3 to 5 exhibit the same behavior as the one obtained for the quarks and the mesons. On one side, it consists to a deformation of the PNJL curves, compared to the NJL ones, towards higher temperatures. At reduced temperatures, it leads to rather constant diquarks masses for the PNJL curves, until $T \approx 200$ MeV. On the other side, NJL and PNJL results coincide at null temperature, whatever the baryonic density. Moreover, the great resemblance between the diquarks and the mesons' curves can be explained by the fact that the equations to be solved are very similar with these two kinds of particles. More precisely, the behavior of the scalar diquarks strongly recalls the one of the pseudo-scalar mesons. In fact, these particles are associated with the $\gamma^5$ interaction channel, and the resemblance between their equations can be easily checked in the appendix D. However, the masses of the diquarks are higher compared to the ones on the pseudo-scalar mesons, leading to an expected result: these diquarks are less stable than these mesons.

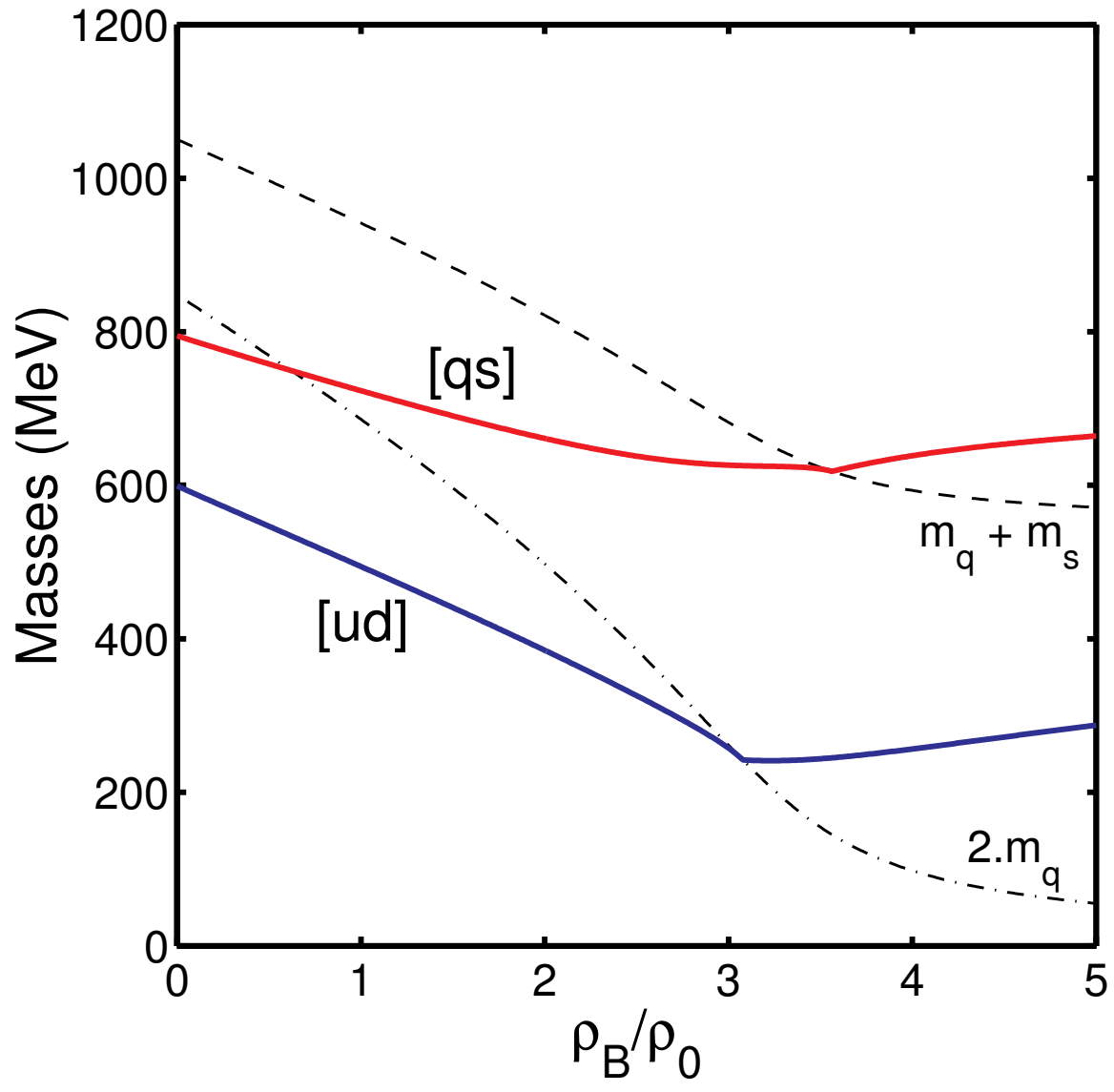

**Figure 4.** Masses of the scalar diquarks according to the baryonic density.

Another difference between these two kinds of particles is the evolution of their masses according to the temperature or to the baryonic density. Indeed, for the scalar diquarks, it is observed in figures 3, 4 a decreasing of the masses when $T$ or $\rho_B$ increases, until their limit of stability. More precisely, a critical temperature and a critical density are present for each diquark, respectively in the figure 3 and in the figure 4. This behavior, especially according to the temperature, recall the one found for vectorial mesons, for which the 't Hooft term is not applied. In fact, a strong baryonic density corresponds to a broad excess of quarks $q$ compared to the antiquarks $\bar{q}$. So, a structure made by quarks, as a diquark, could be "stabilized" by an increase of the baryonic density. This observation may explain the masses' evolution in the figure 4. But, the masses of the quarks that compose our diquarks decrease more rapidly than the diquarks' masses. Thus, the binding energies of these ones, e.g. $m_{[ud]} - 2m_q$, tend to decrease (in absolute values). It can explain the stable/unstable transition observed in the figure 4, for approximately $3\rho_0$ for $[ud]$ and $\approx 3.5\rho_0$ for $[us]$.

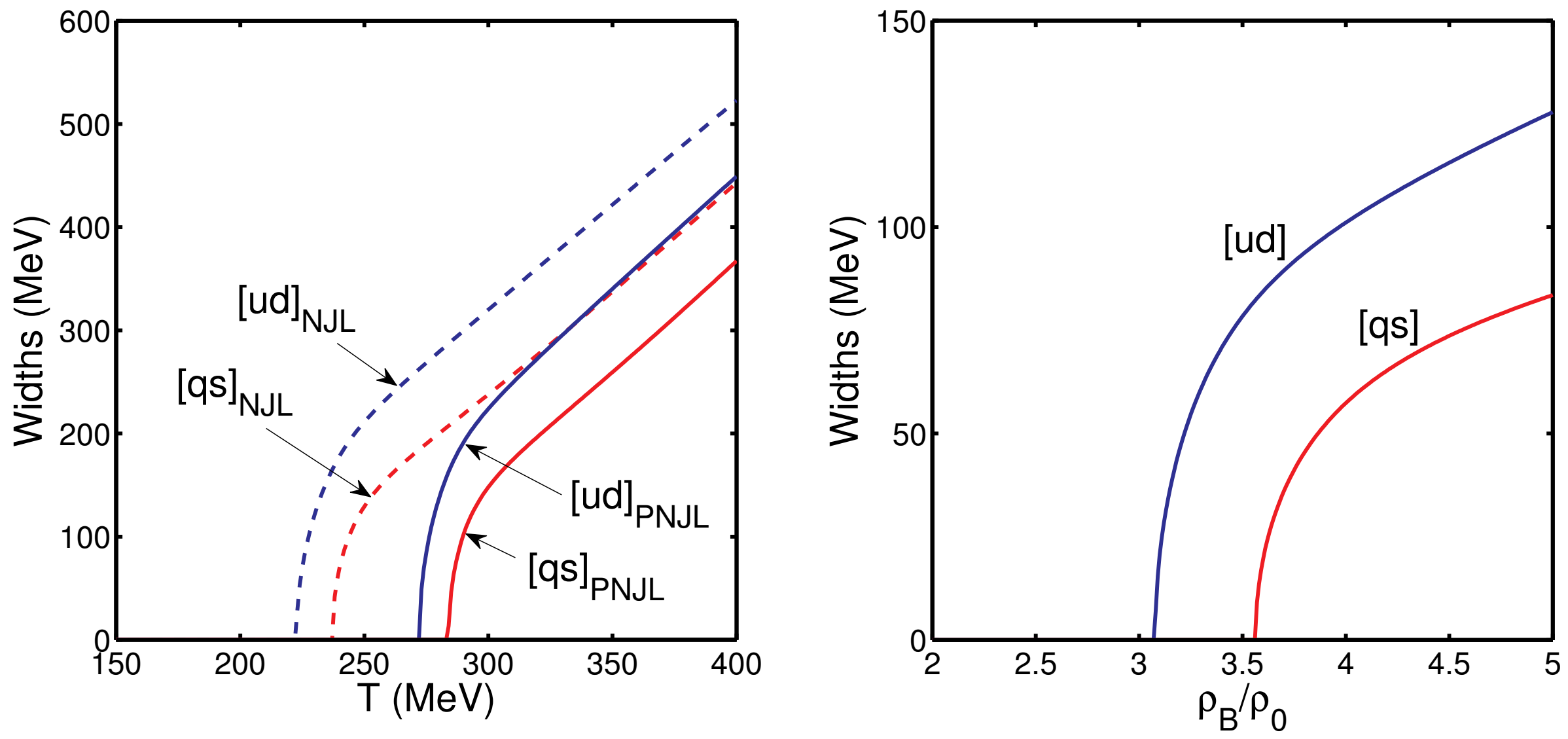

**Figure 5.** Widths of the scalar diquarks according to the temperature and the baryonic density.

The masses of the scalar diquarks seem to be low enough to consider them as good candidate to form baryons. More precisely, the mass of $[ud]$ is about 600 MeV at null temperature and density. Associated with a light quark, whose mass is there about 420 MeV, it is reasonable to envisage to create a nucleon from such a system. It justifies the estimation of the coupling constants of these diquarks, notably to use these data in the baryons' modeling, and then in the associated cross-sections. Our results are proposed in the figures 6 and 7. As with the masses, our pure NJL curves can be compared to the ones of [15, 17], leading to similar remarks.

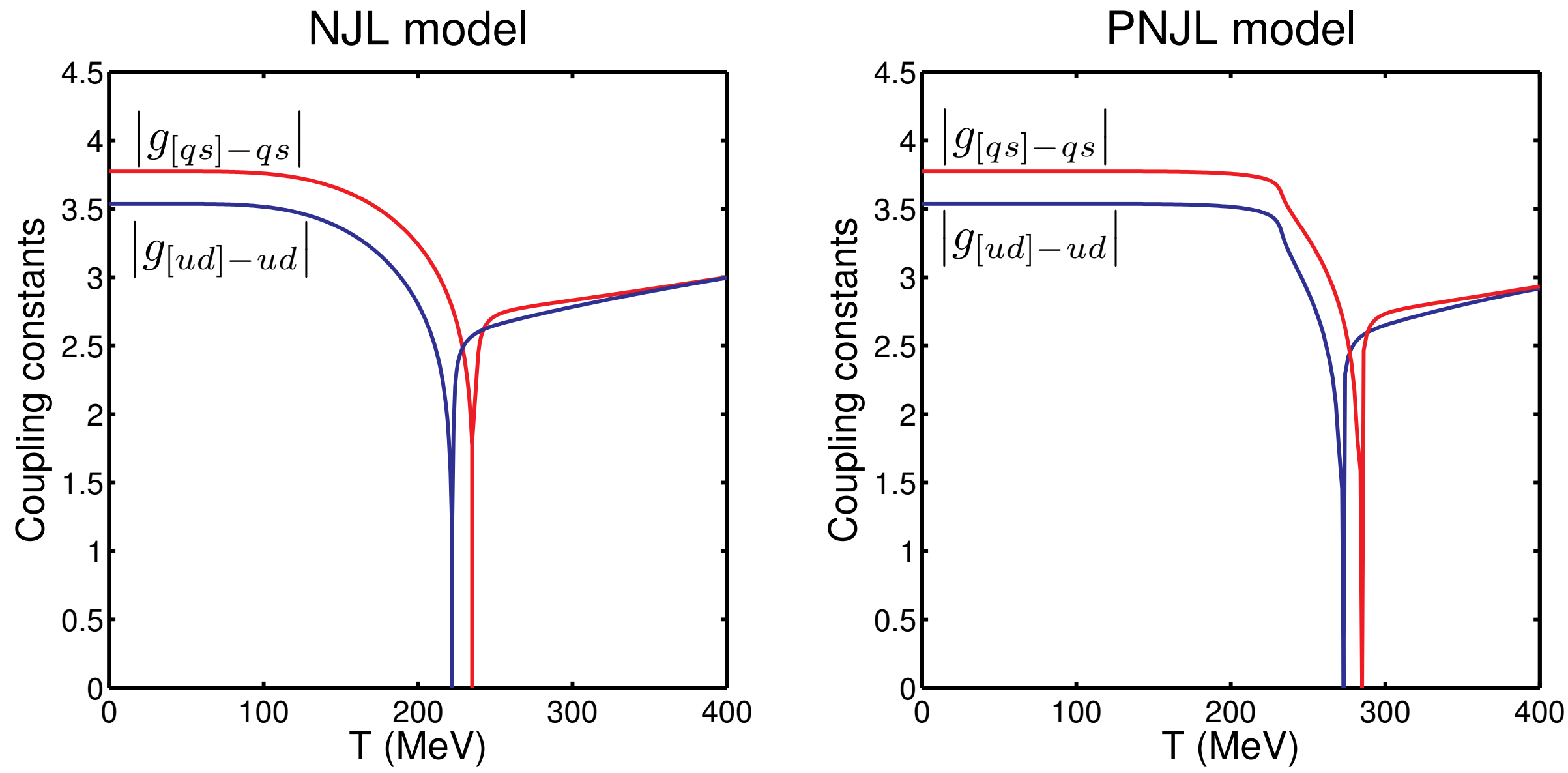

**Figure 6.** Coupling constants of the scalar diquarks according to the temperature.

In fact, the curves strongly recall the ones found for the pions and the kaons, especially according to the temperature, figure 6. Even quantitatively, the values of the coupling constants stay rather close to 4 at low temperatures. Moreover, the strong decrease ( $g \to 0$ ) is also observed with the diquarks. This strong decrease corresponds to a stable/unstable transition, so it confirms our previous observations for these diquarks. The analysis will be extended at negative densities in the subsection 3.5.

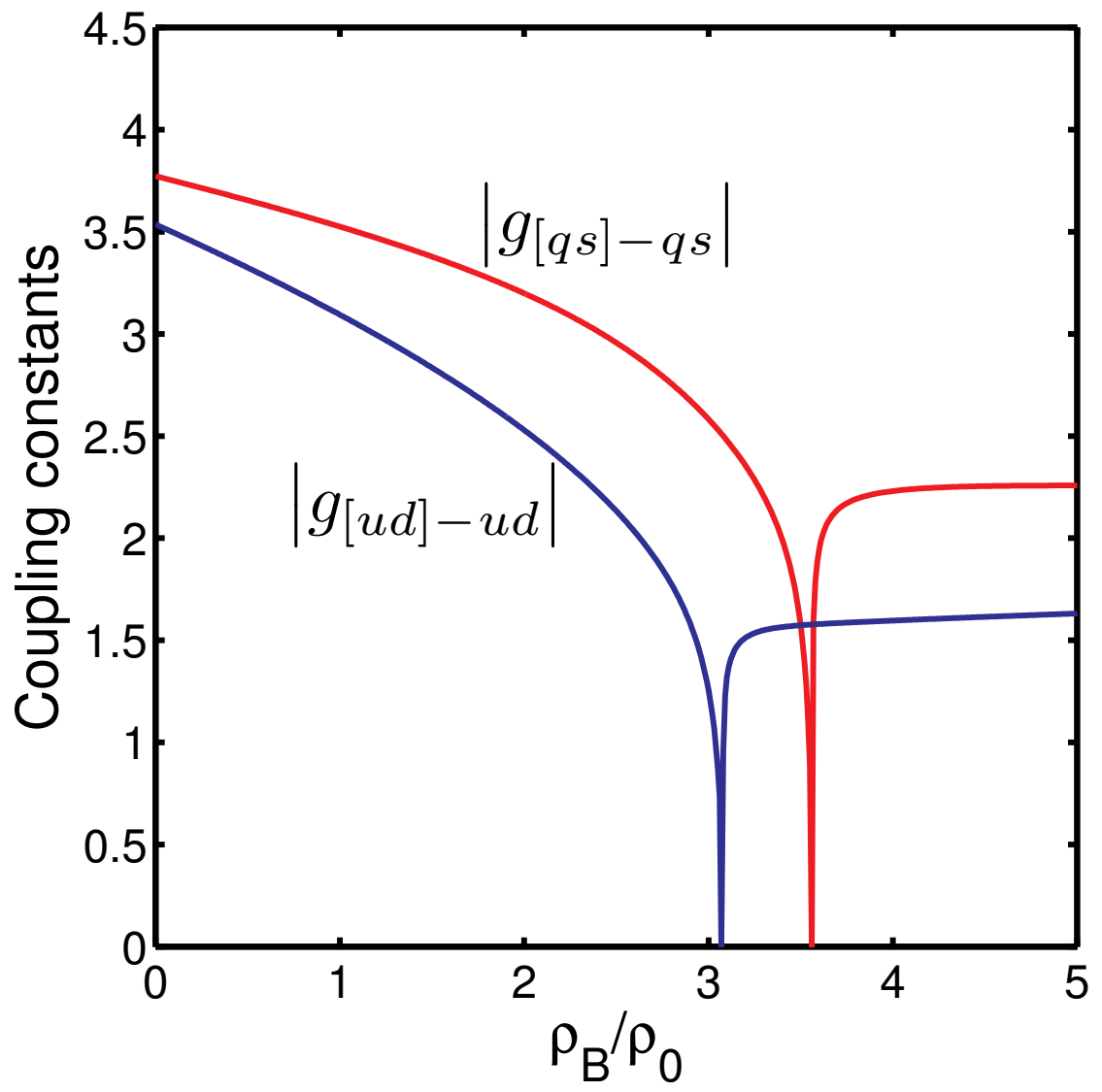

**Figure 7.** Coupling constants of the scalar diquarks according to the baryonic density.

# 3.2 Pseudo-scalar diquarks

The pseudo-scalar and the vectorial diquarks are not treated in the literature. As a consequence, we cannot propose a comparison of our results with other ones. Concerning the pseudo-scalar diquarks, we studied their masses and widths according to the temperature in the figures 8 and 9, and according to the baryonic density in the figure 10. In fact, the behavior of the pseudo-scalar diquarks is qualitatively very close to the one observed for the scalar mesons. Indeed, we found that these particles are always unstable, whatever the temperature or the baryonic density: the masses of the pseudo scalar diquarks are always higher than the masses of the quarks that constitute them, figure 8. Furthermore, the associated widths are always non-null, figure 9. About the evolution of the masses according to the temperature, we note that the masses of these diquarks evolve in a parallel way compared to the ones of the quarks that constitute them, i.e. respectively $2m_q$ for $[ud]$ and $m_q+m_s$ for $[qs]$, until $T \approx 180$ MeV for the NJL model and $T \approx 220$ MeV for the PNJL one. It leads to stable values of the widths, figure 9, until reaching these temperatures. Such a behavior is also present according to the baryonic density, even if it is there less marked.

Quantitatively, the masses of the pseudo-scalar diquarks are relatively strong at low temperatures and densities, about 920 MeV for the $[ud]$ and 1140 MeV for the $[qs]$. In fact, $[ud]$ has a mass close to the one of a nucleon. Thus, in the framework of the nucleon's modeling, we can guess that the contribution of the pseudo-scalar diquarks should not be dominant, and even should be rather reduced. Indeed, if we associate the $[ud]$ pseudo-scalar diquark with a quark, it would require a binding energy higher than 400 MeV to obtain the mass of nucleon close to its actual one, i.e. 938 MeV. Such a value of the binding energy seems to be too strong to be realistic. However, as explained in the introduction of this chapter, the pseudo-scalar diquarks are not useless: they can be considered in cross sections calculations, as propagators, in the framework of reactions involving baryons.

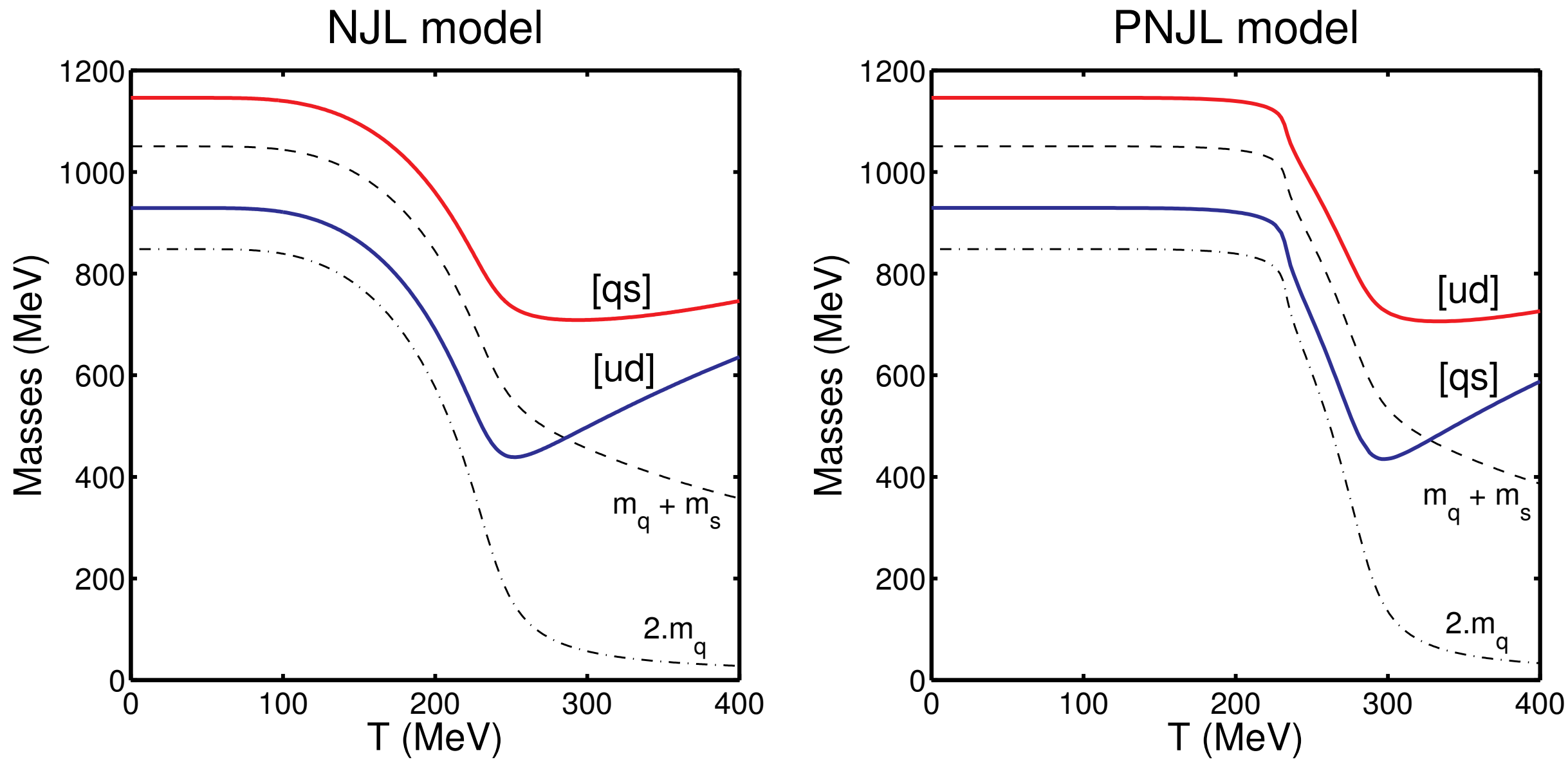

**Figure 8.** Masses of the pseudo scalar diquarks' masses according to the temperature.

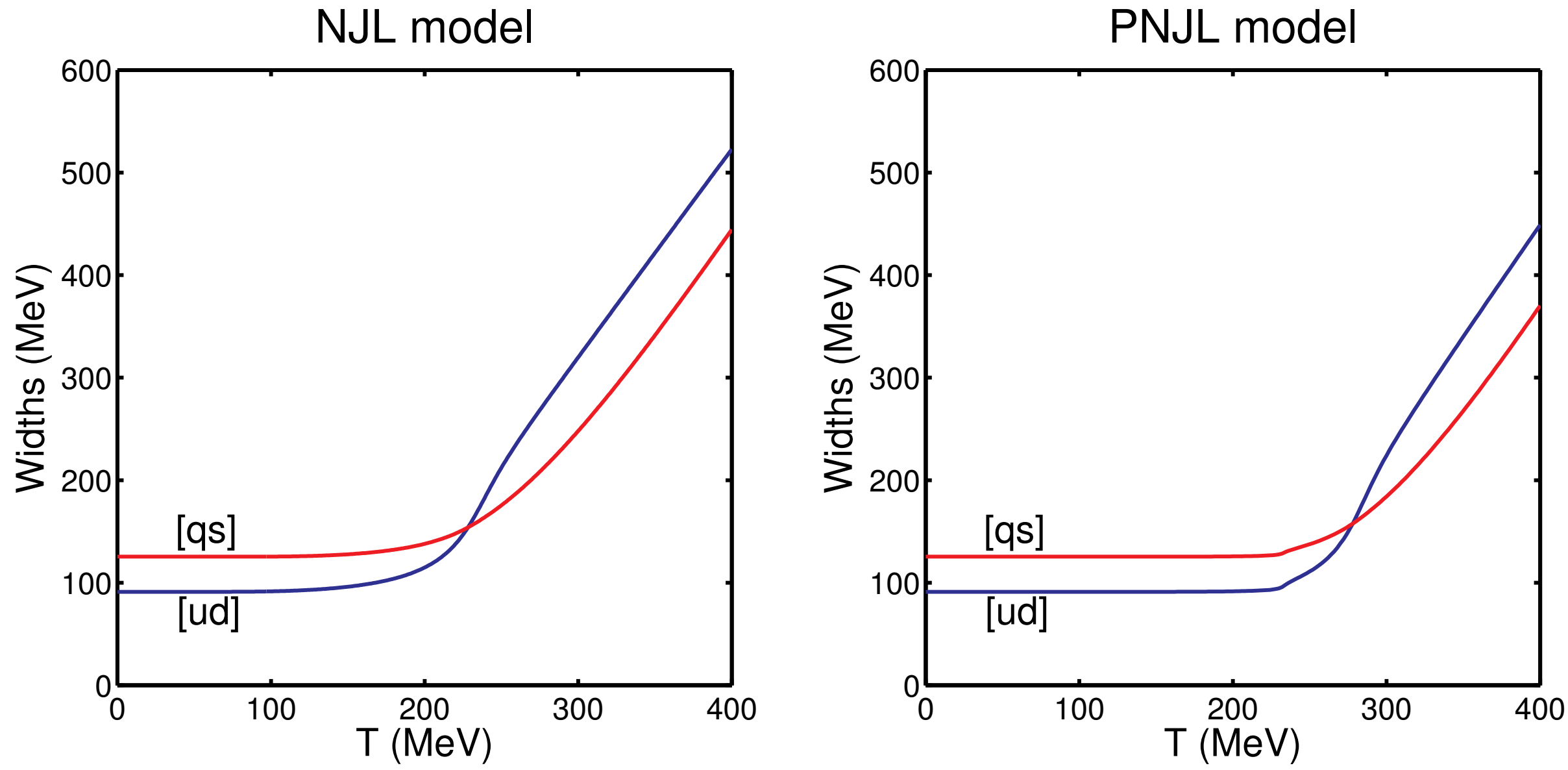

**Figure 9.** Widths of the pseudo scalar diquarks, function of the temperature.

Moreover, as with the mesons, we observe that the $[ud]$ scalar and pseudo-scalar diquarks become degenerate at high temperatures and high densities, figures 3, 8 and 4, 10. This remark is also valid for $[us]$, but the degeneracy occurs for stronger $T, \rho_B$ values. In the same way, this behavior will be observed for the vectorial and axial diquarks $[ud]$ and $[us]$.

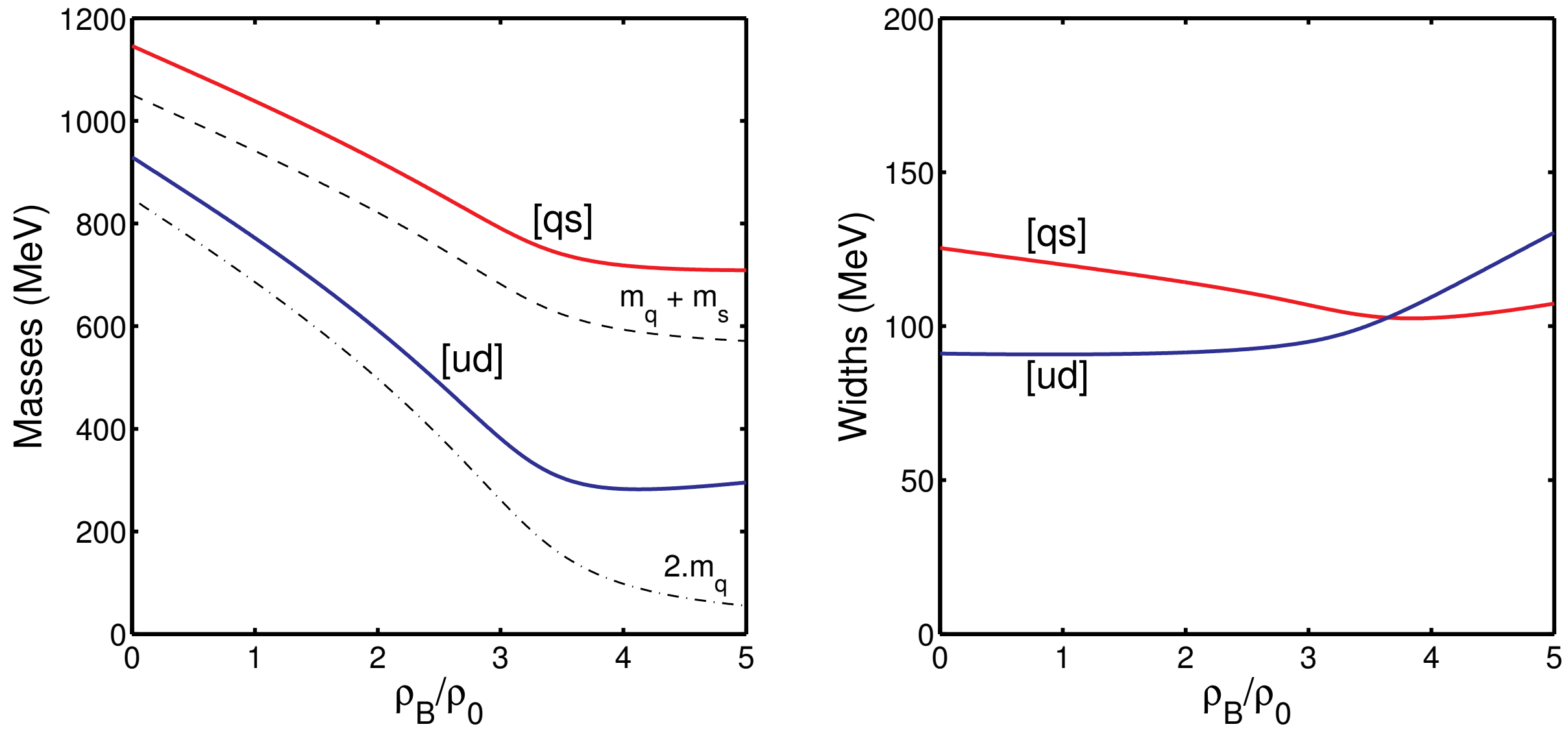

**Figure 10.** Masses and widths of the pseudo scalar diquarks function of the baryonic density.

## 3.3 Vectorial diquarks

The evolution of the vectorial diquarks according to the temperature is represented in the figure 11. The influence of the baryonic density is studied in the figure 12. These results are structurally similar to the ones seen for the pseudo-scalar diquarks. The vectorial diquarks are the heaviest diquarks studied in our work. They strongly recall the axial mesons. As with these particles, the strong masses could suggest that we could be there too at the limit of reliability of our approach. As with the pseudo-scalar diquarks, a possible application of these particles may concern their use as propagators in cross-sections calculations. However, if we really used the pseudo-scalar diquarks for that purpose, it is not the case for the vectorial ones, at least in the framework of the work presented in this thesis.

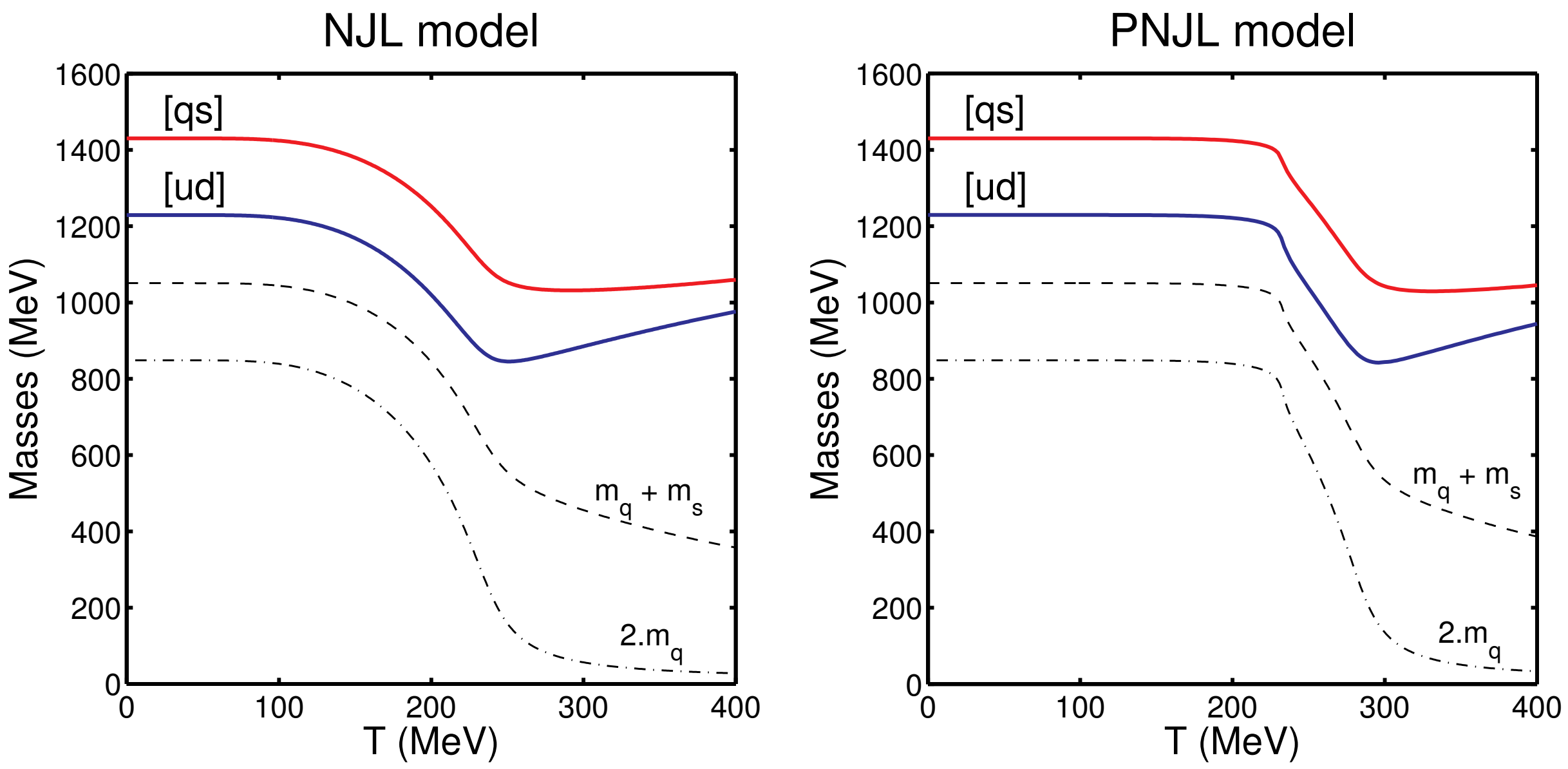

**Figure 11.** Masses of the vectorial diquarks function of the temperature.

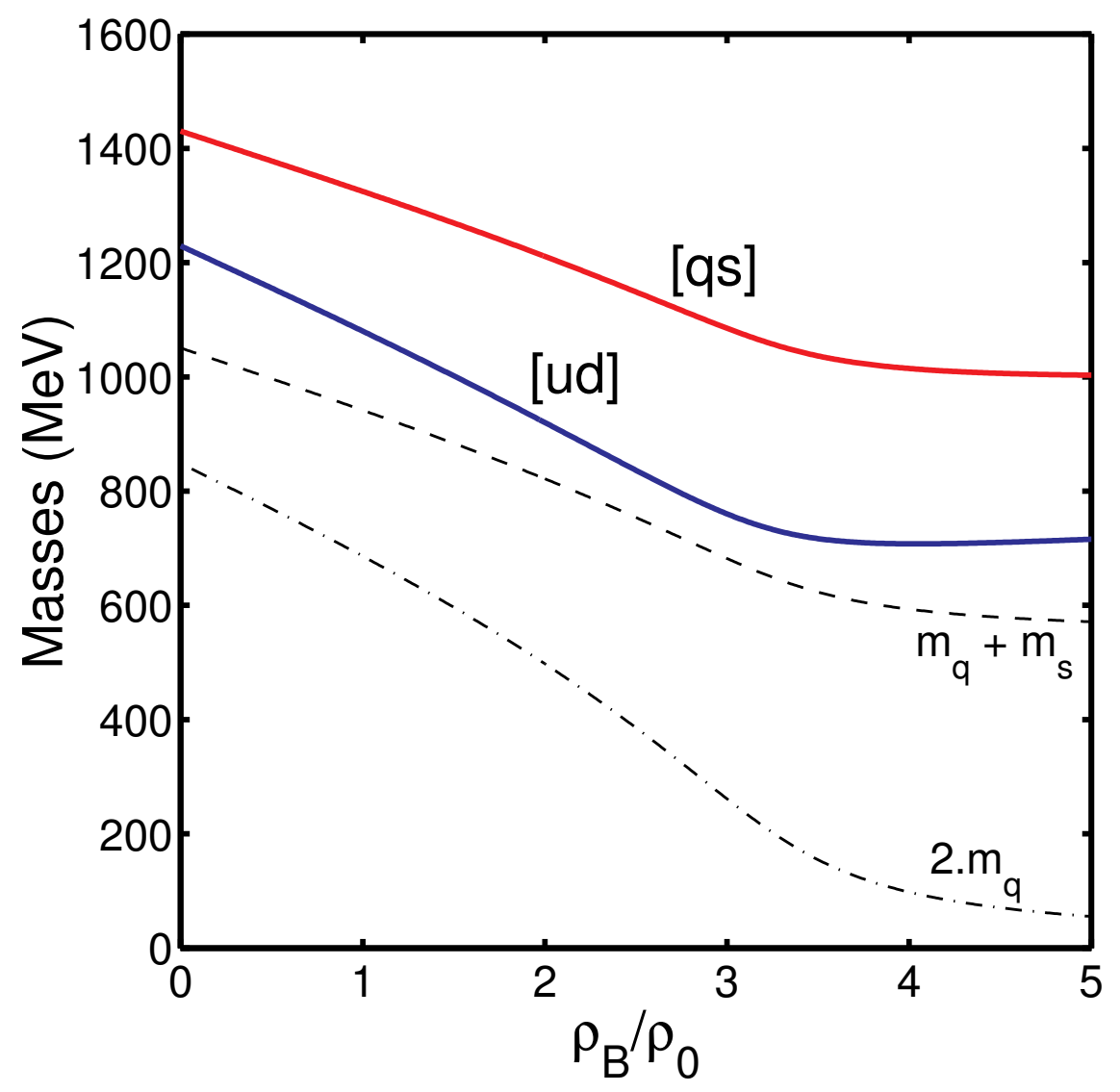

**Figure 12.** Masses of the vectorial diquarks function of the baryonic density.

# 3.4 Axial diquarks

In our work, the axial diquarks are the only ones that can have two quarks with the same flavor. Thanks to this property, these diquarks can intervene in the modeling of baryons as $\Delta$, $\Omega$, or for the axial flavor components of other baryons. As a consequence, they are particularly relevant particles. Our results concerning the evolution of their masses according to the temperature and the baryonic density are presented in the figures 13 and 14. In the literature, we do not have element of comparison, except values estimated at null temperature and null baryonic density, e.g. [11–13].

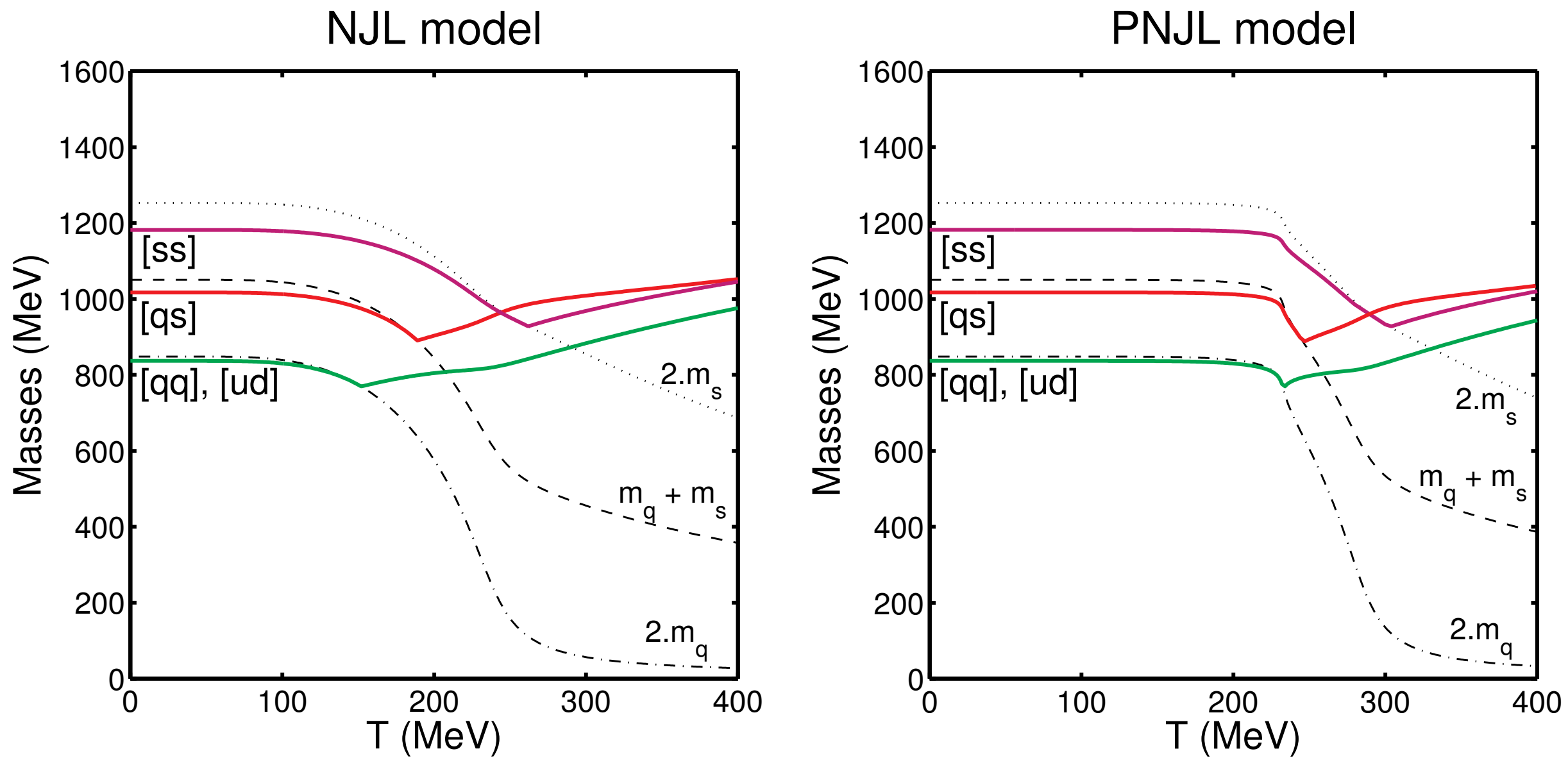

**Figure 13.** Masses of the axial diquarks function of the temperature.

In our two figures, we found that the axial diquarks are stable at reduced temperatures and baryonic densities. Indeed, even if we do not present here the evolution of the associated widths, this information can be easily be found thanks to the behavior of the curves, but also with the results of the figures 15, 16, 18. In fact, qualitatively, the evolution of the axial diquarks' masses recalls the one observed for the scalar ones. However, in their stability zone, the masses of the axial diquarks are very close to the mass of the quark-quark pair that composes them. It wants to say that their binding energy is rather modest (in absolute value). As a consequence, the axial diquarks are less stable than the scalar one. This remark is confirmed in the figure 18, if we compare the stability zone of each diquark. But, this affirmation is not valid for the $[ss]$ diquark. In fact, this particle is composed by two strange quarks, thus $[ss]$ is less sensitive to the temperature and the baryonic density, as the $s$ quark in the chapter 2. Except for $[ss]$ according to the baryonic density, the studied diquarks have a critical temperature and a critical density. Thus, we have a clear separation between their stability and instability zones according to these two parameters.

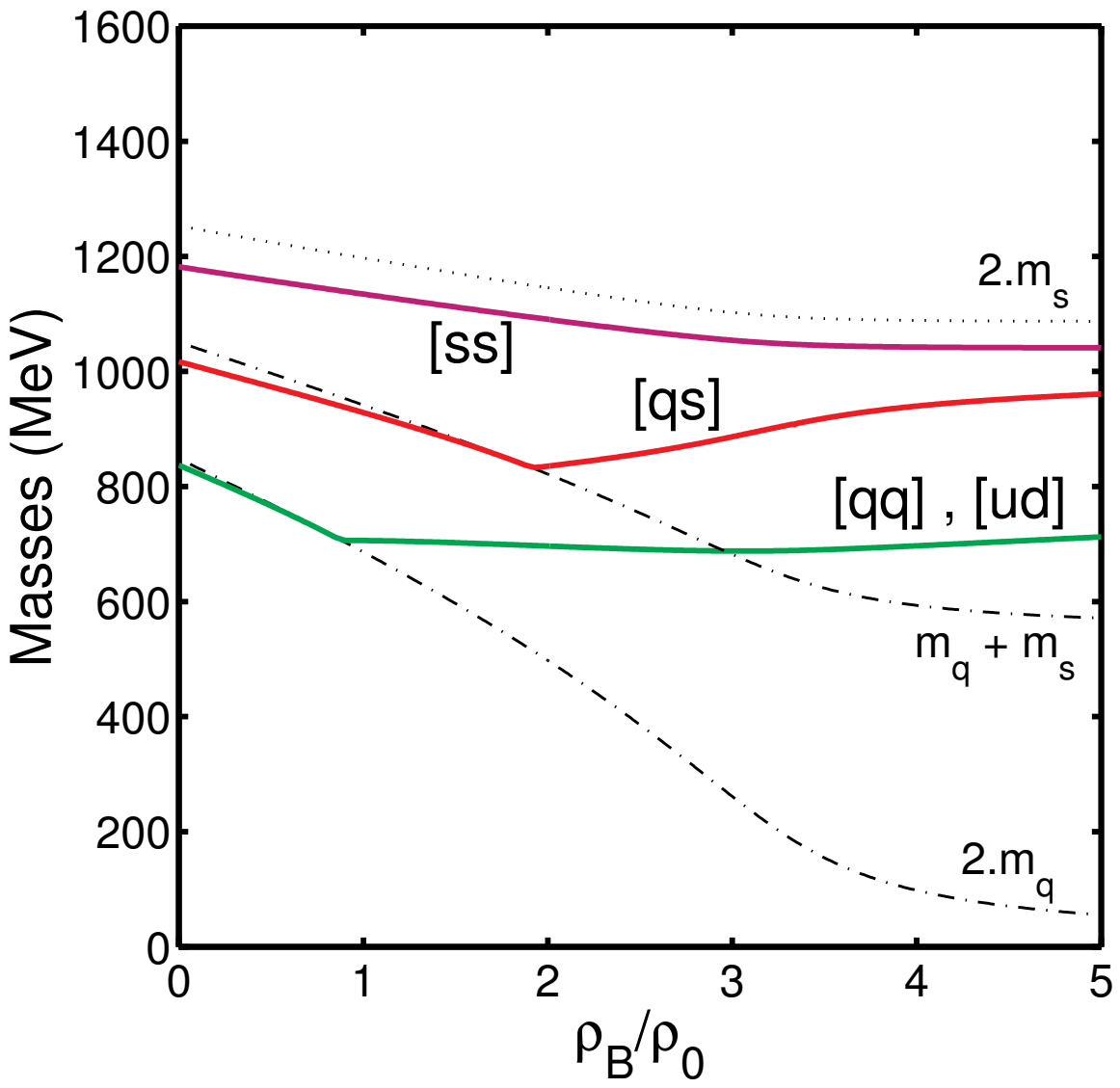

**Figure 14.** Masses of the axial diquarks according to the baryonic density.

Aware of the interest of the axial diquarks, we propose hereafter to study their coupling constants. It corresponds to the figures 15 and 16. Firstly, it can be underlined that their aspect recalls the ones found for the scalar diquarks, or for the mesons. However, some differences are identifiable. It firstly concerns the values at null temperature and density. Except for the $\eta$ meson, we previously found values close to 4, whereas here, we have $\approx 1.73$ for $\left|g_{[qq]-qq}\right|$, 2.32 for $\left|g_{[qs]-qs}\right|$ and 2.85 for $\left|g_{[ss]-ss}\right|$. For the axial diquarks, the lightest ones present the weakest coupling constants at $T=0$ and $\rho_B=0$, and conversely. At temperatures located in the instability zone of the diquarks, the coupling constants become stronger, see figure 15. They exceed the values they had at low temperatures; they are upper than 3.5. In this figure, the curves also give the impression to converge towards very close values. Obviously, these remarks are verified in the NJL and in the PNJL models. On the other hand, at high densities (figure 16), the coupling constants keep very distinct values. But, the curves tend towards a fast stabilization: towards approximately 2.5 for $\left|g_{[ss]-ss}\right|$, 2.8 for $\left|g_{[qq]-qq}\right|$ and 3.1 for $\left|g_{[qs]-qs}\right|$.

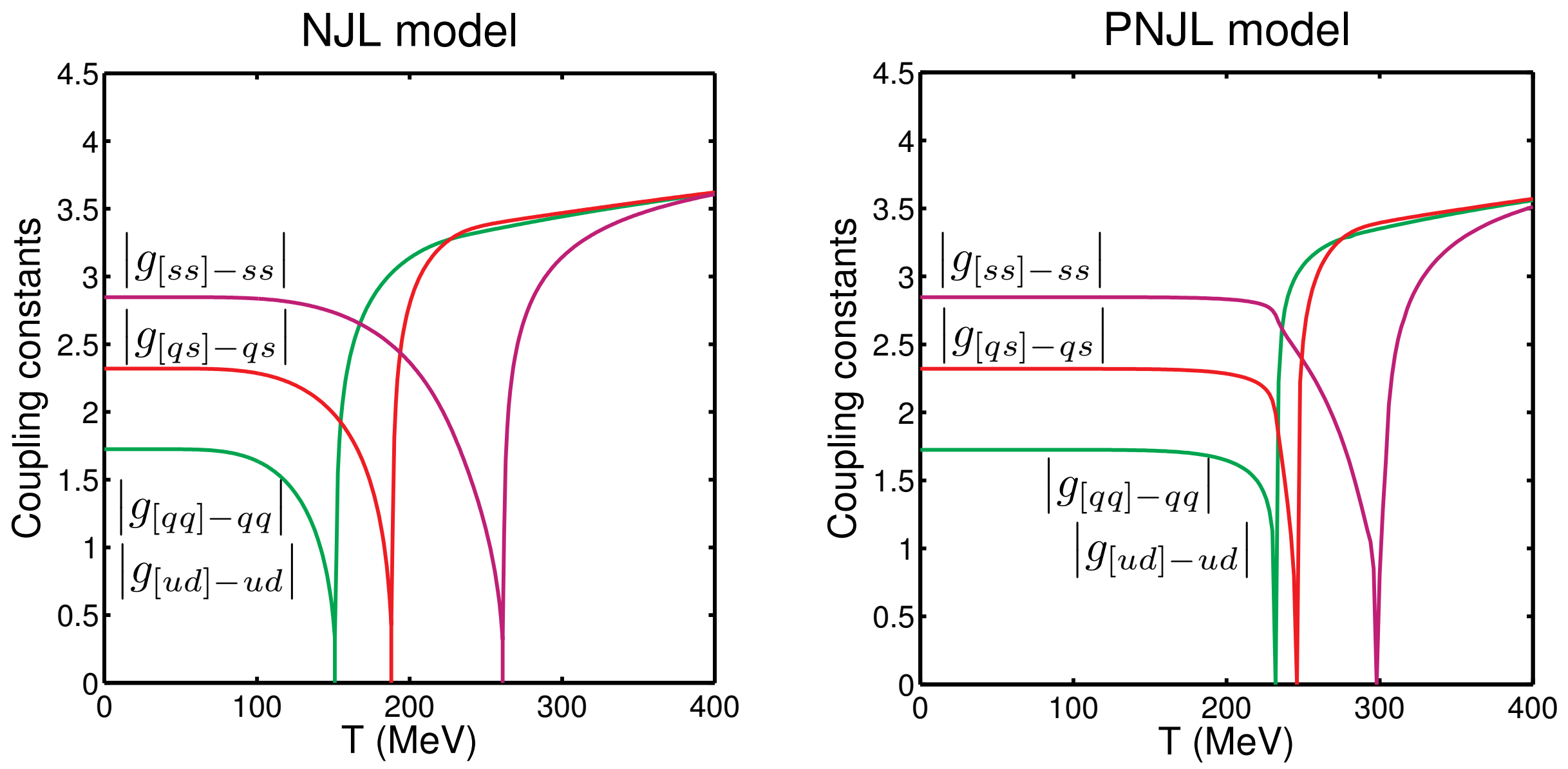

**Figure 15.** Coupling constants of the axial diquarks according to the temperature.

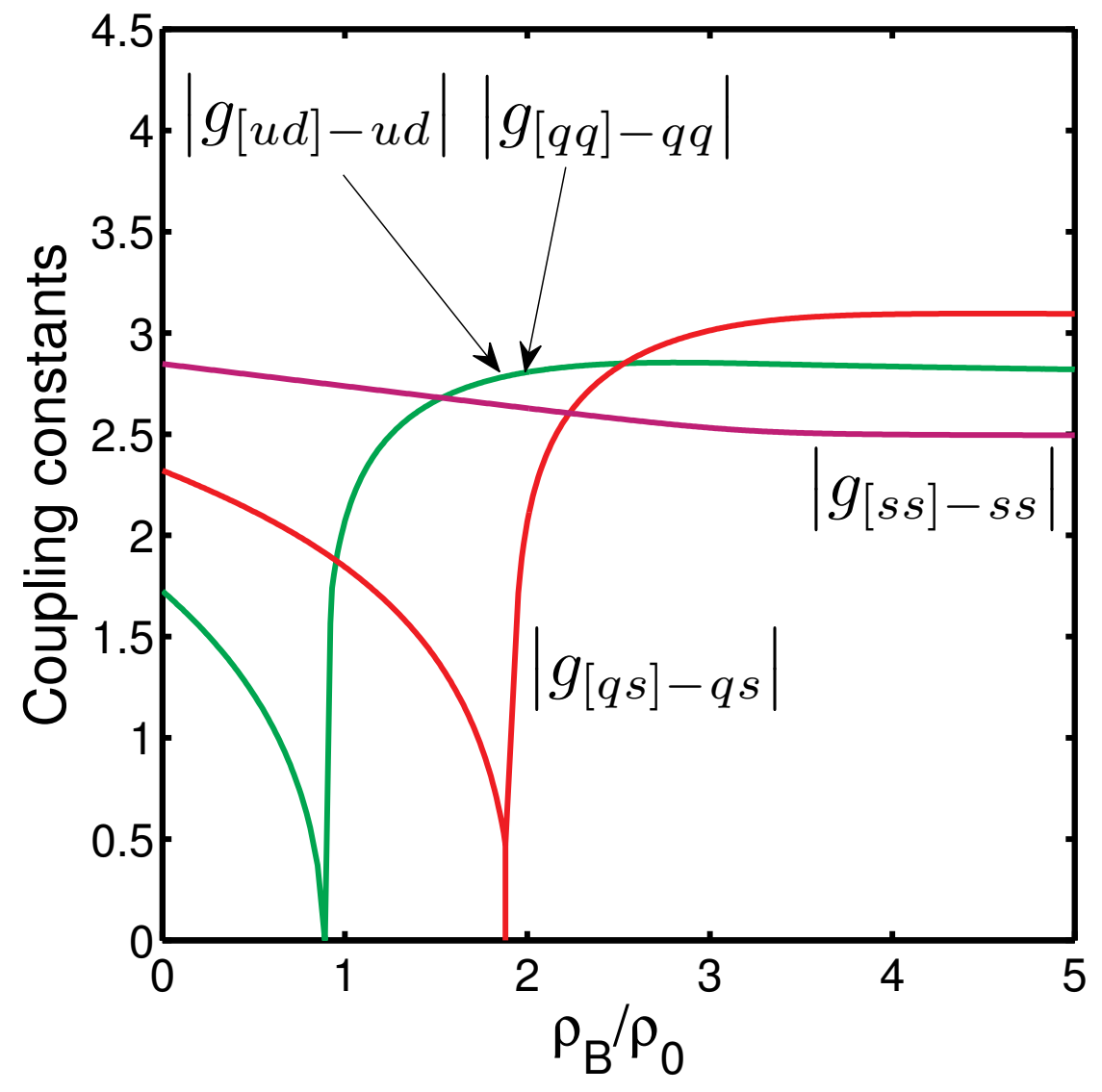

**Figure 16.** Coupling constants of the axial diquarks function of the baryonic density.

# 3.5 Other results

We propose now to extend the results found for the scalar and the axial diquarks to negative densities. More precisely, we studied the behavior of anti-diquarks according to the baryonic density, figure 17, and performed a "phase diagram" extended to negative densities, figure 18.

About the results of the figure 17, our method considers the matter-antimatter symmetry. It consists to say that a particle plunged in a medium in which the density is equal to $-\rho_B$ (negative value, i.e. a medium dominated by the antimatter) acts in the same way as its associated antiparticle plunged in a medium in which the density is equal to $\rho_B$ (i.e. dominated by the matter), and conversely. As a consequence, the anti-diquarks can be simply found by reversing the sign of the baryonic density. With the mesons, these particles can be their own antiparticles, especially when the isospin approximation is employed. Thus, we found that the mesons masses evolved in the same way in positive and negatives densities, except for the kaons.

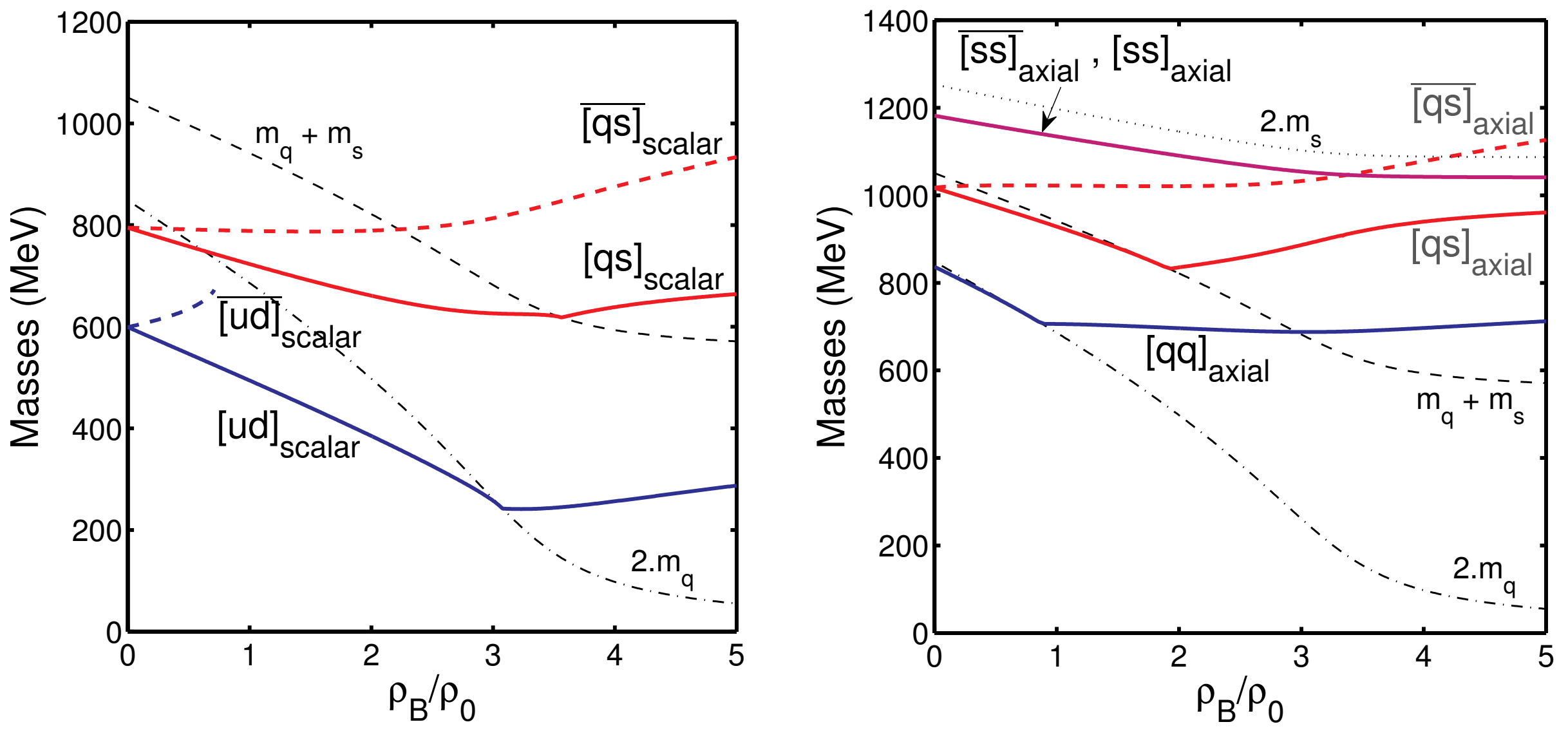

**Figure 17.** Scalar and axial diquarks and their associated anti-diquarks.

Obviously, a diquark and its anti-diquark are expected to act in very different manner, as confirmed in the figure 17. In this figure, the anti-diquarks are represented in dotted lines. The graphs were established at null temperature. Thus, the NJL and the PNJL models strictly coincide. In fact, at null density, the structure of the equations leads to the same masses for each diquark and its associated antiquarks, whatever the temperature. But, at non-null density, the masses of the diquarks/anti-diquarks can be different. This behavior is observed for the diquarks composed by at least one light quark $q$. Indeed, when $\rho_B \neq 0$, the values of the chemical potential $\mu_q$ and $\mu_{\bar{q}}$ are also non-null, and $\mu_{\bar{q}} = -\mu_q$. This sign difference is directly found in the expressions of the diquarks and anti-diquarks loop functions, appendix D, leading to an asymmetry between diquarks and anti-diquarks. It explains the mass splitting. Intuitively, we understand that a couple of diquark/anti-diquark made by two light quarks/antiquarks is more sensitive to this effect than a couple made only by one light quark/antiquark. This is confirmed by the numerical results. Indeed, in the figure 17, we note the difference between the scalar $[ud]$ and $\overline{[ud]}$ masses grows more quickly than for the scalar $[qs]$ and $\overline{[qs]}$ masses. But, this comparison was not possible with their axial counterpart, because the axial $\overline{[qq]}$ could not be calculated numerically. These antiquarks are too sensitive to baryonic density to be modeled in a reliable way. This observation also explain why the scalar $\overline{[ud]}$ curve could not be continued beyond $0.7\rho_0$. At the opposite, the $[ss]$ and $\overline{[ss]}$ are constituted, respectively, by two strange quarks and two strange antiquarks. The strange quarks/antiquarks masses depend on the baryonic density, exactly in the same way. As a consequence, the masses of $[ss]$ and $\overline{[ss]}$ evolve according to $\rho_B$. But, the asymmetry mentioned above cannot intervene for these diquarks. So, there is no mass splitting for them: the curves stay degenerate, whatever the baryonic density.

Moreover, in the figure 18, the evoked asymmetry can be observed also in the diagram of stability/instability of the diquarks, because the $\rho_B = 0$ axis is not here a symmetry axis. Clearly, the diquarks are more stable in a medium in which $\rho_B > 0$ than in a medium dominated by antimatter ( $\rho_B < 0$ ). A physical explanation is diquarks are carrying two quarks. Plunged in a medium where the baryonic density is negative, i.e. in which the antiquarks $\bar{q}$ (confined or not) are in broad excess compared to the quarks $q$, the diquarks tend to liberate their quarks. In other words, they become unstable. These quarks are then able to combine with antiquarks to form mesons, which are clearly more stable. On the other hand, as in the previous paragraph, these explanations cannot be applied to $[ss]$. More precisely, this particle is too stable according to the baryonic density. Thus, we do not represent it in the figure 18.

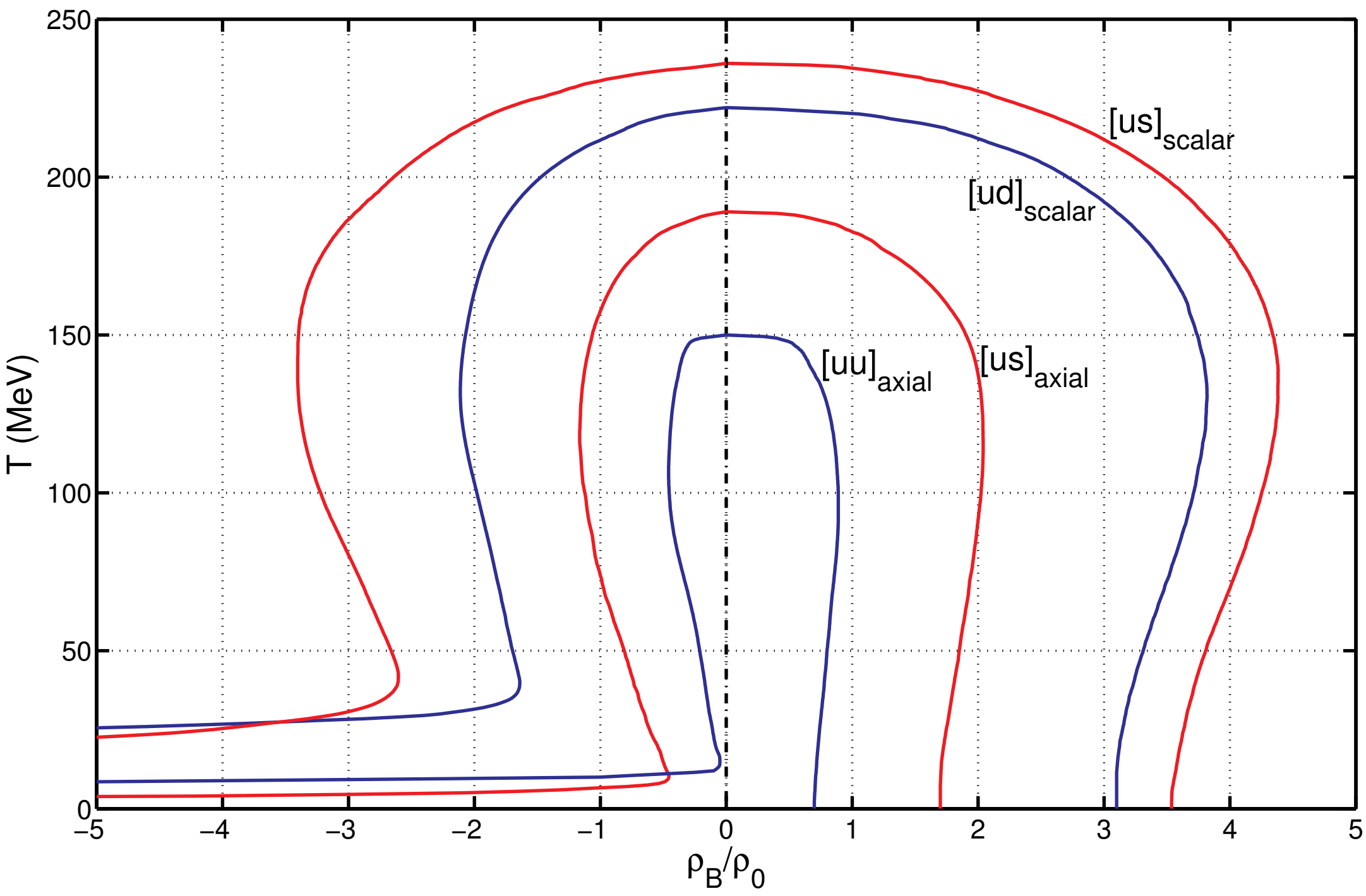

**Figure 18.** NJL diagram of stability/instability for scalar and axial diquarks.

# 3.6 Obtained masses

The table 2 hereafter summarizes the values we found at null temperature and null density. The column associated with the P1 parameter set corresponds to the data described in the previous graphs, i.e. respecting the isospin symmetry. As in the previous chapters, the column related to the EB parameter set does not apply this symmetry. Between these two columns, we inserted values extracted from the literature: from [11] and from [12, 13]. These papers considered other approaches compared to the one used here, and they respected the isospin symmetry. In these references, the widths are null. Indeed, these articles studied the diquarks that appeared as stable. This is why the boxes corresponding to the pseudo-scalar diquarks and vectorial diquarks are empty. For the other diquarks, i.e. the scalar and the axial ones, a very good agreement can be noted between our data and those of these references. Nevertheless, we concede that we used M. Oettel's data [12, 13] in order to fix our $G_{DIQ}$ value. In fact, we consider that $G_{DIQ}/G = 0.705$ in the P1 and EB parameter sets used in this work.

| | | Obtained results (P1 set) | | Values from the literature | | Obtained results (EB set) | |
|---|---|---|---|---|---|---|---|
| Diquarks | | Masses | Widths | Masses from [11] | Masses from [12, 13] | Masses | Widths |
| Scalar | $[ud]$ | 599.14 | 0 | 595 | 598 | 592.82 | 0 |
| | $[us]$ | 794.75 | 0 | 795 | – | 752.43 | 0 |
| | $[ds]$ | 794.75 | 0 | 795 | – | 754.86 | 0 |
| Pseudo scalar | $[ud]$ | 929.37 | 91.11 | – | – | 921.72 | 89.81 |
| | $[us]$ | 1146.25 | 125.40 | – | – | 1100.17 | 118.45 |
| | $[ds]$ | 1146.25 | 125.40 | – | – | 1103.60 | 119.08 |
| Vectorial | $[ud]$ | 1229.39 | 715.76 | – | – | 1222.38 | 715.08 |
| | $[us]$ | 1430.38 | 733.84 | – | – | 1387.24 | 730.39 |
| | $[ds]$ | 1430.38 | 733.84 | – | – | 1390.49 | 730.62 |
| Axial | $[ud]$ | 836.94 | 0 | 835 | 831 | 830.77 | 0 |
| | $[us]$ | 1017.06 | 0 | 1000 | – | 978.27 | 0 |
| | $[ds]$ | 1017.06 | 0 | 1000 | – | 980.81 | 0 |
| | $[uu]$ | 836.94 | 0 | 835 | 831 | 827.94 | 0 |
| | $[dd]$ | 836.94 | 0 | 835 | 831 | 833.59 | 0 |
| | $[ss]$ | 1181.94 | 0 | 1160 | – | 1116.68 | 0 |
| Quarks | $u$ | 424.23 | 0 | 450 | 425 | 419.10 | 0 |
| | $d$ | 424.23 | 0 | 450 | 425 | 422.31 | 0 |
| | $s$ | 626.49 | 0 | 650 | – | 588.17 | 0 |

**Table 2.** Diquarks masses at null density and temperature.

About the values found with P1, we have an good agreement between our scalar $[ud]$ and the one found in [12, 13], but also our scalar $[us]$, $[ds]$ and axial $[ud]$, $[uu]$, $[dd]$ with those of [11]. Concerning the values found with EB, we have similarities between the scalar $[ud]$ and the one of [11], and the axial $[ud]$, $[uu]$, $[dd]$ with those found in [12, 13]. Finally, these comparisons are excellent because the differences do not exceed 3 MeV. About the other values, i.e. the axial $[us]$, $[ds]$, $[ss]$ (unavailable in [12, 13]), our values are of the good order of magnitude compared to those of [11]. But, the differences between the values are slightly stronger than previously.

# 4. Conclusion

In this chapter, we firstly saw that adaptations of the mesons equations can easily leads to obtain the diquarks ones, thanks of the use of the charge conjugation. Indeed, this trick allowed transforming a quark-antiquark loop into a quark-quark loop. As with the mesons, four types of diquarks were considered: the scalar, pseudo-scalar, vectorial and axial diquarks. In fact, it was observed similarities between the diquarks and the mesons, notably as regards the behavior of their masses according to the temperature and the baryonic density. The effect of the Polyakov loop was also studied. It was found that it acts in the same way as observed for mesons, i.e. a shifting of the curves towards higher temperatures.

Moreover, the scalar and the axial diquarks appeared as good candidate to form baryons. We found that these two diquarks types are stable for $T=0$ and $\rho_B=0$. For them, it was then investigated their associated coupling constants, and they behavior at negative baryonic density. It allowed building the diagrams of stability/instability of these diquarks in the $T,\rho_B$ plane. At this occasion, we also studied the possibility to model anti-diquarks, which will permit us to consider anti-baryons. Finally, the masses at null temperature and null density were compared with other theoretical data, using other models. It was found a very good agreement with these references.

# 5. Calculation of the propagator of the charge conjugate quark

This section was largely inspired by Regina Nebauer's notes [23].

The NJL propagator of a charge conjugate quark is written as:

$$\begin{aligned} S^C(\tau-\tau',\vec{x}-\vec{x}') &= -i\cdot\left\langle \mathrm{T}\left(\psi_\xi{}^C(\tau,\vec{x})\bar{\psi}_{\xi'}{}^C(\tau',\vec{x}')\right)\right\rangle \\ &= -i\cdot\left(\theta(\tau-\tau')\cdot\left\langle \psi_\xi{}^C(\tau,\vec{x})\bar{\psi}_{\xi'}{}^C(\tau',\vec{x}')\right\rangle + \theta(\tau'-\tau)\cdot\left\langle \bar{\psi}_{\xi'}{}^C(\tau',\vec{x}')\psi_\xi{}^C(\tau,\vec{x})\right\rangle\right) \end{aligned}, \tag{16}$$

in which T is the time ordering operator, $\theta$ is the Heaviside function and $\tau,\tau'$ correspond to times. By using the decomposition:

$$\begin{cases} \psi_\xi{}^C(\tau,\vec{x}) = \displaystyle\int \frac{d^3k}{(2\pi)^3}\frac{m}{E_k}\sum_\alpha\left(d_\alpha(\tau,k)\cdot u_\xi^\alpha(k)\cdot\exp\left(i\vec{k}\cdot\vec{x}\right) + b_\alpha{}^+(\tau,k)\cdot v_\xi^\alpha(k)\cdot\exp\left(-i\vec{k}\cdot\vec{x}\right)\right) \\ \bar{\psi}_{\xi'}{}^C(\tau',\vec{x}') = \displaystyle\int \frac{d^3k'}{(2\pi)^3}\frac{m}{E_{k'}}\sum_{\alpha'}\left(d_{\alpha'}{}^+(\tau',k')\cdot\bar{u}_{\xi'}^{\alpha'}(k')\cdot\exp\left(-i\vec{k}'\cdot\vec{x}'\right) + b_{\alpha'}(\tau',k')\cdot\bar{v}_{\xi'}^{\alpha'}(k')\cdot\exp\left(i\vec{k}'\cdot\vec{x}'\right)\right) \end{cases}, \tag{17}$$

the propagator is written as:

$$i\cdot S^{C}\left(\tau-\tau',\vec{x}-\vec{x}'\right)$$
$$=\theta(\tau-\tau')\cdot\int d^3\overline{k}\cdot\int d^3\overline{k}'\cdot\sum_{\alpha,\alpha'}\begin{pmatrix}\left\langle d_\alpha(\tau,k)d_{\alpha'}^{+}(\tau',k')\right\rangle\cdot\exp\left(i\vec{k}\cdot\vec{x}-i\vec{k}'\cdot\vec{x}'\right)\cdot u_\xi^\alpha(k)\overline{u}_{\xi'}^{\alpha'}(k')\\+\left\langle b_\alpha^{+}(\tau,k)b_{\alpha'}(\tau',k')\right\rangle\cdot\exp\left(-i\vec{k}\cdot\vec{x}+i\vec{k}'\cdot\vec{x}'\right)\cdot v_\xi^\alpha(k)\overline{v}_{\xi'}^{\alpha'}(k')\end{pmatrix},$$
$$+\theta(\tau'-\tau)\cdot\int d^3\overline{k}\cdot\int d^3\overline{k}'\cdot\sum_{\alpha,\alpha'}\begin{pmatrix}\left\langle d_{\alpha'}^{+}(\tau',k')d_\alpha(\tau,k)\right\rangle\cdot\exp\left(i\vec{k}\cdot\vec{x}-i\vec{k}'\cdot\vec{x}'\right)\cdot\overline{u}_{\xi'}^{\alpha'}(k')u_\xi^\alpha(k)\\+\left\langle b_{\alpha'}(\tau',k')b_\alpha^{+}(\tau,k)\right\rangle\cdot\exp\left(-i\vec{k}\cdot\vec{x}+i\vec{k}'\cdot\vec{x}'\right)\cdot\overline{v}_{\xi'}^{\alpha'}(k')v_\xi^\alpha(k)\end{pmatrix} \qquad (18)$$

$$\text{with } \int d^3\overline{k}\equiv\int\frac{d^3k}{(2\pi)^3}\cdot\frac{m}{E_k}\text{ and }\int d^3\overline{k}'\equiv\int\frac{d^3k'}{(2\pi)^3}\cdot\frac{m}{E_{k'}}. \qquad (19)$$

Now, thanks to the relations,

$$\begin{cases}b_\alpha(\tau,k)=\exp\left(-i(E_k-\mu)\cdot\tau\right)\cdot b_\alpha(k)\\d_\alpha(\tau,k)=\exp\left(-i(E_k+\mu)\cdot\tau\right)\cdot d_\alpha(k)\end{cases}\qquad\begin{cases}b_\alpha^{+}(\tau,k)=\exp\left(i(E_k-\mu)\cdot\tau\right)\cdot b_\alpha^{+}(k)\\d_\alpha^{+}(\tau,k)=\exp\left(i(E_k+\mu)\cdot\tau\right)\cdot d_\alpha^{+}(k)\end{cases}, \qquad (20)$$

we have:

$$i\cdot S^{C}\left(\tau-\tau',\vec{x}-\vec{x}'\right)$$
$$=\theta(\tau-\tau')\cdot\int d^3\overline{k}\cdot\int d^3\overline{k}' \qquad (21)$$
$$\times\sum_{\alpha,\alpha'}\begin{pmatrix}\left\langle d_\alpha(k)d_{\alpha'}^{+}(k')\right\rangle\cdot\exp\left(-i\vec{k}\cdot\vec{x}+i\vec{k}'\cdot\vec{x}'\right)\exp\left(-i(E_k+\mu)\tau\right)\exp\left(i(E_{k'}+\mu)\tau'\right)\cdot u_\xi^\alpha(k)\overline{u}_{\xi'}^{\alpha'}(k')\\+\left\langle b_\alpha^{+}(k)b_{\alpha'}(k')\right\rangle\cdot\exp\left(i\vec{k}\cdot\vec{x}-i\vec{k}'\cdot\vec{x}'\right)\exp\left(i(E_k-\mu)\tau\right)\exp\left(-i(E_{k'}-\mu)\tau'\right)\cdot v_\xi^\alpha(k)\overline{v}_{\xi'}^{\alpha'}(k')\end{pmatrix}.$$
$$+\theta(\tau'-\tau)\cdot\int d^3\overline{k}\cdot\int d^3\overline{k}'$$
$$\times\sum_{\alpha,\alpha'}\begin{pmatrix}\left\langle d_{\alpha'}^{+}(k')d_\alpha(k)\right\rangle\cdot\exp\left(-i\vec{k}\cdot\vec{x}+i\vec{k}'\cdot\vec{x}'\right)\exp\left(i(E_{k'}+\mu)\tau'\right)\exp\left(-i(E_k+\mu)\tau\right)\cdot\overline{u}_{\xi'}^{\alpha'}(k')u_\xi^\alpha(k)\\+\left\langle b_{\alpha'}(k')b_\alpha^{+}(k)\right\rangle\cdot\exp\left(i\vec{k}\cdot\vec{x}-i\vec{k}'\cdot\vec{x}'\right)\exp\left(-i(E_{k'}-\mu)\tau'\right)\exp\left(i(E_k-\mu)\tau\right)\cdot\overline{v}_{\xi'}^{\alpha'}(k')v_\xi^\alpha(k)\end{pmatrix}$$

If we keep only the non-null terms, the propagator is then simplified:

$$i\cdot S^{C}\left(\tau-\tau',\vec{x}-\vec{x}'\right) \qquad (22)$$
$$=\theta(\tau-\tau')\cdot\int d^3\overline{k}\cdot\sum_{\alpha}\begin{pmatrix}\left\langle d_\alpha(k)d_\alpha^{+}(k)\right\rangle\cdot\exp\left(-i\vec{k}\cdot(\vec{x}-\vec{x}')\right)\exp\left(-i(E_k+\mu)(\tau-\tau')\right)\cdot u_\xi^\alpha(k)\overline{u}_\xi^\alpha(k)\\+\left\langle b_\alpha^{+}(k)b_\alpha(k)\right\rangle\cdot\exp\left(i\vec{k}\cdot(\vec{x}-\vec{x}')\right)\exp\left(i(E_k-\mu)(\tau-\tau')\right)\cdot v_\xi^\alpha(k)\overline{v}_\xi^\alpha(k)\end{pmatrix},$$
$$+\theta(\tau'-\tau)\cdot\int d^3\overline{k}\cdot\sum_{\alpha}\begin{pmatrix}\left\langle d_\alpha^{+}(k)d_\alpha(k)\right\rangle\cdot\exp\left(-i\vec{k}\cdot(\vec{x}-\vec{x}')\right)\exp\left(-i(E_k+\mu)(\tau-\tau')\right)\cdot\overline{u}_\xi^\alpha(k)u_\xi^\alpha(k)\\+\left\langle b_\alpha(k)b_\alpha^{+}(k)\right\rangle\cdot\exp\left(i\vec{k}\cdot(\vec{x}-\vec{x}')\right)\exp\left(i(E_k-\mu)(\tau-\tau')\right)\cdot\overline{v}_\xi^\alpha(k)v_\xi^\alpha(k)\end{pmatrix}$$

and:

$$i\cdot S^{C}(\tau-\tau',\vec{x}-\vec{x}') \tag{23}$$

$$=\underbrace{\theta(\tau-\tau')\cdot\int d^3\overline{k}\cdot\sum_{\alpha}\begin{pmatrix}\langle d_{\alpha}(k)d_{\alpha}^{+}(k)\rangle\cdot\exp(-i\vec{k}\cdot(\vec{x}-\vec{x}'))\exp(-i(E_k+\mu)(\tau-\tau'))\cdot\frac{\not{k}+m}{2m}\\+\langle b_{\alpha}^{+}(k)b_{\alpha}(k)\rangle\cdot\exp(i\vec{k}\cdot(\vec{x}-\vec{x}'))\exp(i(E_k-\mu)(\tau-\tau'))\cdot\frac{\not{k}-m}{2m}\end{pmatrix}}_{i\cdot S^{C}(\tau-\tau',\vec{x}-\vec{x}')\big|_{\theta(\tau-\tau')}}$$

$$+\underbrace{\theta(\tau'-\tau)\cdot\int d^3\overline{k}\cdot\sum_{\alpha}\begin{pmatrix}\langle d_{\alpha}^{+}(k)d_{\alpha}(k)\rangle\cdot\exp(-i\vec{k}\cdot(\vec{x}-\vec{x}'))\exp(-i(E_k+\mu)(\tau-\tau'))\cdot\frac{\not{k}+m}{2m}\\+\langle b_{\alpha}(k)b_{\alpha}^{+}(k)\rangle\cdot\exp(i\vec{k}\cdot(\vec{x}-\vec{x}'))\cdot\exp(i(E_k-\mu)(\tau-\tau'))\cdot\frac{\not{k}-m}{2m}\end{pmatrix}}_{i\cdot S^{C}(\tau-\tau',\vec{x}-\vec{x}')\big|_{\theta(\tau'-\tau)}}$$

The finality of the calculation is to express the propagator in energy-momentum space. So, in a first time, a Fourier transform is applied on the time:

$$\begin{aligned}i\cdot S^{C}(i\omega_n,\vec{x}-\vec{x}')=&\int_0^{-i\beta}d(\tau-\tau')\cdot\exp(i(i\omega_n)(\tau-\tau'))\cdot i\cdot S^{C}(\tau-\tau',\vec{x}-\vec{x}')\big|_{\theta(\tau-\tau')}\\&-\int_0^{-i\beta}d(\tau-\tau')\cdot\exp(-i(i\omega_n)(\tau-\tau'))\cdot i\cdot S^{C}(\tau-\tau',\vec{x}-\vec{x}')\big|_{\theta(\tau'-\tau)}\end{aligned} \tag{24}$$

It gives the expression:

$$i\cdot S^{C}(i\omega_n,\vec{x}-\vec{x}')= \tag{25}$$

$$\begin{aligned}&\theta(\tau-\tau')\cdot\int d^3\overline{k}\cdot\left(\langle d(k)d^{+}(k)\rangle\cdot\frac{\not{k}+m}{2m}\cdot\exp(i\vec{k}\cdot(\vec{x}-\vec{x}'))\right)\cdot\left(\int_0^{-i\beta}d(\tau-\tau')\cdot e^{i\cdot(i\omega_n-E_k-\mu)\cdot(\tau-\tau')}\right)\\&+\theta(\tau-\tau')\cdot\int d^3\overline{k}\cdot\left(\langle b^{+}(k)b(k)\rangle\cdot\frac{\not{k}-m}{2m}\cdot\exp(-i\vec{k}\cdot(\vec{x}-\vec{x}'))\right)\cdot\left(\int_0^{-i\beta}d(\tau-\tau')\cdot e^{i\cdot(i\omega_n+E_k-\mu)\cdot(\tau-\tau')}\right)\\&+\theta(\tau'-\tau)\cdot\int d^3\overline{k}\cdot\left(\langle d^{+}(k)d(k)\rangle\cdot\frac{\not{k}+m}{2m}\cdot\exp(i\vec{k}\cdot(\vec{x}-\vec{x}'))\right)\cdot\left(-\int_0^{-i\beta}d(\tau-\tau')\cdot e^{-i\cdot(i\omega_n-E_k-\mu)\cdot(\tau-\tau')}\right)\\&+\theta(\tau'-\tau)\cdot\int d^3\overline{k}\cdot\left(\langle b(k)b^{+}(k)\rangle\cdot\frac{\not{k}-m}{2m}\cdot\exp(-i\vec{k}\cdot(\vec{x}-\vec{x}'))\right)\cdot\left(-\int_0^{-i\beta}d(\tau-\tau')\cdot e^{-i\cdot(i\omega_n+E_k-\mu)\cdot(\tau-\tau')}\right)\end{aligned}$$

About the calculations of the integrals with respect to $\tau-\tau'$, we write:

$$\int_0^{-i\beta}d(\tau-\tau')\cdot\exp(i(i\omega_n-E_k-\mu)(\tau-\tau'))=i\cdot\left(\frac{\exp(-\beta(E_k+\mu))+1}{i\omega_n-E_k-\mu}\right), \tag{26a}$$

$$\int_0^{-i\beta}d(\tau-\tau')\cdot\exp(i(i\omega_n+E_k-\mu)(\tau-\tau'))=i\cdot\left(\frac{\exp(\beta(E_k-\mu))+1}{i\omega_n+E_k-\mu}\right), \tag{26b}$$

$$-\int_0^{-i\beta}d(\tau-\tau')\cdot\exp(-i(i\omega_n-E_k-\mu)(\tau-\tau'))=i\cdot\left(\frac{\exp(\beta(E_k+\mu))+1}{i\omega_n-E_k-\mu}\right), \tag{26c}$$

$$-\int_0^{-i\beta} d(\tau-\tau')\cdot\exp\left(-i\left(i\omega_n+E_k-\mu\right)(\tau-\tau')\right)=i\cdot\left(\frac{\exp\left(-\beta\left(E_k-\mu\right)\right)+1}{i\omega_n+E_k-\mu}\right), \tag{26d}$$

because the Matsubara frequency $i\omega_n$ is fermionic type, which implies:

$$\exp\left(i\omega_n\cdot\beta\right)=\exp\left(i\cdot(2n+1)\cdot\pi\right)=-1. \tag{27}$$

Also, we have:

$$\begin{cases}\left\langle b^+(k)b(k)\right\rangle=\dfrac{1}{\exp\left(\beta\left(E_k-\mu\right)\right)+1}\\ \left\langle b(k)b^+(k)\right\rangle=\dfrac{\exp\left(\beta\left(E_k-\mu\right)\right)}{\exp\left(\beta\left(E_k-\mu\right)\right)+1}\end{cases} \text{ and } \begin{cases}\left\langle d^+(k)d(k)\right\rangle=\dfrac{1}{\exp\left(\beta\left(E_k+\mu\right)\right)+1}\\ \left\langle d(k)d^+(k)\right\rangle=\dfrac{\exp\left(\beta\left(E_k+\mu\right)\right)}{\exp\left(\beta\left(E_k+\mu\right)\right)+1}\end{cases}. \tag{28}$$

We inject now these relations in (25), we group the terms and we obtain:

$$i\cdot S^C\left(i\omega_n,\vec{x}-\vec{x}'\right)=i\int d^3\overline{k}\cdot\frac{\not{k}+m}{2m}\cdot\frac{\exp\left(i\vec{k}\cdot(\vec{x}-\vec{x}')\right)}{i\omega_n-E_k-\mu}+i\int d^3\overline{k}\cdot\frac{\not{k}-m}{2m}\cdot\frac{\exp\left(-i\vec{k}\cdot(\vec{x}-\vec{x}')\right)}{i\omega_n+E_k-\mu}, \tag{29}$$

or:

$$S^C\left(i\omega_n,\vec{x}-\vec{x}'\right)=\int\frac{d^3k}{(2\pi)^3}\cdot\frac{\not{k}+m}{2E_k}\cdot\frac{\exp\left(i\vec{k}\cdot(\vec{x}-\vec{x}')\right)}{i\omega_n-E_k-\mu}+\int\frac{d^3k}{(2\pi)^3}\cdot\frac{\not{k}-m}{2E_k}\cdot\frac{\exp\left(-i\vec{k}\cdot(\vec{x}-\vec{x}')\right)}{i\omega_n+E_k-\mu}. \tag{30}$$

We apply the Fourier transform on the positions:

$$S^C\left(i\omega_n,\vec{p}\right)=\int d^3\left(\vec{x}-\vec{x}'\right)\cdot\exp\left(-i\vec{p}\cdot(\vec{x}-\vec{x}')\right)\cdot S^C\left(i\omega_n,\vec{x}-\vec{x}'\right), \tag{31}$$

it gives:

$$\begin{aligned}S^C\left(i\omega_n,\vec{p}\right)=&\int\frac{d^3k}{(2\pi)^3}\cdot\frac{\not{k}+m}{2E_k}\cdot\frac{1}{i\omega_n-E_k-\mu}\cdot\underbrace{\int d^3\left(\vec{x}-\vec{x}'\right)\cdot\exp\left(i\left(\vec{k}-\vec{p}\right)\cdot(\vec{x}-\vec{x}')\right)}_{(2\pi)^3\cdot\delta^{(3)}\left(\vec{k}-\vec{p}\right)}\\&+\int\frac{d^3k}{(2\pi)^3}\cdot\frac{\not{k}-m}{2E_k}\cdot\frac{1}{i\omega_n+E_k-\mu}\cdot\underbrace{\int d^3\left(\vec{x}-\vec{x}'\right)\cdot\exp\left(-i\left(\vec{k}+\vec{p}\right)\cdot(\vec{x}-\vec{x}')\right)}_{(2\pi)^3\cdot\delta^{(3)}\left(\vec{k}+\vec{p}\right)},\end{aligned} \tag{32}$$

or:

$$S^C\left(i\omega_n,\vec{p}\right)=\frac{\gamma_0\cdot i\omega_n-\vec{\gamma}\cdot\vec{p}+m}{2E_p\cdot\left(i\omega_n-E_p-\mu\right)}+\frac{\gamma_0\cdot i\omega_n+\vec{\gamma}\cdot\vec{p}-m}{2E_p\cdot\left(i\omega_n+E_p-\mu\right)}, \tag{33}$$

and, finally:

$$S^C\left(i\omega_n,\vec{p}\right)=\frac{1}{\not{p}-m-\gamma_0\cdot\mu}. \tag{34}$$

In the framework of the PNJL model, because of the inclusion of the Polyakov loop, the quarks are minimally coupled to this loop, implying a dependence upon the color in the propagator. As evoked before, it leads to the replacement $\mu \to \mu - iA_4$ [21]. The PNJL propagator is then written as [20, 21]:

$$S^{C}{}_{PNJL}\left(\not{p}\right) = \frac{1}{\not{p} - m - \gamma_0 \cdot \left(\mu - iA_4\right)}. \tag{35}$$

Moreover, the expression of the ordinary (non-conjugate) propagator is known within the framework of the Nambu and Jona–Lasinio formalism since a long time [24]. But, it could be confirmed by a calculation similar to the one performed in this section. Indeed, we only need to remake the calculations by replacing the equation (17) by:

$$\begin{cases} \psi_{\xi}\left(\tau,\vec{x}\right) = \displaystyle\int \frac{d^3k}{\left(2\pi\right)^3} \cdot \frac{m}{E_k} \cdot \sum_{\alpha} \left(b_{\alpha}\left(\tau,k\right) \cdot u_{\xi}^{\alpha}\left(k\right) \cdot \exp\left(i\vec{k}\cdot\vec{x}\right) + d_{\alpha}{}^{+}\left(\tau,k\right) \cdot v_{\xi}^{\alpha}\left(k\right) \cdot \exp\left(-i\vec{k}\cdot\vec{x}\right)\right) \\ \bar{\psi}_{\xi'}\left(\tau',\vec{x}'\right) = \displaystyle\int \frac{d^3k'}{\left(2\pi\right)^3} \cdot \frac{m}{E_{k'}} \cdot \sum_{\alpha'} \left(b_{\alpha'}{}^{+}\left(\tau',k'\right) \cdot \bar{u}_{\xi'}^{\alpha'}\left(k'\right) \cdot \exp\left(-i\vec{k}'\cdot\vec{x}'\right) + d_{\alpha'}\left(\tau',k'\right) \cdot \bar{v}_{\xi'}^{\alpha'}\left(k'\right) \cdot \exp\left(i\vec{k}'\cdot\vec{x}'\right)\right) \end{cases}, \tag{36}$$

it gives:

$$S\left(i\omega_n,\vec{p}\right) = \frac{\gamma_0 \cdot i\omega_n - \vec{\gamma}\cdot\vec{p} + m}{2E_p \cdot \left(i\omega_n - E_p + \mu\right)} + \frac{\gamma_0 \cdot i\omega_n + \vec{\gamma}\cdot\vec{p} - m}{2E_p \cdot \left(i\omega_n + E_p + \mu\right)}, \tag{37}$$

and it comes :

$$S\left(i\omega_n,\vec{p}\right) = \frac{1}{\not{p} - m + \gamma_0 \cdot \mu}. \tag{38}$$

In the same way, the transformation $\mu \to \mu - iA_4$ also gives the possibility to find the associated PNJL propagator [20, 21]:

$$S_{\,PNJL}\left(\not{p}\right) = \frac{1}{\not{p} - m + \gamma_0 \cdot \left(\mu - iA_4\right)}. \tag{39}$$

Whatever the model, NJL or PNJL, the only difference between the quark propagator and the charge conjugate one is the sign placed in front of the chemical potential term, i.e. before the $\gamma_0$ matrix. As a consequence, when the chemical potential is equal to zero, the both are equal: $S^{C}\left(i\omega_n,\vec{p}\right) = S\left(i\omega_n,\vec{p}\right)\Big|_{\mu_f=0}$, and idem for the PNJL ones.

# Chapter 5

# Baryons

A part of this chapter was published in *J. Phys. G: Nucl. Part. Phys.* **38** 105003

## 1. Introduction

We saw in the previous chapters that the NJL model can be completed by the inclusion of the Polyakov Loop, forming the PNJL model. As reported in the literature and as observed in our work, these models can allow the modeling of dressed quarks and mesons. The next step is to include baryons in the analysis. In fact, to study the cooling of a quarks-antiquarks plasma, the baryons cannot be neglected, even if a strong mesonization of the system is expected.

However, if the mesons are composite particles formed by a quark-antiquark pair, the baryons are composed by three quarks. It thus requires the modeling of a three-body system. As evoked in the previous chapter, the Faddeev equations have to be considered in such a work [1, 2]. But, it was also shown that a simplification of these equations, i.e. considering the "first order term", leads to consider a baryon as a bound state formed by a quark and a diquark [3–5]. In the literature, studies using this approach were published [6–8]. Furthermore, the baryon modeling was also performed in the framework of the NJL model, notably with this quark-diquark approximation. It leads to various works. During the 1990's, papers as [9–14] can be quoted. After 2000, we have notably [15–21]. Among the performed studies, some of them concerned the estimation of the masses of the baryons at null temperature and density, as in [3]. For example in [15, 19], the baryons were studied at finite densities…

In fact, the quark-diquark model appears to be relevant in an NJL description, notably because of the possibility to use loop functions to treat composites particles in this model. Clearly, we saw in chapter 3 that the mesons were considered by the way of a quark-antiquark loop function. About the diquark, chapter 4, we simply applied a charge conjugation to the antiquark to mimic a quark-quark loop. A loop function made by a quark and a diquark is possible there. However, it is at the price of an extra approximation, known as static approximation [13]. It consists to neglect the momentum of the exchanged quark in front of its mass. Studies applying this idea were performed, as [16–18] that use scalar diquarks to calculate the masses of octet baryons, according to the temperature or the baryonic density. However, it was observable in [16] some limitations of this attempt, in the form of numerical instabilities in some curves. Upon a numerical point of view, it reveals that the equations to be solved require much more numerical considerations compared to the ones used with mesons.

Going back to a physical point of view, it is often considered that the NJL baryon modeling is incompletely treated in the literature. It suggests evolutions and modifications of the already performed works. In addition, the baryons were not treated in framework of the PNJL model. So, it could be interesting to observe the effects of the inclusion of the Polyakov loop. With the quarks, mesons and the diquarks, this modification of the model induced a shifting of the curves towards higher temperatures. About the baryons, it is not trivial to obtain the same result. Furthermore, the baryons masses were mainly studied according to the temperature, more rarely according to the baryonic density, but not in the whole $T, \rho_B$ plane. Clearly, it could be instructive to see the limit of stability of some baryons, as the nucleons. Also, thanks to the work performed in the previous chapter, we have there the possibility to include axial diquarks in the baryon modeling, as evoked e.g. in [11]. This could allow the treatment of decuplet baryons. In the same way, with anti-diquarks, it could be investigated the behavior of anti-baryons.

In this chapter, taking care about these observations, we propose to establish in section 2 the equations devoted to model the baryons. At this occasion, the static approximation is introduced in our work and explained. In section 3, these theoretical calculations focus on the study of each baryon in a systematic way. More precisely, the scalar and/or axial flavor component of these particles are detailed. Then, the section 4 presents our numerical calculations performed at finite temperatures and densities. At this occasion, we notably underline the differences between NJL and PNJL models. This part includes a study of the nucleon's mass in the $T, \rho_B$ plane, diagrams of stability/instability of the studied baryons, and the modeling of anti-baryons. In the section 5, calculations of coupling constants involving baryons are presented. The equations that we used for mesons and diquarks cannot be employed here. As a consequence, we firstly focus on the method to be applied. We present then the obtained results. In the section 6, the masses of the baryons at null temperature and densities are studied, in the framework of the isospin symmetry, and beyond this one. These results are compared to experimental data. It will engage a discussion about the reliability of our approach, which will notably concern the applied approximations. Our method is then compared to other ones. It obviously concerns the already quoted works performed in the framework of the NJL quark-diquark picture, but also the work performed in [22], i.e. not using quarks to describe baryons.

# 2. Presentation of the employed method

The basic idea of our baryon modeling is to come back to a structure close to the one observed for the mesons and diquarks. In other words, the finality is to use the Bethe-Salpeter equation for a quark-diquark scattering. It also wants to say that we must be able to come back to a description using a loop function, involving a quark and a diquark. Our approach is summarized in the figure 1. The finality of this section is to explain the passage from one line to another one in this figure, and to give the associated formulas. In fact, the method described here can be directly applied to "simple" cases, as the scalar flavor component of the nucleon. The adaptations to be done with more complex cases, as with the $\Lambda$ baryon, will be explained in the section 3, when these baryons will be described individually.

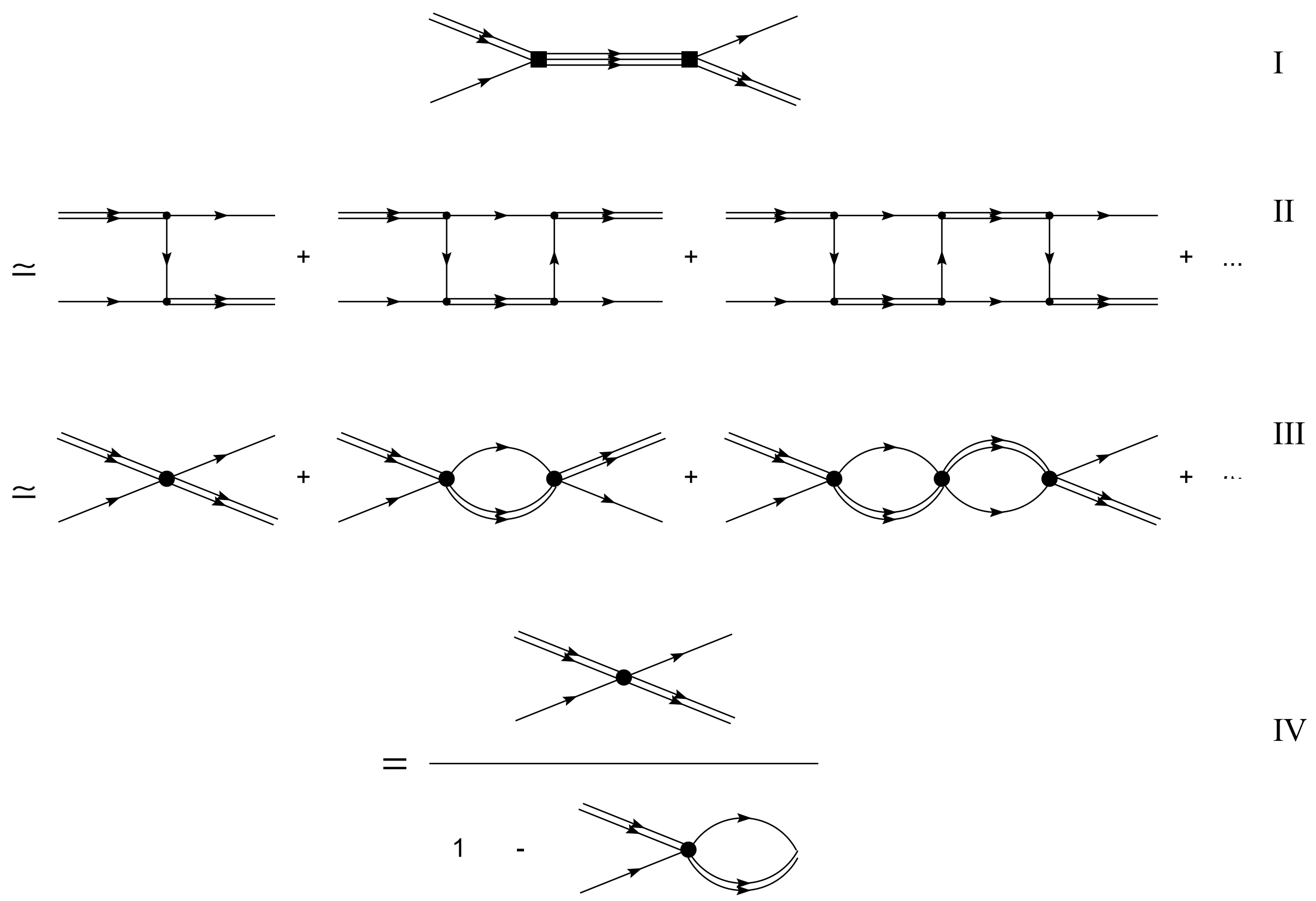

**Figure 1.** Schematization of the baryons' modeling.

## 2.1 Static approximation

The static approximation [13, 16] evoked in the introduction corresponds to the passage from the second to the third line of the figure 1. Thanks to this approximation, we can come back to the wanted loop structure. In practice, it consists to "erase properly" the exchanged quarks visible in the line II in the figure 1.

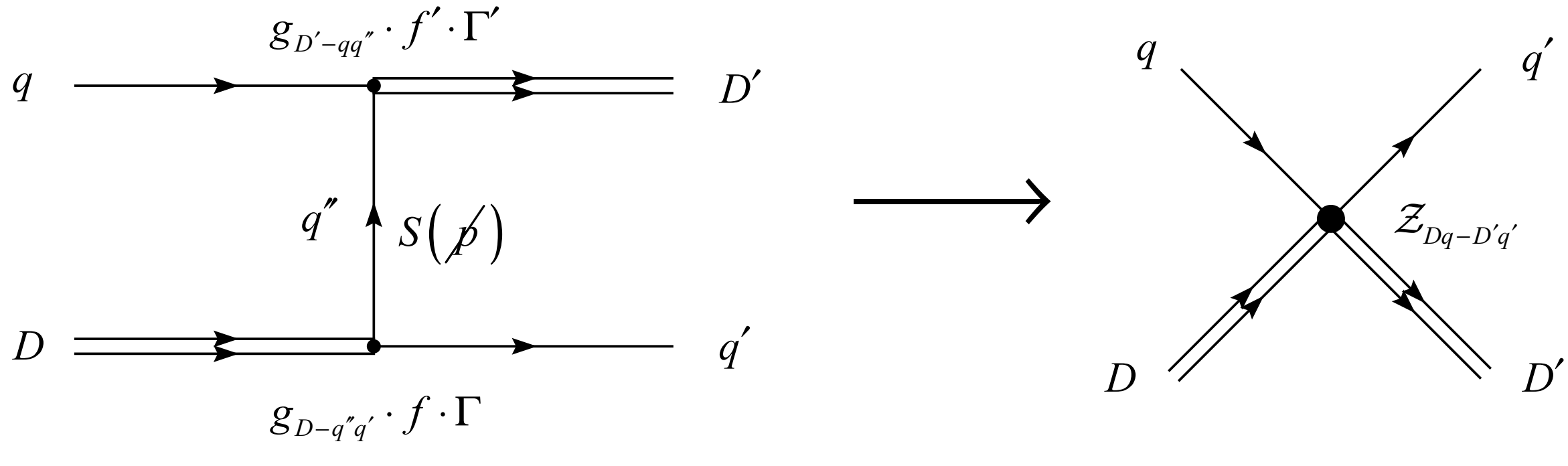

**Figure 2.** The static approximation.

We take together the propagator of the exchanged quark $S(\not{p})$ with its two associated vertices. It firstly leads to the following writing:

$$\left(g_{D-q''q'}\cdot f\cdot\Gamma\right)\cdot i\cdot S(\not{p})\cdot\left(g_{D'-qq''}\cdot f'\cdot\Gamma'\right)=\left(g_{D-q''q'}\cdot f\cdot\Gamma\right)\cdot\left(\frac{i}{\not{p}-m_q}\right)\cdot\left(g_{D'-qq''}\cdot f'\cdot\Gamma'\right). \qquad (1)$$

The two $g$ represent the coupling constants between a diquark and a quark-quark pair. Such terms were studied in the previous chapter. Also, the two $f$ gather terms such as the flavor factor (appendix C), or a factor of color [16, 18]. They also include the two $\Gamma$ terms. These ones correspond to the Dirac matrices that translate the type of interaction, on the level of the vertices. In our study, we consider *scalar* ($\gamma_5$) or *axial* ($\gamma^\mu$) interactions, as in [7, 10, 11]. The other ones are not taken into account, because we saw that pseudo-scalar and vectorial diquarks are not good candidates to form baryons.

In the framework of the static approximation, the four-momentum of the exchanged quark is neglected in front of its mass. This makes it possible to replace the exchanged quark propagator by an effective vertex. This one is materialized by the black spot on the right hand side of the figure 2. We have:

$$\begin{aligned}&\left(g_{D-q''q'}\cdot f\cdot\Gamma\right)\cdot\left(\frac{i}{\not{p}-m_q}\right)\cdot\left(g_{D'-qq''}\cdot f'\cdot\Gamma'\right)\\&\approx\left(g_{D-q''q'}\cdot f\cdot\Gamma\right)\cdot\left(-\frac{i}{m_q}\right)\cdot\left(g_{D'-qq''}\cdot f'\cdot\Gamma'\right)\qquad ,\\&\equiv \mathcal{Z}_{Dq-D'q'}=\begin{cases}g_{D-q''q'}\cdot g_{D'-qq''}\cdot f\cdot f'\cdot\left(-\dfrac{i}{m_q}\right)\cdot 1_4 & \text{if } \Gamma=\Gamma'=\gamma_5\\ 4\cdot g_{D-q''q'}\cdot g_{D'-qq''}\cdot f\cdot f'\cdot\left(-\dfrac{i}{m_q}\right)\cdot 1_4 & \text{if } \Gamma=\Gamma'=\gamma^\mu\end{cases}\end{aligned} \qquad (2)$$

where $1_4$ is the identity $4\times 4$ matrix.

## 2.2 Description of the equations

Now, we focus on the third and the fourth lines of the figure 1. The vertex materialized by a black spot, involving two diquarks and two quarks, corresponds to the two body interaction kernel of the Bethe-Salpeter equation [3, 16]. It is associated with the $\mathcal{Z}$ previously defined for the static approximation:

$$\bullet\equiv\mathcal{Z}\text{ proportional to } -i\cdot\frac{g_{D-q''q'}\cdot g_{D'-qq''}}{m_q}. \qquad (3)$$

In the same way, the diquark/quark loop corresponds to the baryon polarization function, also designated as baryon loop function in our work:

$$\equiv\Pi \qquad (4)$$

The transition matrix T is written thanks to the Bethe-Salpeter equation as $T = \mathcal{Z} + \mathcal{Z} \cdot \Pi \cdot T$. As we did for mesons, we then write, in the general case:

$$T = \mathcal{Z} + \mathcal{Z}\Pi\mathcal{Z} + \mathcal{Z}\Pi\mathcal{Z}\Pi\mathcal{Z} + \mathcal{Z}\Pi\mathcal{Z}\Pi\mathcal{Z}\Pi\mathcal{Z} + \ldots = \frac{\mathcal{Z}}{1 - \Pi\mathcal{Z}} \ . \tag{5}$$

Rigorously, (5) is checked only if:

$$\lim_{n \to \infty} (\Pi\mathcal{Z})^n = 0 \, . \tag{6}$$

Let us pose that $\mathcal{Z}$ is equal to $g^2/m$. Then, $g$ never exceeds 4, see chapter on the diquarks. Furthermore, the quarks effective masses $m$ are higher than 16 in the domain in which the baryons will be studied. In conclusion, $\mathcal{Z}^n$ is necessarily close to zero when $n$ is sufficiently large. About $\Pi$, numerical tests show that the required property is also verified. Then, T is written as:

$$T = \frac{\mathcal{Z}}{1 - \Pi\mathcal{Z}} = \frac{\mathcal{Z}}{\det(1 - \Pi\mathcal{Z})} \cdot {}^{T}\big(\mathrm{com}(1 - \Pi\mathcal{Z})\big) \text{ proportional to } \frac{1}{\det(1 - \Pi\mathcal{Z})} \ . \tag{7}$$

To obtain the baryon's mass, it is necessary that T diverges. Therefore, it requires:

$$\det(1 - \Pi\mathcal{Z}) = 0 \, . \tag{8}$$

For the simple cases, $\Pi\mathcal{Z}$ is a scalar number. So, for a baryon momentum equal to $\vec{k}$, (8) is equivalent to the equation:

$$1 - \Pi\left(k_0, \vec{k}\right) \cdot \mathcal{Z} = 0 \ \Big|_{k_0 = \sqrt{m^2 + \left(\vec{k}\right)^2} \ , \ \vec{k} \text{ fixed}} \ , \tag{9}$$

where $m$ is the mass of the baryon. In more complex cases, $\Pi\mathcal{Z}$ is a matrix. Nevertheless, the relation (8) stays valid.

# 2.3 Baryon loop function

Although structurally close to the mesons or diquarks polarization functions, the baryons loop function $\Pi$ is more delicate to treat for several reasons. Firstly, the loop function does not include one term but two terms. This is due to the asymmetry caused by the fact to consider a quark and a diquark. We can make the choice to take a quark and a charge conjugate diquark. But, we can also consider a diquark and a charge conjugate quark. To write our function, we should build this one as a linear combination of these two possibilities, figure 3 and equations (10, 11). Nevertheless, we showed in the appendix D that the two components are strictly equal. As a consequence, this is a false complication.

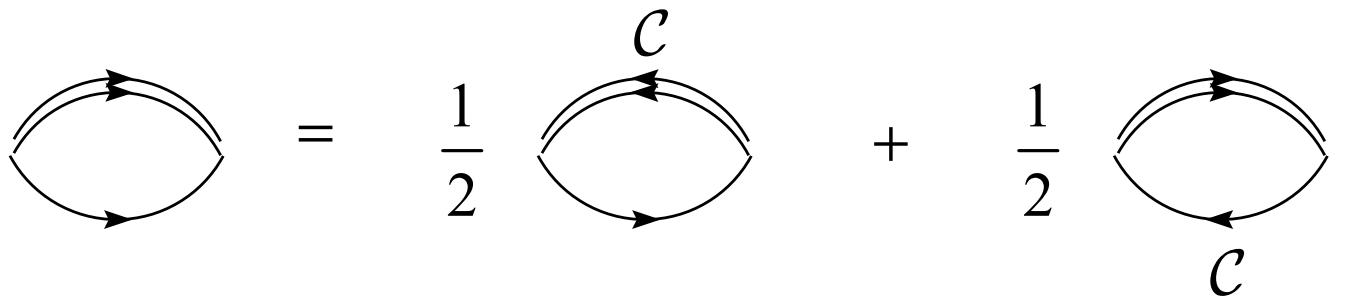

**Figure 3.** The two components of the baryon loop function.

$$-i\cdot\Pi\left(i\cdot\nu_m,\vec{k}\right)=\frac{1}{2}\cdot\left(-i\cdot\Pi^{(1)}\left(i\cdot\nu_m,\vec{k}\right)\right)+\frac{1}{2}\cdot\left(-i\cdot\Pi^{(2)}\left(i\cdot\nu_m,\vec{k}\right)\right)\ , \tag{10}$$

with:

$$\begin{cases} -i\cdot\Pi^{(1)}\left(i\cdot\nu_m,\vec{k}\right)=-\dfrac{i}{\beta}\cdot\sum_n\displaystyle\int\frac{\mathrm{d}^3p}{(2\pi)^3}Tr\left(i\cdot S_q\left(i\cdot\omega_n,\vec{p}\right)\cdot i\cdot S_D{}^C\left(i\cdot\omega_n-i\cdot\nu_m,\vec{p}-\vec{k}\right)\right) & \text{(11a)}\\ -i\cdot\Pi^{(2)}\left(i\cdot\nu_m,\vec{k}\right)=-\dfrac{i}{\beta}\cdot\sum_n\displaystyle\int\frac{\mathrm{d}^3p}{(2\pi)^3}Tr\left(i\cdot S_D\left(i\cdot\omega_n,\vec{p}\right)\cdot i\cdot S_q{}^C\left(i\cdot\omega_n-i\cdot\nu_m,\vec{p}-\vec{k}\right)\right) & \text{(11b)}\end{cases} .$$

Then, we also have to take care that the loop function includes a fermion (the quark) and a boson (the diquark). This imposes to think about the used Matsubara frequencies. In fact, they are not the same in (11a) and (11b). But, for the two components, the total frequency $i\cdot\nu_m$ is the one associated with a baryon. Therefore, $i\cdot\nu_m$ is a fermionic frequency, i.e. an uneven number. In the equation (11a), $i\cdot\omega_n$ is the frequency associated with the quark propagator. Therefore, $i\cdot\omega_n$ is fermionic. The Matsubara frequency of the diquark propagator must then be bosonic. This is checked, since $i\cdot\omega_n-i\cdot\nu_m$ corresponds to this frequency: the sum or the difference of two fermionic frequencies (uneven numbers) necessarily gives a bosonic frequency (even number). In the equation (11b), this time, $i\cdot\omega_n$ is the frequency associated with the diquark propagator. So, it implies that this one is bosonic. The frequency of the quark propagator is fermionic. Indeed, it is equal to $i\cdot\omega_n-i\cdot\nu_m$: the sum or the difference of a bosonic frequency (even number) with a fermionic frequency (odd number) is necessarily a fermionic frequency (odd number).

Concerning the adaptations of the equations due to the inclusion of the Polyakov loop, we continue there to apply the idea evoked in the previous chapters. Clearly, it concerns the adaptation of the Fermi-Dirac statistics in the cases of the quarks and antiquarks, generalizing the method proposed in [23] for the mesons.

# 3. Modeling of each baryon

In this section, we perform a systematic study for each baryon, taking into account the scalar and the axial interaction channels. It gives then two components [7, 11], i.e., respectively, the scalar flavor component and the axial one. For each of them, we propose to write the associated diagrams, as the ones observable in [4, 16]. Then, we study their wave functions, inspiring us from [6, 7, 18]. Finally, we establish the equations to be solved to find the baryons' masses, in the framework of the (P)NJL models.

## 3.1 Nucleons: proton and neutron

The proton scalar flavor component is one of the simplest cases to study, figure 4. Indeed, it concerns one quark $u$ and one scalar diquark $[ud]$. The exchanged quark is the quark *d*.

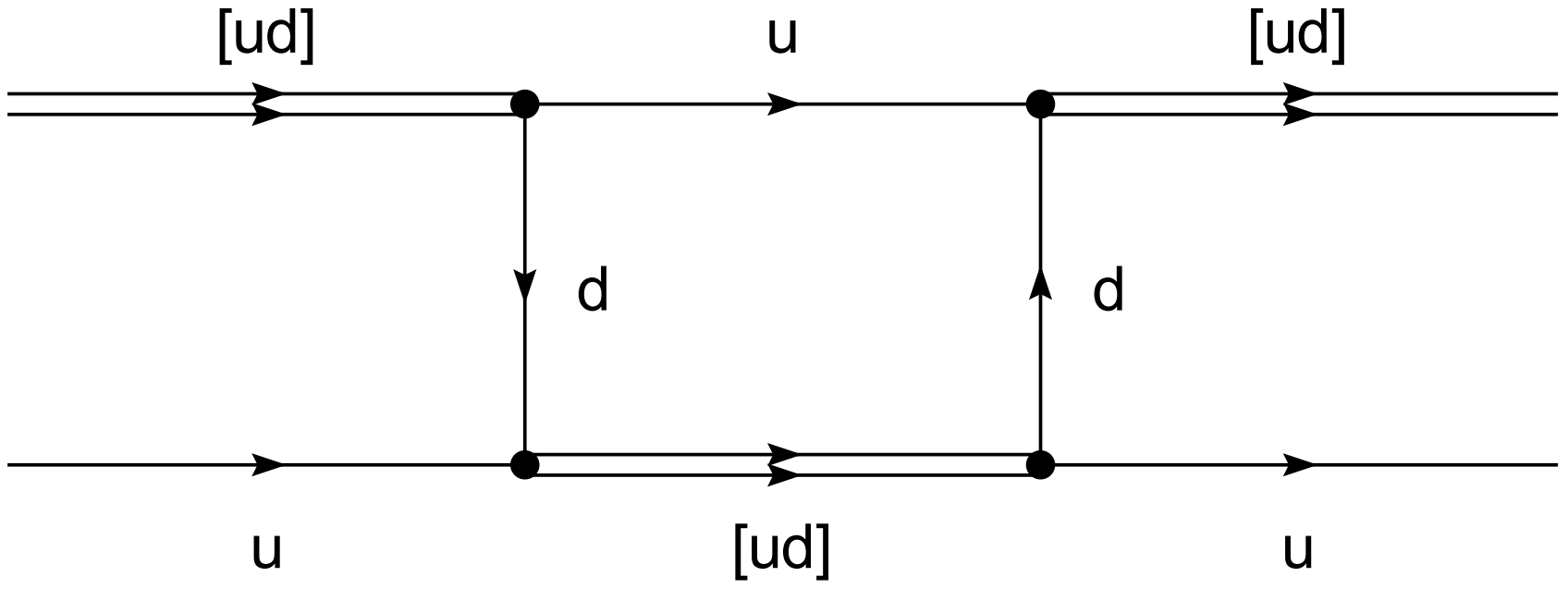

**Figure 4.** Scalar flavor component of the proton.

In the figure 4, a vertex translating the scalar interaction is indicated by a black spot. The wave-function of this component is written as a tensor product that associates a quark $u$ with a diquark $[ud]$ [6, 7, 18]:

$$|p\rangle_{\text{scalar}} = \underbrace{\begin{bmatrix}1\\0\\0\end{bmatrix}}_{u} \otimes \underbrace{\frac{i}{\sqrt{2}} \cdot \lambda_2 \cdot \begin{bmatrix}1\\1\\0\end{bmatrix}}_{[ud]} \ . \tag{12}$$

A mnemotechnic technique to understand this wave-function is to consider that the quarks space is summarized with a column vector. More precisely, the first component is associated with the quark $u$, the second to the quark $d$, and the third to the quark $s$. This idea can be taken up for the diquarks. In this case, we have to take into account two quarks. These ones are of course those that constitute the diquark, so the $\begin{bmatrix}1\\1\\0\end{bmatrix}$ for $[ud]$.

About the matrix $\lambda_2$, the reader is invited to consult the appendix C. More precisely, a parallel can be made between this matrix and the flavor factor for the vertices where $[ud]$ appears [6]. Applying the formula (9) in the case of the scalar flavor component of the proton, we obtain the equation to be solved to find its mass $M_p$. For a proton at rest, it comes:

$$1 - 2 \cdot \frac{-2 \cdot g_{ud}^{\ 2}}{m_d} \cdot \Pi_{[ud],u}\left(M_p, \vec{0}\right) = 0 \ . \tag{13}$$

Now, let us consider the axial flavor component. As indicated in the figure 5, the axial component is made by the state $[ud]_{axial} + u$ and by the state $[uu]_{axial} + d$. In this figure, the vertices materialized by circles indicate axial interactions.

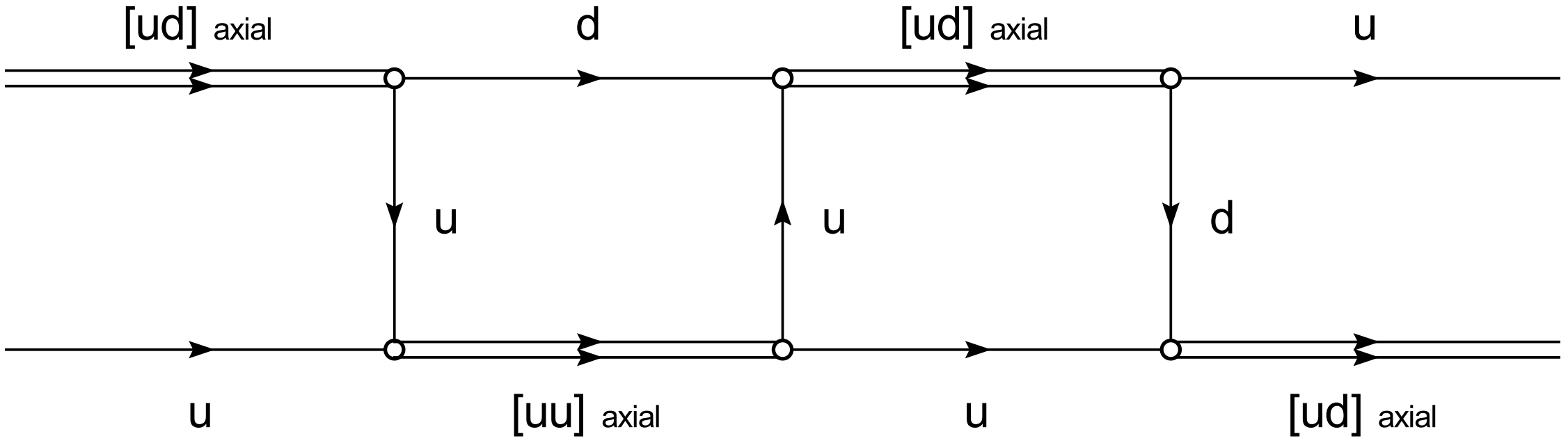

**Figure 5.** Axial flavor component of the proton.

This linear combination between these two states is found in the wave-function expression associated with this axial component, which is written as [6, 7]:

$$|p\rangle_{\text{axial}} = \sqrt{\frac{1}{3}} \cdot \left( \underbrace{\begin{bmatrix} 1 \\ 0 \\ 0 \end{bmatrix}}_{u} \otimes \underbrace{\frac{1}{\sqrt{2}} \cdot \lambda_1 \cdot \begin{bmatrix} 1 \\ 1 \\ 0 \end{bmatrix}}_{[ud]_{\text{axial}}} - \underbrace{\begin{bmatrix} 0 \\ 1 \\ 0 \end{bmatrix}}_{d} \otimes \underbrace{i \cdot \lambda_{+1} \cdot \lambda_2 \cdot \begin{bmatrix} 1 \\ 0 \\ 0 \end{bmatrix}}_{[uu]_{\text{axial}}} \right) , \tag{14}$$

with:

$$\lambda_{\pm 1} = \mp \sqrt{\frac{1}{2}} \cdot \left( \lambda_1 \pm i \cdot \lambda_2 \right) . \tag{15}$$

The method used to obtain the equation to be solved, i.e. the equivalent of (13) for the axial flavor component, will be explained within the framework of the $\Lambda$ baryon, in the subsection 3.2. However, it can be employed without problem with the other baryons. In fact, it will be enough to insert there the quarks and diquarks that correspond to the desired flavor component...

It is possible to remake the work with the neutron. It is only required to replace all the $u$ by $d$ , and conversely, in all the formulas associated with the proton. Obviously, if the isospin symmetry is applied, the results must strictly coincide. Indeed, in this case, the quarks $u$ and $d$ are similar. In the general case, the scalar flavor component of the neutron wave-function is written:

$$|n\rangle_{\text{scalar}} = \underbrace{\begin{bmatrix} 0 \\ 1 \\ 0 \end{bmatrix}}_{d} \otimes \underbrace{\frac{i}{\sqrt{2}} \cdot \lambda_2 \cdot \begin{bmatrix} 1 \\ 1 \\ 0 \end{bmatrix}}_{[ud]} , \tag{16}$$

and the axial flavor component is:

$$|n\rangle_{\text{axial}} = -\sqrt{\frac{1}{3}} \cdot \left( \underbrace{\begin{bmatrix} 0 \\ 1 \\ 0 \end{bmatrix}}_{d} \otimes \underbrace{\frac{1}{\sqrt{2}} \cdot \lambda_1 \cdot \begin{bmatrix} 1 \\ 1 \\ 0 \end{bmatrix}}_{[ud]_{\text{axial}}} - \underbrace{\begin{bmatrix} 1 \\ 0 \\ 0 \end{bmatrix}}_{u} \otimes \underbrace{i \cdot \lambda_{-1} \cdot \lambda_2 \cdot \begin{bmatrix} 0 \\ 1 \\ 0 \end{bmatrix}}_{[dd]_{\text{axial}}} \right) . \tag{17}$$

# 3.2 Λ baryon

The Λ baryon is conceptually rather delicate. Indeed, this baryon flavor wave-function results from the linear combination of three distinct states: $u+[ds]$, $d+[us]$ and $s+[ud]$. The figure 6 hereafter describes the scalar flavor component:

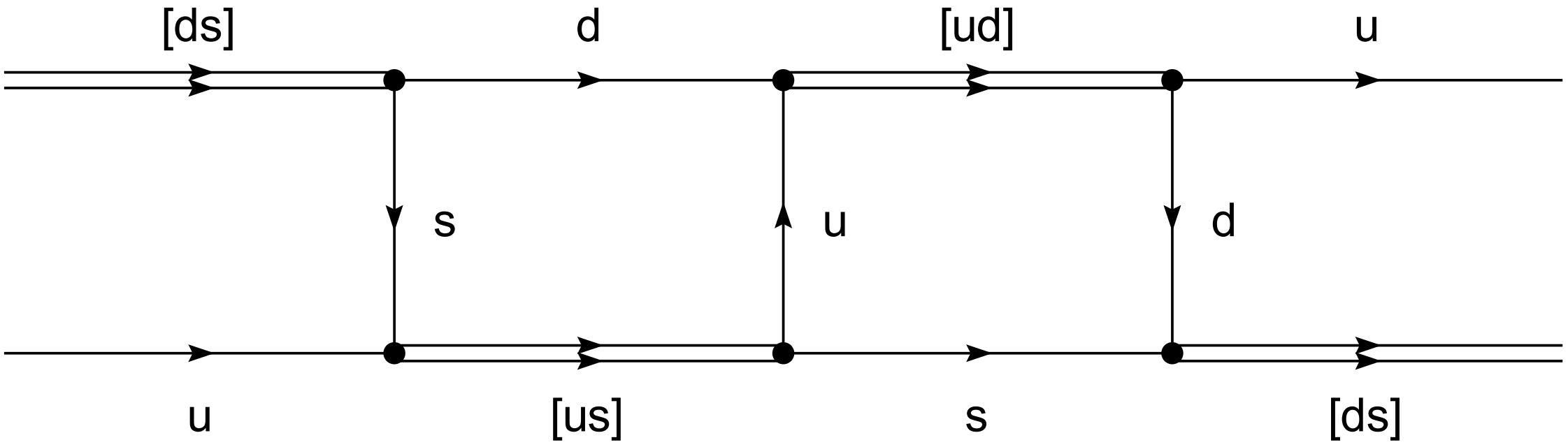

**Figure 6.** Scalar flavor component of the Λ baryon.

The wave-function associated with this figure 6 is written [7, 18]:

$$|\Lambda\rangle_{\text{scalar}} = \frac{1}{\sqrt{12}} \cdot \left( \underbrace{\begin{bmatrix} 1 \\ 0 \\ 0 \end{bmatrix}}_{u} \otimes \underbrace{i\lambda_7 \cdot \begin{bmatrix} 0 \\ 1 \\ 1 \end{bmatrix}}_{[ds]} + \underbrace{\begin{bmatrix} 0 \\ 1 \\ 0 \end{bmatrix}}_{d} \otimes \underbrace{i\lambda_5 \cdot \begin{bmatrix} 1 \\ 0 \\ 1 \end{bmatrix}}_{[us]} - 2 \underbrace{\begin{bmatrix} 0 \\ 0 \\ 1 \end{bmatrix}}_{s} \otimes \underbrace{i\lambda_2 \cdot \begin{bmatrix} 1 \\ 1 \\ 0 \end{bmatrix}}_{[ud]} \right). \tag{18}$$

The interaction vertices between the quarks and diquarks that compose the baryon are gathered in the $\mathcal{Z}$ term, which is now a $3\times3$ matrix. This one takes the form:

$$\mathcal{Z}_\Lambda = \begin{bmatrix} 0 & \mathcal{Z}^{ud} & \mathcal{Z}^{us} \\ \mathcal{Z}^{du} & 0 & \mathcal{Z}^{ds} \\ \mathcal{Z}^{su} & \mathcal{Z}^{sd} & 0 \end{bmatrix} = \begin{bmatrix} 0 & \dfrac{g_{u,s} \cdot g_{d,s}}{m_s} & \dfrac{-2 \cdot g_{u,d} \cdot g_{d,s}}{m_d} \\ \dfrac{g_{u,s} \cdot g_{d,s}}{m_s} & 0 & \dfrac{-2 \cdot g_{u,d} \cdot g_{u,s}}{m_u} \\ \dfrac{-2 \cdot g_{u,d} \cdot g_{d,s}}{m_d} & \dfrac{-2 \cdot g_{u,d} \cdot g_{u,s}}{m_u} & 0 \end{bmatrix}. \tag{19}$$

For each coupling constant $g$, the two quarks put in subscripts are the ones that form the diquark. We can also note the factor –2, which appears in front of the terms including $g_{u,d}$ [16, 18]. This coefficient is a flavor factor. In fact, this factor appears in (13). As with $\mathcal{Z}$, the Π term in (8) is also a $3\times3$ matrix. This one is diagonal:

$$\Pi = \begin{bmatrix} \Pi_u & 0 & 0 \\ 0 & \Pi_d & 0 \\ 0 & 0 & \Pi_s \end{bmatrix}, \tag{20}$$

where $\Pi_u$ is the baryon loop function including the quark $u$ and the diquark $[ds]$. Also, $\Pi_d$ indicates the one including the quark $d$ and the diquark $[us]$. And, $\Pi_s$ is the one including $s$ and the diquark $[ud]$. In this configuration, the transition matrix T is obviously a $3\times 3$ matrix. Nevertheless, T is always defined by the Bethe-Salpeter equation $\mathrm{T}=\mathcal{Z}+\mathcal{Z}\cdot\Pi\cdot\mathrm{T}$. Stricto sensu, (6) is then rewritten as $\mathrm{T}=\mathcal{Z}(1-\Pi\mathcal{Z})^{-1}$, because in the framework of matrices calculations, we cannot divide by a matrix, but we can multiply by its matrix inverse. In all the cases, the equation $\det(1-\Pi\mathcal{Z})=0$ (8) stays valid. Formally, the equation to solve is the same as the one seen previously, i.e. when $\mathcal{Z}$ and $\Pi$ were scalar numbers. But, if we clarify the terms, using (19, 20) in (8), we obtain:

$$1-2\cdot\Pi_u\cdot\Pi_d\cdot\Pi_s\cdot\mathcal{Z}^{ud}\cdot\mathcal{Z}^{us}\cdot\mathcal{Z}^{ds}-\Pi_u\cdot\Pi_d\cdot\left(\mathcal{Z}^{ud}\right)^2-\Pi_u\cdot\Pi_s\cdot\left(\mathcal{Z}^{us}\right)^2-\Pi_d\cdot\Pi_s\cdot\left(\mathcal{Z}^{ds}\right)^2=0 \quad . \qquad (21)$$

The $\Lambda$ baryon can also have an axial flavor component, described by the diagram in the figure 7. It is easy to note the similarity with the figure 6.

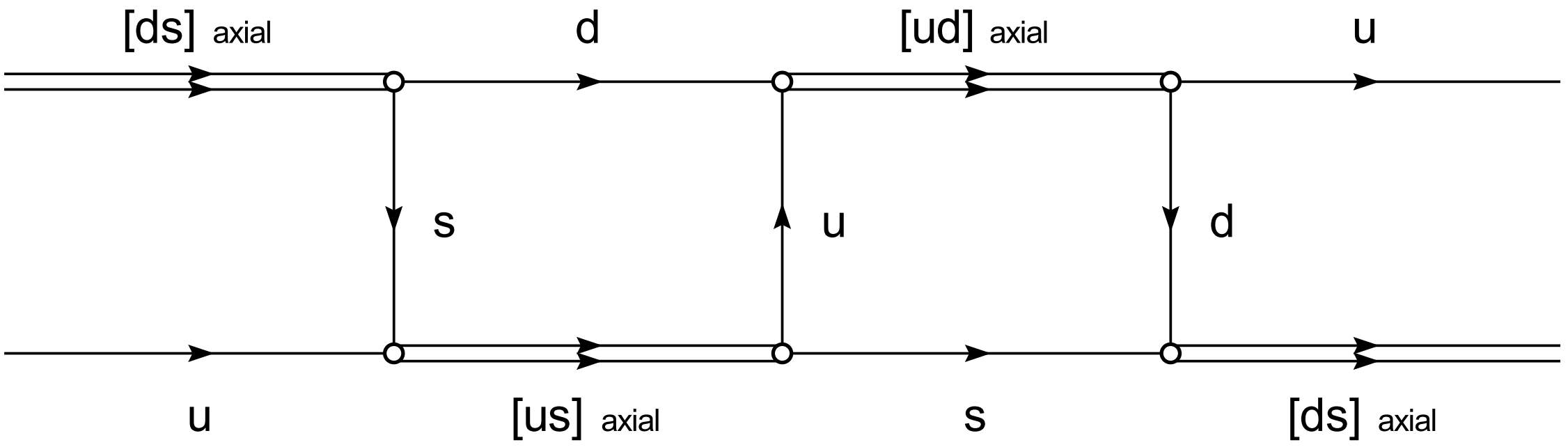

**Figure 7.** Axial flavor component of the $\Lambda$ baryon.

The associated wave-function is written [7]:

$$|\Lambda\rangle_{\text{axial}}=\frac{1}{\sqrt{12}}\cdot\left(\underbrace{\begin{bmatrix}1\\0\\0\end{bmatrix}}_{u}\otimes\underbrace{\lambda_6\cdot\begin{bmatrix}0\\1\\1\end{bmatrix}}_{[ds]_{\text{axial}}}+\underbrace{\begin{bmatrix}0\\1\\0\end{bmatrix}}_{d}\otimes\underbrace{\lambda_4\cdot\begin{bmatrix}1\\0\\1\end{bmatrix}}_{[us]_{\text{axial}}}-2\underbrace{\begin{bmatrix}0\\0\\1\end{bmatrix}}_{s}\otimes\underbrace{\lambda_1\cdot\begin{bmatrix}1\\1\\0\end{bmatrix}}_{[ud]_{\text{axial}}}\right) \quad . \qquad (22)$$

Compared to (18), the scalar diquarks are replaced by the axial ones. Consequently, the nature of the vertices is modified, as the matrices used in (22). Moreover, in the equation to solve to obtain the baryon's mass, i.e. the equivalent of (21), an additional factor 4 should be added at each vertex. Indeed, these ones translate here an axial interaction, see equation (2). Except for these dissimilarities, the equation is structurally identical to (21). So, we will not clarify it.

## 3.3 $\Sigma^0$ baryon

The $\Sigma^0$ baryon is treated separately from the $\Sigma^-$ and $\Sigma^+$, because the way to describe this baryon is very different compared to the two others. Clearly, the structure of the equations describing this baryon is rather close to the one seen for $\Lambda$. The only difference between $\Lambda$ and $\Sigma^0$ is this one does not include the term associated with the loop formed by $s$ and $[ud]$, figure 8. This remark is valid for the scalar and the axial flavor components.

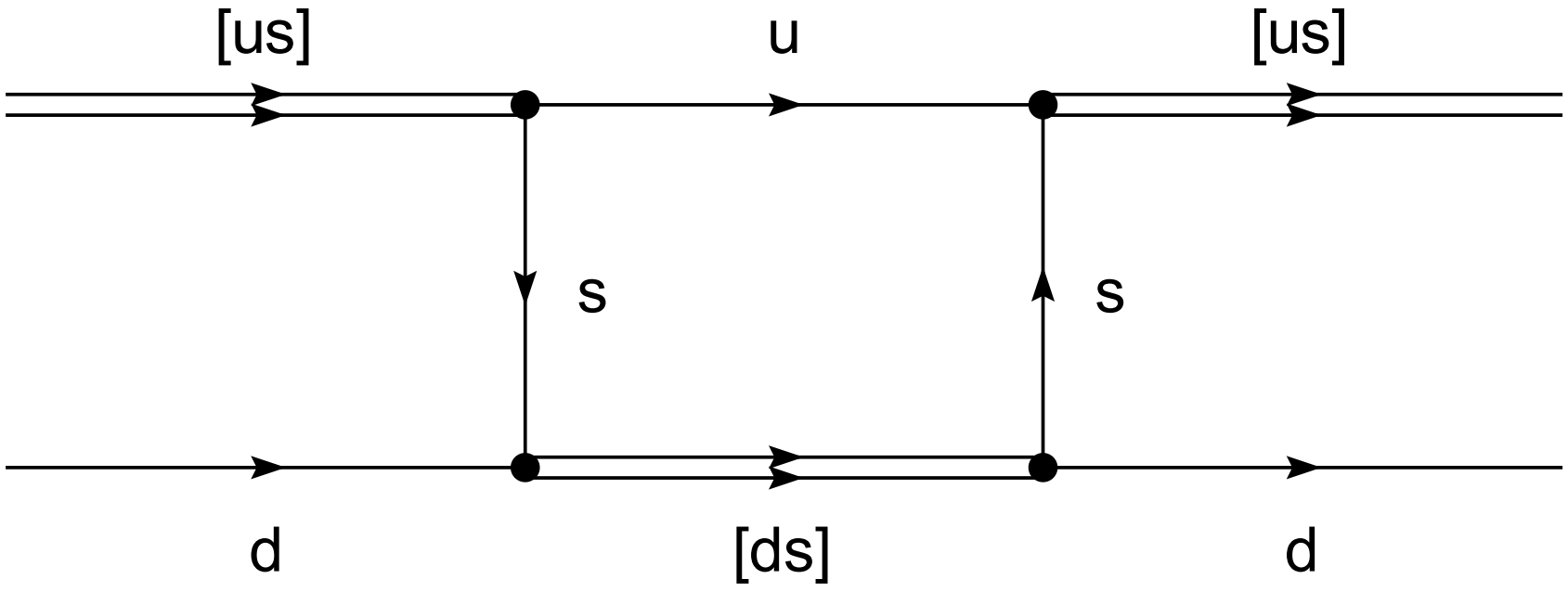

**Figure 8.** Scalar flavor component of the $\Sigma^0$ baryon.

In the other hand, the wave function is rather different compared to (18) [16, 18]:

$$\left|\Sigma^0\right\rangle_{\text{scalar}} = \frac{1}{2}\cdot\left(\underbrace{\begin{bmatrix}1\\0\\0\end{bmatrix}}_{u}\otimes\underbrace{i\lambda_7\cdot\begin{bmatrix}0\\1\\1\end{bmatrix}}_{[ds]}-\underbrace{\begin{bmatrix}0\\1\\0\end{bmatrix}}_{d}\otimes\underbrace{i\lambda_5\cdot\begin{bmatrix}1\\0\\1\end{bmatrix}}_{[us]}\right), \tag{23}$$

because of the normalization term $(1/2)$ and the minus sign in the center of the equation. In spite of that, the equation that gives the baryon mass is easy to obtain. We use what we did for $\Lambda$, and we write [18]:

$$\mathcal{Z}_{\Sigma^0} = \begin{bmatrix}0 & \mathcal{Z}^{ud} & 0\\ \mathcal{Z}^{du} & 0 & 0\\ 0 & 0 & 0\end{bmatrix} = \begin{bmatrix}0 & \dfrac{-g_{u,s}\cdot g_{d,s}}{m_s} & 0\\ \dfrac{-g_{u,s}\cdot g_{d,s}}{m_s} & 0 & 0\\ 0 & 0 & 0\end{bmatrix} \quad \text{and} \quad \Pi = \begin{bmatrix}\Pi_u & 0 & 0\\ 0 & \Pi_d & 0\\ 0 & 0 & 0\end{bmatrix}, \tag{24}$$

so that the equation to be solved is written, starting from $\det(1-\Pi\mathcal{Z})=0$ (8), as :

$$1-\Pi_u\cdot\Pi_d\cdot\left(\mathcal{Z}^{ud}\right)^2=0 \ . \tag{25}$$

The $\Sigma^0$ axial flavor component is obtained in exactly the same way. Indeed, this one is structurally identical to the scalar flavor component, see figure 9 and the wave-function (26).

$$\left|\Sigma^0\right\rangle_{\text{axial}} = \frac{1}{2}\cdot\left(\underbrace{\begin{bmatrix}1\\0\\0\end{bmatrix}}_{u}\otimes\underbrace{\lambda_6\cdot\begin{bmatrix}0\\1\\1\end{bmatrix}}_{[ds]_{\text{axial}}}-\underbrace{\begin{bmatrix}0\\1\\0\end{bmatrix}}_{d}\otimes\underbrace{\lambda_4\cdot\begin{bmatrix}1\\0\\1\end{bmatrix}}_{[us]_{\text{axial}}}\right). \tag{26}$$

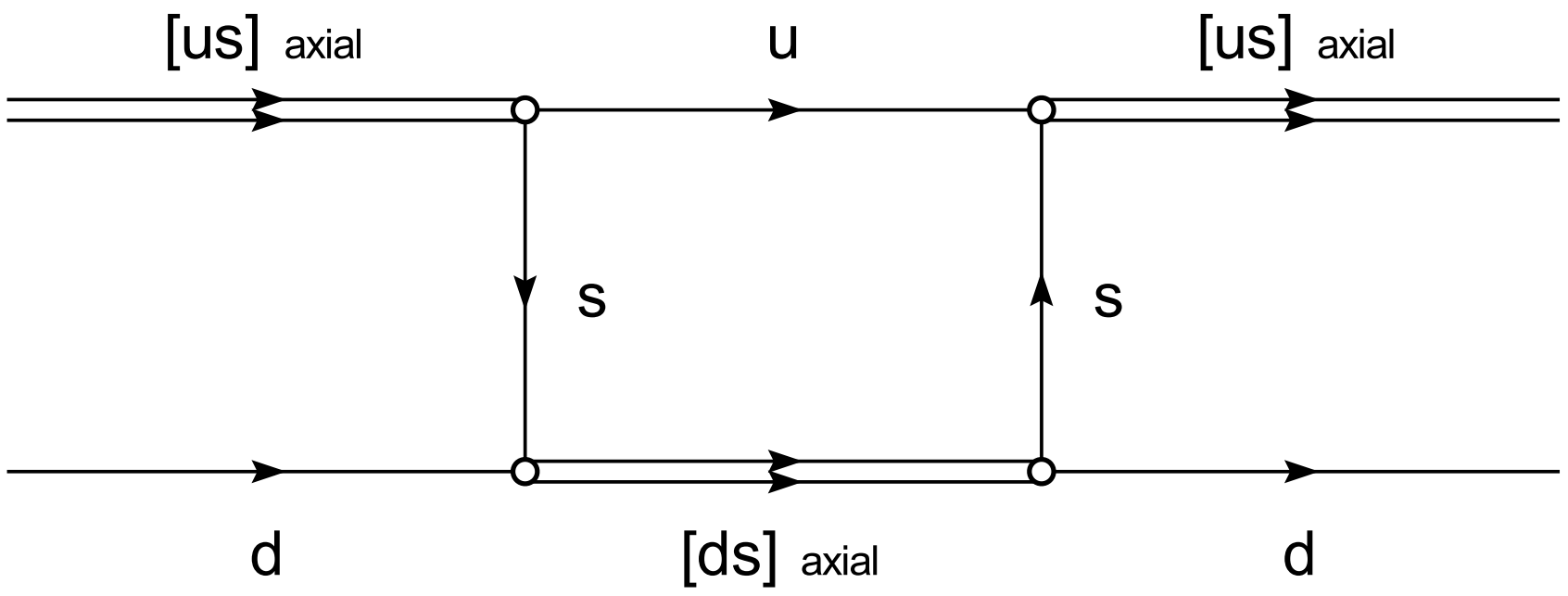

**Figure 9.** Axial flavor component of the $\Sigma^0$ baryon.

## 3.4 $\Sigma^+$ and $\Sigma^-$ baryons

The form of the equations describing these two baryons strongly recalls what was made for the nucleons scalar flavor components. We can see it on the figure 10 for $\Sigma^+$ : it is enough to replace the quark *d* of the proton by the quark *s*.

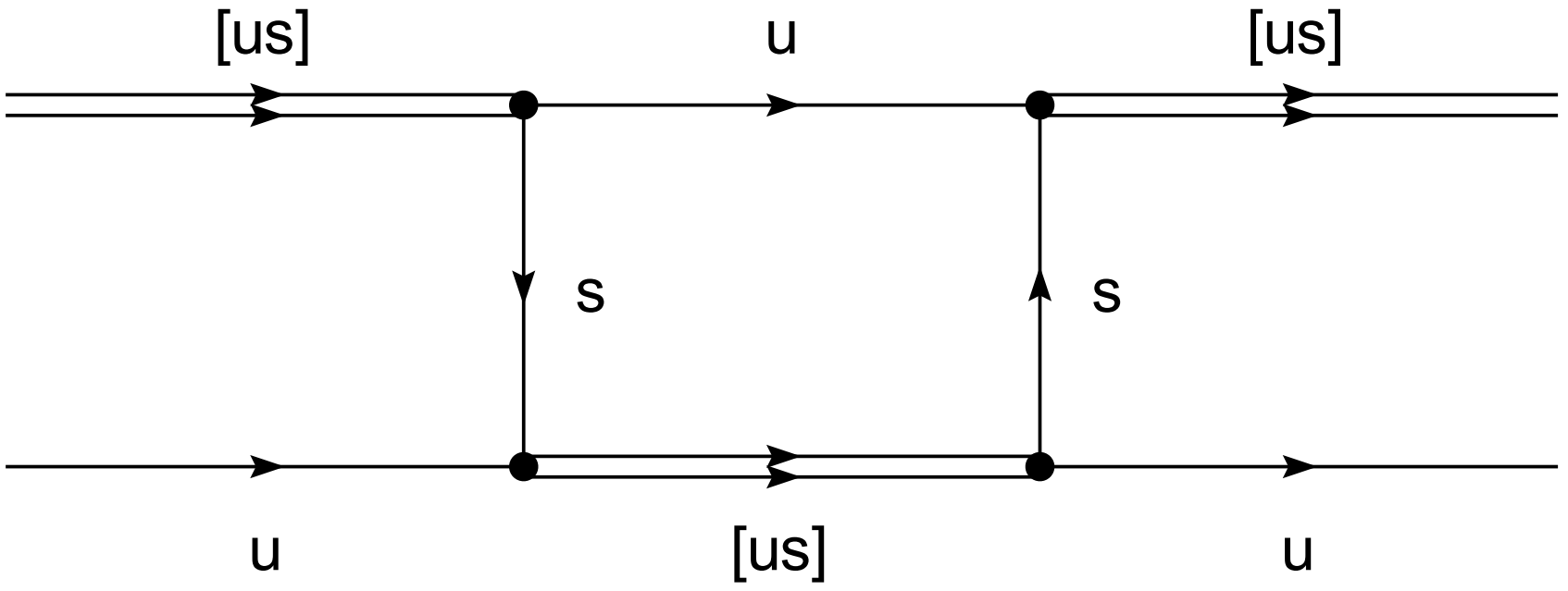

**Figure 10.** Scalar flavor component of the $\Sigma^+$ baryon.

The wave-function is written:

$$\left|\Sigma^+\right\rangle_{\text{scalar}} = \underbrace{\begin{bmatrix}1\\0\\0\end{bmatrix}}_{u} \otimes \underbrace{\frac{i}{\sqrt{2}}\cdot\lambda_5\cdot\begin{bmatrix}1\\0\\1\end{bmatrix}}_{[us]} \quad , \tag{27}$$

and the equation to be solved is:

$$1-2\cdot\frac{-2\cdot g_{us}{}^2}{m_s}\cdot\Pi_{[us],u}\left(M_{\Sigma^+},\vec{0}\right)=0 \quad . \tag{28}$$

An axial flavor component can also be considered. According to figure 11, it has the same structure as its scalar counterpart. The scalar diquarks are only replaced by axial ones; it implies an update of the vertices.

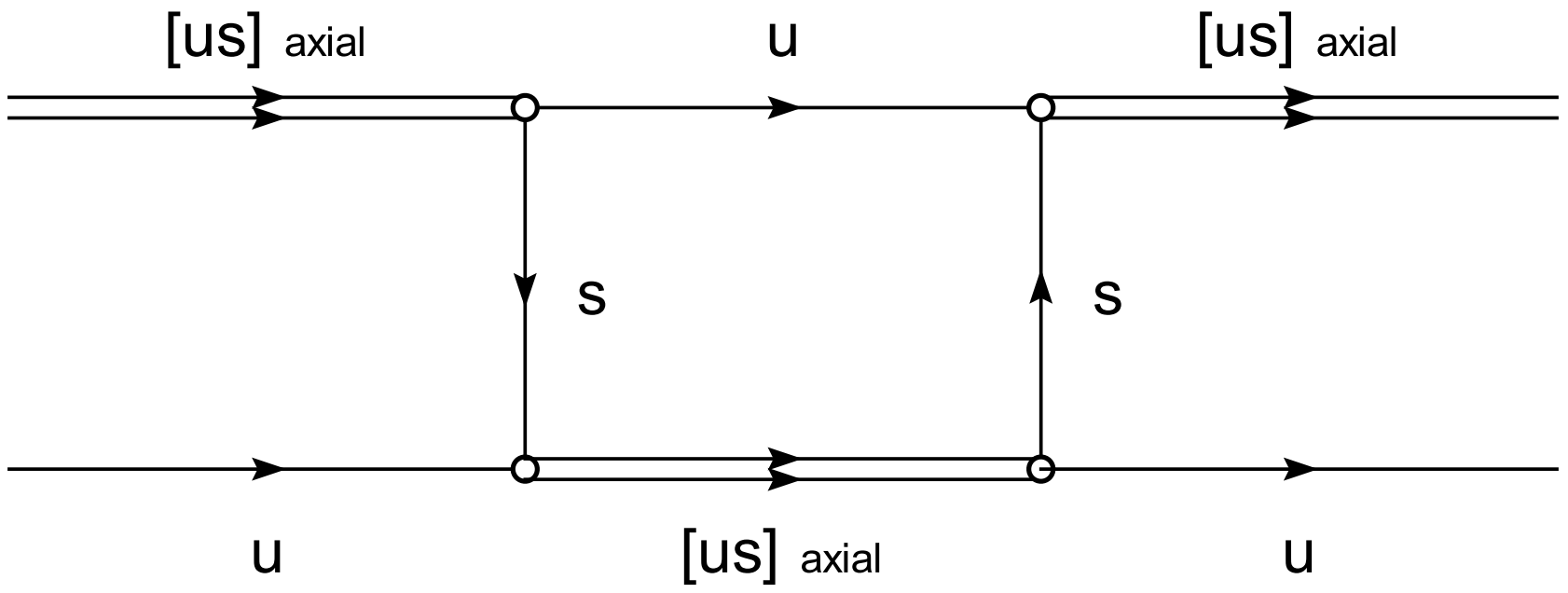

**Figure 11.** Axial flavor component of the $\Sigma^+$ baryon.

Thus, compared to (27), only the flavor matrix term is modified in the expression of the wave function [7]:

$$\left|\Sigma^+\right\rangle_{\text{axial}} = \underbrace{\begin{bmatrix}1\\0\\0\end{bmatrix}}_{u} \otimes \underbrace{\frac{1}{\sqrt{2}} \cdot \lambda_4 \cdot \begin{bmatrix}1\\0\\1\end{bmatrix}}_{[us]_{\text{axial}}} . \tag{29}$$

To obtain $\Sigma^-$, all the $u$ are replaced by $d$. It leads to the two wave-functions that correspond to the two flavor components:

$$\left|\Sigma^-\right\rangle_{\text{scalar}} = \underbrace{\begin{bmatrix}0\\1\\0\end{bmatrix}}_{d} \otimes \underbrace{\frac{i}{\sqrt{2}} \cdot \lambda_7 \cdot \begin{bmatrix}0\\1\\1\end{bmatrix}}_{[ds]} \quad \text{and} \quad \left|\Sigma^-\right\rangle_{\text{axial}} = \underbrace{\begin{bmatrix}0\\1\\0\end{bmatrix}}_{d} \otimes \underbrace{\frac{1}{\sqrt{2}} \cdot \lambda_6 \cdot \begin{bmatrix}0\\1\\1\end{bmatrix}}_{[ds]_{\text{axial}}} . \tag{30}$$

# 3.5 $\Xi$ baryons

The description of these baryons is very close that what we did with the nucleons and $\Sigma^{\pm}$. Indeed, the $\Xi^0$ scalar flavor component is composed by one term, figure 12.

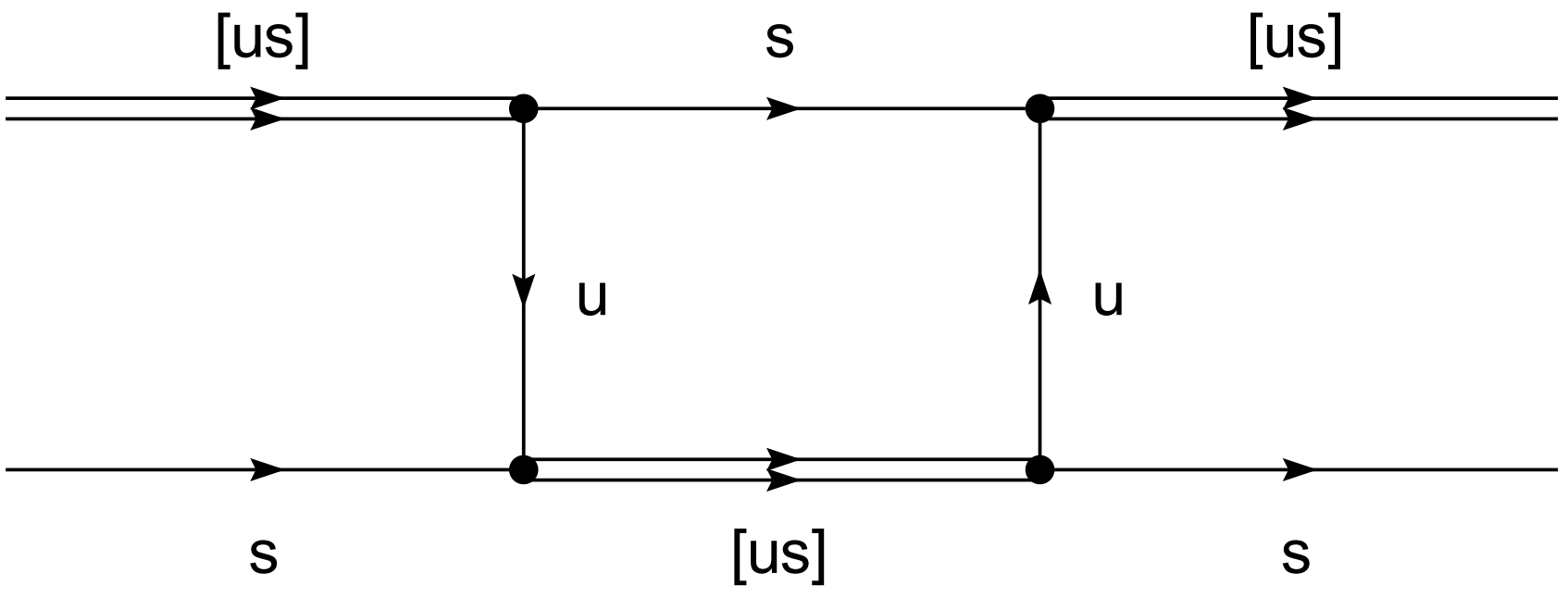

**Figure 12.** Scalar flavor component of the $\Xi^0$ baryon.

As a consequence, the wave-function of the scalar flavor component is written as:

$$\left|\Xi^0\right\rangle_{\text{scalar}} = \underbrace{\begin{bmatrix}0\\0\\1\end{bmatrix}}_{s} \otimes \underbrace{\frac{i}{\sqrt{2}}\cdot\lambda_5\cdot\begin{bmatrix}1\\0\\1\end{bmatrix}}_{[us]} \quad . \tag{31}$$

The equation to be solved to find the mass is:

$$1-\frac{-2\cdot g_{us}{}^2}{m_u}\cdot\Pi_{[us],s}\left(M_{\Xi^0},\vec{0}\right)=0 \quad . \tag{32}$$

Compared to the nucleons and to the $\Sigma^{\pm}$ baryons, we do not have the factor 2 in front of the term $\frac{-2\cdot g_{us}{}^2}{m_u}$. As with the nucleons, the axial flavor component includes two terms, see figure 13.

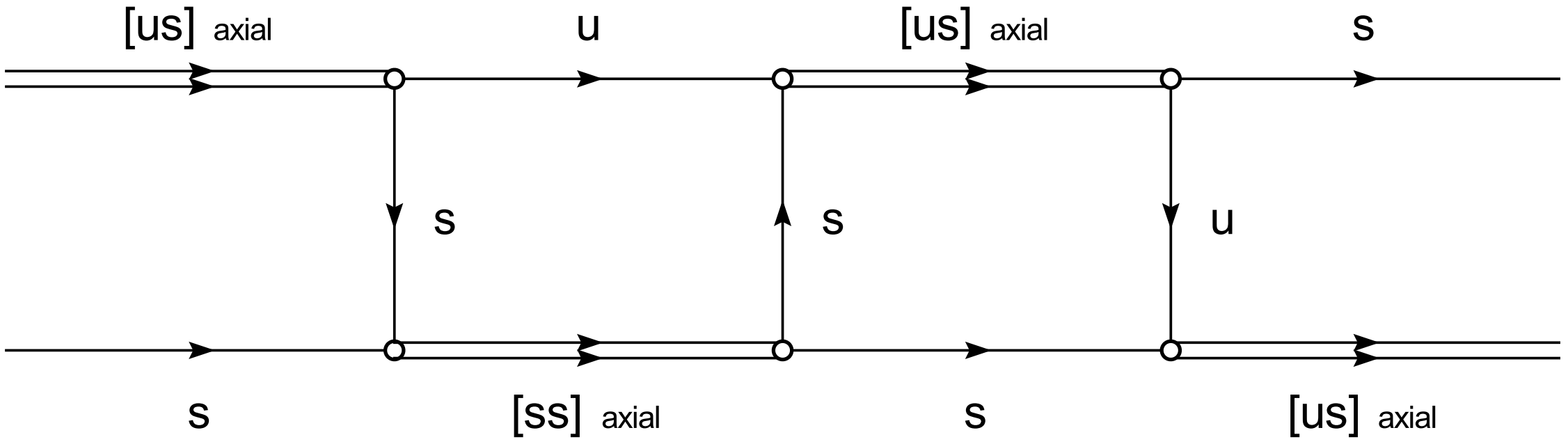

**Figure 13.** Axial flavor component of the $\Xi^0$ baryon.

Therefore, we have [6, 7]:

$$\left|\Xi^0\right\rangle_{\text{axial}} = \sqrt{\frac{1}{3}}\cdot\left(\underbrace{\begin{bmatrix}0\\0\\1\end{bmatrix}}_{s} \otimes \underbrace{\frac{1}{\sqrt{2}}\cdot\lambda_4\cdot\begin{bmatrix}1\\0\\1\end{bmatrix}}_{[us]_{\text{axial}}} - \underbrace{\begin{bmatrix}1\\0\\0\end{bmatrix}}_{u} \otimes \underbrace{\begin{bmatrix}0&&\\&0&\\&&\sqrt{2}\end{bmatrix}\cdot\begin{bmatrix}0\\0\\1\end{bmatrix}}_{[ss]_{\text{axial}}}\right) \quad . \tag{33}$$

About $\Xi^-$, we replace the $u$ by $d$, so that we obtain:

$$\left|\Xi^-\right\rangle_{\text{scalar}} = \underbrace{\begin{bmatrix}0\\0\\1\end{bmatrix}}_{s} \otimes \underbrace{\frac{i}{\sqrt{2}}\cdot\lambda_7\cdot\begin{bmatrix}0\\1\\1\end{bmatrix}}_{[ds]} \quad , \tag{34}$$

and:

$$\left|\Xi^-\right\rangle_{\text{axial}} = -\sqrt{\frac{1}{3}}\cdot\left(\underbrace{\begin{bmatrix}0\\0\\1\end{bmatrix}}_{s}\otimes\underbrace{\frac{1}{\sqrt{2}}\cdot\lambda_6\cdot\begin{bmatrix}0\\1\\1\end{bmatrix}}_{[ds]_{\text{axial}}}-\underbrace{\begin{bmatrix}0\\1\\0\end{bmatrix}}_{d}\otimes\underbrace{\begin{bmatrix}0 & & \\ & 0 & \\ & & \sqrt{2}\end{bmatrix}\cdot\begin{bmatrix}0\\0\\1\end{bmatrix}}_{[ss]_{\text{axial}}}\right). \tag{35}$$

# 3.6 Δ baryons

The $\Delta$ and $\Omega^-$ baryons are treated by only one component, i.e. the axial flavor component. Indeed, if we consider these baryons as a quark-diquark bound state, the diquarks must have flavor symmetrical wave-functions, because we need here to use diquarks as $[uu],[dd],[ss]$. The axial diquarks are the only ones that can satisfy this constraint [7], as observed in the previous chapter. Moreover, if the isospin symmetry is not considered, we have four different $\Delta$ baryons. They are $\Delta^{++}$, $\Delta^-$, $\Delta^+$ and $\Delta^0$. We start with $\Delta^{++}$, see figure 14.

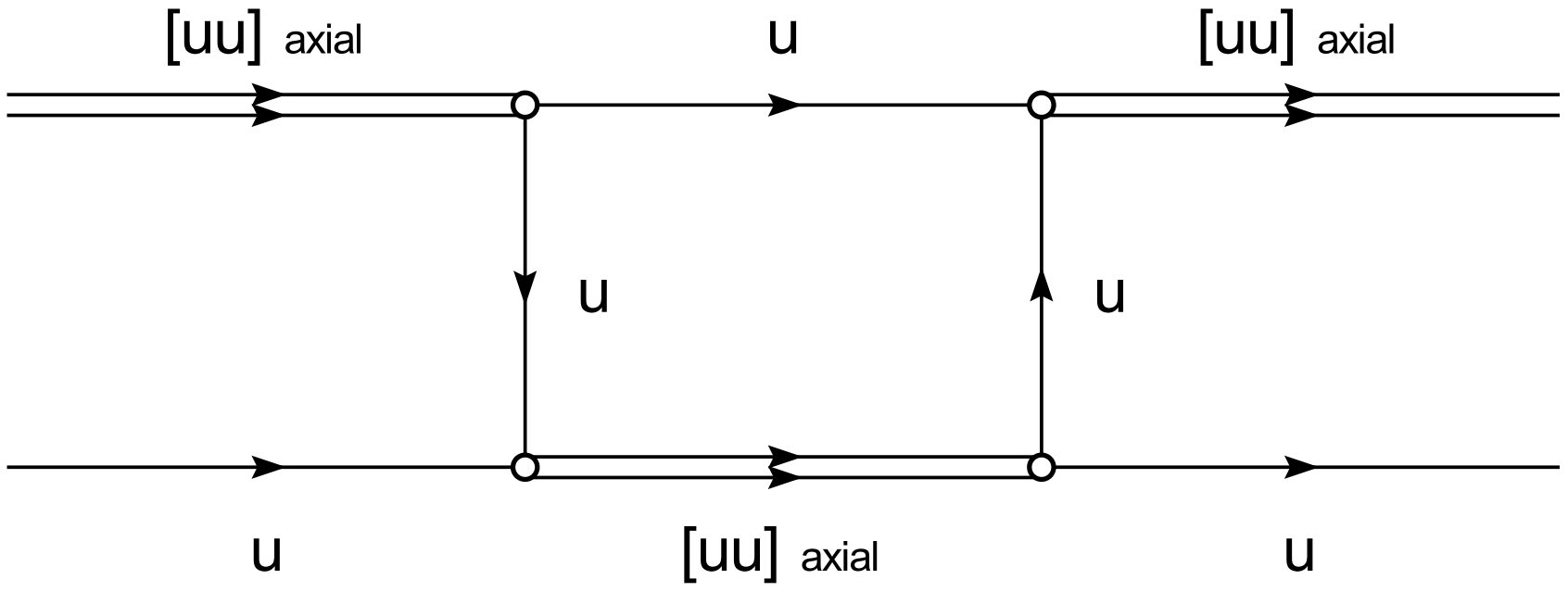

**Figure 14.** Axial flavor component of the $\Delta^{++}$ baryon.

According to what we saw upstream, the associated wave-function is written as [7]:

$$\left|\Delta^{++}\right\rangle = \underbrace{\begin{bmatrix}1\\0\\0\end{bmatrix}}_{u}\otimes\underbrace{\frac{i}{\sqrt{2}}\cdot\lambda_{+1}\cdot\lambda_2\cdot\begin{bmatrix}1\\0\\0\end{bmatrix}}_{[uu]_{\text{axial}}}\ , \tag{36}$$

and the equation used to obtain the baryon mass is:

$$1-2\cdot\frac{-2\cdot 4\cdot g_{uu}{}^2}{m_u}\cdot\Pi_{[uu],u}\left(M_{\Delta^{++}},\vec{0}\right)=0\ . \tag{37}$$

A factor 4 is present, because of the axial interaction channel. The $\Delta^-$ baryon is obtained from what we did for $\Delta^{++}$ by a replacement of the $u$ by $d$. It comes:

$$\left|\Delta^-\right\rangle = \underbrace{\begin{bmatrix}0\\1\\0\end{bmatrix}}_{d} \otimes \underbrace{\frac{i}{\sqrt{2}}\cdot\lambda_{-1}\cdot\lambda_2\cdot\begin{bmatrix}0\\1\\0\end{bmatrix}}_{[dd]_{\text{axial}}} \quad . \tag{38}$$

Concerning $\Delta^+$, we have to consider a linear combination of two terms, figure 15.

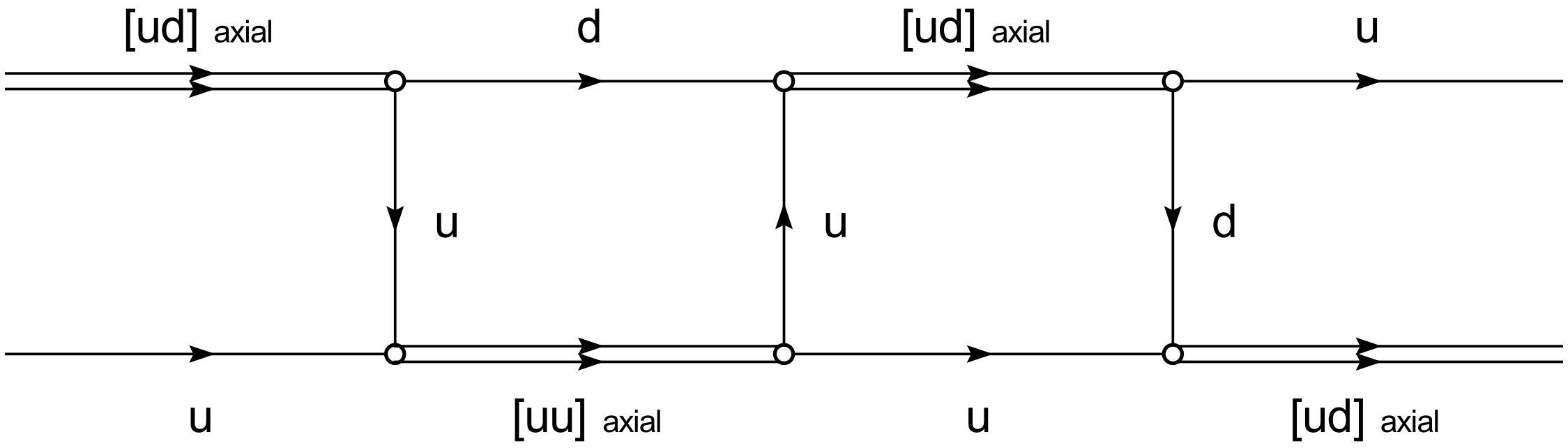

**Figure 15.** $\Delta^+$ baryon.

But, this diagram exactly corresponds to the axial flavor component of the proton. Therefore, the wave-function of the $\Delta^+$ baryon is identical to the one written equation (14).

About the $\Delta^0$ baryon, we replace the $u$ by $d$, and conversely, in the figure 15. It corresponds to the axial flavor component of the neutron, and thus the associated wave-function is identical to (17).

# 3.7 $\Omega^-$ baryon

The baryon $\Omega^-$ can be modeled as an association of a quark $s$ and an axial diquark $[ss]$. Obviously, the exchanged quark is a quark $s$, as represented in the figure 16.

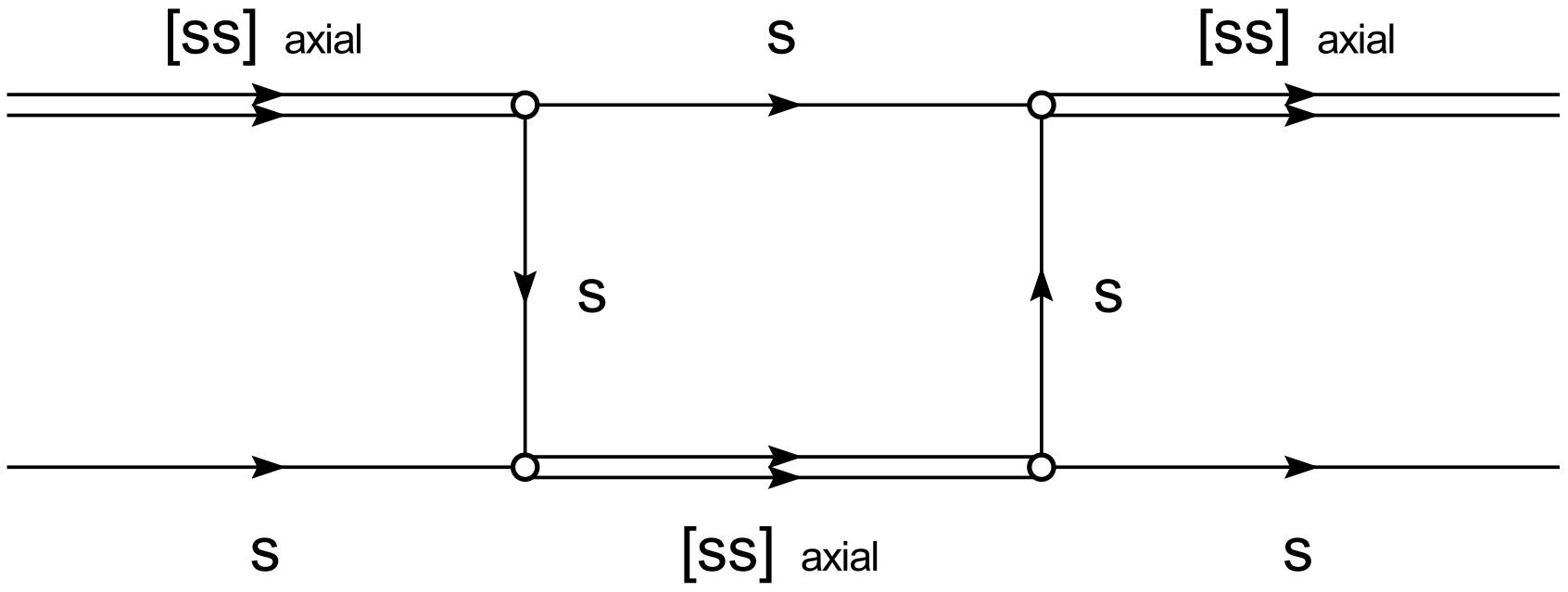

**Figure 16.** $\Omega^-$ baryon.

The wave-function associated with this baryon is written as:

$$\left|\Omega^-\right\rangle = \underbrace{\begin{bmatrix}0\\0\\1\end{bmatrix}}_{s} \otimes \underbrace{\frac{1}{\sqrt{2}} \cdot \begin{bmatrix}0 & & \\ & 0 & \\ & & \sqrt{2}\end{bmatrix} \cdot \begin{bmatrix}0\\0\\1\end{bmatrix}}_{[ss]_{\text{axial}}} \quad , \tag{39}$$

and the equation to be solved is:

$$1 - \frac{-2 \cdot 4 \cdot g_{ss}^{\ 2}}{m_s} \cdot \Pi_{[ss],s}\left(M_{\Omega^-}, \vec{0}\right) = 0 \quad . \tag{40}$$

# 4. Results at finite temperatures and densities

In our numerical results, we noted that the octet baryons can be correctly described only via their scalar flavor component. In fact, even if it was possible to include the axial flavor component, as in [7], our numerical results showed that this contribution can be neglected in the (P)NJL description performed here. In the same way, the decuplet baryons were modeled using their axial flavor component.

Moreover, as in the previous chapters, our calculations at finite temperatures and densities were performed with the P1 parameter set. It wants to say that we considered the isospin symmetry. In fact, with the octet and the decuplet baryons, we thus have to study 18 baryons. Thanks to the isospin symmetry, some baryons' masses are degenerate. As a consequence, the number of curves to be plotted is reduced. Thus, it avoids overloading our graphs.

## 4.1 Octet baryons

As mentioned in the introduction, some works related to the NJL octet baryons at finite temperatures and densities are available in the literature, e.g. [16-18]. These references used the scalar flavor component to describe octet baryons. They did not model axial diquarks, thus they did not include the axial flavor component. About the PNJL results, we recall that the PNJL baryons' modeling was not performed in the literature before our work.

Our results are presented in the figures 17 to 19. In the figure 17, we study the masses of the baryons according to the temperature, at null density, whereas in the figure 18, the masses are calculated at finite densities, with $T = 0$. In these figures, the masses globally decrease when the temperature increases, until the baryons come to their limit of stability. However, according to the density, it was also observed that the mass reaches a minimum, and then increases again. This behavior notably occurs for the nucleons, $\Xi$ and $\Lambda$. In a general way, the octet baryons are more sensitive to the baryonic density than to the temperature. The mass diminution according to this parameter is less than 20%, in the NJL and PNJL models. The rate is about 30% according to the baryonic density. Also, in our description, the mass of the nucleons is equal to 897.5 MeV at null density. At the ordinary nuclear density $\rho_0$, the mass is 724.4 MeV. It leads to a ratio $M_N\left(\rho_B = \rho_0\right)/M_N\left(\rho_B = 0\right) \approx 0.8$, whereas [22] proposes 0.6.

As a consequence, the decrease of the nucleon' mass according to the baryonic density is expected to be stronger than in our approach. Our pure NJL results can be compared to the ones of [16–18]. Especially with the study according to the baryonic density, we note that our curves do not present the defects observable in these references, especially in [16], characterized by strong numerical instabilities. Another difference concerns the obtained masses at reduced temperatures, notably for $\Xi$. This aspect is explainable by the choice of the $G_{DIQ}$ constant in the used parameter set. In fact, our choice seems to be better as regards the behavior of $\Xi$. Indeed, all our baryons have a critical temperature and a critical density, figures 17, 18, whereas it seems not to be the case for the $\Xi$ modeled in [16–18].

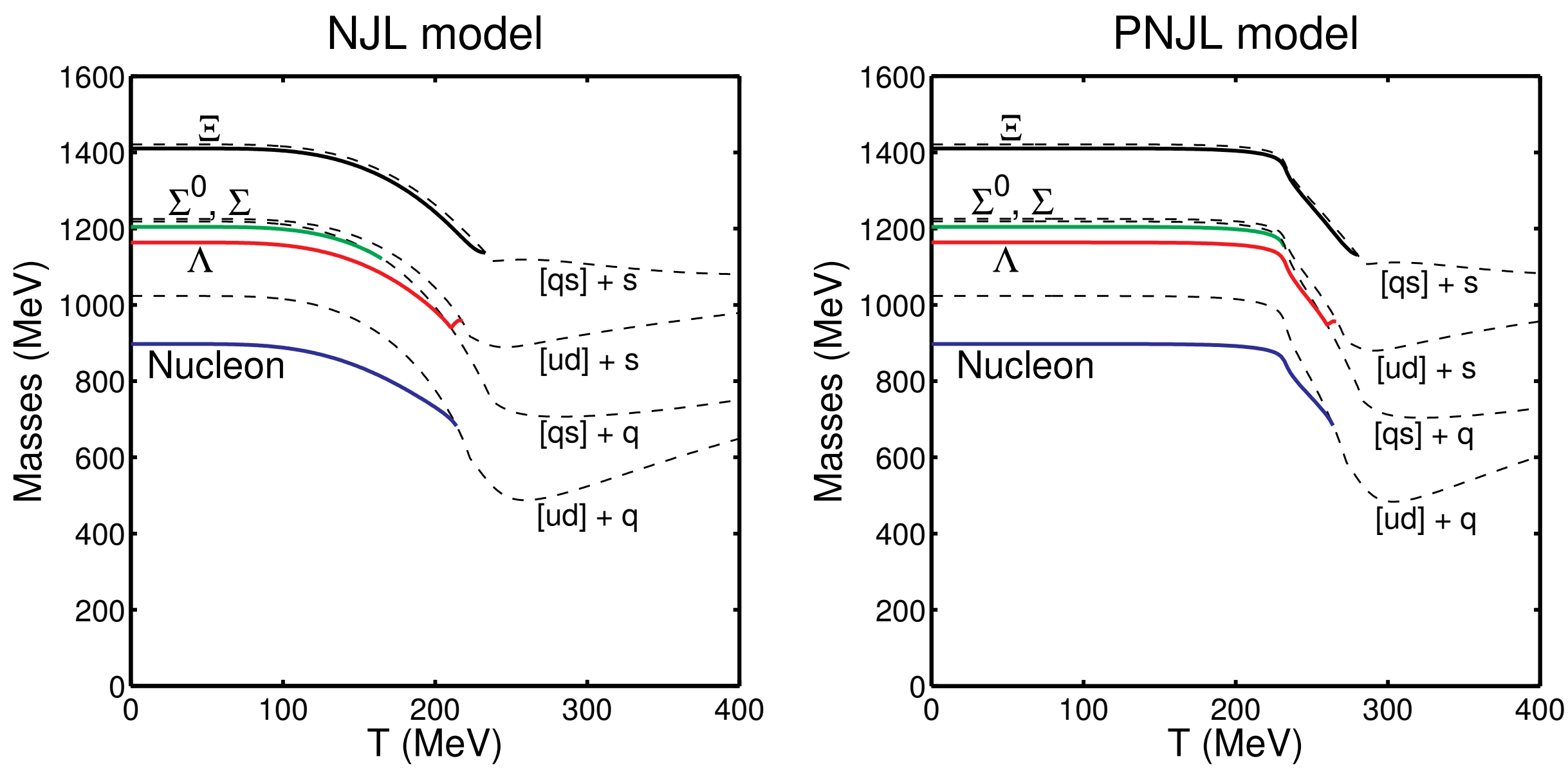

**Figure 17.** Masses of the octet baryons function of the temperature.

More precisely, concerning $\Xi$, we found that these particles are the octet baryons that have the best tolerance as regards the temperature or the baryonic density. Indeed, they have the strongest critical temperature and critical density. The $\Xi$ are *globally* made by two quarks $s$ and by only one light quark. This explains this excess of stability compared to the other baryons. Indeed, the strange quarks $s$ are less sensitive to $T, \rho_B$ than the light quarks. However, the binding energy of $\Xi$ is always weak. We recall that the binding energy can be found if we compare the values given by the $[qs]+s$ and the $\Xi$ curves.

Concerning the $\Sigma$ baryons, the curves associated with $\Sigma^0, \Sigma^\pm$ are strictly degenerate, whatever be the temperature or the baryonic density. In the framework of the isospin symmetry, such a result is perfectly correct. However, the equations used to model $\Sigma^0$ and $\Sigma^\pm$ are different, subsections 3.3 and 3.4. Thus, it was not obvious to *strictly* obtain the same results. This remark will be also applicable to the decuplet baryons $\Sigma^{0*}, \Sigma^*$. Moreover, the binding energies of the $\Sigma$ baryons are always rather reduced.

About the $\Lambda$ baryon, we saw in the subsection 3.2 that its wave-function is described by three distinct states, i.e. $[ud]+s$, $[us]+d$ and $[ds]+u$. In the framework of the isospin symmetry, the two former ones are degenerate, so we have the states $[ud]+s$ and $[qs]+q$. It leads to "two critical temperatures" and "two critical densities" for this baryon. In other words, the phenomenon is present according to the temperature or the baryonic density, respectively, figures 17 and 18. Firstly, when the $\Lambda$ curve crosses the $[qs]+q$ curve, the $\Lambda$ becomes unstable according to this state. But, it stays stable as regards the $[ud]+s$ state, until it crosses the associated curve. After that, the baryon becomes unstable. In the figures, the "double stable/unstable transition" is observable by a fast increase of the $\Lambda$'s mass, between the $[qs]+q$ and the $[ud]+s$ curves.

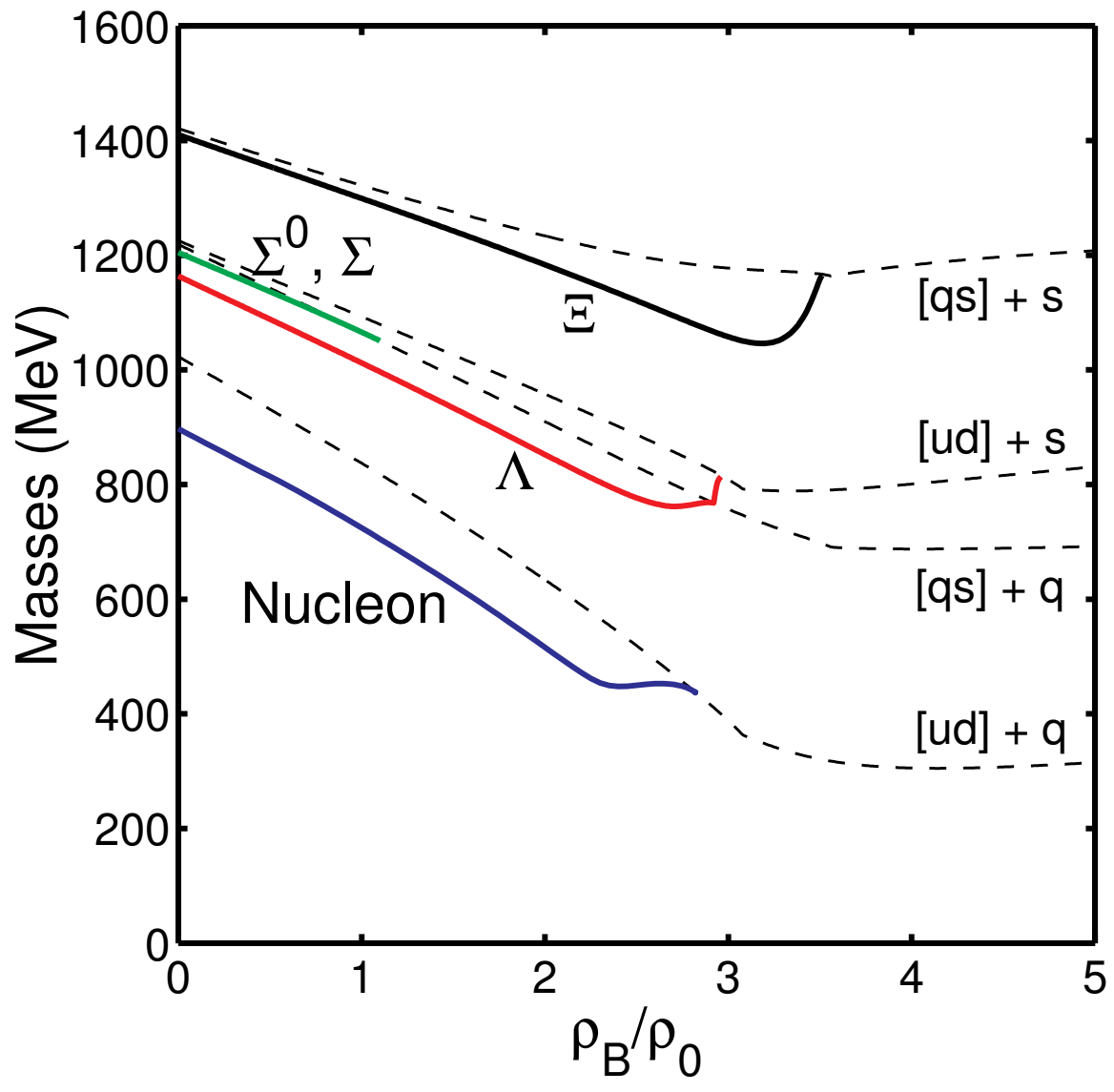

**Figure 18.** Masses of the octet baryons according to the baryonic density.

In the framework of our work, we recall that the instability of a baryon is associated with its disintegration in a quark and a diquark. It does not imply the decay into another baryon. Moreover, in our baryons' results, we focused on a study of their masses in their stability zone, and not in the instability one. In fact, our baryons' equations were constructed using the approximation proposed in [24]. It consists to neglect the imaginary part of the $k_0$ term, see (9), used as argument of the $B_0$ function. With the baryons, this approximation forbids to work in the instability zone of the baryons, as explained in the subsection 5.3 of the appendix D. Clearly, calculations of this kind require performing integrations that use complex numbers. Some versions of our numerical algorithms perform such integrations.

Moreover, the figures 17 show that the inclusion of the Polyakov loop leads to the already observed shifting towards higher temperatures. In fact, we confirm that this curve's distortion does not alter the values: for example, the $\Lambda$'s behavior described upstream can be found in the NJL and in the PNJL results, in the figure 17. In the framework of the PNJL model, at reduced densities, the baryons' masses stay constant in a range of about 200 MeV, as with their constituents (quarks and diquarks). Also, the critical temperatures are shifted towards higher values. Concretely, they are located after 260 MeV for the PNJL curves, i.e. 20 MeV higher than in the NJL description. As a consequence, the baryons' stability zones are enhanced by the inclusion of the Polyakov loop. This observation is confirmed by the figure 19, in which the mass of the nucleon is estimated in the $T, \rho_B$ plane, in its stability zone, for the NJL and the PNJL models. In this figure, we can concretely estimate the extension of the nucleon's stability zone due to the Polyakov loop. It was also confirmed that the NJL and PNJL approaches coincide at null temperature, whatever be the baryonic density. Moreover, as with the quarks in the chapter 2, it was also confirmed that the Polyakov loop only act according to the temperature, and not according to the density. A figure as the figure 19 is also a relevant test as regards the stability of our numerical method. We conclude that the test is very positive, because no defect is observable on the graphs, as discontinuities or other pathological behavior.

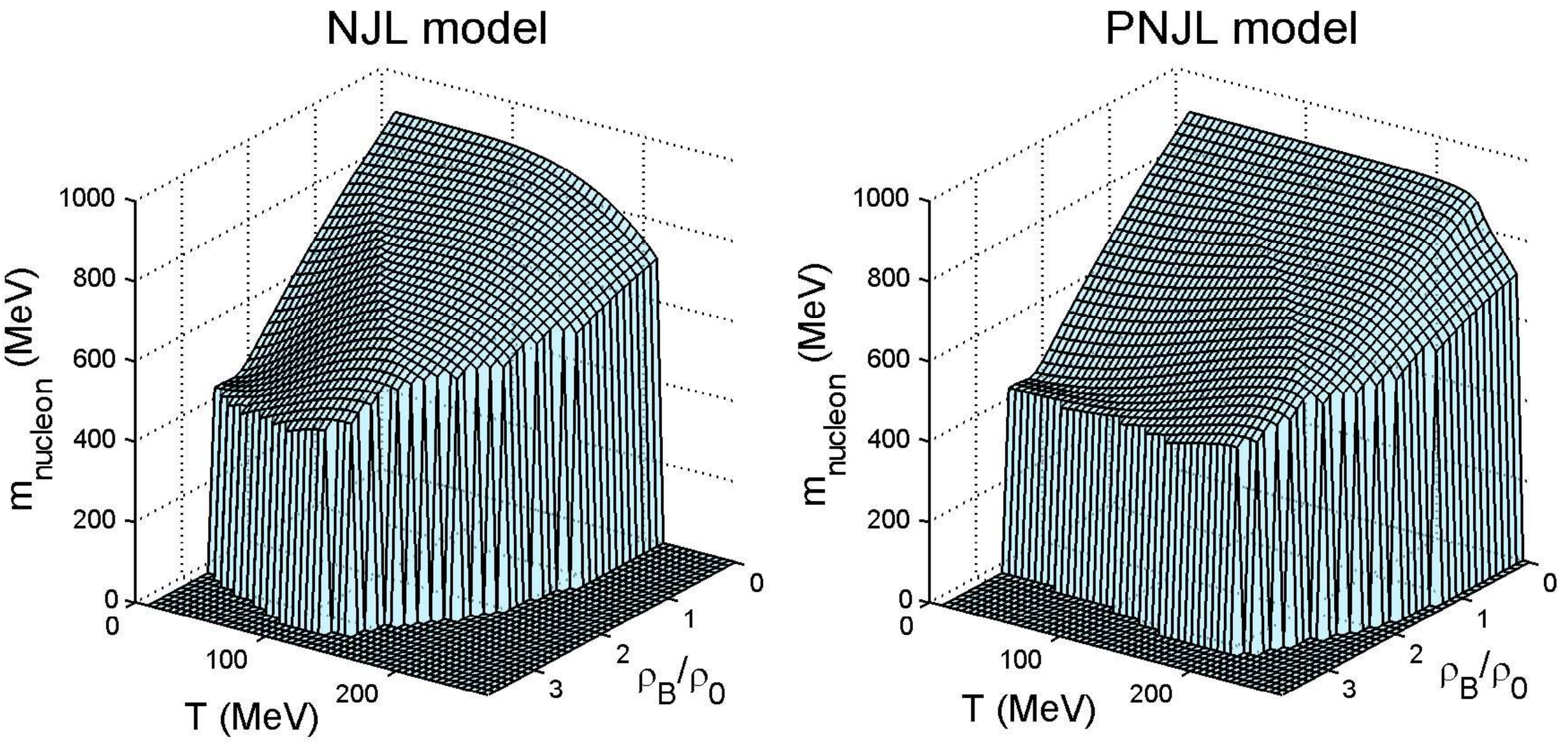

**Figure 19.** Mass of the nucleon in the $T-\rho_B$ plane.

# 4.2 Decuplet baryons

In the (P)NJL models, the study of the decuplet baryons at finite temperatures and densities performed here can be considered as new as regards the literature. In fact, estimations of the masses of these baryons were performed at $T=0, \rho_B=0$, for example in [3]. But, these data will be considered later, in the section 6. Our results are displayed in the figures 20 and 21. Clearly, the decuplet baryons present differences compared to the octet ones. The variations of the masses according to the temperatures are weak, figure 20. The masses of the $\Delta$ baryons tend to increase when the temperature is growing, whereas we found a decrease for all the octet baryons. But, as for these ones, the masses of the decuplet baryons decrease when the baryonic density increases, figure 21. About the NJL results, very disparate values of critical temperatures and densities are found. The critical temperatures of the decuplet baryons are globally weaker compared to the ones of the octet baryons. In a general way, the lightest baryons, as $\Delta$, are clearly the most fragile according to $T, \rho_B$. On the other hand, the $\Omega$ baryon is the most resistant: it critical temperature is comparable to the ones of the octet baryons. Furthermore, this baryon is manifestly too stable to admit a critical density at null temperature, at least in our study domain, see figure 21. The behaviors of these baryons are related to the ones of the axial $[qq]$ and $[ss]$ diquarks that constitute them. Moreover, about the $\Xi^*$ baryons, we observe the same phenomenon of “double transition” as with $\Lambda$, between the curves $[ss]+q$ and $[qs]+s$.

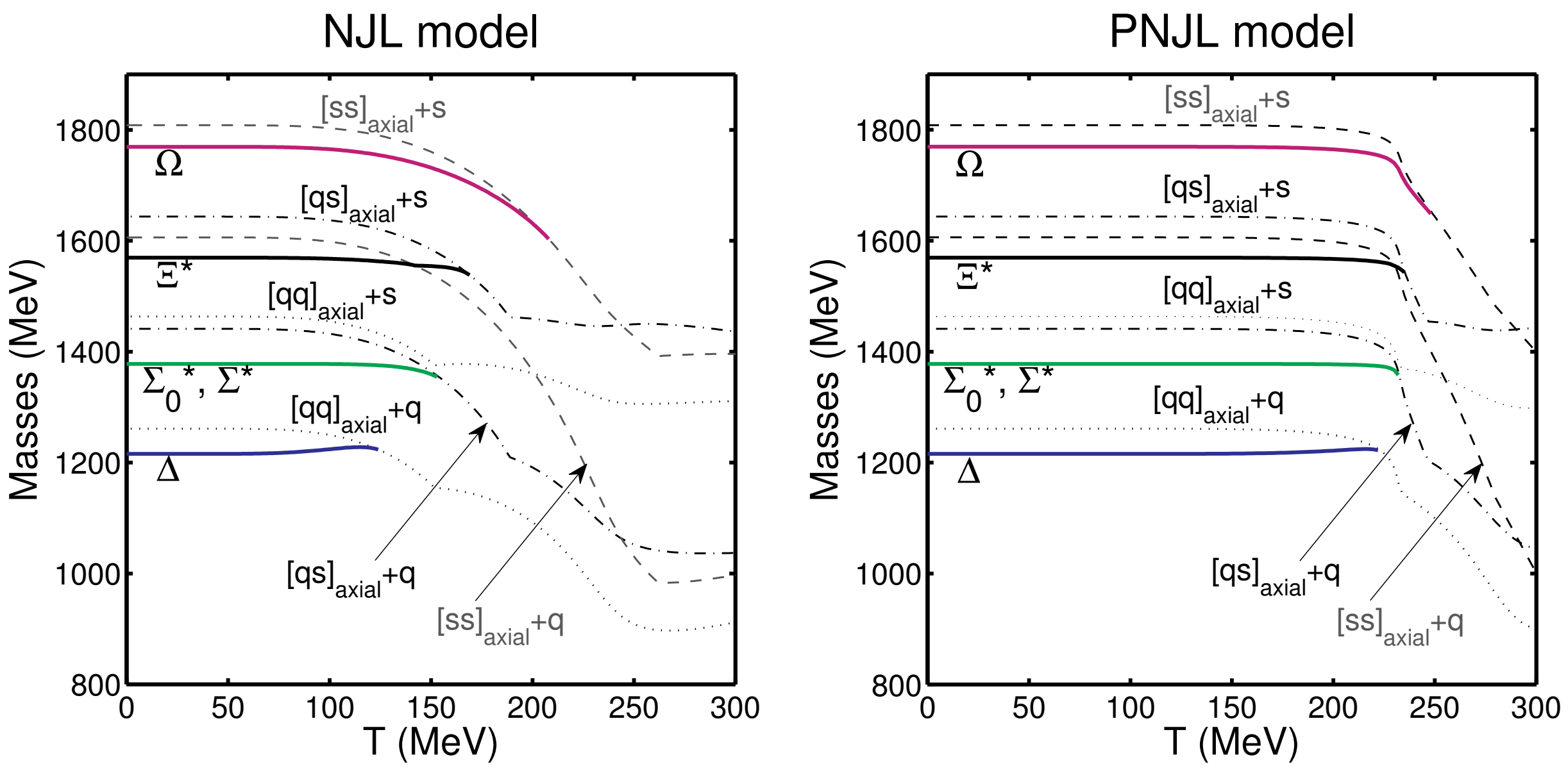

**Figure 20.** Masses of the decuplet baryons according to the temperature.

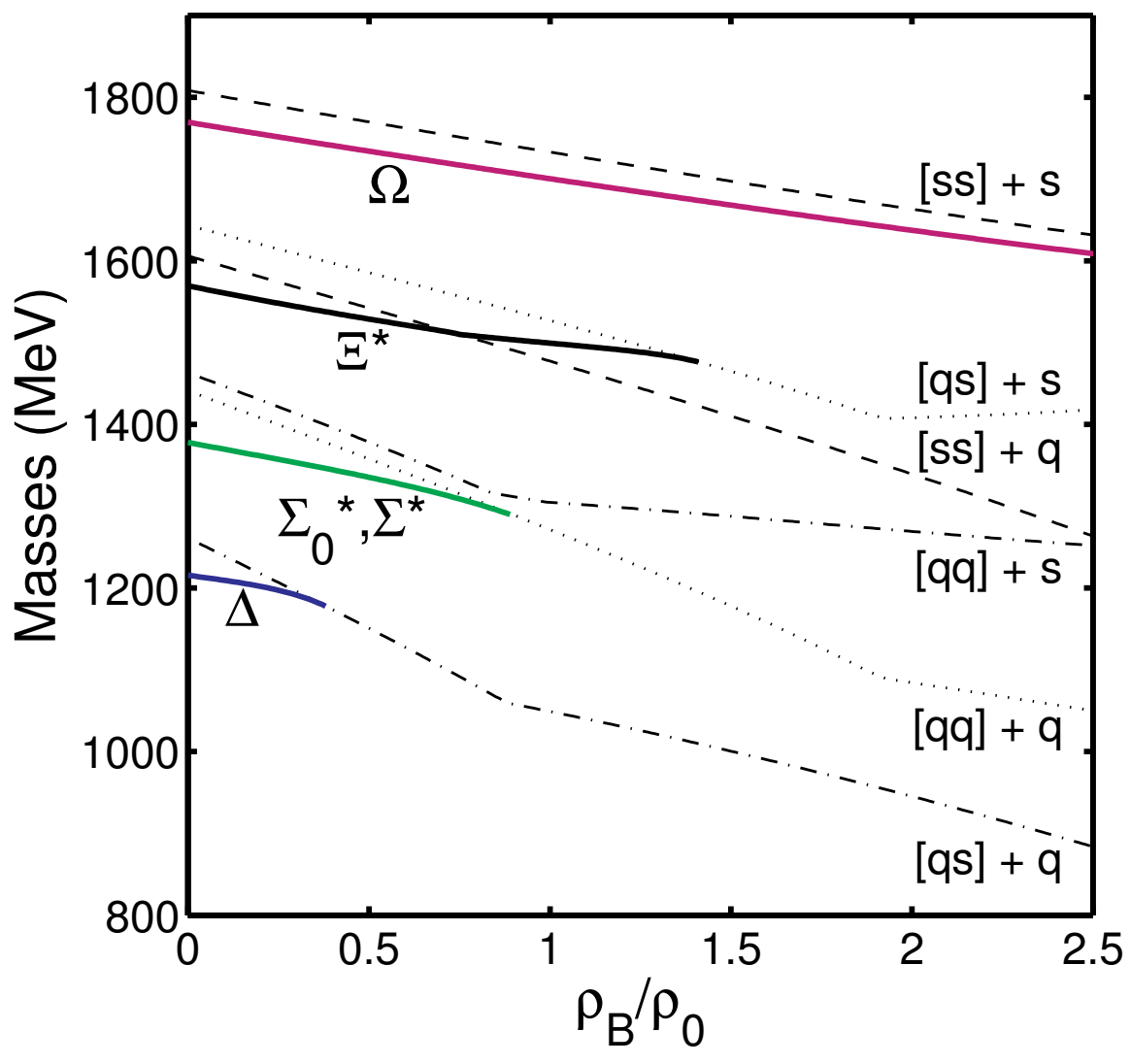

**Figure 21.** Masses of the decuplet baryons function of the baryonic density.

In the PNJL model, apart from the behavior already noted for the previous particles, we can underline the fact that the stability zone of the $\Delta$ baryons is strongly extended by the inclusion of the Polyakov loop. More precisely, for these baryons, we have a critical temperature close to 125 MeV in the NJL approach, against 225 MeV in the PNJL description. Furthermore, if the dispersion of the critical temperatures is important for the treated NJL decuplet baryons, this dispersion is reduced in the framework of the PNJL model.

# 4.3 Other studies

In order to complete the results found in the previous subsections, we propose there to study the anti-baryons and to establish NJL diagrams of stability/instability for the treated baryons. As explained in the previous chapter with the diquarks, the use of the matter-antimatter symmetry makes it possible to study antiparticles. For an anti-baryon, this trick consists to admit that the behavior of a baryon at density equal to $-\rho_B$ is the same as the one of its associated anti-baryon at $\rho_B$, and reversely. Our results for octet and decuplet baryons/anti-baryons are presented in the figure 22. For these particles, the baryons/anti-baryons couples are strictly degenerate at null density. A non-null baryonic density, for all baryons/anti-baryons couples, the mass of the anti-baryon is always higher than the mass of its associated baryon. As expected, it reveals that the anti-baryons are less stable than the baryons at positive densities. In fact, because of the anti-baryons instability in this regime, it explains why our baryonic density domain is so reduced in the figure 22. As a general tendency, the mass difference between a baryon and its anti-baryons quickly grows if the baryon is composed by light quarks. This remark is particularly true with the nucleons and the anti-nucleons. Besides, for this anti-particle, the increase of its mass is so strong that it becomes unstable when $\rho_B \approx 0.1\rho_0$. In the figure 22, the dotted lines indicate that the data were found in its instability zone. At this occasion, we used a numerical method slightly different from the one we usually employed, performing integrals with complex numbers, as evoked before.

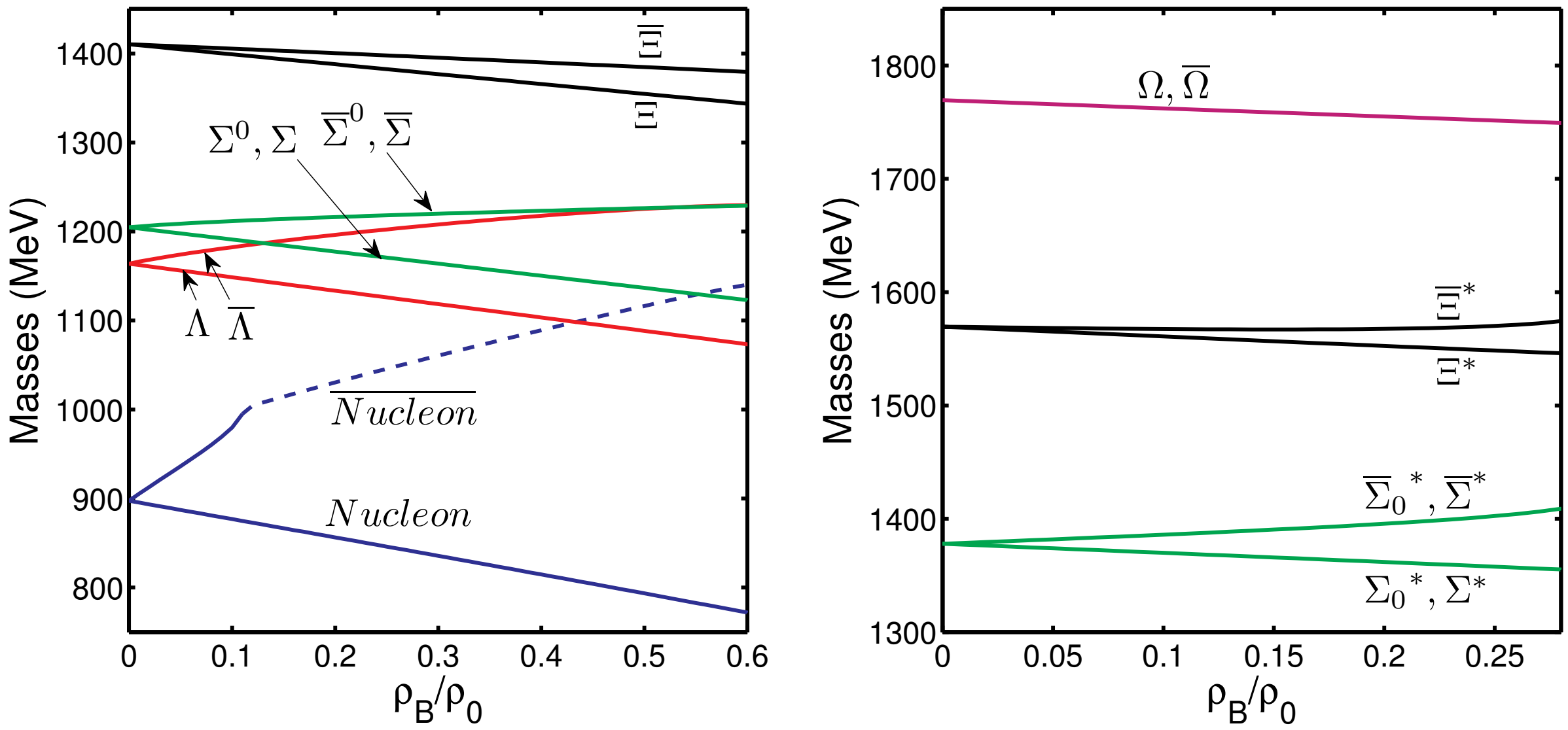

**Figure 22.** Masses of the baryons and anti-baryons function of the baryonic density.

In the previous chapter, we did not model the anti-diquarks $\overline{[qq]}$ at finite densities (for $T = 0$), because they were too instable. Therefore, we did not consider the anti-baryon $\overline{\Delta}$ in the figures 22, 24. At the opposite, the $[ss]$ and $\overline{[ss]}$ are always degenerate, whatever be the baryonic density. This behavior is also observable for the $\Omega, \overline{\Omega}$ baryons/antibaryons, in the figure 22. In the figure 24, it leads to a perfect symmetry of the $\Omega$ curve upon the $\rho_B = 0$ axis. We recall that the strange quarks, and by extension $[ss]$ and $\Omega$, are only affected by the absolute value of the baryonic density.

In the figure 23, we established the NJL diagram of stability/instability for the octet baryons, whereas in the figure 24, we interest us to the NJL diagram of the decuplet baryons, except for $\Delta$. As found with the diquarks, the baryons' curves present an asymmetry according to the $\rho_B = 0$ axis, except for $\Omega$. It confirms the observations done for the figure 22: the baryons are more stable at positive densities than at negative ones. About the baryons that present "several transitions", as the $\Lambda$ and $\Xi^*$, the curve indicates the limit for which *all* the states that compose the baryons are stables, i.e. the "first" transition visible in the figures 17, 18, 20, 21. In the two figures, the dotted curves are associated with the diquarks. They made possible to remark that the baryon's limit of stability is reached before the diquark(s) that compose it become unstable, for all the treated baryons. This point will be discussed later in this chapter.

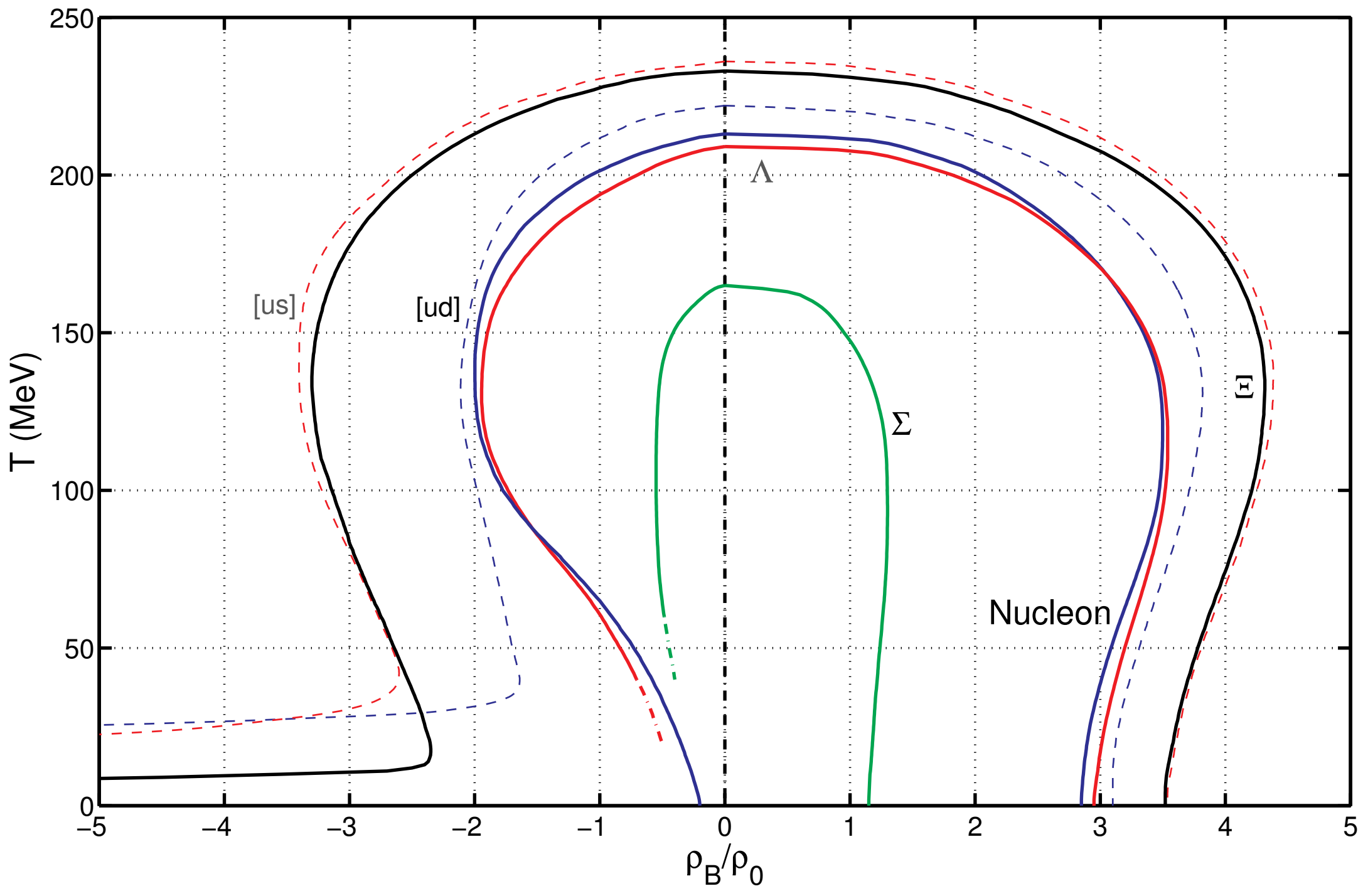

**Figure 23.** NJL diagram of stability/instability for octet baryons.

In fact, in the results obtained for the baryons, we observe that the masses of the baryons, or their limits of stability, are strictly continuous according to the baryonic density, positive or negative. This is a positive sign of the reliability of our approach. As the octet baryons are expected to intervene in the dynamic evolution model performed in the chapter 7, a different result at this stage would have been unacceptable, physically or numerically.

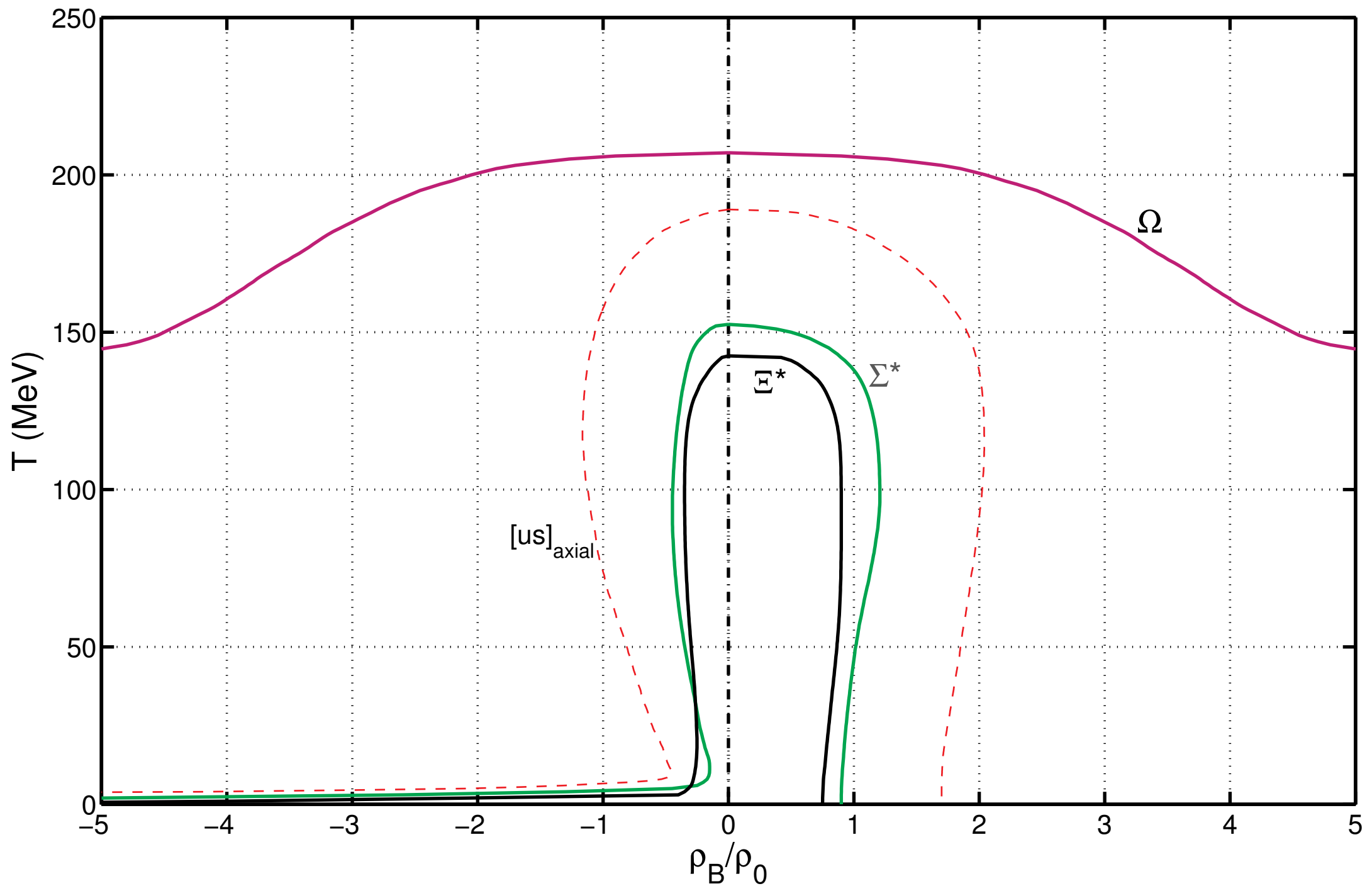

**Figure 24.** NJL diagram of stability/instability for decuplet baryons.

# 5. Coupling constants

## 5.1 Establishment of the baryons' coupling constants

As with the other composite particles studied before, the coupling constants involving a baryon can be considered. However, the method to be applied is more delicate for the baryons. Indeed, we saw with the mesons and the diquarks that the obtained T matrix could be associated with a (P)NJL propagator. This one could be compared to the traditional propagator, i.e. the one found in the framework of the quantum field theory, e.g. $\frac{1}{k^2 - m^2}$ for the mesons. In fact, this former one and the (P)NJL propagator are scalar. With the baryons, we have to remark the scalar nature of the used loop function. As a consequence, the T matrix, and so the (P)NJL propagator are also scalar, whereas the quantum field theory baryon propagator (free propagator) should be a four-vector, i.e. $\frac{\not{k} + m}{k^2 - m^2}$. However, the divergence is also verified with this former propagator. It allows us to validate (8). This aspect does not have consequence on the reliability of the method [16–18], but it imposes some precautions for the calculations associated with the coupling constants.

Several solutions are possible to establish the expression of the baryons' coupling constants. But, in all the cases, it seems necessary to proceed to an approximation. For example, a simple method considers the trace of the baryon's free propagator, and then to establish the equivalence with the (P)NJL propagator. For a nucleon, it leads to the expression:

$$\left.\frac{-2\cdot\dfrac{g^2}{m_q}}{1+4\cdot\dfrac{g^2}{m_q}\cdot\Pi\left(k_0,\vec{k}\right)}\right|_{k^2=m_B{}^2} \approx iG\cdot\left(Tr\left(\frac{\not{k}+m_B}{k^2-m_B{}^2}\right)\Bigg|_{k^2=m_B{}^2}\right)\cdot iG, \qquad (41)$$

in which $m_B$ is the mass of the nucleon and $G$ is the wanted coupling constant. Then, we apply the same method as in the previous chapters. Clearly, we firstly invert the equation (41), we derive according to $k$, and finally we pose $k_0 = m$, $\vec{k} = \vec{0}$. We obtain the relation:

$$G = \sqrt{\frac{1}{4\cdot\left.\dfrac{\partial\Pi\left(k_0,\vec{0}\right)}{\partial k_0}\right|_{k_0=m_B}}} \qquad (42)$$

In this formula, the derivative of $\Pi$ according to $k_0$ was numerically found as negative, leading to $G\in\mathbb{C}$. But, only $|G|$ is required in the calculations. Therefore, this behavior is without consequence on the results. Concerning the other baryons, we only need to take again (41) and to replace $\mathcal{D}$ by the one of the considered baryon. For the baryons described by a linear combination between several distinct states, as $\Lambda$, it is necessary to consider each state separately.

## 5.2 Results

We obtained the data presented in the figures 25 to 28. The figures 25 and 26 concern the octet baryons, whereas the figures 27 and 28 are associated with the decuplet ones. For the mesons and the diquarks, except for $\eta$, it was observed that the coupling constants are rather constant at reduced temperatures. Then, the curves decrease, they tend towards zero for the critical temperature, and then they increase again when the particle becomes unstable. A similar behavior was also observed according to the baryonic density when the particle presents a stable/unstable transition according to this parameter. Concerning the baryons, we recall that we studied them until they reach their limits of stability. Therefore, only a part of the curve is displayed in our results. According to the temperature, some of the baryons' curves correspond to this description, i.e. a decreasing until zero of the coupling constants. However, as with $\left|g_{\Lambda-[ud]s}\right|$, $\left|g_{\Xi-[qs]s}\right|$, the curves have a different behavior. For the first one, the curve does not tend towards zero. About the second one, the curve admits a maximum before reaching the critical temperature. Such a behavior is also observed for some octet baryons, according to the baryonic density, figure 26. This remark is particularly true for the $\left|g_{nucleon-[ud]q}\right|$, for which the maximum is spectacular. In fact, these maximums correspond to the densities for which the masses of the baryons are stable, as visible in the figure 18. More precisely, a coupling constant admits a maximum when the mass of the baryon presents a minimum according to $\rho_B$, or at least a stabilization of the mass.

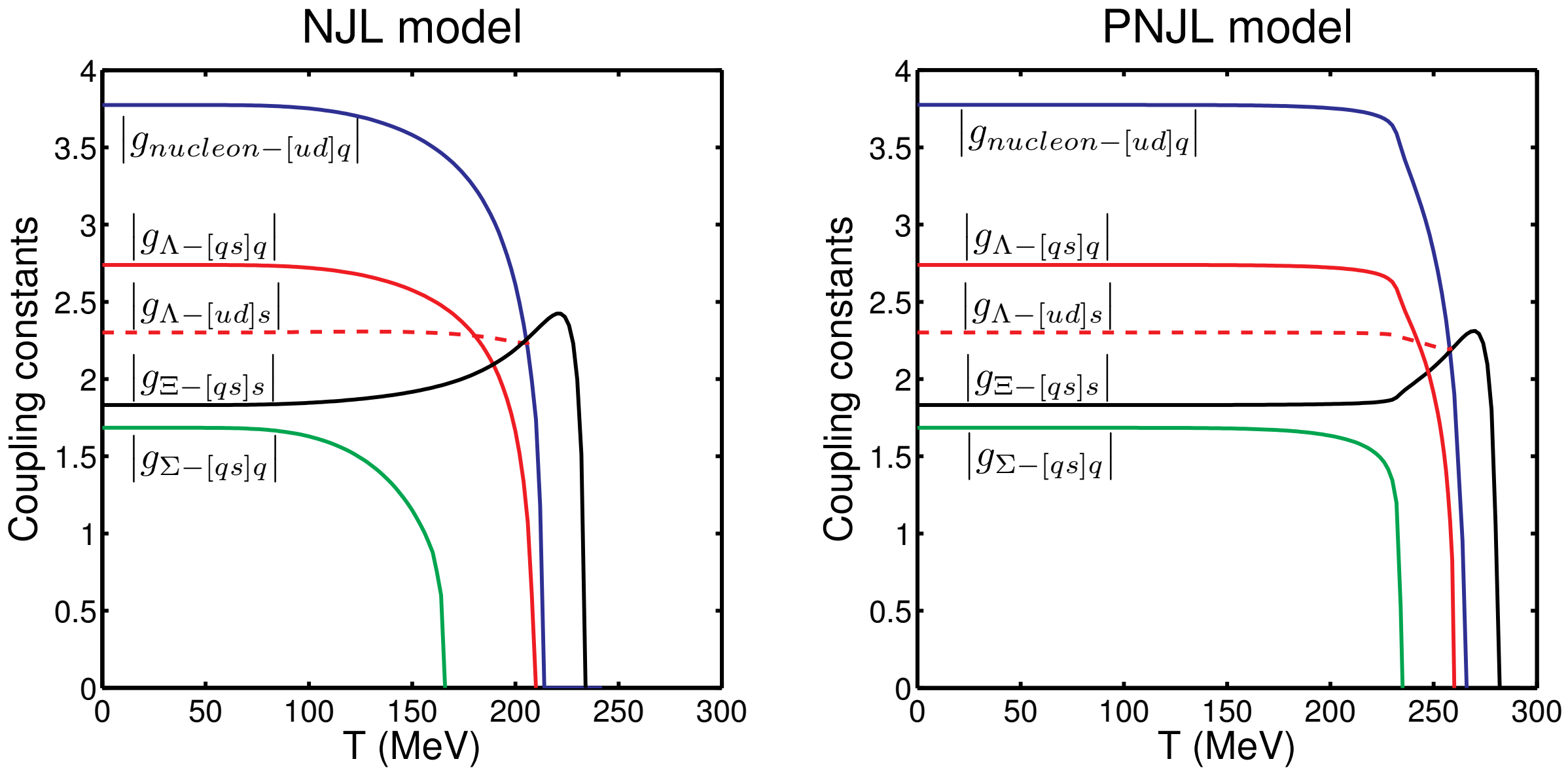

**Figure 25.** Coupling constants of the octet baryons function of the temperature.

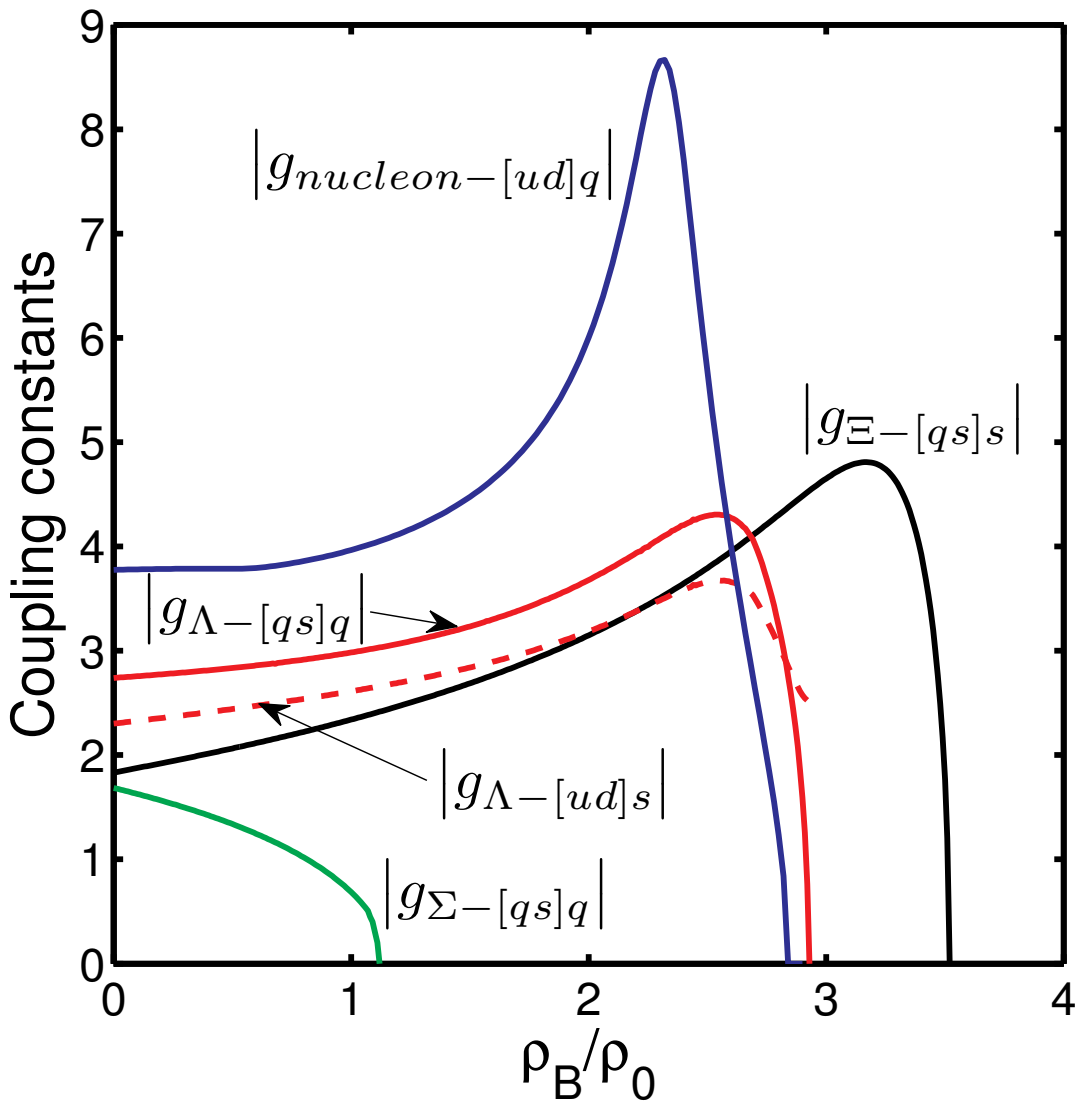

**Figure 26.** Coupling constants of the octet baryons following the baryonic density.

Concerning the decuplet baryons, we can underline that the figure 27 confirms our observations done in the subsection 4.2, i.e. the effect of the Polyakov loop leads to the shifting of the critical temperatures, but it also tends to gather the various critical temperatures. Moreover, the $\left|g_{\Xi^*-[ss]q}\right|$ curves present an interesting aspect, according to the temperature or the baryonic density. Indeed, it is the only baryon for which we obtained values after $g \to 0$, i.e. when the curves increase. Clearly, we found there a similar behavior as the one observed for the mesons and diquarks in their instability regime. Such a result is visible for the $\Xi^*$ because of its "double stable/unstable transition" associated with its two states $[ss]+q$ and $[qs]+s$, see figures 20, 21. Furthermore, the gap between the transitions is large enough, according to $T$ and $\rho_B$, to make visible this phenomenon in the figures 27 and 28. The same behavior should be observed for $\Lambda$. But, it is not the case in the figures 25 and 26, because the gap between the two transitions is there too reduced, according to the temperature and the density.

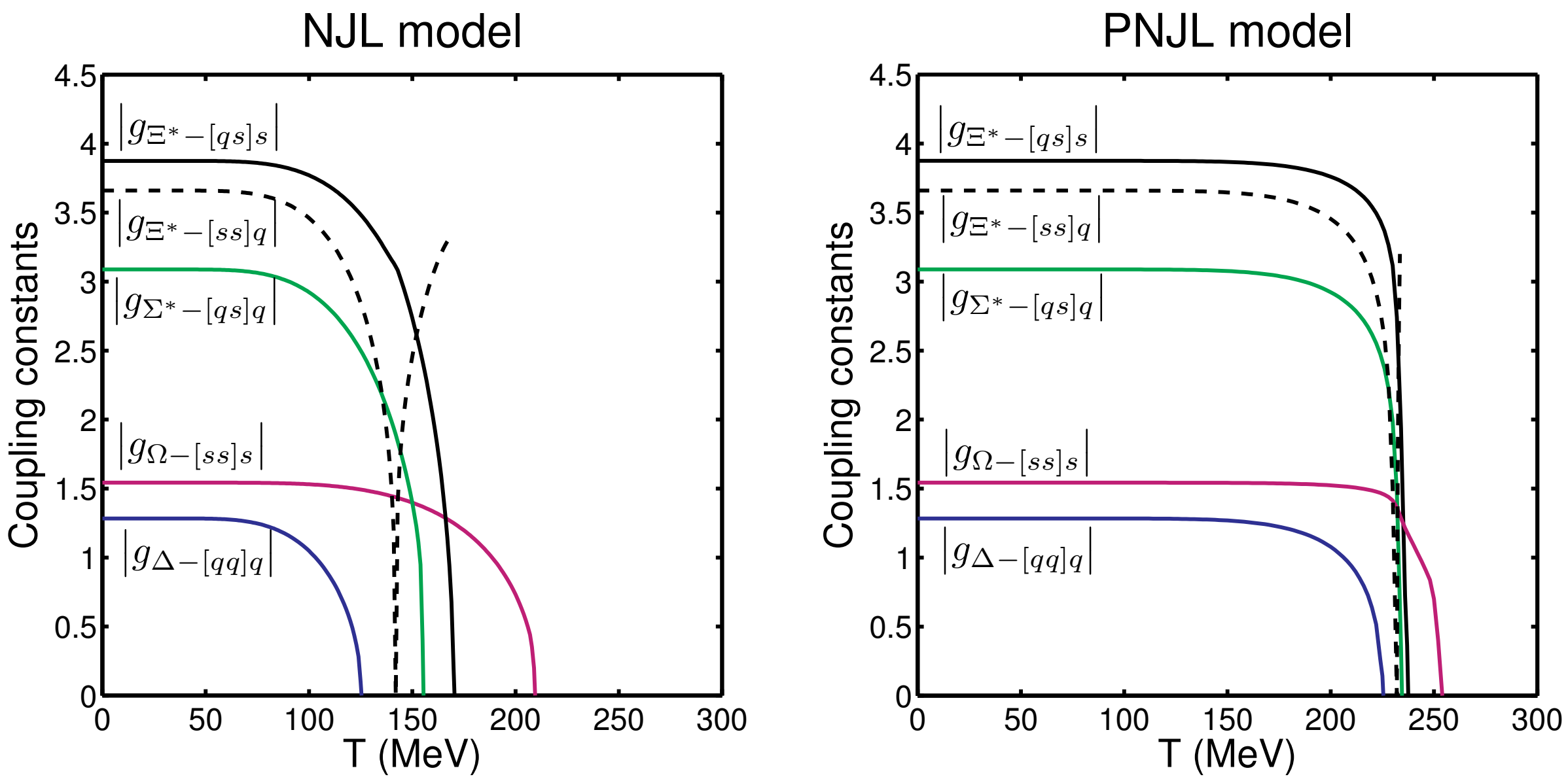

**Figure 27.** Coupling constants of the decuplet baryons function of the temperature.

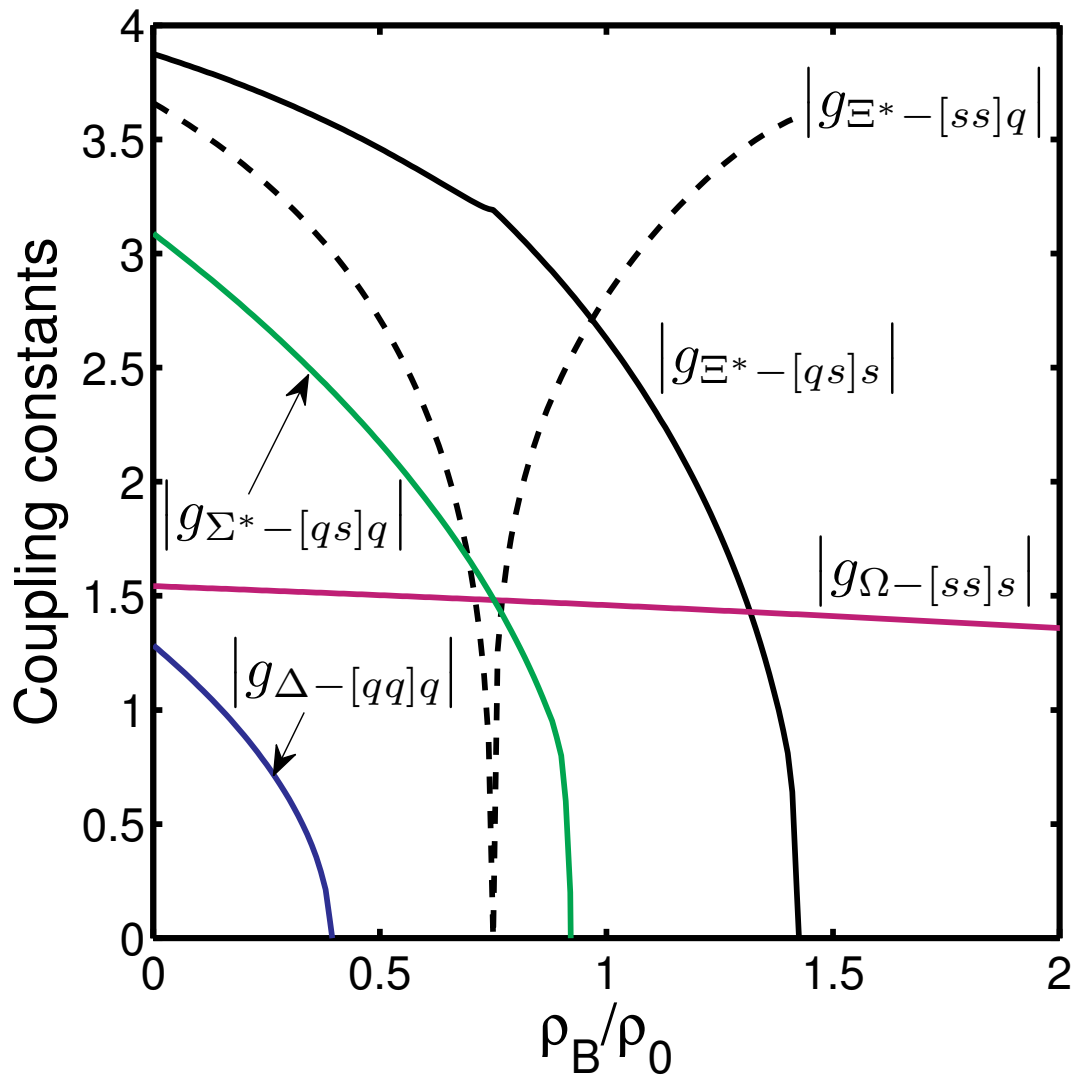

**Figure 28.** Coupling constants of the decuplet baryons function of the baryonic density, with $T = 0$ .

# 6. Masses at null temperature and density

## 6.1 Presentation of the results

In order to compare our results with the ones of the literature, we gather in table 1 the masses of the baryons found at null temperature and densities. As done with the mesons and the diquarks, we used the P1 parameter set to calculate our data in the framework of the isospin symmetry, and the EB parameter set that not respect this symmetry. Concerning the data found in the literature, we mention [3, 16] that used an NJL description. Other theoretical studies that modeled baryons are for example [6, 7, 22]. About the isospin symmetry, the values proposed in the framework of the Quantum Molecular Dynamics model (QMD) [27] also constitute an interesting source of data. These ones are reproduced in the appendix A. Moreover, experimental values can be found in [25, 26]. They are visible in the associated column in the table 1.

In the framework of our numerical results, we made the choice to treat the octet baryons by their scalar flavor component, and the decuplet ones by their axial flavor component. The obtained data prove that this choice was judicious, because of the good agreement with the ones of other studies, or with the experimental data.

| | | | Obtained results (P1 set) | Obtained results (EB set) | Experimental values | |
|---|---|---|---|---|---|---|
| Baryons | | | Masses | Masses | Masses | Widths |
| Octet | proton | *uud* | 897.46 | 887.11 | 938.3 | 0 |
| | neutron | *udd* | 897.46 | 884.76 | 939.6 | |
| | $\Lambda$ | *uds* | 1163.83 | 1119.45 | 1116 | |
| | $\Sigma^+$ | *uus* | 1204.72 | 1148.91 | 1189 | |
| | $\Sigma^0$ | *uds* | 1204.72 | 1151.22 | 1193 | |
| | $\Sigma^-$ | *dds* | 1204.72 | 1153.52 | 1197 | |
| | $\Xi^0$ | *uss* | 1410.44 | 1332.36 | 1315 | |
| | $\Xi^-$ | *dss* | 1410.44 | 1335.37 | 1321 | |
| Decuplet | $\Delta^{++}$ | *uuu* | 1215.64 | 1211.62 | 1232 | 120 |
| | $\Delta^+$ | *uud* | 1215.64 | 1212.55 | 1232 | 120 |
| | $\Delta^0$ | *udd* | 1215.64 | 1213.34 | 1232 | 120 |
| | $\Delta^-$ | *ddd* | 1215.64 | 1214.19 | 1232 | 120 |
| | $\Sigma^{*+}$ | *uus* | 1377.96 | 1336.08 | 1383 | 36 |
| | $\Sigma^{*0}$ | *uds* | 1377.96 | 1336.96 | 1384 | 36 |
| | $\Sigma^{*-}$ | *dds* | 1377.96 | 1337.81 | 1387 | 39 |
| | $\Xi^{*0}$ | *uss* | 1569.55 | 1512.74 | 1532 | 9 |
| | $\Xi^{*-}$ | *dss* | 1569.55 | 1514.73 | 1535 | 10 |
| | $\Omega^-$ | *sss* | 1769.54 | 1674.88 | 1672 | |

**Table 1.** Masses of the baryons, at null temperature and null densities, in MeV.

About our P1 values, we observe that these results are of the good order of magnitude. We do not obtain any aberrant value. These data are comparable to the ones of [3]. In order to quantify the exactitude of these results, we can use the Gell-Mann–Okubo relations, improved according to [28]:

$$\begin{cases} 2\cdot\left(m_N+m_\Xi\right)=3m_\Lambda+m_\Sigma+\dfrac{2}{13}\cdot\left(m_N+m_\Sigma-2m_\Lambda\right) \\ m_\Omega-m_\Delta=3\cdot\left(m_{\Xi^*}-m_{\Sigma^*}\right) \end{cases}. \quad (43)$$

We obtain a variation lower than 1% for the first relation, and less than 4% for the second. In addition, if we compare the data with the ones of [27], the variations are, respectively, less than 0.4% and less than 0.91%. However, the nucleons' mass is underestimated in our approach, whereas the masses of the $\Xi$ seem to be overestimated. At the opposite, the agreement is better for decuplet baryons.

Concerning the EB parameter set, the abandon of the isospin symmetry leads to an improvement of the precision for some baryons. This enhancement is particularly visible for the octet baryons: the overestimation of $\Xi$ by the P1 parameter set is corrected by the EB one. However, the P1 values are slightly better than the EB ones for some decuplet baryon, as the $\Sigma^*$, except for $\Omega^-$. Indeed, the mass of this baryon obtained with the EB parameter set is very close to the experimental value: the error is less than 0.2%.

Moreover, the abandon of the isospin symmetry leads to the apparition of a "hierarchy" between baryons that were found as degenerate when $m_u=m_d$. This mass hierarchy is notably visible for the $\Delta$ baryons. Even if this phenomenon is not observed experimentally, it stays rather consistent. Indeed, we have $m_u<m_d$, thus it appears physically admissible to find $m_{\Delta^{++}}<m_{\Delta^+}<m_{\Delta^0}<m_{\Delta^-}$, because of their respective composition in *u* and *d* quarks, see table 1. This reasoning also works with $\Sigma^+,\Sigma^0,\Sigma^-$, with $\Xi^0,\Xi^-$, and with their decuplet partners. In these cases, the mass hierarchy is confirmed by experimental data. Nevertheless, this reasoning does not work for the proton and the neutron. For them, our EB results still underestimate the experimental data. But, the most important aspect to be underlined is the fact that our results indicate that the proton is heavier than the neutron. Of course, this aspect is not in agreement with experimental facts. The explanation of this behavior is detailed in the next subsection.

# 6.2 Explanation of the mass inversion

In order to explain the mass inversion between the proton and the neutron, we write again the equation (13) to be solved for these two baryons:

$$\begin{cases} 1+4\cdot\dfrac{g_{ud}^{2}}{m_d}\cdot\Pi_{[ud],u}\left(M_p,\vec{0}\right)=0 \\ 1+4\cdot\dfrac{g_{ud}^{2}}{m_u}\cdot\Pi_{[ud],d}\left(M_n,\vec{0}\right)=0 \end{cases}. \quad (44)$$

The first equation concerns the proton of mass $M_p$, the second is for the neutron of mass $M_n$. The second argument of the $\Pi$ functions, fixed to zero, is the baryon's momentum, i.e. the baryons are considered at rest. In the following explanations, we will drop this argument. We note that the coupling constant $g_{ud}{}^2$ is the same for the two baryons. The only difference between these ones is the loop function $\Pi$ and the exchanged quark, expressed via its mass: $m_d$ for the proton, $m_u$ for the neutron. When the two equations (44) are satisfied, we formally write:

$$\frac{\Pi_{[ud],u}\left(M_p\right)}{m_d}=\frac{\Pi_{[ud],d}\left(M_n\right)}{m_u} \quad . \tag{45}$$

Certainly, the $\Pi_{[ud],u}(M_B)$ function is different compared to the $\Pi_{[ud],d}(M_B)$ function, but we can reasonably neglect this difference, even if we drop the isospin symmetry. So, we have:

$$\Pi_{[ud],u}(M_B)\approx\Pi_{[ud],d}(M_B)\equiv\Pi(M_B) \ , \tag{46}$$

where $M_B$ is a baryon mass, used as argument of the $\Pi$ loop function. We rewrite equation (45) as:

$$\Pi\left(M_p\right)\cdot m_u=\Pi\left(M_n\right)\cdot m_d \ . \tag{47}$$

We studied the behavior of the $\Pi$ baryon loop function. In a wide range of the baryon's mass, we found that the function is a negative and decreasing function. By multiplying the equation (47) by a minus one, we have:

$$-\Pi\left(M_p\right)\cdot m_u=-\Pi\left(M_n\right)\cdot m_d \ . \tag{48}$$

Since we have $m_d>m_u$ outside the framework of the isospin symmetry, we deduce from (48) that:

$$-\Pi\left(M_p\right)>-\Pi\left(M_n\right) \ . \tag{49}$$

The function $-\Pi(M_B)$ is a growing function, so we finally find:

$$M_p>M_n \ . \tag{50}$$

In conclusion, the equations' structure, via the static approximation, is the responsible of this incorrect deduction. On the other hand, we do not observe this phenomenon with the flavor axial components or with the other baryons. Indeed, the proton and neutron flavor scalar components are the only ones that use the same coupling constant $g_{ud}{}^2$ with two different exchanged quark mass $m_d$ and $m_u$.

# 6.3 Discussion

At the light of our results, we can now discuss about the validity of our approach, and its possible limitations. Firstly, the NJL quark-diquark picture proposed by [3] constitutes a first approximation. However, this reference showed the validity of this description. Outside of the NJL model, we can quote the work performed in [29–31] that investigate the possibility to go beyond this approximation. It fact, they proved that the quark-diquark picture leads to a variation of about 5 % compared to a three-quark description. So, this approximation appears to be well validated.

On the other hand, as indicated with the mesons, we manipulate here heavy particles. Because of the use of a cut-off in the numerical integrations, the (P)NJL descriptions can present some limitations to describes heavy baryons. The good agreement obtained with our massive baryons lead us to consider the results as reliable, even with this limitation. Another aspect concerns the fact that we do not integrate in the description the baryon disintegration, which necessary would have required the inclusion of mesons into the modeling. More precisely, an improvement of the present work is to include the decays of heavy baryons into lightest ones. Such a work may leads to modifications concerning these heavy baryons, notably as regards their stability zones, figures 23, 24.

Then, an important point of the baryon modeling concerns the use of the static approximation. Indeed, this approximation is suspected to be at the origin of imprecisions in our results. We can refer to [7] that not used this approximation, but not in the NJL description. Also, in the framework of the NJL model, some discussions have already been made in the literature upon this aspect, as in [16–18], which notably refer to the works performed in [13, 15]. In fact, our results confirm the discussion performed in [18]. The static approximation consists to neglect the four-momentum of the exchanged quark in front of its mass. Therefore, this simplification is well validated for the heavy quarks, i.e. here the strange quarks, when they are used as exchanged quarks. This observation is a possible explanation of the good agreement found for the heavy baryons, and notably for $\Omega$. In addition, we showed that the strange quarks are not sensible to the temperature and the baryonic density. It wants to say that the static approximation stays applicable for the heavy baryons, for a wide range of temperatures and densities.

At the opposite, the light quarks present low masses, and tend to reach the values of their naked masses at high temperatures/densities. There, the static approximation is expected to be less trustable. Even at null temperature/density, we can be tempted to evoke the static approximation to explain the fact that some of our results underestimate the experimental data. Also, the work carried out in [15] concluded that the use of this approximation leads the nucleon's mass to decrease more quickly compared to an approach not using this one. Anyway, we showed in the previous subsection that the unphysical mass inversion found for the proton and the neutron can be related to this approximation. Another explanation of the underestimation of the octet baryons' masses, including the nucleons, could be the non-inclusion of the axial flavor component. But, this argument is not applicable to the decuplet baryons, as $\Delta$.

Another aspect to be mentioned concerns the equations (11a) and (11b) used to define the baryon loop function $\Pi$. In this one, as in works as [16–18], the used propagators to model the diquarks are the free propagators, i.e. the quantum field propagators, and not the (P)NJL ones. In fact, we recall that these two propagators are strictly equivalent when the particle is on mass shell. In this case, the couplings $g$ intervening in the $\mathcal{Z}$ term (3) are constant for a given temperature and density. It thus justifies the "coupling constant" naming that is frequently found in the literature. Clearly, these ones were obtained in the chapter associated with the diquarks, thanks to the equivalence between the two propagators when $k^2 \to m^2$.

But, for arbitrary momenta, it was shown that the couplings become momentum dependent, as explained in [32, 33]. In fact, this observation does not invalidate our results, but our treatment of the $g$ constitutes an approximation. Taking into account this momentum dependence should constitute a future improvement of our work.

Another property of our study is visible in the figures 23, 24. Whatever the temperature or the baryonic density, it is observed that the baryons are stable only when the diquarks that compose them are stable. In other words, we do not model stable baryons with diquarks in unstable states. This observation does not contradict works as [11, 15, 18], which considered the Faddeev equations (and their simplification), as done in our work. However, it is explained in [20] the modeling of NJL stable baryons composed by a quark, and by a diquark that can be stable or unstable. In this description, the baryon's behavior is compared to a Borromean (or Efimov) state. In fact, this result does not necessarily contradict our modeling, because our equations do not forbid the creation of stable baryons formed by unstable diquarks, even if we did not observe it. It could be an interesting extension of our work to verify this aspect in the framework of our (P)NJL descriptions. However, even if [20] uses the NJL model, the approach performed in this paper seems to be different compared to the ones of [11, 15, 18]. Anyway, the topic of this paper can also suggest another improvement of our work, which consists to investigate the behavior of the baryons near the color-superconducting phase.

# 7. Conclusion

In this chapter, we detailed the method that we used to include baryons in our (P)NJL description. By a simplification of the Faddeev equations, we saw that the baryon can be considered as a bound state of a quark and a diquark. In this modeling, we come back to a structure close to the one described in the previous chapters. In other word, we used the Bethe-Salpeter equation to find the baryon propagator. It leads to consider a loop function made by a quark-diquark pair. This modeling also implied the use of approximations, as the static approximation. Then, we analyzed the equations associated with the octet and decuplet baryons.

Concerning our results, we showed that the octet baryons can be modeled in a reliable way by using only its scalar flavor component, whereas the decuplet baryons were treated by their axial flavor component. We investigated the behavior of these baryons at finite temperatures and densities. About the difference between the NJL and the PNJL results, we observed the same behavior as in the previous chapters, i.e. a distortion of the curves according to the temperature. It leads to a significant extension of the stability zones of some baryons. Other studies concerned the modeling of the anti-baryons or the estimation of the coupling constants involving baryons. Then, we focused on a study on the results at null temperature and density, in order to compare them to other studies, or to experimental data. Even if our modeling can be considered as rather simple, we obtained good results. We also noted that the abandon of the symmetry isospin leads to an improvement of the precision of our data.

In a last part, we discussed on the reliability of our approach. Indeed, if the modeling of the mesons and diquarks is rather standard, the baryons' modeling involves various considerations and approximations. Among these ones, we particularly analyzed the effects of the static approximation. Indeed, this one is suspected to be at the origin of some defects in our results. Moreover, the other simplifications and limitations of our modeling were described. They can suggest several future developments of our work.

# Chapter 6

# Cross sections

---

## 1. Introduction

To model the cooling of a quaks/antiquarks system, the cross sections calculations [1–5] appear as an essential stage of the study. Indeed, the knowledge of the interactions between the particles is crucial to characterize the dynamics of the system. Such a study can be divided in two parts, corresponding to the two types of cross sections: the inelastic and the elastic ones. During the cooling, the inelastic reactions allow the formation of composite particles starting from the quarks/antiquarks. The cross-sections of these reactions are strongly related to the creation rate of the composite particles. Moreover, the elastic reactions are responsible of heat transfers from hot zones towards colder ones.

The Nambu and Jona-Lasinio model showed its relevancy to model particles as quarks, mesons or baryons, as done in the previous chapters. Furthermore, it allows cross sections calculations involving these particles. More precisely, it was reported in the literature the possibility to evaluate cross sections reactions producing mesons starting from a quark-antiquark pair $q+\overline{q}\rightarrow M+M$ [6, 7]. Very large cross-sections were observed, especially near to the kinematic threshold of the reactions. It can allow a massive mesonization of the quarks-antiquarks plasma during its cooling, as observed in high energies collisions. In addition, elastic cross-sections were also calculated in the NJL model. They concerns elastic scattering between two quarks $q+q\rightarrow q+q$, and between a quark-antiquark pair $q+\overline{q}\rightarrow q+\overline{q}$ [8, 9]. More recently, some studies were initiated, firstly to consider again the processes described above [10], but also to try to evaluate the cross-sections of baryonization reactions [11, 12]. As a whole, reactions involving two incoming particles and two outgoing ones are treated, except with studies as [13]. Indeed, three-body reactions are expected to be too rare to intervene in a notable way.

Moreover, the quoted works globally concern cross sections calculations as a function of the Mandelstam variable $\sqrt{s}$, for several temperatures. Some studies considered the influence of the baryonic chemical potential, but more rarely the baryonic density. Indeed, [7, 8] or [14, 15] for example considered the temperature, [16] the baryonic density, and [17] the both. In fact, even if processes as $q+\overline{q}\rightarrow q+\overline{q}$ or $q+\overline{q}\rightarrow M+M$ are crucial to correctly describe the mesonization of a quarks/antiquarks system, they have not yet been treated at finite densities. In the framework of dynamical studies, neglecting the influence of the density in these reactions may leads to miss some important aspects of the cooling, especially in physical systems for which the density is positive. Furthermore, it was reported the limitations of the NJL approach, due to its main defect: the absence of confinement [18]. In the cross sections calculations, this aspect can limit the reliability of the obtained results.

Thus, several ways are possible to improve and develop cross-sections calculations involving the quoted particles. Firstly, it consists to investigate the influence of the baryonic density on the cross-sections, especially with the processes described by [7, 8]. Moreover, thanks to the inclusion of the Polyakov loop, it can be interesting to see the induced modifications on the results. More precisely, what are the consequences of the confinement mechanism simulated by the PNJL model, compared to a pure NJL one? We saw in the previous chapters that the masses of the PNJL particles are shifted towards higher temperatures compared to NJL ones. In the framework of the cross-sections, the modifications induced by the inclusion of the Polyakov loop acts at several levels of the required calculations. As a consequence, the final result is not obvious. Such comparisons of NJL-PNJL cross-sections are not treated in the literature, and should be performed. In addition, thanks to the baryon modeling performed in the chapter 6, we are able to include baryonization reactions in our study. We inspire us by the reactions mentioned in [11, 12], but we also add new ones. Indeed, the list of NJL reactions already treated in the literature is interesting, but not exhaustive. In order to prepare the dynamical study performed in the next chapter, this list should be completed. It notably concerns elastic reactions. Clearly, these ones can compete with inelastic ones, thus they intervene in the dynamics of the system. Also, the role played by the diquarks is to be clearly specified. More precisely, we should investigate if they mainly act as propagators in the baryonization reactions or if they act as intermediate particles in the system, i.e. two quarks should associate themselves to form a diquark, and then a quark should react with it to form a baryon.

The work described in this chapter was performed while considering the evolutions proposed in the previous paragraph. In the section 2, we recall the general methods used to perform cross-sections calculations. Firstly, we treat inelastic reactions. In the section 3, we consider again the mesonization reactions [7], in order to see the effect of the densities and the Polyakov loop. In the section 4, reactions inducing the formation of diquarks are treated, whereas section 5 focuses on reactions creating baryons. In fact, the performed calculations in these two sections require delicate mathematical developments, notably spinors calculations implying different momenta. Such calculations are detailed in the appendix B, in which we use the reference [19]. Then, we focus on elastic reactions. In the section 6, quark-quark scattering and quark-antiquark scattering shown in [8] are calculated again. As in section 2, one objective is to compare NJL-PNJL results, and to extend to calculations to finite densities. In section 7, elastic reactions involving mesons and diquarks are considered. Some reactions involving baryons are then treated in section 8.

# 2. Calculation methods

The method required to perform the cross-sections calculations is appreciably always the same, whatever the reactions that we will describe in this chapter. First, we have to list the possible channels, labeled as $s$, $s'$, $t$, $u$ … We will see thereafter in the concrete cases what are these channels. Each of them corresponds to a Feynman diagram and a matrix element $\mathcal{M}_i$. These matrix elements are summed, the square of the absolute value is then evaluated. Also, we sum over final states and we average over initial states, as explained e.g. in [7], according to the degrees of freedom of spin and color:

$$\frac{1}{4N_c^2}\sum_{s,c}\left|\mathcal{M}_{\text{total}}\right|^2 \quad , \text{ with } \quad \mathcal{M}_{\text{total}} = \sum_{i \text{ channels}} \pm\mathcal{M}_i \, . \tag{1}$$

It implies to calculate squared terms $\left|\mathcal{M}_i\right|^2$ and mixed terms $\mathcal{M}_i \cdot \mathcal{M}_j^*$, also designated as interference terms in the literature. In the appendix B, the method used to determine the squared and mixed terms is explained. In the right-hand side of (1), the signs ± placed in front of the matrix elements are closely related to the symmetrization or the anti-symmetrization of the wave-functions implied in the diffusion process, notably in the case of identical particles. In order to explain it, we consider the example of the elastic scattering between two identical diquarks. The interaction can use a channel named $t$ and a channel $u$, figure 39. This one differs from the channel $t$ only by the exchange of the two outgoing particles. If the considered interaction uses the two channels, it leads to add the transition amplitudes and we have $\mathcal{M}_{\text{total}} = \mathcal{M}_t + \mathcal{M}_u$. Indeed, the mesons are bosons: the wave-function describing the outgoing way must be symmetrical by exchange of these two particles. This explains the plus sign. At the opposite, concerning the elastic reaction using two identical quarks, figure 30, we have $\mathcal{M}_{\text{total}} = \mathcal{M}_t - \mathcal{M}_u$. Indeed, the associated wave-function must be antisymmetric by exchange of the two quarks, because the quarks are fermions. This explains the minus sign.

Moreover, the matrix elements can have the following structure, extracted from section 3:

$$-i\mathcal{M}_s = f_s \;\delta_{c_1,c_2}\; \bar{v}(p_2)\; ig_1\; i\mathcal{D}_s^S\; \Gamma\; ig_2\; u(p_1) \tag{2}$$

The $u(p_1)$ and $\bar{v}(p_2)$ designate spinors. They are obtained with the Feynman rules [4, 5]. The $\mathcal{D}$ refers to a (P)NJL propagator. We can remark the difference in the notation with the $S$ used to designate free propagators, as the ones seen in the previous chapters, notably in the loop functions $\Pi$. Clearly, in this chapter, the (P)NJL propagators are used for the mesons and diquarks, whereas the free propagators concerns the quarks. Moreover, in (2), $f_s$ is a flavor factor term [7]. These ones are detailed in the appendix C. The $g_1$ and $g_2$ are coupling constants at the level of vertices. These ones were studied in the previous chapters. The other terms will be explained later. Then, the differential cross section in the centre of mass reference frame of the two incoming particles is written, e.g. [7, 8]:

$$\frac{\partial\sigma}{\partial t} = \frac{1}{64\pi\, s\, \left\|\vec{p}_1^{\,*}\right\|^2} \cdot \frac{1}{4N_c^2}\sum_{s,c}\left|\mathcal{M}_{\text{total}}\right|^2 \quad . \tag{3}$$

The appendix F recalls some notion about the Mandelstam variables $s,t,u$ used in this relation. It also proposes to define our writing conventions in kinematics. Notably, the incoming particles are labeled particles 1 and 2, whereas outgoing ones are labeled particles 3 and 4. To estimate the cross section $\sigma$, the integration is performed according to the Mandelstam variable $t$. Then, two *blocking factors* are inserted [20]. They take into account that the two produced particles appear in a medium where other identical particles already exist. If the particle 3 or 4 is a fermion, its blocking factor is $1 - f_F\left(\beta\cdot\left(E_{3,4}^* - \mu_{3,4}\right)\right)$. In the case of a boson, it is $1+f_B\left(\beta\cdot\left(E^*_{3,4}-\mu_{3,4}\right)\right)$. The $f_F$ and $f_B$ indicate, respectively, the Fermi-Dirac and the Bose-Einstein statistics. The signs in front of the chemical potentials $\mu_{3,4}$ are adapted in the cases of anti-fermions or anti-bosons. As explained in the previous chapters, the Fermi-Dirac statistics are modified for the quarks and antiquarks if the calculations are performed in the PNJL approach, due to the Polyakov field [21]. The cross section is then written as [7, 8]:

$$\sigma(s,T)=\left(1\pm f_{F,B}\left(E^*_3-\mu_3\right)\right)\cdot\left(1\pm f_{F,B}\left(E^*_4-\mu_4\right)\right)\cdot\int_{t_-}^{t_+} dt\cdot\frac{\partial\sigma(s,T)}{\partial t} \quad . \tag{4}$$

In practice, the cross sections are non-null according to $\sqrt{s}$ after a *kinematic threshold*, defined as [7, 11]:

$$\sqrt{s}_{\text{threshold}}=\max\left[\left(m_1+m_2\right),\left(m_3+m_4\right)\right] \quad , \tag{5}$$

in which $m_{1,2}$ are the masses of the incoming particles, and $m_{3,4}$ the masses of the outgoing ones. If $m_1+m_2>m_3+m_4$, the kinematic threshold corresponds to $m_1+m_2$. At this threshold, $\left\|\vec{p}^*_1\right\|$ and $\left\|\vec{p}^*_2\right\|$ tend towards zero, see appendix F. According to (3), the differential cross section diverges, except of course if the matrix elements give a null value at this moment.

With the inelastic reactions, it could be useful to consider reverse reactions, e.g. $M+M\rightarrow q+\overline{q}$ [10]: the particles 3 and 4 produce the particles 1 and 2. Obviously, the blocking factors are adapted. Then, a similar reasoning could be applied for $\left\|\vec{p}^*_3\right\|$ or $\left\|\vec{p}^*_4\right\|$, which replace $\left\|\vec{p}^*_1\right\|$ in the relation (3). So, for a reverse reaction, there is possibility of divergence at the threshold if $m_3+m_4>m_1+m_2$.

Moreover, it could be relevant to introduce the transition rate $\omega$, as proposed in [7]. This quantity is associated with the cross section $\sigma$ by the relation [2, 7]:

$$\omega(s,T)=v_{\text{rel}}\cdot\sigma(s,T) \quad \text{with} \quad v_{\text{rel}}=\frac{\left\|\vec{p}^*_1\right\|}{E^*_1}+\frac{\left\|\vec{p}^*_2\right\|}{E^*_2}. \tag{6}$$

In the equation (6), $v_{\text{rel}}$ is the relative velocity of the incoming particles, in their center of mass reference frame.

# Inelastic reactions

## 3. Mesonization reactions

Firstly, we consider $q + \bar{q} \rightarrow M + M$ : a quark and an antiquark give two pseudo-scalar mesons. This is a dominating reaction in hadronization processes of a quarks/antiquarks plasma. It is true especially with pseudo-scalar mesons, because they are the lightest ones. As explained in the introduction of this chapter, this reaction was proposed in [7], in which the cross-sections calculations were performed at finite temperatures. We propose to recover these results, to see the differences with the PNJL description, and to enlarge the calculations at non-null baryonic densities.

The possible channels are presented in figure 1 by their Feynman diagrams. For each of them, their corresponding matrix elements are written in (7).

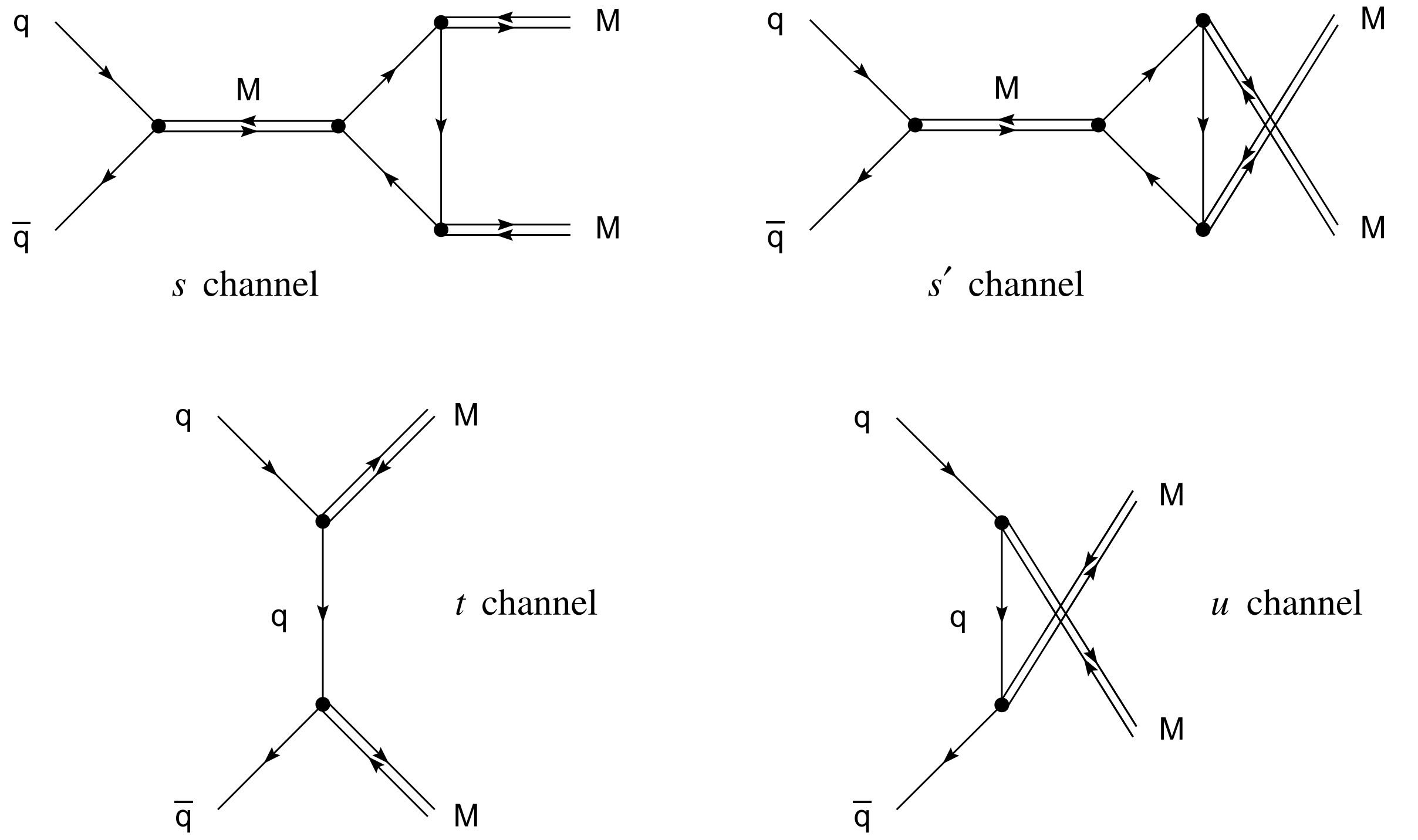

**Figure 1.** Feynman diagrams.

$$\begin{aligned}
-i\mathcal{M}_s &= f_s\ \delta_{c_1,c_2}\ \bar{v}(p_2)\ ig_1\ i\mathcal{D}_s^S(p_1+p_2)\ \Gamma(p_1+p_2\ ,\ p_3)\ ig_2\ u(p_1)\\
-i\mathcal{M}_{s'} &= f_{s'}\ \delta_{c_1,c_2}\ \bar{v}(p_2)\ ig_1\ i\mathcal{D}_{s'}^S(p_1+p_2)\ \Gamma(p_1+p_2\ ,\ p_4)\ ig_2\ u(p_1)\\
-i\mathcal{M}_t &= f_t\ \delta_{c_1,c_2}\ \bar{v}(p_2)\ i\gamma_5\ ig_1\ iS_F(p_1-p_3)\ i\gamma_5\ ig_2\ u(p_1)\\
-i\mathcal{M}_u &= f_u\ \delta_{c_1,c_2}\ \bar{v}(p_2)\ i\gamma_5\ ig_1\ iS_F(p_1-p_4)\ i\gamma_5\ ig_2\ u(p_1)
\end{aligned} \quad . \qquad (7)$$

In the $s$ and $s'$ channels, the incoming quarks/antiquarks temporarily form scalar mesons, whose (P)NJL propagators are noted as $\mathcal{D}^{S}_{s,s'}$. As seen in the chapter 3 and in [7], we can write them, for $a_0$ or $K_0^*$, by the following writing:

$$\mathcal{D}^{S}_{s,s'}\left(k_0,\vec{k}\right)=\frac{2K^{-}_{i\bar{i}}}{1-4K^{-}_{i\bar{i}}\cdot\Pi^{S}\left(k_0,\vec{k}\right)} \quad . \tag{8}$$

These mesons are linked to the "quark-triangle" structure, described by the function $\Gamma$, as in [7]:

$$\begin{aligned}\Gamma\left(i\nu_m,\vec{k};i\alpha_l,\vec{p}\right)&=-N_c\cdot\frac{i}{\beta}\cdot\sum_n\int\frac{\mathrm{d}^3q}{(2\pi)^3}\\ &\times Tr\left[iS_{f_1}\left(i\omega_n,\vec{q}\right)\cdot i\gamma_5\cdot iS_{f_2}\left(i\omega_n-i\alpha_l,\vec{q}-\vec{p}\right)\cdot i\gamma_5\cdot iS_{f_3}\left(i\omega_n-i\nu_m,\vec{q}-\vec{k}\right)\right]\end{aligned} \quad . \tag{9}$$

As explained in the appendix D, this $\Gamma$ requires the use of the $C_0$ function [22]. The $\Gamma$ structure allows producing two outgoing pseudo-scalar mesons. As visible in the figure 1, the difference between the $s$ and $s'$ channels is these mesons are exchanged in the $s'$ channel compared to the $s$ one.

The rule concerning the coupling constants terms $g_1, g_2$ in (7) is to include them at each vertex connected to external lines, or to internal lines that are not (P)NJL propagators (i.e. quarks propagators in practice). In fact, coupling constants are included by construction in the (P)NJL propagators [23]. Moreover, the channels $t$ and $u$ use the quark propagator labeled as $S_F$ in (7). This one was defined for example in the chapter 4. For the $t$ and $u$ channels, the $i\gamma_5$ matrices are included because of the pseudo-scalar nature of the outgoing mesons. Also, the rules associated with trace calculations impose an even number of such matrices in (9), see appendix B, which justifies scalar mesons as propagators. Clearly, pseudo-scalar mesons are not possible as propagators in this case. Also, the $f_i$ terms are flavor factors, appendix C. Furthermore, the mesons are non-colored objects, within the meaning of the QCD. It imposes restrictions on the choice of the quarks colors, are indicated by the Kronecker symbol $\delta$.

The existence of the four channels presented in the figure 1 depends on the involved particles in the reaction. We propose in the table 1 a list of the most used reactions in the framework of $\mathrm{q}+\bar{\mathrm{q}}\rightarrow \mathrm{M}+\mathrm{M}$. This list is not exhaustive. More precisely, reactions involving heavy mesons as outgoing mesons, as $\eta'$, have weaker cross sections than the other ones [7]. So, they can be omitted.

Concerning the numerical calculations, we turn our attention to the reaction $u+\bar{u}\rightarrow\pi^{+}+\pi^{-}$. As observed in [7, 10], this example generates the strongest cross sections of the $\mathrm{q}+\bar{\mathrm{q}}\rightarrow \mathrm{M}+\mathrm{M}$ process. The results are gathered in the figures 2 to 7. Firstly, a comparison between NJL and PNJL models is possible with the figure 2, 3. In fact, I have already used a figure similar to the figure 2 in the reference [10]. As with the particles' masses and the coupling constants, the inclusion of the Polyakov loop leads also to a shifting of the values towards higher temperatures. The found values are not modified, just shifted. Indeed, we use the same values to legend the figures 2 and 3. Clearly, the inclusion of the Polyakov loop does not lead here to higher and lower cross-sections. Numerically, this loop intervenes in the gap equations, and in some other relations by the replacement of the Fermi-Dirac statistics with the modified ones [21], as done in the previous chapters. It concerns the equations used to find the masses of the mesons, and some specific calculations, as the $\Gamma$ function (9).

| reactions | propagated meson(s) (*s* channel) | channel | | | |
|---|---|---|---|---|---|
| | | $s$ | $s'$ | $t$ | $u$ |
| $u+\bar{u}\to\pi^++\pi^-$ | $a_0^0, f_0, f_0'$ | √ | √ | √ | |
| $u+\bar{u}\to\pi^0+\pi^0$ | $a_0^0, f_0, f_0'$ | √ | √ | √ | √ |
| $u+\bar{u}\to K^++K^-$ | $\left(a_0^0\right), f_0, f_0'$ | √ | √ | √ | |
| $u+\bar{u}\to\pi^0+\eta$ | $a_0^0, f_0, f_0'$ | √ | √ | √ | √ |
| $u+\bar{d}\to\pi^++\pi^0$ | $a_0^+$ | √ | √ | √ | √ |
| $u+\bar{d}\to K^++\bar{K}^0$ | $a_0^+$ | √ | | √ | |
| $u+\bar{d}\to\pi^++\eta$ | $a_0^+$ | √ | √ | √ | √ |
| $u+\bar{s}\to\pi^++K^0$ | $K_0^{*+}$ | √ | | √ | |
| $u+\bar{s}\to\pi^0+K^+$ | $K_0^{*+}$ | √ | | √ | |
| $u+\bar{s}\to\eta+K^+$ | $K_0^{*+}$ | √ | √ | √ | √ |
| $s+\bar{s}\to K^-+K^+$ | $f_0, f_0'$ | √ | √ | √ | |
| $s+\bar{s}\to\bar{K}^0+K^0$ | $f_0, f_0'$ | √ | √ | √ | |
| $s+\bar{s}\to\eta+\eta$ | $f_0, f_0'$ | √ | | √ | |

**Table 1.** Some mesonization reactions.

As explained in the section 2, a divergence at the threshold can be present if the incoming particles are heavier than the outgoing ones. With $u+\bar{u}\to\pi^++\pi^-$, the quarks are heavier than the pions at moderate temperatures and densities, so divergences are observable in this case, as confirmed in the figure 6. As noted in [10], these divergences lead to high cross sections values. At the level of the threshold, a tiny range according to $\sqrt{s}$ can sometimes exceeds 40 mb, or more. In the NJL model, the divergence is found until $T=240$ MeV. The maximal cross-sections are found just before this temperature. This one corresponds to the critical temperature of the pions at null density, as observed in the chapter 3. With the PNJL approach, this temperature is higher, i.e. $T=290$ MeV. As a consequence, the divergence (the darkest colored zone) exists for a wider range of temperature.

The study of the influence of the baryonic densities is new in the framework of $q+\bar{q}\to M+M$. As a consequence, we do not have element of comparison in the literature. To investigate this aspect, we firstly propose to consider the figures 4 and 5. In the figure 4, the calculations were performed at null temperature, whereas we consider $T=200$ MeV for the figure 5. As a whole, the baryonic density acts in a comparable way compared to the temperature. Concerning the figure 4, we note that the zone for which $\sigma>10$ mb is wider than in the results found in the figures 2 and 3. Furthermore, at about $2.5\rho_0$, this zone presents its maximal width. It leads there to extreme values, of about few barns locally, just after the threshold. At the opposite, for $\rho_B>3\rho_0$, the cross sections becomes weaker. As observed in the chapter 3, the pion does not have stable/unstable transition according to the baryonic density when $T=0$. However, $\rho_B > 3\rho_0$ corresponds to the zone for which the pion becomes heavier than the quark-antiquark pair that composite it.

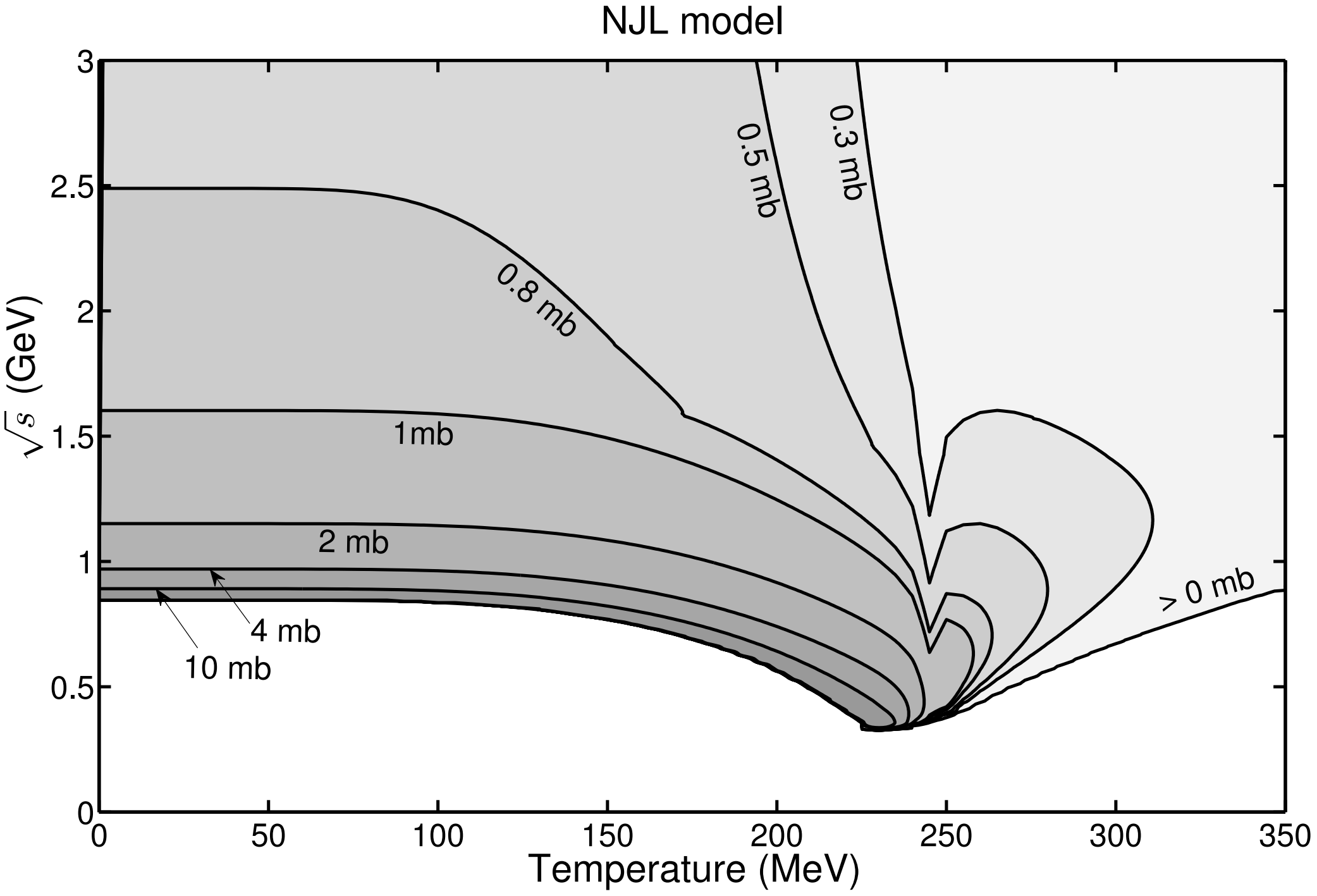

**Figure 2.** Cross sections of $u+\bar{u} \to \pi^+ + \pi^-$ in the $T-\sqrt{s}$ plane, for $\rho_B = 0$, using the NJL model.

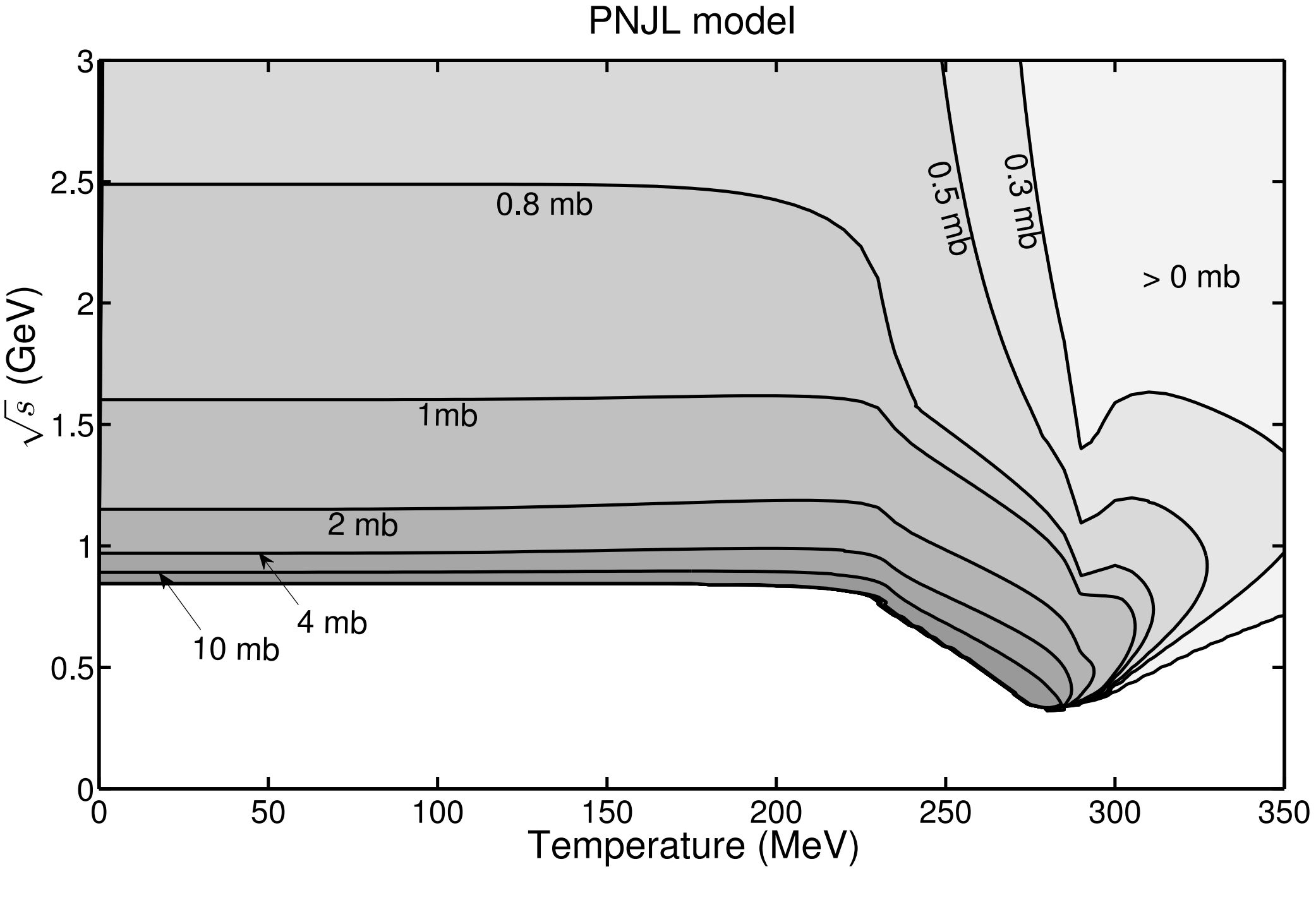

**Figure 3.** Cross sections of $u+\bar{u} \to \pi^+ + \pi^-$ in the $T-\sqrt{s}$ plane, for $\rho_B = 0$, using the PNJL model.

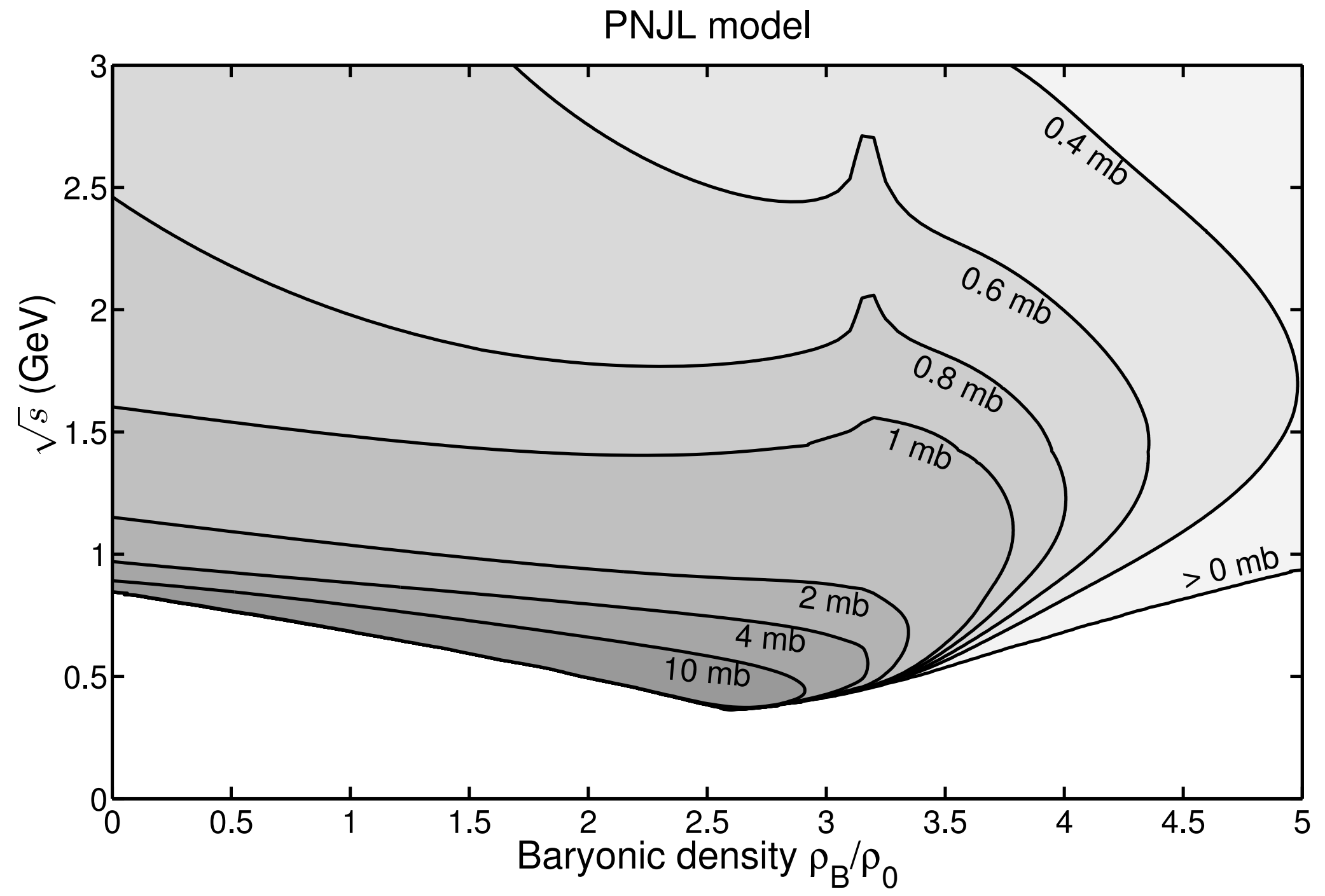

**Figure 4.** Cross sections of $u+\bar{u} \rightarrow \pi^{+}+\pi^{-}$ in the $\rho_B - \sqrt{s}$ plane, at null temperature.

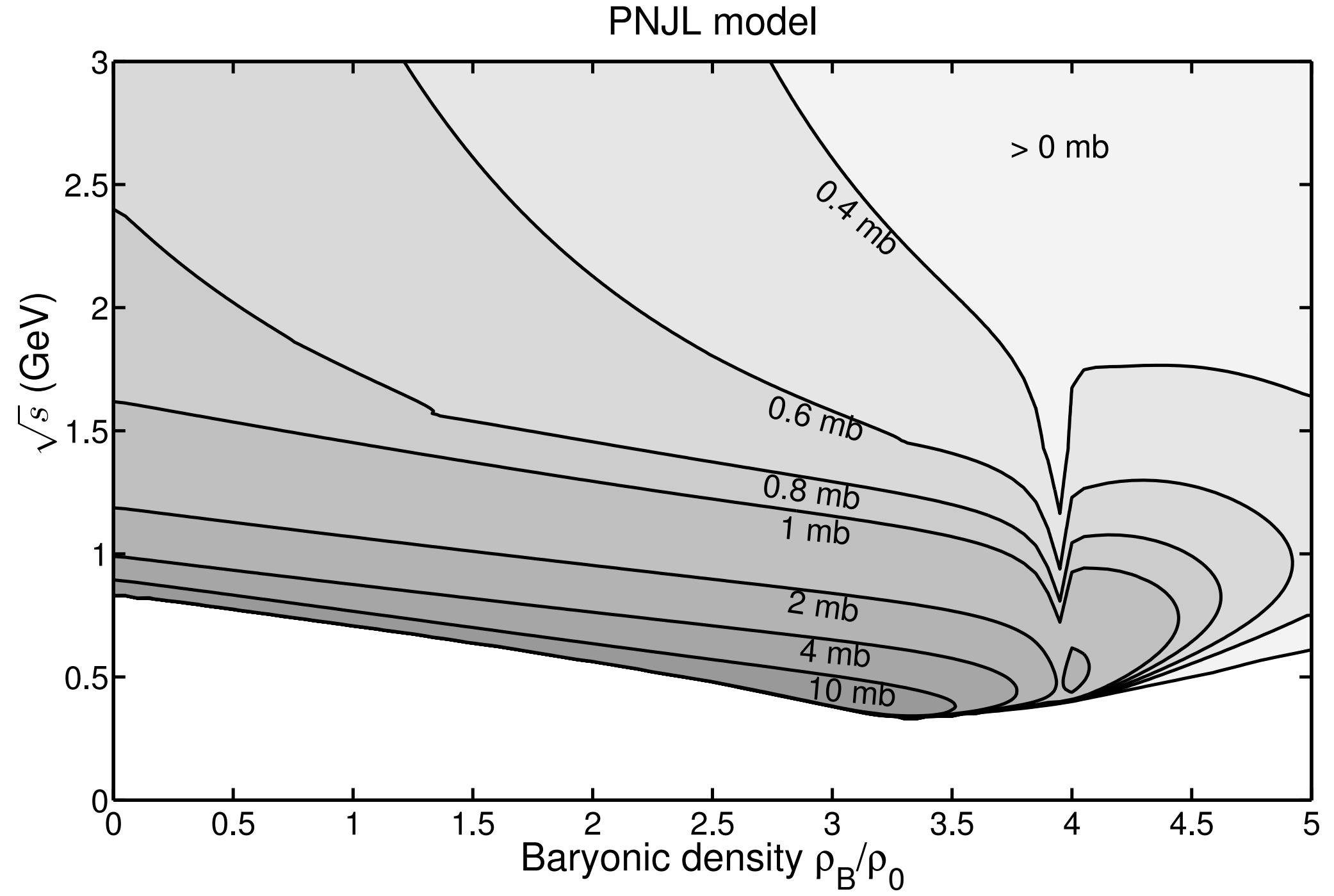

**Figure 5.** Cross sections of $u+\bar{u} \rightarrow \pi^{+}+\pi^{-}$ in the $\rho_B - \sqrt{s}$ plane, at $T = 200$ MeV .

About the figure 5, some differences can be noted compared to the figure 4. Firstly, the $\sigma > 10$ mb zone is extended towards higher densities, because this zone exists until $3.5\rho_0$. Nevertheless, its width is more reduced than in the figure 4. Another difference concerns the fact that at $T = 200$ MeV , the pion becomes unstable by a stable/unstable transition. In the PNJL model, this transition occurs just before $4\rho_0$. It corresponds to the structures visible in the figure 5, along the $\rho_B = 4\rho_0$ vertical line. This phenomenon is also present according to the temperature, but it seems less marked in the figures 2 and 3. It is explained by the cancellation of the pion coupling constant, as seen in the chapter 3.

We propose to complete this analysis with the figures 6 to 8. In the figure 6, the cross-sections are studied according to $\sqrt{s}$ for several temperatures for $\rho_B = 0$ (left hand side of the figure), and for several densities for $T = 250$ MeV. In fact, the choice of this temperature is motivated by the remark performed in the chapter 2 about the color-superconductivity. As explained, at finite densities and reduced temperatures, the color-superconductivity state is expected to intervene [24, 25] and it can affect our cross-sections calculations. However, as argued, temperatures above 200 MeV are definitely not concerned by this phenomenon. Moreover, to study the cooling of a quarks/antiquarks system, it is worthwhile to work at temperatures near to the one of the phase transition. Clearly, it justifies our choice to perform our finite densities calculations at such temperatures in these figures, as in the rest of this chapter.

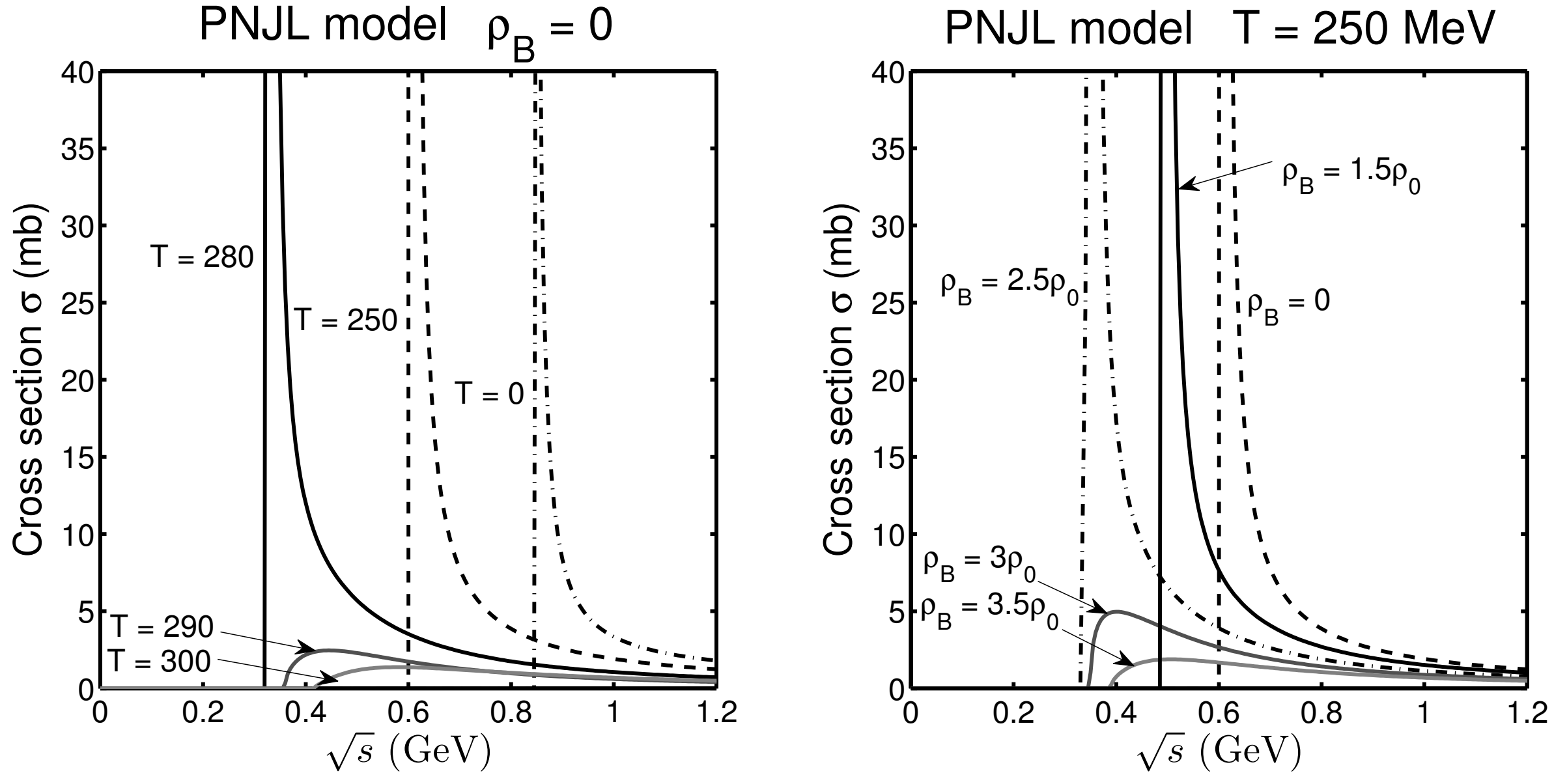

**Figure 6.** Cross sections of $u + \bar{u} \to \pi^+ + \pi^-$ function of the temperature and the baryonic density.

The figure 6 is also completed by the figure 7, in which the transition rate $\omega$, defined in the section 2, is estimated in the same conditions as in the figure 6. The advantage of the transition rate is to attenuate the divergence at the threshold. Indeed, when $T \le 280$ MeV, the value of the threshold corresponds to the mass sum of the incoming quark and antiquark. There, the cross section tends towards the infinity, but the relative speed between them is very close to zero, see (3–6) and the appendix F. For the figures 6 and 7, the global behavior confirms the observations of the figure 3: $\sigma$ and $\omega$ tend towards higher values near to the threshold when the temperature increases, until the pions reach their critical temperature. After this temperature, the divergences disappear, explained by the fact that the outgoing pions are then heavier than the incoming quark-antiquark pairs.

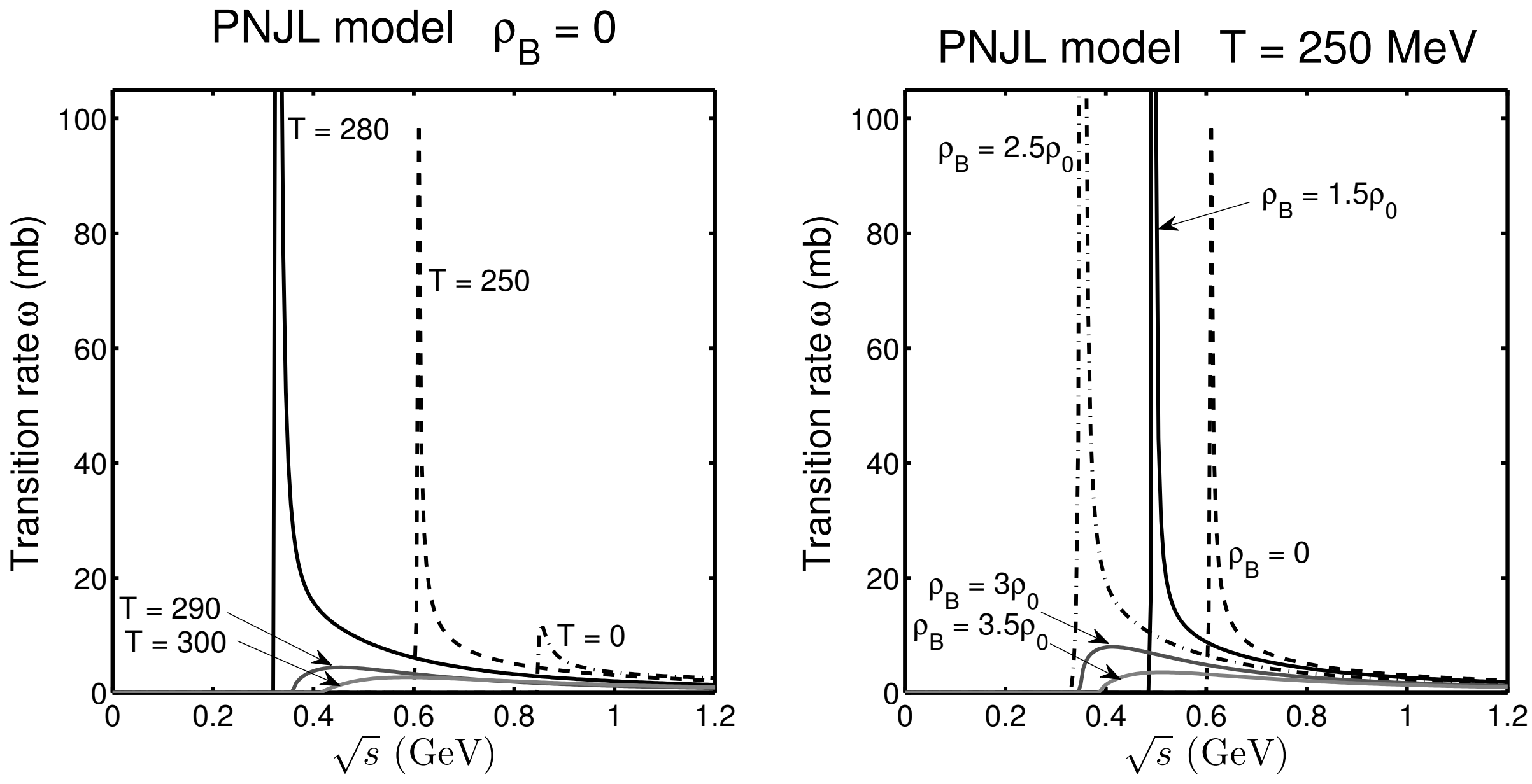

**Figure 7.** Transition rates of $u+\bar{u}\rightarrow\pi^{+}+\pi^{-}$.

The figure 8 shows the transition rate obtained with the reaction $\pi^{+}+\pi^{-}\rightarrow u+\bar{u}$, i.e. the reverse reaction of $u+\bar{u}\rightarrow\pi^{+}+\pi^{-}$ studied in the previous figures. The left hand side of this figure can be compared to the results of [10], in which the NJL cross sections of these reverse reactions were studied at finite temperatures and null density. Moreover, if we also compare the left hand side of the figure 8 with the one of the figure 7, we conclude that the found values are manifestly weaker for the reverse reaction compared to the direct one.

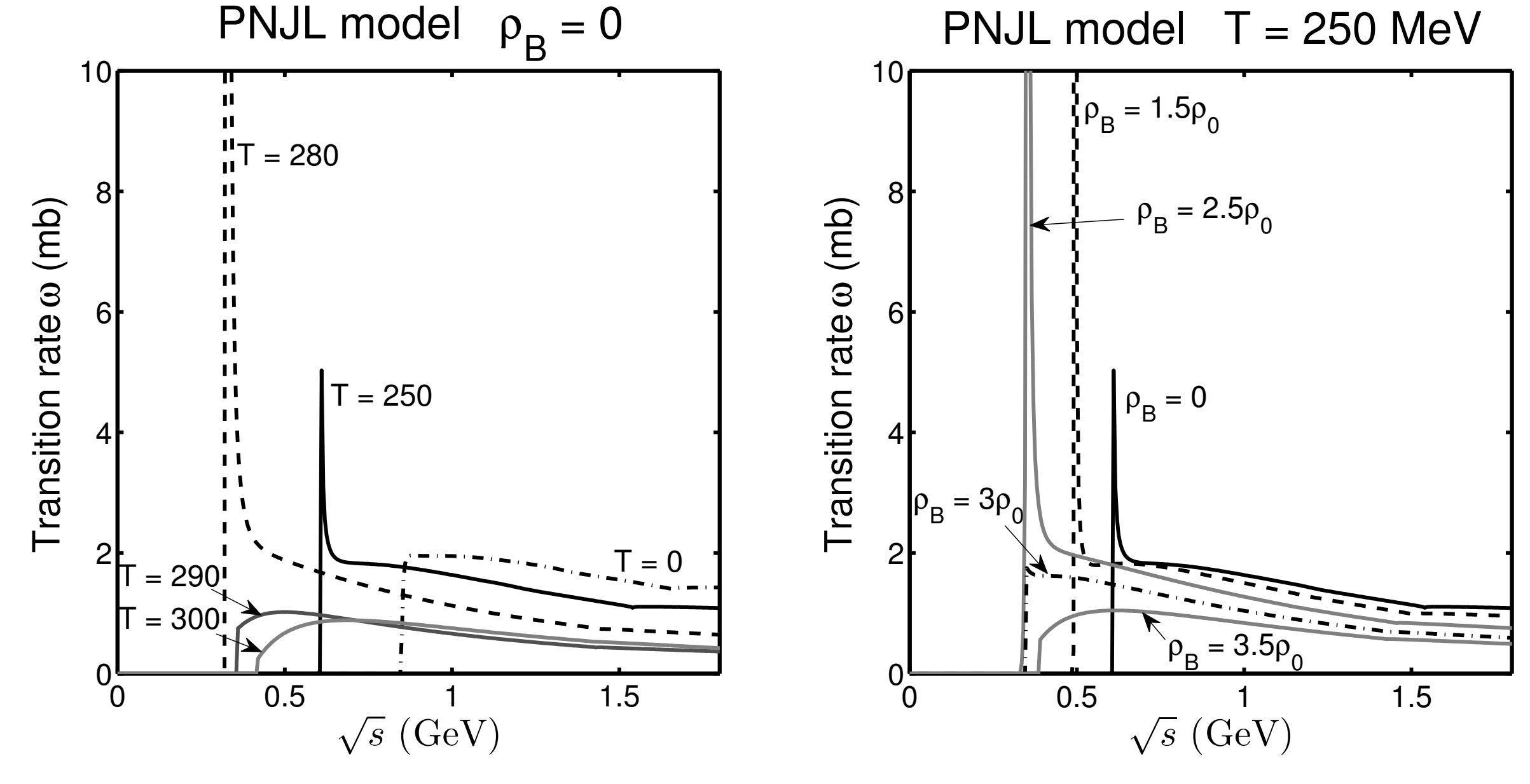

**Figure 8.** Transition rates of $\pi^{+}+\pi^{-}\rightarrow u+\bar{u}$.

Now, we turn our attention to the right hand side of the figures 6 to 8. At $T=250$ MeV and at $\rho_B=0$, the pions are still lighter than the quark-quark pairs that compose them, i.e. they are still stable. For reduced densities, $\sigma$ and $\omega$ are still increasing when the density is growing. At this temperature, the disappearance of the divergence at the threshold is located between $2.5\rho_0$ and $3\rho_0$ in the PNJL model. At $2.5\rho_0$, $\sigma$ and $\omega$ locally explode. After $3\rho_0$, the evolutions of $\sigma$ and $\omega$ strongly recall what was observed for high temperatures. In the same way, the values of the transition rate for the reverse reaction are rather weak according to the density, right hand side of figure 8. Even if the curves have some peaks at the threshold, they stay reduced compared to the ones observable for the direct reaction, figure 7. Thus, as a whole, the reverse reaction is not expected to disturb the direct reaction, at least in the conditions of the mesonization.

To summarize this analysis, we see that the cross-sections (and the transition rates) tend to explode just before the pion becomes unstable, or at least when it mass is stronger than the one of the quark-antiquark pair that composes it. As indicated in [10], this behavior and the extension of the $\sigma > 10$ mb zone in the $T - \rho_B$ plane suggest an overwhelming mesonization of a quark/antiquark plasma, when the temperature and/or the density decrease enough.

# 4. Reactions involving diquark

## 4.1 Feynman rules

The Feynman diagrams and the associated matrix elements presented previously correspond well to the techniques already used in the quantum field theory [1–5, 19], adapted to the framework of the (P)NJL models, as described in [7, 8, 14, 15]. But, to be able to handle matrix elements implying diquarks, some precisions are required [11, 12]. The mesons are described in the (P)NJL models by a loop composed by a quark going towards the future, and a quark going to the past. According to the Feynman point of view[1], this second quark should be understood as an antiquark. Clearly, the meson is made by a quark and an antiquark. Concerning the diquark, we saw in the chapter 4 that applying a charge conjugation to the antiquark gives the possibility to *mimic* the behavior of a quark. This trick is also relevant to write the matrix elements. In fact, the traditional Feynman rules as we use them are unaware about the charge conjugation. In another words, it is necessary to find which is (are) the charge conjugate quark(s)/antiquark(s), in order to find which particle(s) is (are) *truly* quark(s) or antiquark(s). Logically, there could be ambiguity only on the level of the vertices where a diquark/anti-diquark takes part. In the cases treated below, there are sometimes “several solutions”. Therefore, several channels for the same Feynman graph could be possible. In the following examples, only scalar diquarks are used as incoming or outgoing particles. These examples were inspired from [11, 12]. However, we observed that the cross-sections results of these references are not really in agreement with the ones of the literature, or with ours. It notably includes a disagreement between some results of [11] with the ones of [8], whereas in our side, we confirmed the results presented in [8].

[1] Feynman R P 1949 The Theory of Positrons *Phys. Rev.* **76** 749–59

## 4.2 $\overline{\mathrm{q}}+\mathrm{D}\rightarrow\mathrm{M}+\mathrm{q}$ reactions

Firstly, we consider the $\overline{\mathrm{q}}+\mathrm{D}\rightarrow\mathrm{M}+\mathrm{q}$ process and its reverse one, i.e. $\mathrm{M}+\mathrm{q}\rightarrow\overline{\mathrm{q}}+\mathrm{D}$. Clearly, this one is particularly interesting in the framework of an eventual diquark production. Moreover, these reactions are the simplest we treated in this section. Indeed, only one channel is considered, the *t* channel, described by figure 9 and by its matrix element, equation (18). This is a first occasion to use the method described in the previous subsection. The higher vertex of the Feynman diagram is not connected to a diquark. So, the antiquark in position *1*, i.e. in top left position, should be really treated as an antiquark, implying the $\overline{v}(p_1)$ spinor. On the other hand, the lower vertex uses a diquark, thus the quark in position *4* (right bottom position) should be seen as a charge conjugate antiquark. Therefore, this one is represented by $v(p_4)$. In the equation (18), the $i\gamma_5$ matrices are associated with the pseudo-scalar meson and the scalar diquark. Furthermore, the completely antisymmetric tensor $\varepsilon^{c_2,c_t,c_4}$ is used to recall the color constraints between the diquark, the propagated quark and the quark in position *4*. More precisely, the color of the diquark must be "the sum" of the ones of the two quarks. This term $\varepsilon^{c_2,c_t,c_4}$, also used in [11, 12], can be found also in the writing of the diquark Lagrangian or in the associated conserved currents, see chapter 4. We recall that the scalar diquarks are antisymmetric color diquarks.

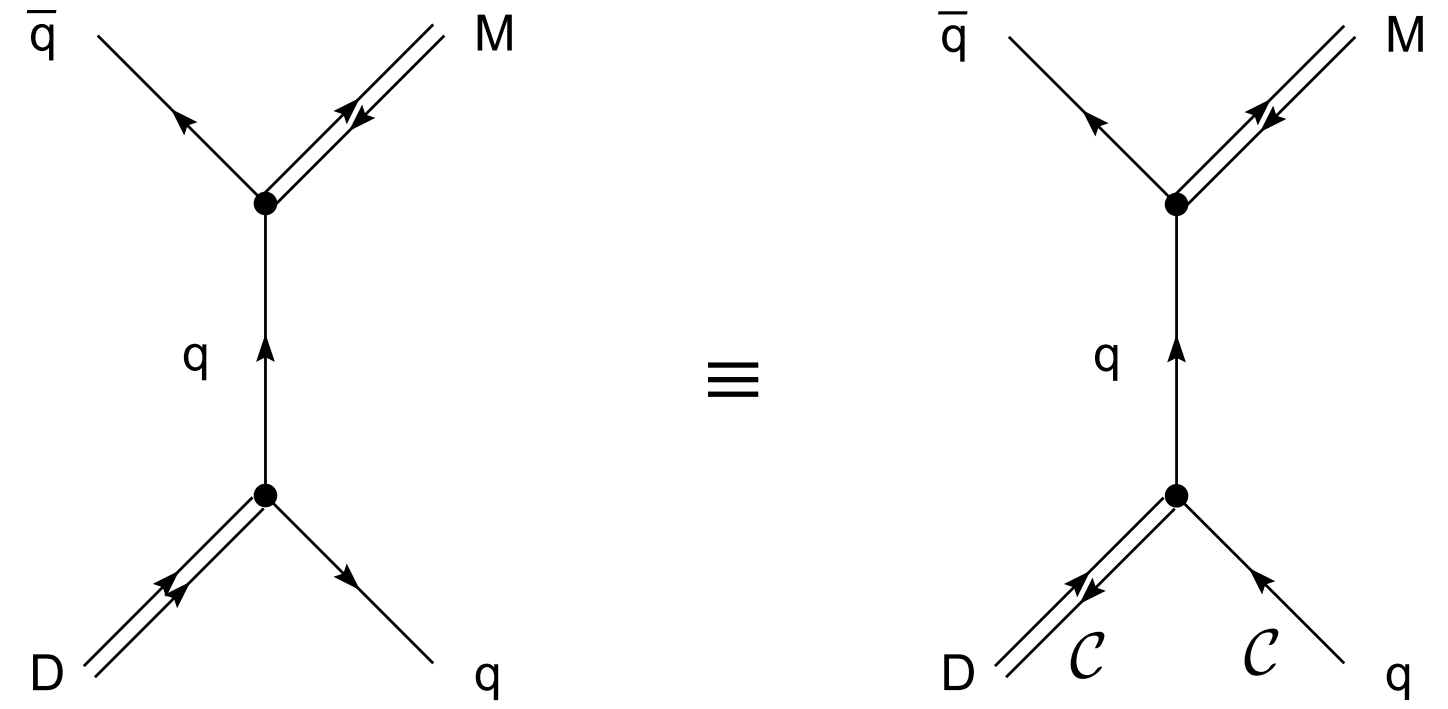

**Figure 9.** *t* channel.

$$-i\mathcal{M}_t = f_t \ \delta_{c_t,c_1} \ \varepsilon^{c_2,c_t,c_4} \ \overline{v}(p_1) \ i\gamma_5 \ ig_1 \ iS_F(p_3-p_1) \ i\gamma_5 \ ig_2 \ v(p_4) \ . \tag{10}$$

The results found for the reaction $\overline{u}+[ud]\rightarrow\pi^-+u$ are proposed in the figure 10. The shape of the curves recalls some of the one of the figure 6. The reaction can present divergences at the threshold because the mass sum of the incoming particles is stronger than the one of the outgoing particles. Indeed, the $[ud]$ diquark is more massive than the pion. When the temperature increases, the divergence at the threshold is still present, at the opposite of the behavior found in the section 3. Also, when the masses of the diquark begins to decrease, from a temperature of $T=250$ MeV, the threshold is shifted towards lower values along the $\sqrt{s}$ axis. Then, before the pion and the diquark reach their critical temperature, the cross sections are minimal. After $T=270$ MeV, the cross sections increase strongly, and tend to explode after $T=300$ MeV.

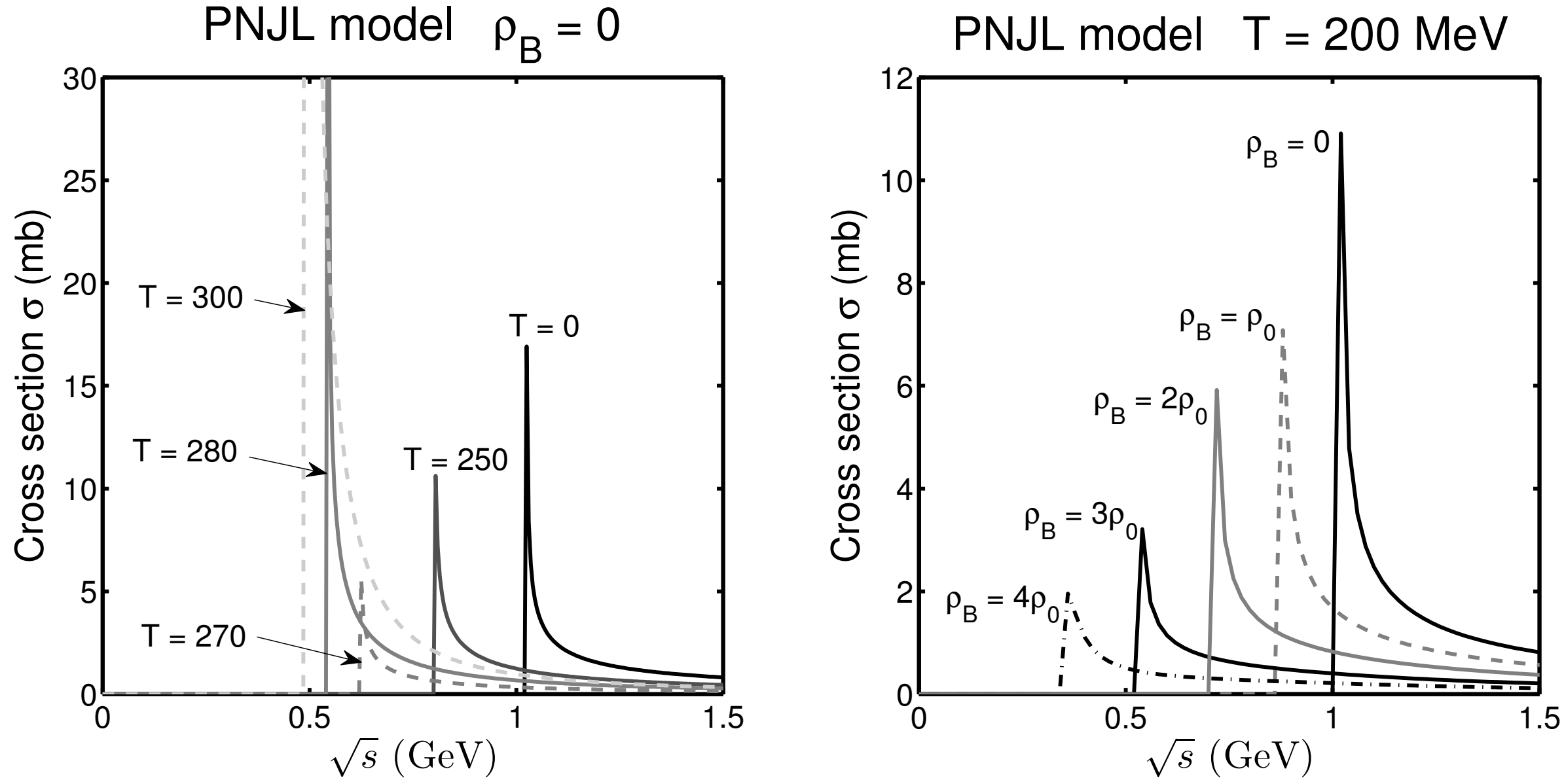

**Figure 10.** Cross sections of $\bar{u} + [ud] \to \pi^- + u$ .

The figure 11 exhibits the results associated with the reverse reaction $\pi^- + u \to \bar{u} + [ud]$. It does not present divergence at the threshold as the direct reaction. In the left hand side of the figure 10, the cross sections reach values of few millibarns at low temperatures. At the level of the threshold, the values are found to be about 10 times weaker compared to the direct reaction. On the other hand, the cross sections of the reverse reaction vary in the same way as for the direct reaction, i.e. a diminution until $T = 270$ MeV , and then they quickly increase again.

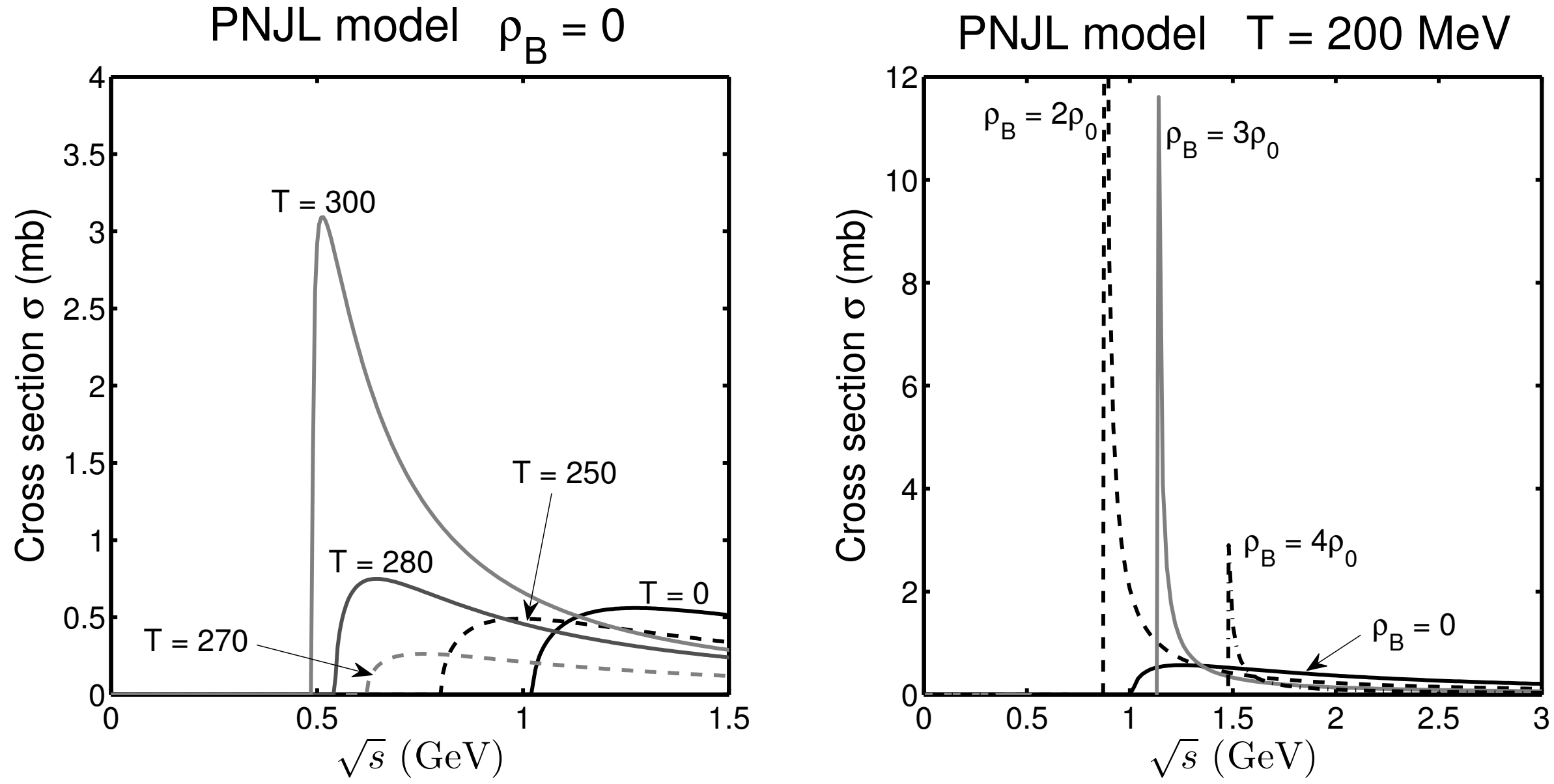

**Figure 11.** Cross sections of $\pi^- + u \to \bar{u} + [ud]$ .

According to the baryonic density, right hand side of the figures 10 and 11, an increase of this parameter leads to a progressive diminution of the cross sections, for the $\bar{u}+[ud]\to\pi^-+u$ reaction. Concerning the reverse reaction, the cross sections increase to become optimal for densities of $\rho_B \approx 2\rho_0$, and decrease after this density. In fact, if we generalize these results to all the reactions as $\mathrm{M}+\mathrm{q}\to\bar{\mathrm{q}}+\mathrm{D}$, they are expected to produce diquarks in such conditions. Nevertheless, except in this case, the process $\bar{\mathrm{q}}+\mathrm{D}\to\mathrm{M}+\mathrm{q}$ globally presents stronger cross sections than it reverse process. As a consequence, the diquarks are certainly consumed, but not produced. Clearly, $\mathrm{M}+\mathrm{q}\to\bar{\mathrm{q}}+\mathrm{D}$ is not expected to be a good candidate to form diquarks. However, as remarked in [12], $\bar{\mathrm{q}}+\mathrm{D}\to\mathrm{M}+\mathrm{q}$ requires an antiquark to occur. So, a way to interpret these results is to imagine that a potential formation of diquarks, and by extension baryons, cannot intervene until the massive mesonization had strongly consumed the antiquarks, to "neutralize" them into mesons.

## 4.3 $\mathrm{q}+\bar{\mathrm{q}}\to\mathrm{D}+\bar{\mathrm{D}}$ reactions

This kind of reactions can be modeled by adapting the works performed for $\mathrm{q}+\bar{\mathrm{q}}\to\mathrm{M}+\mathrm{M}$. As indicated in the figure 12, we applied a charge conjugation on some quarks and antiquarks of the figure 1. As a consequence, the associated matrix elements (11) are structurally identical to the ones of (7). In practice, the same $\Gamma$ function as the one seen for mesons was used, only a charge conjugation of one of the quarks was applied, as indicated in the figure 12. Moreover, the mesons used as propagators in the *s* channel are scalar ones, as in the section 3.

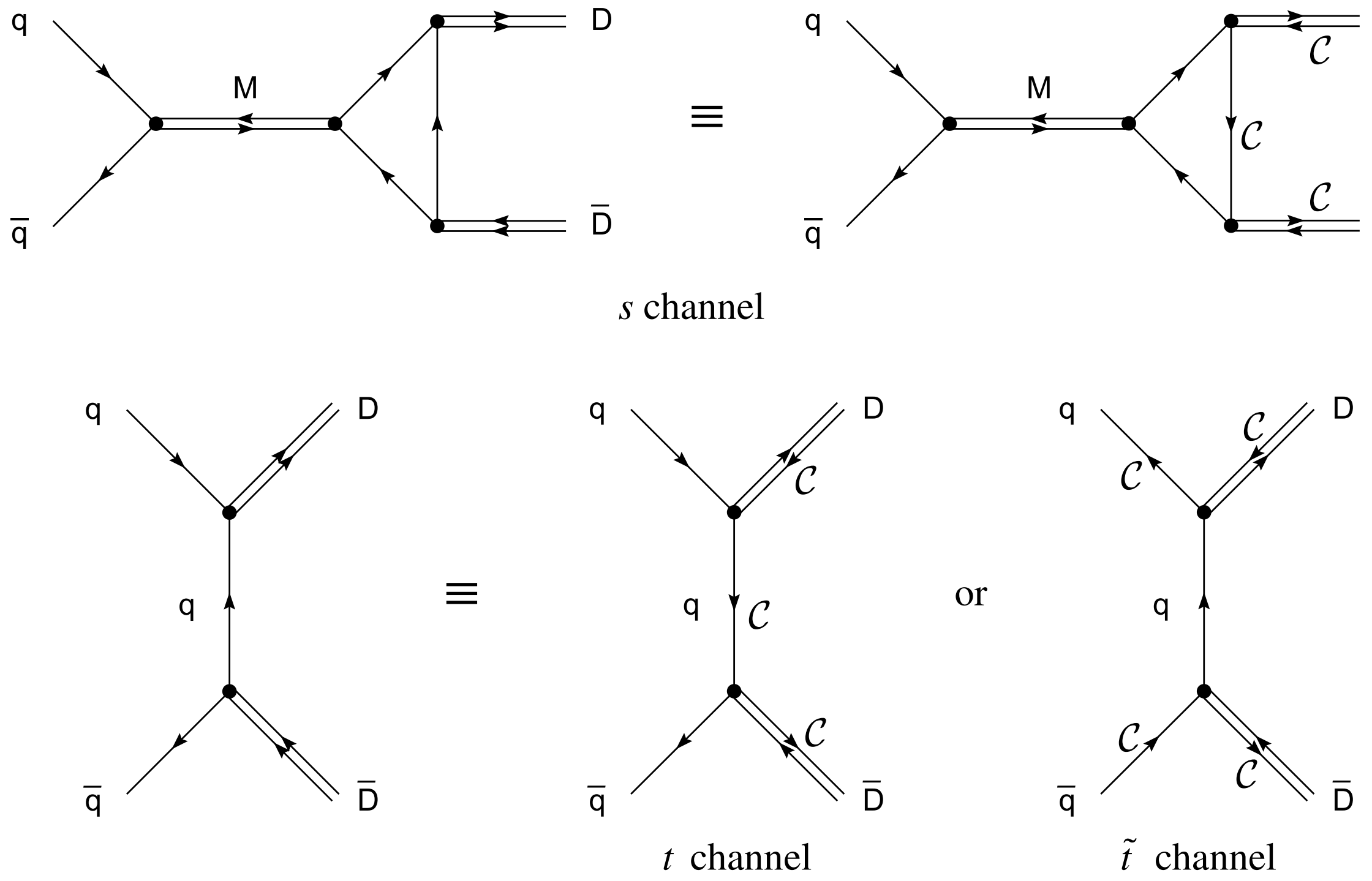

**Figure 12.** Channels of the reaction.

$$
\begin{aligned}
-i\mathcal{M}_s &= f_s\ \delta_{c_1,c_2}\ \bar{v}(p_2)\ u(p_1)\ i\mathcal{D}_s^S(p_1+p_2)\ \Gamma(p_1+p_2\ ,\ p_3)\ ig_1\ ig_2 \\
-i\mathcal{M}_t &= f_t\ \bar{v}(p_2)\ i\gamma_5\ \varepsilon^{c_2,c_t,c_4}\ ig_1\ iS_F(p_1-p_3)\ i\gamma_5\ \varepsilon^{c_t,c_1,c_3}\ ig_2\ u(p_1) \quad . \\
-i\mathcal{M}_{\tilde{t}} &= -f_t\ \bar{v}(p_1)\ i\gamma_5\ \varepsilon^{c_1,c_t,c_3}\ ig_1\ iS_F(p_3-p_1)\ i\gamma_5\ \varepsilon^{c_t,c_2,c_4}\ ig_2\ u(p_2)
\end{aligned} \tag{11}
$$

In the subsection 4.2, we saw that we had only one $t$ channel. Here, two "solutions" can be proposed for this channel. Indeed, we have two possibilities as regards the identification of the charge conjugate particles. The channel labeled as $t$ stipulates that the quark and the anti-quark are really which they are. On the other hand, for the $\tilde{t}$ channel, these two incoming particles are in fact charge conjugate ones. In practice, these two channels must be treated as two distinct ones. A cross term as $\mathcal{M}_t\ \mathcal{M}_{\tilde{t}}^*$ are thus possible. They leads to calculations involving terms as $\bar{v}(p_2)v(p_1)$ and $u(p_1)\bar{u}(p_2)$, i.e. spinors with two different momenta, see appendix B.

In the figure 13, we present the results found for the reaction $u+\bar{u}\rightarrow[ud]+\overline{[ud]}$, and for its reverse one $[ud]+\overline{[ud]}\rightarrow u+\bar{u}$. The associated cross-sections were estimated at finite temperatures, and for a null density. Indeed, as observed in the chapter 4, the anti-diquark $\overline{[ud]}$ is described in our model only when the baryonic density is modest, largely lower than $\rho_0$. So, the calculations were not performed a non-null densities. The direct reaction does not give strong cross sections, whatever the temperature. In fact, there is no divergence at the threshold, and the values never exceed 0.40 mb. So, this reaction is certainly dominated by reactions as $\mathrm{q}+\bar{\mathrm{q}}\rightarrow \mathrm{M}+\mathrm{M}$. Moreover, the cross-sections of the reverse reaction $[ud]+\overline{[ud]}\rightarrow u+\bar{u}$ are slightly higher than the ones of the direct reaction. But, in the same way, they do not exceed 0.6 mb. As a consequence, the process $\mathrm{q}+\bar{\mathrm{q}}\rightarrow \mathrm{D}+\bar{\mathrm{D}}$ and it reverse one are not expected to intervene in a notable way.

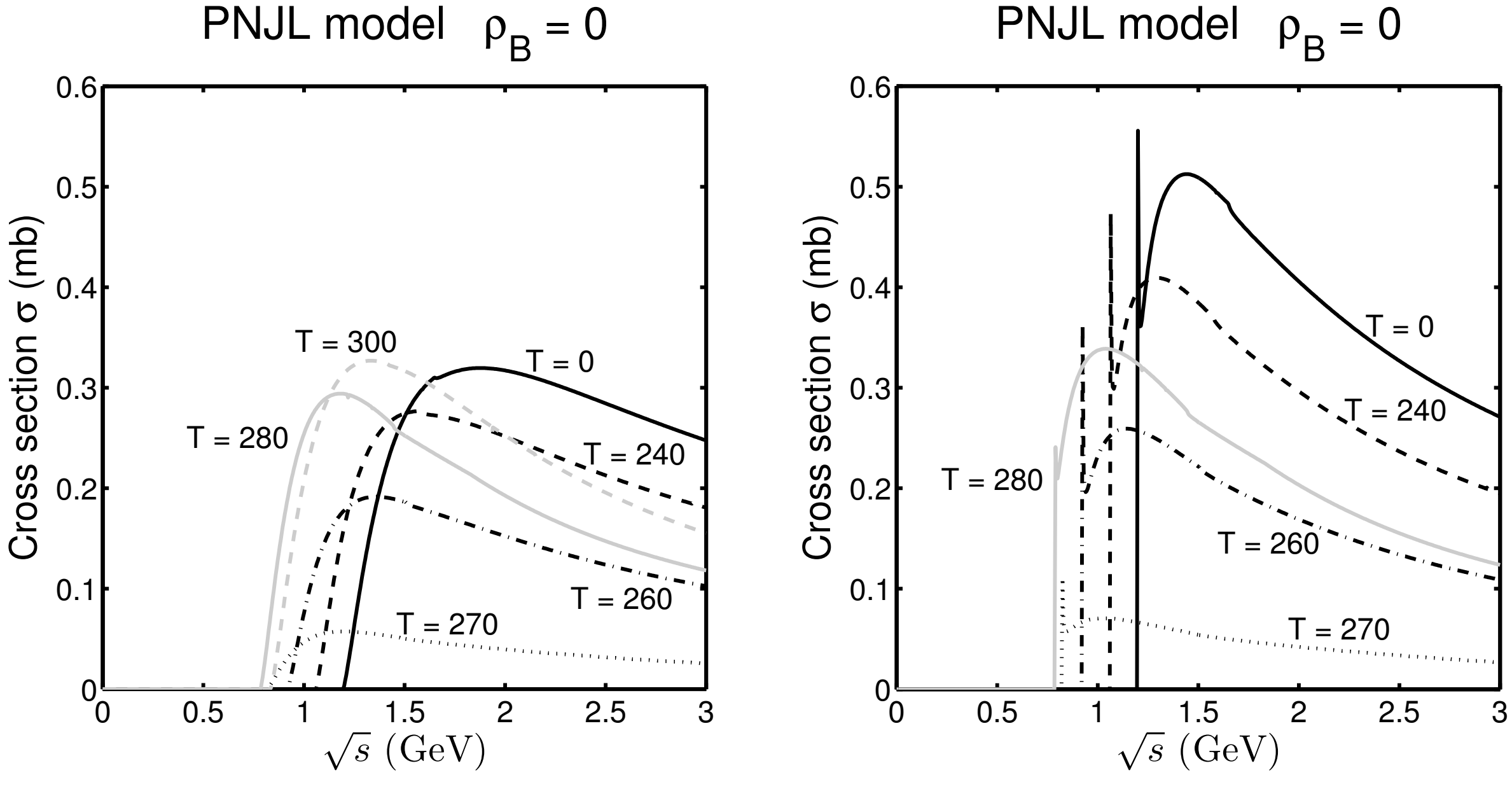

**Figure 13** Cross sections of $u+\bar{u}\rightarrow[ud]+\overline{[ud]}$ and $[ud]+\overline{[ud]}\rightarrow u+\bar{u}$ at null density.

## 4.4 $q+q \rightarrow D+M$ reactions

Like with the previous reactions, we firstly identify the particles that are charge conjugate ones. In the Feynman diagram associated with the *t* channel, figure 14, the quark in position *1* is a charge conjugate antiquark. Its associated spinor is $\overline{v}(p_1)$, as written in the matrix element equation (12). This process also admits a channel *u*. In this case, the vertex in which the diquark intervenes is the one in bottom of the graph. So, the quark in position *2* is a charge conjugate antiquark. This explains the spinor $\overline{v}(p_2)$ in (12).

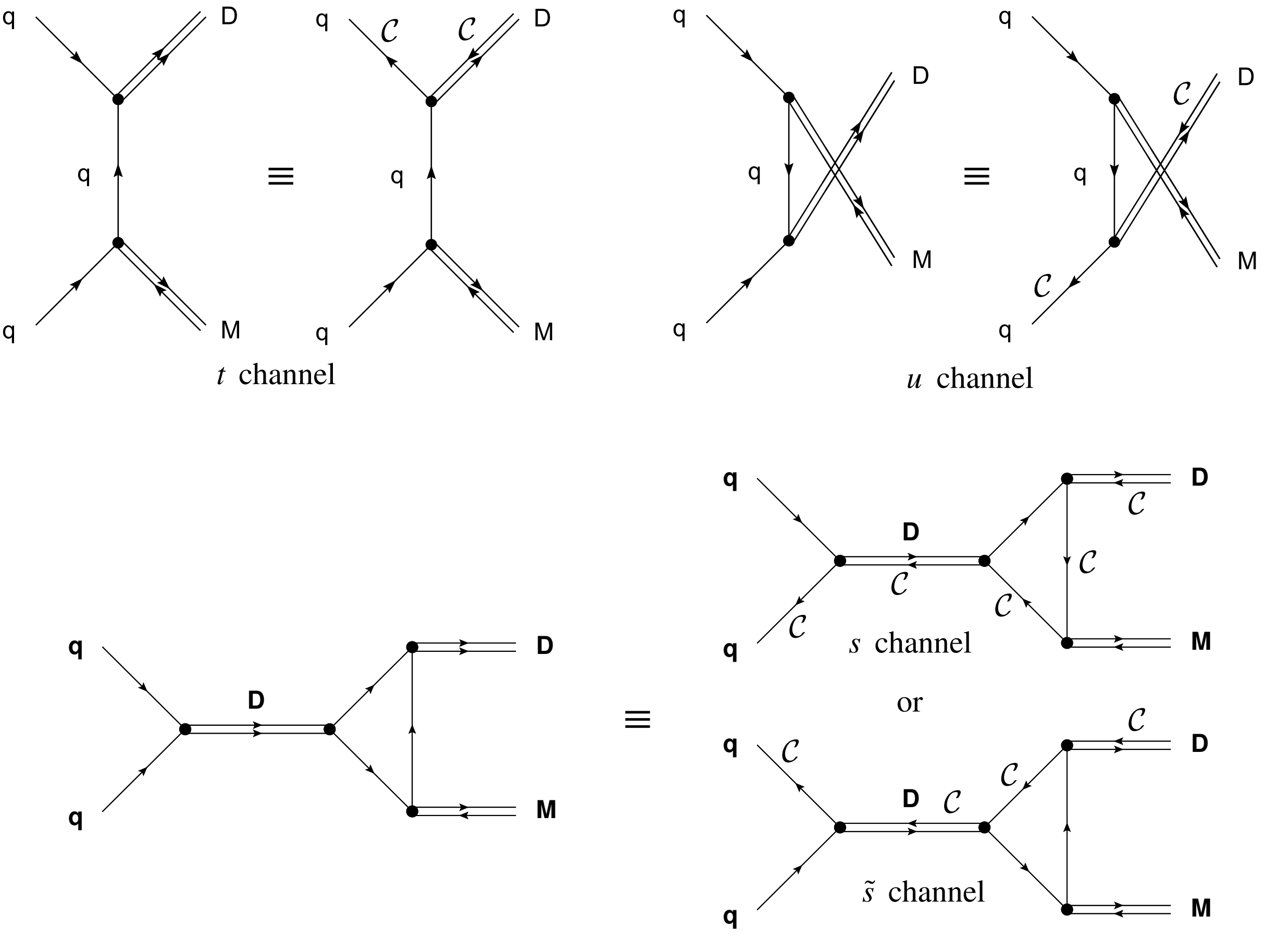

**Figure 14.** s, *t* and *u* channels.

Moreover, two *s* channels are considered. As in the section 3 and in the subsection 4.3, they include a $\Gamma$ function (9). This one takes into account the two $i\gamma_5$ matrices associated with the outgoing meson and diquark. As a consequence, the diquark used as propagator must be a pseudo-scalar one: $1_4$ contribution, i.e. a $4\times4$ identity matrix. On the other hand, scalar diquarks are not possible here as propagators. Indeed, the trace in the $\Gamma$ function must have an even number of $i\gamma_5$ matrices to be non-null. As in the subsection 4.3, the effect of the charge conjugation on the quarks in the $\Gamma$ function consists to reverse the sign of their chemical potential. Furthermore, by a study of the charge conjugated quarks/antiquarks, two possible solutions are obtained. The first keeps the name of "*s* channel" in the figure 14. In the writing of the associated matrix element, it has the same structure as the *s* channel used for $q+\overline{q}\rightarrow M+M$. The second solution is labeled as "$\tilde{s}$ channel". In the same way, the study also included $s'$ channels. They were established as in (7): they consist in an exchange of the outgoing particles at the exit of the triangle modeled by $\Gamma$.

The calculations of the squared terms do not present difference compared to the ones treated in the literature [7, 8]. Nevertheless, as in subsection 4.3, some mixed terms, as $\mathcal{M}_t\ \mathcal{M}_u^*$ and $\mathcal{M}_u\ \mathcal{M}_t^*$, imply calculations involving $\bar{v}(p_1)v(p_2)$ and $u(p_2)\bar{u}(p_1)$. In the appendix B, we present our method with the example of the $\mathcal{M}_u \cdot \mathcal{M}_t^*$ term.

$$\begin{aligned}
-i\mathcal{M}_t &= f_t\ \delta_{c_t,c_2}\ \varepsilon^{c_1,c_t,c_3}\ \bar{v}(p_1)\ i\gamma_5\ ig_1\ iS_F(p_3-p_1)\ i\gamma_5\ ig_2\ u(p_2) \\
-i\mathcal{M}_u &= f_u\ \delta_{c_u,c_1}\ \varepsilon^{c_2,c_u,c_3}\ \bar{v}(p_2)\ i\gamma_5\ ig_1\ iS_F(p_1-p_4)\ i\gamma_5\ ig_2\ u(p_1) \\
-i\mathcal{M}_s &= f_s\ \varepsilon^{c_1,c_2,c_D}\ \bar{v}(p_2)\ ig_1\ i\mathcal{D}_s^{PS}(p_1+p_2)\ \Gamma(p_1+p_2\ ,\ p_3)\ ig_2\ u(p_1) \\
-i\mathcal{M}_{\tilde{s}} &= f_s\ \varepsilon^{c_1,c_2,c_D}\ \bar{v}(p_1)\ ig_1\ i\mathcal{D}_{\tilde{s}}^{PS}(p_1+p_2)\ \Gamma(p_1+p_2\ ,\ p_4)\ ig_2\ u(p_2) \\
-i\mathcal{M}_{s'} &= f_{s'}\ \varepsilon^{c_1,c_2,c_D}\ \bar{v}(p_2)\ ig_1\ i\mathcal{D}_s^{PS}(p_1+p_2)\ \Gamma(p_1+p_2\ ,\ p_4)\ ig_2\ u(p_1) \\
-i\mathcal{M}_{\tilde{s}'} &= f_{s'}\ \varepsilon^{c_1,c_2,c_D}\ \bar{v}(p_1)\ ig_1\ i\mathcal{D}_{\tilde{s}}^{PS}(p_1+p_2)\ \Gamma(p_1+p_2\ ,\ p_3)\ ig_2\ u(p_2)
\end{aligned} \quad . \qquad (12)$$

We consider the reaction $u+d \to [ud]+\pi^0$ and its reverse one, i.e. $[ud]+\pi^0 \to u+d$. Our associated results are presented, respectively, in the figures 15 and 16. In the figure 15, the cross sections show divergence at the threshold. Indeed, the two incoming quarks can be heavier than the produced diquark and pion. However, the divergences are less marked than for the mesonization reactions. According to the results function of the temperature, the cross sections tend to increase when the temperature grows, until $T = 200$ MeV, left hand side of the figure 15. For higher values, they decrease. At $T = 300$ MeV and after, the divergence at the threshold disappears and the cross sections become highly negligible. The maximum values reached by this reaction stay rather modest: they cannot exceed 2 mb at null density. The contribution of the $s$ and $s'$ channels seems to be optimal near to the kinematic threshold. The mass of the $[ud]$ pseudo scalar diquark used as propagator can explain this observation. Indeed, at null density, its mass is about 930 MeV until $T = 200$ MeV, so its resonance when $\sqrt{s}$ is close to this value permits this contribution for reduced temperatures.

According to the baryonic density, the cross sections increase until $\rho_B \approx 2\rho_0$, density for which the cross sections can punctually exceed 10 mb, see the right hand side of the figure 15. As a whole, the cross sections are affected in a non-negligible way by the blocking factors (4), especially at non-null densities. Indeed, an increasing of the baryonic density leads to variations of the chemical potentials. These ones intervene in the Fermi-Dirac and Bose-Einstein statistics used by the blocking factors. This is one explanation of the very different curve shapes observed for the direct reaction $u+d \to [ud]+\pi^0$ and for the reverse reaction, figure 16, at non null baryonic density. Clearly, the direct reaction uses bosonic blocking factors, whereas the reverse process uses fermionic ones.

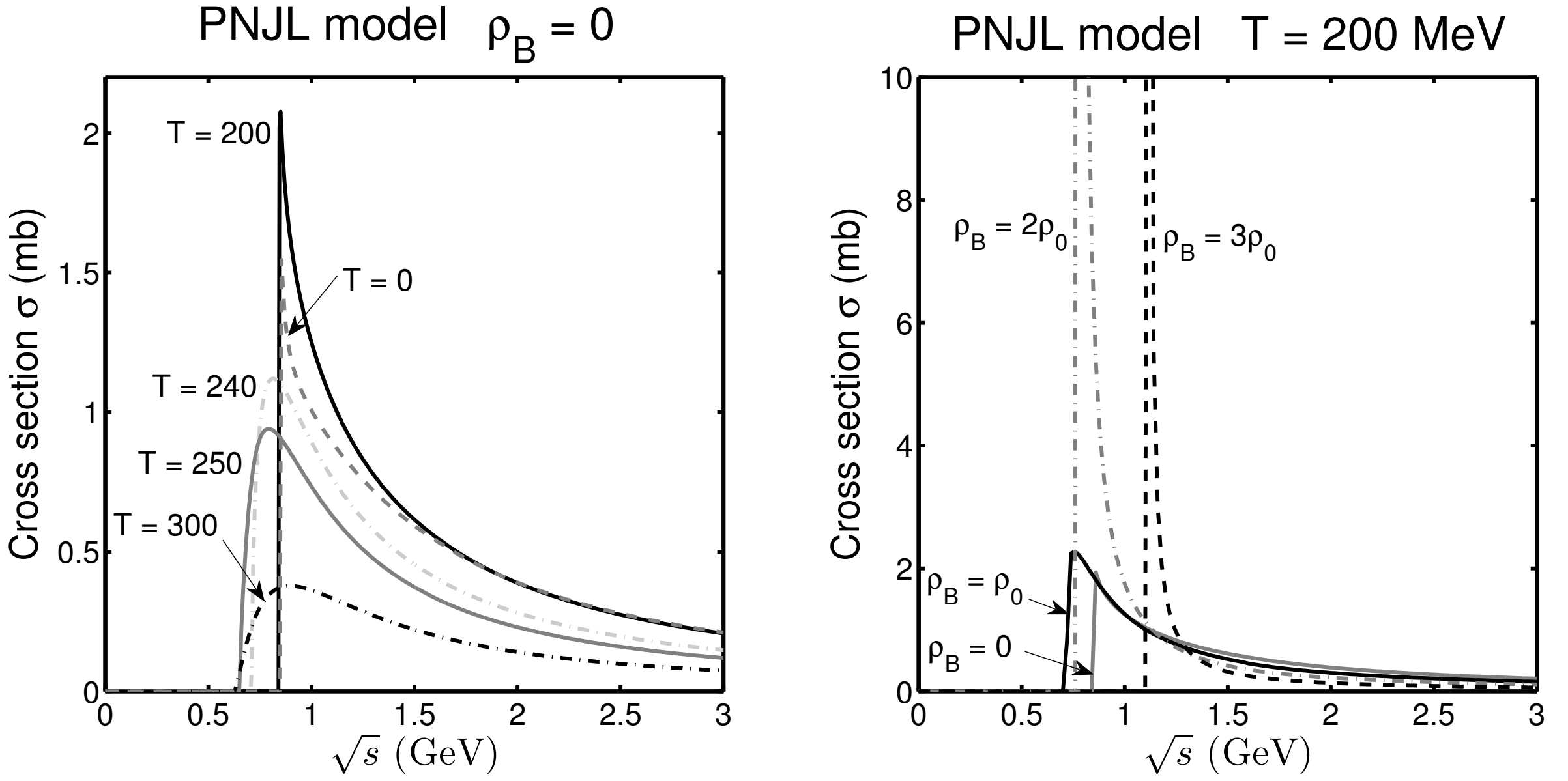

**Figure 15.** Cross sections of $u+d \to [ud]+\pi^0$.

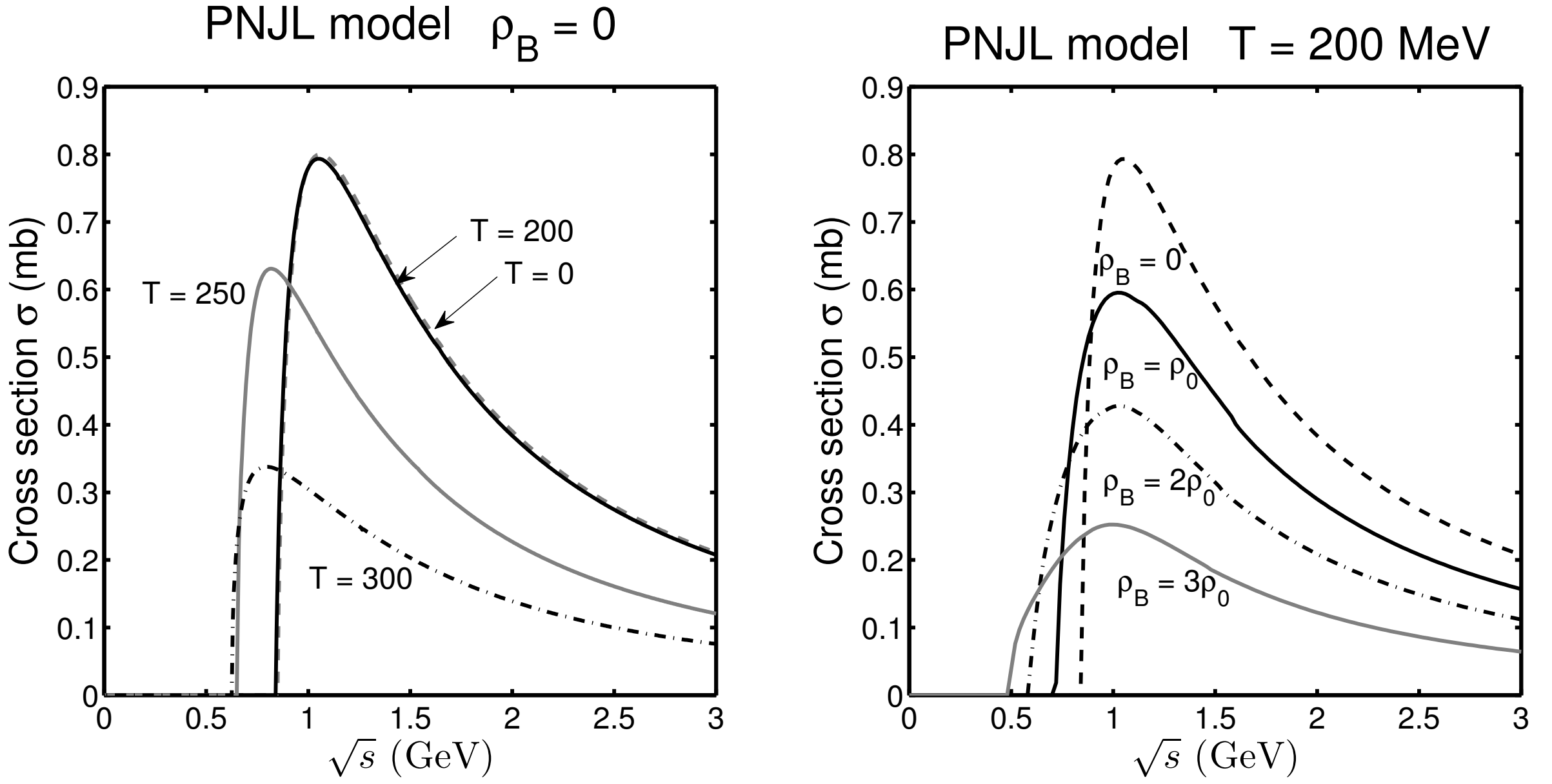

**Figure 16.** Cross sections of $[ud]+\pi^0 \to u+d$.

The values found with the $u+d \to [ud]+\pi^0$ reaction are globally stronger than the ones of $[ud]+\pi^0 \to u+d$. However, the difference is not very marked at null density. The diquark production seems to be rather reduced. Its optimal conditions could be near to $T = 200$ MeV and $\rho_B \approx 2\rho_0$. For such temperatures or higher, if the mesonization processes consumed enough antiquarks, these ones would not be able to destroy the diquarks, for example via $\overline{\mathrm{q}}+\mathrm{D} \to \mathrm{M}+\mathrm{q}$. With this hypothesis, some diquarks might be formed.

# 5. Baryonization reactions

We treat now inelastic reactions in which a baryon is produced, starting from quarks and/or diquarks. As in the previous section, we consider some of the processes described in [11, 12]. It concerns $\overline{\mathrm{q}}+\mathrm{D}\to\overline{\mathrm{D}}+\mathrm{B}$, $\mathrm{M}+\mathrm{D}\to\overline{\mathrm{q}}+\mathrm{B}$, $\mathrm{q}+\mathrm{q}\to\mathrm{B}+\overline{\mathrm{q}}$, $\mathrm{q}+\mathrm{D}\to\mathrm{M}+\mathrm{B}$, but we add the process $\mathrm{D}+\mathrm{D}\to\mathrm{B}+\mathrm{q}$. For all of these processes, we focus on the production of nucleons. More precisely, as explained in the previous chapter, these baryons were described by their scalar flavor component, and not by their axial one. This simplification will be used again in the descriptions performed in this section. Indeed, a vertex involving a nucleon and a diquark will thus translate a scalar interaction. Anyway, we had seen with the diquarks that some precautions are required in the writing of our matrix elements. It was the case especially for the identification of the charge conjugate quarks/anti-quarks. This method will be considered here, even if we no longer make appear our equivalent Feynman graphs, which revealed in an explicit way the particles that were charge conjugate ones.

Moreover, thanks to the works performed in the previous chapter, we remark that the mass of the nucleon is very close to the mass sum of its constituents, i.e. a light quark and a $[ud]$ diquark. It leads to a rather reduced binding energy. More precisely, at null temperature and density, this one is less than $150\ \mathrm{MeV}$, in absolute values. In the framework of the nucleon production, it wants to say that we cannot model processes in which the incoming particles are heavier than the outgoing ones, if one of the outgoing particles is a nucleon. As noticed before, this behavior is translated by the absence of divergence at the level of the kinematic threshold. In addition, it can induce stronger cross sections for the reverse reactions, precisely those that tend to destroy baryons. Clearly, for each of the treated examples, this property is able to limit the field of application of these baryonization processes. But, we will study in what conditions the associated reverse reactions are expected to not intervene.

## 5.1 $\overline{\mathrm{q}}+\mathrm{D}\to\overline{\mathrm{D}}+\mathrm{B}$ reactions

Whatever the concerned particles, the *t* channel is considered for these reactions. This one is described by the Feynman diagram in the figure 17, and by the associated matrix element equation (13). The vertex involving the anti-diquark (placed in top on the right of the graph) and the two anti-quarks uses a $\gamma_5$ matrix, because the anti-diquark is a scalar one. On the other hand, the vertex of the bottom in the diagram is connected to a baryon, a scalar diquark and a quark. This vertex indicates an interaction of scalar type. This is translated by a $1_4$ matrix (identity matrix), in the same spirit of the equation (2) seen in the previous chapter. This $1_4$ matrix does not appear explicitly in (13). The "rule" at the level of such vertices is to consider that the quark is always a real quark. In other words, this quark is not a charge conjugate antiquark. We conclude that the only solution is the anti-quark in position *1* is in fact a charge conjugate quark: its associated spinor is $u(p_1)$ in the equation (13).

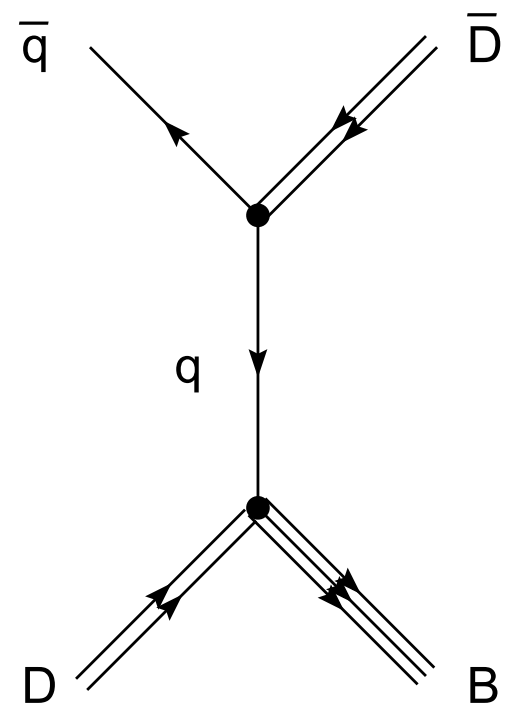

**Figure 17.** *t* channel.

$$-i\mathcal{M}_t = f_t \; ig_1 \; \bar{u}(p_4) \; iS_F(p_4 - p_2) \; i\gamma_5 \; ig_2 \; \varepsilon^{c_t, c_1, c_3} \; u(p_1) \quad . \tag{13}$$

As an example, we consider the reaction $\bar{u} + [ud] \to \overline{[ud]} + n$. The neutron is symbolized by $n$. Since this reaction uses an anti-diquark, the study of the cross sections according to the baryonic density was not performed, as with the reaction $u + \bar{u} \to [ud] + \overline{[ud]}$ seen previously. The results, available in the figure 18, reveal that the cross sections are rather weak. In fact, at null baryonic density, they never exceed 0.25 mb. We can then extrapolate with all the reactions $\bar{\mathrm{q}} + \mathrm{D} \to \bar{\mathrm{D}} + \mathrm{B}$. In addition, at null or positive densities, we predict that the anti-diquarks are so badly tolerated particles by a physical system that their existence would be sources of destruction of the baryons already present in the medium, precisely by the reverse reactions of the ones treated here, i.e. $\bar{\mathrm{D}} + \mathrm{B} \to \bar{\mathrm{q}} + \mathrm{D}$.

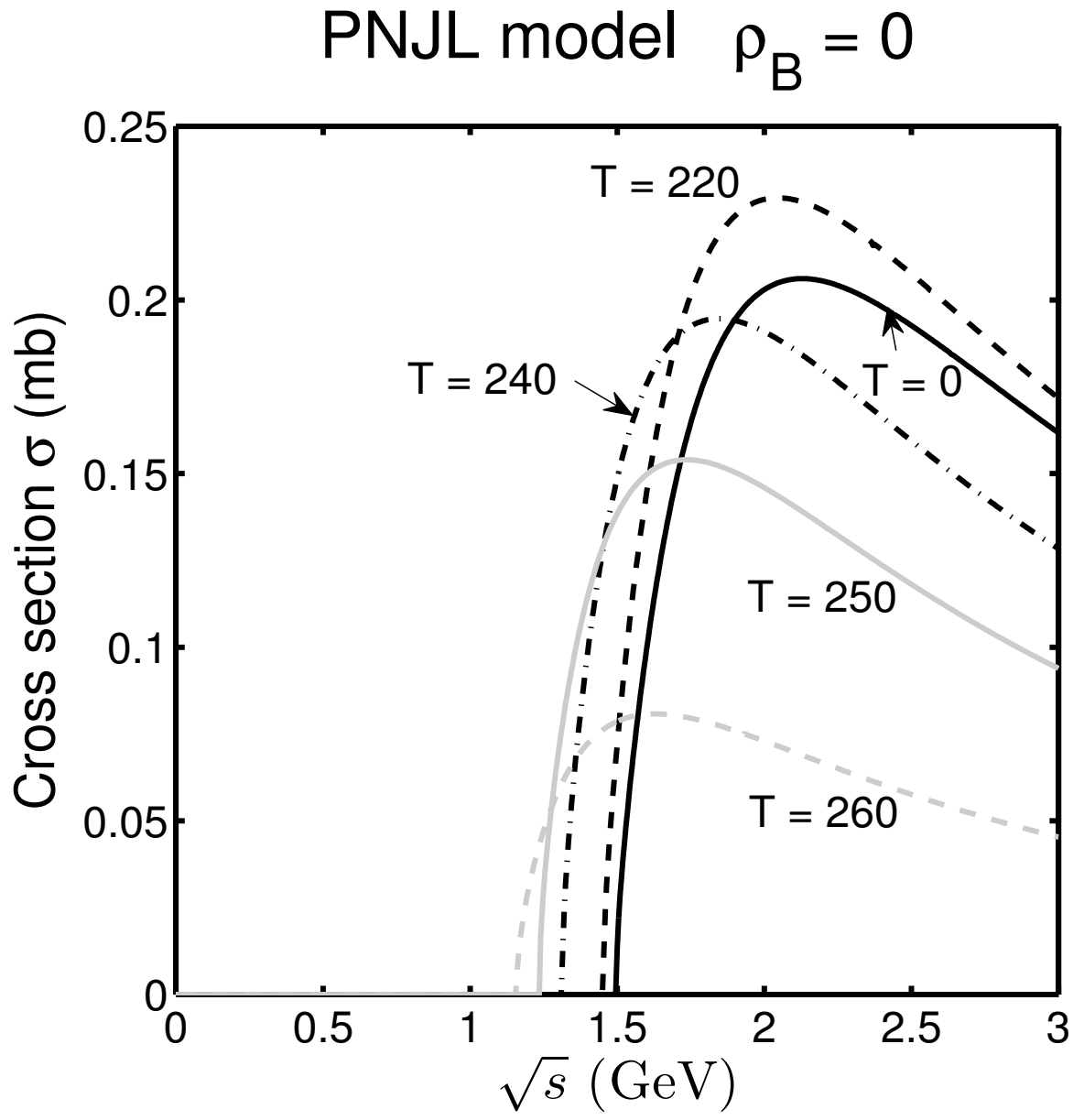

**Figure 18.** Cross sections of the reaction $\bar{u} + [ud] \to \overline{[ud]} + n$ for several temperatures.

As a consequence, we can consider that this reaction does not really contribute to the baryonization of the system, and can be neglected. As a whole, as $u+\bar{u} \rightarrow [ud]+\overline{[ud]}$, reactions producing anti-diquarks are not expected to intervene in a significant way.

## 5.2 $\mathrm{M}+\mathrm{D} \rightarrow \bar{\mathrm{q}}+\mathrm{B}$ reactions

We test here a process that is able to consume a diquark and to create a baryon. This reaction requires the presence of a meson. Because of the strong mesonization caused by the reactions $\mathrm{q}+\bar{\mathrm{q}} \rightarrow \mathrm{M}+\mathrm{M}$, this aspect does not seem to be a problem. In our description, the $\mathrm{M}+\mathrm{D} \rightarrow \bar{\mathrm{q}}+\mathrm{B}$ reactions are described by the *t* channel, see figure 19 and equation (14). The vertex at the bottom in the diagram is connected to a baryon. Thus, this vertex indicates a scalar type interaction, so uses a $1_4$ matrix. According to the "rule" established in the previous subsection, the quark involved in this vertex is a real quark. As a consequence, the antiquark in position *3* is a real antiquark.

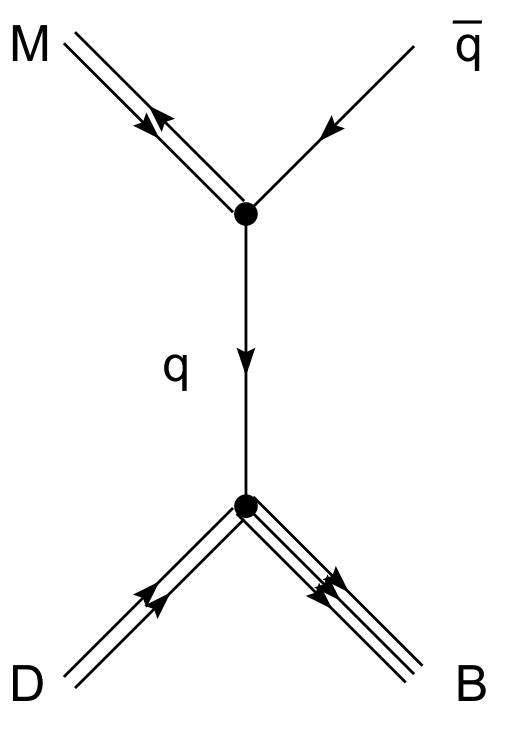

**Figure 19.** *t* channel.

$$-i\mathcal{M}_t = f_t \; ig_1 \; \bar{u}(p_4) \; iS_F(p_4 - p_2) \; i\gamma_5 \; ig_2 \; v(p_3) \quad . \tag{14}$$

The figure 20 describes the cross sections found with the reaction $\pi^- + [ud] \rightarrow \bar{u} + n$. As a whole, these cross sections are reduced. They do not exceed one millibarn in the presented results. At null baryonic density, the cross sections tend to decrease when the temperature is growing. At $T = 200$ MeV, the same behavior is observed when we vary the baryonic density. Moreover, the outgoing particles are heavier than the incoming ones. As a consequence, no divergence at the threshold is observed. Another consequence is the reverse reaction has stronger cross-sections. However, as indicated in the subsection 4.4, if the antiquarks tend to be rarer, this reverse reaction cannot occur. In parallel, some diquarks could be formed, to allow the formation of few baryons, for example via this process $\mathrm{M}+\mathrm{D} \rightarrow \bar{\mathrm{q}}+\mathrm{B}$. The antiquarks produced in the same time might be consumed preferentially by the mesonization reactions. Anyway, $\mathrm{M}+\mathrm{D} \rightarrow \bar{\mathrm{q}}+\mathrm{B}$ seems to be too limited to assume alone the baryonization of the quark system.

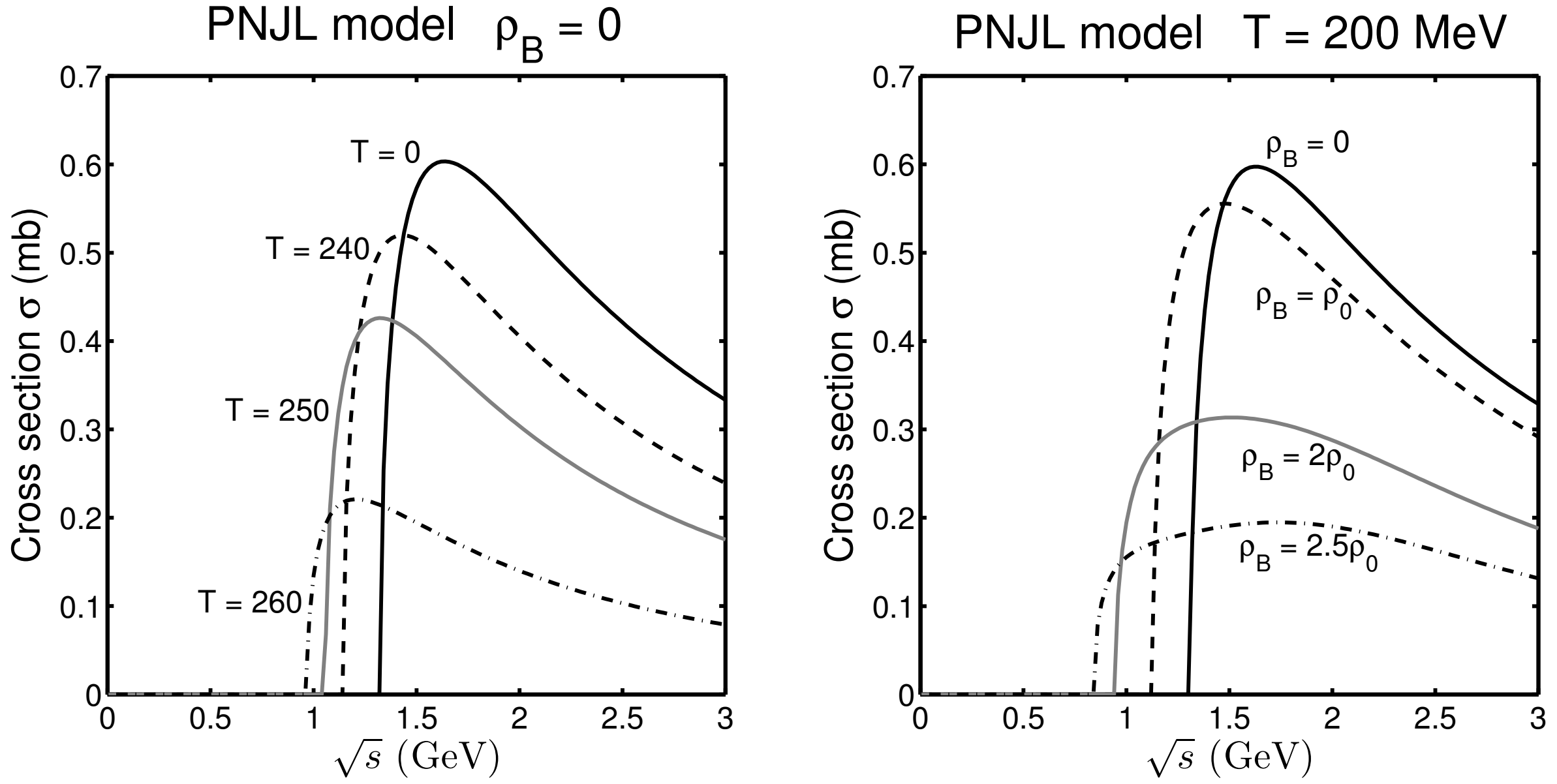

**Figure 20.** Cross sections of $\pi^- + [ud] \rightarrow \bar{u} + n$ .

## 5.3 $\mathrm{D} + \mathrm{D} \rightarrow \mathrm{B} + \mathrm{q}$ reactions

We describe the reactions $\mathrm{D} + \mathrm{D} \rightarrow \mathrm{B} + \mathrm{q}$ by two channels, figure 21 and equation (15). There is no ambiguity at the level of the baryon vertices, but with the ones involving a diquark and two quarks. The quark at the position *4* is a charge conjugate antiquark for the two channels. It is translated in (15) by the $v(p_4)$ spinor.

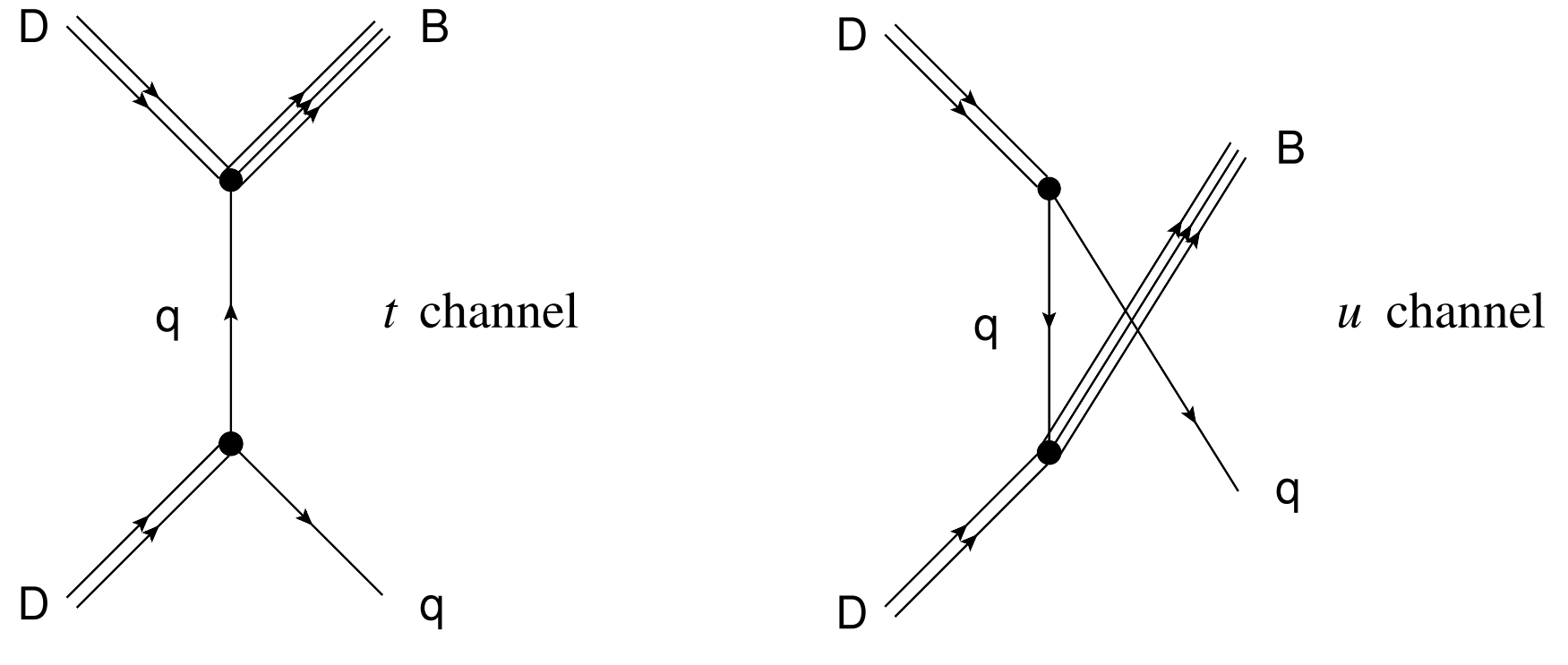

**Figure 21.** Involved channels.

$$\begin{aligned} -i\mathcal{M}_t &= f_t \ \varepsilon^{c_4,c_t,c_2} \ ig_1 \ v(p_4) \ i\gamma_5 \ iS_F(p_3 - p_1) \ ig_2 \ \bar{u}(p_3) \\ -i\mathcal{M}_u &= f_u \ \varepsilon^{c_4,c_u,c_1} \ ig_1 \ v(p_4) \ i\gamma_5 \ iS_F(p_1 - p_4) \ ig_2 \ \bar{u}(p_3) \end{aligned} \quad . \tag{15}$$

The numerical results correspond to the reaction $[ud] + [ud] \rightarrow p + d$, in which $p$ is a proton. The figure 22 indicates that the temperature has a modest influence on the cross sections, until $T = 250$ MeV. After this temperature, the cross sections brutally decrease. Similar observations can be done according to the densities, for $T = 200\,\text{MeV}$. At densities above $2\rho_0$, the cross sections cannot exceed one millibarn.

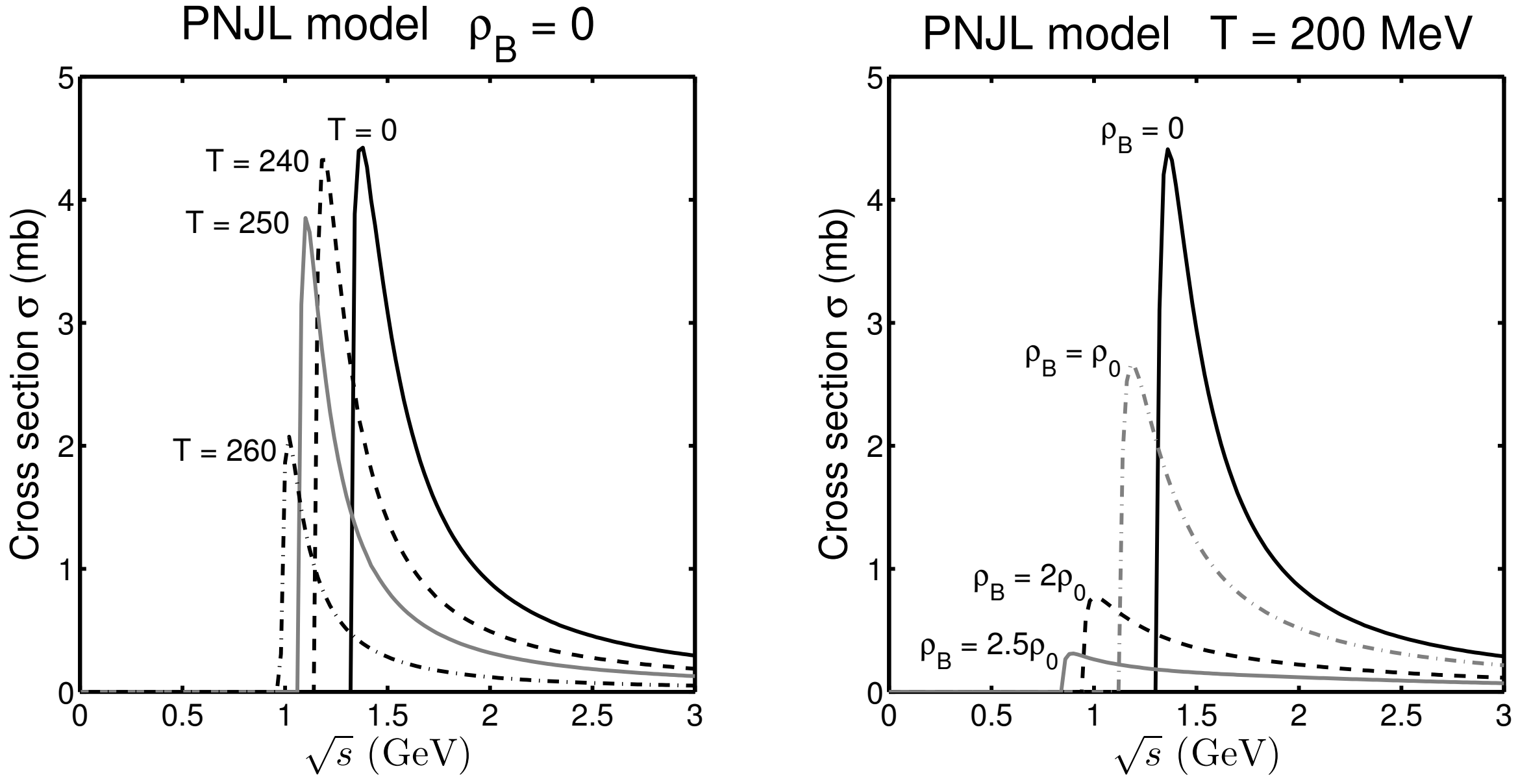

**Figure 22.** Cross sections of the reaction $[ud]+[ud] \to p+d$ for several temperatures and baryonic densities.

The reaction is able to have cross sections close to 4.5 mb at the maximum. This is better than the reactions treated at the previous subsections. Furthermore, we will see hereafter the other cross-sections associated with baryonization reactions will not give higher values. Nevertheless, it stays rather weak compared to the values found for the mesonization reactions. In addition, this inelastic collision between two diquarks supposes a relative important production of these particles. The results found in the section 4 do not seem to confirm this hypothesis, especially at null density. In fact, $\mathrm{D+D \to B+q}$ is expected to act during the baryonization in a modest way. Clearly, the reverse reaction cannot really intervene in Nature, excepted in a very hot or dense medium. There is no real chance that a stable baryon, hit by a quark, disintegrates into two colored objects as diquarks in “normal” conditions, i.e. moderate temperatures and densities.

## 5.4 $\mathrm{q+q \to B+\bar{q}}$ reactions

At the opposite of the reactions $\mathrm{D+D \to B+q}$, the reactions $\mathrm{q+q \to B+\bar{q}}$ are an alternative to the scenario saying that the diquarks intervene as intermediate particles. A reduced production of diquarks can comfort the study of reactions that are not concerned by these particles. Indeed, the process described in this subsection creates baryons starting directly from quarks. The diquarks only play the role of intermediates, because they act as propagators in the treated channels, figure 23. They correspond to the $\mathcal{D}^D$ (P)NJL propagators in the matrix elements (16). The used diquarks are scalar ones, so they imply $i\gamma_5$ at the level of the vertices involving also two quarks/antiquarks. The structure of the Feynman diagrams has similarities with the diagrams used to model the elastic quark/antiquark scatterings, see [8] or section 6. About the $t/\tilde{t}$ channels, we have an ambiguity on the level of the quark in position *2* and the antiquark in position *4*. One of the two is a charge conjugate one. For the $t$ channel, the antiquark is in fact a charge conjugate quark. For the $\tilde{t}$ channel, the quark is a charge conjugate antiquark. Then, for the $u/\tilde{u}$ channels, the ambiguity is at the level of the quark in position *1* and the antiquark in position *4*. If the antiquark is a charge conjugate quark, we obtain the $u$ channel. Otherwise, we have the $\tilde{u}$ channel. About the $s/\tilde{s}$ channels, the ambiguity is related to the two incoming quarks. One of them is a charge conjugate antiquark…

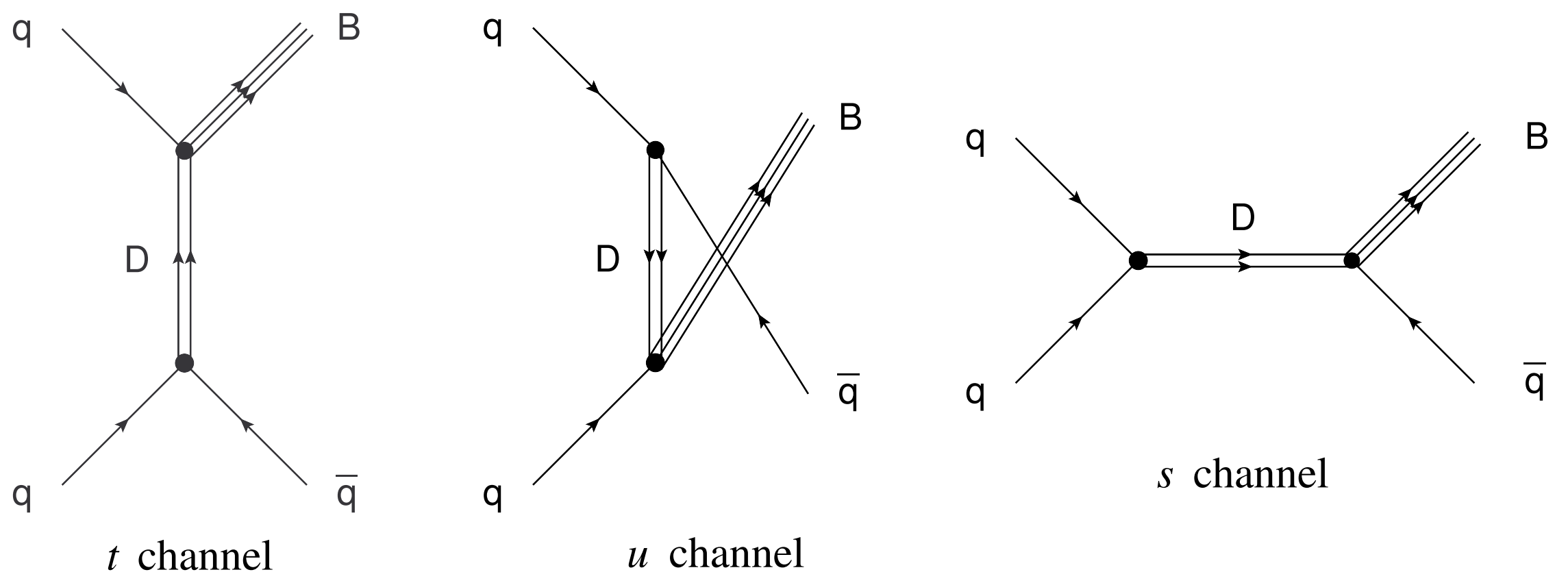

**Figure 23.** Involved channels.

$$
\begin{aligned}
-i\mathcal{M}_s &= f_s\ \varepsilon^{c_2,c_1,c_s}\ \overline{v}(p_2)\ i\gamma_5\ u(p_1)\ i\mathcal{D}_s^D(p_1+p_2)\ \overline{u}(p_3)\ v(p_4) \\
-i\mathcal{M}_{\tilde{s}} &= f_s\ \varepsilon^{c_2,c_1,c_s}\ \overline{v}(p_1)\ i\gamma_5\ u(p_2)\ i\mathcal{D}_s^D(p_1+p_2)\ \overline{u}(p_3)\ v(p_4) \\
-i\mathcal{M}_t &= f_t\ \varepsilon^{c_4,c_2,c_t}\ \overline{u}(p_4)\ i\gamma_5\ u(p_2)\ i\mathcal{D}_t^D(p_3-p_1)\ \overline{u}(p_3)\ u(p_1) \\
-i\mathcal{M}_{\tilde{t}} &= f_t\ \varepsilon^{c_2,c_4,c_t}\ \overline{v}(p_2)\ i\gamma_5\ v(p_4)\ i\mathcal{D}_t^D(p_3-p_1)\ \overline{u}(p_3)\ u(p_1) \\
-i\mathcal{M}_u &= f_u\ \varepsilon^{c_4,c_1,c_u}\ \overline{u}(p_4)\ i\gamma_5\ u(p_1)\ i\mathcal{D}_u^D(p_3-p_2)\ \overline{u}(p_3)\ u(p_2) \\
-i\mathcal{M}_{\tilde{u}} &= f_u\ \varepsilon^{c_1,c_4,c_u}\ \overline{v}(p_1)\ i\gamma_5\ v(p_4)\ i\mathcal{D}_u^D(p_3-p_2)\ \overline{u}(p_3)\ u(p_2)
\end{aligned}
\quad . \qquad (16)
$$

The results obtained with $u+u \to p+\overline{d}$ are presented in the figure 24. The $s/\tilde{s}$ channels are not available in this example. Indeed, $[uu]$ diquarks, i.e. flavor symmetrical, are axial diquarks, but not the scalar ones. Furthermore, at the contrary of some processes evoked in this work, notably $\mathrm{q}+\overline{\mathrm{q}} \to \mathrm{M}+\mathrm{M}$ [7], the contribution of the $s/\tilde{s}$ channels is not important in the case of the reactions $\mathrm{q}+\mathrm{q} \to \mathrm{B}+\overline{\mathrm{q}}$. Moreover, the choice of scalar diquarks, instead of pseudo scalar ones, was motivated by the fact that scalar ones allow obtaining stronger cross sections. Concerning the evolution of the cross sections, the temperature and the baryonic density act in the same manner. Indeed, these parameters shift the threshold towards low $\sqrt{s}$ values. However, the temperature does not really modify the obtained cross-sections, whereas an increase of the density leads to a reduction of the found values.

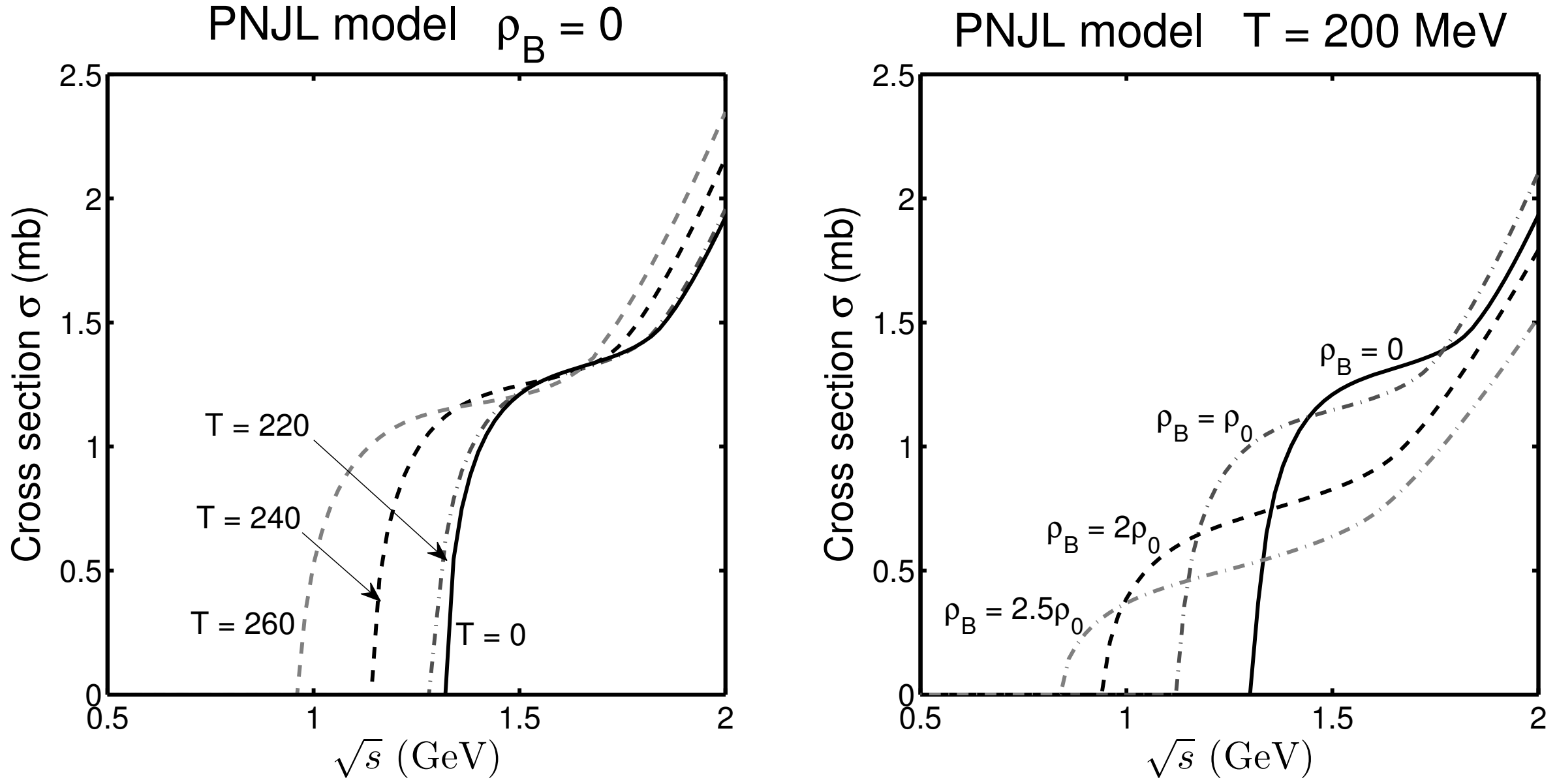

**Figure 24.** Cross sections of the reaction $d+u \to n+\bar{d}$ .

In the figure 24, the values stay rather low: they do not exceed 1.5 mb when $\sqrt{s}$ is lower than 1.7 GeV. Furthermore, $\mathrm{q+q \to B+\bar{q}}$ is able to compete with $\mathrm{q+q \to D+M}$ or with the quark elastic scattering [8]. These three reactions have cross sections of equal importance, even if $\mathrm{q+q \to D+M}$ allow the formation of diquarks, so it can permit the production of baryons. Concerning the reverse reaction of $\mathrm{q+q \to B+\bar{q}}$, it is expected to be stronger than our reaction, when it is supposed to be realizable. But, as indicated previously, the disappearance of the antiquarks (thanks to a massive mesonization) can neutralize the process $\mathrm{B+\bar{q} \to q+q}$. In addition, the mesonization is expected to continue when the baryonization starts, by capturing all the antiquarks that could be formed during this phase. Clearly, the process $\mathrm{q+q \to B+\bar{q}}$ can be considered as relevant in the framework of the baryonization of the system. It has reduced cross-sections, but it is totally independent of the diquark production.

## 5.5 $\mathrm{q+D \to M+B}$ reactions

In these reactions, the outgoing particles are not colored objects, whereas it is the case for the incoming ones, i.e. the diquark and the quark. Also, $\mathrm{q+D \to M+B}$ has the advantage that the reverse reactions cannot intervene, except in extreme conditions. In a scenario in which few diquarks would be produced, $\mathrm{q+D \to M+B}$ is imagined to be the final process that could combine the formed diquarks with the remaining free quarks. At this stage, the antimatter, i.e. notably the antiquarks, is supposed to be already consumed by previous reactions to form mesons. After the $\mathrm{q+D \to M+B}$ reactions, all the formed particles are expected to be observables ones. In our work, this process is described by the $t$ channel, figure 25. The associated matrix element is written equation (17). There is no vertex implying a diquark and two quarks. Thus, there is no ambiguity on the quarks and anti-quarks that are present. Therefore, the $t$ channel is not subdivided in another $\tilde{t}$ channel.

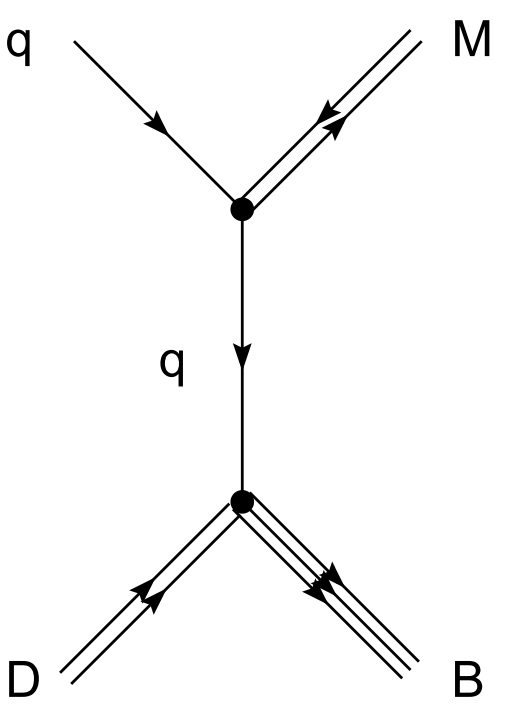

**Figure 25.** *t* channel.

$$-i\mathcal{M}_t = f_t \; ig_1 \; \bar{u}(p_4) \; iS_F(p_1 - p_3) \; i\gamma_5 \; ig_2 \; u(p_1). \quad (17)$$

The reaction $u + [ud] \to \pi^+ + n$ is considered as an example. The associated results, figure 26, indicate that the cross sections are weak. However, even if we note a resemblance between the equations (17) and (14), the values are here slightly higher than the ones observed in the subsection 5.2. Indeed, the reaction $\mathrm{q} + \mathrm{D} \to \mathrm{M} + \mathrm{B}$ is able to reach 1.2 mb in some conditions. The behavior according to the temperature, at null density, is described in the left hand side of the figure 26. It is shown that the cross sections increase very slowly when the temperature is growing, until $T = 220$ MeV. After that, the values quickly drop until the nucleon's limit of stability. About the evolution according to the density, right hand side of the figure, we suggest to compare it with the figure 20 to qualitatively observe a similar behavior between the two reactions. Moreover, even if one *s* channel was imagined by [11, 12], no divergence at the threshold is expected for this reaction, even with it. Maybe the inclusion of such a channel is able to increase the cross-sections, but it can be only in a rather reduced way.

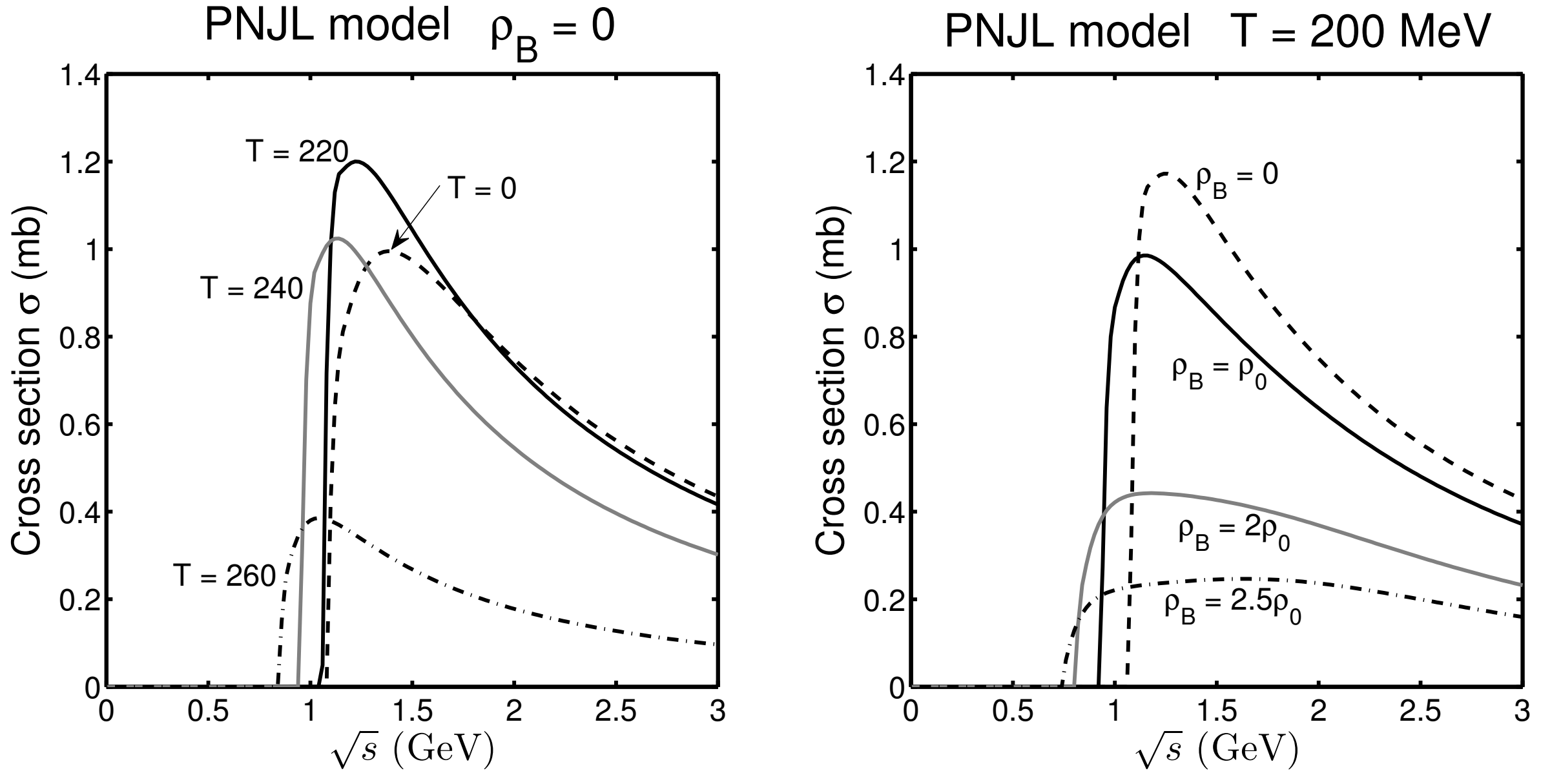

**Figure 26.** Cross sections of the reaction $u + [ud] \to \pi^+ + n$.

# Elastic reactions

We turn our attention now to elastic reactions. These ones were less encountered in the NJL literature compared to inelastic ones, except for the quark-quark and quark/antiquark elastic scatterings [8]. About these former ones, as explained in the introduction of this chapter, the objectives are to confirm the results found in this reference, extend them at non-null densities, and perform such calculations in the framework of the PNJL model. Notably with $\mathrm{q}+\overline{\mathrm{q}} \rightarrow \mathrm{q}+\overline{\mathrm{q}}$, another goal is to be able to compare them with the values found with $\mathrm{q}+\overline{\mathrm{q}} \rightarrow \mathrm{M}+\mathrm{M}$, to see if the elastic scattering is able to disturb the mesonization, i.e. if it can be a serious source of competition. Such a study is then extendable to the other elastic reactions treated in this second part of the chapter. Thus, the objective is to see the elastic processes that can intervene in a notable way, compared to the inelastic ones.

# 6. Elastic collisions with quarks and antiquarks

## 6.1 Quark–antiquark scattering

As observed in [8], the elastic scattering between a quark and an antiquark is described with $s$ and $u$ channels, figure 27 and equation (18). The used propagators concern scalar and pseudo-scalar mesons, that are respectively represented by the NJL/PNJL propagators $\mathcal{D}^S_{t,u}$ and $\mathcal{D}^P_{t,u}$. The $T$ symbols refer to flavor factor terms.

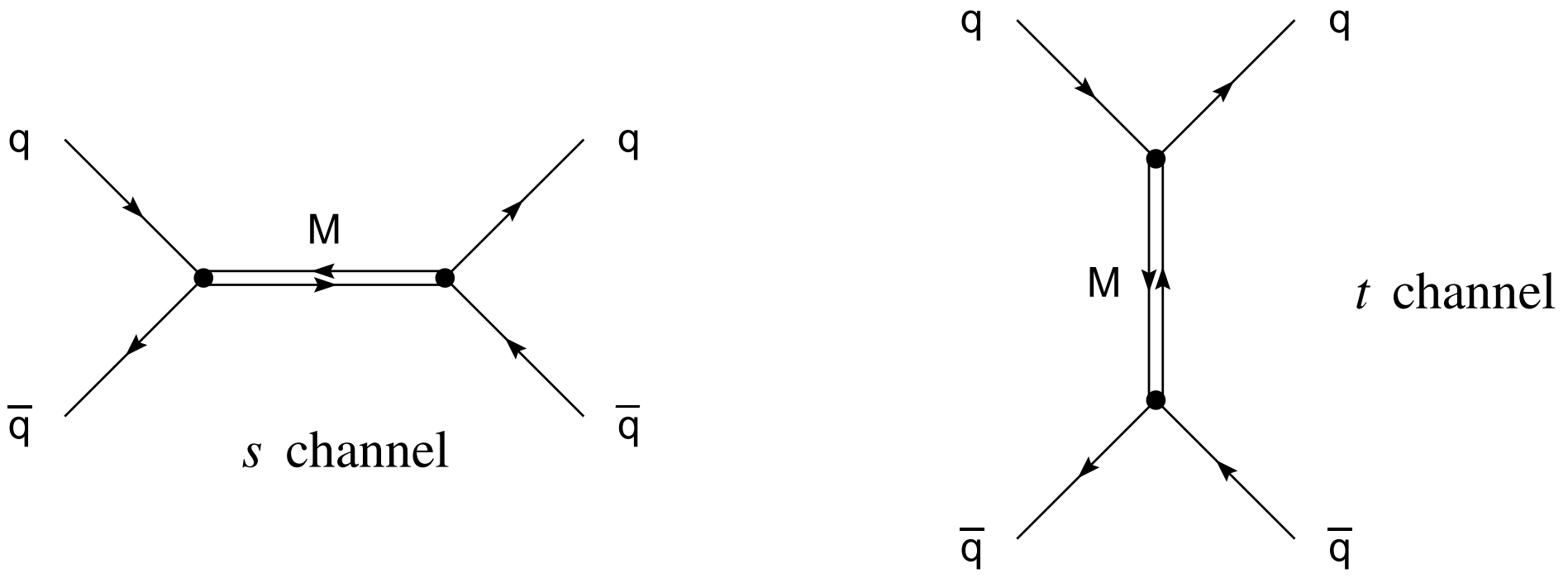

**Figure 27.** $s$ and $t$ channels.

$$\begin{aligned}
-i\mathcal{M}_s &= \delta_{c_1,c_2}\delta_{c_3,c_4}\ \overline{v}(p_2)\ T\ u(p_1)\ i\mathcal{D}^S_s(p_1+p_2)\ \overline{u}(p_3)\ T\ v(p_4) \\
&\quad + \delta_{c_1,c_2}\delta_{c_3,c_4}\ \overline{v}(p_2)\ i\gamma_5\ T\ u(p_1)\ i\mathcal{D}^P_s(p_1+p_2)\ \overline{u}(p_3)\ i\gamma_5\ T\ v(p_4) \\
-i\mathcal{M}_t &= \delta_{c_1,c_3}\delta_{c_2,c_4}\ \overline{u}(p_3)\ T\ u(p_1)\ i\mathcal{D}^S_t(p_1-p_3)\ v(p_4)\ T\ \overline{v}(p_2) \\
&\quad + \delta_{c_1,c_3}\delta_{c_2,c_4}\ \overline{u}(p_3)\ i\gamma_5\ T\ u(p_1)\ i\mathcal{D}^P_t(p_1-p_3)\ v(p_4)\ i\gamma_5\ T\ \overline{v}(p_2)
\end{aligned} \quad . \tag{18}$$

We evaluated the cross-sections of $u+\overline{u} \rightarrow u+\overline{u}$. The results are presented figures 28 and 29. The left hand side of the figure shows our NJL results, whereas the other part concerns PNJL values. For all these graphs, as mentioned in [8], the calculations are restricted by a limit value according to $\sqrt{s}$ :

$$\sqrt{s}_{\text{limit}} = 2\sqrt{\Lambda^2 + m_f^{\,2}}\Big|_{f=u,d,s} \,, \tag{19}$$

in which $\Lambda$ is the upper limit of the integrals used to calculate the masses of the concerned particles, see chapter 2.

Concerning the NJL cross sections, we observe a good agreement with the results of [8]. For the NJL and PNJL results of the figure 28, the cross-sections are modest at low temperatures. Then, they brutally increase, and they from structures close to the kinematic threshold. After that, at high temperatures, the values decrease when the temperature is growing. As with $\mathrm{q}+\bar{\mathrm{q}} \to \mathrm{M}+\mathrm{M}$, the inclusion of the Polyakov loop leads to a shifting of the curves towards higher temperatures. More precisely, with the NJL model, the strongest cross sections are obtained at $T = 250\ \mathrm{MeV}$, in agreement with [8], whereas in the PNJL approach, the optimal cross-sections are found for temperatures close to $300\ \mathrm{MeV}$. In fact, whatever the model, NJL or PNJL, the optimal temperature corresponds to the critical temperatures (Mott temperatures) of some pseudo-scalar mesons used as propagators.

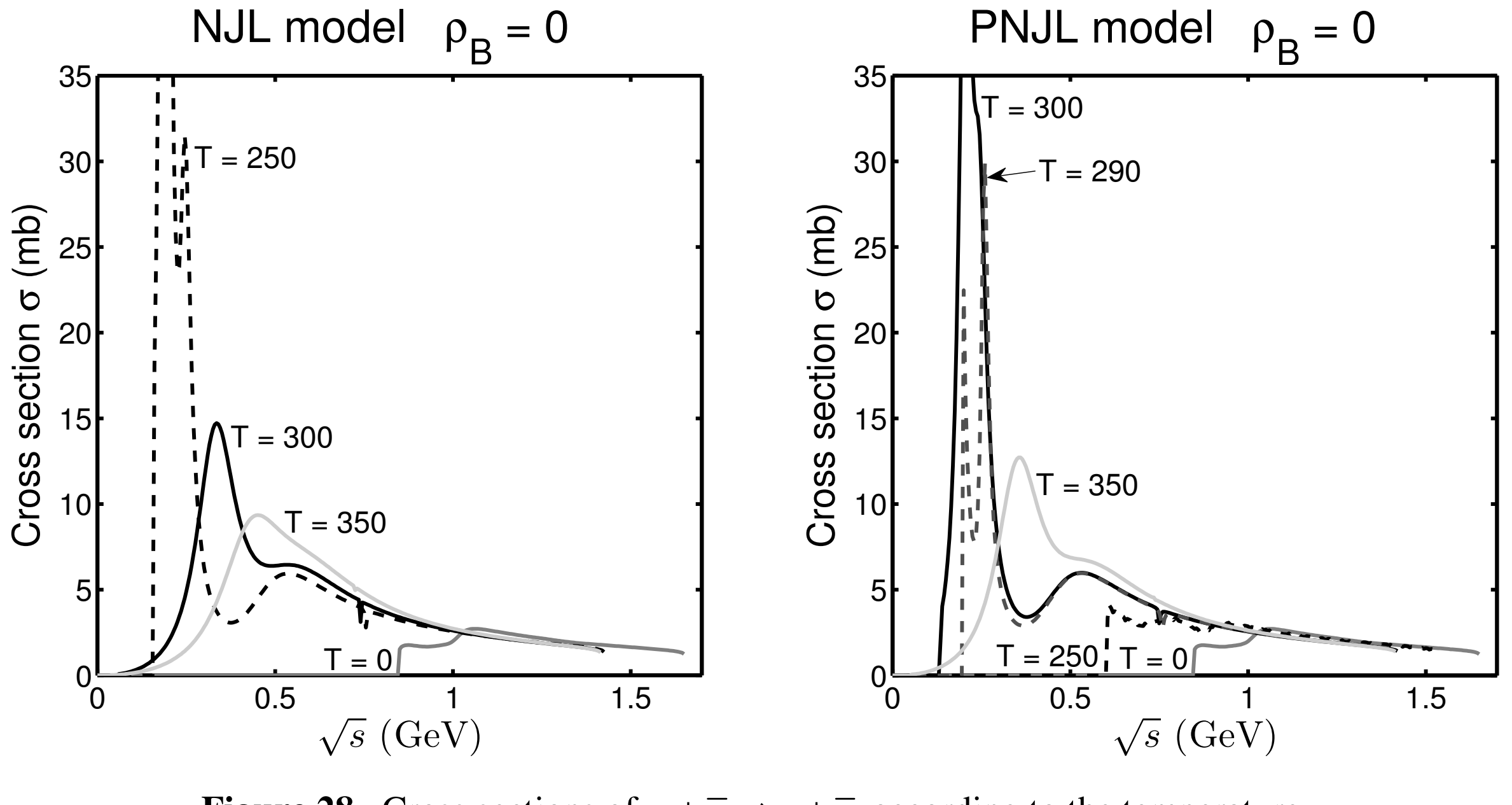

**Figure 28.** Cross sections of $u+\bar{u} \to u+\bar{u}$ according to the temperature.

About the study at non null densities, figure 29, we estimated the evolution of the PNJL cross sections at $T = 250\ \mathrm{MeV}$. Indeed, for this temperature, the effect of the density is rather spectacular. In fact, the behavior of the cross-sections is comparable to the one found with the temperature. The values are weak at low densities. They increase and form structures similar to the ones of the figure 28, and then they decrease again at high densities, i.e. here $\rho_B \approx 5\rho_0$.

A global explanation consists to say that the cross sections become stronger when $\sqrt{s}$ is comparable to the mass of the lightest pseudo-scalar mesons used as propagators in the $s$ channel, i.e. the pion and $\eta$. In that way, we checked that the $s$ channel dominates, especially when the cross sections are strong. In the figure 28, except for the divergence at the threshold, the two other maximums in the NJL curve at $T = 250$ MeV and in the PNJL at $T = 290$ MeV translate the resonances of these quoted mesons when $\sqrt{s}$ is equal to their masses.

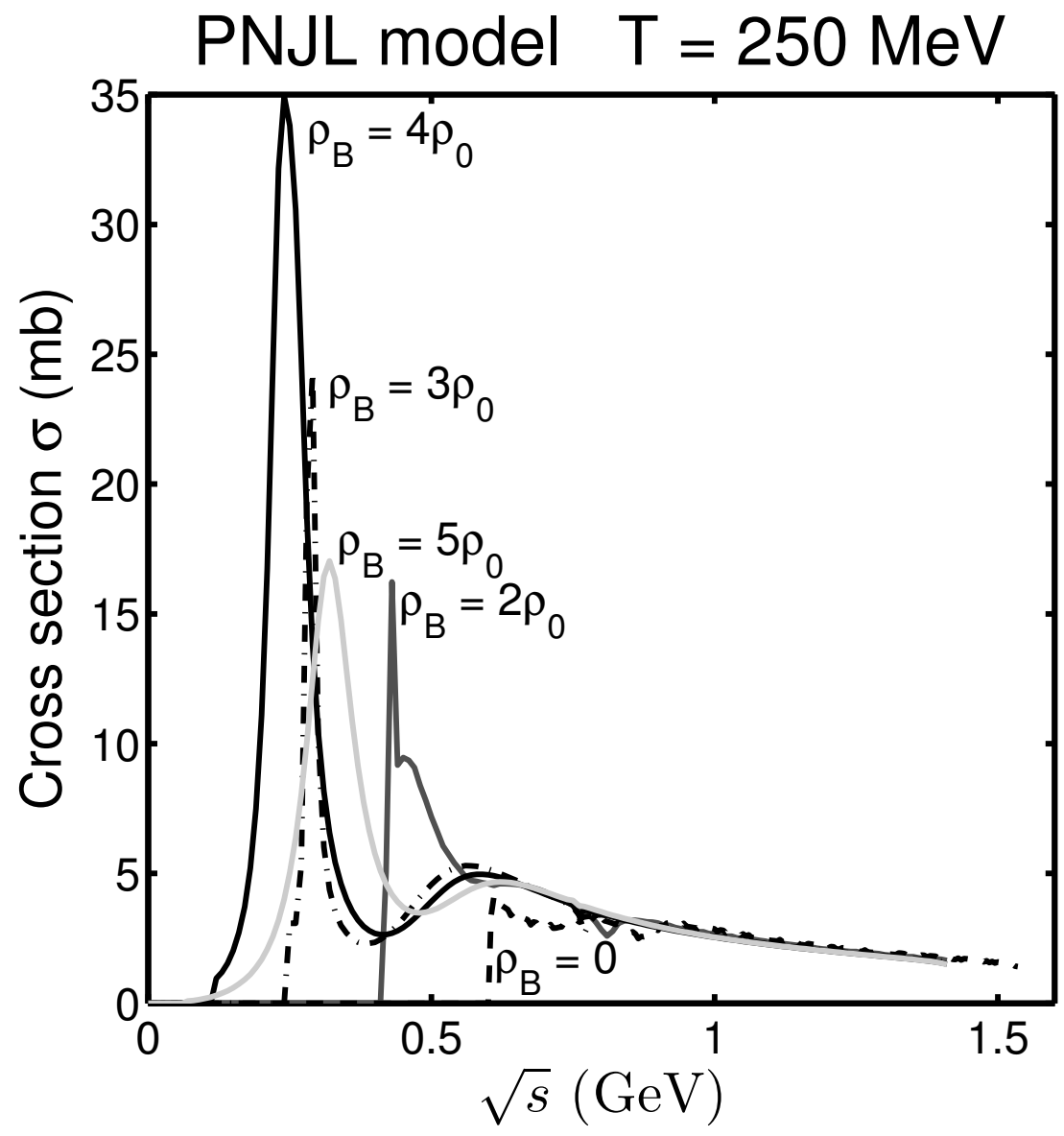

**Figure 29.** Cross sections of $u + \bar{u} \to u + \bar{u}$ according to the baryonic density.

We consider now the differences between $\mathrm{q} + \bar{\mathrm{q}} \to \mathrm{q} + \bar{\mathrm{q}}$ and $\mathrm{q} + \bar{\mathrm{q}} \to \mathrm{M} + \mathrm{M}$. At null density, the maximum cross-sections for the elastic process $u + \bar{u} \to u + \bar{u}$ are found at a temperature close to $T = 300$ MeV, whereas the optimal temperature for $u + \bar{u} \to \pi^+ + \pi^-$ was found in the figure 6 at $T = 280$ MeV. Such a difference could already been noted with the NJL description. In this model, the optimal temperatures are $T = 230$ MeV for $u + \bar{u} \to \pi^+ + \pi^-$ (see figure 2), versus $T = 250$ MeV for $u + \bar{u} \to u + \bar{u}$ [10]. Going back to the PNJL results, at $T = 250$ MeV, the cross section of $u + \bar{u} \to \pi^+ + \pi^-$ explodes for a density of about $2\rho_0$ or $2.5\rho_0$, but lower than $3\rho_0$. The figure 29 indicates that the optimal density for the $u + \bar{u} \to u + \bar{u}$ reaction is higher, because we obtained $4\rho_0$. Extrapolating at the whole $T, \rho_B$ plane, we can guess that these two quoted reactions should not interfere in a notable way, because they might occur at close but different conditions. The elastic scattering is able to intervene for highest temperatures and/or densities. During the cooling of a quarks/antiquarks plasma, a thermalization of the system, ruled by the elastic process, is expected to occur just before a massive mesonization.

## 6.2 Quark–quark scattering

In the framework of the evolution of a quark system, the quark-quark scattering is expected to play an important role, especially in the first moments of its expansion. Inspiring us from [8], we propose to estimate the cross-sections by the *t* and *u* channels represented in the figure 30. Their corresponding matrix elements are written is (20). As with the quark-antiquark scattering, scalar and pseudo-scalar mesons are used as propagators. They are noted, respectively, $i\mathcal{D}^S_{t,u}$ and $i\mathcal{D}^P_{t,u}$. In the same way, $T$ refers also to flavor factor terms, using the same notations as in the reference [8].

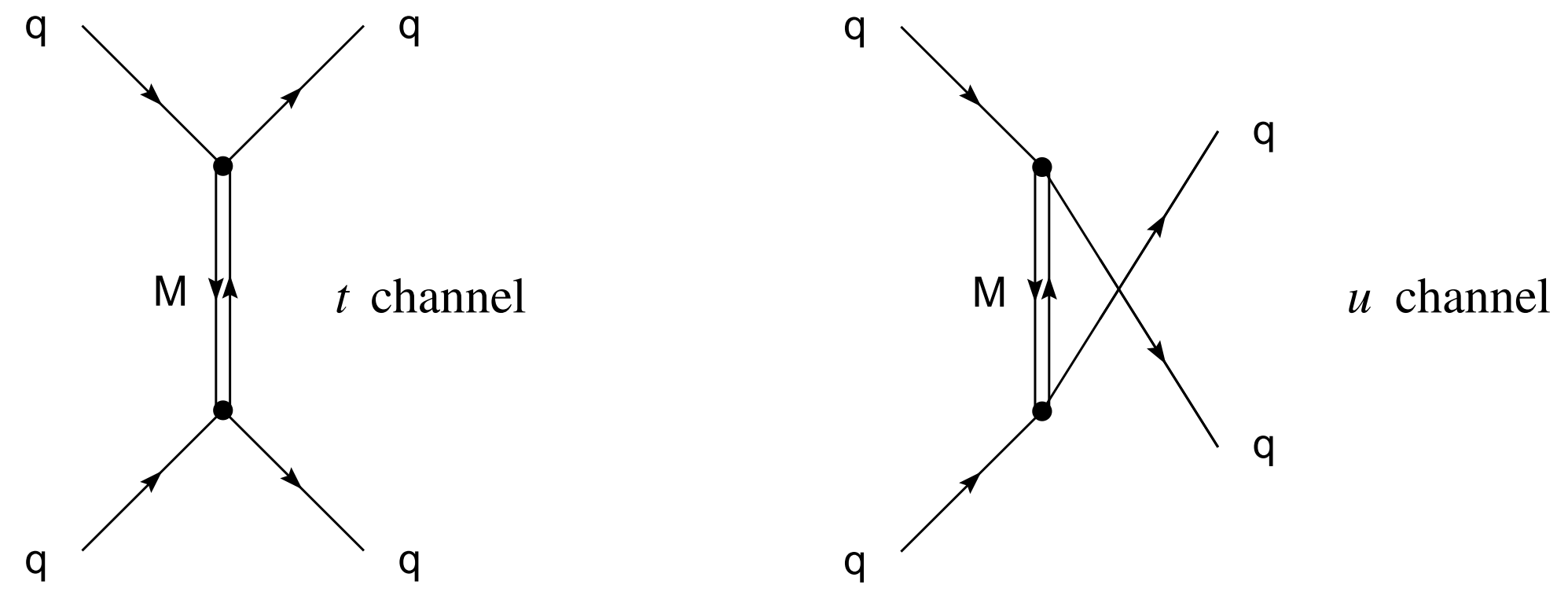

**Figure 30.** $t$ and $u$ channels.

$$
\begin{aligned}
-i\mathcal{M}_t &= \delta_{c_1,c_3}\delta_{c_2,c_4}\ \bar{u}(p_3)\ T\ u(p_1)\ i\mathcal{D}^S_t(p_1-p_3)\ \bar{u}(p_4)\ T\ u(p_2)\\
&\quad+\delta_{c_1,c_3}\delta_{c_2,c_4}\ \bar{u}(p_3)\ i\gamma_5\ T\ u(p_1)\ i\mathcal{D}^P_t(p_1-p_3)\ \bar{u}(p_4)\ i\gamma_5\ T\ u(p_2)\\
-i\mathcal{M}_u &= \delta_{c_1,c_4}\delta_{c_2,c_3}\ \bar{u}(p_4)\ T\ u(p_1)\ i\mathcal{D}^S_u(p_1-p_4)\ \bar{u}(p_3)\ T\ u(p_2)\\
&\quad+\delta_{c_1,c_4}\delta_{c_2,c_3}\ \bar{u}(p_4)\ i\gamma_5\ T\ u(p_1)\ i\mathcal{D}^P_u(p_1-p_4)\ \bar{u}(p_3)\ i\gamma_5\ T\ u(p_2)
\end{aligned}
\quad . \qquad (20)
$$

Our results are proposed in the figures 31 and 32. They concern the study of the elastic scattering of two $u$ quarks. The figure 31 gather the data found at null baryonic density, for the NJL and PNJL models. The NJL ones are in agreement with [8]. As with the other comparisons between the NJL and the PNJL models, it was also observed a shifting according to the temperature. As an example, the NJL curve found at $T = 250$ MeV strongly resembles to the PNJL one found at $T = 300$ MeV. These two values correspond to the temperature for which the cross-sections are optimal, for each model. In fact, focusing on $\sqrt{s}$ values lower than 1 GeV, the cross sections tend to grow when the temperature increases, until these quoted temperatures, but very slowly. After that, they begin to decrease. The propagated mesons are $\pi_0, \eta, \eta', a_0, f_0, f_0'$ for this scattering. The presence of the pseudo-scalar mesons is a possible explanation of the observed behavior. Because of their low masses, they could intervene for rather low $\sqrt{s}$ values. The absolute maximum observed for the reaction $u+u \rightarrow u+u$ would correspond rather well to the critical temperatures of these mesons, i.e. $T = 250$ MeV (NJL) and $T = 300$ MeV (PNJL). And, the disappearance of this maximum beyond these temperatures could be associated with their instability.

According to the baryonic density, the calculations were performed at $T = 250$ MeV, with the PNJL model. The results are shown in the figure 32. They indicate that the cross sections tend to decrease when the baryonic density grows. This evolution is expected to occur at other temperatures. At high temperatures or baryonic densities, the cross sections have a very regular aspect, almost linear according to $\sqrt{s}$. As a whole, the elastic scattering between two quarks seems to be rather weak whatever the applied parameters, at least at low $\sqrt{s}$. Indeed, whatever the used model, the cross sections stay relatively low, below 2.5 mb. Such a values are comparable to the ones of $q+q \to D+M$ and $q+q \to B+\overline{q}$, i.e. the direct source of competition of the quark elastic scattering.

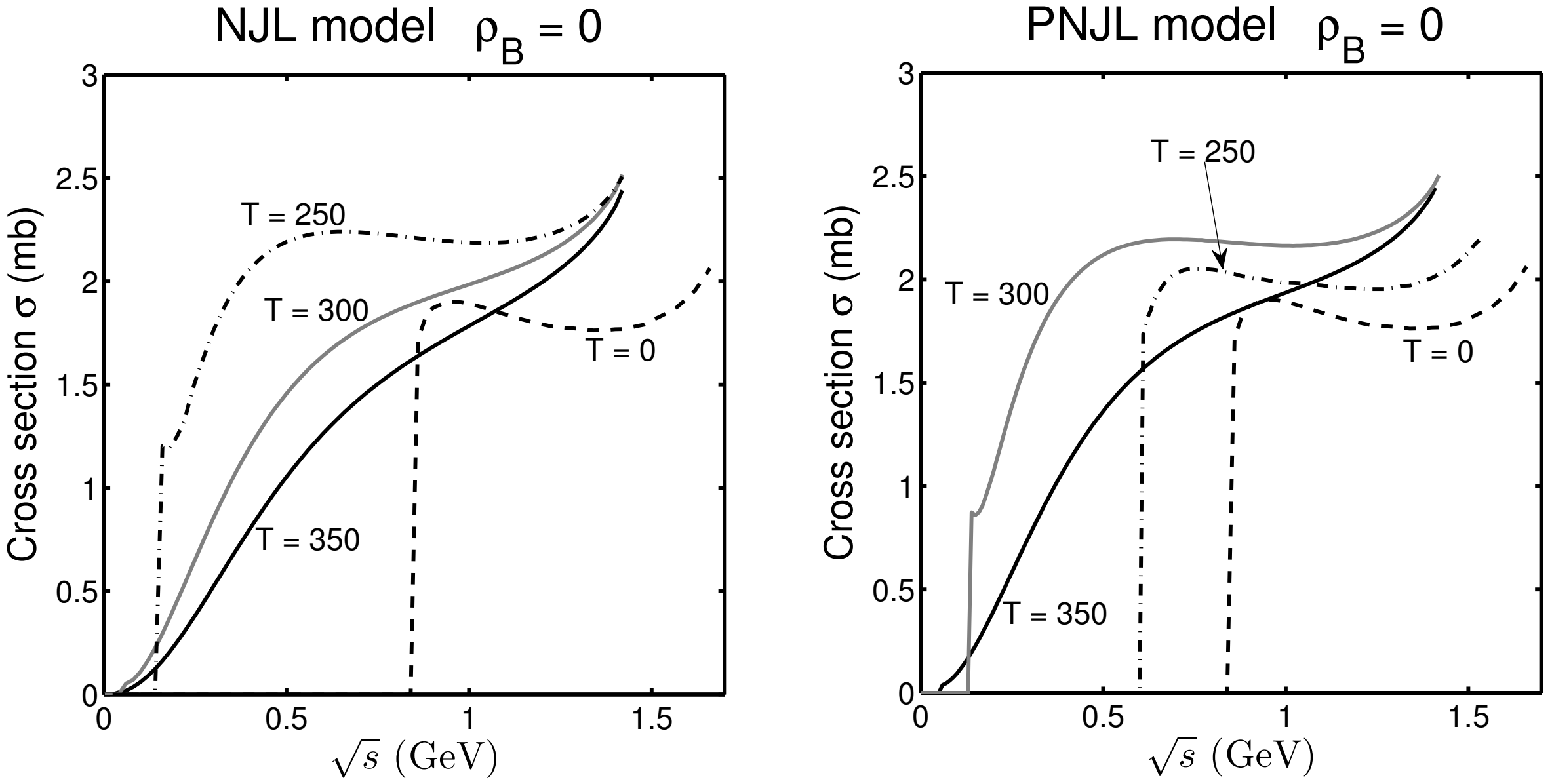

**Figure 31.** Cross sections of $u+u \to u+u$ according to the temperature.

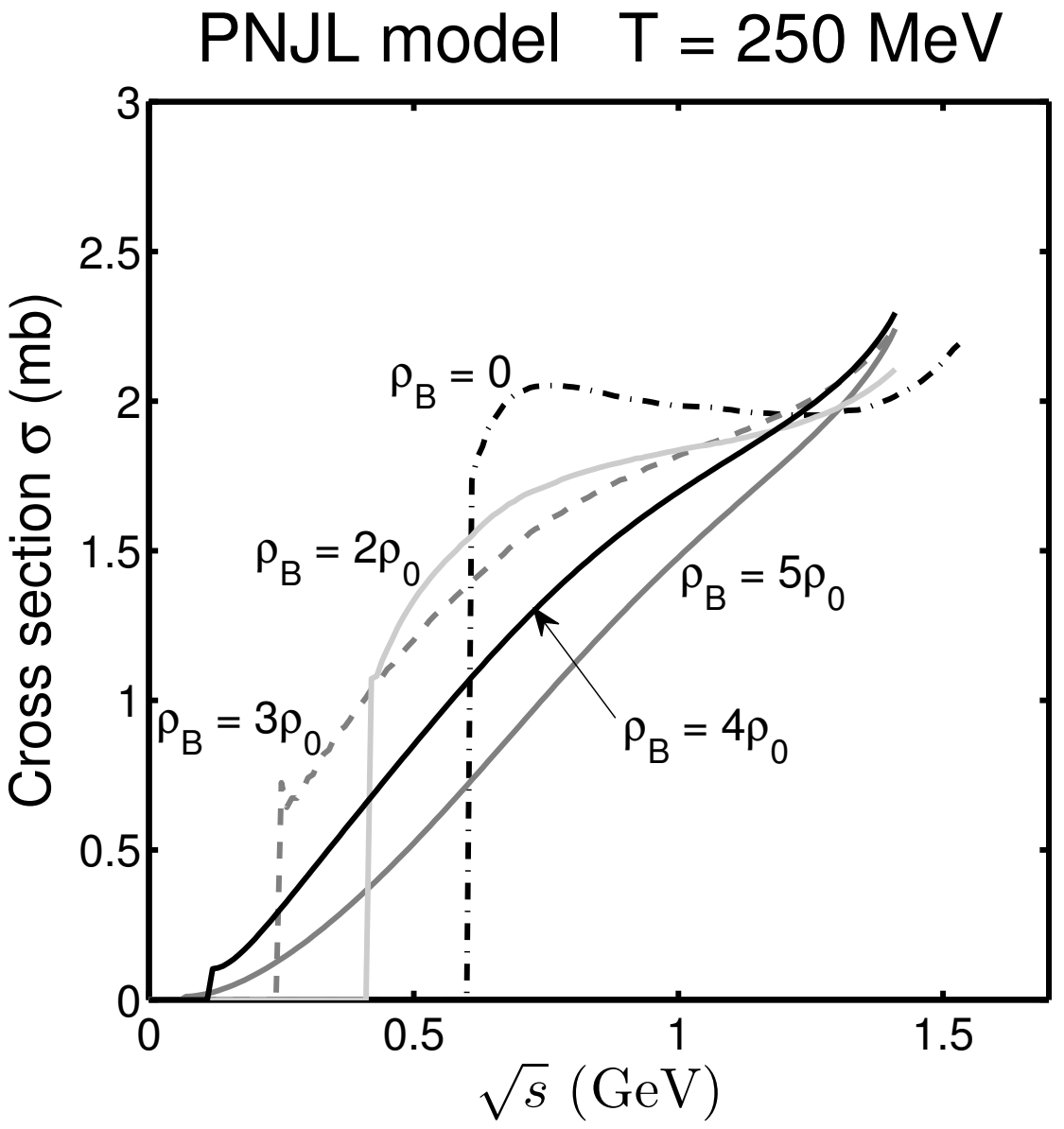

**Figure 32.** Cross sections of $u+u \to u+u$ according to the baryonic density.

Moreover, a possible extension of the modeling of this elastic scattering could be to take into account an $s$ channel. This one could employ scalar or pseudo scalar diquarks as propagators, as in the subsections 4.4 and 5.4. Their contributions on the cross-sections are then to be evaluated. Another extension of this work also concerns the study of antiquark-antiquark scattering, with reactions as $\overline{u}+\overline{u} \to \overline{u}+\overline{u}$. But, at null density, the cross-sections associated with $\overline{q}+\overline{q} \to \overline{q}+\overline{q}$ are identical to the ones of $q+q \to q+q$. At non-null densities, we can use the matter-antimatter symmetry evoked in the previous chapters. We recall that this one consists to say that an antiquark plunged in a medium with positive density will act as a quark plunged in a medium with negative density. But, as observed in the chapter 2, the quarks are only affected by the absolute value of the density. With $\bar{u}+\bar{u}\rightarrow\bar{u}+\bar{u}$, this observation can be extended to all the terms of (20), including the mesons propagators. Finally, the only difference between $\bar{u}+\bar{u}\rightarrow\bar{u}+\bar{u}$ and $u+u\rightarrow u+u$ is finally due only to the blocking factors (4), i.e. $\left(1-f_F(E_3^*-\mu_3)\right)\cdot\left(1-f_F(E_4^*-\mu_4)\right)$. Clearly, the difference between a quark and an antiquark is precisely the sign of their chemical potential. This one is positive for a quark, negative for an antiquark (at positive density). Naturally, it induces different results at the level of the Fermi-Dirac distributions.

# 7. Elastic cross sections involving mesons or diquarks

## 7.1 Meson–quark scattering

In the beginning of this chapter, it was seen the strong cross sections found for the reactions $\mathrm{q}+\bar{\mathrm{q}}\rightarrow\mathrm{M}+\mathrm{M}$, leading to consider a massive mesonization. These mesons are expected to be formed early in the evolution of the physical system. So, their collisions with the remaining quarks or antiquarks are to be considered. To treat the $\mathrm{q}+\mathrm{M}\rightarrow\mathrm{M}+\mathrm{q}$ reactions, we consider the *t* channel described by the figure 33 and by its matrix element (21). A possible evolution of this work could include other channels, but they are not expected to induce important modifications. Moreover, the $\bar{\mathrm{q}}+\mathrm{M}\rightarrow\mathrm{M}+\bar{\mathrm{q}}$ process can be deduced from the $\mathrm{q}+\mathrm{M}\rightarrow\mathrm{M}+\mathrm{q}$ one. More precisely, its Feynman diagram differ from (21) by a replacement of $u(p_1)$ by $\bar{v}(p_1)$, $S_F(p_1-p_3)$ by $S_F(p_3-p_1)$ and $\bar{u}(p_4)$ by $v(p_4)$, recalling the work done in the subsection 4.2. In fact, it can be checked that the squared term $|\mathcal{M}_t|^2$ is equal for these two processes. Only the blocking factors are different at non-null baryonic densities.

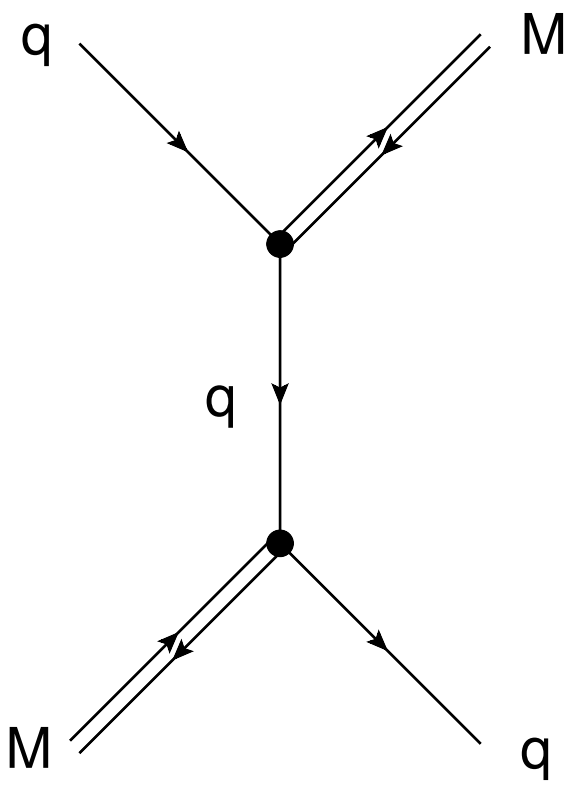

**Figure 33.** *t* channel of $\mathrm{q}+\mathrm{M}\rightarrow\mathrm{M}+\mathrm{q}$.

$$-i\mathcal{M}_t = f_t\ \delta_{c_t,c_1}\ \delta_{c_4,c_t}\ u(p_1)\ i\gamma_5\ ig_1\ iS_F(p_1-p_3)\ i\gamma_5\ ig_2\ \bar{u}(p_4)\ . \qquad (21)$$

Our results consider the elastic scattering between a pion $\pi^0$ and a quark $u$, in the figure 34. In the left and side of the figure, we observe that the cross-sections increase when the temperature is growing. Indeed, the values are less than one millibarn at reduced temperatures. On the other hand, they exceed few millibarns for $T = 280$ MeV. Such values cannot be neglected. According to the baryonic density, right hand side of the figure, we note that the cross-sections tend to decrease when the baryonic density increases, except at the level of the threshold. Clearly, the divergence at the threshold becomes significant.

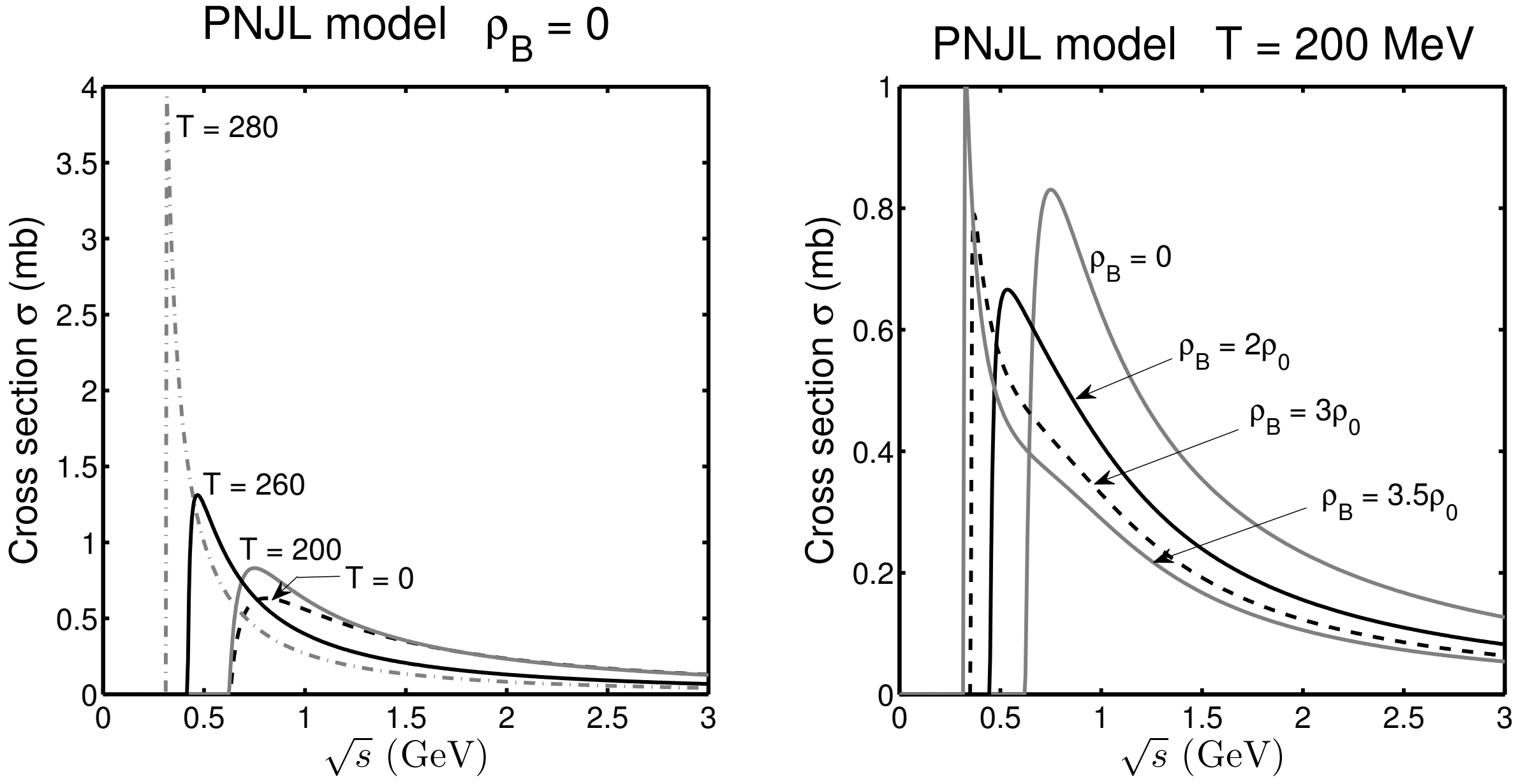

**Figure 34.** Cross sections of the reaction $u + \pi^0 \to u + \pi^0$.

The meson–quark elastic scattering has only one competitor, i.e. $\mathrm{M} + \mathrm{q} \to \overline{\mathrm{q}} + \mathrm{D}$ studied in the subsection 4.2. In fact, the results of figures 11 and 34 are rather comparable according to the temperature, even if the elastic reaction seems to be slightly favored. However, at non-null densities, $\mathrm{M} + \mathrm{q} \to \overline{\mathrm{q}} + \mathrm{D}$ have higher cross-sections, notably via the strong divergences at the threshold observed for $\rho_B \approx 2 - 3\rho_0$.

# 7.2 Diquark–quark scattering

Inspiring us from the subsection 5.5, we propose in the figure 35 a Feynman diagram to model the elastic reactions $\mathrm{q} + \mathrm{D} \to \mathrm{D} + \mathrm{q}$. It could be argued that an $s$ channel involving a baryon as propagator should be included. However, as in the subsection 5.5, we do not include it, notably to obtain similar cross-sections between $\mathrm{q} + \mathrm{D} \to \mathrm{D} + \mathrm{q}$ and $\mathrm{q} + \mathrm{D} \to \mathrm{M} + \mathrm{B}$. The two possible matrix elements are written in the equation (22): we have a $t$ and a $\tilde{t}$ channel. In fact, there is ambiguity at the level of the two vertices in the figure 35, explaining these two channels. The first solution consists to say that only the propagated quark is a charge conjugate one. It corresponds to the matrix element of the $t$ channel. The other possibility consists to say the opposite: the quarks associated with the external lines in the Feynman diagram are in fact charge conjugate anti-quarks. On the other hand, the propagated quark is not affected. It gives the $\tilde{t}$ channel.

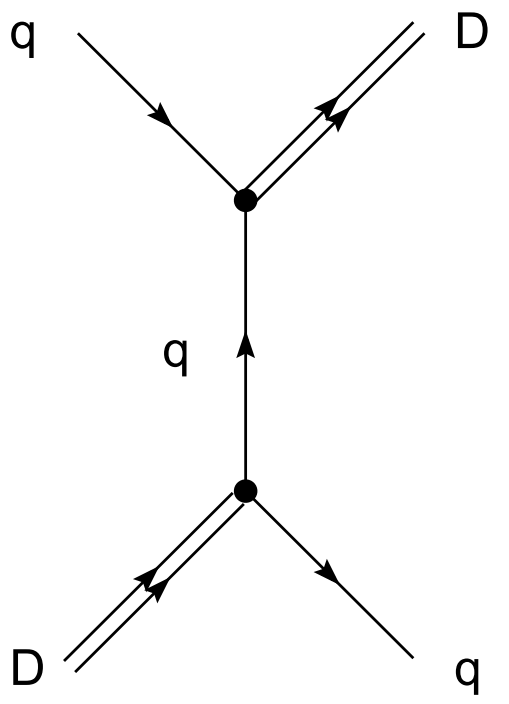

**Figure 35.** $t/\tilde{t}$ channels.

$$\begin{aligned} -i\mathcal{M}_t &= f_t \ \varepsilon^{c_1,c_3,c_t} \ \bar{u}(p_4) \ i\gamma_5 \ ig_1 \ iS_F(p_1 - p_3) \ i\gamma_5 \ \varepsilon^{c_1,c_4,c_t} \ ig_2 \ u(p_1) \\ -i\mathcal{M}_{\tilde{t}} &= f_t \ \varepsilon^{c_1,c_3,c_t} \ v(p_4) \ i\gamma_5 \ ig_1 \ iS_F(p_2 - p_4) \ i\gamma_5 \ \varepsilon^{c_1,c_4,c_t} \ ig_2 \ \bar{v}(p_1) \end{aligned} . \quad (22)$$

As an example, we consider the reaction $u + [ud] \to [ud] + u$. The results are gathered in the figures 36. At finite temperatures and null density, the values are less than 2.5 mb. In addition, more the temperature increases, more the cross-sections are weak. On the other hand, according to the baryonic density, we observe that a divergence at the threshold appears when the baryonic density is close to $2\rho_0$. There, the cross-sections can largely exceed 6 mb. Above this density, the divergence at the threshold is still present, but the cross-sections become rather reduced.

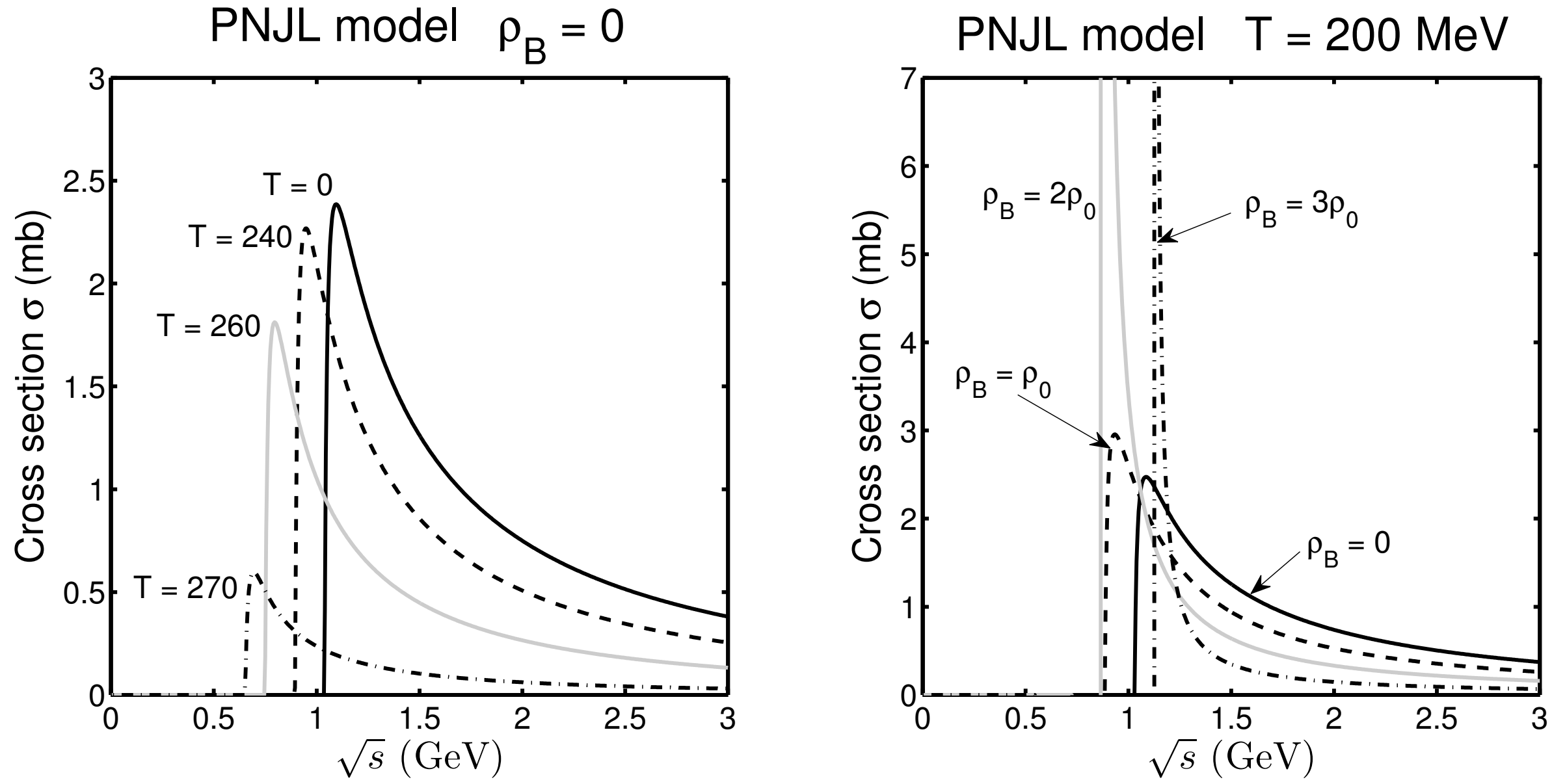

**Figure 36.** Cross sections of the reaction $u + [ud] \to [ud] + u$.

If we compare $q+D \to D+q$ and $q+D \to M+B$ at null density, we note that the elastic process globally has cross sections that are twice compared to the ones of the inelastic one, figures 26 and 36. At finite densities, $q+D \to D+q$ largely dominates $q+D \to M+B$. As a consequence, the elastic process is able to limit the baryon production by the way of the $q+D \to M+B$. The most favored conditions as regards this baryon production seem to be at reduced densities.

# 7.3 Diquark–antiquark scattering

The $\bar{q}+D \to \bar{q}+D$ process is described in our works by the $s$ and $\tilde{s}$ channels, figure 37 and the equation (23). Clearly, because of the two vertices involving a diquark and an antiquark, it leads to two distinct solutions on the choice of the charge conjugate quarks. More precisely, for the $s$ channel, the incoming and the outgoing antiquarks are in fact charge conjugate quarks, leading respectively to the $u(p_1)$ and $\bar{u}(p_3)$ spinors. Concerning the $\tilde{s}$ channels, the quark used as propagator is in fact a charge conjugate antiquark.

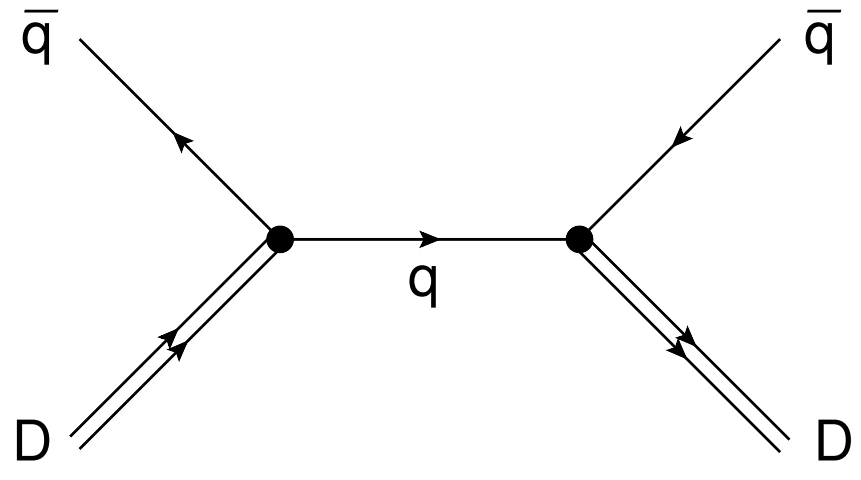

**Figure 37.** $s/\tilde{s}$ channels.

$$\begin{aligned} -i\mathcal{M}_s &= f_s\ \varepsilon^{c_1,c_2,c_s}\ u(p_1)\ i\gamma_5\ ig_1\ iS_F(p_1+p_2)\ i\gamma_5\ \varepsilon^{c_3,c_s,c_s}\ ig_2\ \bar{u}(p_3) \\ -i\mathcal{M}_{\tilde{s}} &= f_s\ \varepsilon^{c_1,c_2,c_s}\ \bar{v}(p_1)\ i\gamma_5\ ig_1\ iS_F(-p_1-p_2)\ i\gamma_5\ \varepsilon^{c_1,c_4,c_t}\ ig_2\ v(p_3) \end{aligned} \quad . \tag{23}$$

The results associated with this process are proposed in the figure 38, in which we consider the $\bar{u}+[ud] \to \bar{u}+[ud]$ reaction. Whatever the temperature and the baryonic density, the cross-sections stay reduced, because they seem unable to exceed 1 mb, except maybe for divergences at the threshold, for densities as $3-4\rho_0$. In fact, these divergences are not found according to the temperature, left hand side of figure 38.

Two other kinds of processes use an antiquark and a diquark as incoming particles. They are $\bar{q}+D \to q+M$ and $\bar{q}+D \to \bar{D}+B$. Firstly, thanks to the figure 10, we conclude that $\bar{q}+D \to \bar{q}+D$ cannot really constitute a real source of competition compared to $\bar{q}+D \to q+M$. Indeed, the cross-sections of this process are largely higher than the ones of $\bar{q}+D \to \bar{q}+D$. On the other hand, the values found with $\bar{q}+D \to \bar{D}+B$, figure 18, are comparable to the ones of our elastic process. In fact, it wants to say that the reactions associated with $\bar{q}+D \to \bar{D}+B$ and $\bar{q}+D \to \bar{q}+D$ are negligible compared to $\bar{q}+D \to q+M$. This remark is true according to the temperature, and according to the baryonic density.

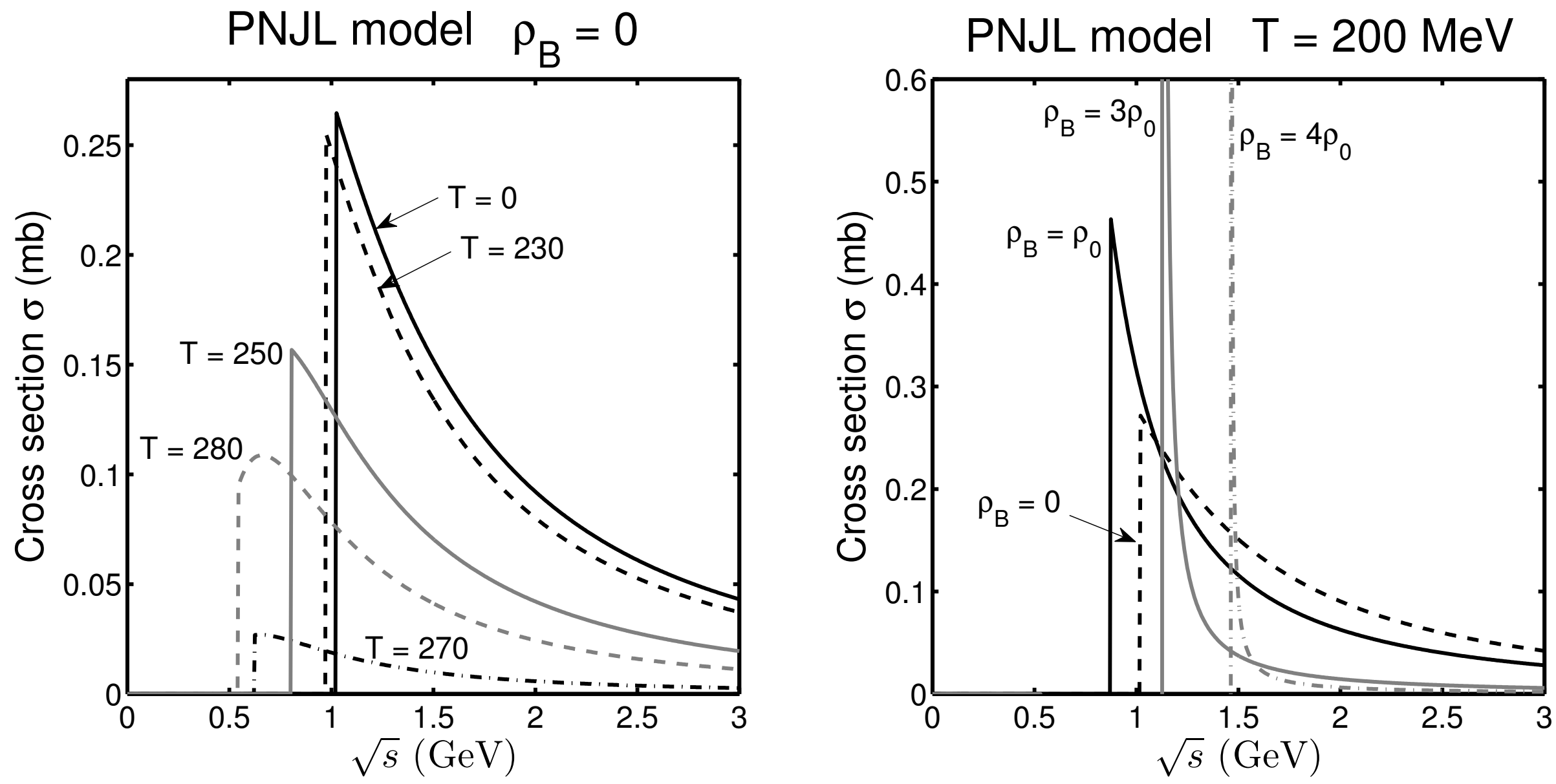

**Figure 38.** Cross sections of the reaction $\bar{u}+[ud]\rightarrow\bar{u}+[ud]$.

# 7.4 Diquark–diquark scattering

Studies of cross-sections between two composite particles are rare in the framework of the (P)NJL models. It is explained by the complications caused by such calculations. However, we can notably quote the works performed in [14–16]. These papers permit us to consider the diquark-diquark scattering, by using Feynman diagrams, figure 39, close to the ones proposed in these references. The triangle structures for the $t$ and $u$ channels correspond to the $\Gamma$ function (9) used to model $\mathrm{q}+\bar{\mathrm{q}}\rightarrow\mathrm{M}+\mathrm{M}$. In the writing of the matrix elements, equation (24), they are designated as $\Gamma_{t,u}^{up}$ or $\Gamma_{t,u}^{down}$. Between them, a meson (P)NJL propagator is used, corresponding to $\mathcal{D}_{t,u}^{S}$. As recalled by the $S$, only scalar mesons are used for this purpose, as in [15]. Moreover, the figure 39 also includes a new channel, the "box channel" [15, 16]. This channel cannot be described with the generic functions $A, B_0$ described in the appendix D. As a consequence, we do not include it in our calculations.

As an example, we consider the elastic scattering between two diquarks $[ud]$. The results are proposed in the figure 40. We remark the relative weakness of the found values. At null density, they never exceed the millibarn, in spite of the presence of divergence at the threshold, whatever the temperature. In fact, the cross-sections tend to decrease when the temperature is growing. At $T=200\text{ MeV}$, the baryonic density acts in a reverse way: the cross-sections are stronger at $\rho_B=\rho_0$ than at null density, especially at the kinematic threshold.

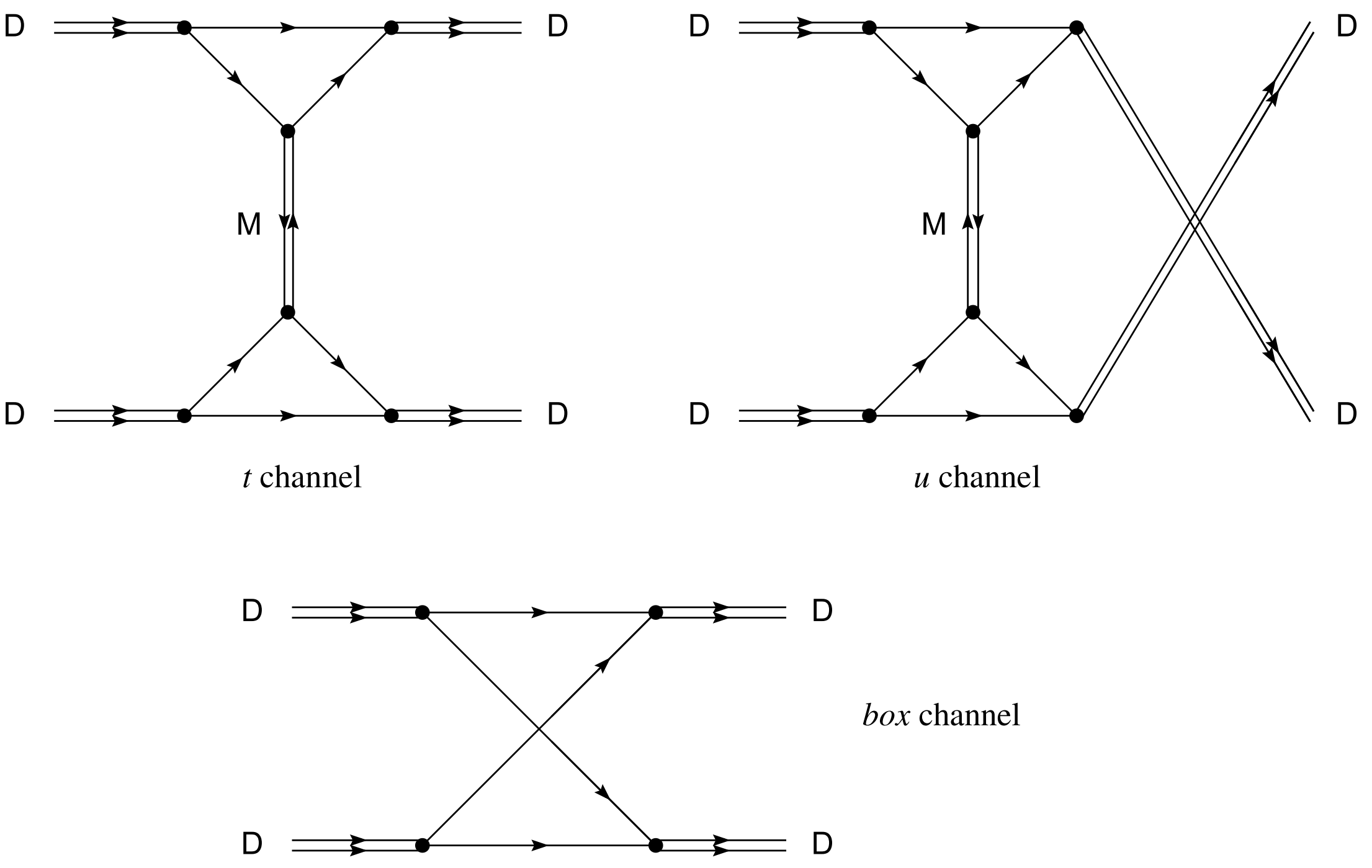

**Figure 39.** Possible channels.

$$\begin{aligned} -i\mathcal{M}_t &= f_t\ ig_1\ ig_3\ \Gamma_t^{up}(p_1-p_3,p_1)\ i\mathcal{D}_t^S(p_1-p_3)\ \Gamma_t^{down}(p_1-p_3,p_4)\ ig_2\ ig_4 \\ -i\mathcal{M}_u &= f_u\ ig_1\ ig_4\ \Gamma_u^{up}(p_1-p_4,p_1)\ i\mathcal{D}_u^S(p_1-p_4)\ \Gamma_u^{down}(p_1-p_4,p_3)\ ig_2\ ig_3 \end{aligned} . \qquad (24)$$

An interest of this study is to compare the results with the ones of the other reactions that use two diquarks as incoming particles, i.e. $D+D \to q+B$. According to the results found in the figure 22, we see that the inelastic reaction largely dominates the elastic one at null density. Indeed, we found cross-sections that can exceed 4 mb for $D+D \to q+B$. At finite densities, the cross-sections of this process decrease when the density is growing. As a consequence, the domination of $D+D \to q+B$ compared to $D+D \to D+D$ is there less manifest. Nevertheless, we conclude that the baryon production by the way of reactions as $D+D \to q+B$ may not be really disturbed by the diquark-diquark scattering. As seen previously, the main limitation of $D+D \to q+B$ is in fact the lack of diquarks…

An extension of the work performed in this subsection is obviously to extend them to elastic cross-sections involving two mesons, a meson and a diquark, or baryon-baryon scattering. In fact, $M+D \to M+D$ cross-sections calculations were done in the framework of our work, leading to results comparable to the ones found in the figure 40, but higher values were obtained. For the meson-meson cross-sections, a complication appears to model the scattering between two pions. Indeed, as explained in [14], such a modeling requires using $\rho$ meson as propagator, in a $s$ channel. These calculations constitute an interesting evolution of our work. Moreover, in the same way, it is possible to imagine the description of the baryon-baryon scattering with the (P)NJL models, via Feynman diagrams as the ones of the figure 39. However, because of the fermionic nature of the baryons, such a study induces complications at the level of the matrix elements calculations. As a consequence, we consider the data and formulas supplied by [26, 27], to estimate the cross-sections of the $B+M \to B+M$ and $B+B \to B+B$ processes.

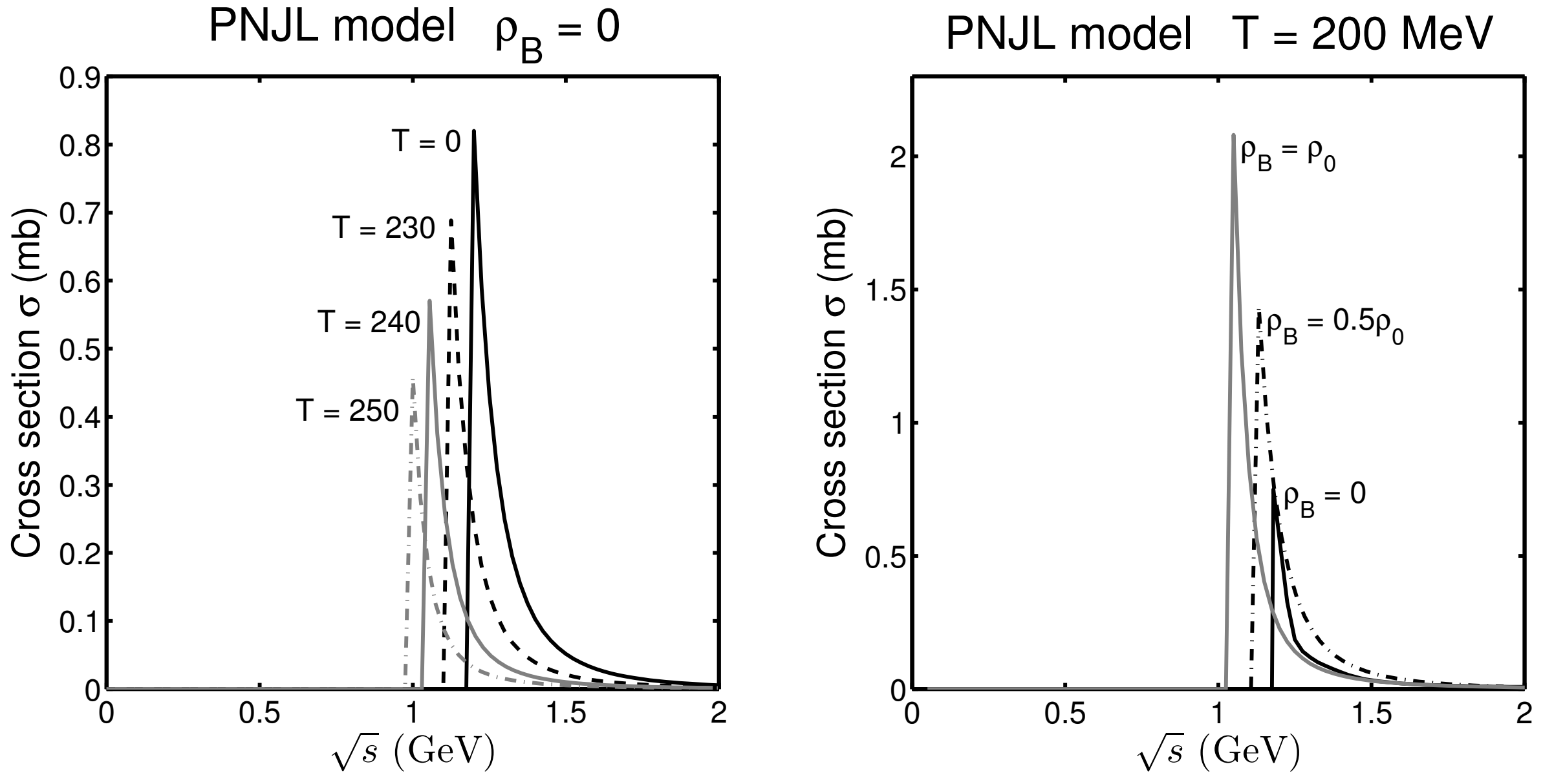

**Figure 40.** Cross sections of $[ud]+[ud]\to[ud]+[ud]$.

# 8. Elastic cross sections involving baryons

## 8.1 Baryon–quark scattering

To model the elastic scattering between a baryon and a quark, we consider the $t$ channel presented in the figure 41 and equation (25). The diquarks used as propagators, via the $\mathcal{D}_t$ term, are scalar ones. In fact, we concede that our description may be completed by more complex channels, notably by box channels.

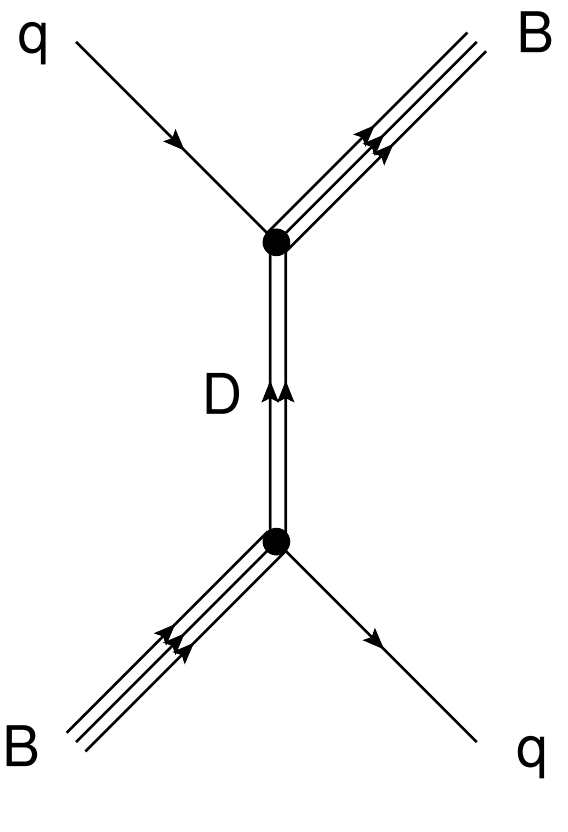

**Figure 41.** $t$ channel.

$$-i\mathcal{M}_t = f_t\ \bar{u}(p_4)u(p_2)\ i\mathcal{D}_t(p_3-p_1)\ \bar{u}(p_3)u(p_1)\quad . \tag{25}$$

Concerning our numerical results, we chose to study $p+u \to p+u$, i.e. the elastic scattering between a proton and a $u$ quark. The associated results are presented in the figure 42. As a whole, the found values are not negligible, because they can exceed 10 mb at reduced temperatures and densities. In fact, these two parameters act in the same way: when the temperature or the baryonic density increase, a diminution of the cross-sections is observed. At the limit of stability of the proton according to the temperature, the cross-sections can still reach 4 mb. According to the baryonic density, the diminution is stronger. At $\rho_B = 2\rho_0$ and $T = 200$ MeV, the values are close to 2 mb.

Moreover, the other process involving a baryon and a quark as incoming particles is $\mathrm{q+B \to D+D}$. Clearly, the associated reactions are the reverse ones of $\mathrm{D+D \to q+B}$, studied subsection 5.3. Even if we do not present the results associated with these reverse reactions, they are stronger than the ones of $\mathrm{D+D \to q+B}$. Indeed, the total mass of two diquarks is lower than the mass sum of a baryon and a quark, thus divergence at the threshold are not observed for $\mathrm{D+D \to q+B}$. But, the divergences are present for $\mathrm{q+B \to D+D}$, allowing this process to reach high cross-sections. In practice, at the level of this divergence, the values frequently exceed 20 mb. As a consequence, we conclude that $\mathrm{q+B \to q+B}$ and $\mathrm{q+B \to D+D}$ have rather equivalent cross-sections, even if the inelastic process dominates, thanks to its divergence at the threshold. Anyway, an interesting aspect of the elastic process is its ability to disturb $\mathrm{q+B \to D+D}$, i.e. to reduce an unwanted baryon's destruction by the quarks.

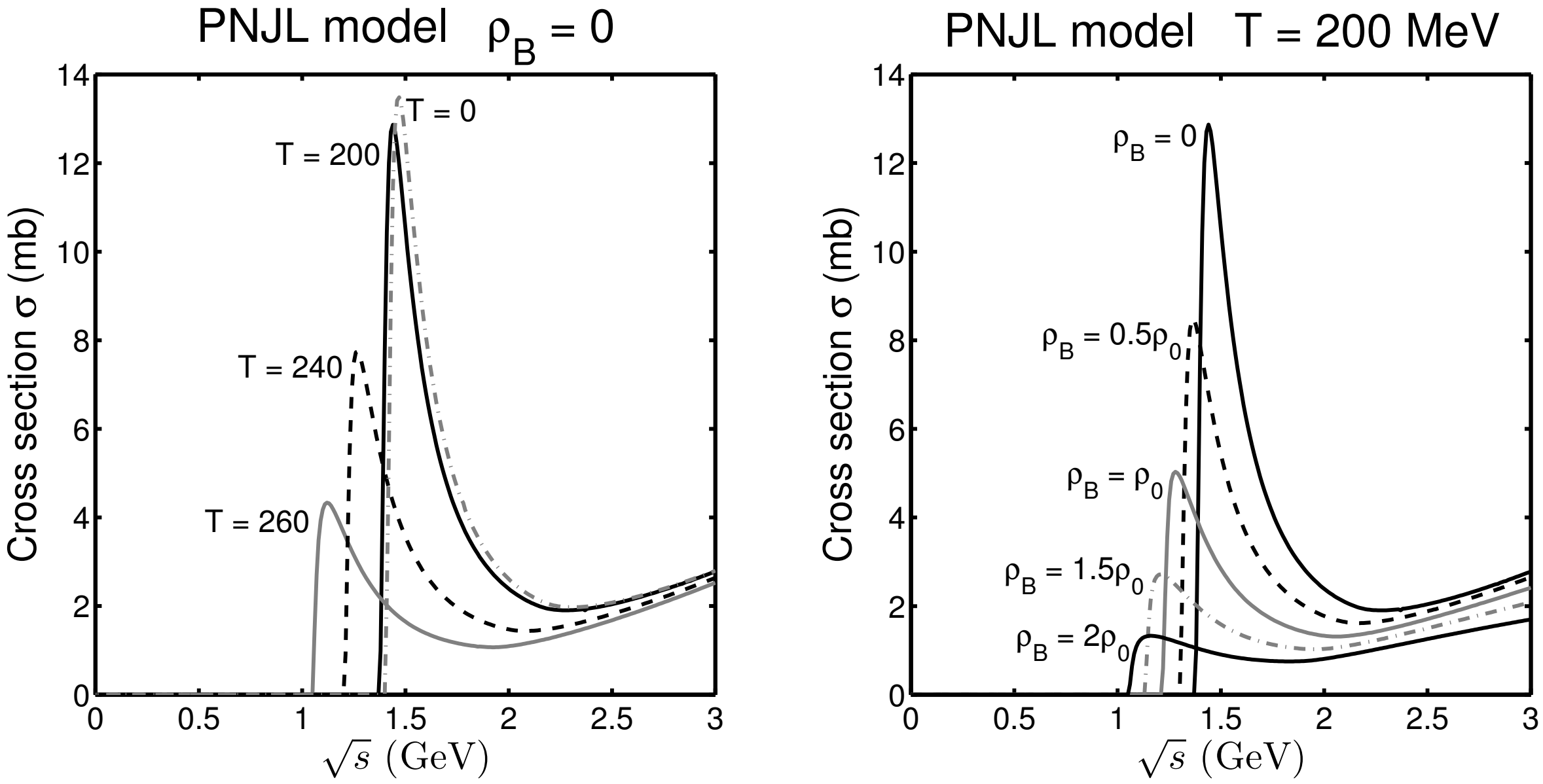

**Figure 42.** Cross sections of the reaction $u + p \to u + p$.

Moreover, the study performed in this subsection can easily be adapted to the reactions $\mathrm{\bar{q}+\bar{B} \to \bar{q}+\bar{B}}$, using the matter-antimatter symmetry, as explained upstream for other reactions. However, it is true that anti-baryons are not expected in a physical system in which the matter largely dominates the antimatter.

# 8.2 Baryon–antiquark scattering

A possible modeling of the elastic collisions between a baryon and an antiquark includes the $s$ channel, presented in the figure 43 and equation (26). In this channel, scalar diquarks are used as propagators. They correspond to the $\mathcal{D}_s$ term.

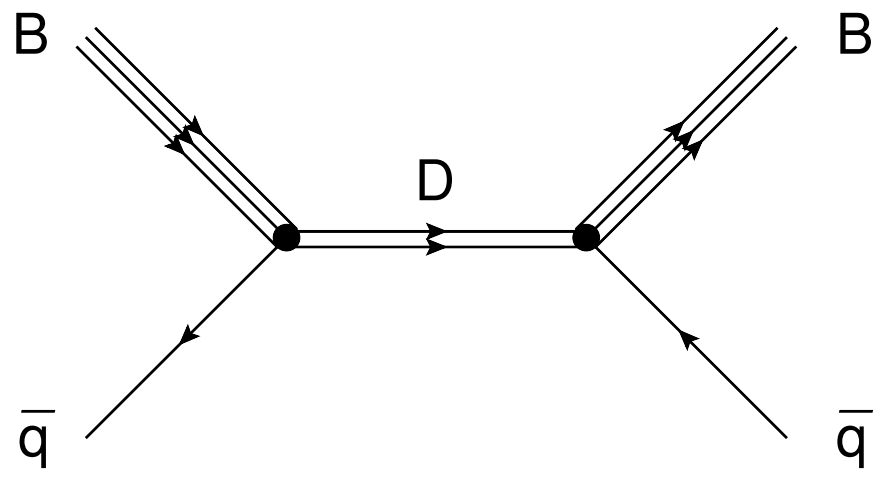

**Figure 43.** $s$ channel.

$$-i\mathcal{M}_s = f_s\ \bar{v}(p_2)u(p_1)\ i\mathcal{D}_s(p_1+p_2)\ \bar{u}(p_3)v(p_4)\ . \tag{26}$$

As an example, we estimated the cross-sections of the $p+\bar{u} \to p+\bar{u}$ scattering, and gathered our results in the figure 44. The obtained curves have a very different behavior compared to the ones previously described. Clearly, they have no similarities with the cross-sections found for example with $\mathrm{q+B \to q+B}$. The results show that $p+\bar{u} \to p+\bar{u}$ is not able to intervene before $\sqrt{s} < 1{,}6$ GeV. After this value, the cross-sections strongly increase, but they stay lower than one millibarn for $\sqrt{s} = 2$ GeV, whatever the temperature or the baryonic density.

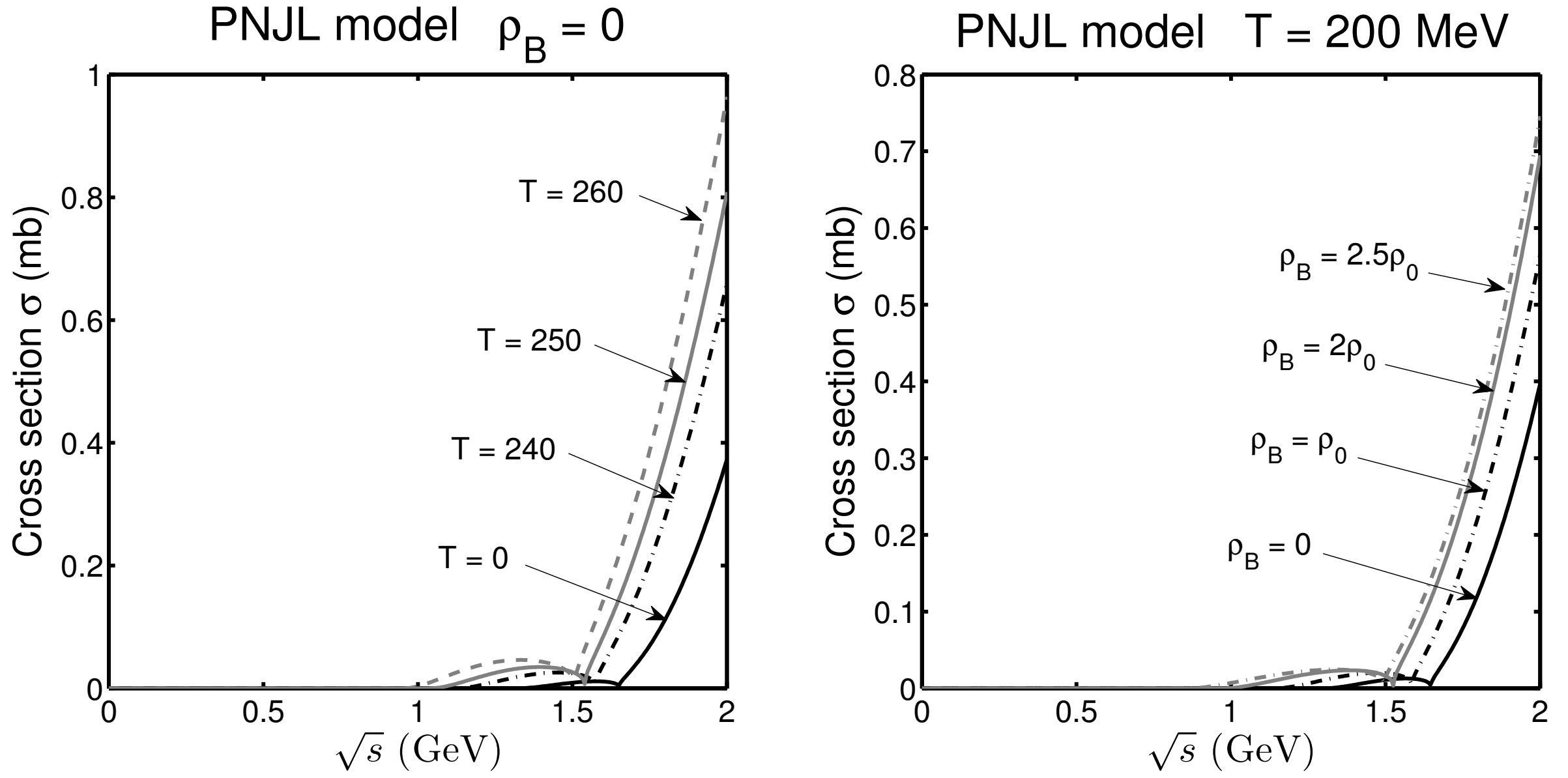

**Figure 44.** Cross sections of the reaction $p+\bar{u} \to p+\bar{u}$ .

Moreover, two other processes consider a baryon and an antiquark as incoming particles: they are $\mathrm{\bar{q}+B \to q+q}$ and $\mathrm{\bar{q}}+B \to \mathrm{M+D}$. They corresponds to the reverse processes of baryonization reactions seen in the subsection 5.2 and 5.4. Also, $\mathrm{\bar{q}+B \to q+q}$ and $\overline{q}+B \to M+D$ are *exothermic* reactions (incoming particles heavier than outgoing ones). So, they have divergence at the threshold, and they present stronger cross-sections than the ones of $M+D \to \overline{q}+B$ and $q+q \to \overline{q}+B$. It wants to say that $\overline{q}+B \to \overline{q}+B$ is highly negligible for $\sqrt{s} < 2\ \text{GeV}$. This elastic process is thus unable to avoid the destruction of baryons via reactions as $\overline{q}+B \to q+q$ or $\overline{q}+B \to M+D$. As remarked upstream in this chapter, it suggests a scenario that imagines that the baryon production may be effective only if the population of the antiquarks is reduced enough.

# 8.3 Baryon–diquark scattering

The last process described in this chapter concerns the elastic reactions between a baryon and a diquark. We propose to study this process with a *t* channel, figure 45. The associated matrix element is written equation (27). In our modeling, a quark is used as propagator. We are particularly aware that this description can be completed by the inclusion of more complex channels, as a box channel in which the diquark and the baryon exchange one quark, inspiring us with the one of the figure 39.

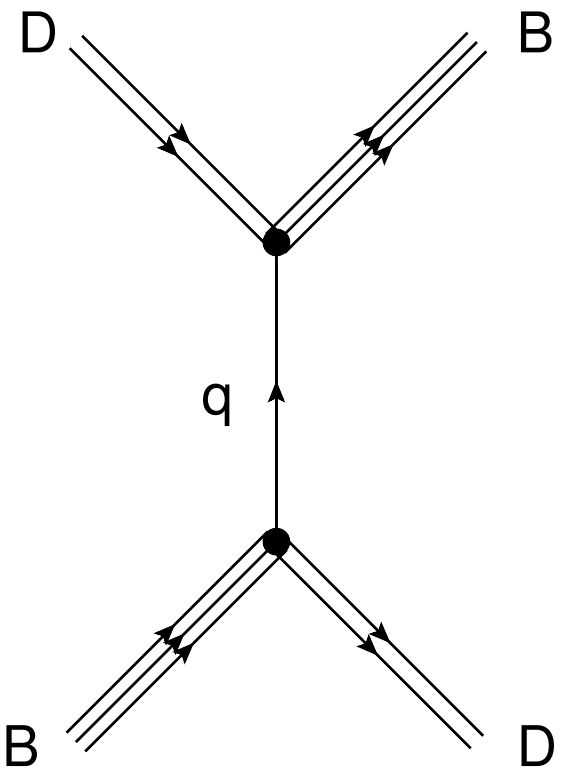

**Figure 45.** *t* channel.

$$-i\mathcal{M}_t = f_t\ u(p_2)\ ig_2\ iS_F(p_3 - p_1)\ ig_1\ \bar{u}(p_3) \quad . \tag{27}$$

As an example, the reaction $p+[ud] \to p+[ud]$ is considered. Our results are exhibited in the figure 46. Firstly, we observe that the obtained cross-sections can exceed few millibarns, i.e. more than 6 mb at some occasions. On the left hand side of the figure, we note that the values slightly increase when the temperature is growing, until $T = 230\ \text{MeV}$. Then, the cross-sections decrease, until the limit of stability of the baryon, for $T \approx 260\ \text{MeV}$. According to the baryonic density, right hand side of the figure, the peak at the threshold becomes important when the density is strong enough. But, in the same time, the cross-sections become reduced far from the threshold, i.e. when $\sqrt{s} > 2\ \text{GeV}$. As a whole, thanks to the found values, the reaction cannot be considered as negligible. Furthermore, $D+B \to B+D$ is the only process in which we have a baryon and a diquark as incoming particles. Thus, there is no competition for such a process. As a consequence, it is probable that $D+B \to B+D$ occurs during the cooling of the quarks/antiquarks plasma.

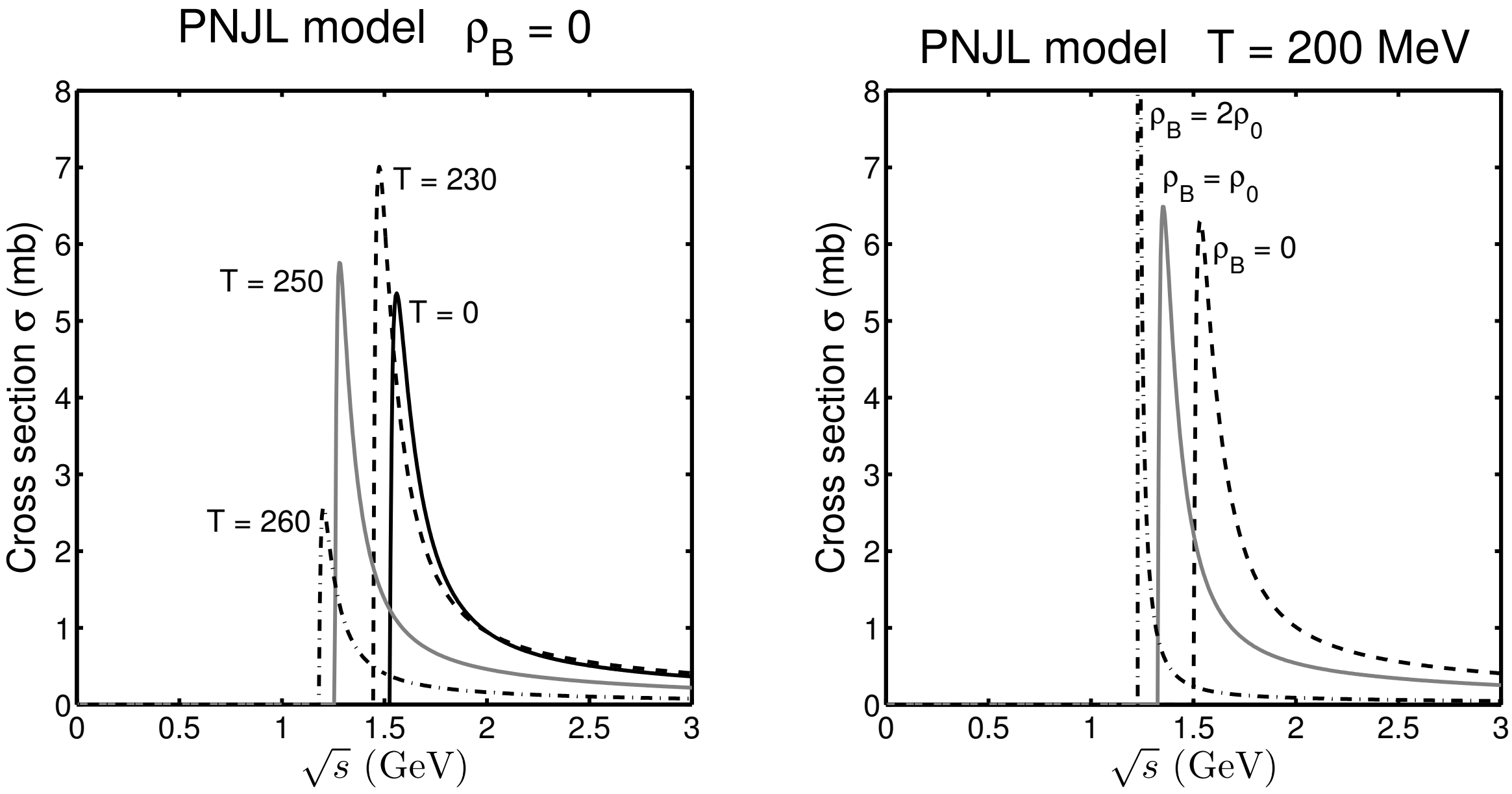

**Figure 46.** Cross sections of the reaction $[ud]+p \rightarrow [ud]+p$.

# 9. Conclusion

In this chapter, we studied some cross-sections involving particles modeled in the framework of our work, i.e. quarks/antiquarks, pseudo scalar mesons, scalar diquarks and octet baryons. More precisely, we restrained our calculations to the reactions involving light particles, i.e. we did not include the strangeness. Indeed, particles as $q$ quarks, pions, $[ud]$ and the nucleons allow obtaining the strongest cross sections. In fact, our calculations are fully applicable to strange particles, but the presentation of the associated results would have overloaded this chapter.

A first objective was to consider again some reactions already treated in the NJL literature. Firstly, it concerned the process $\mathrm{q}+\overline{\mathrm{q}} \rightarrow \mathrm{M}+\mathrm{M}$ [7]. As noted in [10], we confirmed that the divergence at the kinematic threshold leads to very strong cross-sections for these reactions, i.e. greater than $100\ \mathrm{mb}$. This behavior was also confirmed according to the baryonic density. Furthermore, we also investigated the effects of the inclusion of the Polyakov loop on the obtained results. As with the masses of the particles, we observed that the PNJL curves are shifted towards higher temperatures compared to the NJL ones. Quantitatively, the optimal cross sections for the $u+\overline{u} \rightarrow \pi^{+}+\pi^{-}$ reaction were found at $T=230$ MeV (for $\rho_B=0$) with NJL, whereas it is of about $T=280$ MeV in a PNJL description. As a whole, we saw that these optimal conditions, according to the temperature or the baryonic density, were found just before the produced pions become heavier than the $q/\overline{q}$ pairs that compose them, i.e. just before their limit of stability. Moreover, it is interesting to note that $T=280$ MeV is very comparable to the value of the critical deconfinement temperature in a pure gauge theory, expected to be $T_0=270$ MeV [28], chapter 2.

In the same way, we also proceeded to similar works as regards the elastic reactions involving quarks/antiquarks, i.e. $\mathrm{q+q \rightarrow q+q}$ and $\mathrm{q+\overline{q} \rightarrow q+\overline{q}}$, following the method detailed in [8]. We confirmed the results of this reference according to the temperature, but we also extended them at finite densities. As with $\mathrm{q+\overline{q} \rightarrow M+M}$, we also observed the effects of the Polyakov loop in these results. Clearly, for $u+u \rightarrow u+u$ and $u+\overline{u} \rightarrow u+\overline{u}$, we found optimal cross-sections for $T = 250$ MeV in the NJL model, versus $T = 300$ MeV with the PNJL description. By a comparison of these optimal temperatures with the ones of $u+\overline{u} \rightarrow \pi^{+}+\pi^{-}$, it was imagined that, during the cooling of the quarks/antiquarks system, $\mathrm{q+\overline{q} \rightarrow q+\overline{q}}$ should strongly act notably to allow a thermalization of the system, before an overwhelming mesonization via $\mathrm{q+\overline{q} \rightarrow M+M}$. In other words, $\mathrm{q+\overline{q} \rightarrow q+\overline{q}}$ and $\mathrm{q+\overline{q} \rightarrow M+M}$ are not expected to intervene in the same time.

Moreover, other works performed in this chapter concerned the modeling of reactions involving diquarks, and then reactions involving baryons. These calculations were performed in the framework of the PNJL model, according to the temperature or to the baryonic density. Inspiring us by the processes proposed in [11, 12], and adding new ones as $\mathrm{D+D \rightarrow B+q}$, the objective was thus to see the dominant reactions that can allow the formation of baryons, and the role played by the diquarks. As a general tendency, the cross-sections was found as rather low compared to the ones of $\mathrm{q+\overline{q} \rightarrow M+M}$. They have values of few millibarns. According to our results, a reduced diquark production is expected, allowed only if the mesonization processes had already consumed the great majority of the antiquarks initially present in the system. It appeared that the dominant process that can allow this creation of diquarks is $\mathrm{q+q \rightarrow D+M}$. The optimal conditions of this diquark production seems to be at moderate densities, i.e. about two times the nuclear density $\rho_0$, and below the optimal mesonization temperature.

In fact, our results lead us to develop a scenario describing the cooling of the quarks/antiquarks plasma. After a domination of $\mathrm{q+\overline{q} \rightarrow q+\overline{q}}$ at high temperatures ($\approx 300$ MeV in the PNJL model), and after a strong mesonization via $\mathrm{q+\overline{q} \rightarrow M+M}$ at temperatures close to $T = 280$ MeV, the baryonization is supposed to intervene, notably when the temperature is reduced enough. Furthermore, an overlap between the mesonization and baryonization is expected, in order to “suppress” all the antiquarks, formed e.g. via $\mathrm{q+q \rightarrow B+\overline{q}}$, that could destroy the produced baryons. With this hypothesis, the relative weakness of the baryonization’s cross sections is not a real limitation of our modeling. Indeed, except with elastic scatterings, the remaining quarks might only react themselves by reactions that could form diquarks and/or baryons. In the framework of the production of baryons, we saw that $\mathrm{q+q \rightarrow B+\overline{q}}$, $\mathrm{q+D \rightarrow M+B}$, $\mathrm{D+D \rightarrow B+q}$ are particularly relevant. Obviously, the former one is strongly linked to the diquark production.

We saw that the reactions described in this part required interpreting the Feynman diagrams correctly, by identifying the quarks/antiquarks that are in fact charge conjugate ones. It sometimes leads to a subdivision of the involved channels, with e.g. $t/\tilde{t}$ ones. Also, it leads to treat uncommon calculations, because they imply spinors with different momenta. It required establishing a method, described in the appendix B, to perform such calculations. Obviously, the method does not seem to be limited to the NJL/PNJL models, but may be adapted to other cases involving a consequent number of four-vectors. In addition, the results found with this method finally stay rather close to the ones obtained with “standard” spinors calculations, as the ones described in [7, 8]. This observation was far from being trivial before managing the calculations. In fact, these similarities can be explained by the method developed in [29]. More precisely, the approach explained in this paper proposes transformations that allow rewriting matrix elements, in order to greatly simplify the calculations. More precisely, as showed in the appendix B, the $\mathcal{M}_u \cdot \mathcal{M}_t^*$ term of $\mathrm{q}+\mathrm{q} \to \mathrm{D}+\mathrm{M}$ (subsection 4.4) becomes equivalent after transformation to the one of the process $\mathrm{q}+\overline{\mathrm{q}} \to \mathrm{M}+\mathrm{M}$. Clearly, this alternative approach confirms our calculations. Moreover, it allows validating the signs of the mixed terms, designated as RSIF (Relative Sign of Interfering Feynman graphs) in the reference.

Another aspect of this chapter concerns the modeling of elastic reactions involving mesons, diquarks and baryons. Except with the meson-meson scattering [14–16], the (P)NJL literature is poor as regards such reactions. Some of the treated processes presented interesting cross-sections, as $\mathrm{M}+\mathrm{q} \to \mathrm{M}+\mathrm{q}$, $\mathrm{q}+\mathrm{D} \to \mathrm{q}+\mathrm{D}$ and $\mathrm{B}+\mathrm{q} \to \mathrm{B}+\mathrm{q}$. It was shown that these reactions are able to constitute a source of competition compared to the inelastic reactions involving the same incoming particles. For example, we can mention the rivalry between $\mathrm{q}+\mathrm{D} \to \mathrm{q}+\mathrm{D}$ and $\mathrm{q}+\mathrm{D} \to \mathrm{M}+\mathrm{B}$.

Moreover, we saw that the description of some of these processes may be completed by the inclusion of new channels, as the box channels. This aspect may constitute a possible future development of our work. In fact, other evolutions can be evoked, as a complete modeling of the meson-meson scattering [14–16] in the framework of the (P)NJL models, especially to describe the scattering between two pions. In the continuity of this idea, as suggested in [11], we may also consider the treatment of elastic reactions as $\mathrm{B}+\mathrm{B} \to \mathrm{B}+\mathrm{B}$ or $\mathrm{B}+\mathrm{M} \to \mathrm{B}+\mathrm{M}$. Other possibility of evolution may also concern the study of the influence of the color-superconductivity [24, 25] on the cross-sections, at low temperatures and finite densities.

# Acknowledgment

We wish to thank the authors of the *Jaxodraw* software [30, 31], notably used in this chapter to realize our Feynman diagrams.

# Chapter 7

# Time evolution of a system

This chapter was published in *Phys. Rev.* C **89** 065204.

## 1. Introduction

In the previous chapters, we saw the possibilities offered by the (P)NJL models, in order to model quarks, mesons, diquarks and baryons. This work allows estimating the mass of each of the quoted particles according to the temperature $T$ and the baryonic density $\rho_B$. In the chapter 6, we also studied the cross-sections associated with reactions involving these particles, as a function of $\sqrt{s}$, but also taking account the influence of $T$ and $\rho_B$. Therefore, at this stage of the work, we have the required tools to focus on the last part of this thesis, i.e. to model the evolution of the system according to the time. It concerns there the cooling of a quark-antiquark plasma, and its hadronization.

In the literature devoted to nuclear or particles physics, various dynamical models were proposed. Firstly, we can quote hydrodynamics and the QMD/UrQMD models. Relativistic hydrodynamics [1–9] was proposed in 1953 by Landau to describe strongly interacting systems. In this approach, the matter is modeled as a continuous fluid. This method is applicable if local thermodynamic equilibrium is satisfied: it requires low variations of the temperature and the pressure, according to the distance and the time. In fact, the Bjorken scenario [10], seen in the chapter 1, plans the existence of a hydrodynamic evolution, since the (rapid) thermalization of the QGP phase, until the freeze out of the hadronic phase. Experimental results seem to confirm this hypothesis, validating that this evolution seems to satisfy the local equilibrium criterion. In practice, relativistic hydrodynamics can be used to model the evolution of the QGP phase and the one of the hadronic phase. The UrQMD (Ultra-Relativistic Quantum Molecular Dynamics) [11, 12] is a microscopic approach. It wants to say that it models the evolution of particles, in highly relativistic regime. More precisely, this method considers covariant transport equations and treats interactions between particles via cross-sections. UrQMD can be employed in practice to describe the hadronic phase, e.g. [13, 14]. In some UrQMD versions, Lund strings [15] are considered to describe high energies collisions between hadrons. They allow the formation of new particles (mesons), while respecting the confinement of the quarks inside the hadrons. Recent approaches can also be quoted, as the PHSD (Parton Hadron String Dynamics) [16–19], the BAMPS (Boltzmann Approach for Multi-Parton Scattering) [20], etc.

These works are often considered as a basis in the framework of such studies. For example, the collision criterion proposed in [11, 12] is found in various works, as the relativistic equations of motion presented in [11]. These aspects, included in an NJL model, already gave some results available in the literature, as [21–24]. As a whole, the temperature appears as a crucial parameter that rules the cooling of the quark system. This cooling can be treated following two points of view. The first one considers the use of an external thermostat that imposes a programmed decrease of the temperature according to the time. The second one treats the temperature as a local parameter [24]. It wants to say that the temperature felt by each particle depends on its close environment. If the first method seems to be adapted to model the behavior of infinite system, the second one is able to describe the mutual influence of the particles, and to treat local effects, as core-corona interactions.

The NJL dynamical works evoked in the previous paragraph often focus on a phase transition between quark-antiquark plasma and a mesonic phase. Indeed, for example in [24], reactions involving quarks, antiquarks and mesons were included. This reference indicates a “quasi-complete” hadronization of the system, so it supposes an initial system composed by quarks and antiquarks in equal quantities, i.e. a null mean density. Thanks to the inclusion of the baryons in the (P)NJL models, this description should be extended to quark-antiquark systems in which we have more quarks than antiquarks, i.e. a positive mean density. Furthermore, if study as [23] proposed an attempt using baryons in such work, the number of possible reactions should be increased. The finality is to be able to treat the possible collisions between two unspecified particles (quarks, mesons, diquarks, baryons and their antiparticles), by elastic reactions, and/or possibly by inelastic ones.

In the previous chapter, it was proposed a scenario taking into account the obtained cross-sections. Notably, at high temperatures, $q+\overline{q}\rightarrow q+\overline{q}$ is expected to dominate $q+\overline{q}\rightarrow M+M$, whereas the situation is reversed at moderate and reduced temperatures. It lets foresee a massive mesonization, i.e. a strong consumption of quarks and antiquarks. In the same time, it was reported that the reactions forming diquarks and baryons present modest cross-sections. It was thus envisaged that the baryonization could act after the mesonization, notably when most of the antiquarks would have disappeared, in order to avoid parasite reactions between baryons and antiquarks. Clearly, by the way of a dynamical study, it could be interesting to validate or not this scenario. At this occasion, the role played by the diquarks should be investigated: is their production enough, during the evolution of the system, to really intervene? In the same time, the particles and the cross sections depend in our work on the temperature and the density. Thus, the influence of the density should also be evaluated.

Furthermore, because of the relative weakness of the cross-sections associated with the baryonization reactions, at least compared to mesonization ones, it should be considered the potential consequences on the simulations. More precisely, because of the system expansion, we can imagine that some quarks cannot have the time to interact via inelastic reactions before leaving the system. Indeed, the NJL model does not include confinement, thus a quark (or a diquark) can appear as free outside of the conditions that are expected for the quark gluon plasma, i.e. it can be found insolated from the other quarks. Obviously, as noted in the chapter 1, this behavior is physically not acceptable: a quark cannot be observed as free. This aspect is an important constraint in the framework of a dynamical study that must be fully taken into account. As a consequence, the NJL model can be considered as insufficient to model the system evolution reliably. In the previous chapters, we saw that the inclusion of the Polyakov loop allows simulating a confinement mechanism in the NJL model, forming the PNJL one. It was previously noted its effect on the masses of the included particles, and on the cross-sections. At this stage of the work, it is interesting now to evaluate the effect of the Polyakov loop on the system dynamics. Thus, a crucial question that motivates the work performed in this chapter is to see if the inclusion of this loop could allow avoiding free quarks/antiquarks in the final state of the simulations.

In this chapter, we start our analysis in section 2 by explaining our method, and establishing the required equations to model the system's evolution. We begin this section with a description of the global algorithm. The rest of the section focuses then on a study of each steps of this algorithm. Firstly, it concerns the evaluation of the external parameters felt by each particle, i.e. the densities and the temperature. About the temperature, it is explained how this statistical parameter, well mastered in thermodynamics, has to be adapted in the framework of relativistic systems. Also, an important part of the section 2 concerns the treatment of the collisions in our modeling. The devoted procedure is presented and then a list of all the possible collisions included in the program is proposed. We continue the description by detailing the relativistic equations of motion. These ones were inspired by [11, 24]. At this moment, a discussion is initiated on the manner to interpret them. Notably, it is introduced the notion of remote interaction between the particles. In the section 3, we present some preliminary results. Some of them are associated with the discussion started in section 2 about the remote interaction. Another test concerns the study of a close system, in order to simulate a system of hot quarks. Notably, a finality of this work is to characterize the quarks' trajectories. At this occasion, we want to see if their behavior can be described by a relativistic Brownian motion [25] or not. In the section 4, the results of simulations were presented. A first objective is to compare NJL and PNJL results, in order to see if one of them can allow a full hadronization of the quark-antiquark plasma. In section 5, we describe the results of a complete simulation, including a study of the evolution of some observables, in order to describe the various phases of the cooling. Also, we notably focus on an analysis of the collisions that occurred during the simulations, in order to distinguish the dominant ones.

# 2. Setting in equation of the dynamical model

## 2.1 The global algorithm

In the framework of the work presented in this chapter, we developed a stand alone computer program. The program, and by extension the algorithm, can be designated by the software's unofficial name, i.e. ARCHANGE. Please note that this name does not correspond to an acronym. The global algorithm used is not really different compared to some already described in the literature, as [23, 24]. Indeed, we consider the following steps:

1. System initialization. In this step, the initial particles are created and added into the system. In the framework of the work described in this chapter, it concerns the pseudo-scalar mesons, the scalar diquarks and the octet baryons. Apart from the nature of these particles, we supply only the positions and momenta of each particle as input data. Indeed, the program determines the environment of each of them, i.e. it calculates the values of the external parameters $(T, \rho_f)$ in the vicinity of each particle. It allows estimating the initial mass of each of them.

2. Treatment of the collisions. This step is devoted to investigate all the possibility of collisions between each couple of particles that could be considered. The applied collisions are determined by the program. This one treats then the particles' replacement into the system, in the case of collisions that modify them.

3. Treatment of the movements. There, we apply equations of motion in order to periodically update the position and the momentum of each particle.

4. Return to the point #2 for the next time iteration, until the end of the simulation.

## 2.2 Determination of the densities

Our approach considers local parameters. As seen in the previous subsection, the determination of these external parameters is used in the first step, i.e. the system initialization. However, in practice, the update of these values is performed in almost all the other steps, notably after a collision or after the treatment of the motions. In our work, these external parameters are the temperature *T* and the densities. These ones are noted $\rho_f$, with $f = u,d,s$, i.e. a density for each quark flavor. In the framework of the isospin symmetry, $\rho_u = \rho_d$, and the baryonic density $\rho_B$ can be found using the relation $\rho_B = {}^2\!/_3\,\rho_q$ [26]. However, in the simulations performed hereafter, this symmetry is not considered. Clearly, we used the EB parameter set defined in the chapter 2. The temperature and the densities are evaluated for each particle in the laboratory reference frame. It could be considered that this reference frame always coincides with the center of mass reference frame of the whole system.

Concerning the calculation of the densities, we firstly remark that our choice to use them is different to the one of [24] that uses the chemical potential. Our choice is explained by the behavior obtained in the figure 6 of the chapter 2. More precisely, the relation between $\rho_B$ and $\mu_q$ ($q \equiv u,d$) is not trivial in the (P)NJL models. In fact, $\mu_q$ is a function of $\rho_B$ but also of the temperature *T*, especially at low temperatures. This remark is fully extendable to the $\rho_f$ and to the associated chemical potentials $\mu_f$. However, each $\mu_f$ is solution of the equation set used to find the quarks' masses (chapter 2), so the chemical potentials can be calculated by the algorithm. They are then written in the output data supplied by the program.

We consider now an unspecified particle, labeled as *i*. The density $\rho_f$ felt by this particle is determined by the formula:

$$\left(\rho_f\right)_i = \frac{1}{V} \cdot \sum_{j \neq i} w(i,j) \cdot \left( \left(n_f\right)_j - \left(n_{\bar{f}}\right)_j \right), \tag{1}$$

where *f* is a quark flavor ($u,d,s$). Also, $V = 4/3 \cdot \pi \cdot R^3$ is the volume of a fictitious sphere centered on the studied particle $i$. This sphere defines the vicinity of the particle. The *j* summation in (1) is performed over the particles forming the system. Also, $\left(n_f\right)_j$ and $\left(n_{\bar{f}}\right)_j$ designates, respectively, the number of flavor *f* quarks and anti-flavor $\bar{f}$ antiquarks "contained" in the particle *j*. For example, in the case of a proton, $n_u = 2$, $n_{\bar{u}} = 0$, $n_d = 1$, $n_{\bar{d}} = 0$, $n_s = 0$ and $n_{\bar{s}} = 0$. In the same way, with the $\bar{u}$ antiquark, $n_{\bar{u}} = 1$, and zero for the other counters. As a consequence, the counting does not consider the states of the quarks, i.e. confined or not confined. But, in order to take into account the distance between the *i* and *j* particles, a coefficient is applied during the calculation. This coefficient is supplied by a function that [24] named *weighting function*. As in this reference, we use a Gaussian function:

$$w(i,j) = \exp\left(-\frac{d_{ij}^{\,2}}{2D^2}\right), \tag{2}$$

where $d_{ij}$ is the distance between the *i* and *j* particles, and *D* is linked to the sphere radius *R*. In practice, we chose $D = 1.75\ \text{fm}$ in our simulations. The relation (2) allows considering the densities as local parameters, as expected.

# 2.3 Estimation of the temperature

In [24], a link was established between the densities and the temperature. In our approach, temperature and densities are independent parameters, without correlation between them. It allows considering various system configurations: hot or/and dense systems. In statistical physics, the use of the equipartition theorem is a relevant approach to estimate the temperature. In our case, the complications come to the adaptation of this theorem to the relativistic framework and the necessity to work at thermal equilibrium. About the relativistic treatment, some approaches were proposed in order to introduce the notion of relativistic temperature [27–34]. We propose here to explain their reasoning.

Firstly, in a non-relativistic case, we recall that the temperature of an ideal gas can be obtained, at the thermal equilibrium, by the standard equipartition theorem:

$$\langle E_K \rangle_{NR} = \frac{3}{2} \cdot k_B \cdot T, \tag{3}$$

where $\langle E_K \rangle_{NR}$ is the expectation value of the non-relativistic kinetic energy, i.e. the average kinetic energy per particle. $k_B$ is the Boltzmann constant, considered as equal to 1 with our unities. In fact, the equipartition theorem can be extended. A general formulation explains that for each degree of freedom $\phi_i$, i.e. for each quadratic variable in the energy writing, the following relation is satisfied [27]:

$$\left\langle \phi_i \cdot \frac{\partial E}{\partial \phi_i} \right\rangle = k_B \cdot T. \tag{4}$$

In a relativistic formulation, the mean energy of a particle is written as $\langle E \rangle = \left\langle \sqrt{p_x^{\,2} + p_y^{\,2} + p_z^{\,2} + m^2} \right\rangle$. We have three configurations:

- Non-relativistic regime. Clearly, in this case $(\vec{p})^2 \ll m^2$, leading to the well known approximation:

$$\langle E \rangle \approx \left\langle \frac{p_x^{\,2} + p_y^{\,2} + p_z^{\,2}}{2m} + m \right\rangle \equiv \langle E_K \rangle_{NR} + \langle m \rangle. \tag{5}$$

Using the equations (4, 5), we obtain $\left\langle p_i \cdot \frac{\partial E}{\partial p_i} \right\rangle = \left\langle \frac{p_i^2}{m} \right\rangle = k_B \cdot T$, where $i = x, y, z$. Obviously, we find (3).

- Ultra-relativistic regime [34]. There, the mass is negligible in front of the momentum: $(\vec{p})^2 \gg m^2$. In this case, we have:

$$\langle E \rangle \approx \left\langle \sqrt{p_x^2 + p_y^2 + p_z^2} \right\rangle \equiv \langle E_K \rangle_{UR}. \tag{6}$$

In the same way, we find $\left\langle p_i \cdot \frac{\partial E}{\partial p_i} \right\rangle = \left\langle \frac{p_i^2}{\sqrt{p_x^2 + p_y^2 + p_z^2}} \right\rangle = k_B \cdot T$, and it comes:

$$\langle E_K \rangle_{UR} = \left\langle \frac{p_x^2 + p_y^2 + p_z^2}{\sqrt{p_x^2 + p_y^2 + p_z^2}} \right\rangle = \sum_{i=x,y,z} \left\langle p_i \cdot \frac{\partial E}{\partial p_i} \right\rangle \text{ and } \langle E_K \rangle_{UR} = 3 \cdot k_B \cdot T. \tag{7}$$

- General case. Using the same method as in the previous cases, we have again $\langle E \rangle = \left\langle \sqrt{p_x^2 + p_y^2 + p_z^2 + m^2} \right\rangle$, and we write:

$$\left\langle p_i \cdot \frac{\partial E}{\partial p_i} \right\rangle = \left\langle \frac{p_i^2}{\sqrt{p_x^2 + p_y^2 + p_z^2 + m^2}} \right\rangle = k_B \cdot T. \tag{8}$$

We sum the contributions of the three components $x, y, z$, and we obtain [27, 33, 34]:

$$\left\langle \frac{(\vec{p})^2}{E} \right\rangle = 3k_B \cdot T \text{ or } \left\langle \frac{(\vec{p})^2}{2E} \right\rangle = \frac{3}{2} \cdot k_B \cdot T. \tag{9}$$

The equation (9) presents the advantage to be satisfied in all the cases, because no approximation was used as with the two regimes described upstream. We considered this relation (9) in our calculations. In this work, the notion of local temperature is used, thus the thermal equilibrium should be considered locally. In practice, the $\langle \ \rangle$ operator concerns the momenta $\vec{p}$ and the energy $E$ of the particles on the vicinity of the studied particle. As with the densities, the weighting function (2) is considered in this calculation. At this occasion, we can underline the important role played by $D$ in this function, because this parameter defines what is considered as the vicinity of each particle. It has an influence on the calculations of the temperature and the densities.

Moreover, the energy $E = \sqrt{\vec{p}^2 + m^2}$ of a particle depends on its mass $m$. As a consequence, the mass of a particle $j$ intervenes via (9) on the calculation of the temperature felt by a particle $i$. We saw in the chapters 2 to 4 that the masses of all the treated particles depend on the temperature in the (P)NJL models. So, the temperature of the particle $i$ has an incidence on the mass of this particle $i$. This mass variation has consequences on the temperature felt by the particle $j$. This temperature has an incidence on the mass of this particle $j$… Clearly, there is interdependence between masses and temperatures. Numerically, this aspect is treated by the mean of successive iterations, until convergence. This is notably the case for the step #1 of the subsection 2.1, i.e. the system initialization.

## 2.4 Treatment of the collisions

The collision algorithm firstly consists to take into account a particle in particular. We label it as *i*. Then, the program establishes a list of all the couples that can be formed using this particle *i* and another particle of its vicinity, labeled with *j*. Each couple $(i, j)$ represents a possibility of incoming particles that can interact.

The momenta of the particles are expressed in the laboratory reference frame in the data manipulated by the program. The (P)NJL cross-sections available in the literature and by extension the ones calculated in our model (chapter 6) are estimated in the center of mass reference frame of the two incoming particles $(i, j)$. As a consequence, the procedure consists to calculate $\sqrt{s}$ for the system formed by these two particles, and then to apply a Lorentz boost to this system, in order to express the two particle's energy and momentum in their center of mass reference frame.

For each couple $(i, j)$, the impact parameter $b^*_{i,j}$ is estimated in their center of mass reference frame. Some couples can be invalidated, notably if the program determinates that the particles are moving away themselves, or if the impact parameter is physically too high, i.e. incompatible with the collision criterion (10) whatever the $\sigma$ values physically admissible. Also, the program proceeds to an extrapolation of the trajectories in order to validate the couple $(i, j)$ only if the distance between *i* and *j* is minimal, following the idea mentioned in [24]. In other words, the collision does not occur at a time *t* if the program estimates that the particles will be still approaching at the next time iteration.

For each remaining couple $(i, j)$, the program establishes a sub-list of the reaction types that can occur between *i* and *j*. In our modeling, collisions involving two incoming particles and two outgoing ones are considered. For example, with a quark/antiquark pair $(q,\bar{q})$, the possible reaction types are $q+\bar{q} \to q+\bar{q}$, $q+\bar{q} \to M+M$ and $q+\bar{q} \to D+\bar{D}$. Then, for each reaction type, the algorithm foresees the possible outgoing particles that are creatable by each process. If this step is trivial for reactions as $q+q \to q+q$, the things are more complicated in other cases. The table 1 proposes a list in the case of the quark/antiquark couple $(u,\bar{u})$.

The notion of *elastic reactions* should be understood as reactions in which the types of the particles (quark, meson, etc.) are not modified. But, the particles themselves may be replaced, as in $u+\bar{u} \to d+\bar{d}$. About $q+\bar{q} \to M+M$, we do not included reactions in which $\eta'$ could appear. Indeed, the cross sections of the mesonization reactions implying $\eta'$ are negligible in front of the mesonization processes that form lighter mesons, as pions [35].

| Reaction type | Detail on the possible reactions |
|---|---|
| $q+\overline{q}\rightarrow q+\overline{q}$ | $u+\overline{u}\rightarrow u+\overline{u}$<br>$u+\overline{u}\rightarrow d+\overline{d}$<br>$u+\overline{u}\rightarrow s+\overline{s}$ |
| $q+\overline{q}\rightarrow M+M$ | $u+\overline{u}\rightarrow \pi^0+\pi^0$<br>$u+\overline{u}\rightarrow \pi^0+\eta$<br>$u+\overline{u}\rightarrow \eta+\eta$<br>$u+\overline{u}\rightarrow \pi^+ +\pi^-$<br>$u+\overline{u}\rightarrow K^+ +K^-$ |
| $q+\overline{q}\rightarrow D+\overline{D}$ | $u+\overline{u}\rightarrow [ud]+\overline{[ud]}$<br>$u+\overline{u}\rightarrow [us]+\overline{[us]}$ |

**Table 1.** List of the reactions when one $u$ quark interacts with one $\overline{u}$ anti-quark.

At this stage of the procedure, the program evaluates the cross-sections of all the listed reactions. For each couple $(i,j)$ and for each reaction (labeled with $k$), the obtained cross section $\sigma_{i,j,k}$ is compared to the impact parameter $b^*_{i,j}$ of the couple. More precisely, the collision criterion mentioned in [11, 12] is applied: it stipulates that a reaction is possible if its cross section satisfies the inequality (10), and schematized in the figure 1.

$$b^*_{i,j} \leq \sqrt{\sigma_{i,j,k}/\pi} \,. \tag{10}$$

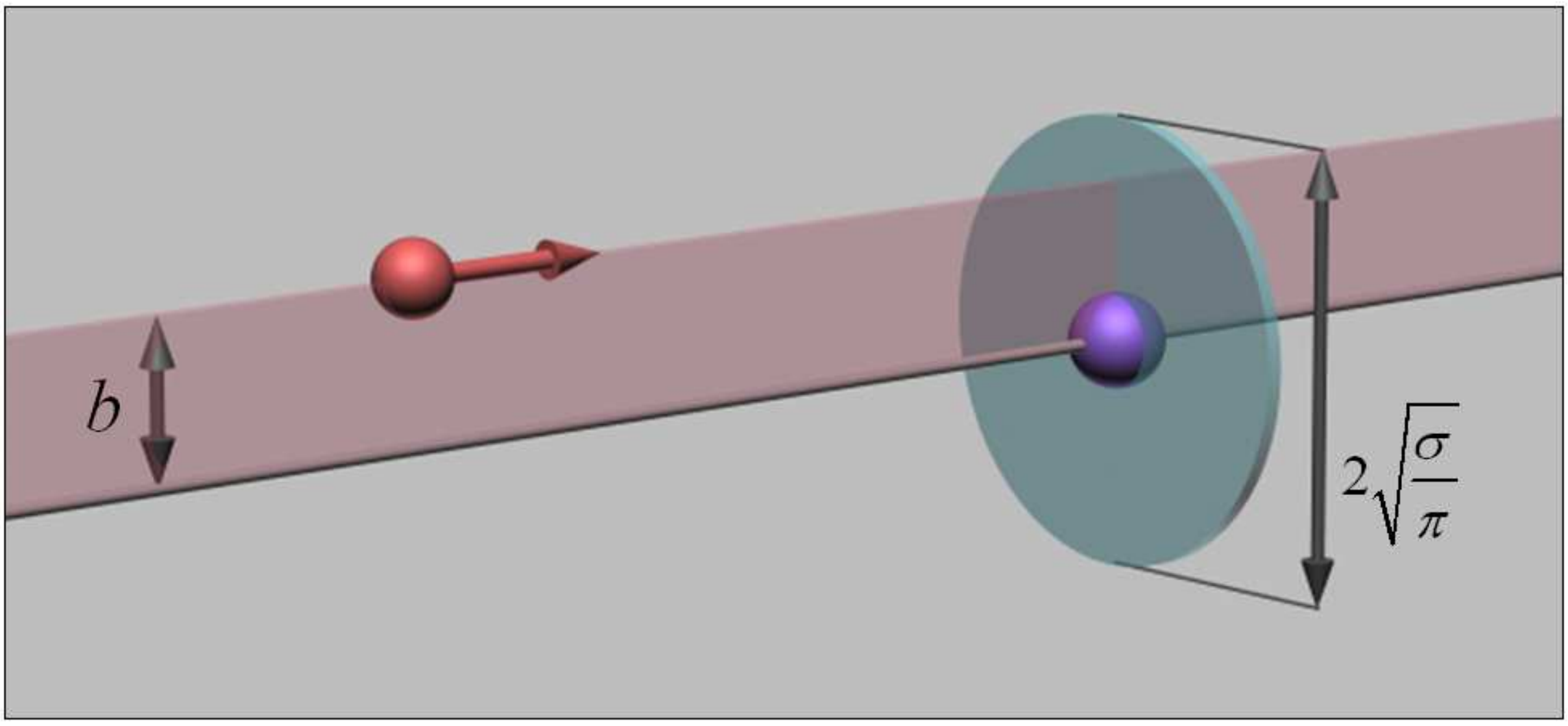

**Figure 1.** Schematization of the geometrical criterion of collision with two particles.

Among all the reactions that verify this criterion, the program chooses the reaction $k$ associated with the couple $(i,j)$ that will be considered. This choice is stochastically determined, but the probability of each reaction $(i,j,k)$ is weighted by its associated cross section. In other words, the reaction that has the strongest cross section at this moment is the one that has the strongest chance to occur. At the opposite, a reaction presenting null cross-sections at this moment cannot be chosen.

Once the reaction and the associated couple are determined, i.e. it was found the particle $j$ that will react with the particle $i$ via the reaction $k$, the program estimates the scattering angle and the momenta of the outgoing particles. When these particles are inserted into the system, their masses are beforehand estimated in the exact conditions (temperature and densities) felt by the incoming particles $(i, j)$ before their replacement. The formulas used to treat the collision upon its geometrical aspects (Lorentz boosts, impact parameter, scattering angle) are gathered in the appendix F.

Whatever the reaction that occurred, elastic or inelastic, these new particles will no longer interact before the next time iteration: they are provisory "deactivated" by the procedure. This is done to avoid that the same particles interact indefinitely together during the same time iteration. In the same way, if the program judges that the considered particle $i$ cannot interact at this moment (too far compared to the other ones, etc.), the particle $i$ is also deactivated until the next iteration.

Finally, the program considers then another particle $i$, and proceeds again as explained in this subsection, until all the particles are reviewed, i.e. "deactivated".

# 2.5 List of the type of reactions included in our model

The table 2 hereafter gathers the reaction types implemented in the program. It reveals that an important number of possible reactions is to be considered, especially because our simulations do not apply the isospin symmetry. We recall that the included particles in the program are the quarks $q$, the pseudo-scalar mesons $M$, the scalar diquarks $D$, the octet baryons $B$ and their associated anti-particles.

Furthermore, as observed previously in this thesis, the cross-sections depend on $\sqrt{s}$, the temperature $T$ and the densities (or the chemical potentials). As a consequence, using pre-calculated values is not the best solution, notably by the use of a cross-sections database, as in [24]. Certainly this solution is applicable, but at the price of too strong approximations. Indeed, for example with $q+\overline{q} \to M+M$, a strong mesonization is expected, but in some precise $\sqrt{s}$, $T$, $\rho_f$ conditions. In fact, this aspect is important to study the cooling of a quark system. During the simulations, the cross-sections are *real time* calculated, i.e. they are estimated when the program requires it, taking care about $\sqrt{s}, T, \rho_f$. The method to estimate these cross-sections with the (P)NJL models was detailed in the previous chapter. Notably, about the types of reactions described in the table 2, we used the method described in [36, 37] to calculate cross-sections as $D+D \to D+D$. Nevertheless, we used the formulas proposed in [12] to treat reactions as $M+M \to M+M$, $M+B \to M+B$ and $B+B \to B+B$. This remark is obviously valid for reactions involving their antiparticles, as $\overline{B}+\overline{B} \to \overline{B}+\overline{B} \ldots$

| Incoming particles | Possible reaction types | | |
|---|---|---|---|
| $q+\bar{q}$ | $q+\bar{q}\to q+\bar{q}$ | $q+\bar{q}\to M+M$ | $q+\bar{q}\to D+\bar{D}$ |
| $q+q$ | $q+q\to q+q$ | $q+q\to M+D$ | $q+q\to \bar{q}+B$ |
| $\bar{q}+\bar{q}$ | $\bar{q}+\bar{q}\to \bar{q}+\bar{q}$ | $\bar{q}+\bar{q}\to M+\bar{D}$ | $\bar{q}+\bar{q}\to q+\bar{B}$ |
| $q+M$ | $q+M\to q+M$ | $q+M\to \bar{q}+D$ | |
| $\bar{q}+M$ | $\bar{q}+M\to \bar{q}+M$ | $\bar{q}+M\to q+\bar{D}$ | |
| $\bar{q}+D$ | $\bar{q}+D\to \bar{q}+D$ | $\bar{q}+D\to q+M$ | $\bar{q}+D\to \bar{D}+B$ |
| $q+\bar{D}$ | $q+\bar{D}\to q+\bar{D}$ | $q+\bar{D}\to \bar{q}+M$ | $q+\bar{D}\to D+\bar{B}$ |
| $q+D$ | $q+D\to q+D$ | $q+D\to M+B$ | |
| $\bar{q}+\bar{D}$ | $\bar{q}+\bar{D}\to \bar{q}+\bar{D}$ | $\bar{q}+\bar{D}\to M+\bar{B}$ | |
| $\bar{q}+B$ | $\bar{q}+B\to \bar{q}+B$ | $\bar{q}+B\to q+q$ | $\bar{q}+B\to M+D$ |
| $q+\bar{B}$ | $q+\bar{B}\to q+\bar{B}$ | $q+\bar{B}\to \bar{q}+\bar{q}$ | $q+\bar{B}\to M+\bar{D}$ |
| $q+B$ | $q+B\to q+B$ | $q+B\to D+D$ | |
| $\bar{q}+\bar{B}$ | $\bar{q}+\bar{B}\to \bar{q}+\bar{B}$ | $\bar{q}+\bar{B}\to \bar{D}+\bar{D}$ | |
| $M+M$ | $M+M\to M+M$ | $M+M\to q+\bar{q}$ | |
| $M+D$ | $M+D\to M+D$ | $M+D\to q+q$ | $M+D\to \bar{q}+B$ |
| $M+\bar{D}$ | $M+\bar{D}\to M+\bar{D}$ | $M+\bar{D}\to \bar{q}+\bar{q}$ | $M+\bar{D}\to q+\bar{B}$ |
| $M+B$ | $M+B\to M+B$ | $M+B\to q+D$ | |
| $M+\bar{B}$ | $M+\bar{B}\to M+\bar{B}$ | $M+\bar{B}\to \bar{q}+\bar{D}$ | |
| $D+\bar{D}$ | $D+\bar{D}\to D+\bar{D}$ | $D+\bar{D}\to q+\bar{q}$ | |
| $D+D$ | $D+D\to D+D$ | $D+D\to q+B$ | |
| $\bar{D}+\bar{D}$ | $\bar{D}+\bar{D}\to \bar{D}+\bar{D}$ | $\bar{D}+\bar{D}\to \bar{q}+\bar{B}$ | |
| $\bar{D}+B$ | $\bar{D}+B\to \bar{D}+B$ | $\bar{D}+B\to \bar{q}+D$ | |
| $D+\bar{B}$ | $D+\bar{B}\to D+\bar{B}$ | $D+\bar{B}\to q+\bar{D}$ | |
| $D+B$ | $D+B\to D+B$ | | |
| $\bar{D}+\bar{B}$ | $\bar{D}+\bar{B}\to \bar{D}+\bar{B}$ | | |
| $B+\bar{B}$ | $B+\bar{B}\to B+\bar{B}$ | | |
| $B+B$ | $B+B\to B+B$ | | |
| $\bar{B}+\bar{B}$ | $\bar{B}+\bar{B}\to \bar{B}+\bar{B}$ | | |

**Table 2.** List of the reaction types included in the model.

# 2.6 Equations of motion

In the subsection 2.3, we underlined the necessity to work in our model in a relativistic regime. This aspect must appear in an explicit way in the equations of motion. In the literature, relativistic theories were developed, e.g. [11, 12, 38, 39]. As noted in [11] or in [24], the writing of the starting equations strongly recalls the Hamilton equations. However, as explained in these references, the equations to be used in a relativistic regime are different. In our work, we consider the equations used for example in [24]:

$$\begin{cases} \dfrac{d\left(r_i^{\mu}\right)}{d\tau} = \dfrac{p_i^{\mu}}{E_i} \\ \dfrac{d\left(p_i^{\mu}\right)}{d\tau} = -\sum\limits_{j\neq i} \dfrac{1}{2E_j} \cdot \dfrac{\partial V_j}{\partial r_{i\,\mu}} = -\sum\limits_{j\neq i} \dfrac{m_j}{E_j} \cdot \dfrac{\partial m_j}{\partial r_{i\,\mu}} \end{cases} . \tag{11}$$

They correspond to classical (non-quantum) equations of motion. In these relations, $E=\sqrt{\vec{p}^2+m^2}$ is the energy of the concerned particle, i.e. the zero component of the associated energy-momentum four-vector. As mentioned in [24], the rewriting of the second equation of (11), as a function of the masses $m_j$, is justified by the fact that we do not have an explicit potential $V_j$ in the framework of the (P)NJL models. However, the mass of each particle depends on parameters as the local temperature and the densities. Clearly, these parameters are calculated for each particle via its vicinity, i.e. the other particles. So, it is possible to consider that the potential is "hidden" in the masses of the particles, making the replacement $\dfrac{\partial V_j}{\partial r_{i\,\mu}} = \dfrac{\partial m_j^2}{\partial r_{i\,\mu}}$. The $\sum\limits_{j\neq i} \dfrac{m_j}{E_j} \cdot \dfrac{\partial m_j}{\partial r_{i\,\mu}}$ is thus interpretable as a *remote interaction* between the particles.

Also, following [24], the derivative $\dfrac{\partial m_j}{\partial r_{i\,\mu}}$ can be developed as:

$$\frac{\partial m_j}{\partial r_{i\,\mu}} = \frac{\partial m_j}{\partial T_j} \cdot \frac{\partial T_j}{\partial r_{i\,\mu}} + \sum_{f=u,d,s} \frac{\partial m_j}{\partial \rho_{f\,j}} \cdot \frac{\partial \rho_{f\,j}}{\partial r_{i\,\mu}} . \tag{12}$$

Nevertheless, another interpretation of this term can be given: we can say that it directly represents the variation of the masses of the particles *j* induced by the variation of the position of the particle *i*. Our reasoning is illustrated by the figure 2.

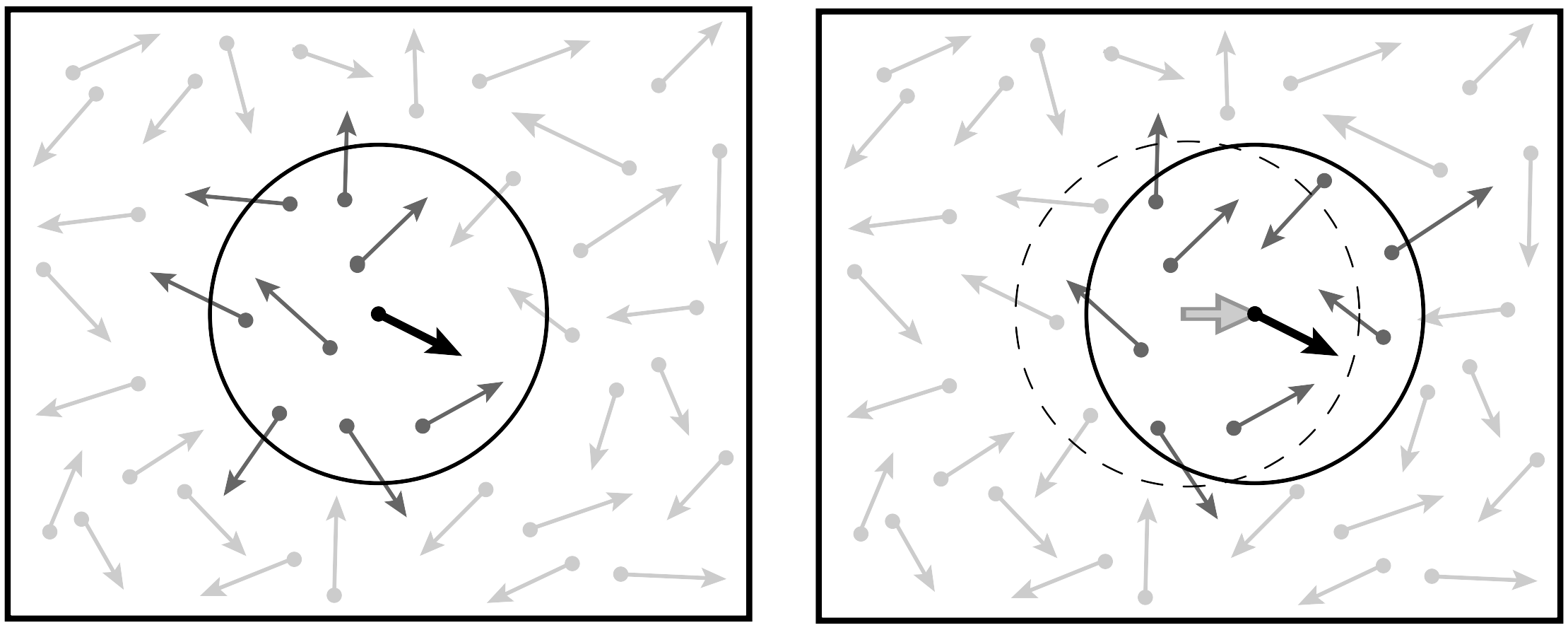

**Figure 2.** Schematization of the effect of one particle's motion.

More precisely, the displacement of a particle induces perturbations on the particles located in its vicinity. Firstly, the densities felt by the other particles are modified, especially for the particles in its close neighborhood. Secondly, the displacement also induces a modification of the local temperature. In fact, these two effects can be understood by the way we estimate the densities and the temperature, subsections 2.2, 2.3, especially by the use of the weighing function (2). More precisely, every displacement induces a variation of the values returned by this function, so it operates modifications in the counting.

We use the temperature and the densities as parameters to calculate the mass of the particle. Thus, it explains why the displacement of one particle is able to modify the mass of the other particles. Clearly, these observations show the necessity to update the masses, temperatures and densities after the step #3 seen in the subsection 2.1 (treatment of the motion). Concerning the densities, this parameter is linked to the positions of the particles, so the calculation is easy. But, as noted in the subsection 2.3, the interdependence between the masses and the local temperature requires performing several iterations in the calculation of these quantities, until convergence. However, in practice, this convergence is quickly reached. The conservation of total energy and total momenta are two constraints imposed to the algorithm during the described procedures. In fact, the conservation of the energy is compatible with how we use (11). Clearly, it is true that the determination of the external parameters $T, \rho_f$ is imperfect, in the sense that they depend on the arbitrary $D$ introduced in the equation (2). However, firstly, these parameters are calculated in the same way for all the particles present in the studied system. Secondly, we remark that the second equation of (11) allow variations of the kinetic energy of a particle by variations of its potential energy. But, the total energy of the studied particle does not vary, if this one does not participate to a collision.

# 3. Preliminary results

## 3.1 Remote interaction between particles

In this subsection, we propose to investigate numerically the properties of the remote interaction evoked in the subsection 2.6. The finality is to try to answer an interrogation we can have since some chapters. Let us summarize: the strong interaction, described by Quantum Chromodynamics, governs the interactions between the quarks and/or antiquarks via the exchange of gluons. This interaction is very attractive because it is responsible of the confinement of the quarks and antiquarks inside the hadrons. When we consider the NJL model, we lose gluons and the confinement, even if the inclusion of the Polyakov loop mimics a mechanism of confinement. The question is finally to see if we still have an attractive potential between quarks and/or antiquarks. Also, is it enough to mimic the strong interaction in a realistic way, at least in these conditions?

Firstly, two tests were performed, using the PNJL model. In the first one, two low $u$ quarks are considered. The initial momenta are directed in opposite directions, but with $\|\vec{p}\| = 70$ MeV for the two quarks. In a second test, the two $u$ quarks have got the same momentum, with also $\|\vec{p}\| = 70$ MeV. For these tests, the collision procedure is deactivated. An external thermostat imposes a temperature of 200 MeV to the quarks, added to the one produced by the particles themselves by the way of (9). The trajectories in the laboratory reference frame are represented in the form of a chronogram in the figure 3. The trajectories of the quarks are periodically marked with dots. The dots with the darkest tone correspond to the more recent quarks' positions. For these tests, as with the ones described in figure 4, it was verified that the total energy and the total momentum are strictly conserved, at each time of the simulations.

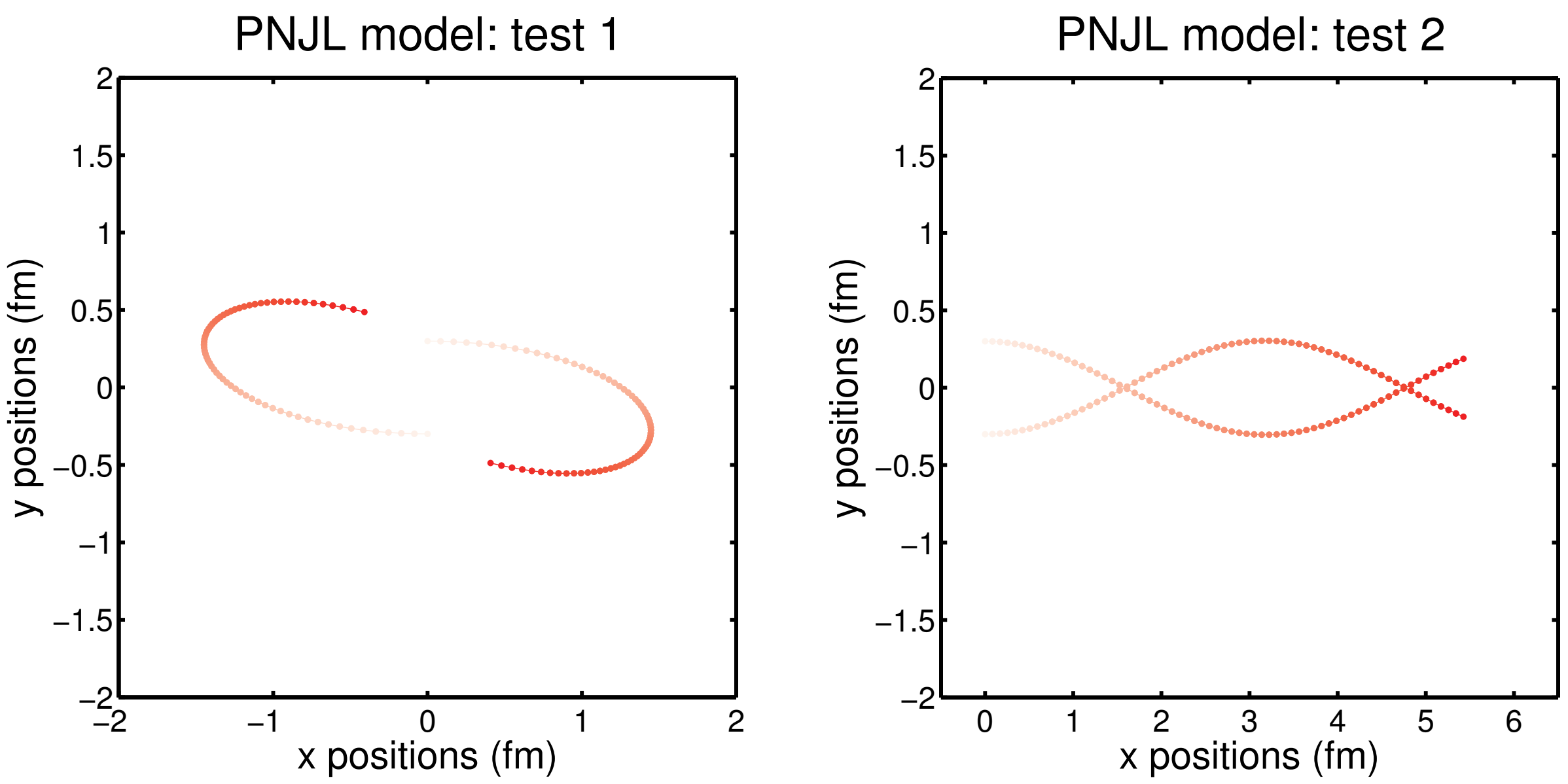

**Figure 3.** Trajectories of two $u$ quarks, at $T \approx 200$ MeV .

The results of figure 3 show a remote interaction between the quarks. This interaction is clearly attractive. Following the discussion started in the subsection 2.6, this behavior is explained by the second equation of (11). A general tendency is a particle modifies its trajectory in order to tend to minimize the masses of the particles in its vicinity. So, the two quarks tend to be attracted themselves. Indeed, for one quark $j$, the approach of another quark $i$ induces that the temperature and the density $\rho_u$ felt by the quark $j$ increase, so its mass decreases, as confirmed in the figure 3 of the chapter 2. The same reasoning is valid for the quark $i$ submitted to the action of the quark $j$, in a symmetric way. In the $T, \rho_B$ zone in which evolves the two quarks, i.e. $\rho_B \approx 0$ and $T \approx 200$ MeV , the derivative with respect to the $\rho_u$ density seems to be upper than the one upon the temperature. This remark is extendable to zones in which the densities are reduced and $T < 200$ MeV . In the tests of figure 3, the observed quark behavior seems to come mainly from the density variations felt by theses two particles, see equations (11, 12).

More precisely, we interpret the $-\sum_{j \neq i} \frac{m_j}{E_j} \cdot \frac{\partial m_j}{\partial r_{i\,\mu}}$ term in the second line of (11) as a force. It should be stated that the remote interaction is clearly non linear. Indeed, if one multiplies the number of quarks, the attractive effect is not necessarily multiplied by the same factor. It should be seen as an N-body interaction. Furthermore, the observed interaction has a limited range, in opposition with the QCD quark-quark potential. In fact, the (P)NJL remote interaction is strongly linked to the variations of the $T$ and densities induced by a particle on its vicinity. Obviously, the weighting function (2) plays a central role in the behavior of this interaction, especially as regards its range. In a dynamical evolution, (P)NJL models appears to be able to mimic short range phenomena described by QCD, but could present some limitations to model long range ones.

On the other hand, the masses of the quarks tend to their naked values ($m_{0f}$) at high temperatures and densities, as seen in the chapter 2. As a consequence, in these extreme conditions, the mass of a quark will no longer be influenced by these parameters, so by its environment. Thus, the quarks become free in this regime. It coincides with the asymptotic freedom phenomenon, treated by the perturbative QCD evoked in the chapter 1.

Now, we investigate the differences between the NJL and the PNJL models as regards the observed interaction. In other words, we try to see the influence of the Polyakov loop on this phenomenon. These differences are evaluated via two tests, in which two $u$ quarks interact themselves in identical conditions, with initially $\|\vec{p}\| = 30$ MeV and for a temperature close to 250 MeV, imposed by an external thermostat. The only difference between the two tests is the used model: NJL or PNJL. The results presented in figure 4 indicate that the PNJL remote interaction is more intense than the NJL one, at least in the framework of these simulations. Indeed, the trajectories show that the deviations induced by the remote interaction are more important when the PNJL model is used. In the NJL test, the attraction was not enough strong to hold the quarks together: they went away in oppose directions.

A first explanation of these observed behaviors is given by the figure 3 of the chapter 2: the effect of the Polyakov loop upon the quarks' masses is to shift the observed values towards higher temperature, applying a distortion effect on the curves. According to the explanations given in the previous paragraphs, the quark-quark interaction is expected to be directly related to the mass variations upon $T$ and $\rho_f$. The mass variation is stronger in the PNJL approach compared to the NJL one, in a region for which the baryonic density is lower than $3\rho_0$ and the temperature is roughly between 200 and 300 MeV. This zone is particularly interesting, because it corresponds to the conditions in which the hadronization is expected to intervene.

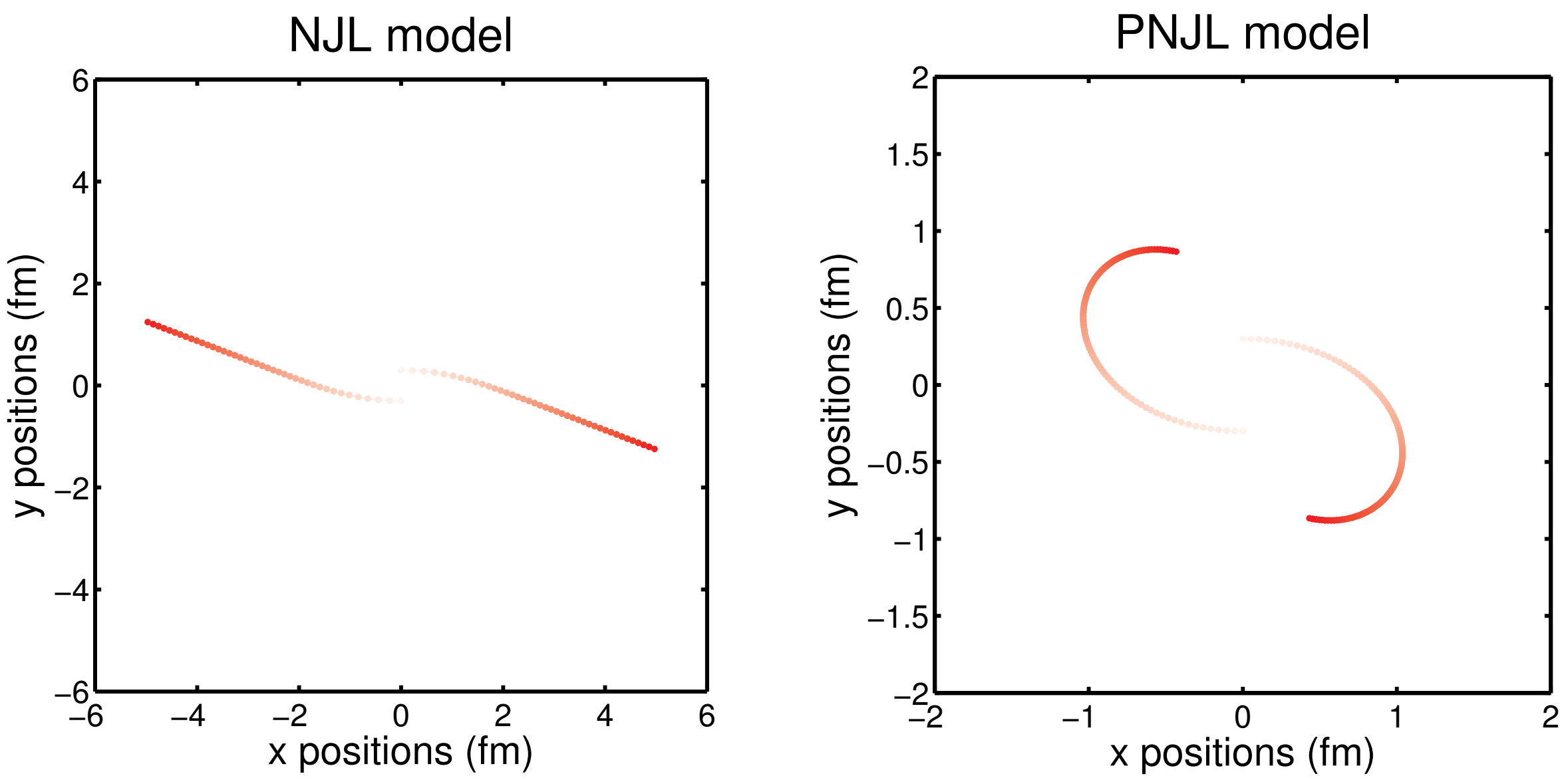

**Figure 4.** Comparison of the results found with the NJL and PNJL models, at $T \approx 250$ MeV .

There, the remote quark–quark interaction could be stronger in the PNJL model, compared to the NJL one. Following this hypothesis, it could induce a provisory collapse of the quark system, which can optimize the hadronization processes. Moreover, at reduced densities, the zone for which $\frac{\partial m}{\partial \rho_f}$ is important is wider in the PNJL model than in the NJL one. Indeed, we find, respectively, $T < 220\text{ MeV}$ against $T < 150\text{ MeV}$ in figure 3 of the chapter 2. Aware of the results of figure 3 above, explained by the way of the mass variations according to the density, it lets foresee that the attractive effect can be more present during the cooling of the quark system thanks to the inclusion of the Polyakov loop.

## 3.2 Relativistic Brownian motion?

If we consider the remote interaction highlighted in the previous subsection and the collisions described in the subsections 2.4, 2.5, we will see now if it could be a dominant behavior between them. In other words, if the collisions largely rule the dynamics, the system could show a behavior close to the one of a relativistic Brownian motion [25]. At the opposite, if the remote interaction is strong enough, it might induce the collapse motion evoked in the previous subsection.

In order to try to answer to this question, we performed another simulation with the PNJL description. We gathered 6 $u$ quarks and 6 $d$ quarks in a cube of size 2 fm, in the conditions of a hot and dense system. More precisely, these 12 quarks in a volume equal to $8\text{ fm}^3$ correspond to a baryonic density close to $3\rho_0$. Moreover, the average momentum of each quark is chosen to be close to 780 MeV. In agreement with (9) and aware of the light quarks masses in these conditions, it corresponds to a temperature near 250 MeV. In order to simulate the behavior of an infinite system, the cube's walls are perfectly reflective for the contained quarks. Clearly, the momentum direction is modified by the rebound, but not its modulus. As a consequence, the total energy conservation can be satisfied. We checked that it was the case. Moreover, the inelastic scattering processes are not included in this simulation. It wants to say that our quarks are not modified in this test. Also, as with the following simulations, no external thermostat was used.

The results are presented in the figures 5 to 8. In the figure 5 and 6, the evolution of some relevant physical quantities is represented: the mean mass and the mean momentum of the quarks, the mean temperature and the mean Polyakov field $\Phi$ felt by these particles. Concerning this former one, we observed in the chapter 2 that $\bar{\Phi} \approx \Phi$ whatever the temperature or the baryonic density. It explains why we only represented $\Phi$ in our graphs.

The plotted quantities present fluctuations according to the time, but no deviation are observed. The mass oscillates around an average value close to 220 MeV. The average temperature is near to 250 or 260 MeV, as expected. Also, the average momentum seems to be about 780 MeV. Moreover, we note that the variations of the temperature and the momentum have some similarities. These ones are explainable with (9): roughly speaking, the temperature can be seen as an averaging of the momenta's modulus. Furthermore, the variations of the mass and the ones of the temperature seem to be reversed. In other words, when the temperature is maximal, the mass admits a minimum, and conversely. This aspect is explained by the behavior of the quarks' masses according to the temperature. More precisely, for the observed temperatures, the mass of a light quark $q$ decreases almost linearly when the temperature is growing, chapter 2.

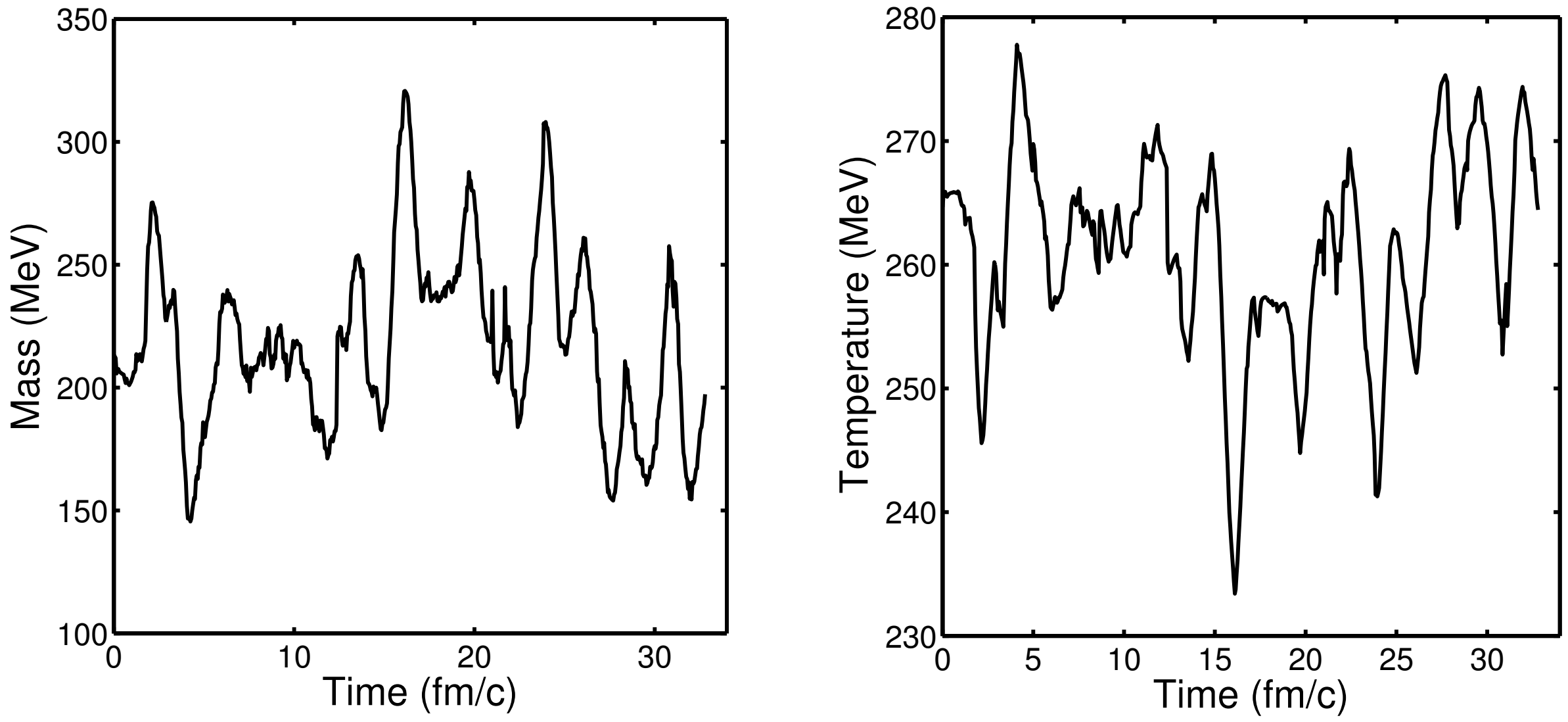

**Figure 5.** Evolution of the mean mass and the mean temperature according to the time.

About the Polyakov field $\Phi$, it is often above 0.5. Stricto sensu, the "deconfined" regime corresponds to $\Phi \to 1$. However, because of the used effective potential, i.e. the one of [40, 41], such a value is only reached at infinite temperature, by construction. In practice, we observed that $\Phi \approx 0.8$ at $T \approx 400$ MeV. Furthermore, the value of $T_0$, i.e. the critical deconfinement temperature in a pure gauge theory [42], was chosen to be equal to 270 MeV. As a consequence, the values of $T$ and $\Phi$ found in the figures 5, 6 suggest that we are close to the "deconfinement transition" in this test. So, our simulation corresponds to the description of a rather hot system. It may be associated with the conditions of a cooling system, when the hadronization is expected to occur.

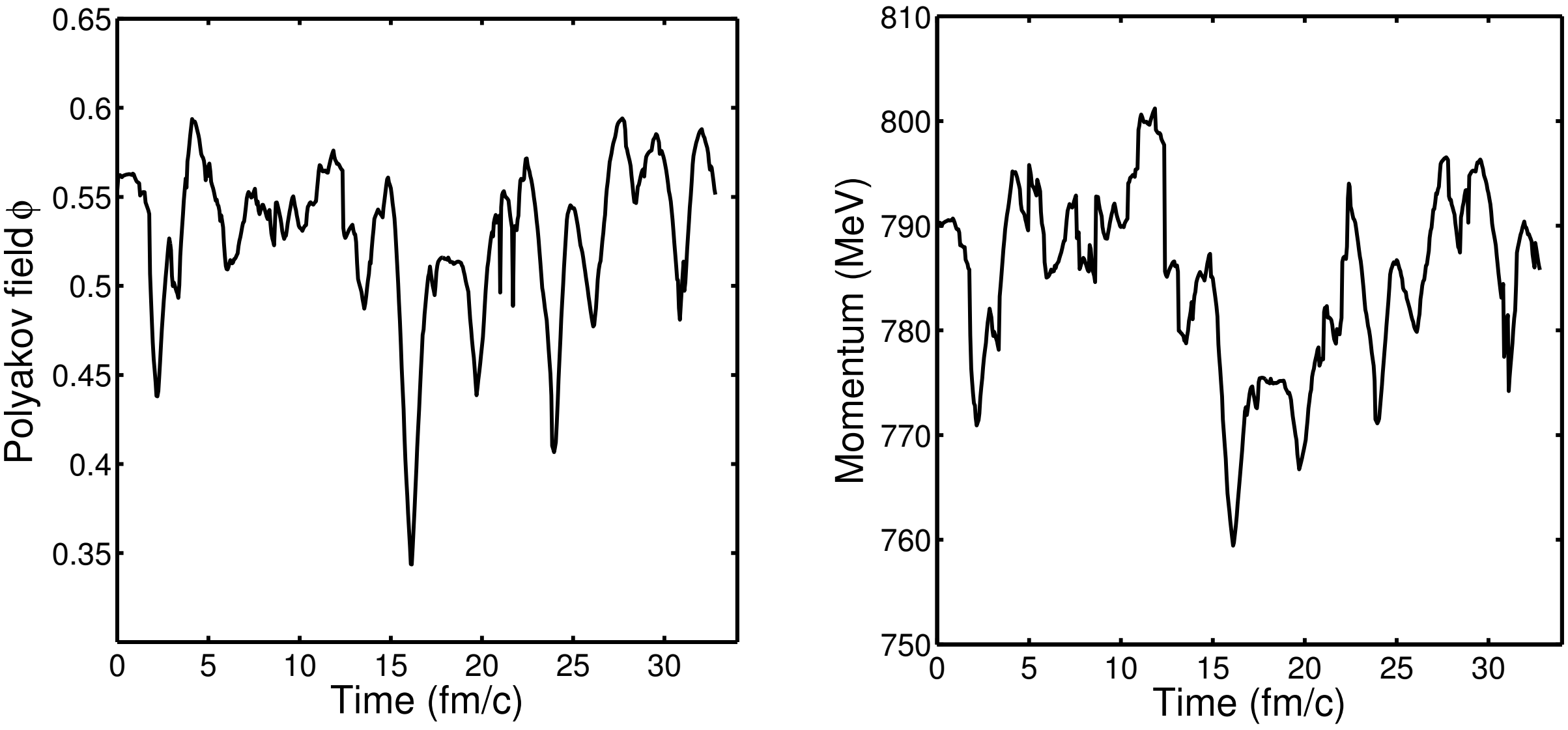

**Figure 6.** Mean Polyakov field and mean momentum of the particles, according to the time.

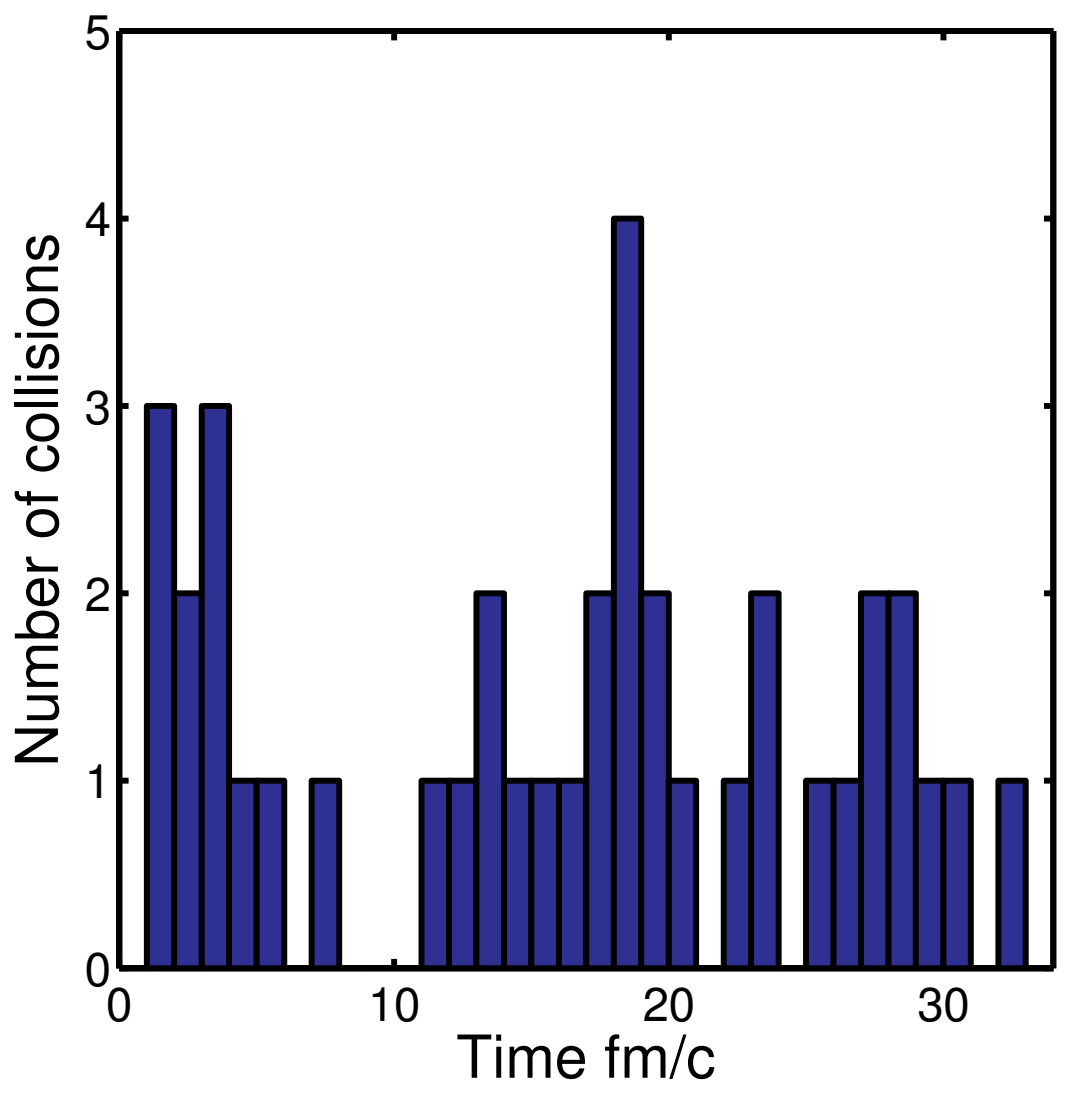

**Figure 7.** Number of collisions according to the time.

The evolution of the number of collisions according to the time is plotted in the figure 7. The collision rate is similar to the one observed in [24], in close conditions. Also, these results are in agreement with the ones of the figures 5 and 6. Indeed, even if we observe variations, the results seem to oscillate around an average one. It confirms that our results describe the evolution of a system at equilibrium. This behavior is perfectly explainable for such an isolated and closed system. In fact, this equilibrium seems to be reached since the beginning of the simulation. Upon numerical aspects, a deviation of the values would have been the sign of a possible anomaly in our algorithms, or a badly chosen iteration time. More precisely, in this test and the following ones, we consider an iteration time $\Delta t = 5\times10^{-2}$ fm/c .

To justify this choice of $\Delta t$, we consider the standard method used in statistical physics to estimate the mean free path $\lambda$. This quantity corresponds to the average distance traveled by a particle between two collisions. We interpret it as the height of a fictitious cylinder, as schematized in the figure 8 hereafter.

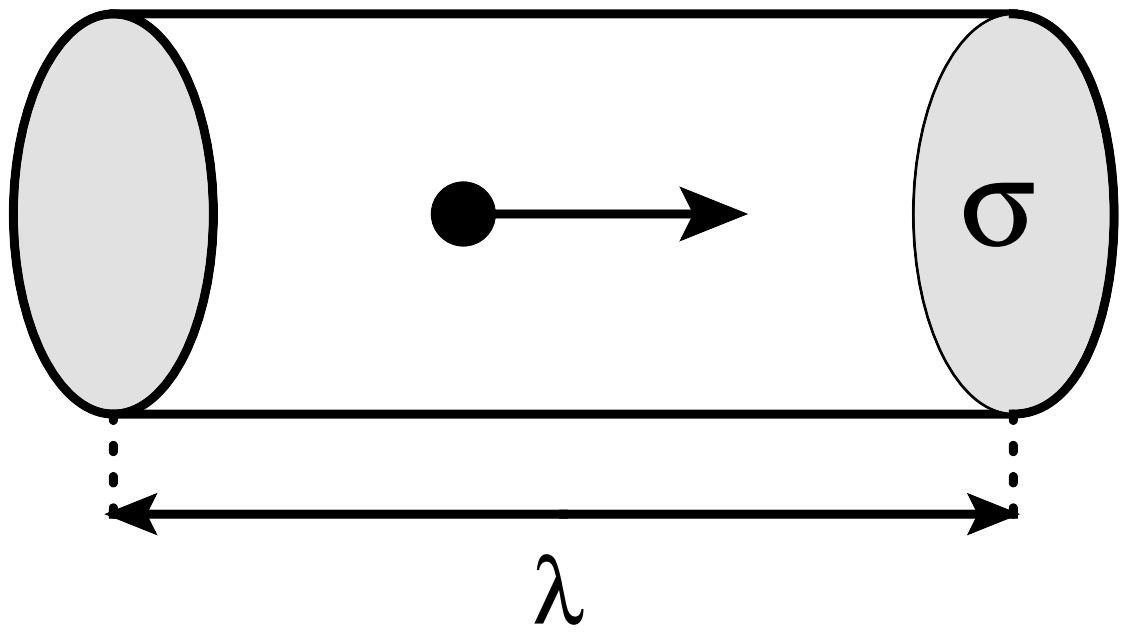

**Figure 8.** Method to evaluate the mean free path $\lambda$.

The area of the base of this cylinder corresponds to the cross section $\sigma$ associated with the collisions between the particle and the other ones of its vicinity. The volume $V$ of the cylinder is the maximum one that the particle can occupy alone. If we consider a density $\rho$ defined by the number of particles/antiparticles divided by the volume ($\neq \rho_f$ …), we have $\rho \cdot V = 1$. Moreover, $V = \sigma \cdot \lambda$, so that the mean free path is given by:

$$\lambda = \frac{1}{\rho \cdot \sigma} \quad . \tag{13}$$

In our test, $\rho = 12/2^3$ and $\sigma \approx 2\text{ mb} = 0.2\text{ fm}^2$ (quark-quark scattering), so $\lambda \approx 3\text{ fm}$. Moreover, thanks to the mean free path, we estimate the mean time $\Delta\tau$ between two collisions. Indeed, if we note $v$ the velocity of the particle, it comes $\Delta\tau = (\rho \cdot \sigma \cdot v)^{-1}$. As done implicitly in [24], we take $v \equiv 1$ (the speed of light). It leads to minimize $\Delta\tau$, so we consider the most unfavorable case. Also, we will see that this approximation will be justified with other simulations, figure 28. As a consequence, we propose:

$$\Delta\tau = \frac{1}{\rho \cdot \sigma} . \tag{14}$$

In this test, $\Delta\tau \approx 3\text{ fm/c}$. Our iteration time is $\Delta t = 5\times10^{-2}\text{ fm/c}$, thus we conclude that the condition $\Delta t < \Delta\tau$ is satisfied, i.e. the iteration time was correctly chosen. In the other simulations, higher cross-sections are considered, notably $\sigma \approx 100\text{ mb} = 10\text{ fm}^2$ for $q + \bar{q} \to M + M$ in some conditions. If we keep the value of $\rho$ used in the previous calculations (it overestimates the values really found in these simulations), we find now $\Delta\tau = 7\times10^{-2}\text{ fm/c}$, so that our choice on $\Delta t$ is still valid …

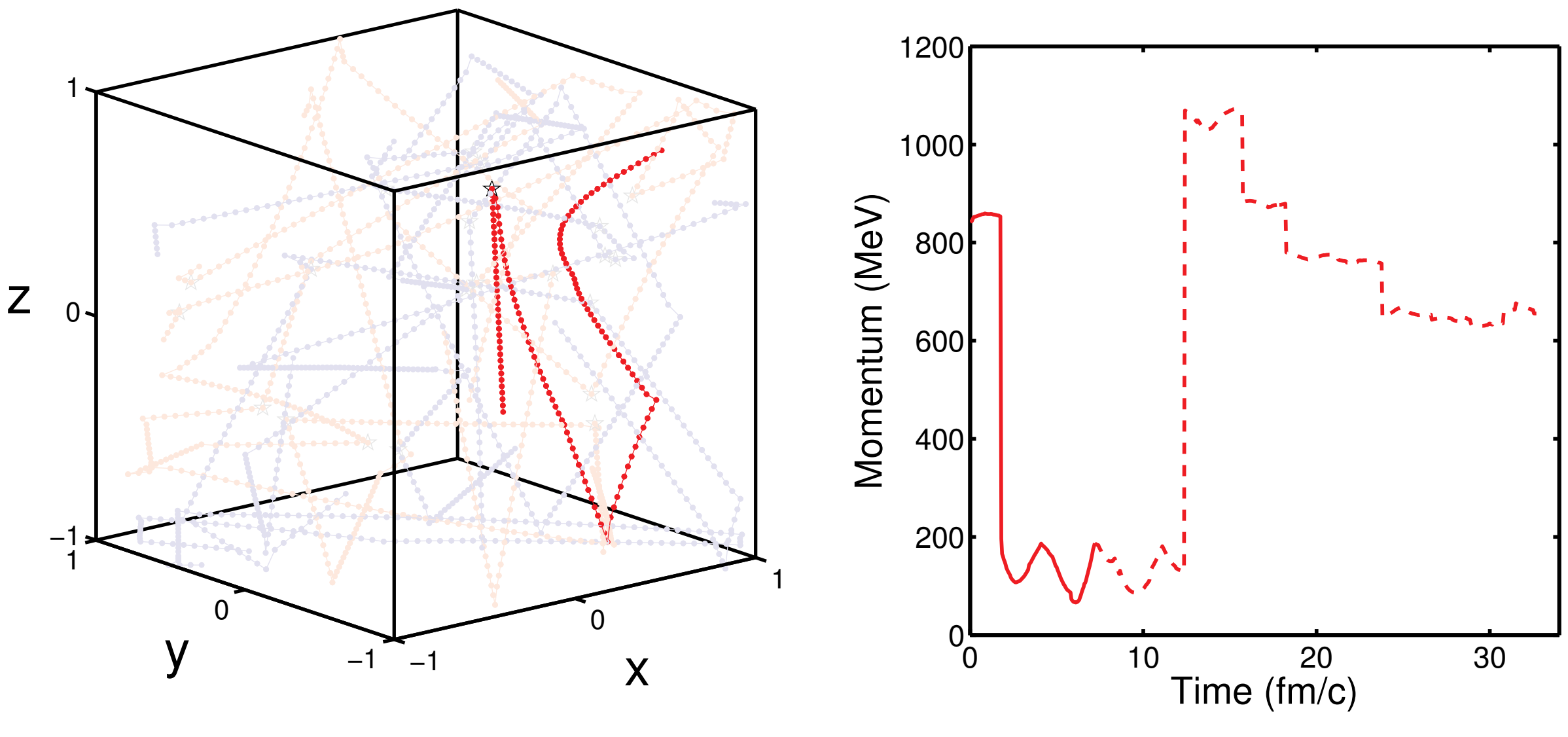

**Figure 9.** Simulation of 12 light quarks in a cube, and detail on the momentum of one of these quarks.

Now, we turn our attention to the figure 9. In the left hand side of this figure, the trajectories of the quarks are represented, in the first moments of the simulation, i.e. $0 \leq t \leq 7.2\text{ fm/c}$. We highlighted the trajectory of one of the quarks, in order to facilitate the reading of the graph. Moreover, the collisions between the quarks are marked with stars, but not the rebounds of these particles against the walls of the cube. In the right hand side of the figure 9, the momentum of the highlighted quark is plotted according to the time. The part of the curve plotted with a solid line corresponds to the simulation times that are really represented in the left hand side of the figure. At the opposite, the part of the curve in dotted line is associated with the evolution of the quark's momentum at the other moments of the simulation.

As a whole, we remark that the trajectories are straight lines. It suggests that the remote interaction identified in the previous subsection does not intervene in this simulation. It can be explained by several reasons, as the high temperatures of the medium, as explained in the subsection 3.1. More precisely, the particles' momentum is higher than in the figures 3 and 4. Clearly, the remote interaction does not have the time to really influence the quarks' motion. Then, another argument is the abundance of quarks in a reduced volume, inducing a rather constant environment. As mentioned previously, the remote interaction really intervenes only when the external parameters felt by the particles present enough variations.

Nevertheless, the underlined quark is an exception of this global behavior. In fact, its trajectory is strongly curved, especially in the right part of the tridimensional graph. This portion of the trajectory corresponds to simulation times between 3 and 7.2 fm/c. For $t \approx 7.2$ fm/c, the quark goes towards the upper right corner of the graph. The deviation of the quark was induced by the other quarks present in its vicinity, in agreement with the behavior described in the figures 3 and 4. In fact, in the right hand side of the figure 9, for $3 < t < 7.2$ fm/c, the quark's momentum is largely reduced compared to the ones of the other quarks, see figure 6, because it is less than 200 MeV. This observation confirms our previous observations: the remote interaction between quarks may really act only on slow quarks. In the framework the cooling of a quark/antiquarks plasma, involving (at least initially) high temperatures, so rapid quarks, the influence of this remote interaction appears to be rather limited. As a consequence, the collisions seem to rule the quarks' dynamics, and dominate the effects of the remote interaction. According to these results, the quarks' motion can be compared to a relativistic Brownian motion.

# 4. First simulations

## 4.1 Comparison between NJL and PNJL results

Now, we proceed to complete simulations of open systems. In other words, we do not consider a box as in the previous subsection that confines the particles. We consider a spherical system initially composed by light quarks and antiquarks, table 3 and left hand side of the figure 10. The matter dominates the antimatter. Indeed, the quarks/antiquarks ratio is close to 1.5. Also, our system does not initially content strange quarks/antiquarks. The sphere is inhomogeneous according to the quarks' momentum, as visible in the right hand side of the figure 10. This profile is comparable to the ones of [24]. High momenta are synonymous of high temperatures, so the objective is clearly to obtain a layer structure, with a hot core and colder external layers. The figure 13 shows that the obtained temperature profiles are in agreement with this description.

| Particles | $u$ | $\bar{u}$ | $d$ | $\bar{d}$ |
|---|---|---|---|---|
| Number | 30 | 20 | 36 | 25 |

**Table 3.** Initial composition of the system.

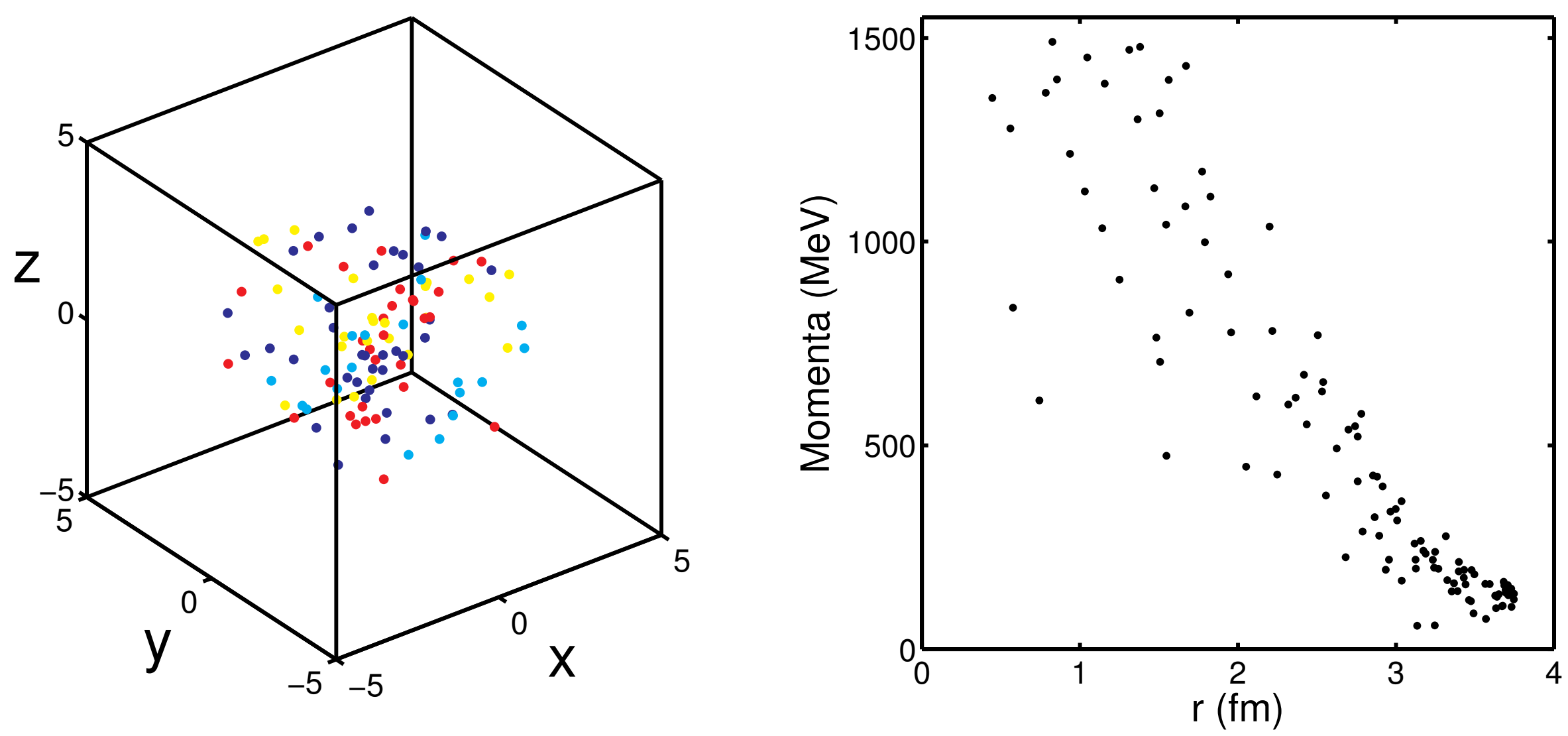

**Figure 10.** Left hand side: initial position of the quarks. Right hand side: their initial momenta according to their distance from the center of the system.

The simulation of the expansion/hadronization of this system was performed in the NJL and the PNJL models, using each time exactly the initials conditions described above. The associated results are presented in the figures 11, 12 and in the table 4. Concerning the colored versions of this thesis, these figures use the color convention explained in the appendix G. The evolution of the number of quarks/antiquarks according to the time is presented in the figure 11, whereas the figure 12 concerns mesons. These data correspond to two simulations, but we checked that other tests gave very close results. More precisely, in the used collision algorithm described subsection 2.4, the collisions are determined using stochastic considerations, even if each reaction are taken into account, and each one is weighted by is associated cross-sections in this "choice". It leads to statistic fluctuations. However, we verified that the variations concern few particles in the results displayed hereafter, typically less than 5 quarks/mesons. Moreover, about the collision procedure, aware of the relative fragility of composite anti-particles, i.e. $\bar{D}$ and $\bar{B}$, these objects were not included in these simulations. The reactions presented in table 2 that include these anti-particles were deactivated. Upon numerical aspects, the reactions were taken into account in the algorithm, but the program returned null cross-sections for them.

In these simulations, we saw that the quark remote interaction acts in a negligible way on the results. It confirms our conclusion formulated in the subsection 3.2. Clearly, as visible in the figure 10, the quarks initially located in the most external layers seems to be slow enough to undergo this interaction in a notable way. In fact, it leads to some modifications of the quarks' trajectories. However, the cross-sections can be rather high, especially with reactions as $q+\bar{q} \to M+M$ . As a consequence, the deviations of the trajectories are not strong enough to intervene on the collisions. Furthermore, the quarks from the core or from the hot layers are definitely too rapid to undergo the remote interaction. As a consequence, rectilinear trajectories were found for them.

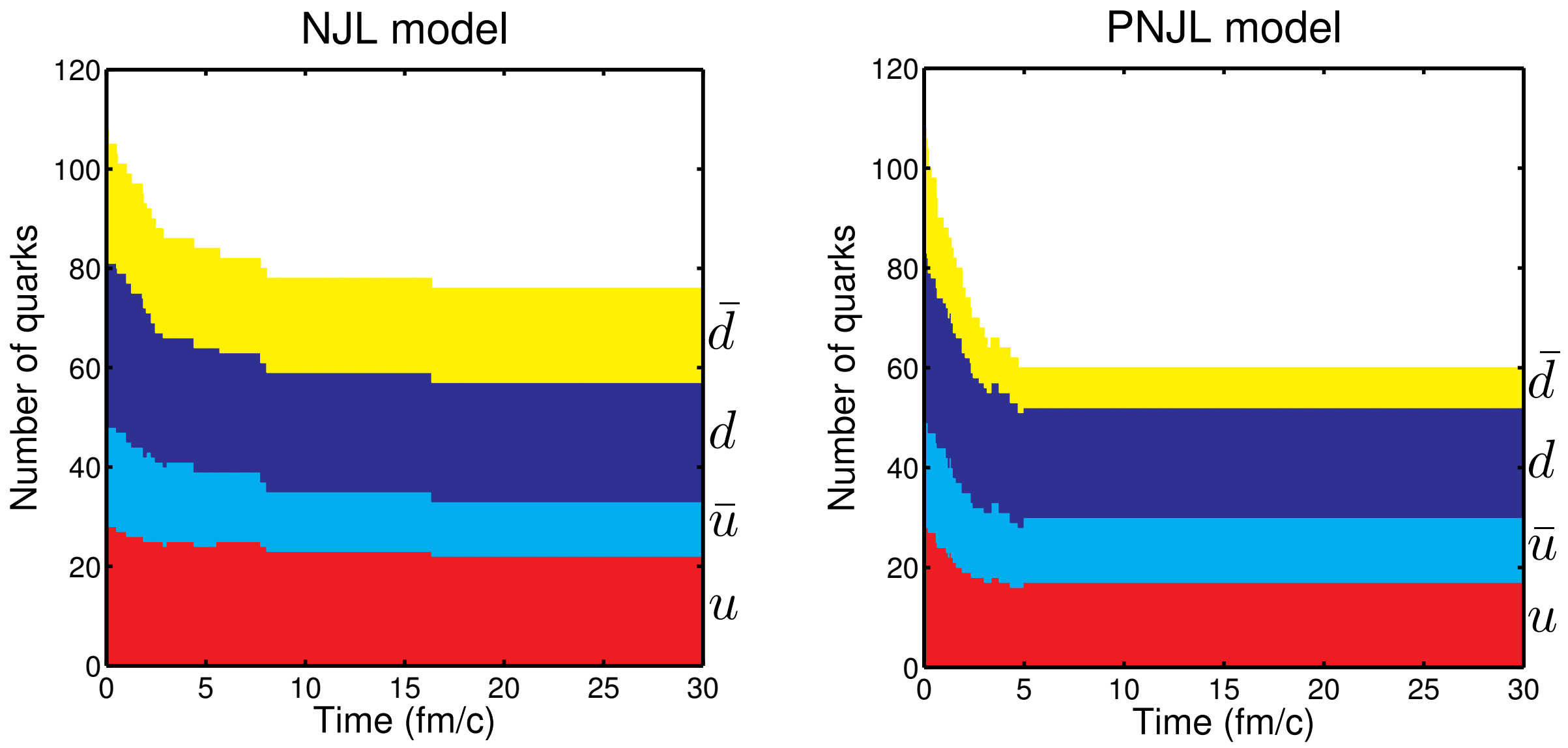

**Figure 11.** Comparison of the quarks-antiquarks consumption in the NJL and PNJL models.

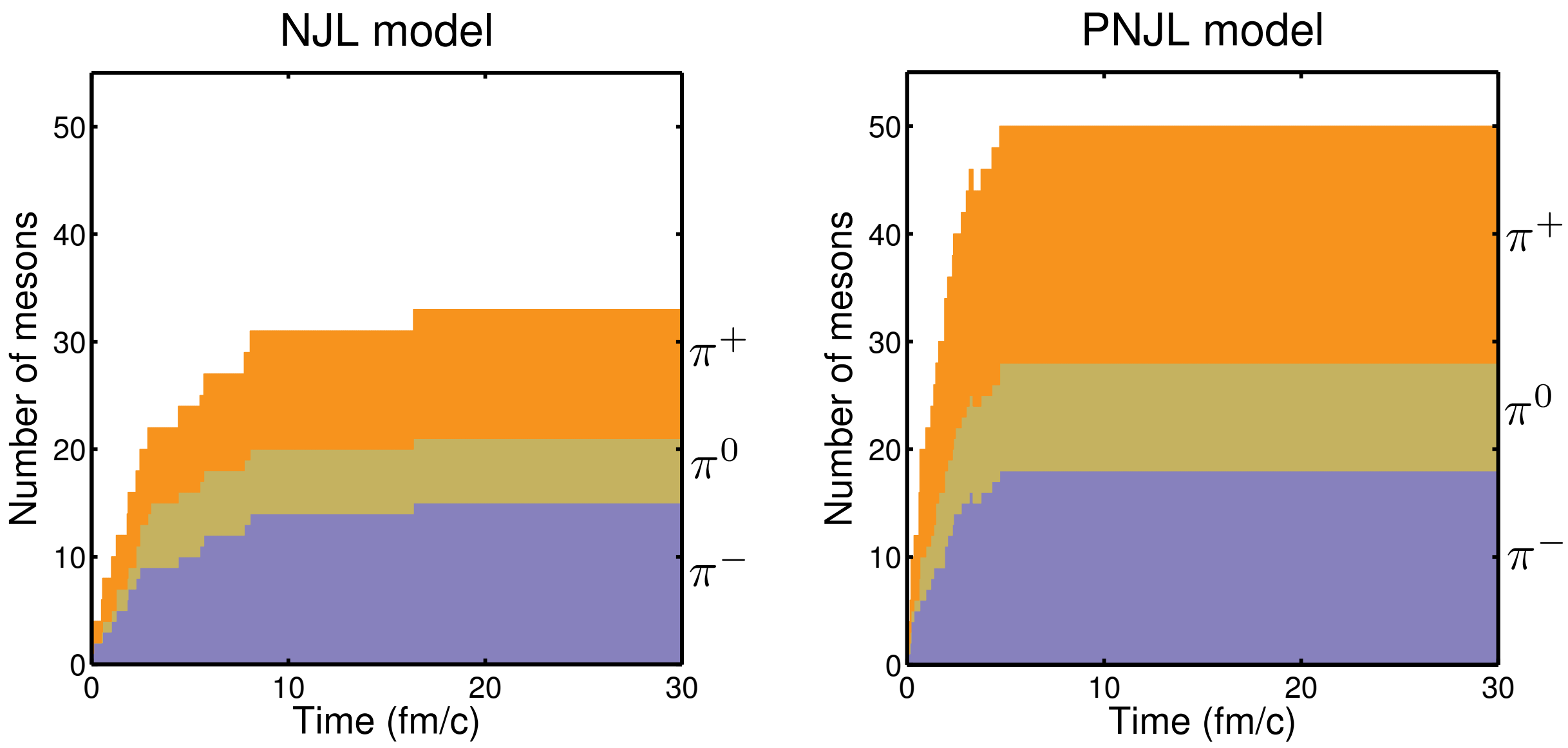

**Figure 12.** Comparison of the production of mesons in the NJL and PNJL models.

Qualitatively, the production of mesons described in the figure 12 only concerns pions. Our simulations does not consider the isospin symmetry, so $\pi^-$, $\pi^0$ and $\pi^+$ are independent. But, the results do not show production of $\eta$ or kaons. In the same way, in the figure 11, no production of strange quark or antiquark is observed. In fact, even if we do not include strange particles/antiparticles in the initial composition, they can appear in the system by the way of collisions, see table 2. However, as show in [43] in the case of $u+\bar{u}\to s+\bar{s}$ and in [35] for $u+\bar{d}\to K^+ + \bar{K}^0$ or $u+\bar{u}\to K^- + K^+$, the cross-sections of reactions producing strange matter are reduced compared to the ones that exclusively use light particles. It explains the absence of strange matter in these simulations, even if it was theoretically possible to observe strange particles. In fact, in some tests similar to the ones presented here, we noted the production of few kaons.

About the evolution of the population of quarks/antiquarks and mesons, we firstly note that their populations do not seem to vary after $t > 20$ fm/c. This remark is valid for the NJL or the PNJL models. For the both, the simulations ended at $t = 30$ fm/c. So, this simulation time corresponds to the "final state" of the system. The table 4 shows its composition at this simulation time. In these simulations the production of diquarks and baryons are strongly reduced. Therefore, the evolution of the system can be described only via the figures 12 and 13, i.e. via the quarks/antiquarks and the mesons. As a whole, the production of the mesons is optimal in the first moments of the simulation. It induces a diminution of the quarks/antiquarks' population in a symmetrical way. This strong mesonization is explained by the high concentration of quarks and antiquarks, leading to an important collision rate. As seen previously, the attractive quark interaction described in the subsection 3.2 does not really intervene in a notable way. As a consequence, the system extends spatially. The system is open, thus such an expansion is without limitation. It leads to a dilution of the particles, so a diminution of the collision rate. When the expansion becomes too strong, the particles do no longer interact. It explains the observed stagnation when $t > 20$ fm/c.

| Particles | Quarks | Mesons | Diquarks | Baryons |
|---|---|---|---|---|
| NJL | 76 | 33 | 1 | 1 |
| PNJL | 60 | 50 | 0 | 1 |

**Table 4.** Composition of the system at $t = 30$ fm/c.

Quantitatively, differences are found between the NJL and the PNJL models. It constitutes an important aspect of our results. The mesons' production is more rapid and more efficient in the PNJL model than in a pure NJL one. More precisely, the number of mesons stagnates in the PNJL model for $t > 5$ fm/c, whereas the stagnation intervenes after 15 fm/c in the NJL model. In addition, the production of mesons is equal to 50 in the PNJL description, against 33 with the NJL one. Such a difference can be explained with the figures 13 and 14 hereafter. The figure 13 plots the initial temperature felt by each quark/antiquark. The figure 14 studies their initial masses. The two figures allow a comparison between the two models. Clearly, even if the initial positions and momenta were strictly equal for the NJL and the PNJL simulations, figure 10, the estimation of the quarks' masses and the temperatures does not give the same results in the two models. Moreover, in our dynamical model, we recall the interdependence between the mass and the temperature: the quarks' masses are estimated at a given temperature (see chapter 2) and the temperatures require the quarks' masses in the evaluation of the energies (9). So, these two quantities cannot be studied separately. As a consequence of the inclusion of the Polyakov loop, the initial temperatures are lower in the PNJL model than in the NJL one. At the opposite, the masses are stronger in the PNJL description than in the NJL one. These differences are particularly visible in the core of the system, i.e. $r < 1.5$ fm.

In fact, it can be noted that the mass difference between the (P)NJL models intervenes in the first equation of (11), via the energy. For identical momentum, a PNJL quark is heavier, so the term $p/E$ is expected to be more reduced for this quark. As a consequence, $\frac{dr}{d\tau}$ should be more reduced. Therefore, the velocity of the PNJL quark should be more reduced than the NJL one. So, it should reduce the velocity of the expansion. However, most of the quarks are highly relativistic. So, this explanation cannot explain the found differences.

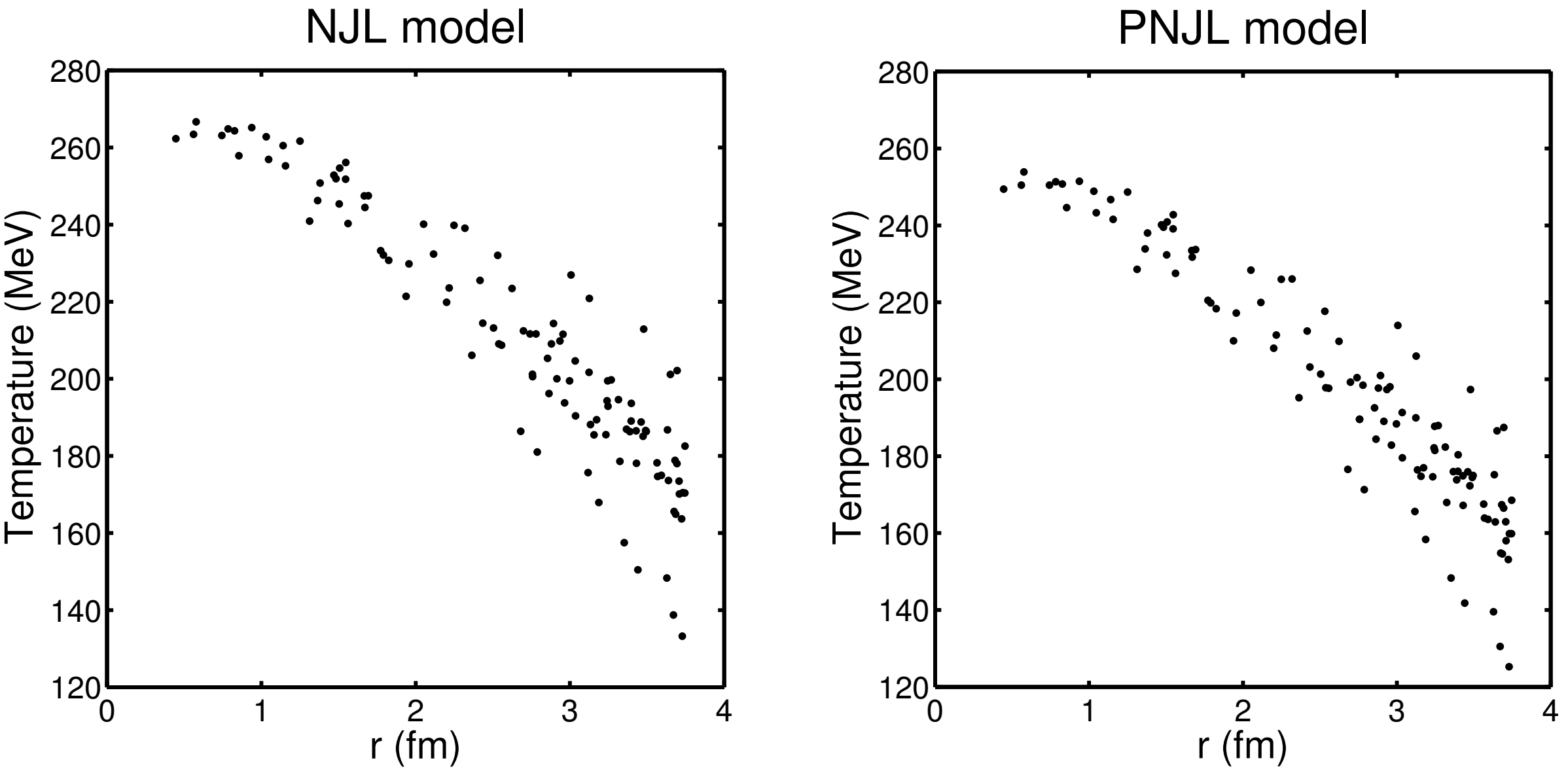

**Figure 13.** Initial temperatures in the NJL and PNJL models.

Clearly, a more relevant explanation directly concerns the cross sections values. According to the work performed in the previous chapter, the optimal temperature (at null density) of the mesonization process, via $u+\bar{u}\rightarrow\pi^{+}+\pi^{-}$, seems to be close to 230 MeV for the NJL model, against 280 MeV for the PNJL one. In fact, $u+\bar{u}\rightarrow\pi^{+}+\pi^{-}$ is the dominant reaction of the ones written as $q+\bar{q}\rightarrow M+M$, and these results can be extrapolated to the other mesonization reactions of this kind, involving light quarks and pions [35]. In addition, the elastic reaction $u+\bar{u}\rightarrow u+\bar{u}$ is optimal for a temperature 20 MeV above the ones found for the mesonization process, in the NJL and PNJL descriptions. At the light of these data and with the left hand side of figure 13, for the NJL simulation, we conclude that a significant part of the system is initially too hot to undergo the mesonization process. Clearly, in the core, the elastic reactions between the quarks and/or the antiquarks should have dominated the inelastic mesonization reactions, at least in the first instants of the expansion. It leads to a reduced production of mesons, which mainly occurred in the external layers. At the opposite, for the PNJL simulation, right hand side of the figure 13, all the system is initially below $T=280\text{ MeV}$. Even if the mesonization is not optimal at modest temperatures, we still have rather strong cross-sections values, especially near the kinematic threshold [35, 44]. As a consequence, in the PNJL simulation, the mesonization can start in all the system since the beginning of the simulation, leading to a more rapid and a more efficient meson production.

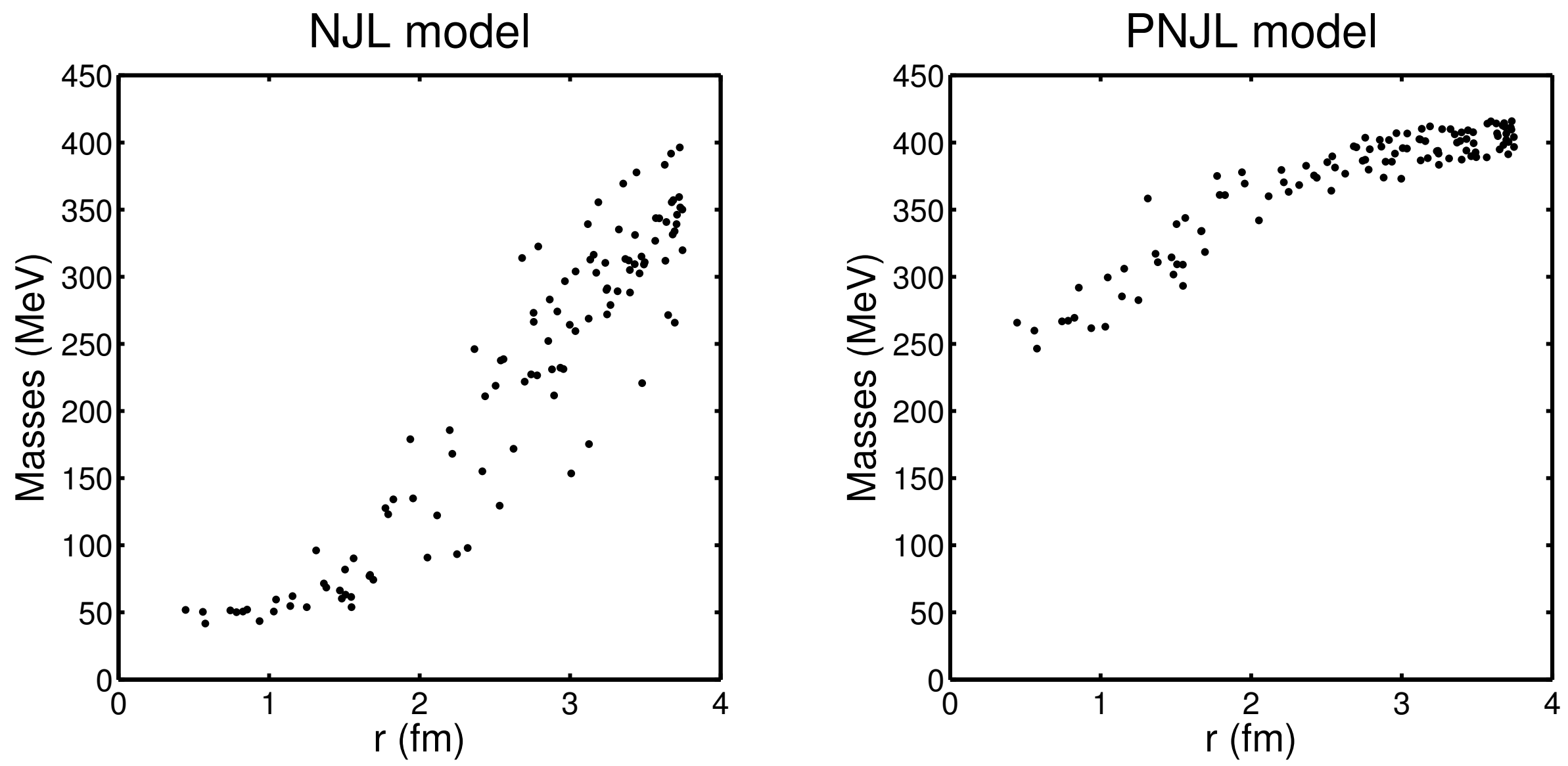

**Figure 14.** Initial masses in the NJL and PNJL models.

However, even the PNJL model did not reach a complete hadronization in these simulations. More precisely, 68 % of the quarks/antiquarks are still free in the end of the simulation with the NJL model, and 54 % with the PNJL one. Even if it is imaginable to enhance these results, notably at the level of the system composition, geometry, initial momenta, etc., the (P)NJL models alone do not seem to be able to allow a full hadronization of such a system. In fact, an explanation of this phenomenon could be associated with the relative weakness of the cross-sections of the baryonization processes, notably compared to the cross-sections found for $q+\bar{q} \rightarrow M+M$ . So, it is realistic to reach high hadronization yields with systems composed equitably by quarks and antiquarks. In these systems, a total hadronization is possible exclusively via a mesonization, i.e. by neglecting baryonization reactions, as in [24]. In fact, we saw previously the possibility to model a short range interaction between quarks. But, this one proved to be too limited. As a consequence, our results show the necessity of a long range interaction between the quarks/antiquarks, as the one observed in the framework of the QCD. Clearly, the finality of such an interaction is to reduce the velocity of the system expansion, at least for the quarks/antiquarks, in order to allow the quarks to combine themselves, to form diquarks and then baryons.

## 4.2 A solution to allow a complete hadronization

To solve the problem evoked in the previous subsection, several solutions are possible. The finality is to find a mechanism that can model the mentioned long range interaction. It could lead to a modification of the used (P)NJL models. However, in this work, we propose to add a sphere that will confine the quarks-antiquarks system described in the figure 10 in a "QGP phase", and so it mimics the behavior of a long range spring-like force. Like the box in the subsection 3.2, the wall of the sphere is reflective according to the Snell-Descartes' law on the reflection. However, the only concerned particles are quarks, antiquarks and the diquarks gathered in this sphere. At the opposite, the non-colored particles, i.e. mesons and baryons, can leave the sphere freely. The populations of quarks, antiquarks and diquarks are obviously expected to decrease according to the time. Thus, the radius of the sphere is updated for each iteration time. More precisely, the volume of the sphere is proportional to the total number of quarks/antiquarks, free or combined into diquarks. This QGP phase is expected to present a spherical symmetry during the simulation. So, the center of the sphere coincides with the center of the system.

The evolution of the populations of each particle's type is proposed in the figures 15 to 17. In this simulation, we used the PNJL model. It is observable that the effect of the sphere leads to the complete hadronization, because the total number of quarks/antiquarks converges towards zero (and reaches zero), and there is no diquark in the end the simulation. In fact, the hadronization was fully completed at the time 86.6 fm/c. This time is longer than the one expected in references as [16, 24]. However, the evolution of the number of quarks/antiquarks and mesons observed in the figures 15 and 16 recalls the one predicted in [16]. Indeed, the mesons production is strong in the first moments of the simulation, until $t \approx 8$ fm/c, via $q+\bar{q} \to M+M$ reactions. Then, the production begins to be less rapid, until $t \approx 25$ fm/c. After that, the variations of the number of mesons are slow, but rather regular. But, in this last phase, the production of mesons is ruled by reactions as $q+D \to M+B$, and not by $q+\bar{q} \to M+M$. In fact, even if the sphere avoids that the quarks/antiquarks leave the QGP phase, the collision rate tends to decrease according to the time. As observed in the figure 25 with another simulation, this diminution cannot be associated with a modification of the conditions in the QGP phase. Indeed, in was found that an indirect effect of the sphere is to stabilize the temperature and the densities of this phase. The reason of the diminution of the collision rate is related to kinematic considerations. Clearly, with reactions as $q+\bar{q} \to M+M$, the cross sections are maximum near the kinematic threshold. It wants to say that the probability of creating mesons is optimal if the momenta of the incoming quark and antiquark are reduced in the center of reference frame of these two particles, appendix F. In the beginning of the simulation, the great number of quarks/antiquarks makes highly probable the satisfaction of this kinematic condition. But, as soon as the number of quarks and antiquarks decrease, the probability decreases also. This behavior can be compared to the processes ruled by the exponential distribution. Furthermore, rapid quarks/antiquarks are clearly not favored as regards this kinematic condition. They may stay in the QGP phase for long time before their hadronization. However, elastic reactions allow decreasing their momenta, and they lead these particles to react more easily via inelastic reactions.

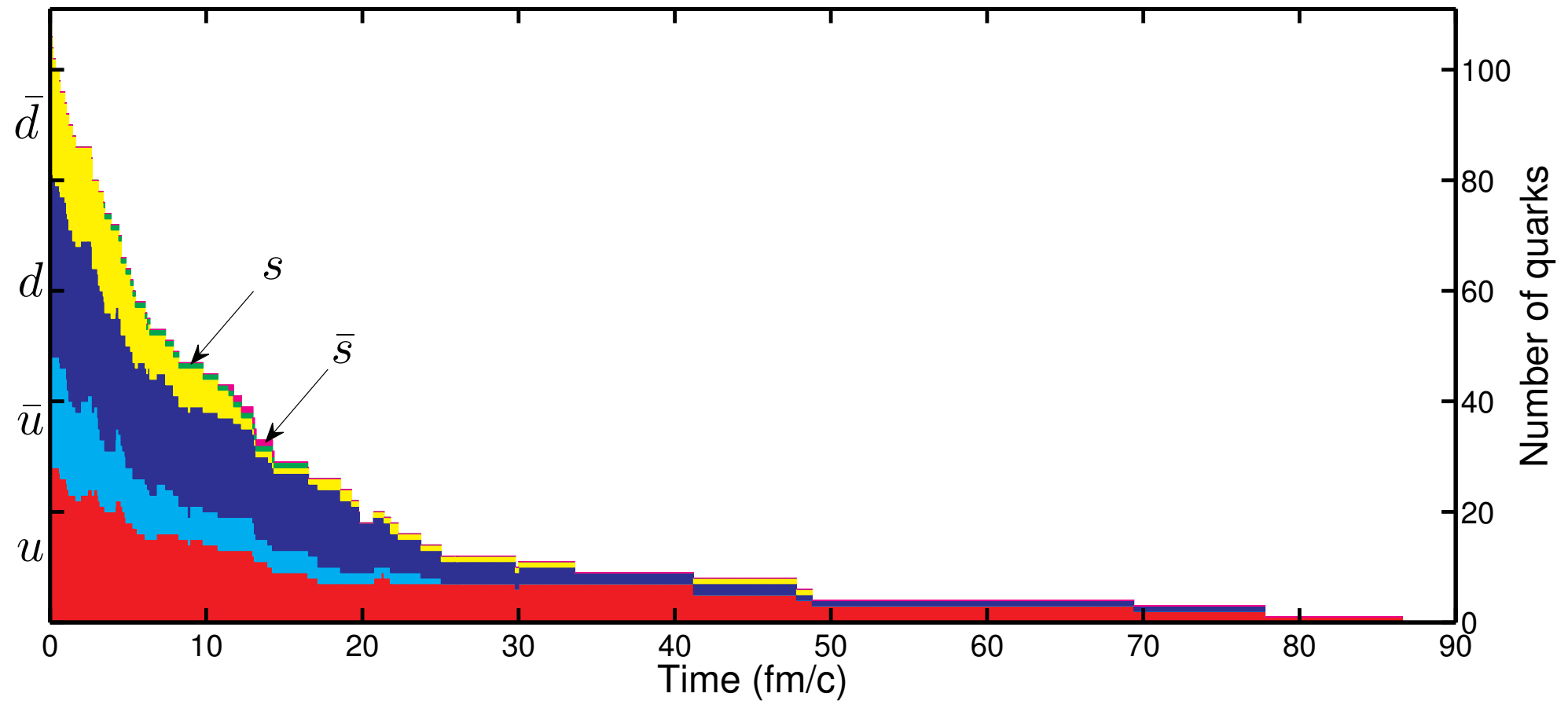

**Figure 15.** Evolution of the quarks-antiquarks population according to the time.

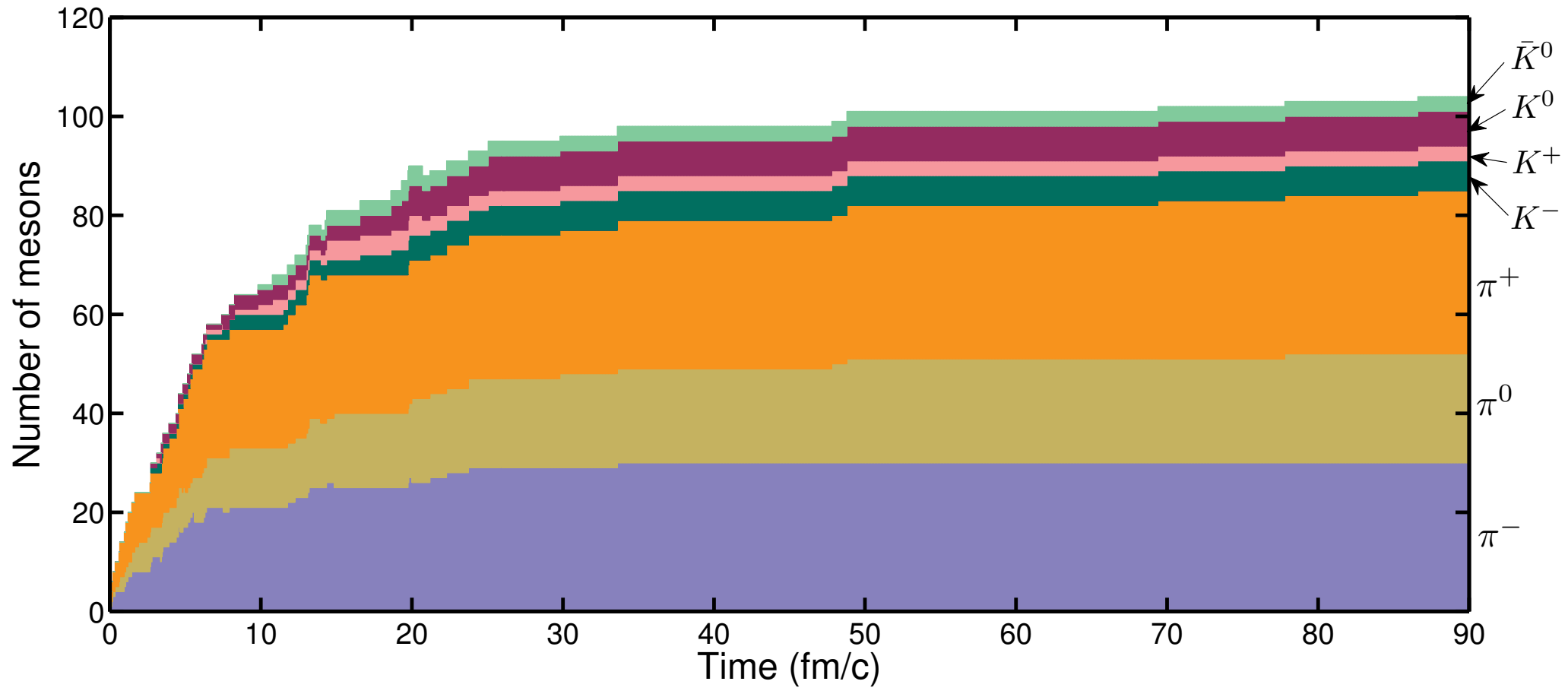

**Figure 16.** Mesons' production according to the time.

Moreover, rapid quarks/antiquarks can also be good candidates to react via reactions involving strange matter. As seen previously, we quote $u+\bar{u} \rightarrow s+\bar{s}$, $u+\bar{d} \rightarrow K^{+}+\bar{K}^{0}$ or $u+\bar{u} \rightarrow K^{-}+K^{+}$. These reactions have reduced cross-sections, but their kinematic thresholds are higher compared to reactions involving exclusively light particles. They cannot be neglected for moderate $\sqrt{s}$ values [35, 43]. The inclusion of the sphere avoids that the rapid quarks/antiquarks quit the system, and in the same time it allows increasing the number of collisions during the whole simulation. As a consequence, the figure 15 shows that strange quarks and antiquarks were produced, e.g. via $q+\bar{q} \rightarrow s+\bar{s}$. Furthermore, a production of kaons was observed, figure 36 and table 5. As expected, this production is reduced compared to the one of pions. In the same way, figure 37 and table 5, a $\Sigma^{+}$ was formed.

| Particles | $\pi^-$ | $\pi^0$ | $\pi^+$ | $K^-$ | $K^+$ | $K^0$ | $\bar{K}^0$ | *Neutron* | *Proton* | $\Sigma^+$ |
|---|---|---|---|---|---|---|---|---|---|---|
| Number | 30 | 22 | 33 | 6 | 3 | 7 | 3 | 4 | 2 | 1 |

**Table 5.** Final composition of the system.

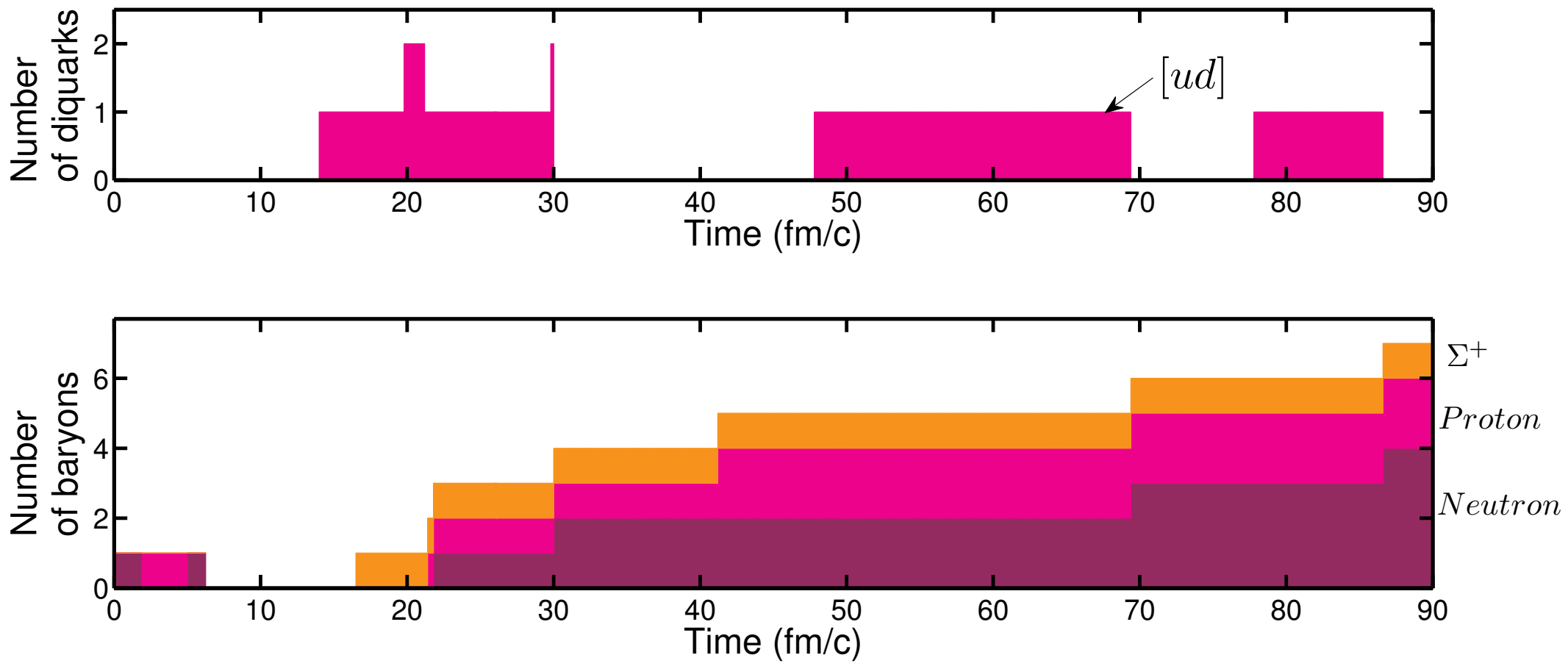

**Figure 17.** Number of diquarks and baryons according to the time.

Concerning the baryons, their production started later in this simulation, i.e. after 16 fm/c. In this description, we do not take into account the nucleons formed in the first moment of the simulation, because they were rapidly destroyed by inelastic processes. These observations confirm our scenario that imagined that the mesonization would occur before the baryonization, in order to "consume" enough antiquarks to block inelastic process that can destroy baryons, as $\bar{q}+B \to q+q$ and $\bar{q}+B \to M+D$. Moreover, no free antiquark (not combined into mesons) was found in the system firstly for $t \approx 33{,}6$ fm/c and then definitively from $t \approx 48{,}8$ fm/c. Clearly, the mesonization, via $q+\bar{q} \to M+M$, was completed earlier than the baryonization.

# 5. Complete study of a simulation

The previous simulations involved a rather reduced number of particles. As a consequence, the modest diquark production observed in the figure 17 cannot be considered as a general result. So, we performed another PNJL simulation, involving 279 particles. The initial composition is presented in the table 6. Compared to the previous simulations, table 3, the asymmetry between matter and anti-matter is enhanced. More precisely, the quarks/antiquarks ratio is now equal to 2. As previously, no strange quark/antiquark is initially present in the system. The initial positions and mometa of the particles are represented in the figure 18. In the previous simulation, the initial radius was close to 3.8 fm. Now, this one is about 4.5 fm. Moreover, as visible in the figure 18, the maximal momenta can largely exceed 1500 MeV, i.e. more than the momenta visible in figure 10.

| Particles | $u$ | $\bar{u}$ | $d$ | $\bar{d}$ |
|---|---|---|---|---|
| Number | 86 | 43 | 100 | 50 |

**Table 6.** Initial composition of the system.

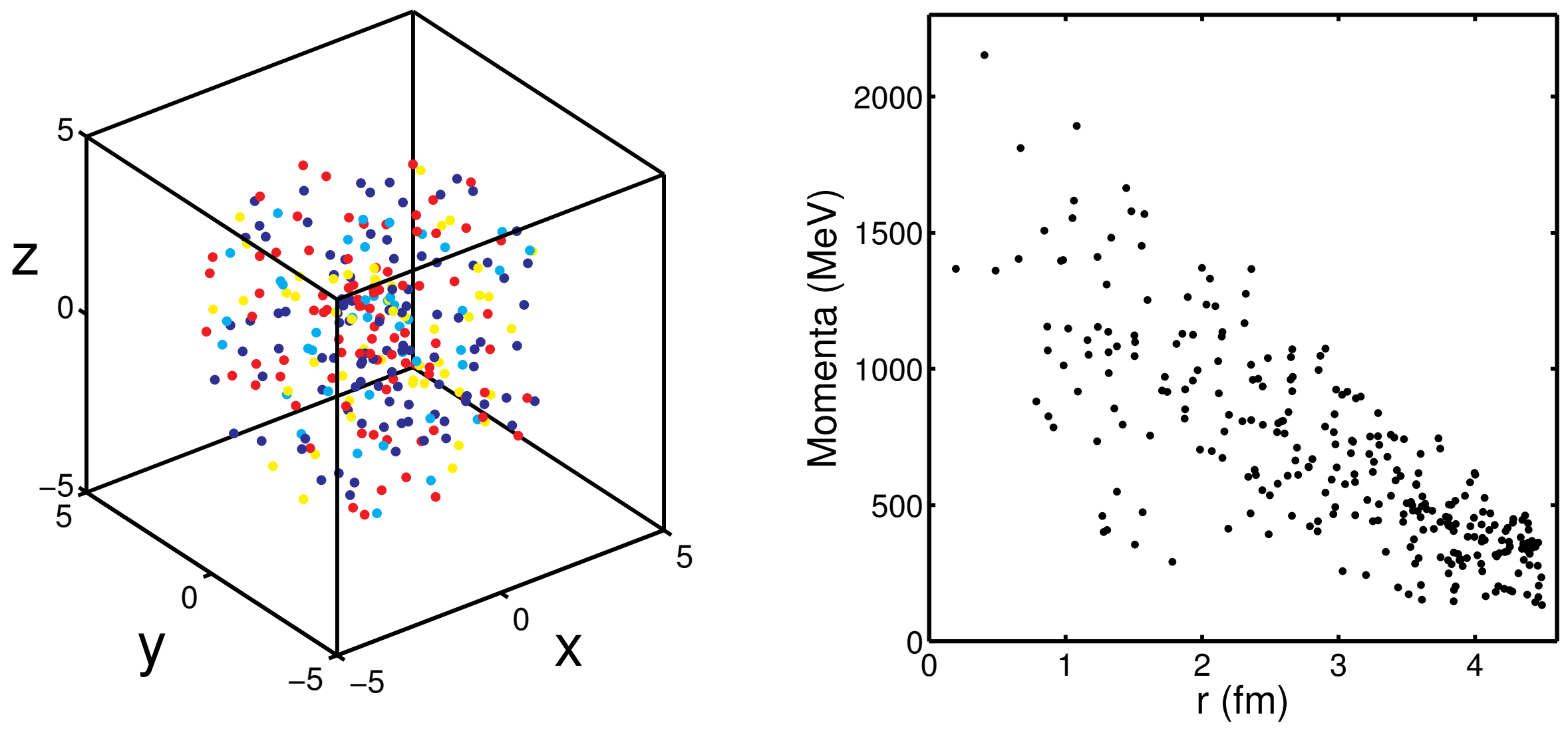

**Figure 18.** Initial position and momenta of the quarks-antiquarks forming our system.

These initial conditions lead to a hot and dense system. This affirmation is confirmed by the figures 19–21. Firstly, in the right hand side of the figure 19, the temperature in the core can exceed 300 MeV. It wants to say that we are there above the optimal mesonization temperature $T = 280\,\text{MeV}$. The temperature regularly decreases as soon as the radius is growing up. In the most external layers, the values of the temperature are greater than 180 MeV. In the left hand side of the figure 19, the masses of the quarks/antiquarks appear to be reduced in the core, but stay greater than their naked masses. Indeed, in the framework of the PNJL model, the masses of the naked quarks can only be reached for temperatures greater than or equal to 400 MeV, chapter 2. However, the values of the Polyakov field $\Phi$ are close to 0.7 in the core, figure 21. So, we can consider that the quarks/antiquarks are there in a "deconfined" regime.

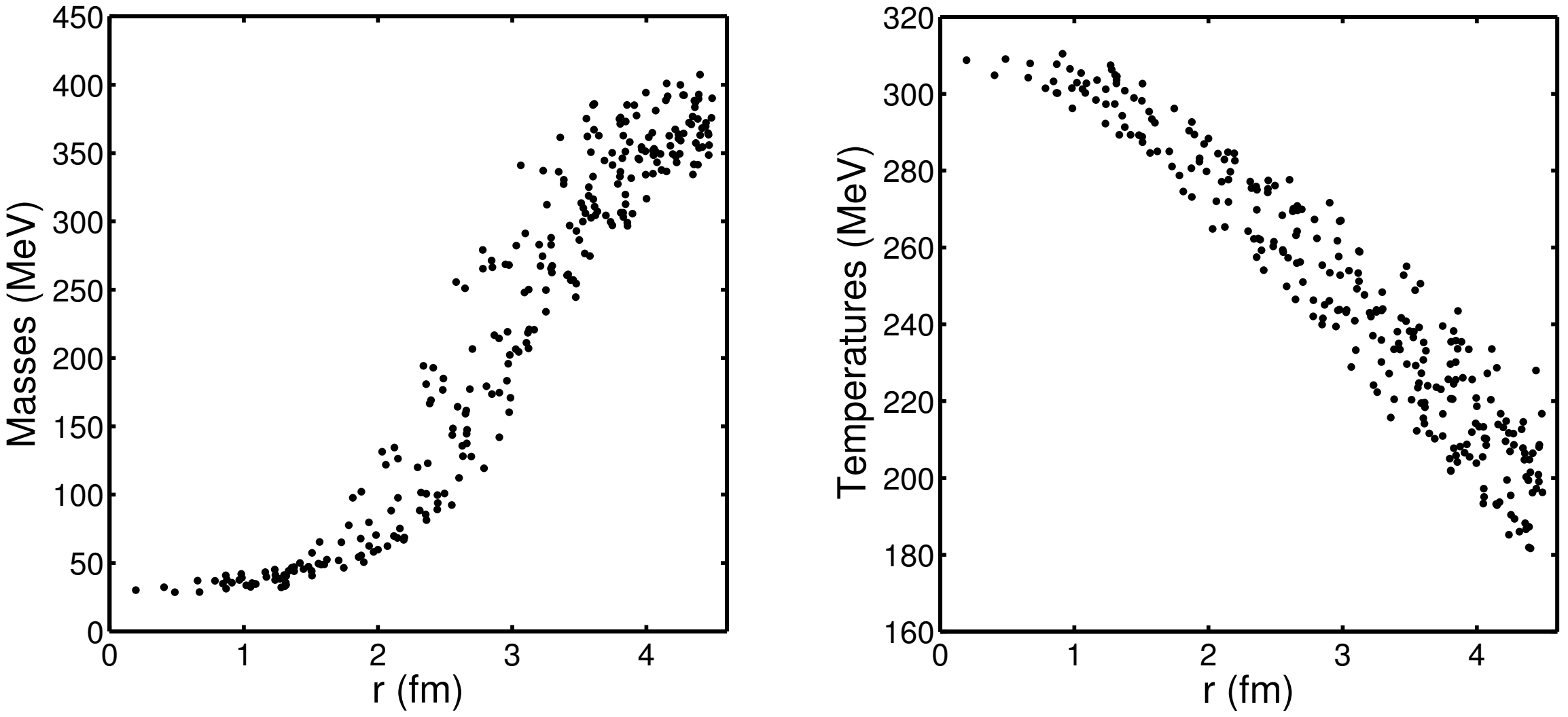

**Figure 19.** Initial masses and initial temperatures of the quarks-antiquarks.

Moreover, in the previous simulations, the densities were not taken into account in our descriptions. But, in the framework of this simulation, the densities are found to be strong, as proved by the figure 20. The profiles of the densities $\rho_u$ and $\rho_d$ are similar, but the densities found for $\rho_d$ are slightly upper than the values of $\rho_u$, because of the excess of $d$ quarks compared to $u$ ones, table 6. In fact, thanks to the relation $\rho_B = 2/3\,\rho_q$ [26] used in the framework of the isospin symmetry with $\rho_s \equiv 0$, we conclude that the baryonic density is close to $3\rho_0$ in the core.

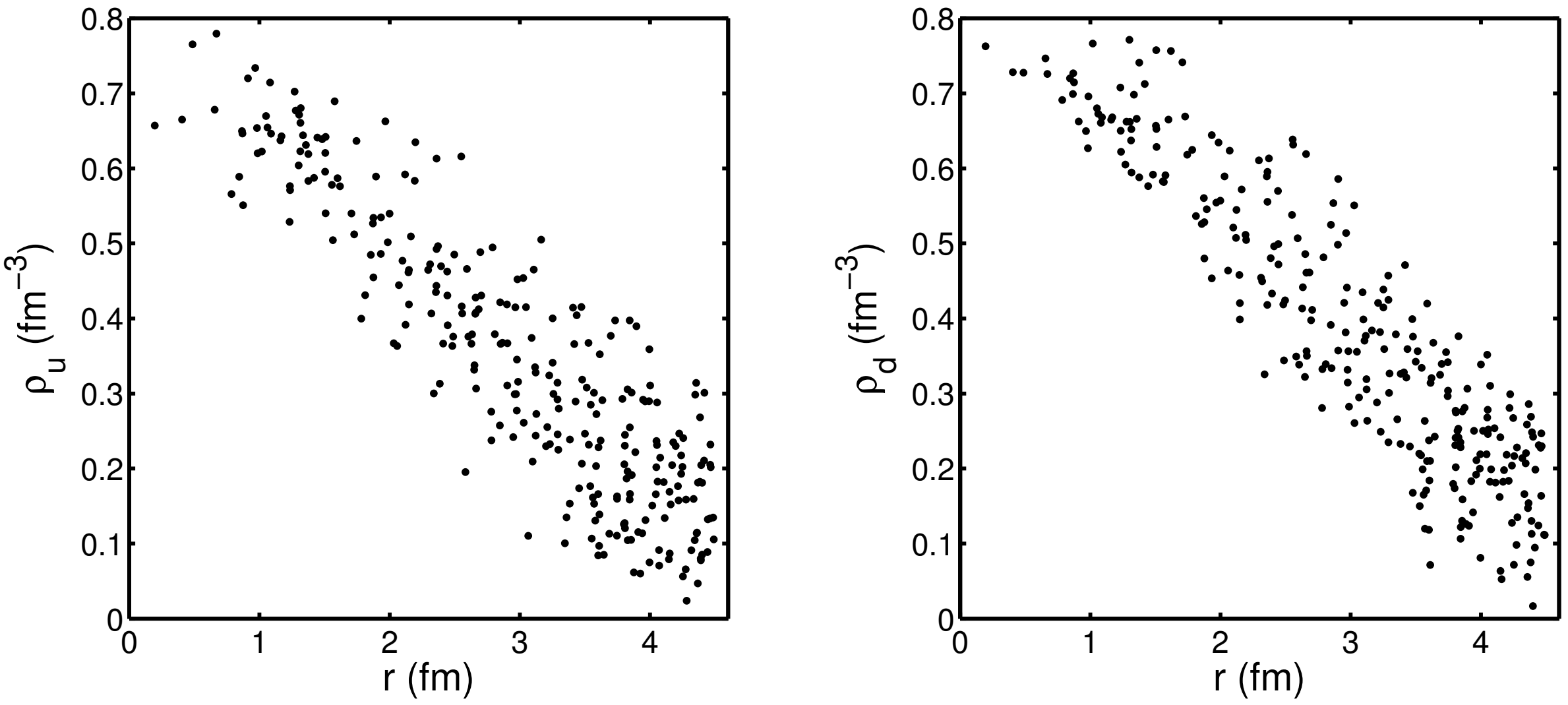

**Figure 20.** Initial densities.

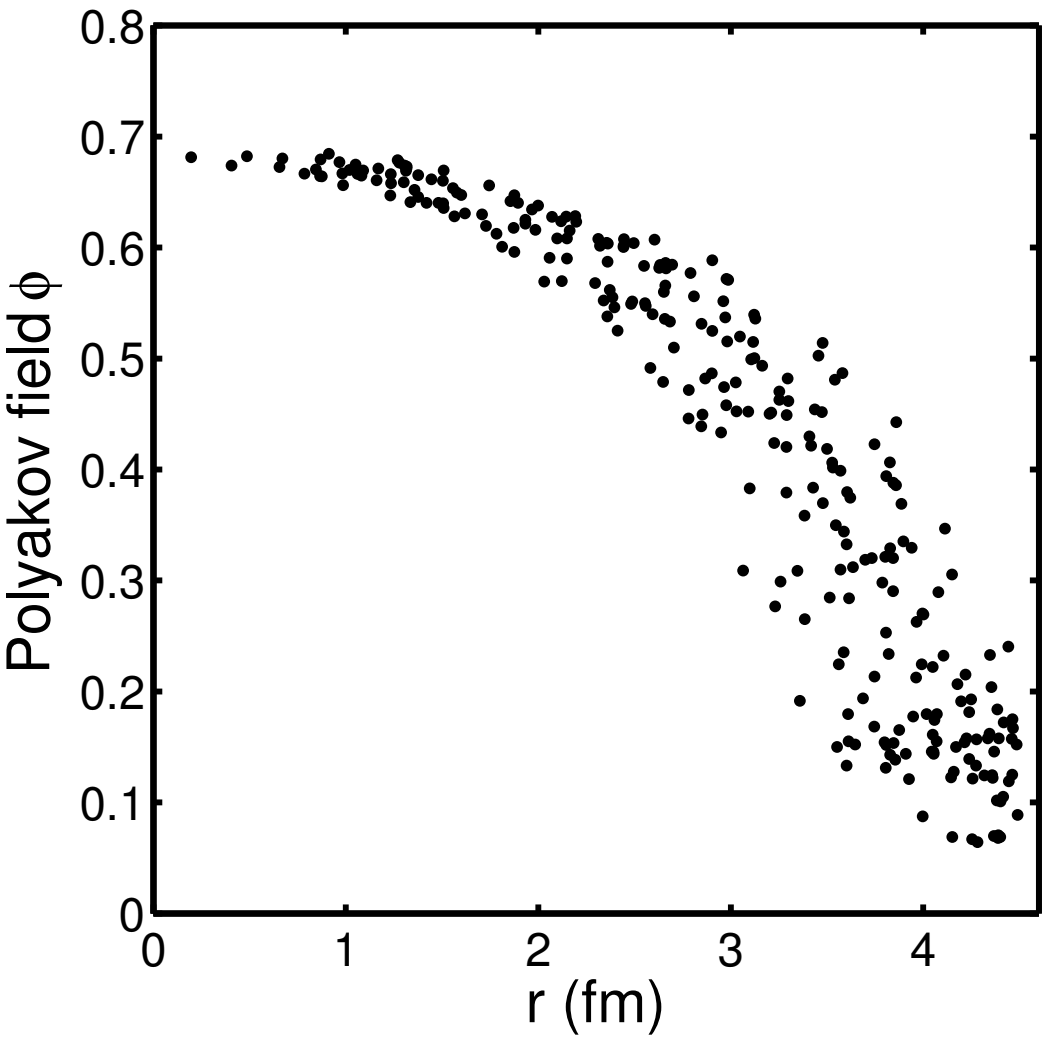

**Figure 21.** Initial values of the Polyakov field felt by the quarks and antiquarks.

The appendix G proposes a visualization of the particles' positions according to the time. Moreover, the evolution of the populations of the particles is represented in the figure 22. Concerning the quarks/antiquarks and the mesons, the observed evolutions strongly recall the ones found in the previous simulation, figures 15 and 16. Roughly speaking, the population of quarks/antiquarks is decreasing exponentially. However, about the diquarks and the baryons, differences are observable compared to the figure 17. In the former simulation, the diquarks were punctually produced. Here, a massive production of diquarks occurs in the first moments of the expansion, until $t \approx 10$ fm/c. During about 15 fm/c, the number of diquarks stagnates, and then decreases exponentially. About the baryons, their formation really starts at 3.7 fm/c. At first, the number of baryons is reduced, until $t \approx 25$ fm/c. Then, the production becomes stronger. This behavior is directly explained by the consumption of the diquarks, in order to form the baryons. After 60 fm/c, the production begins to be more reduced, and stagnates after 102 fm/c. In fact, the evolution of the diquarks and baryons' populations strongly recall what are observed in chemistry. More precisely, the diquarks perfectly play the role of intermediate states. Clearly, by their non-negligible production, they allow a more efficient formation of baryons. It is notably true for $30 < t < 60$ fm/c. In this simulation, this behavior is explainable by the temperatures and densities obtained in some layers of the system, allowing reactions as $q + q \to M + D$. Indeed, it was seen in the chapter 6 that these reactions have optimal cross sections for densities close to $2 - 3\rho_0$ and temperatures close to 200 MeV, or slightly more.

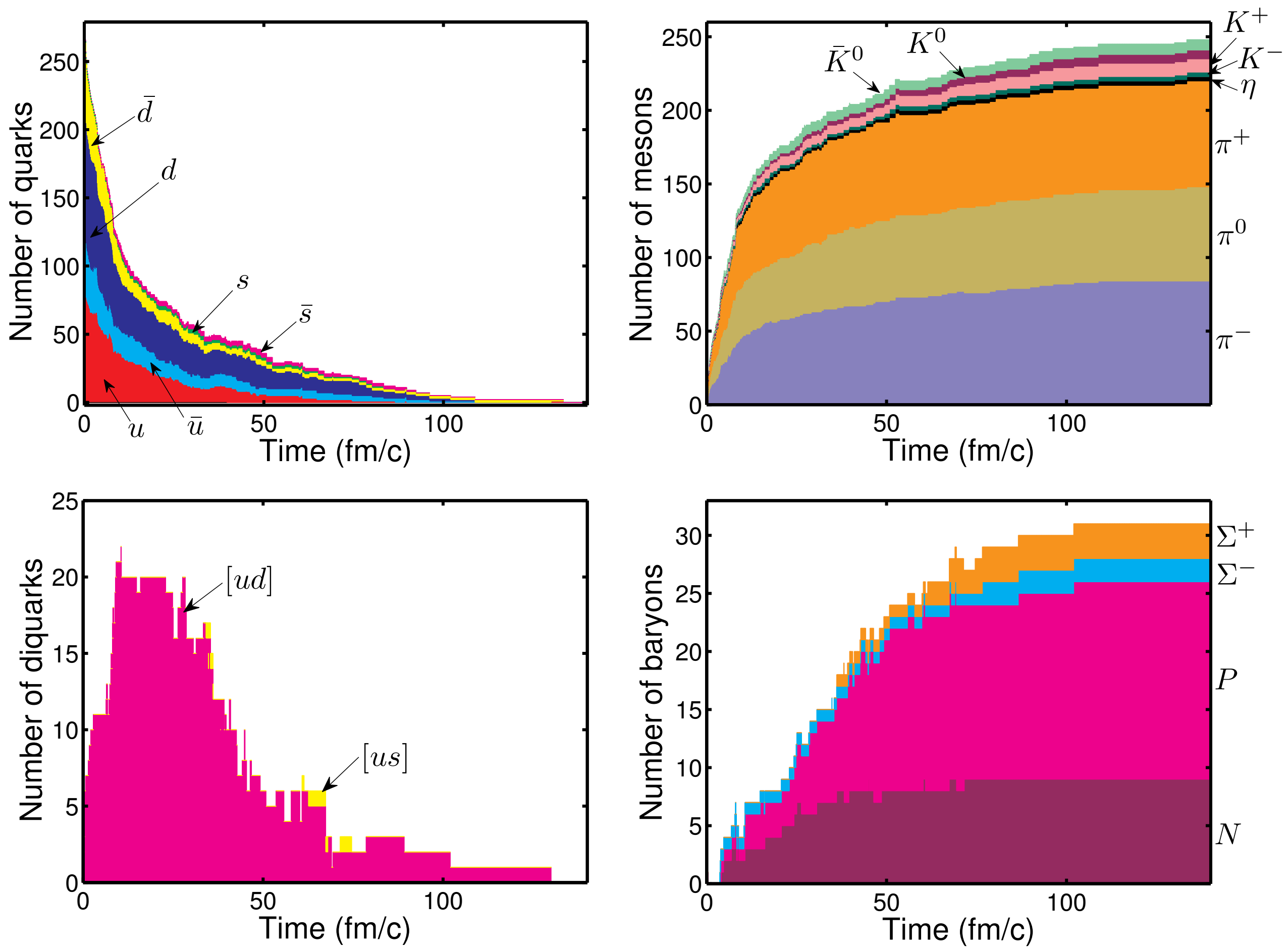

**Figure 22.** Populations of the involved particles according to the time (*N*: neutron, *P*: proton).

Qualitatively, the figure 22 is completed by the table 7. This one describes the final composition of the system, once the hadronization is complete, i.e. for $t = 133.4$ fm/c. As with the table 5, the production of pions is strong. Clearly, they represent about 79 % of the particles found in the end of this simulation. Also, a production of 26 nucleons is observable, i.e. it represents slightly less than 10 %. In addition, strange particles were also produced. In fact, we managed to produce rare particles because we considered more particles and because we reached higher temperatures than in the previous simulation. Clearly, thanks to the conditions met in this simulation, $s, \bar{s}$ pairs have been produced. It notably allowed the formation of a diquark as $[us]$. In the final state, we remark the presence of $\eta$ mesons and hyperons $\Sigma^-, \Sigma^+$. These particles are formed by reactions that present very limited cross-sections.

| Particles | $\pi^-$ | $\pi^0$ | $\pi^+$ | $\eta$ | $K^-$ | $K^+$ | $K^0$ | $\bar{K}^0$ | *Neutron* | *Proton* | $\Sigma^-$ | $\Sigma^+$ |
|---|---|---|---|---|---|---|---|---|---|---|---|---|
| Number | 84 | 64 | 72 | 3 | 3 | 9 | 6 | 7 | 9 | 17 | 2 | 3 |

**Table 7.** Final composition of the system.

We consider now the figure 23. This one shows the evolution of the temperature according to the radius (i.e. the distance from the center of the system) and the time. In this graph, we also plotted the radius of the sphere introduced in the simulation described subsection 4.2. The zone to the left of this curve is the interior of the sphere. It represents the phase in which the quarks/antiquarks and the diquarks are present, i.e. the QGP phase. Until 20 fm/c, a massive production of mesons occurred, leading to a strong consumption of quarks/antiquarks. As a consequence, the decrease in the radius of the sphere is strong in the beginning of the simulation. Then, between 20 and 110 fm/c, this decrease is less strong, and seems to be almost linear. This behavior is explained by the diminution of the quarks/antiquarks consumption. After 110 fm/c, the radius stagnates, until the total hadronization about 20 fm/c later.

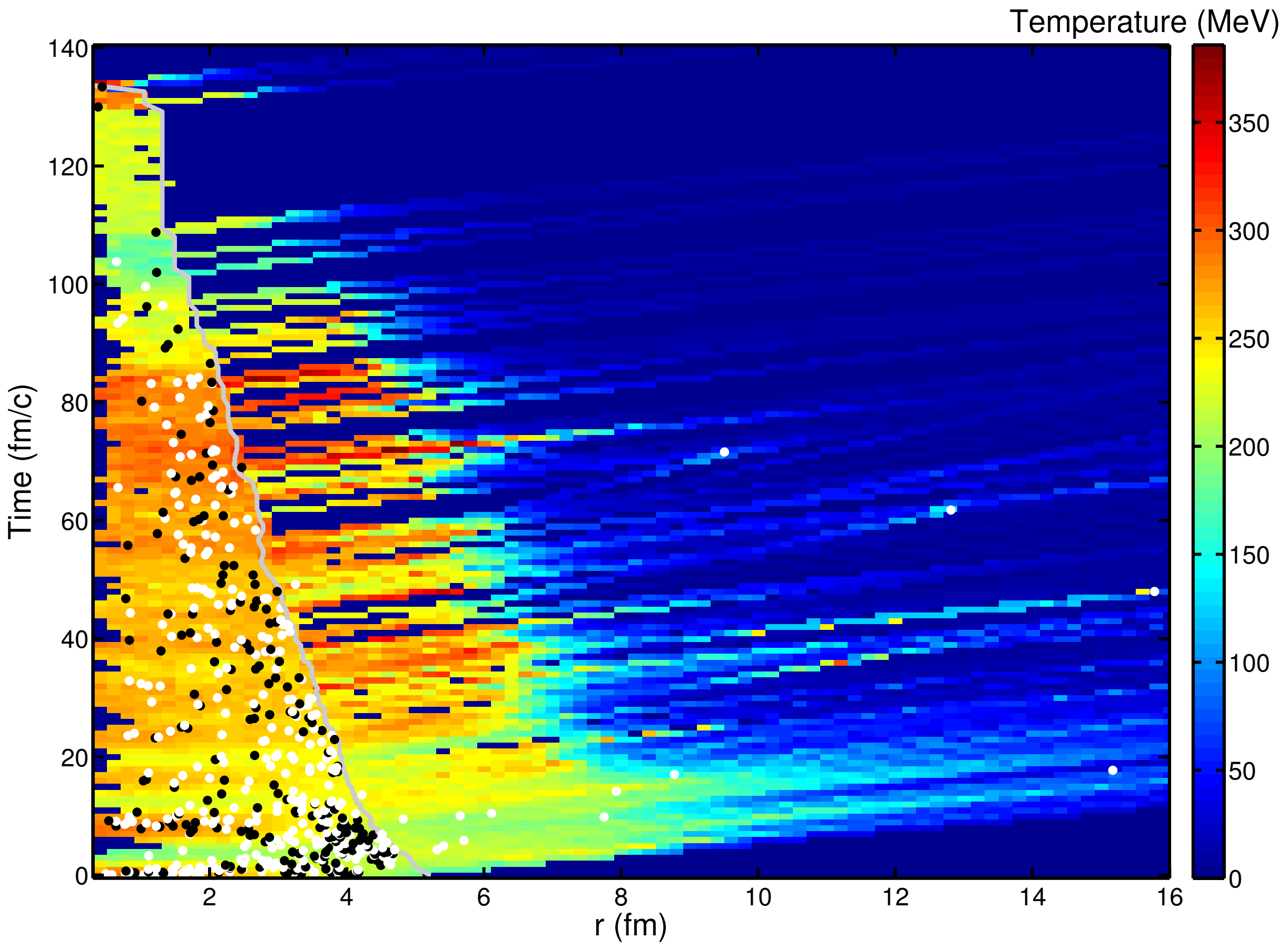

**Figure 23.** Temperature according to the distance from the center of the system, and according to the time. The gray curve materializes the limit of the QGP phase. The white dots correspond to elastic collisions, and the black dots correspond to inelastic ones.

Moreover, during all the hadronization of the QGP phase, its temperature is rather constant, as attested by the figure 25. According to the right hand side of this figure, this remark is also valid for the densities $\rho_u$ and $\rho_d$. The average temperature in the QGP phase is close to 250 MeV, even if variations are observable. Clearly, the inclusion of the sphere leads to this behavior. At the opposite, the dynamics of the system outside the sphere is completely different. More precisely, this part of the system is composed by the mesons and baryons that quitted the QGP phase. The reactions $q+\bar{q}\rightarrow M+M$ involving light particles are particularly *exothermic* [35]. It wants to say that the incoming particles are heavier than the outgoing ones. Especially with the pions produced by such reactions, these particles are expected to have strong velocities, as confirmed by the figure 28. These strong velocities imply strong temperatures around the QGP phase, forming a “hot corona” of pions. This corona explains the slight increase in the temperature of the $q/\bar{q}$ plasma, visible in the figure 23, notably between 20 and 70 fm/c. Consequently, the expansion of the mesons/baryons phase is very rapid. It leads to the right hand side of the graph presented in the figure 23. As observed, these particles gradually cool when $r$ and the time increase, because of the dilution. As checked in the figure 25, it leads to an exponential decrease in the mean temperature and densities $\rho_{u,d}$ of the system.

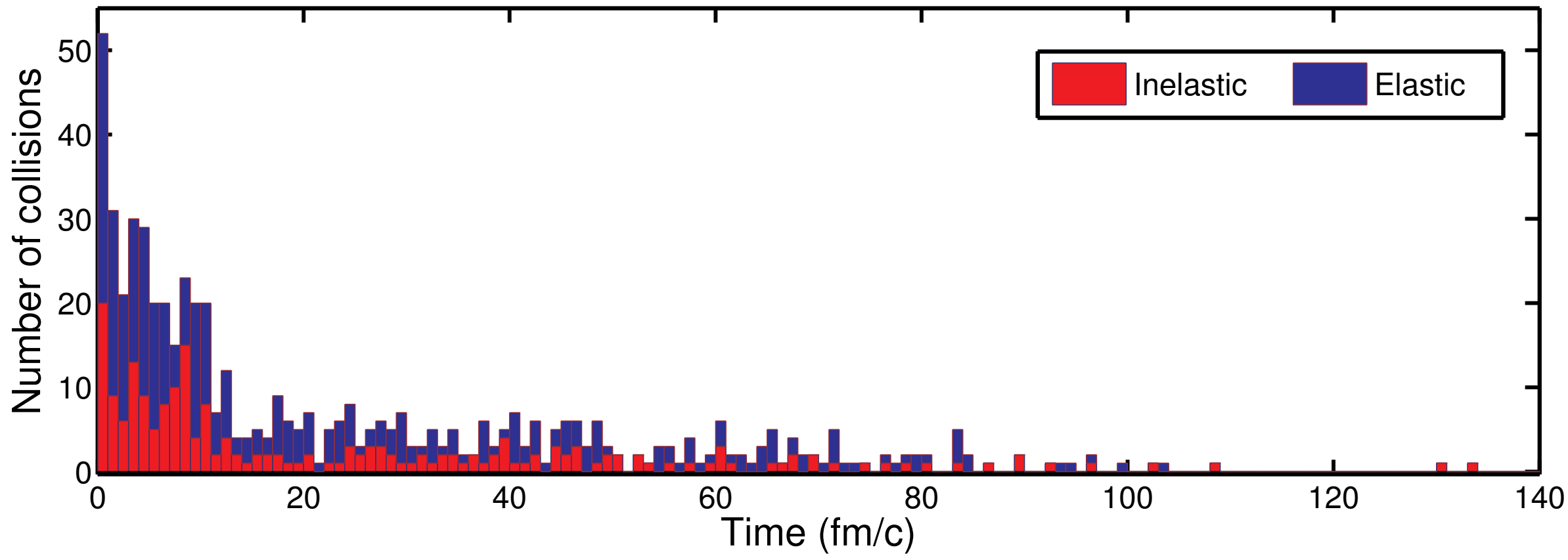

**Figure 24.** Elastic and inelastic collisions according to the time.

Concerning the collisions, the figures 23, 24 and the table 8 permit to study them upon several aspects. In this simulation, we counted 559 collisions. The elastic reactions represent about $63\ \%$, against $37\ \%$ for the inelastic ones. According to the figure 24, the collisions preferentially occur in the first moments of the simulation, until 10 fm/c. There, the elastic collisions dominate the inelastic ones. This behavior is explainable by the strong temperatures. It allows reactions as $q+\bar{q}\rightarrow q+\bar{q}$, $q+q\rightarrow q+q$ and $\bar{q}+\bar{q}\rightarrow\bar{q}+\bar{q}$ to dominate, figures 23, 25. According to the work described in the previous chapter, we recall that strong temperatures allow these reactions to present higher cross sections than $q+\bar{q}\rightarrow M+M$. However, this type of inelastic reactions is also very present, and it permits the massive mesonization visible in the figure 22. Between 10 and 80 fm/c, because of the diminution of the population in the QGP phase, the collision rate strongly decreases, but stay rather constant. As seen previously, the temperature of the QGP phase is almost constant during the simulation. So, as with $t<10$ fm/c, the elastic reactions also dominate the inelastic ones for $10\text{ fm/c}<t<80\text{ fm/c}$. After 80 fm/c, the number of particles forming the QGP phase is low. It leads to a strong reduction of the probability of interaction between two particles in the QGP phase, as with the simulation described in the subsection 4.2.

These observations are in agreement with the data visible in the figure 23. This one also gives information concerning the spatial distribution of the collisions. As a whole, the inelastic collisions (black dots) preferentially occur in the QGP phase, but near to its surface. In this zone, the temperature is more reduced than in the center: it corresponds to more favorable conditions for inelastic reactions as $q+\overline{q}\to M+M$. However, *endothermic* inelastic reactions, as $q+q\to\overline{q}+B$, are not concerned by this remark. They are associated with the inelastic reactions visible in the figure 23 near the center of the system. Moreover, the elastic collisions are observed in the whole QGP phase. It mainly concerns the elastic reactions between quarks and/or antiquarks. But, elastic reactions are also observable outside of this phase. These ones concern processes as $M+M\to M+M$, and secondly $M+B\to M+B$ and $B+B\to B+B$, table 8. The domination of $M+M\to M+M$ compared to the two other ones is explained by the fact that the mesons' production largely dominates the one of baryons, figure 22. As a consequence, the meeting between two mesons is highly more probable than the $M,B$ or the $B,B$ meeting, independently of the cross-sections associated with these processes.

Moreover, the table 8 presents the occurrence of all the processes treated in the simulation. Among the 559 collisions, the $q+\overline{q}\to q+\overline{q}$ reactions represent more than a quarter. As observed previously, these elastic scatterings dominate the inelastic mesonization reactions $q+\overline{q}\to M+M$. Indeed, these ones correspond to less than 20 % of the observed reactions. We saw in the subsection 4.2 that the mesonization ended early. But, this observation is not confirmed in this simulation. Indeed, antiquarks were found in the system until the total hadronization. As a consequence, the high temperatures met in this simulation affect the $q+\overline{q}\to M+M$ process in a non-negligible way. Nevertheless, these inelastic reactions were not really disturbed by their reverse ones, i.e. $M+M\to q+\overline{q}$, because these ones occurred only one time in all the simulation. Indeed, the cross-sections found with the reactions $q+\overline{q}\to M+M$ are globally stronger than the ones of $M+M\to q+\overline{q}$. Obviously, $q+\overline{q}\to M+M$ is the privileged way to produce mesons starting from the quark-antiquark system, largely before $q+q\to M+D$. But, this conclusion can depend on the initial ratio between quarks and antiquarks. Moreover, the mesons mainly interact with other particles via elastic reactions, as $M+M\to M+M$, $q+M\to q+M$ or $\overline{q}+M\to\overline{q}+M$, but more rarely by inelastic ones, as $q+M\to\overline{q}+D$ or $M+D\to\overline{q}+B$. It explains the constant growth of the number of mesons, figure 22.

| | | | | | |
|---|---|---|---|---|---|
| $q+\overline{q}\to q+\overline{q}$ | 147 | $q+D\to M+B$ | 14 | $D+D\to D+D$ | 3 |
| $q+\overline{q}\to M+M$ | 106 | $\overline{q}+M\to\overline{q}+M$ | 14 | $M+B\to M+B$ | 3 |
| $q+q\to q+q$ | 59 | $q+B\to D+D$ | 10 | $B+B\to B+B$ | 2 |
| $\overline{q}+\overline{q}\to\overline{q}+\overline{q}$ | 36 | $q+M\to\overline{q}+D$ | 7 | $\overline{q}+B\to q+q$ | 1 |
| $q+D\to q+D$ | 31 | $\overline{q}+B\to M+D$ | 7 | $M+M\to q+\overline{q}$ | 1 |
| $M+M\to M+M$ | 24 | $\overline{q}+D\to q+M$ | 5 | $\overline{q}+D\to\overline{q}+D$ | 0 |
| $q+q\to M+D$ | 22 | $D+B\to D+B$ | 4 | $\overline{q}+B\to\overline{q}+B$ | 0 |
| $q+M\to q+M$ | 22 | $q+B\to q+B$ | 3 | $M+D\to q+q$ | 0 |
| $D+D\to q+B$ | 17 | $M+D\to\overline{q}+B$ | 3 | $M+B\to q+D$ | 0 |
| $q+q\to\overline{q}+B$ | 15 | $M+D\to M+D$ | 3 | | |

**Table 8.** Occurrence of each type of collision.

About other reactions evoked in the table 8, even if $q+\overline{q}\to q+\overline{q}$ intervenes in a notable way, the other elastic reactions involving quark/antiquark, i.e. $q+q\to q+q$ and $\overline{q}+\overline{q}\to\overline{q}+\overline{q}$, are also present. Their occurrence is several times lower compared to the quark-antiquark scattering, because their cross-sections are globally more reduced than the ones of $q+\bar{q} \to q+\bar{q}$ [43]. On one side, the cross-sections of $q+q \to q+q$ and $\bar{q}+\bar{q} \to \bar{q}+\bar{q}$ are similar at moderate densities. On the other side, the initial ratio between matter and antimatter is equal to two. Thus, it explains why the occurrence of $q+q \to q+q$ is about twice compared to the one of $\bar{q}+\bar{q} \to \bar{q}+\bar{q}$.

Concerning the diquarks, their contribution cannot be neglected in this simulation, figure 22. As mentioned above, among the inelastic processes studied in the previous chapter, the reactions $q+q \to M+D$ are clearly the ones that produce diquarks in the most notable manner, as visible in the table 8. Furthermore, the reverse reactions $M+D \to q+q$ did not occur during the simulation. On the other hand, the contribution of the reactions $q+M \to \bar{q}+D$ is reduced, notably because of low cross-sections. Another reason is the reverse reactions $\bar{q}+D \to q+M$ present stronger cross-sections.

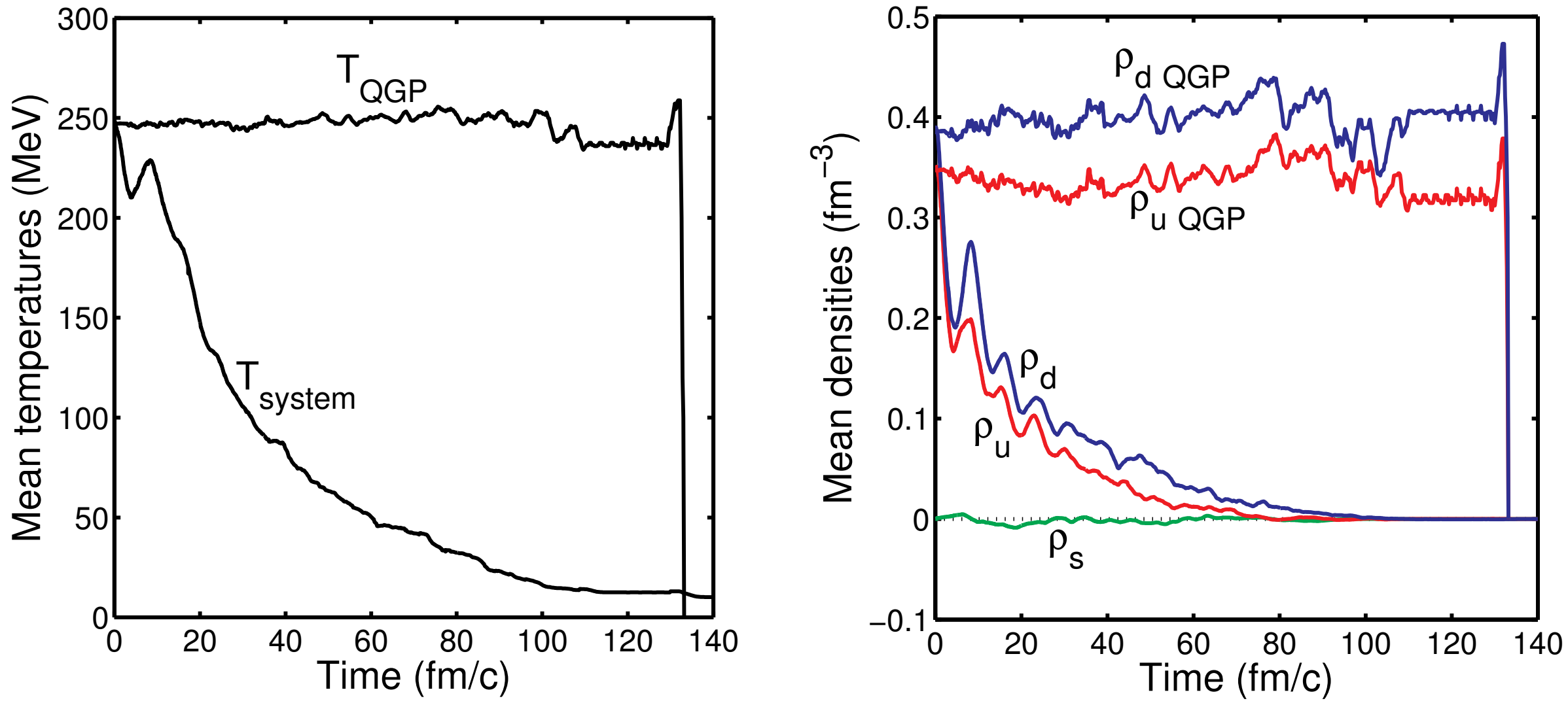

**Figure 25.** Evolution of the temperature and the densities.

About the formation of the baryons, we confirm the hypothesis formulated in the previous chapter that $M+D \to \bar{q}+B$ can be neglected, because of low cross-sections values, and because the ones of the reverse reactions $\bar{q}+B \to M+D$ are always stronger. Clearly, as expected, the three dominant processes allowing the baryonization are $D+D \to q+B$, $q+D \to M+B$ and $q+q \to \bar{q}+B$. In the framework of this simulation, they gave similar contributions. The baryons' production via $D+D \to q+B$ was permitted by the non-negligible diquark production, figure 22, by interesting cross-sections, and by the relative high temperatures in the QGP phase. Indeed, these endothermic reactions require such temperatures to intervene. The argument of the temperature is also valid with the $q+q \to \bar{q}+B$ reactions. But, concerning $D+D \to q+B$ and $q+q \to \bar{q}+B$, we firstly note that $q+B \to D+D$ is not negligible: it lowers the efficiency of the baryon's production starting from two diquarks. In addition, even if $\bar{q}+B \to q+q$ can be neglected according to our results, it is not the case for $\bar{q}+B \to M+D$, leading also to some limitations of the efficiency of the production of baryons via $q+q \to \overline{q}+B$. At the opposite, $q+D \to M+B$ is not limited by reverse reactions. Indeed, the mesons do not really take part to the inelastic reactions once they are produced. As a consequence $M+B \to q+D$ was not observed in our simulation, whereas $M+B \to M+B$ occurred three times. However, it is true that $q+D \to q+D$ appears as a non negligible source of competition of $q+D \to M+B$, even if this elastic process can only slow down the baryon production.

Now, we turn our attention to the figure 26 hereafter. Its finality is to check the validity of our results. This graph displays the evolution of the total energy, according to the time. Physically, the total energy is expected to be constant, but due to numerical rounding, some variations are observable. The values presented in the figure 26 can be considered as the most unfavorable of the simulations described in this chapter. However, the variations always stay less than 0.8 %. Furthermore, after the variations observed until 40 fm/c, we note stabilization towards 0.5%. Such a value is acceptable, and it is slightly higher than the variations announced in [24]. Clearly, in the framework of our simulations, these variations can be decreased if we enhance the precision of the numerical calculations, but at the prize of an extension of the calculation time. We checked that the inclusion of the sphere does not have a direct incidence on these variations.

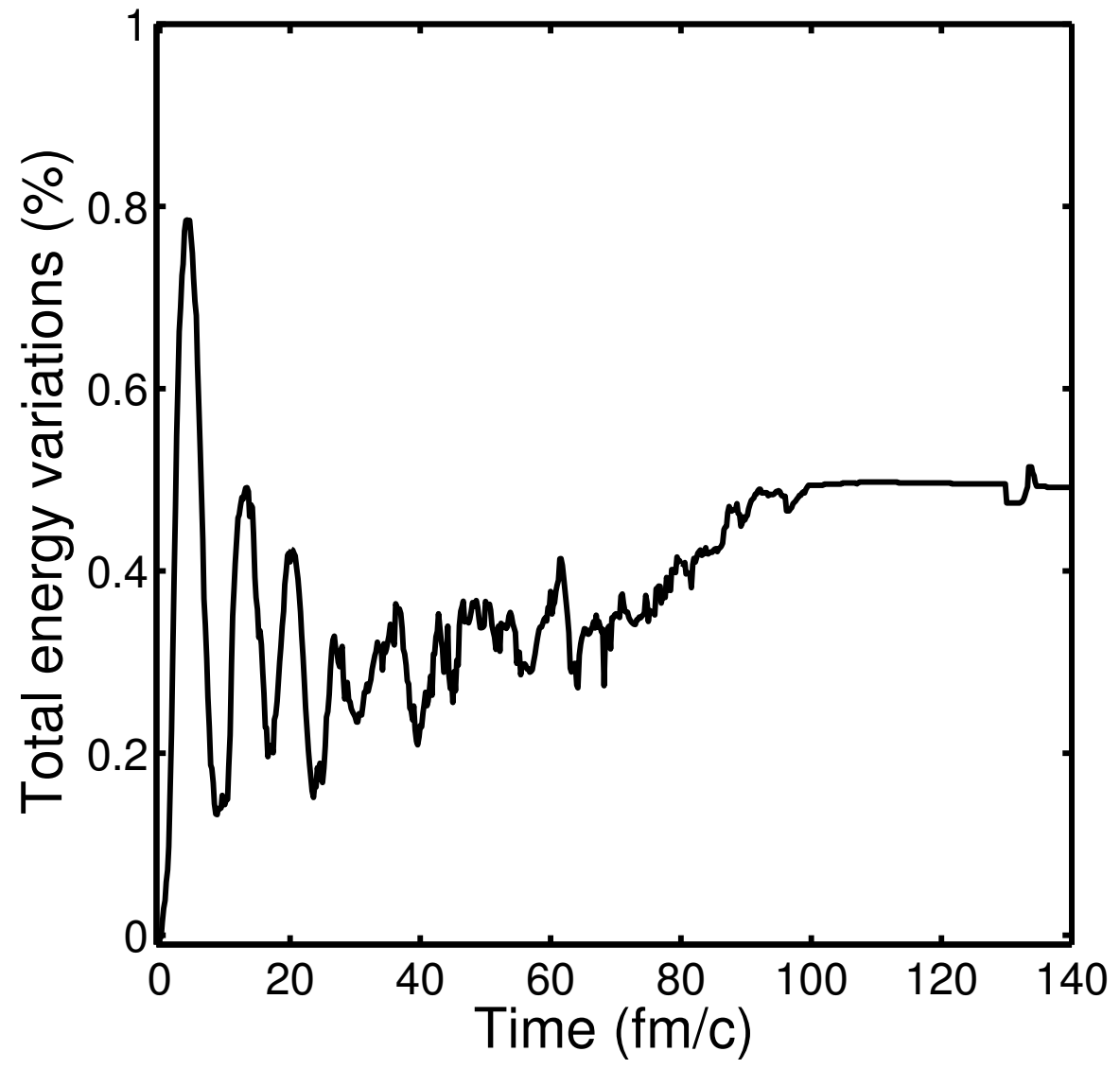

**Figure 26.** Fluctuations of the system's total energy according to the time.

To continue this analysis, we consider the figures 27 and 28 that describe the system at the end of the simulation, i.e. for a time $t \approx 140$ fm/c. The figure 27 shows the masses and the momenta of the particles. Obviously, in agreement with the table 7, these particles are necessarily mesons or baryons. At the end of the simulation, the temperature and the densities are reduced, figure 25. So, the masses of the particles, found in the left hand side of the figure 27, are very close to the ones found for $T=0$ and $\rho_f=0$, chapters 3 and 5. As a consequence, we observe four distinct layers in this graph. Two of them concern the mesons: one for the light one, and one for strange ones (kaons and $\eta$). In the same way, the two other ones are associated with the light baryons (nucleons) and strange ones, i.e. $\Sigma^{\pm}$ in the present simulation. We also remark a strong concentration of pions for radii between 100 and 150 fm. They testify of the system's massive mezonization in the first moments of the simulation.

Moreover, these particles are the ones that travelled the greater distance. Indeed, they have high velocities, because of the exothermic behavior of $q+\bar{q}\rightarrow M+M$ with these particles. We can also invoke the fact that they were produced early, as confirmed by the figure 22. In the right hand side of the figure 27, we plotted the momenta of the particles. The highest momentum is found to be close of 2 GeV, whereas the lowest is slightly upper than 70 MeV. As a consequence, very disparate values are observable. However, the dots seem to be more concentrated in the lower right corner of the graph. According to our previous explanations, these dots mainly concern the mesons, and notably the pions.

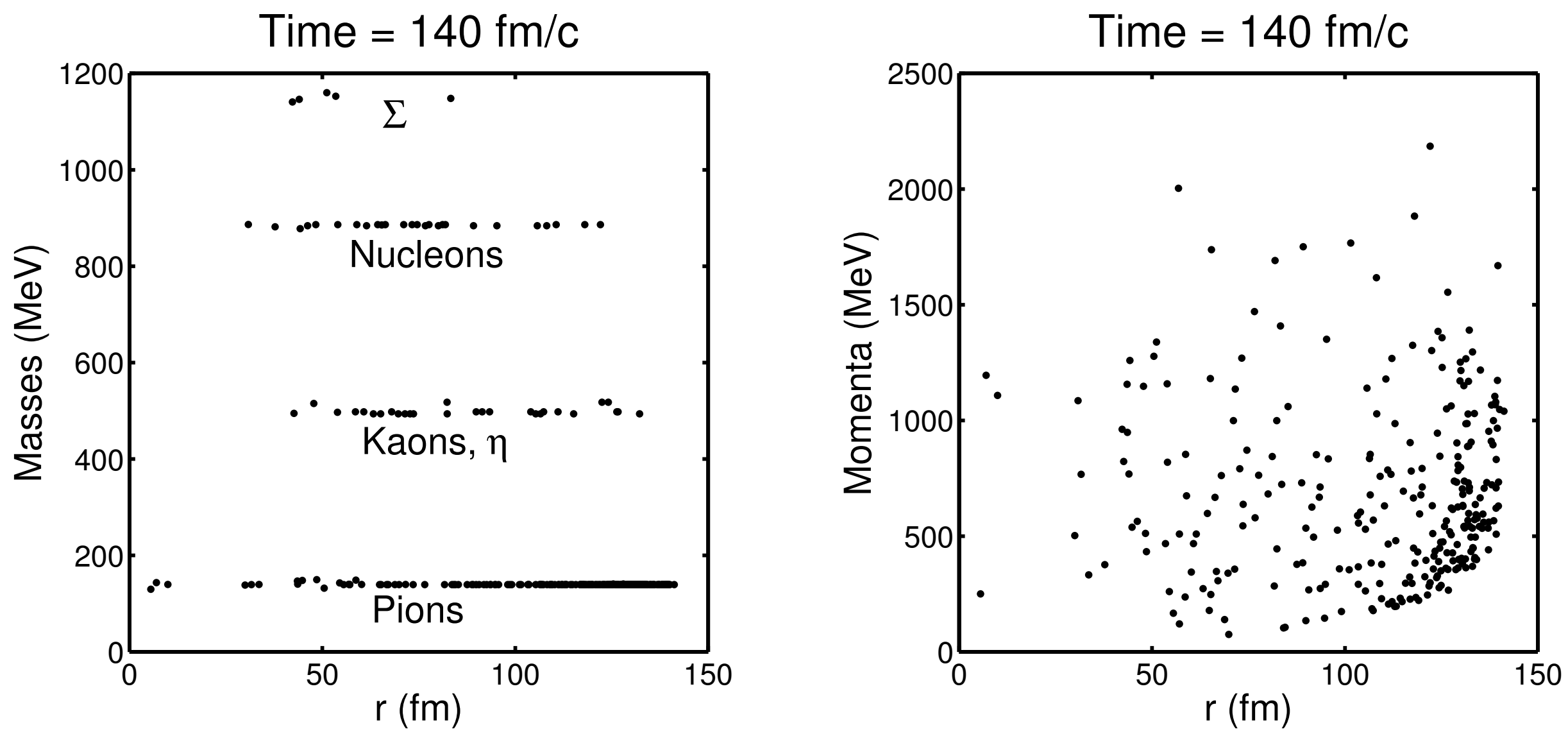

**Figure 27.** Masses of the particles and their momenta at the end of the simulation.

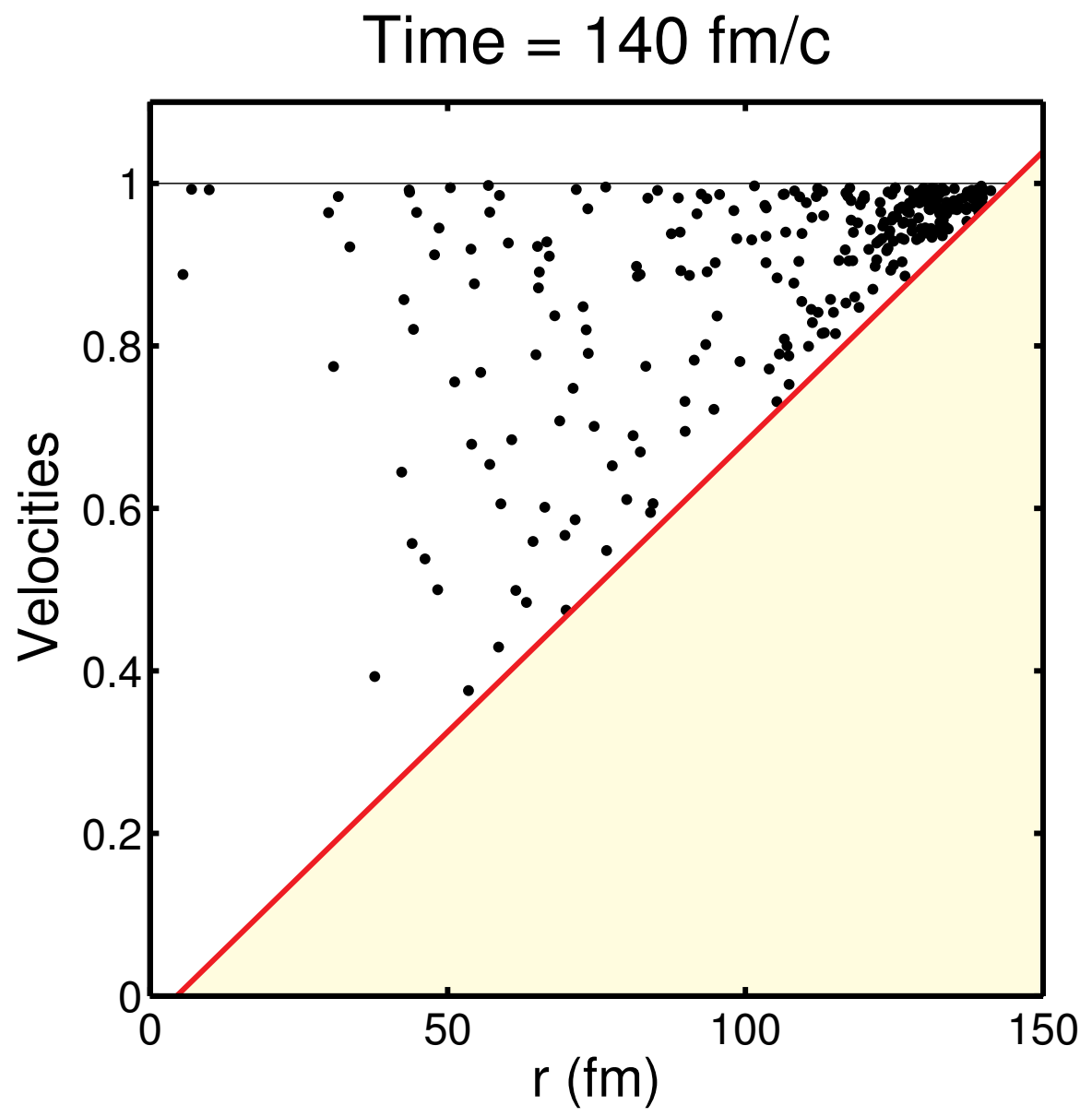

**Figure 28.** Velocities of the particles at the end of the simulation.

In order to conclude this description, the figure 28 shows the velocities of the particles according to the distance from the center of the system, at $t=140$ fm/c . These velocities were obtained by the data supplied by the figure 27, using the formula $\vec{v}=\vec{p}\cdot c^2/E$. In the figure, the horizontal line, for which the velocity is equal to 1, materializes the speed of light $c$. A significant part of the particles present strong velocities, close to 1, because the momenta were initially strong, and because of the exothermic reactions. Obviously, no tachyon was observed… Moreover, we saw that the remote interaction between the particles can be neglected. This remark is true outside of the QGP phase, because of the limited range of this interaction. In other words, once the final particles (mesons and baryons) are produced, their velocities are not expected to vary until the end of the simulation. It leads to consider them as free particles. Taking into account this hypothesis, a diagonal line is traced in the figure 28: it materializes the limit of causality of the graph. The shifting of this line according to the $r$ axis takes into account the initial radius of the system, about $4.5\ \text{fm}$. Clearly, particles on the right of this line would be non-causal, because they would be at too great distances from the center of the system in comparison to their velocities. Such unphysical behavior was not observed in our results. It would have been the sign of failures of the algorithm.

# 6. Discussion and conclusions

In this chapter, we presented a model devoted to dynamically study the cooling of a quarks/antiquarks plasma. It was an occasion to gather the works performed in the previous chapters in our program devoted to perform the simulation; this one is named ARCHANGE. Indeed, we saw that such a program included the calculations of the masses of particles in the (P)NJL models. These particles were here the $u,d,s$ quarks, the associated antiquarks, the pseudo scalar mesons, the scalar diquarks and the octet baryons, without the isospin symmetry. These masses were calculated according to the temperature $T$ and the densities $\rho_{u,d,s}$. More precisely, these *external parameters* were considered in our modeling as local parameters. In other words, for each particle, they translate the influence of by the vicinity, i.e. the other particles of the close neighborhood.

Furthermore, in order to describe the collisions, our algorithm included the calculations of cross-sections, using the work performed in the chapter 6. Notably, 60 different types of reactions were implemented, in order to allow the treatment of the possible collisions between quarks, mesons, diquarks, baryons and their associated anti-particles. In addition, the associated cross-sections were *real time calculated*, taking into account $\sqrt{s}$, but also the local temperature and densities. We showed that this approach allow obtaining some interesting results, as a non-negligible production of diquarks via reactions as $q+q \rightarrow D+M$. Indeed, these reactions can have strong cross-sections, but in some precise conditions according to the temperature and the densities.

In addition, we considered classical relativistic equations of motion. It gave us the occasion to interpret them in the framework of the (P)NJL models. In this way, we highlighted an attractive remote interaction between the particles, notably between quarks, by the way of their mass. As a consequence, some of our simulations concerned the study of this remote interaction. We saw that the attractive effect was found to be stronger in the PNJL description, compared to an NJL one. However, this interaction appeared to be negligible in the framework of hot systems, because involving rapid particles. Also, this interaction has a limited range, i.e. few femtometers in our descriptions.

Then, we performed complete simulations. We firstly noted that the PNJL model supplied better results than a pure NJL one, notably as regards the quarks/antiquarks consumption and the mesons production. These enhancements of the PNJL model were explained by the shifting of the temperatures. More precisely, it concerned the modifications of the optimal temperature of the $q+\overline{q} \to M+M$ reactions involving light particles: 230 MeV for a pure NJL model, against 280 MeV in the PNJL description. As a consequence, the mesonization process can act at higher temperatures in the PNJL description. So, the mesonization is more efficient in this model.

However, the (P)NJL models did not allow the complete hadronization of our open system. Even the mechanism of confinement proposed by the PNJL model was not enough in the framework of these simulations. We evoked the relative weakness of the baryonization reactions to explain these results. Also, this incomplete hadronization is a possible sign that the modeling requires long range interactions, to aggregate the quarks/antiquarks. In our work, we proposed to mimic this behavior by the way of a sphere. Of course, this trick constitutes a first step, and future developments of this work must concern the modeling of this long range interaction.

This sphere allowed performing two simulations, for which the hadronization was complete. The evolution of the particles present some similarities with the ones described e.g. in [16]. But, we also note that the required time to obtain full hadronization was higher compared to the one expected in this reference, and in general to the one expected in the literature, notably in the framework of the Bjorken scenario [10]. The absence of gluons in the modeling as dynamical particles can be evoked to try to explain this aspect. Clearly, reactions as $g+g \to q+\overline{q}$ are expected to be non negligible during the evolution of the QGP, chapter 1. More precisely, such a process can lead to an increase in the quarks/antiquarks population, so an increase in the collision rate. The inclusion of gluons as real dynamical particles could lead to the wanted long range interaction between quarks/antiquarks. However, such an evolution cannot concern (P)NJL models … Moreover, the influence on the hadronization time of reactions involving more than two incoming/outgoing particles should be investigated. These processes can intervene when the system is dense enough, i.e. potentially in the beginning of the simulation. Neglecting them may lead to underestimate the collision rate.

However, for the two simulations, some interesting aspects were observed. Firstly, we confirmed the scenario formulated in the previous chapter. It explained that at high temperatures, the system is dominated by elastic reactions between quarks/antiquarks. But when the temperature is low enough, the reactions $q+\overline{q} \to M+M$ make possible an overwhelming mesonization. The strong consumption of antiquarks allows then the formation of baryons. With the second simulation, we showed a non-negligible formation of diquark. They really acted as an intermediate state to boost the baryons' production. In this configuration, we saw that the baryons were preferentially formed by reactions as $D+D \to q+B$, $q+q \to \overline{q}+B$ and $q+D \to M+B$. Furthermore, it was observed a slight advantage for $q+D \to M+B$. Moreover, in the simulations, another interesting aspect is the creation of strange particles starting for light quarks/antiquarks.

Indeed, we saw in the chapter 1 that the strangeness production is one of the possible signatures of the QGP. But, our modeling presents limitations to study the other signatures. It is true especially with the emission of dileptons and thermal photons, because these particles are not considered in our approach. About the $J/\psi$ production, it requires the inclusion of $c$ quarks. It was saw in [45] that this improvement is possible in the PNJL model. About the jet quenching, we could study the variation of the quarks' energy before their hadronization. However, a major limitation is the absence of gluons, preventing to study gluon bremsstrahlung. Moreover, because of the choice of our initial conditions (presenting a spherical symmetry), we did not study $v_2$. As a consequence, an improvement of our approach is to consider initial conditions that can be compatible with the ones expected in heavy ion collisions, for which $b \neq 0$. For that purpose, we can quote e.g. the Glauber model [6] used in various simulations.

Apart from the improvements mentioned above, we can propose other possible evolutions. It may concern for example the inclusion of vectorial mesons, decuplet baryons, etc. Clearly, such a development requires performing the associated cross-sections calculations for these particles. Furthermore, as remarked in the previous chapters for heavy particles, a correct description of the decays of these particles seems to be necessary, as done in [24] for pseudo scalar mesons (when they are in their instability zone). Upon numerical aspects, some developments can be done following two objectives. The first is to reduce the numerical rounding, in order to minimize the variations of the total energy. The second concerns a reduction of the time required to perform our simulations, in order to be able to treat systems involving more particles.

# Conclusions

In this thesis, the main objective was to study the quarks physics, and notably the phase transition between the quark gluon plasma and the hadronic matter. To reach this goal, we firstly drawn a rapid overview of the current knowledge associated with this topic. We recalled some notions about the group theory. Indeed, we saw throughout this thesis the relevancy of such a theory in our work. Moreover, we described the Quantum Chromodynamics (QCD), notably its Lagrangian and its properties. We saw that this theory was not applicable in our description, even if it constitutes the most sophisticated tool to model the quarks. Indeed, QCD cannot be solved in the general case. Calculations are possible in the framework of the Lattice QCD. They are notably considered as references, but present some limitations at finite densities, because of the fermion sign problem. As a consequence, we turned our attention to an effective model, the Nambu and Jona Lasinio model. In this approach, the description of the interaction between quarks is simplified, by considering "frozen" gluons. More precisely, these ones are treated via effective terms, and then finally disappear in this modeling as dynamical degrees of freedom. We saw that a direct consequence of this treatment is the confinement is absent in a pure NJL description. In order to correct this limitation, we considered an evolution of the NJL description recently proposed in the literature [1]. It consists to couple the quarks/antiquarks to a Polyakov loop, in order to mimic a mechanism of confinement. It forms the Polyakov Nambu Jona-Lasinio model (PNJL).

During three chapters of the thesis, we modeled particles with the NJL model, and compared them to the ones found in the PNJL description. These particles were the $u,d,s$ quarks, the mesons, the diquarks and the baryons. Our modeling notably concerned the study of the masses of these particles in the $T,\rho_B$ plane. We also investigated the stability of the treated composites particles, i.e. mesons, diquarks and baryons. As a general tendency, we found that the inclusion of the Polyakov loop leads to a distortion of the curves towards higher temperatures. In other words, the found masses values were not modified, but shifted towards upper temperatures, compared to a pure NJL treatment. This effect was not observed according to the density. Concerning the quarks, we confirmed the results of the NJL [2] and PNJL [3] literature, and extended these calculations to the $T,\rho_B$ plane. We also proposed to study the expectation values of the Polyakov field $\Phi$ and its conjugate one $\bar{\Phi}$ in this plane. In the same way, such an analysis was done for the light chemical potential $\mu_q$. We saw that the relation between $\mu_q$, the temperature $T$ and the baryonic density $\rho_B$ is not trivial in the (P)NJL approaches. About the mesons, we also found results in agreement with [3, 4]. It notably concerns the pseudo-scalar and scalar mesons. We also studied the axial and the vectorial ones. Concerning the stable mesons (pseudo-scalar and vectorial ones), we presented diagrams of stability/instability. Moreover, our results at null temperature and density were very close to experimental data, especially when we the isospin symmetry was not considered. Then, about the baryons, we saw that it was possible to describe them as a bound state of a quark and a diquark. It thus motive us to model diquarks in the framework of the (P)NJL models. We concluded that scalar diquarks are usable to model octet baryons, whereas axial diquarks can allow describing the decuplet ones. In fact, even of some important simplifications in the baryon's modeling, as the static approximation, our results appeared to be close to the ones found in other theoretical approaches [2, 6, 7] or experiment data. However, we saw that our modeling can be subject to various evolutions, as the inclusion of decay processes for heavy baryons.

Then, an important chapter of this thesis concerned the calculation of the cross-sections of reactions involving the evoked particles. A part of this work focused on reactions as $q+\overline{q} \to M+M$, $q+\overline{q} \to q+\overline{q}$ and $q+q \to q+q$. These ones, calculated initially in [4, 5], were estimated according to $\sqrt{s}, T, \rho_B$ in our work. We also investigated the effect of the Polyakov loop in such cross-sections. As with the masses, the values of the PNJL cross-sections were shifted towards higher temperatures compared to the NJL ones. More precisely, an optimal mesonization via $q+\overline{q} \to M+M$ is expected for $T = 280\,\text{MeV}$ in the PNJL description, against 230 MeV in the NJL one. Moreover, inspiring us by the reactions proposed in [6], we also studied the PNJL inelastic cross-sections of reactions involving diquarks and/or baryons. We found reduced values for these processes. On one side, only the reactions $q+q \to D+M$ seem to be able to produce diquarks in a non-negligible way, but in some precise conditions. On the other side, three types of reactions appeared to be good candidates to allow the formation of baryons: $D+D \to B+q$ , $q+D \to M+B$ and $q+q \to B+\overline{q}$. To describe these processes, we saw the necessity to develop calculation methods, notably to be able to perform uncommon theoretical calculations, as the ones implying spinors with different momenta. It implies complex developments, detailed in the appendix B. Then, we also considered elastic reactions less or not treated in the literature, as $q+M \to M+q$, $q+D \to D+q$, $q+B \to q+B$, $D+B \to B+D$. The finality was to see if they can intervene as a potential source of competition of the inelastic ones. We observed that it could be the case, as with $q+D \to D+q$ compared to $q+D \to M+B$. We saw that our description should be completed, notably by the inclusion of more channels in the modeling of some reactions, as the *box* channels. Furthermore, the work performed for $D+D \to D+D$, inspired from [8, 9], should also be extended to meson-meson scattering, or even to $B+M \to B+M$ or $B+B \to B+B$.

A last aspect of our work was associated with our dynamical model. The finality was to simulate the evolution of a quark/antiquark plasma according to the time, to observe its cooling and its hadronization. We firstly described the various stages required to construct such a dynamical model. In our modeling, the particles can be influenced at the level of their masses, via external parameters. These ones are the temperature and the densities according to the $u,d,s$ flavors. More precisely, $T, \rho_f$ were treated as local parameters. According to the equations of motion, we saw that the influence of these parameters on the masses allowed the creation of a remote interaction between particles. We investigated the properties of this short range interaction, but we observed its limitations to really act on the dynamics of a system, notably if rapid particles are involved. Moreover, we described our algorithm devoted to treat the collisions. We used the collision criterion used in [10] that compares the cross-section with the impact parameter. We noted that a great advantage of our description is to estimate the cross-sections in the conditions met by the two incoming particles, i.e. taking into account $\sqrt{s}, T, \rho_f$, and not using database. Clearly, most of the work used in the previous chapters was implemented in this dynamical model, to estimate the masses or the cross-sections. Then, we focused on a description of the performed simulations. They involved $u,d,s$ quarks, their antiquarks, pseudo-scalar mesons, scalar diquarks and octet baryons. We explained that the inclusion of heavier particles in our dynamical model, as decuplet baryons, can be considered as a possible and interesting evolution of our work. But it requires an extension of the cross-sections calculations to these particles, and the treatment of the decays. Moreover, we concluded that the inclusion of the Polyakov loop allowed obtaining better results than in a pure NJL one. Indeed, the mesonization in the PNJL model is more rapid and more efficient than in the NJL model, thanks to the shifting of the temperatures mentioned above. Nevertheless, we do not succeeded to obtain a full hadronization of an open system with the PNJL model. Because of the relative weakness of the cross sections baryonization reactions, it requires the inclusion of a long range interaction between the particles to hold enough the quarks in the QGP phase of the system. As a consequence, an important evolution of our description concerns the modeling of the evoked long range force. However, we managed to mimic this one by the inclusion of a sphere that confined the colored particles until their hadronization into mesons or baryons. Thanks to this trick, we observed a complete hadronization for two different simulations. We compared our results with the ones of other approaches [11, 12]. Our data are qualitatively in agreement with these references. However, we found a required time to obtain a complete hadronization higher than the one expected in the literature. Such a difference may be notably explained by the cross-sections values and the absence of gluons as dynamical particles. Moreover, we noted the non-negligible contribution of the diquarks in the baryons formation, but in precise conditions upon the temperature and the densities. We also established and verified a scenario, which planned a domination of quark/antiquark scattering at high temperatures, a massive mesonization via $q+\overline{q} \to M+M$ when $T \leq 280$ MeV (PNJL model), and then a production of baryons, when the population of antiquarks will be reduced enough.

As a consequence, throughout this thesis, we showed the relevance of the NJL model to describe the quarks physics, upon several aspects, and notably to model the cooling/hadronization of a quark/antiquark plasma dynamically. Furthermore, by the inclusion of the Polyakov loop, to form the PNJL model, we confirmed that this enhancement of the model is very promising, as announced in the associated literature. Indeed, by the mechanism of confinement simulated by the model, we showed the modifications on the results, and its advantages, notably in the dynamical simulations. However, we concede that we agree with the conclusions formulated in [13]: the PNJL mimics some aspect of the confinement behavior, but there is not *true confinement* in this model. The confinement is not still fully mastered. Its description via the PNJL model is interesting, but not complete. Certainly, future evolutions of PNJL model will take into account this present limitation. About the current enhancements proposed in the literature, it would be instructive to consider for example the Entangled Polyakov Nambu Jona-Lasino model (EPNJL) [14], and to investigate the implied modifications in our results (masses, cross-sections, simulations). Moreover, another possible evolution of our work concerns the treatment of the color-superconductivity phases [15] with the PNJL model. As argued previously, this phenomenon is not expected to occur in the $T,\rho_f$ conditions of our dynamical simulations. As a consequence, it does not constitute a limitation of our performed work. But, the treatment of the color-superconductivity can allow the comprehension of the whole $T,\rho_B$ plane, or at least the bad known domain involving reduced temperatures and strong baryonic densities. In addition, the color-superconductivity presents applications in astrophysics, in order to describe the deep layers of cold and dense objects, as neutron stars. It thus constitutes a possible extension of our work, taking care of course about the modifications to be applied in the model, by the use of the Nambu-Gorkov formalism [16] …

# Appendix A

# Particles data

## 1. Quarks

| flavor | name | mass (MeV) | electric charge | quantum number |
|---|---|---|---|---|
| $u$ | up | 1.8–3.0 | $2/3\ e$ | $I_z = +1/2$ |
| $d$ | down | 4.5–5.3 | $-1/3\,e$ | $I_z = -1/2$ |
| $s$ | strange | 90–100 | $-1/3\,e$ | strangeness $S = -1$ |
| $c$ | charm | 1.250–1.300 GeV | $2/3\ e$ | charm $C = +1$ |
| $b$ | bottom | 4.15–4.69 GeV | $-1/3\,e$ | bottomness $B = -1$ |
| $t$ | top | $173.07 \pm 1.24$ GeV | $2/3\ e$ | topness $T = +1$ |

**Table 1.** Description of the quarks.

In the column labeled quantum number, only the non-zero values are indicated (except $I_z$). For example, the strangeness of a quark other than $s$ is inevitably equal to zero. $I_z$ indicates projection according to $z$ of $I$ isospin. The quarks are fermions. Their spin is equal to ½. The data were extracted from [1, 2].

# 2. Mesons

| | $0^{+-}$ | $1^{--}$ | $0^{++}$ | $1^{++}$ | $1^{+-}$ | $2^{++}$ | $1^{--}$ | $1^{--}$ |
|---|---|---|---|---|---|---|---|---|
| name | $\pi$ | $\rho$ | $a_0$ | $a_1$ | $b_1$ | $a_2$ | $\rho$ | $\rho$ |
| mass | 138 | 769 | 984 | 1230 | 1235 | 1318 | 1465 | 1700 |
| width | 0 | 151 | 100 | 400 | 142 | 107 | 310 | 323 |
| isospin | 1 | 1 | 1 | 1 | 1 | 1 | 1 | 1 |
| strangeness | 0 | 0 | 0 | 0 | 0 | 0 | 0 | 0 |
| name | $K$ | $K$* | $K_0$* | $K_1$* | $K_1$ | $K_2$* | $K$* | $K$* |
| mass | 495 | 893 | 1429 | 1273 | 1400 | 1430 | 1410 | 1680 |
| width | 0 | 50 | 287 | 90 | 174 | 100 | 227 | 235 |
| isospin | 1/2 | 0 | 1/2 | 1/2 | 1/2 | 1/2 | 1/2 | 1/2 |
| strangeness | ±1 | ±1 | ±1 | ±1 | ±1 | ±1 | ±1 | ±1 |
| name | $\eta$ | $\omega$ | $f_0$ | $f_1$ | $h_1$ | $f_2$ | $\omega$ | $\omega$ |
| mass | 547 | 782 | 980 | 1282 | 1170 | 1275 | 1419 | 1662 |
| width | 0 | 8.43 | 100 | 24 | 360 | 185 | 174 | 280 |
| isospin | 0 | 0 | 0 | 0 | 0 | 0 | 0 | 0 |
| strangeness | 0 | 0 | 0 | 0 | 0 | 0 | 0 | 0 |
| name | $\eta$' | $\phi$ | $f_0$* | $f_1$' | $h_1$' | $f_2$' | $\phi$ | $\phi$ |
| mass | 958 | 1019 | 1370 | 1512 | 1380 | 1525 | 1680 | 1900 |
| width | 0.201 | 4.43 | 200 | 350 | 80 | 76 | 150 | 400 |
| isospin | 0 | 0 | 0 | 0 | 0 | 0 | 0 | 0 |
| strangeness | 0 | 0 | 0 | 0 | 0 | 0 | 0 | 0 |

**Table 2.** The mesons.

Concerning the first line of table 2, the number indicates the spin of the meson; the first ± is associated with the behavior of its wave function according to the parity. The second ± refers to the charge conjugation. These data correspond to the QMD/URQMD ones, extracted from [3]. They respect the isospin approximation. The masses and the widths are in MeV.

# 3. Baryons (particles and resonances)

Nucleon : *uud* or *udd*
$S = 0,\ I = 1/2$

| Mass | Width | Spin |
|---|---|---|
| 0938 | 000 | 1/2 |
| 1440 | 200 | 1/2 |
| 1520 | 125 | 3/2 |
| 1535 | 150 | 1/2 |
| 1650 | 150 | 1/2 |
| 1675 | 140 | 5/2 |
| 1680 | 120 | 5/2 |
| 1700 | 100 | 3/2 |
| 1710 | 110 | 1/2 |
| 1720 | 150 | 3/2 |
| 1900 | 500 | 3/2 |
| 1990 | 550 | 7/2 |
| 2080 | 250 | 3/2 |
| 2190 | 550 | 7/2 |
| 2200 | 550 | 9/2 |
| 2250 | 470 | 9/2 |

$\Delta$ : *uuu*, *uud*, *udd* or *ddd*
$S = 0,\ I = 3/2$

| Mass | Width | Spin |
|---|---|---|
| 1232 | 115 | 3/2 |
| 1600 | 200 | 3/2 |
| 1620 | 180 | 1/2 |
| 1700 | 300 | 3/2 |
| 1900 | 240 | 1/2 |
| 1905 | 280 | 5/2 |
| 1910 | 250 | 1/2 |
| 1920 | 150 | 3/2 |
| 1930 | 250 | 5/2 |
| 1950 | 250 | 7/2 |

$\Lambda$ : *uds*
$S = -1,\ I = 0$

| Mass | Width | Spin |
|---|---|---|
| 1116 | 000 | 1/2 |
| 1405 | 050 | 1/2 |
| 1520 | 016 | 3/2 |
| 1600 | 150 | 1/2 |
| 1670 | 035 | 1/2 |
| 1690 | 060 | 3/2 |
| 1800 | 300 | 1/2 |
| 1810 | 150 | 1/2 |
| 1820 | 080 | 5/2 |
| 1830 | 095 | 5/2 |
| 1890 | 100 | 3/2 |
| 2100 | 200 | 7/2 |
| 2110 | 200 | 5/2 |

$\Sigma$ : 1 quark *s*
$S = -1,\ I = 1$

| Mass | Width | Spin |
|---|---|---|
| 1192 | 000 | 1/2 |
| 1385 | 036 | 3/2 |
| 1660 | 100 | 1/2 |
| 1670 | 060 | 3/2 |
| 1750 | 090 | 1/2 |
| 1775 | 120 | 5/2 |
| 1915 | 120 | 5/2 |
| 1940 | 220 | 3/2 |
| 2030 | 180 | 7/2 |

$\Xi$ : *uss* or *dss*
$S = -2,\ I = 1/2$

| Mass | Width | Spin |
|---|---|---|
| 1315 | 00 | 1/2 |
| 1530 | 09 | 3/2 |
| 1690 | 50 | 3/2 |
| 1820 | 24 | 3/2 |
| 1950 | 60 | 3/2 |
| 2030 | 20 | 5/2 |

$\Omega$ : *sss*
$S = -3,\ I = 0$

| Mass | Width | Spin |
|---|---|---|
| 1672 | 00 | 3/2 |

**Table 3.** The baryons.

The mass of the particle corresponds to the lowest value for each sub-table. The other ones are resonances. The masses and the widths are in MeV. The data were extracted from [3].

# Appendix B

# Field theory formulary and trace calculations

The sections 3 and 4 were published in *J. Phys. G: Nucl. Part. Phys.* **39** 105003

## 1. Lie groups generators

### 1.1 *SU(2)* generators

$SU(2)$ corresponds the 2-dimension unitary matrices that have a determinant equal to 1. The generators of this group are the Pauli matrices. Their usual representation is:

$$\sigma_1 = \begin{bmatrix} 0 & 1 \\ 1 & 0 \end{bmatrix} \qquad \sigma_2 = \begin{bmatrix} 0 & -i \\ i & 0 \end{bmatrix} \qquad \sigma_3 = \begin{bmatrix} 1 & 0 \\ 0 & -1 \end{bmatrix} \tag{1}$$

### 1.2 *SU(3)* generators

The $SU(3)$ group designates the 3-dimension unitary matrices that have a determinant equal to 1. The generators of $SU(3)$ are the Gell-Mann matrices. A possible representation of these generators is [1]:

$$\begin{gathered} \lambda_1 = \begin{bmatrix} 0 & 1 & 0 \\ 1 & 0 & 0 \\ 0 & 0 & 0 \end{bmatrix} \qquad \lambda_2 = \begin{bmatrix} 0 & -i & 0 \\ i & 0 & 0 \\ 0 & 0 & 0 \end{bmatrix} \qquad \lambda_3 = \begin{bmatrix} 1 & 0 & 0 \\ 0 & -1 & 0 \\ 0 & 0 & 0 \end{bmatrix} \\ \lambda_4 = \begin{bmatrix} 0 & 0 & 1 \\ 0 & 0 & 0 \\ 1 & 0 & 0 \end{bmatrix} \qquad \lambda_5 = \begin{bmatrix} 0 & 0 & -i \\ 0 & 0 & 0 \\ i & 0 & 0 \end{bmatrix} \qquad \lambda_6 = \begin{bmatrix} 0 & 0 & 0 \\ 0 & 0 & 1 \\ 0 & 1 & 0 \end{bmatrix} \\ \lambda_7 = \begin{bmatrix} 0 & 0 & 0 \\ 0 & 0 & -i \\ 0 & i & 0 \end{bmatrix} \qquad \lambda_8 = \frac{1}{\sqrt{3}} \cdot \begin{bmatrix} 1 & 0 & 0 \\ 0 & 1 & 0 \\ 0 & 0 & -2 \end{bmatrix} \end{gathered} \tag{2}$$

These matrices are linked by the commutation relation:

$$\left[\frac{\lambda^a}{2},\frac{\lambda^b}{2}\right]=i\cdot f^{abc}\cdot\frac{\lambda^c}{2}, \tag{3}$$

where the $f^{abc}$ are the structure constants of $SU(3)$. We have: $f^{123}=1$, $f^{147}=1/2$, $f^{156}=-1/2$, $f^{246}=1/2$, $f^{678}=\sqrt{3}/2$, the others are equal to zero.

In the framework of the Nambu and Jona-Lasinio model, one often introduces $\lambda_0$. Of course, this matrix is not a 9th generator; it is defined as:

$$\lambda_0=\sqrt{\frac{2}{3}}\cdot\begin{bmatrix}1&0&0\\0&1&0\\0&0&1\end{bmatrix}. \tag{4}$$

# 2. Dirac matrices

## 2.1 Usual representations

### Dirac (standard)

$$\gamma_D^0=\begin{bmatrix}1_2&0\\0&-1_2\end{bmatrix}\qquad \gamma_D^{\mathrm{i}}=\begin{bmatrix}0&\sigma^{\mathrm{i}}\\-\sigma^{\mathrm{i}}&0\end{bmatrix}\qquad \gamma_D^5=\begin{bmatrix}0&1_2\\1_2&0\end{bmatrix} \tag{5}$$

$$\mathrm{i}=1,2,3$$

where $1_2$ is the $2\times 2$ identity matrix.

### Chiral (spinorial)

$$\gamma^0=\begin{bmatrix}0&1_2\\1_2&0\end{bmatrix}\qquad \gamma^{\mathrm{i}}=\begin{bmatrix}0&-\sigma^{\mathrm{i}}\\\sigma^{\mathrm{i}}&0\end{bmatrix}\qquad \gamma^5=\begin{bmatrix}1_2&0\\0&-1_2\end{bmatrix} \tag{6}$$

### Majorana

$$\gamma^0=\begin{bmatrix}0&\sigma_2\\\sigma_2&0\end{bmatrix}\qquad \gamma^5=\begin{bmatrix}\sigma_2&0\\0&-\sigma_2\end{bmatrix}$$

$$\gamma^1=\begin{bmatrix}i\cdot\sigma_3&0\\0&i\cdot\sigma_3\end{bmatrix}\qquad \gamma^2=\begin{bmatrix}0&-\sigma_2\\\sigma_2&0\end{bmatrix}\qquad \gamma^3=\begin{bmatrix}-i\sigma_1&0\\0&-i\sigma_1\end{bmatrix} \tag{7}$$

# 2.2 Dirac matrices properties

## Clifford algebra

Whatever the representation, the following relations are always valid [2]:

$$\begin{cases} \{\gamma^\mu,\gamma^\nu\} = \gamma^\mu\gamma^\nu + \gamma^\nu\gamma^\mu = 2\eta^{\mu\nu}\ 1_4 \\ \{\gamma^\mu,\gamma^5\} = 0 \\ \left(\gamma^5\right)^2 = 1_4 \\ \gamma^5 = i\ \gamma^0\gamma^1\gamma^2\gamma^3 \\ \left(\gamma^0\right)^\dagger = \gamma^0\ ,\ \left(\gamma^i\right)^\dagger = -\gamma^i = \gamma^0\gamma^i\gamma^0\ ,\ \left(\gamma^5\right)^\dagger = \gamma^5 \end{cases} . \tag{8}$$

More precisely, (8) defines the Clifford algebra, with:

$$\eta = \begin{bmatrix} 1 & & & \\ & -1 & & \\ & & -1 & \\ & & & -1 \end{bmatrix}, \tag{9}$$

which is the Minkowski metric and $1_4$ the $4\times4$ identity matrix. We can also introduce:

$$\sigma^{\mu\nu} = \frac{i}{2}\left[\gamma^\mu,\gamma^\nu\right] = \frac{i}{2}\cdot\left(\gamma^\mu\cdot\gamma^\nu - \gamma^\nu\cdot\gamma^\mu\right). \tag{10}$$

## Other properties

$\gamma_\mu\gamma^\mu = 4$

(implicit summation over the $\mu$)

$\gamma_\mu \not{a}\ \gamma^\mu = -2\not{a}$

$\gamma_\mu \not{a}\not{b}\ \gamma^\mu = 4ab$

$\gamma_\mu \not{a}\not{b}\not{c}\ \gamma^\mu = -2\not{c}\not{b}\not{a}$

$\gamma_\mu\, \sigma_{\gamma\nu}\, \gamma^\mu = 0$

$Tr\left(\not{a}\not{b}\right) = 4ab$

$Tr\left(\not{a}\not{b}\gamma^5\right) = 0$

$$\begin{aligned} & Tr(1_4) = 4 \\ & Tr\left(\gamma^5\right) = 0 \\ & Tr\left(\gamma^\mu\gamma^\nu\ldots\gamma^\alpha\gamma^\beta\right) = Tr\left(\gamma^\beta\gamma^\mu\gamma^\nu\ldots\gamma^\alpha\right) \\ & \qquad = Tr\left(\gamma^\beta\gamma^\alpha\ldots\gamma^\nu\gamma^\mu\right) \\ & Tr\left(\text{odd number of } \gamma^\mu\right) = 0 \\ & Tr\left(\gamma^\mu\gamma^\nu\right) = 4\eta^{\mu\nu} \\ & Tr\left(\gamma^\mu\gamma^\nu\gamma^5\right) = 0 \\ & Tr\left(\gamma^\mu\gamma^\nu\gamma^\alpha\gamma^\beta\right) = 4\left(\eta^{\mu\nu}\eta^{\alpha\beta} + \eta^{\mu\beta}\eta^{\nu\alpha} - \eta^{\mu\alpha}\eta^{\nu\beta}\right) \end{aligned} \tag{11}$$

where $a_\mu$ is the $\mu^{\text{th}}$ component of a four-vector, and $a_\mu\cdot\gamma^\mu \equiv \not{a}$ (Feynman slash). Idem for *b* and *c*.

# 3. Standard trace calculations

The vocation of the sections 3 and 4 is to show how to calculate the matrix elements that appear in the cross-section calculations presented in chapter 6. As a whole, such calculations concern the evaluation of traces, in which terms as $\not{a}$ are present. The section 3 concerns relatively classic cases. The section 4 treats more delicate cases, which imply spinors with different momenta.

## 3.1 First example

As a first example, we evaluate the squared term of the $s$ channel of the $\mathrm{q}+\overline{\mathrm{q}}\to \mathrm{M}+\mathrm{M}$ process. The matrix element is written as:

$$-i\mathcal{M}_s = f_s \;\delta_{c_1,c_2}\; \overline{v}(p_2)\; 1_4 \; ig_1 \; i\mathcal{D}_s^S(p_1+p_2)\;\Gamma(p_1+p_2\;,p_3)\; ig_2\; u(p_1)\,. \tag{12}$$

The squared term is calculated by a summation over colors and spins:

$$\begin{aligned}&\frac{1}{4N_c^{\,2}}\cdot\sum_{\substack{\text{spin}\\ \text{color}}}\mathcal{M}_s\cdot\mathcal{M}_s^{\,*}\\ &=\frac{f_s^{\,2}}{4N_c}\cdot\left|g_1g_2\right|^2\cdot\left|\mathcal{D}_s^S\right|^2\cdot\left|\Gamma\right|^2\cdot\sum_{s_1,s_2}\left(\overline{v}(p_2,s_2)\cdot 1_4\cdot u(p_1,s_1)\right)\cdot\left(\overline{v}(p_2,s_2)\cdot 1_4\cdot u(p_1,s_1)\right)^*\end{aligned}\,, \tag{13}$$

with $s_1,s_2$ are the spins of the incoming quark/antiquark. The summation over spins gives:

$$\begin{aligned}&\sum_{s_1,s_2}\overline{v}(p_2,s_2)\cdot 1_4\cdot u(p_1,s_1)\cdot\overline{u}(p_1,s_1)\cdot 1_4\cdot v(p_2,s_2)\\ &=\sum_{s_2}v(p_2,s_2)\cdot\overline{v}(p_2,s_2)\cdot 1_4\times\sum_{s_1}u(p_1,s_1)\cdot\overline{u}(p_1,s_1)\cdot 1_4\\ &=\sum_{s_2}v_i(p_2,s_2)\cdot\overline{v}_j(p_2,s_2)\cdot[1_4]_{jk}\times\sum_{s_1}u_k(p_1,s_1)\cdot\overline{u}_l(p_1,s_1)\cdot[1_4]_{li}\end{aligned}\,. \tag{14}$$

In the last line of (14), the spinor indices $i,j,k,l$ are indicated by applying the Einstein summation convention. Using the completeness relations for spinors:

$$\begin{cases}\displaystyle\sum_{s_1=\pm}u_k(p_1,s_1)\cdot\overline{u}_l(p_1,s_1)=\left(\not{p}_1\right)_{kl}+m_1\cdot\delta_{kl}\\ \displaystyle\sum_{s_2=\pm}v_i(p_2,s_2)\cdot\overline{v}_j(p_2,s_2)=\left(\not{p}_2\right)_{ij}-m_2\cdot\delta_{ij}\end{cases}, \tag{15}$$

(14) is rewritten as:

$$\begin{aligned}&\left(\left(\not{p}_2\right)_{ij}-m_2\cdot\delta_{ij}\right)\cdot[1_4]_{jk}\cdot\left(\left(\not{p}_1\right)_{kl}+m_1\cdot\delta_{kl}\right)\cdot[1_4]_{li}\\ &=Tr\left(\left(\not{p}_2-m_2\right)\cdot\left(\not{p}_1+m_1\right)\right)=4\cdot\left(p_2\cdot p_1-m_2\cdot m_1\right)\end{aligned}\,. \tag{16}$$

$2p_1 \cdot p_2 = s - m_1^2 - m_2^2$, appendix F. So, we recover the result of [3]:

$$\frac{1}{4N_c^2} \cdot \sum_{s,c} \left|\mathcal{M}_s\right|^2 = \frac{f_s^2}{2N_c} \cdot \left|g_1 g_2\right|^2 \cdot \left|\mathcal{D}_s^S \cdot \Gamma\right|^2 \cdot \left(s - (m_1 + m_2)^2\right). \tag{17}$$

# 3.2 Second example

We consider now the calculation associated with the mixed term between the $u$ and $t$ channels, for $\mathrm{q} + \overline{\mathrm{q}} \to \mathrm{M} + \mathrm{M}$. The matrix elements are:

$$\begin{array}{l} -i\mathcal{M}_t = f_t \;\; \delta_{c_1,c_2} \;\; \overline{v}(p_2) \;\; i\gamma_5 \;\; ig_1 \;\; iS_F(p_1 - p_3) \;\; i\gamma_5 \;\; ig_2 \;\; u(p_1) \\ -i\mathcal{M}_u = f_u \;\; \delta_{c_1,c_2} \;\; \overline{v}(p_2) \;\; i\gamma_5 \;\; ig_1 \;\; iS_F(p_1 - p_4) \;\; i\gamma_5 \;\; ig_2 \;\; u(p_1) \end{array}. \tag{18}$$

We write:

$$\begin{array}{l} \dfrac{1}{4N_c^2} \cdot \sum\limits_{\substack{\text{spin} \\ \text{color}}} \mathcal{M}_t \cdot \mathcal{M}_u^* = \dfrac{f_t f_u \left|g_1 g_2\right|^2}{4N_c} \cdot \dfrac{1}{\left(t - m_t^2\right)\left(u - m_u^2\right)} \\ \times Tr\left(\left(\not{p}_2 - m_2\right)\gamma_5\left(\not{p}_1 - \not{p}_3 + m_t\right)\gamma_5\left(\not{p}_1 + m_1\right)\gamma_5\left(\not{p}_1 - \not{p}_4 + m_u\right)\gamma_5\right) \end{array}. \tag{19}$$

The trace calculation is more delicate than in (16). Taking into account the relations presented in the subsection 2.2, we propose the following formula:

$$\begin{array}{l} Tr\left(\left(\not{p}_a + m_a\right)\gamma_5\left(\not{p}_b + m_b\right)\gamma_5\left(\not{p}_c + m_c\right)\gamma_5\left(\not{p}_d + m_d\right)\gamma_5\right) \\ = \dfrac{1}{4} \cdot \begin{pmatrix} Tr\left(\left(\not{p}_a + m_a\right)\gamma_5\left(\not{p}_b + m_b\right)\gamma_5\right) \cdot Tr\left(\left(\not{p}_c + m_c\right)\gamma_5\left(\not{p}_d + m_d\right)\gamma_5\right) \\ + Tr\left(\left(\not{p}_b + m_b\right)\gamma_5\left(\not{p}_c + m_c\right)\gamma_5\right) \cdot Tr\left(\left(\not{p}_d + m_d\right)\gamma_5\left(\not{p}_a + m_a\right)\gamma_5\right) \\ - Tr\left(\left(\not{p}_a + m_a\right)\gamma_5\left(\not{p}_c + m_c\right)\gamma_5\right) \cdot Tr\left(\left(\not{p}_b + m_b\right)\gamma_5\left(\not{p}_d + m_d\right)\gamma_5\right) \end{pmatrix} \end{array}. \tag{20}$$

The relation (20) is valid with four $\gamma_5$ matrices. In cases involving two or zero $\gamma_5$ matrices, the formula is adaptable, using the property $\gamma_5 \cdot \gamma^\mu \cdot \gamma_5 = -\gamma^\mu$. In fact, by this relation, the idea is to split the trace into smallest ones (comparable to the trace that intervenes in (16)), whose general form is:

$$Tr\left(\left(\not{p}_a + m_a\right)\gamma_5\left(\not{p}_b + m_b\right)\gamma_5\right) = 4\left(-p_a \cdot p_b + m_a \cdot m_b\right), \tag{21}$$

where $p_a \cdot p_b$ is a dot product of two four-vectors. This product is function of the Mandelstam variables, as listed in the Appendix F. So, thanks to (20), we obtain the result of [3]:

$$\frac{1}{4N_c^{\,2}}\cdot\sum_{\substack{\text{spin}\\ \text{color}}}\mathcal{M}_t\cdot\mathcal{M}_u^{\,*}=\frac{f_t f_u\left|g_1 g_2\right|^2}{4N_c}\cdot\frac{1}{\left(t-m_t^{\,2}\right)\left(u-m_u^{\,2}\right)}$$

$$\times\begin{pmatrix}-\left(m_2^2-m_3^2+u\right)\left(m_1^2-m_3^2+t\right)-\left(m_1^2-m_4^2+u\right)\left(m_2^2-m_4^2+t\right)\\ +2m_1m_u\cdot\left(m_2^2-m_4^2+t\right)+2m_2m_u\cdot\left(m_1^2-m_3^2+t\right)\\ +2m_1m_t\cdot\left(m_2^2-m_3^2+u\right)+2m_2m_t\cdot\left(m_1^2-m_4^2+u\right)\\ +\left(s-\left(m_1-m_2\right)^2\right)\left(-m_1^2-m_2^2+m_3^2+m_4^2\right)+\left(s-\left(m_1+m_2\right)^2\right)\left(2m_um_t\right)\end{pmatrix}. \tag{22}$$

Naturally, $\mathcal{M}_u\cdot\mathcal{M}_t^{\,*}$ leads to the same result.

In the same spirit of (20), we also propose the relation:

$$\begin{aligned}&Tr\left(\left(\not{p}_a+m_a\right)\gamma^5\left(\not{p}_b+m_b\right)\gamma^5\left(\not{p}_c+m_c\right)\right)\\ =\;& m_a\cdot Tr\left(\not{p}_b\gamma^5\not{p}_c\gamma^5\right)+m_b\cdot Tr\left(\left(\not{p}_a+m_a\right)\left(\not{p}_c+m_c\right)\right)+m_c\cdot Tr\left(\not{p}_a\gamma^5\not{p}_b\gamma^5\right)\end{aligned}, \tag{23}$$

in the cases of traces involving thee terms as $\not{p}_i+m_i$.

# 4. Dirac spinors terms using different momenta

## 4.1 Preliminary calculations

We start from the relations described in [4]. From this reference, we extract the formulas involving *u* spinors with two momenta $p_1$ and $p_2$, which are associated, respectively, to the masses $m_1$ and $m_2$:

$$\begin{cases}u\left(p_1,\pm\right)\cdot\bar{u}\left(p_2,\pm\right)=\left(j_+\cdot\dfrac{1\pm\gamma_5}{2}-j_-\cdot\dfrac{1\mp\gamma_5}{2}+m_1\cdot\not{k}_2-m_2\cdot\not{k}_1\right)\cdot\dfrac{\not{\omega}_\pm}{\sqrt{2}}\\ u\left(p_1,\pm\right)\cdot\bar{u}\left(p_2,\mp\right)=\left(j_+\cdot\not{k}_1+m_1\right)\cdot\not{k}_2\cdot\dfrac{1\pm\gamma_5}{2}+\left(j_-\cdot\not{k}_2+m_2\right)\cdot\not{k}_1\cdot\dfrac{1\mp\gamma_5}{2}\end{cases}, \tag{24}$$

where $\pm$ is linked to the spin, and:

$$\not{k}_1=\frac{1}{2\Delta}\cdot\left(j_+\not{p}_1-\frac{m_1}{m_2}j_-\not{p}_2\right)\qquad \not{k}_2=\frac{1}{2\Delta}\cdot\left(j_+\not{p}_2-\frac{m_2}{m_1}j_-\not{p}_1\right)\;, \tag{25}$$

$$j_\pm=\frac{\sqrt{p_1\cdot p_2+m_1\cdot m_2}\pm\sqrt{p_1\cdot p_2-m_1\cdot m_2}}{\sqrt{2}}\qquad \Delta=\sqrt{\left(p_1\cdot p_2\right)^2-\left(m_1\cdot m_2\right)^2}\;, \tag{26}$$

in which $k_1,k_2,\omega_+,\omega_-$ designate light-like four-vectors, as defined in [4]. In practice, we do not have to detail the $\omega_\pm$ terms in the calculations described in this appendix. Furthermore, $k_1,k_2,\omega_+,\omega_-$ satisfy the following properties:

$$\begin{cases} k_i^{\,2}=0 \\ k_i\cdot k_j=\dfrac{1}{2} \end{cases} \qquad \begin{cases} \omega_\pm\cdot\omega_\pm=0 \\ \omega_\pm\cdot\omega_\mp=-1 \\ \omega_\pm\cdot k_i=0 \end{cases} \qquad \left| \begin{matrix} i,j=1,2 \\ i\neq k \end{matrix} \right. . \tag{27}$$

We try now to express terms as $v(p_1,\pm)\cdot\bar{v}(p_2,\pm)$. To reach this goal, we employ the relations:

$$\not{\omega}_- u(p,+)=\sqrt{2}v(p,+) \qquad \not{\omega}_+ u(p,-)=-\sqrt{2}v(p,-)\,, \tag{28}$$

and we write:

$$\begin{aligned} v(p_1,\pm)\cdot\bar{v}(p_2,\pm) &= \frac{\not{\omega}_- u(p_1,+)}{\sqrt{2}}\cdot\left(\frac{\not{\omega}_- u(p_2,+)}{\sqrt{2}}\right)^\dagger\cdot\gamma_0 \\ &= \frac{1}{2}\cdot\not{\omega}_- u(p_1,+)\cdot\left(u(p_2,+)\right)^\dagger\cdot\gamma_0\cdot\not{\omega}_-\cdot\gamma_0\cdot\gamma_0=\frac{1}{2}\cdot\not{\omega}_- u(p_1,+)\cdot\bar{u}(p_2,+)\not{\omega}_- \end{aligned} . \tag{29}$$

If we write the $u(p_1,+)\cdot\bar{u}(p_2,+)$ term thanks to (24), it gives:

$$v(p_1,+)\cdot\bar{v}(p_2,+)=\frac{1}{2}\cdot\not{\omega}_-\cdot\left(j_+\cdot\frac{1+\gamma_5}{2}-j_-\cdot\frac{1-\gamma_5}{2}+m_1\cdot\not{k}_2-m_2\cdot\not{k}_1\right)\cdot\frac{\not{\omega}_+}{\sqrt{2}}\cdot\not{\omega}_- \,. \tag{30}$$

$\not{\omega}_\pm\cdot\not{k}_i+\not{k}_i\cdot\not{\omega}_\pm=2\cdot\omega_\pm\cdot k_i=0$, thus we deduce that $\not{\omega}_\pm\cdot\not{k}_i=-\not{k}_i\cdot\not{\omega}_\pm$. In addition, using $\{\gamma_\mu,\gamma_5\}=0$, the equation (30) becomes:

$$v(p_1,+)\cdot\bar{v}(p_2,+)=\left(j_+\cdot\frac{1-\gamma_5}{2}-j_-\cdot\frac{1+\gamma_5}{2}-m_1\cdot\not{k}_2+m_2\cdot\not{k}_1\right)\cdot\frac{\not{\omega}_-\cdot\not{\omega}_+}{2\cdot\sqrt{2}}\cdot\not{\omega}_- \,. \tag{31}$$

Also, $\not{\omega}_-\cdot\not{\omega}_++\not{\omega}_+\cdot\not{\omega}_-=-2$, so that $\not{\omega}_-\cdot\not{\omega}_+\cdot\not{\omega}_-=-2\not{\omega}_--\not{\omega}_+\cdot\not{\omega}_-\cdot\not{\omega}_-=-2\not{\omega}_-$. The method is identical with $v(p_1,-)\cdot\bar{v}(p_2,-)$. It leads to write the relation:

$$v(p_1,\pm)\cdot\bar{v}(p_2,\pm)=\left(-j_+\cdot\frac{1\mp\gamma_5}{2}+j_-\cdot\frac{1\pm\gamma_5}{2}+m_1\cdot\not{k}_2-m_2\cdot\not{k}_1\right)\cdot\frac{\not{\omega}_\mp}{\sqrt{2}}\,. \tag{32}$$

In parallel, with $v(p_1,+)\cdot\bar{v}(p_2,-)$, we obtain:

$$\begin{aligned} v(p_1,+)\cdot\bar{v}(p_2,-) &= -\frac{1}{2}\cdot\not{\omega}_- u(p_1,+)\cdot\bar{u}(p_2,-)\not{\omega}_+ \\ &= -\frac{1}{2}\cdot\not{\omega}_-\left(\left(j_+\cdot\not{k}_1+m_1\right)\cdot\not{k}_2\cdot\frac{1+\gamma_5}{2}+\left(j_-\cdot\not{k}_2+m_2\right)\cdot\not{k}_1\cdot\frac{1-\gamma_5}{2}\right)\not{\omega}_+ \\ &= \left(\left(-j_+\cdot\not{k}_1+m_1\right)\cdot\left(-\not{k}_2\right)\cdot\frac{1-\gamma_5}{2}+\left(-j_-\cdot\not{k}_2+m_2\right)\cdot\left(-\not{k}_1\right)\cdot\frac{1+\gamma_5}{2}\right)\frac{\not{\omega}_-\cdot\not{\omega}_+}{-2} \\ &= \left(\left(j_+\cdot\not{k}_1-m_1\right)\cdot\not{k}_2\cdot\frac{1-\gamma_5}{2}+\left(j_-\cdot\not{k}_2-m_2\right)\cdot\not{k}_1\cdot\frac{1+\gamma_5}{2}\right)\frac{-2-\not{\omega}_+\cdot\not{\omega}_-}{-2} \end{aligned} . \tag{33}$$

The terms formed by $\not{\omega}_+\cdot\not{\omega}_-$ vanish because they end with $\not{k}_2(1-\gamma_5)\not{\omega}_+$ or $\not{k}_1(1+\gamma_5)\not{\omega}_+$. Indeed, if we rewrite them as $\not{k}_2\cdot\not{\omega}_+(1+\gamma_5)$ and $\not{k}_1\cdot\not{\omega}_+(1-\gamma_5)$, [4] indicates that such expressions are null. The same method is applied to find the expression associated with $v(p_1,-)\cdot\bar{v}(p_2,+)$.

Finally, if we gather all these relations, we have:

$$\begin{cases} u(p_1,\pm)\cdot\bar{u}(p_2,\pm) = \left( j_+ \cdot \dfrac{1\pm\gamma_5}{2} - j_- \cdot \dfrac{1\mp\gamma_5}{2} + m_1 \cdot \not{k}_2 - m_2 \cdot \not{k}_1 \right) \cdot \dfrac{\not{\omega}_\pm}{\sqrt{2}} \\ u(p_1,\pm)\cdot\bar{u}(p_2,\mp) = \left( j_+ \cdot \not{k}_1 + m_1 \right) \cdot \not{k}_2 \cdot \dfrac{1\pm\gamma_5}{2} + \left( j_- \cdot \not{k}_2 + m_2 \right) \cdot \not{k}_1 \cdot \dfrac{1\mp\gamma_5}{2} \\ v(p_1,\pm)\cdot\bar{v}(p_2,\pm) = \left( -j_+ \cdot \dfrac{1\mp\gamma_5}{2} + j_- \cdot \dfrac{1\pm\gamma_5}{2} + m_1 \cdot \not{k}_2 - m_2 \cdot \not{k}_1 \right) \cdot \dfrac{\not{\omega}_\mp}{\sqrt{2}} \\ v(p_1,\pm)\cdot\bar{v}(p_2,\mp) = \left( j_+ \cdot \not{k}_1 - m_1 \right) \cdot \not{k}_2 \cdot \dfrac{1\mp\gamma_5}{2} + \left( j_- \cdot \not{k}_2 - m_2 \right) \cdot \not{k}_1 \cdot \dfrac{1\pm\gamma_5}{2} \end{cases} . \tag{34}$$

# 4.2 Application

As an example to apply (34), we calculate the mixed term $\mathcal{M}_u \cdot \mathcal{M}_t^*$ of the process $\mathrm{q}+\mathrm{q} \to \mathrm{D}+\mathrm{M}$. The *u* and *t* channels are expressed by the following matrix elements:

$$\begin{aligned} -i\mathcal{M}_t &= f_t \ \delta_{c_t,c_2} \ \varepsilon^{c_1,c_t,c_3} \ \bar{v}(p_1) \ i\gamma_5 \ ig_1 \ iS_F(p_3-p_1) \ i\gamma_5 \ ig_2 \ u(p_2) \\ -i\mathcal{M}_u &= f_u \ \delta_{c_u,c_1} \ \varepsilon^{c_2,c_u,c_3} \ \bar{v}(p_2) \ i\gamma_5 \ ig_1 \ iS_F(p_1-p_4) \ i\gamma_5 \ ig_2 \ u(p_1) \end{aligned} . \tag{35}$$

We pose $m_t$ and $m_u$ as the masses of the propagated quarks according to the two channels *t* and *u*. Assuming that the $\varepsilon$ terms lead to one $+1$ term, the mixed term is written as:

$$\begin{aligned} &\frac{1}{4{N_c}^2} \cdot \sum_{\substack{\text{spin}\\ \text{color}}} \mathcal{M}_u \cdot \mathcal{M}_t^* = \frac{f_u f_t \left|g_1 g_2\right|^2}{4N_c} \cdot \frac{1}{\left(t-{m_t}^2\right)\left(u-{m_u}^2\right)} \\ &\times \sum_{s_1,s_2} v(p_1,s_1)\bar{v}(p_2,s_2)\cdot\gamma_5\left(\not{p}_1-\not{p}_4+m_u\right)\gamma_5 \cdot u(p_1,s_1)\bar{u}(p_2,s_2)\cdot\gamma_5\left(\not{p}_3-\not{p}_1+m_t\right)\gamma_5 \end{aligned} . \tag{36}$$

Here, the summations associated with the spins $s_1, s_2$ cannot be uncoupled as in section 3. So, it gives birth to four cases, i.e. $s_1=+, s_2=+$ ; $s_1=+, s_2=-$ ; $s_1=-, s_2=+$ ; $s_1=-, s_2=-$, labeled respectively as $T^{++}, T^{+-}, T^{-+}, T^{--}$. Firstly, concerning $T^{++}$, the use of (34) allows writing the term as:

$$T^{++} = Tr\left( \begin{aligned} &\left( -j_+ \cdot \frac{1-\gamma_5}{2} + j_- \cdot \frac{1+\gamma_5}{2} + m_1 \cdot \not{k}_2 - m_2 \cdot \not{k}_1 \right) \cdot \frac{\not{\omega}_-}{\sqrt{2}} \cdot \left( -\not{p}_1 + \not{p}_4 + m_u \right) \\ &\times \left( j_+ \cdot \frac{1+\gamma_5}{2} - j_- \cdot \frac{1-\gamma_5}{2} + m_1 \cdot \not{k}_2 - m_2 \cdot \not{k}_1 \right) \cdot \frac{\not{\omega}_+}{\sqrt{2}} \cdot \left( -\not{p}_3 + \not{p}_1 + m_t \right) \end{aligned} \right) . \tag{37}$$

Aware of the pseudo scalar behavior of the $\gamma_5$ matrices, we cut the expression (37) in two parts. The 6 terms that compose the first one are labeled for reason of convenience in the developments performed hereafter. The $\gamma_5$ matrices of the second part are treated with the relations of the Clifford algebra described subsection 2.2. We have:

$$T^{++} = Tr\left(\begin{array}{l} \overbrace{\left(-j_+\cdot\frac{1}{2}+j_-\cdot\frac{1}{2}+m_1\cdot \not{k}_2-m_2\cdot\not{k}_1\right)}^{1}\cdot\overbrace{\frac{\not{\omega}_-}{\sqrt{2}}}^{2}\cdot\overbrace{\left(-\not{p}_1+\not{p}_4+m_u\right)}^{3} \\ \times\underbrace{\left(j_+\cdot\frac{1}{2}-j_-\cdot\frac{1}{2}+m_1\cdot \not{k}_2-m_2\cdot\not{k}_1\right)}_{4}\cdot\underbrace{\frac{\not{\omega}_+}{\sqrt{2}}}_{5}\cdot\underbrace{\left(-\not{p}_3+\not{p}_1+m_t\right)}_{6}\end{array}\right) \tag{38}$$
$$+Tr\left(\begin{array}{l}\left(j_+\cdot\frac{\gamma_5}{2}+j_-\cdot\frac{\gamma_5}{2}\right)\cdot\frac{\not{\omega}_-}{\sqrt{2}}\cdot\left(-\not{p}_1+\not{p}_4+m_u\right)\\ \times\left(j_+\cdot\frac{\gamma_5}{2}+j_-\cdot\frac{\gamma_5}{2}\right)\cdot\frac{\not{\omega}_+}{\sqrt{2}}\cdot\left(-\not{p}_3+\not{p}_1+m_t\right)\end{array}\right).$$

The first trace of (38) presents a similar structure as the one of (20). The difference is (38) has got 6 terms, whereas (20) involved 4 terms. The idea is now to extend the method. Firstly, we consider a relation as in (20), but using $\gamma_5\cdot\gamma^\mu\cdot\gamma_5=-\gamma^\mu$ to "remove" the $\gamma_5$ matrices in this relation. We establish that:

$$\begin{array}{l} Tr\left(\left(\not{p}_1+m_1\right)\cdot\left(\not{p}_2+m_2\right)\cdot\left(\not{p}_3+m_3\right)\cdot\left(\not{p}_4+m_4\right)\right) \\ =\frac{1}{4}\cdot\left(\begin{array}{l} Tr\left(\left(\not{p}_1+m_1\right)\cdot\left(\not{p}_2+m_2\right)\right)\cdot Tr\left(\left(\not{p}_3+m_3\right)\cdot\left(\not{p}_4+m_4\right)\right) \\ +Tr\left(\left(\not{p}_1+m_1\right)\cdot\left(\not{p}_4+m_4\right)\right)\cdot Tr\left(\left(\not{p}_2+m_2\right)\cdot\left(\not{p}_3+m_3\right)\right) \\ -Tr\left(\left(\not{p}_1-m_1\right)\cdot\left(\not{p}_3+m_3\right)\right)\cdot Tr\left(\left(\not{p}_2-m_2\right)\cdot\left(\not{p}_4+m_4\right)\right)\end{array}\right).\end{array} \tag{39}$$

In a symbolic way, we rewrite (39) as:

$$Tr(1,2,3,4)=\frac{1}{4}\cdot\left(Tr(1,2)\cdot Tr(3,4)+Tr(1,4)\cdot Tr(2,3)-Tr\left(\overline{1,3}\right)\cdot Tr\left(\overline{2,4}\right)\right). \tag{40}$$

Each number is associated with a term involving $\not{p}_i+m_i$, i.e. a four-vector that uses the Feynman slash, and a scalar (a mass in (39)). The $Tr\left(\overline{i,j}\right)$ notation recalls that the sign of *one* of the scalars of $Tr\left(\overline{i,j}\right)$ must be reversed, as shown in (39). This method is then extended to trace expressions involving more than 4 of such terms. The condition is the number of terms must be even. Obviously, the method can be also derived in the case of uneven terms.

The general idea of this trace splitting method is to associate the terms by group of two, in order to form Lorentz invariants, which the trace is easily evaluable. In the case of a trace with 4 terms, 3 permutations are possible, as in (40). The sign placed in front of some traces, as with $-Tr\left(\overline{1,3}\right)\cdot Tr\left(\overline{2,4}\right)$ in (40), corresponds to the signature of the applied permutation. More precisely, the $Tr\left(\overline{i,j}\right)$ traces concern couples formed by terms with the same parity (even/even or uneven/uneven). This complication is motivated by our wish to gather the four-vectors and the scalars together, in order to avoid the difficulty to expand the whole expression completely. Also, the $1/4$ factor in front of the expression, in (40), recalls that a trace calculation of a Lorentz invariant adds a factor equal to 4, subsection 2.2. Such a factor is required to express a trace by a product of two traces. In the case of $n$ terms, the value of the factor is $4^{\frac{-n}{2}+1}$. We recall that $n$ is considered here as an even number. In addition, $(n-1)!!$ permutations are possible, where !! designates the double factorial, i.e. the product of uneven positive integers less than $n$. With (40), we have $n=4$: we check that we have $(3)!!=1\times 3=3$ permutations.

We apply this method to the 6-term trace of the equation (38). It forms a total of 15 possibilities of permutations. Most of them vanish, using the properties (27). Among the 5 remaining permutations, 3 of them concern the trace $Tr(1,4)$. After some simplifications, these ones can be expressed in the following way:

$$-\frac{1}{16}\cdot\frac{1}{2}\cdot\left(j_+^{\,2}+j_-^{\,2}\right)\begin{pmatrix} Tr\left(\not{\omega}_-\cdot\left(\not{p}_1-\not{p}_4\right)\right)\cdot Tr\left(\not{\omega}_+\cdot\left(\not{p}_3-\not{p}_1\right)\right) \\ +Tr\left(\not{\omega}_+\cdot\left(\not{p}_1-\not{p}_4\right)\right)\cdot Tr\left(\not{\omega}_-\cdot\left(\not{p}_3-\not{p}_1\right)\right) \\ +16\left(\left(\not{p}_1-\not{p}_4\right)\cdot\left(\not{p}_3-\not{p}_1\right)+m_u m_t\right) \end{pmatrix}. \tag{41}$$

However, the second part of (38), which gathered the $\gamma_5$ matrices, can be rewritten as:

$$\begin{aligned} &Tr\begin{pmatrix} \left(j_+\cdot\frac{\gamma_5}{2}+j_-\cdot\frac{\gamma_5}{2}\right)\cdot\frac{\not{\omega}_-}{\sqrt{2}}\cdot\left(-\not{p}_1+\not{p}_4+m_u\right) \\ \times\left(j_+\cdot\frac{\gamma_5}{2}+j_-\cdot\frac{\gamma_5}{2}\right)\cdot\frac{\not{\omega}_+}{\sqrt{2}}\cdot\left(-\not{p}_3+\not{p}_1+m_t\right) \end{pmatrix}, \\ &=-\frac{1}{2}\cdot\frac{1}{4}\cdot\left(j_+^{\,2}+j_-^{\,2}\right)\cdot Tr\left(\not{\omega}_-\cdot\left(\not{p}_1-\not{p}_4+m_u\right)\times\not{\omega}_+\cdot\left(-\not{p}_3+\not{p}_1+m_t\right)\right) \end{aligned} \tag{42}$$

and it gives the opposite. In other words, (41) and (42) vanish themselves.

Then, it only remains 2 permutations, $Tr(1,6)\cdot Tr(2,5)\cdot Tr(3,4)$ and $Tr\left(\overline{1,3}\right)\cdot Tr(2,5)\cdot Tr\left(\overline{4,6}\right)$. These ones are found as strictly equals, and lead to the expression:

$$\begin{aligned} T^{++}=&-\left(-m_1 2k_2\cdot\left(p_1-p_4\right)+m_2 2k_1\cdot\left(p_1-p_4\right)+m_u\left(j_+-j_-\right)\right) \\ &\times\left(-m_1 2k_2\cdot\left(p_3-p_1\right)+m_2 2k_1\cdot\left(p_3-p_1\right)-m_t\left(j_+-j_-\right)\right) \end{aligned}. \tag{43}$$

By symmetry, the $s_1=-, s_2=-$ case, i.e. $T^{--}$, is equal to the result of (43). We write:

$$\begin{aligned} T^{++}+T^{--}=&-2\left(-m_1 2k_2\cdot\left(p_1-p_4\right)+m_2 2k_1\cdot\left(p_1-p_4\right)+m_u\left(j_+-j_-\right)\right) \\ &\times\left(-m_1 2k_2\cdot\left(p_3-p_1\right)+m_2 2k_1\cdot\left(p_3-p_1\right)-m_t\left(j_+-j_-\right)\right) \end{aligned}. \tag{44}$$

For the moment, we will keep this expression, in particular the $k_{i,j}$ four-vectors defined in (25). On the other hand, it is interesting to note that it was not necessary to explicit the $\omega_\pm$ four-vectors [4].

Now, we consider the $s_1=+, s_2=-$ case, written as:

$$T^{+-}=Tr\begin{pmatrix} \left(\left(j_+\not{k}_1-m_1\right)\not{k}_2\frac{1-\gamma_5}{2}+\left(j_-\not{k}_2-m_2\right)\not{k}_1\frac{1+\gamma_5}{2}\right)\left(-\not{p}_1+\not{p}_4+m_u\right) \\ \times\left(\left(j_+\not{k}_1+m_1\right)\not{k}_2\frac{1+\gamma_5}{2}+\left(j_-\not{k}_2+m_2\right)\not{k}_1\frac{1-\gamma_5}{2}\right)\left(-\not{p}_3+\not{p}_1+m_t\right) \end{pmatrix}. \tag{45}$$

Expanding this equation, we obtain the following expression:

$$
\begin{aligned}
T^{+-} &= Tr\left(\left((j_+\not{k}_1 - m_1)\not{k}_2\frac{1-\gamma_5}{2}\right)(-\not{p}_1 + \not{p}_4 + m_u)\left((j_+\not{k}_1 + m_1)\not{k}_2\frac{1+\gamma_5}{2}\right)(-\not{p}_3 + \not{p}_1 + m_t)\right) \\
&+ Tr\left(\left((j_-\not{k}_2 - m_2)\not{k}_1\frac{1+\gamma_5}{2}\right)(-\not{p}_1 + \not{p}_4 + m_u)\left((j_+\not{k}_1 + m_1)\not{k}_2\frac{1+\gamma_5}{2}\right)(-\not{p}_3 + \not{p}_1 + m_t)\right) \\
&+ Tr\left(\left((j_+\not{k}_1 - m_1)\not{k}_2\frac{1-\gamma_5}{2}\right)(-\not{p}_1 + \not{p}_4 + m_u)\left((j_-\not{k}_2 + m_2)\not{k}_1\frac{1-\gamma_5}{2}\right)(-\not{p}_3 + \not{p}_1 + m_t)\right) \\
&+ Tr\left(\left((j_-\not{k}_2 - m_2)\not{k}_1\frac{1+\gamma_5}{2}\right)(-\not{p}_1 + \not{p}_4 + m_u)\left((j_-\not{k}_2 + m_2)\not{k}_1\frac{1-\gamma_5}{2}\right)(-\not{p}_3 + \not{p}_1 + m_t)\right)
\end{aligned}
. \qquad (46)
$$

As with (38), the $\frac{1\pm\gamma_5}{2}$ terms lead to split each line of (46) in two parts. After some manipulations, we found that the first and the fourth traces of (46) give together the result:

$$
\left(j_+^{\,2} + j_-^{\,2}\right)\cdot\begin{pmatrix} -2k_1\cdot(p_3 - p_1)\times k_2\cdot(p_1 - p_4) \\ -2k_1\cdot(p_1 - p_4)\times k_2\cdot(p_3 - p_1) \\ +(p_1 - p_4)\cdot(p_3 - p_1) \end{pmatrix}, \qquad (47)
$$

whereas the second and the third lead to:

$$
2m_1m_2\cdot\begin{pmatrix} -2k_1\cdot(p_3 - p_1)\times k_2\cdot(p_1 - p_4) \\ -2k_1\cdot(p_1 - p_4)\times k_2\cdot(p_3 - p_1) \\ +(p_1 - p_4)\cdot(p_3 - p_1) \end{pmatrix}. \qquad (48)
$$

In other words, only the pre-factor differs between (47) and (48). We also observe that $T^{-+} = T^{+-}$. It allows to write, using the relation $j_+^{\,2} + j_-^{\,2} = 2p_1\cdot p_2$,

$$
T^{+-} + T^{-+} = (2p_1\cdot p_2 + 2m_1m_2)\cdot\begin{pmatrix} -2k_1\cdot(p_3 - p_1)\times 2k_2\cdot(p_1 - p_4) \\ -2k_1\cdot(p_1 - p_4)\times 2k_2\cdot(p_3 - p_1) \\ +2(p_1 - p_4)\cdot(p_3 - p_1) \end{pmatrix}. \qquad (49)
$$

We gather all the cases $T^{++}, T^{+-}, T^{-+}$ and $T^{--}$, and after some simplifications, it comes:

$$
\begin{aligned}
&T^{++} + T^{+-} + T^{-+} + T^{--} = \\
&-2p_2\cdot(p_1 - p_4)\times 2p_1\cdot(p_3 - p_1) - 2p_1\cdot(p_1 - p_4)\times 2p_2\cdot(p_3 - p_1) \\
&+2p_1\cdot p_2\times 2(p_1 - p_4)\cdot(p_3 - p_1) + 2m_1m_2 2(p_1 - p_4)\cdot(p_3 - p_1) \\
&+2m_um_1 2p_2\cdot(p_3 - p_1) - 2m_um_2 2p_1\cdot(p_3 - p_1) \\
&-2m_tm_1 2p_2\cdot(p_1 - p_4) + 2m_tm_2 2p_1\cdot(p_1 - p_4) \\
&+2m_um_t 2p_1\cdot p_2 - 4m_1m_2m_um_t
\end{aligned}
. \qquad (50)
$$

Finally, the complete expression of our mixed term $\mathcal{M}_u \cdot \mathcal{M}_t^*$ is:

$$\frac{1}{4N_c^2} \cdot \sum_{\substack{\text{spin}\\ \text{color}}} \mathcal{M}_u \cdot \mathcal{M}_t^* = \frac{f_u f_t |g_1 g_2|^2}{4N_c} \cdot \frac{1}{\left(t - m_t^2\right)\left(u - m_u^2\right)}$$
$$\times \begin{pmatrix} -\left(m_2^2 - m_3^2 + u\right)\left(m_1^2 - m_3^2 + t\right) - \left(m_1^2 - m_4^2 + u\right)\left(m_2^2 - m_4^2 + t\right) \\ +2m_1 m_u \cdot \left(m_2^2 - m_4^2 + t\right) + 2m_2 m_u \cdot \left(m_1^2 - m_3^2 + t\right) \\ +2m_1 m_t \cdot \left(m_2^2 - m_3^2 + u\right) + 2m_2 m_t \cdot \left(m_1^2 - m_4^2 + u\right) \\ +\left(s - (m_1 - m_2)^2\right)\left(-m_1^2 - m_2^2 + m_3^2 + m_4^2\right) + \left(s - (m_1 + m_2)^2\right)(2m_u m_t) \end{pmatrix}. \tag{51}$$

The calculations done with $\mathcal{M}_t \cdot \mathcal{M}_u^*$ give the same results. Furthermore, (51) is equal to the expression found in (22). This remark can be explained by the method detailed in [5]. More precisely, according to the approach suggested in this paper, we can choice an arbitrary flow that allows rewriting our matrix elements. In the case of (35), the flow going from the particle 1 towards the particle 2 does not modify $-i\mathcal{M}_u$. But, with $-i\mathcal{M}_t$, it leads to the replace $S_F(p_3 - p_1)$ by $S_F(p_1 - p_3)$, $\bar{v}(p_1)$ by $\bar{v}(p_2)$ and $u(p_2)$ by $u(p_1)$. Clearly, except for the $\varepsilon$ terms, we obtain the same matrix elements as the ones of (18). So, it explains why (51) and (22) are equivalent.

# 4.3 Other configurations

Concerning other (and simplest) calculations of this kind, we established that:

$$\begin{aligned} &\sum_{s_1,s_2} u(p_1,s_1)\bar{u}(p_2,s_2) \cdot (\not{p} + m) \cdot v(p_1,s_1)\bar{v}(p_2,s_2) \\ &= 2m(2p_1 \cdot p_2 - 2m_1 m_2) - 2m_1\, 2p_2 \cdot p + 2m_2\, 2p_1 \cdot p \end{aligned}, \tag{52}$$

$$\begin{aligned} &\sum_{s_1,s_2} v(p_1,s_1)\bar{v}(p_2,s_2) \cdot (\not{p} + m) \cdot u(p_1,s_1)\bar{u}(p_2,s_2) \\ &= 2m(2p_1 \cdot p_2 - 2m_1 m_2) + 2m_1\, 2p_2 \cdot p - 2m_2\, 2p_1 \cdot p \end{aligned}, \tag{53}$$

and:

$$\sum_{s_1,s_2} u(p_1,s_1)\bar{u}(p_2,s_2) \cdot v(p_1,s_1)\bar{v}(p_2,s_2) = 2 \cdot \left(s - (m_1 + m_2)^2\right). \tag{54}$$

Equation (54) coincides with the result obtained in equations (16, 17) with:

$$\sum_{s_1,s_2} u(p_1,s_1)\bar{u}(p_1,s_1) \cdot v(p_2,s_2)\bar{v}(p_2,s_2). \tag{55}$$

# Appendix C

# Flavor factors

## 1. General method

A flavor factor is a scalar constant that notably intervenes in the calculation of the cross sections. It is determined at each vertex of the considered Feynman diagram. The method to calculate a flavor factor is the same whatever the studied particles. In the framework of the determination of a flavor factor, the isospin approximation cannot be applied, i.e. the *u* and the *d* quarks must be considered there as different particles.

Each vertex materializes the meeting between three particles: a composite particle (meson, diquark, and baryon) and two other “elementary” particles that give the composite particle when they are joined together. For each composite particle, we write a matrix made with one or with a linear combination of two *SU(3)* generators (Appendix B). In the obtained matrix, each line is associated with a first "elementary" particle. Each column corresponds to the second “elementary” particle. In the case of mesons, the lines are associated by convention with the quarks, the columns with the anti-quarks. For the diquarks, the lines correspond to a first quark, the columns with a second quark. For the baryons, the lines are associated with the diquarks and the columns with the quarks.

## 2. Mesons

| | $\bar{u}$ | $\bar{d}$ | $\bar{s}$ |
|---|---|---|---|
| $u$ | | | |
| $d$ | | | |
| $s$ | | | |

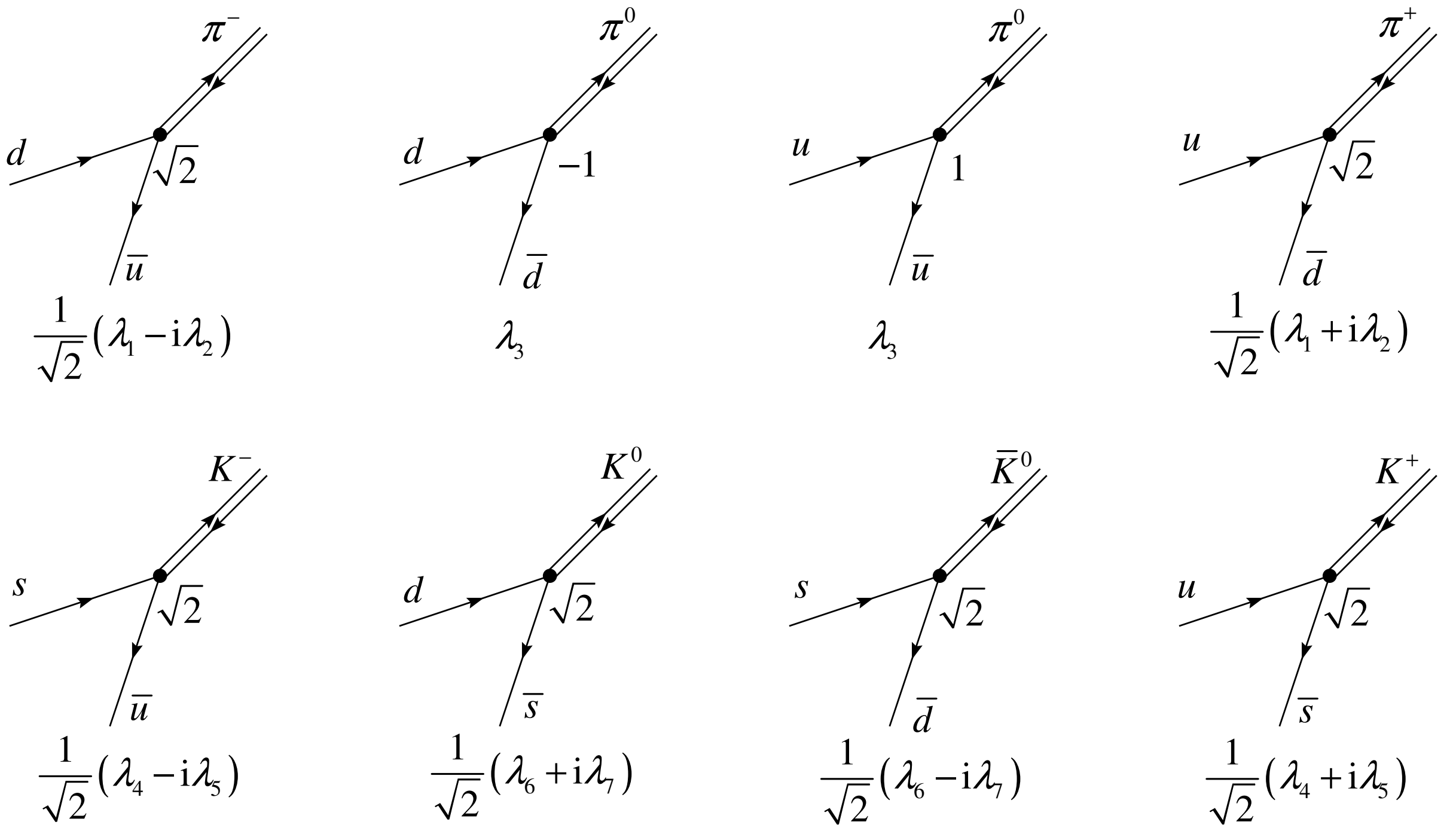

About $\eta$ and $\eta'$, the mixing angle $\theta$ must be used. It was presented in the chapter 3, associated with the mesons. Formally, we write, as in [1]:

$$\begin{bmatrix}\eta\\ \eta'\end{bmatrix}=\begin{bmatrix}\cos(\theta) & -\sin(\theta)\\ \sin(\theta) & \cos(\theta)\end{bmatrix}\begin{bmatrix}\eta_8\\ \eta_0\end{bmatrix}, \tag{1}$$

where:

$$\eta_8 \Leftrightarrow \lambda_8=\frac{1}{\sqrt{3}}\cdot\begin{bmatrix}1 & & \\ & 1 & \\ & & -2\end{bmatrix} \text{ and } \eta_0 \Leftrightarrow \lambda_0=\sqrt{\frac{2}{3}}\cdot 1_3=\sqrt{\frac{2}{3}}\cdot\begin{bmatrix}1 & & \\ & 1 & \\ & & 1\end{bmatrix}. \tag{2}$$

It gives:

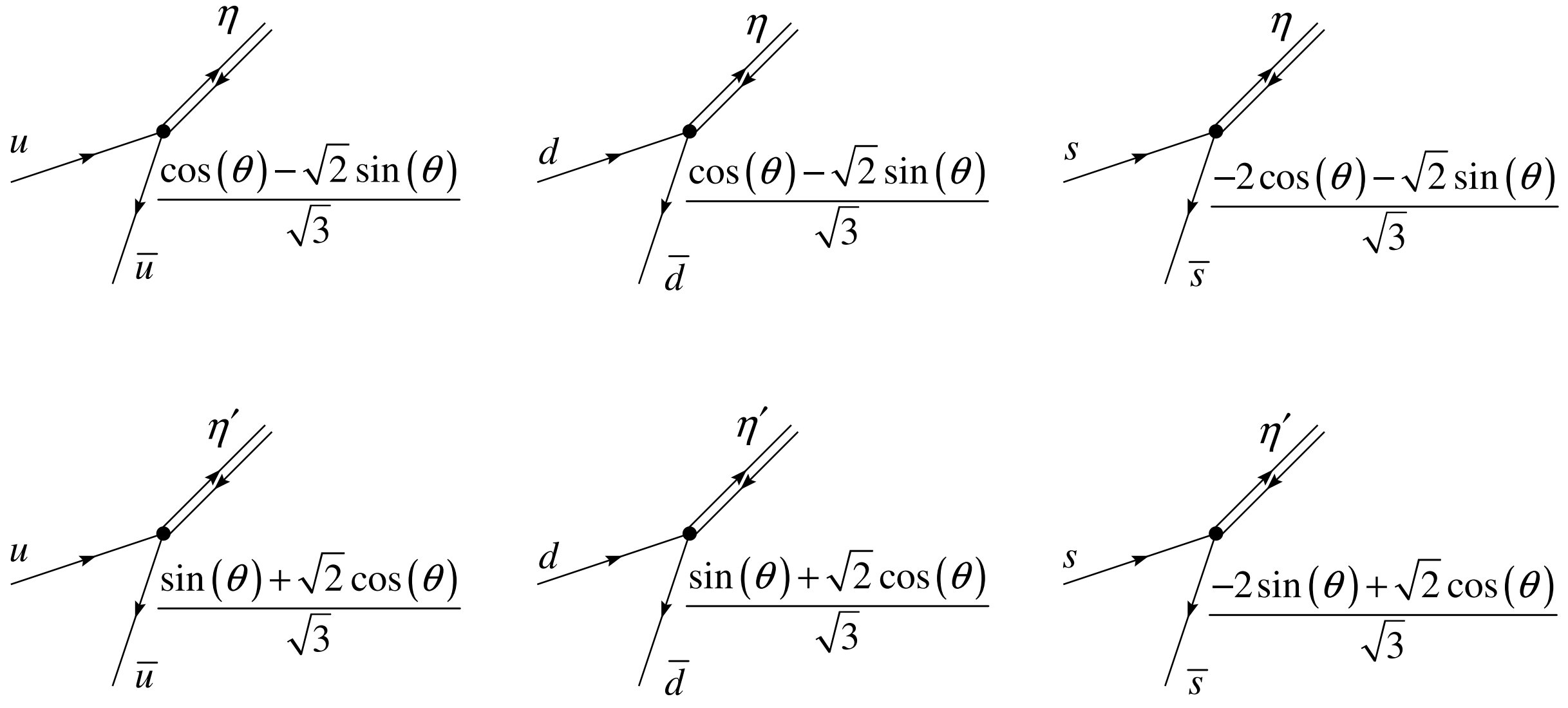

# 3. Diquarks

| | $u$ | $d$ | $s$ |
|---|---|---|---|
| $u$ | | | |
| $d$ | | | |
| $s$ | | | |

## 3.1 Scalar diquarks

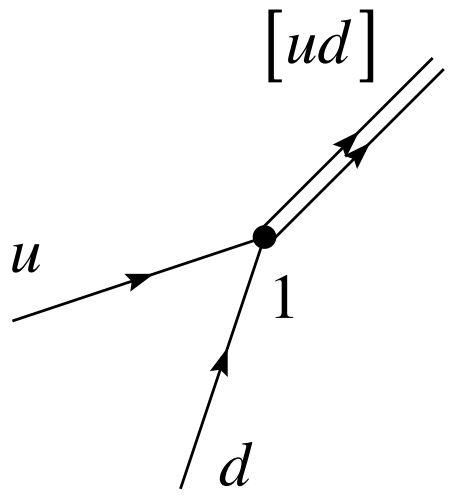

$$\frac{\left.\frac{-\mathrm{i}}{\sqrt{2}}\cdot\lambda_2\right|_{ud}-\left.\frac{-\mathrm{i}}{\sqrt{2}}\cdot\lambda_2\right|_{du}}{-\sqrt{2}}$$

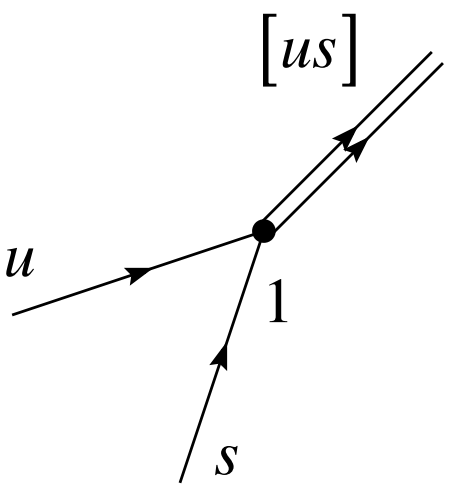

$$\frac{\left.\frac{-\mathrm{i}}{\sqrt{2}}\cdot\lambda_5\right|_{us}-\left.\frac{-\mathrm{i}}{\sqrt{2}}\cdot\lambda_5\right|_{su}}{-\sqrt{2}}$$

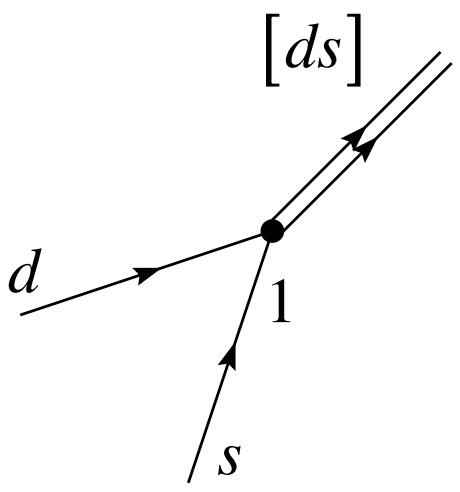

$$\frac{\left.\frac{-\mathrm{i}}{\sqrt{2}}\cdot\lambda_7\right|_{ds}-\left.\frac{-\mathrm{i}}{\sqrt{2}}\cdot\lambda_7\right|_{sd}}{-\sqrt{2}}$$

This writing can be explained by saying that the diquarks $[ud]$, $[us]$, $[ds]$ are in fact linear combinations of two possible associations of two quarks. More precisely, the scalar diquarks wave functions are antisymmetric in flavor. For example, with $[ud]$, we write [2]:

$$[ud]=-\sqrt{\frac{1}{2}}\cdot(ud-du)\,. \tag{3}$$

The things are identical with $[us]$ and $[ds]$.

# 3.2 Axial diquarks

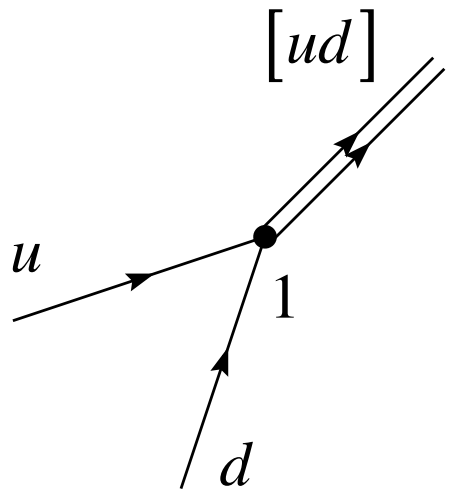

$$\frac{\frac{1}{\sqrt{2}}\cdot\lambda_1\Big|_{ud}+\frac{1}{\sqrt{2}}\cdot\lambda_1\Big|_{du}}{\sqrt{2}}$$

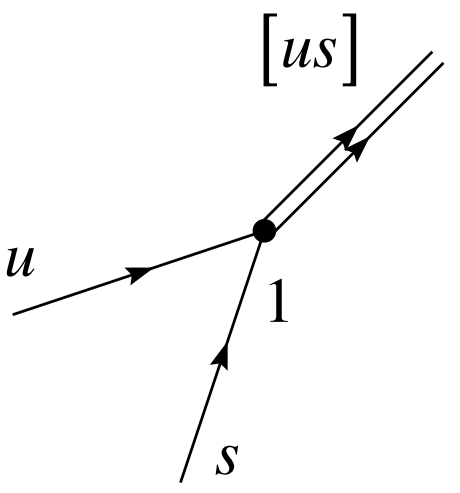

$$\frac{\frac{1}{\sqrt{2}}\cdot\lambda_4\Big|_{us}+\frac{1}{\sqrt{2}}\cdot\lambda_4\Big|_{su}}{\sqrt{2}}$$

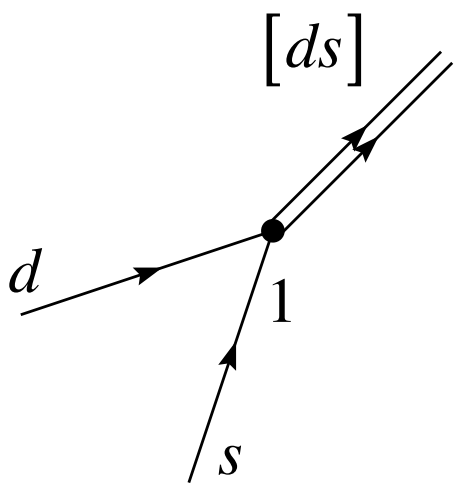

$$\frac{\frac{1}{\sqrt{2}}\cdot\lambda_6\Big|_{ds}+\frac{1}{\sqrt{2}}\cdot\lambda_6\Big|_{sd}}{\sqrt{2}}$$

The reasoning is here identical compared to the scalar diquarks. Only the matrices and the linear combination [2] are different, since the axial diquarks wave functions are symmetrical in flavor:

$$[ud]=\sqrt{\frac{1}{2}}\cdot(ud+du)\,. \tag{4}$$

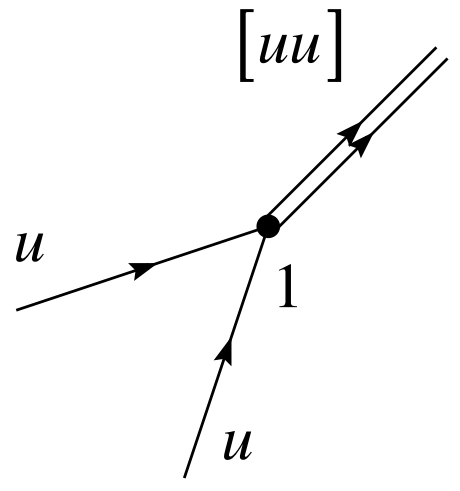

$1_3$

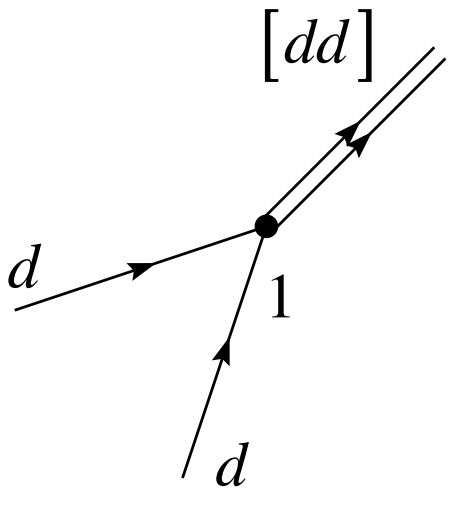

$1_3$

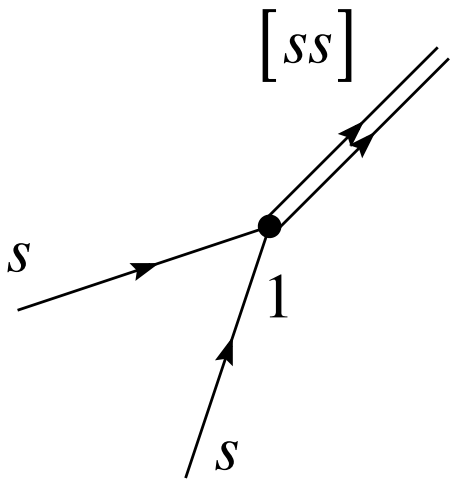

$1_3$

$1_3$ indicates the $3\times 3$ identity matrix.

# 4. Baryons

| | $u$ | $d$ | $s$ |
|---|---|---|---|
| $[ds]$ | | | |
| $[us]$ | | | |
| $[ud]$ | | | |

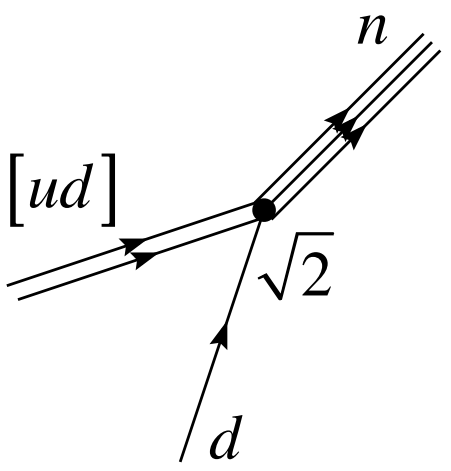

$$\frac{1}{\sqrt{2}}(\lambda_6 - i\lambda_7)$$

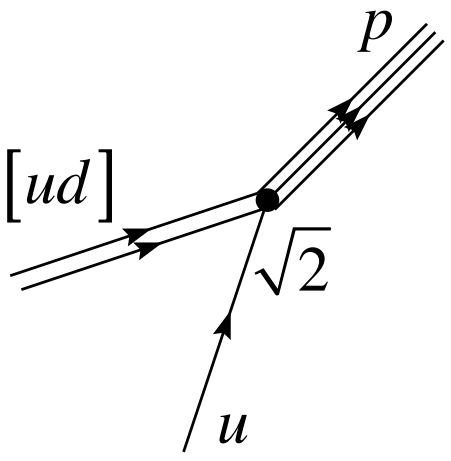

$$\frac{1}{\sqrt{2}}(\lambda_4 - i\lambda_5)$$

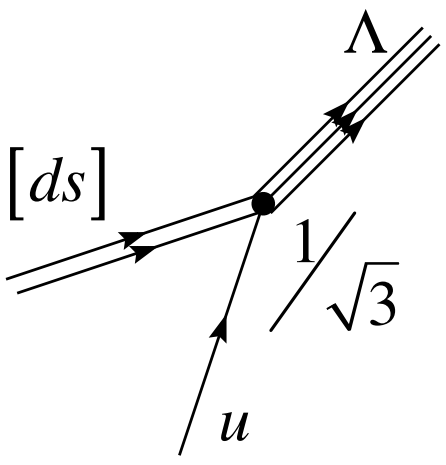

$\lambda_8$

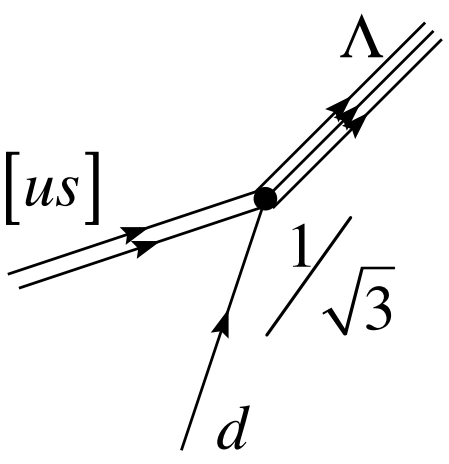

$\lambda_8$

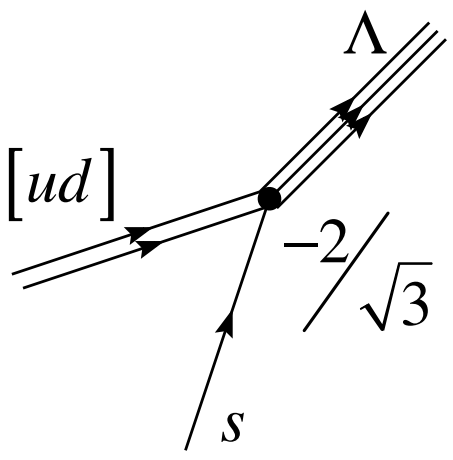

$\lambda_8$

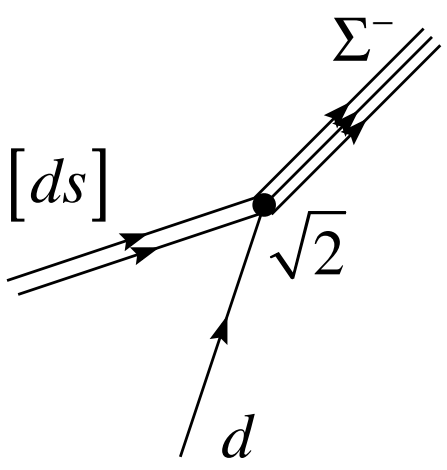

$$\frac{1}{\sqrt{2}}(\lambda_1 + i\lambda_2)$$

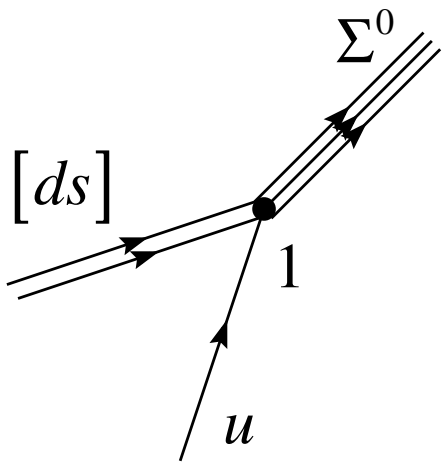

$\lambda_3$

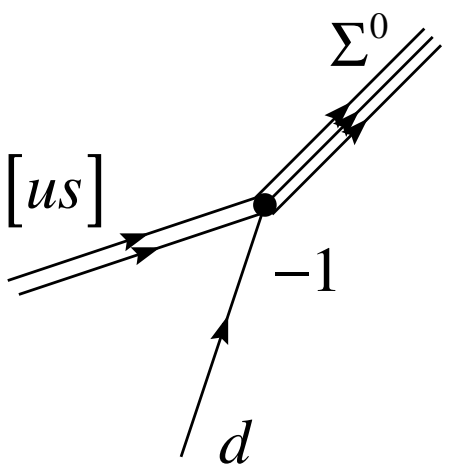

$\lambda_3$

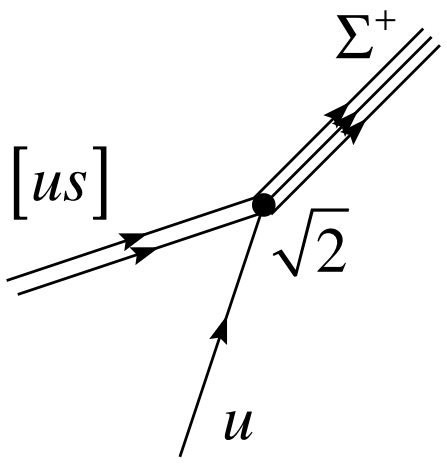

$$\frac{1}{\sqrt{2}}(\lambda_1 - i\lambda_2)$$

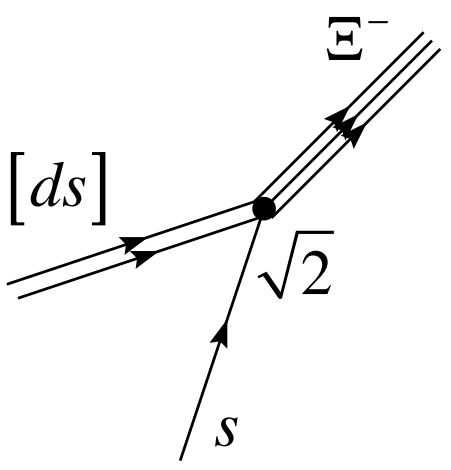

$$\frac{1}{\sqrt{2}}(\lambda_4 + i\lambda_5)$$

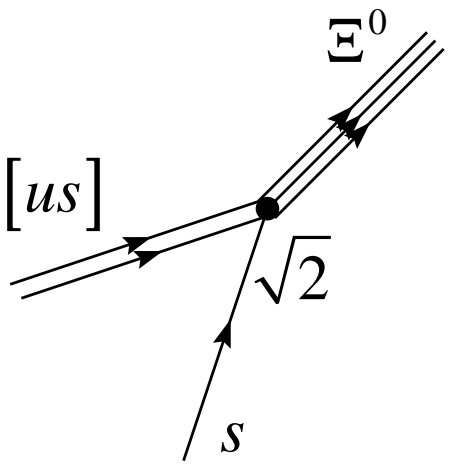

$$\frac{1}{\sqrt{2}}(\lambda_6 + i\lambda_7)$$

# Appendix D

# Loop and triangle functions

A part of this appendix was published in *J. Phys. G: Nucl. Part. Phys.* **38** 105003

## 1. Preliminary calculations

Before describing the used functions, we establish here some general formulas, in order to perform Matsubara summations. The calculations detailed in the following lines are strongly inspired from [1], which we adapted in our case. The goal is to evaluate the following quantity:

$$\frac{1}{\beta}\sum_{n=-\infty}^{\infty}\phi(i\cdot\omega_n), \tag{1}$$

in which $n$ is an integer, $\beta = 1/T$ and $i\cdot\omega_n$ is a Matsubara frequency that can be:

- fermionic: $$\omega_n^{FD} = \frac{(2n+1)\cdot\pi}{\beta} \tag{2a}$$
- bosonic : $$\omega_n^{BE} = \frac{2n\cdot\pi}{\beta} \tag{2b}$$

We suppose that the function $\phi$ has only simple poles, which we note $z_e$. The $i\cdot\omega_n$ never belong to these poles. Also, with the NJL approach, we consider:

$$f^{BE}(z) = \frac{1}{e^{\beta\cdot z}-1} \quad \text{and} \quad f^{FD}(z) = \frac{1}{e^{\beta\cdot z}+1}, \tag{3}$$

i.e. the Fermi-Dirac and Bose-Einstein statistics. We check that $i\cdot\omega_n^{FD}$ and $i\cdot\omega_n^{BE}$ are, respectively, the poles of $f^{FD}(z)$ and of $f^{BE}(z)$. In the framework of the PNJL model, $f^{FD}$ must be updated, as mentioned in chapter 2 and in [2], but the results stay valid.

We consider now the following theorem, quite useful for a residue calculation:

*if a function* $\psi(z)$ *is written on the form* $\psi(z) = \dfrac{P(z)}{Q(z)}$, *then* $res(\psi, z_0) = \dfrac{P(z_0)}{\left.\dfrac{\partial Q}{\partial z}\right|_{z=z_0}}$. (4)

From (4), we deduce that:

$$res\left(f^{BE}, i\cdot\omega_n^{BE}\right) = \frac{1}{\beta} \quad \text{and} \quad res\left(f^{FD}, i\cdot\omega_n^{FD}\right) = -\frac{1}{\beta}. \tag{5}$$

Therefore, the function $f^{FD}\cdot\phi$ (respectively $f^{BE}\cdot\phi$) admits $z_e$ and $i\cdot\omega_n^{FD}$ (respectively $i\cdot\omega_n^{BE}$) as poles. We integrate these functions in the complex plane, on a circular contour $\mathcal{C}$ with radius $R\to\infty$, i.e. which includes all the poles. We have, according to the Residue Theorem:

$$\underbrace{\int_{\mathcal{C}}\frac{dz}{2i\pi}\cdot f^{FD/BE}(z)\cdot\phi(z)}_{0}=res\left(f^{FD/BE},i\cdot\omega_n^{FD/BE}\right)\cdot\sum_n\phi\left(i\cdot\omega_n^{FD/BE}\right)+res(\phi,z_e)\cdot\sum_e f^{FD/BE}(z_e)\,. \quad (6)$$

This gives the two following relations, as in [1]:

$$\begin{cases}\dfrac{1}{\beta}\sum_n\phi\left(i\cdot\omega_n^{FD}\right)=\sum_e f^{FD}(z_e)\cdot res(\phi,z_e) & (7a)\\ -\dfrac{1}{\beta}\sum_n\phi\left(i\cdot\omega_n^{BE}\right)=\sum_e f^{BE}(z_e)\cdot res(\phi,z_e) & (7b)\end{cases}.$$

# 2. Generic functions

Except for the baryons loop function, the functions used in our calculations are built from three generic functions $A$, $B_0$ and $C_0$. They were introduced by S. Klevansky's group [3]. They intervene respectively in loops with 1, 2, 3 fermions. Here, these fermions are quarks.

## 2.1 *A* function (one-fermion loop)

*A* is a real function that only admits real arguments. It is defined by:

$$A(m,\mu,\beta,\Lambda)=\frac{16\pi^2}{\beta}\cdot\sum_n\int\frac{d^3p}{(2\pi)^3}\frac{1}{(i\cdot\omega_n+\mu)^2-E^2}\,, \quad (8)$$

with:

$$E=\sqrt{\vec{p}^2+m^2}\,. \quad (9)$$

The fraction present in the right part of (8) is associated with a quark (fermion). Thus, $i\cdot\omega_n$ is fermionic:

$$\omega_n=\frac{(2n+1)\cdot\pi}{\beta}\,. \quad (10)$$

The poles are:

$$\begin{cases}i\cdot\omega_n=E-\mu\\ i\cdot\omega_n=-E-\mu\end{cases}. \quad (11)$$

So, using (7a):

$$\begin{aligned} A(m,\mu,\beta,\Lambda) &= 16\pi^2 \cdot \int \frac{\mathrm{d}^3 p}{(2\pi)^3} \cdot \frac{1}{2E} \cdot \left( f^{FD}(E-\mu) - f^{FD}(-E-\mu) \right) \\ &= 16\pi^2 \cdot \int \frac{\mathrm{d}^3 p}{(2\pi)^3} \cdot \frac{1}{2E} \cdot \left( -1 + f^{FD}(E+\mu) + f^{FD}(E-\mu) \right) \end{aligned} . \tag{12}$$

We also establish the relation:

$$A(m,\mu,\beta,\Lambda) = A(m,-\mu,\beta,\Lambda) \equiv A(m,|\mu|,\beta,\Lambda) . \tag{13}$$

## 2.2 $B_0$ function (two-fermion loop)

$B_0$ is a complex function that only admits real arguments:

$$\begin{aligned} & B_0\left(\vec{k}, m_1, \mu_1, m_2, \mu_2, i \cdot \nu_m, \beta, \Lambda\right) \\ & = \frac{16\pi^2}{\beta} \cdot \sum_n \int \frac{\mathrm{d}^3 p}{(2\pi)^3} \frac{1}{(i\cdot\omega_n + \mu_1)^2 - E_1^2} \cdot \frac{1}{(i\cdot\omega_n - i\cdot\nu_m + \mu_2)^2 - E_2^2} \end{aligned} , \tag{14}$$

with:

$$E_1 = \sqrt{\vec{p}^2 + m_1^2} \quad \text{and} \quad E_2 = \sqrt{\left(\vec{p} - \vec{k}\right)^2 + m_2^2} . \tag{15}$$

$i \cdot \omega_n$ is a fermionic frequency, like for $A$:

$$\omega_n = \frac{(2n+1)\cdot\pi}{\beta} . \tag{16a}$$

$B_0$ expresses a loop between two quarks (fermions), so the loop itself will be bosonic. It explains why $B_0$ intervenes in the construction of the mesons. Its argument $i \cdot \nu_m$ is thus a bosonic frequency:

$$\nu_m = \frac{2m\cdot\pi}{\beta} . \tag{16b}$$

Each fraction of (14) corresponds to a quark (fermion), so the total Matsubara frequency for each fraction is then inevitably fermionic. Indeed, the fraction on the left-hand side has $i \cdot \omega_n$ as frequency, which is fermionic, and the fraction on the right-hand side has $i \cdot \omega_n - i \cdot \nu_m$ as frequency, which is also fermionic. Indeed, adding or subtracting an odd number with an even number necessarily gives an odd number.

The poles are:

$$\begin{cases} i\cdot\omega_n = E_1 - \mu_1 \\ i\cdot\omega_n = -E_1 - \mu_1 \\ i\cdot\omega_n = E_2 - \mu_2 + i\cdot\nu_m \\ i\cdot\omega_n = -E_2 - \mu_2 + i\cdot\nu_m \end{cases} . \tag{17}$$

We use (7a), and we obtain:

$$B_0\left(\vec{k},m_1,\mu_1,m_2,\mu_2,i\cdot\nu_m,\beta,\Lambda\right)$$
$$=16\pi^2\cdot\int\frac{d^3p}{(2\pi)^3}\cdot\left(\begin{array}{l}\frac{f^{FD}(E_1-\mu_1)}{2E_1}\cdot\frac{1}{(E_1-\mu_1+\mu_2-i\cdot\nu_m)^2-E_2{}^2}\\-\frac{f^{FD}(-E_1-\mu_1)}{2E_1}\cdot\frac{1}{(-E_1-\mu_1+\mu_2-i\cdot\nu_m)^2-E_2{}^2}\\+\frac{f^{FD}(E_2-\mu_2)}{2E_2}\cdot\frac{1}{(E_2+\mu_1-\mu_2+i\cdot\nu_m)^2-E_1{}^2}\\-\frac{f^{FD}(-E_2-\mu_2)}{2E_2}\cdot\frac{1}{(-E_2+\mu_1-\mu_2+i\cdot\nu_m)^2-E_1{}^2}\end{array}\right). \tag{18}$$

The equation (18) enables to verify the relation:

$$B_0\left(\vec{k},m_1,\mu_1,m_2,\mu_2,-i\cdot\nu_m,\beta,\Lambda\right)=B_0\left(\vec{k},m_2,\mu_2,m_1,\mu_1,i\cdot\nu_m,\beta,\Lambda\right). \tag{19}$$

# 2.3 $C_0$ function (three-fermion loop)

$C_0$ is a complex function with real variables:

$$C_0\left(\vec{k},\vec{q},\delta_{\vec{k},\vec{q}},m_1,\mu_1,m_2,\mu_2,i\cdot\nu_m,m_3,\mu_3,i\cdot\alpha_l,\beta,\Lambda\right)=$$
$$\frac{16\pi^2}{\beta}\cdot\sum_n\int\frac{d^3p}{(2\pi)^3}\frac{1}{(i\cdot\omega_n+\mu_1)^2-E_1^2}\cdot\frac{1}{(i\cdot\omega_n-i\cdot\nu_m+\mu_2)^2-E_2^2}\cdot\frac{1}{(i\cdot\omega_n-i\cdot\alpha_l+\mu_3)^2-E_3^2}, \tag{20}$$

with:

$$E_1=\sqrt{\vec{p}^2+m_1^2},\quad E_2=\sqrt{\left(\vec{p}-\vec{k}\right)^2+m_2^2}\quad\text{and}\quad E_3=\sqrt{(\vec{p}-\vec{q})^2+m_3^2}. \tag{21}$$

The poles are:

$$\begin{cases}i\cdot\omega_n=E_1-\mu_1\\i\cdot\omega_n=-E_1-\mu_1\\i\cdot\omega_n=E_2-\mu_2+i\cdot\nu_m\\i\cdot\omega_n=-E_2-\mu_2+i\cdot\nu_m\\i\cdot\omega_n=E_3-\mu_3+i\cdot\alpha_l\\i\cdot\omega_n=-E_3-\mu_3+i\cdot\alpha_l\end{cases}. \tag{22}$$

As previously, the fermionic nature of $i\cdot\omega_n$ brings us to use (7a):

$$C_0\left(\vec{k},\vec{q},\delta_{\vec{k},\vec{q}},m_1,\mu_1,m_2,\mu_2,i\cdot\nu_m,m_3,\mu_3,i\cdot\alpha_l,\beta,\Lambda\right)=$$
$$16\pi^2\cdot\int\frac{\mathrm{d}^3p}{(2\pi)^3}\cdot\left(\begin{array}{l}\frac{f^{FD}(E_1-\mu_1)}{2E_1}\cdot\frac{1}{\left((\lambda_1-E_1)^2-E_2{}^2\right)}\cdot\frac{1}{\left((\lambda_2-E_1)^2-E_3{}^2\right)}\\-\frac{f^{FD}(-E_1-\mu_1)}{2E_1}\cdot\frac{1}{\left((\lambda_1+E_1)^2-E_2{}^2\right)}\cdot\frac{1}{\left((\lambda_2+E_1)^2-E_3{}^2\right)}\\+\frac{f^{FD}(E_2-\mu_2)}{2E_2}\cdot\frac{1}{\left((\lambda_1+E_2)^2-E_1{}^2\right)}\cdot\frac{1}{\left((\lambda_3+E_2)^2-E_3{}^2\right)}\\-\frac{f^{FD}(-E_2-\mu_2)}{2E_2}\cdot\frac{1}{\left((\lambda_1-E_2)^2-E_1{}^2\right)}\cdot\frac{1}{\left((\lambda_3-E_2)^2-E_3{}^2\right)}\\+\frac{f^{FD}(E_3-\mu_3)}{2E_3}\cdot\frac{1}{\left((\lambda_2+E_3)^2-E_1{}^2\right)}\cdot\frac{1}{\left((\lambda_3-E_3)^2-E_2{}^2\right)}\\-\frac{f^{FD}(-E_3-\mu_3)}{2E_3}\cdot\frac{1}{\left((\lambda_2-E_3)^2-E_1{}^2\right)}\cdot\frac{1}{\left((\lambda_3+E_3)^2-E_2{}^2\right)}\end{array}\right),\tag{23}$$

where:

$$\begin{cases}\lambda_1=i\cdot\nu_m+\mu_1-\mu_2\\\lambda_2=i\cdot\alpha_l+\mu_1-\mu_3\\\lambda_3=i\cdot\nu_m-i\cdot\alpha_l-\mu_2+\mu_3\end{cases}.\tag{24}$$

# 3. Mesons loop functions

Pseudo-scalar mesons:

$$\Pi^P_{q_1,\bar{q}_2}\left(k_0,\vec{k}\right)=-\frac{N_c}{8\pi^2}\cdot\left[\begin{array}{l}A(m_1,\mu_1)+A(m_2,\mu_2)\\+\left((m_1-m_2)^2-(k_0+\mu_1-\mu_2)^2+\vec{k}^2\right)\cdot B_0\left(\vec{k},m_1,\mu_1,m_2,\mu_2,\mathcal{R}e(k_0)\right)\end{array}\right].\tag{25}$$

Scalar mesons:

$$\Pi^S_{q_1,\bar{q}_2}\left(k_0,\vec{k}\right)=-\frac{N_c}{8\pi^2}\cdot\left[\begin{array}{l}A(m_1,\mu_1)+A(m_2,\mu_2)\\+\left((m_1+m_2)^2-(k_0+\mu_1-\mu_2)^2+\vec{k}^2\right)\cdot B_0\left(\vec{k},m_1,\mu_1,m_2,\mu_2,\mathcal{R}e(k_0)\right)\end{array}\right].\tag{26}$$

Axial mesons:

$$\Pi^{A}_{q_1,\bar{q}_2}\left(k_0,\vec{k}\right) = -2\cdot\frac{N_c}{8\pi^2}\cdot\begin{bmatrix} A(m_1,\mu_1)+A(m_2,\mu_2) \\ +\left(m_1{}^2+m_2{}^2+4m_1m_2-(k_0+\mu_1-\mu_2)^2+\vec{k}^2\right)\cdot B_0\left(\vec{k},m_1,\mu_1,m_2,\mu_2,\mathcal{R}e(k_0)\right) \end{bmatrix}. \quad (27)$$

Vectorial mesons:

$$\Pi^{V}_{q_1,\bar{q}_2}\left(k_0,\vec{k}\right) = -2\cdot\frac{N_c}{8\pi^2}\cdot\begin{bmatrix} A(m_1,\mu_1)+A(m_2,\mu_2) \\ +\left(m_1{}^2+m_2{}^2-4m_1m_2-(k_0+\mu_1-\mu_2)^2+\vec{k}^2\right)\cdot B_0\left(\vec{k},m_1,\mu_1,m_2,\mu_2,\mathcal{R}e(k_0)\right) \end{bmatrix}\cdot \quad (28)$$

About the loop functions mentioned here, the arguments of the functions $A$ and $B_0$ are necessarily real numbers. When $k_0$ is a complex number, an approximation is applied, in which we only take the real part of $k_0$, noted $\mathcal{R}e(k_0)$, as an argument of $B_0$ [3]. However, the $k_0$ placed in the factor in front of $B_0$ stays complex. In fact, this $k_0$ enables us to work in the zone where the mesons are unstable. This remark is also valid for the diquarks.

# 4. Diquarks loop functions

## 4.1 Expression of the diquarks loop functions

Scalar diquarks:

$$\Pi^{S}_{q_1,q_2}\left(k_0,\vec{k}\right) = -\frac{1}{\pi^2}\cdot\begin{bmatrix} A(m_1,\mu_1)+A(m_2,-\mu_2) \\ +\left((m_1-m_2)^2-(k_0+\mu_1+\mu_2)^2+\vec{k}^2\right)\cdot B_0\left(\vec{k},m_1,\mu_1,m_2,-\mu_2,\mathcal{R}e(k_0)\right) \end{bmatrix}\cdot \quad (29)$$

Pseudo-scalar diquarks:

$$\Pi^{P}_{q_1,q_2}\left(k_0,\vec{k}\right) = -\frac{1}{\pi^2}\cdot\begin{bmatrix} A(m_1,\mu_1)+A(m_2,-\mu_2) \\ +\left((m_1+m_2)^2-(k_0+\mu_1+\mu_2)^2+\vec{k}^2\right)\cdot B_0\left(\vec{k},m_1,\mu_1,m_2,-\mu_2,\mathcal{R}e(k_0)\right) \end{bmatrix}\cdot \quad (30)$$

Vectorial diquarks:

$$\Pi^{V}_{q_1,q_2}\left(k_0,\vec{k}\right) = -\frac{2}{\pi^2}\cdot\begin{bmatrix} A(m_1,\mu_1)+A(m_2,-\mu_2) \\ +\left(m_1{}^2+m_2{}^2+4m_1m_2-(k_0+\mu_1+\mu_2)^2+\vec{k}^2\right)\cdot B_0\left(\vec{k},m_1,\mu_1,m_2,-\mu_2,\mathcal{R}e(k_0)\right) \end{bmatrix}\cdot \quad (31)$$

Axial diquarks:

$$\Pi^{A}_{q_1,q_2}\left(k_0,\vec{k}\right) = -\frac{2}{\pi^2}\cdot\left[\begin{array}{l} A(m_1,\mu_1)+A(m_2,-\mu_2) \\ +\left(m_1^{\,2}+m_2^{\,2}-4m_1m_2-(k_0+\mu_1+\mu_2)^2+\vec{k}^2\right)\cdot B_0\left(\vec{k},m_1,\mu_1,m_2,-\mu_2,\mathcal{R}e(k_0)\right)\end{array}\right]. \tag{32}$$

# 4.2 Exchange of the two quarks

About the mesons, $\Pi_{q_1,\bar{q}_2}\neq\Pi_{q_2,\bar{q}_1}$ in the general case, except if $q_1\equiv q_2$ (coupled mesons $\pi_0,\eta,\eta'$). About the diquarks, we expect to have $\Pi_{q_1,q_2}=\Pi_{q_2,q_1}$. In other words, the exchange of the two quarks must leave invariant the loop functions. In fact, thanks to the writing of the diquarks polarization functions (29–32), we note that the exchange of the quarks $q_1$ and $q_2$ only intervenes by the exchange of the masses $m_1$, $m_2$ and by the chemical potentials $\mu_1$, $\mu_2$. The factor placed in front of $B_0$ is invariant by this exchange, whatever the diquarks loop function. About the $A$ functions, with (13), we write:

$$A(m_1,\mu_1)+A(m_2,-\mu_2)=A(m_2,\mu_2)+A(m_1,-\mu_1). \tag{33}$$

Thus, we only need to show that the exchange of the two quarks leaves invariant $B_0$ itself. Firstly, we write:

$$B_0\left(\vec{k},m_1,\mu_1,m_2,-\mu_2,i\cdot\nu_m,\beta,\Lambda\right) = 16\pi^2\cdot\int\frac{d^3p}{(2\pi)^3}\cdot\left(\begin{array}{l}\frac{f^{FD}(E_1-\mu_1)}{2E_1}\cdot\frac{1}{(E_1-\lambda)^2-E_2^{\,2}}-\frac{f^{FD}(-E_1-\mu_1)}{2E_1}\cdot\frac{1}{(-E_1-\lambda)^2-E_2^{\,2}} \\ +\frac{f^{FD}(E_2+\mu_2)}{2E_2}\cdot\frac{1}{(E_2+\lambda)^2-E_1^{\,2}}-\frac{f^{FD}(-E_2+\mu_2)}{2E_2}\cdot\frac{1}{(-E_2+\lambda)^2-E_1^{\,2}}\end{array}\right), \tag{34}$$

with here:

$$\lambda=i\cdot\nu_m+\mu_1+\mu_2. \tag{35}$$

In parallel, we have:

$$B_0\left(\vec{k},m_2,\mu_2,m_1,-\mu_1,i\cdot\nu_m,\beta,\Lambda\right) = 16\pi^2\cdot\int\frac{d^3p}{(2\pi)^3}\cdot\left(\begin{array}{l}\frac{f^{FD}(E_2-\mu_2)}{2E_2}\cdot\frac{1}{(E_2-\lambda)^2-E_1^{\,2}}-\frac{f^{FD}(-E_2-\mu_2)}{2E_2}\cdot\frac{1}{(-E_2-\lambda)^2-E_1^{\,2}} \\ +\frac{f^{FD}(E_1+\mu_1)}{2E_1}\cdot\frac{1}{(E_1+\lambda)^2-E_2^{\,2}}-\frac{f^{FD}(-E_1+\mu_1)}{2E_1}\cdot\frac{1}{(-E_1+\lambda)^2-E_2^{\,2}}\end{array}\right), \tag{36}$$

which gives:

$$B_0\left(\vec{k},m_2,\mu_2,m_1,-\mu_1,i\cdot\nu_m,\beta,\Lambda\right) \tag{37}$$

$$=16\pi^2\cdot\int\frac{\mathrm{d}^3p}{(2\pi)^3}\cdot\left(\begin{array}{l}\frac{1-f^{FD}(-E_2+\mu_2)}{2E_2}\cdot\frac{1}{(E_2-\lambda)^2-E_1^{\ 2}}-\frac{1-f^{FD}(E_2+\mu_2)}{2E_2}\cdot\frac{1}{(-E_2-\lambda)^2-E_1^{\ 2}}\\+\frac{1-f^{FD}(-E_1-\mu_1)}{2E_1}\cdot\frac{1}{(E_1+\lambda)^2-E_2^{\ 2}}-\frac{1-f^{FD}(E_1-\mu_1)}{2E_1}\cdot\frac{1}{(-E_1+\lambda)^2-E_2^{\ 2}}\end{array}\right).$$

In (37), we identify $B_0\left(\vec{k},m_1,\mu_1,m_2,-\mu_2,i\cdot\nu_m,\beta,\Lambda\right)$ written in (34):

$$B_0\left(\vec{k},m_2,\mu_2,m_1,-\mu_1,i\cdot\nu_m,\beta,\Lambda\right)=B_0\left(\vec{k},m_1,\mu_1,m_2,-\mu_2,i\cdot\nu_m,\beta,\Lambda\right)$$
$$+16\pi^2\cdot\int\frac{\mathrm{d}^3p}{(2\pi)^3}\cdot\left(\begin{array}{l}\frac{1}{2E_2}\cdot\frac{1}{(-E_2+\lambda)^2-E_1^{\ 2}}-\frac{1}{2E_2}\cdot\frac{1}{(E_2+\lambda)^2-E_1^{\ 2}}\\+\frac{1}{2E_1}\cdot\frac{1}{(-E_1-\lambda)^2-E_2^{\ 2}}-\frac{1}{2E_1}\cdot\frac{1}{(E_1-\lambda)^2-E_2^{\ 2}}\end{array}\right), \tag{38}$$

and:

$$B_0\left(\vec{k},m_2,\mu_2,m_1,-\mu_1,i\cdot\nu_m,\beta,\Lambda\right)=B_0\left(\vec{k},m_1,\mu_1,m_2,-\mu_2,i\cdot\nu_m,\beta,\Lambda\right)$$
$$+16\pi^2\cdot\int\frac{\mathrm{d}^3p}{(2\pi)^3}\cdot\left(\begin{array}{l}\frac{2\lambda}{\left((-E_2+\lambda)^2-E_1^{\ 2}\right)\cdot\left((E_2+\lambda)^2-E_1^{\ 2}\right)}\\+\frac{-2\lambda}{\left((-E_1-\lambda)^2-E_2^{\ 2}\right)\cdot\left((E_1-\lambda)^2-E_2^{\ 2}\right)}\end{array}\right). \tag{39}$$

It comes:

$$B_0\left(\vec{k},m_2,\mu_2,m_1,-\mu_1,i\cdot\nu_m,\beta,\Lambda\right)=$$
$$B_0\left(\vec{k},m_1,\mu_1,m_2,-\mu_2,i\cdot\nu_m,\beta,\Lambda\right) \tag{40}$$
$$+16\pi^2\cdot\int\frac{\mathrm{d}^3p}{(2\pi)^3}\cdot\left(2\lambda\cdot\frac{\overbrace{\left((-E_1-\lambda)^2-E_2^{\ 2}\right)\cdot\left((E_1-\lambda)^2-E_2^{\ 2}\right)-\left((-E_2+\lambda)^2-E_1^{\ 2}\right)\cdot\left((E_2+\lambda)^2-E_1^{\ 2}\right)}^{0}}{\left((-E_2+\lambda)^2-E_1^{\ 2}\right)\cdot\left((E_2+\lambda)^2-E_1^{\ 2}\right)\cdot\left((-E_1-\lambda)^2-E_2^{\ 2}\right)\cdot\left((E_1-\lambda)^2-E_2^{\ 2}\right)}\right)$$

Thus, we obtain:

$$B_0\left(\vec{k},m_2,\mu_2,m_1,-\mu_1,i\cdot\nu_m,\beta,\Lambda\right)=B_0\left(\vec{k},m_1,\mu_1,m_2,-\mu_2,i\cdot\nu_m,\beta,\Lambda\right). \tag{41}$$

This enables to conclude that:

$$\Pi_{q_1,q_2}=\Pi_{q_2,q_1}. \tag{42}$$

# 5. Baryons loop functions

To build a baryon, a quark-diquark loop is considered, where the charge conjugation is applied to the diquark, or to the quark. It corresponds to two distinct writings of the baryon loop function. The nature of the concerned Matsubara frequencies that appear in these writings are explained in the chapter 5 devoted to baryons.

## 5.1 First term $\Pi^{(1)}$

If the charge conjugation is applied to the diquark, the first term of the baryons loop function is written as:

$$-i\cdot\Pi^{(1)}\left(i\cdot\nu_m,\vec{k}\right)=-\frac{i}{\beta}\cdot\sum_n\int\frac{\mathrm{d}^3p}{(2\pi)^3}Tr\left(i\cdot S_q\left(i\cdot\omega_n,\vec{p}\right)\cdot i\cdot S_D{}^C\left(i\cdot\omega_n-i\cdot\nu_m,\vec{p}-\vec{k}\right)\right), \tag{43}$$

where:

$$S_q\left(i\cdot\omega_n,\vec{p}\right)\equiv S_q\left(\not{p}\right)=\frac{1}{\not{p}+\gamma_0\cdot\mu_q-m_q}, \tag{44}$$

$$S_D{}^C\left(i\cdot\omega_n-i\cdot\nu_m,\vec{p}-\vec{k}\right)\equiv S_D{}^C\left(\not{k}\right)=\frac{1}{\left(k_0-\mu_D\right)^2-\vec{k}_D{}^2-m_D{}^2}, \tag{45}$$

respectively designate the propagators of the $q$ flavor quark and the charge conjugation diquark (=anti-diquark). The equation (44) must be updated in the framework of the PNJL model, as explained in the chapter 4, by the replacement $\mu_q\to\mu_q-iA_4$. It leads to an adaptation of the Fermi-Dirac distributions [2], e.g. in (55), but our results are still applicable.

The trace of (43) is then expressed as:

$$Tr\left(i\cdot S_q\left(\not{p}\right)\cdot i\cdot S_D{}^c\left(\not{k}\right)\right)=\frac{\overbrace{Tr\left(\not{p}+m_q\right)}^{4m_q}}{\left(\left(i\cdot\omega_n+\mu_q\right)^2-E_q{}^2\right)\cdot\left(\left(i\cdot\omega_n-i\cdot\nu_m-\mu_D\right)^2-E_D{}^2\right)}, \tag{46}$$

with:

$$E_q=\sqrt{\vec{p}^2+m_q{}^2}\quad\text{and}\quad E_D=\sqrt{\left(\vec{p}-\vec{k}\right)^2+m_D{}^2}. \tag{47}$$

(43) is re-written as:

$$-i\cdot\Pi^{(1)}\left(i\cdot\nu_m,\vec{k}\right)=\frac{i}{\beta}\cdot\sum_n\int\frac{\mathrm{d}^3p}{(2\pi)^3}\cdot 4m_q\cdot\frac{1}{\left(\left(i\cdot\omega_n+\mu_q\right)^2-E_q{}^2\right)\cdot\left(\left(i\cdot\omega_n-i\cdot\nu_m-\mu_D\right)^2-E_D{}^2\right)}. \tag{48}$$

Then, it comes :

$$-i\cdot\Pi^{(1)}\left(i\cdot\nu_m,\vec{k}\right)$$

$$=\frac{i}{\beta}\cdot\sum_n\int\frac{\mathrm{d}^3p}{(2\pi)^3}\cdot 4m_q\cdot\left(\begin{array}{c}\frac{1}{\left(i\cdot\omega_n+\mu_q-E_q\right)\cdot\left(i\cdot\omega_n+\mu_q+E_q\right)}\\ \cdot\frac{1}{\left(i\cdot\omega_n-i\cdot\nu_m-\mu_D-E_D\right)\cdot\left(i\cdot\omega_n-i\cdot\nu_m-\mu_D+E_D\right)}\end{array}\right). \tag{49}$$

The poles are:

$$\begin{cases} i\cdot\omega_n=-\mu_q+E_q\\ i\cdot\omega_n=-\mu_q-E_q\\ i\cdot\omega_n=\mu_D+E_D+i\cdot\nu_m\\ i\cdot\omega_n=\mu_D-E_D+i\cdot\nu_m\end{cases}. \tag{50}$$

$i\cdot\omega_n$ is a fermionic frequency, so we use (7a) and we obtain:

$$-i\cdot\Pi^{(1)}\left(i\cdot\nu_m,\vec{k}\right) \tag{51}$$

$$=i\cdot\int\frac{\mathrm{d}^3p}{(2\pi)^3}\cdot 4m_q\cdot\left(\begin{array}{c}\frac{f^{FD}\left(-\mu_q+E_q\right)}{2E_q\cdot\left(\lambda-E_q+E_D\right)\cdot\left(\lambda-E_q-E_D\right)}-\frac{f^{FD}\left(-\mu_q-E_q\right)}{2E_q\cdot\left(\lambda+E_q+E_D\right)\cdot\left(\lambda+E_q-E_D\right)}\\ +\frac{f^{FD}\left(\mu_D+E_D+i\cdot\nu_m\right)}{2E_D\cdot\left(\lambda-E_q+E_D\right)\cdot\left(\lambda+E_q+E_D\right)}-\frac{f^{FD}\left(\mu_D-E_D+i\cdot\nu_m\right)}{2E_D\cdot\left(\lambda-E_q-E_D\right)\cdot\left(\lambda+E_q-E_D\right)}\end{array}\right),$$

where:

$$\lambda=i\cdot\nu_m+\mu_q+\mu_D. \tag{52}$$

$i\cdot\nu_m$ is a fermionic frequency: $$\nu_m=\frac{(2m+1)\cdot\pi}{\beta}, \tag{53}$$

therefore: $$f^{FD}\left(z+i\cdot\nu_m\right)=\frac{1}{1+\underbrace{e^{i\cdot\nu_m\cdot\beta}}_{-1}\cdot e^{\beta\cdot z}}=-f^{BE}(z). \tag{54}$$

Finally, we have:

$$-i\cdot\Pi^{(1)}\left(i\cdot\nu_m,\vec{k}\right) \tag{55}$$

$$=4m_q\cdot i\cdot\int\frac{\mathrm{d}^3p}{(2\pi)^3}\cdot\left(\begin{array}{c}\frac{f^{FD}\left(-\mu_q+E_q\right)}{2E_q\cdot\left(\lambda-E_q+E_D\right)\cdot\left(\lambda-E_q-E_D\right)}-\frac{f^{FD}\left(-\mu_q-E_q\right)}{2E_q\cdot\left(\lambda+E_q+E_D\right)\cdot\left(\lambda+E_q-E_D\right)}\\ +\frac{-f^{BE}\left(\mu_D+E_D\right)}{2E_D\cdot\left(\lambda-E_q+E_D\right)\cdot\left(\lambda+E_q+E_D\right)}-\frac{-f^{BE}\left(\mu_D-E_D\right)}{2E_D\cdot\left(\lambda-E_q-E_D\right)\cdot\left(\lambda+E_q-E_D\right)}\end{array}\right).$$

## 5.2 Second term $\Pi^{(2)}$

Now, if the charge conjugation is applied to the quark, the second term is written as:

$$-i\cdot\Pi^{(2)}\left(i\cdot\nu_m,\vec{k}\right)=-\frac{i}{\beta}\cdot\sum_n\int\frac{\mathrm{d}^3p}{(2\pi)^3}Tr\left(i\cdot S_D\left(i\cdot\omega_n,\vec{p}\right)\cdot i\cdot S_q^{\,C}\left(i\cdot\omega_n-i\cdot\nu_m,\vec{p}-\vec{k}\right)\right),\tag{56}$$

where:

$$S_D\left(i\cdot\omega_n,\vec{p}\right)\equiv S_D\left(\not{p}\right)=\frac{1}{\left(p_0+\mu_D\right)^2-\vec{p}_D^{\,2}-m_D^{\,2}},\tag{57}$$

$$S_q^{\,C}\left(i\cdot\omega_n-i\cdot\nu_m,\vec{p}-\vec{k}\right)\equiv S_q^{\,C}\left(\not{k}\right)=\frac{1}{\not{k}-\gamma_0\cdot\mu_q-m_q},\tag{58}$$

respectively designate the propagator of the diquark and of the charge conjugation quark (=anti-quark). As with the equation (44), (58) is modified in the PNJL model. Then, (56) is written as:

$$-i\cdot\Pi^{(2)}\left(i\cdot\nu_m,\vec{k}\right)=\frac{i}{\beta}\cdot\sum_n\int\frac{\mathrm{d}^3p}{(2\pi)^3}\cdot 4m_q\cdot\left(\begin{array}{c}\dfrac{1}{\left(i\cdot\omega_n-i\cdot\nu_m-\mu_q-E_q\right)\cdot\left(i\cdot\omega_n-i\cdot\nu_m-\mu_q+E_q\right)}\\ \cdot\dfrac{1}{\left(i\cdot\omega_n+\mu_D-E_D\right)\cdot\left(i\cdot\omega_n+\mu_D+E_D\right)}\end{array}\right).\tag{59}$$

The poles are:

$$\begin{cases}i\cdot\omega_n=-\mu_D+E_D\\ i\cdot\omega_n=-\mu_D-E_D\\ i\cdot\omega_n=\mu_q+E_q+i\cdot\nu_m\\ i\cdot\omega_n=\mu_q-E_q+i\cdot\nu_m\end{cases}.\tag{60}$$

$i\cdot\omega_n$ is now bosonic, so we use (7b), and:

$$\begin{aligned}&-i\cdot\Pi^{(2)}\left(i\cdot\nu_m,\vec{k}\right)\\ &=i\cdot\int\frac{\mathrm{d}^3p}{(2\pi)^3}\cdot 4m_q\cdot\left(\begin{array}{l}\dfrac{-f^{BE}\left(-\mu_D+E_D\right)}{2E_D\cdot\left(\lambda-E_D+E_q\right)\cdot\left(\lambda-E_D-E_q\right)}-\dfrac{-f^{BE}\left(-\mu_D-E_D\right)}{2E_D\cdot\left(\lambda+E_D+E_q\right)\cdot\left(\lambda+E_D-E_q\right)}\\ +\dfrac{-f^{BE}\left(\mu_q+E_q+i\cdot\nu_m\right)}{2E_q\cdot\left(\lambda-E_D+E_q\right)\cdot\left(\lambda+E_D+E_q\right)}-\dfrac{-f^{BE}\left(\mu_q-E_q+i\cdot\nu_m\right)}{2E_q\cdot\left(\lambda-E_D-E_q\right)\cdot\left(\lambda+E_D-E_q\right)}\end{array}\right).\end{aligned}\tag{61}$$

$i\cdot\nu_m$ is a fermionic frequency, thus:

$$f^{BE}\left(z+i\cdot\nu_m\right)=\frac{1}{\underbrace{e^{i\cdot\nu_m\cdot\beta}}_{-1}\cdot e^{\beta\cdot z}-1}=-f^{FD}\left(z\right).\tag{62}$$

We have:

$$
\begin{aligned}
&-i\cdot\Pi^{(2)}\left(i\cdot\nu_m,\vec{k}\right) \\
&= 4m_q\cdot i\cdot\int\frac{\mathrm{d}^3p}{(2\pi)^3}\cdot\left(\begin{array}{l}\dfrac{-f^{BE}(-\mu_D+E_D)}{2E_D\cdot\left(\lambda-E_D+E_q\right)\cdot\left(\lambda-E_D-E_q\right)}-\dfrac{-f^{BE}(-\mu_D-E_D)}{2E_D\cdot\left(\lambda+E_D+E_q\right)\cdot\left(\lambda+E_D-E_q\right)}\\+\dfrac{f^{FD}\left(\mu_q+E_q\right)}{2E_q\cdot\left(\lambda-E_D+E_q\right)\cdot\left(\lambda+E_D+E_q\right)}-\dfrac{f^{FD}\left(\mu_q-E_q\right)}{2E_q\cdot\left(\lambda-E_D-E_q\right)\cdot\left(\lambda+E_D-E_q\right)}\end{array}\right).
\end{aligned}
\tag{63}
$$

Thanks to the relations,

$$f^{FD}(-x)=1-f^{FD}(x)\quad\text{and}\quad -f^{BE}(-x)=1+f^{BE}(x),\tag{64}$$

the equation (63) is re-written as:

$$
\begin{aligned}
&-i\cdot\Pi^{(2)}\left(i\cdot\nu_m,\vec{k}\right) \\
&= 4m_q\cdot i\cdot\int\frac{\mathrm{d}^3p}{(2\pi)^3}\cdot\left(\begin{array}{l}\dfrac{f^{BE}(\mu_D-E_D)+1}{2E_D\cdot\left(\lambda-E_D+E_q\right)\cdot\left(\lambda-E_D-E_q\right)}-\dfrac{f^{BE}(\mu_D+E_D)+1}{2E_D\cdot\left(\lambda+E_D+E_q\right)\cdot\left(\lambda+E_D-E_q\right)}\\+\dfrac{1-f^{FD}\left(-\mu_q-E_q\right)}{2E_q\cdot\left(\lambda-E_D+E_q\right)\cdot\left(\lambda+E_D+E_q\right)}-\dfrac{1-f^{FD}\left(-\mu_q+E_q\right)}{2E_q\cdot\left(\lambda-E_D-E_q\right)\cdot\left(\lambda+E_D-E_q\right)}\end{array}\right).
\end{aligned}
\tag{65}
$$

Thanks to (55), we note that a part of (65) corresponds to $-i\cdot\Pi^{(1)}\left(i\cdot\nu_m,\vec{k}\right)$:

$$
\begin{aligned}
&-i\cdot\Pi^{(2)}\left(i\cdot\nu_m,\vec{k}\right)=-i\cdot\Pi^{(1)}\left(i\cdot\nu_m,\vec{k}\right) \\
&+4m_q\cdot i\cdot\int\frac{\mathrm{d}^3p}{(2\pi)^3}\cdot\left(\begin{array}{l}\dfrac{1}{2E_D\cdot\left(\lambda-E_D+E_q\right)\cdot\left(\lambda-E_D-E_q\right)}-\dfrac{1}{2E_D\cdot\left(\lambda+E_D+E_q\right)\cdot\left(\lambda+E_D-E_q\right)}\\+\dfrac{1}{2E_q\cdot\left(\lambda-E_D+E_q\right)\cdot\left(\lambda+E_D+E_q\right)}-\dfrac{1}{2E_q\cdot\left(\lambda-E_D-E_q\right)\cdot\left(\lambda+E_D-E_q\right)}\end{array}\right),
\end{aligned}
\tag{66}
$$

$$
\begin{aligned}
&-i\cdot\Pi^{(2)}\left(i\cdot\nu_m,\vec{k}\right)=-i\cdot\Pi^{(1)}\left(i\cdot\nu_m,\vec{k}\right) \\
&+4m_q\cdot i\cdot\int\frac{\mathrm{d}^3p}{(2\pi)^3}\cdot\underbrace{\left(\begin{array}{l}\dfrac{\lambda}{\left(\lambda-E_D+E_q\right)\cdot\left(\lambda-E_D-E_q\right)\cdot\left(\lambda+E_D+E_q\right)\cdot\left(\lambda+E_D-E_q\right)}\\-\dfrac{\lambda}{\left(\lambda-E_D+E_q\right)\cdot\left(\lambda+E_D+E_q\right)\cdot\left(\lambda-E_D-E_q\right)\cdot\left(\lambda+E_D-E_q\right)}\end{array}\right)}_{0}.
\end{aligned}
\tag{67}
$$

Thus, we obtain:

$$-i\cdot\Pi^{(2)}\left(i\cdot\nu_m,\vec{k}\right)=-i\cdot\Pi^{(1)}\left(i\cdot\nu_m,\vec{k}\right).\tag{68}$$

## 5.3 Complete expression of the baryon loop function

In order to take into account the two components, we formally write:

$$-i\cdot\Pi\left(i\cdot\nu_m,\vec{k}\right)=\frac{1}{2}\cdot\left(-i\cdot\Pi^{(1)}\left(i\cdot\nu_m,\vec{k}\right)\right)+\frac{1}{2}\cdot\left(-i\cdot\Pi^{(2)}\left(i\cdot\nu_m,\vec{k}\right)\right)$$

$$=4m_q\cdot i\cdot\int\frac{\mathrm{d}^3p}{(2\pi)^3}\cdot\left(\begin{array}{l}\dfrac{1/2\cdot\left(f^{FD}\left(-\mu_q+E_q\right)-f^{FD}\left(\mu_q-E_q\right)\right)}{2E_q\cdot\left(\lambda-E_q+E_D\right)\cdot\left(\lambda-E_q-E_D\right)}\\ -\dfrac{1/2\cdot\left(f^{FD}\left(-\mu_q-E_q\right)-f^{FD}\left(\mu_q+E_q\right)\right)}{2E_q\cdot\left(\lambda+E_q+E_D\right)\cdot\left(\lambda+E_q-E_D\right)}\\ -\dfrac{1/2\cdot\left(f^{BE}\left(\mu_D+E_D\right)-f^{BE}\left(-\mu_D-E_D\right)\right)}{2E_D\cdot\left(\lambda-E_q+E_D\right)\cdot\left(\lambda+E_q+E_D\right)}\\ +\dfrac{1/2\cdot\left(f^{BE}\left(\mu_D-E_D\right)-f^{BE}\left(-\mu_D+E_D\right)\right)}{2E_D\cdot\left(\lambda-E_q-E_D\right)\cdot\left(\lambda+E_q-E_D\right)}\end{array}\right)\cdot \tag{69}$$

If we take into account (68) and we clarify all the arguments of $\Pi$, it can also be written:

$$-i\cdot\Pi\left(\vec{k},m_q,\mu_q,m_D,\mu_D,i\cdot\nu_m,\beta,\Lambda\right) \tag{70}$$

$$=4m_q\cdot i\cdot\int\frac{\mathrm{d}^3p}{(2\pi)^3}\cdot\left(\begin{array}{l}\dfrac{f^{FD}\left(-\mu_q+E_q\right)}{2E_q\cdot\left(\lambda-E_q+E_D\right)\cdot\left(\lambda-E_q-E_D\right)}\\ -\dfrac{f^{FD}\left(-\mu_q-E_q\right)}{2E_q\cdot\left(\lambda+E_q+E_D\right)\cdot\left(\lambda+E_q-E_D\right)}\\ -\dfrac{f^{BE}\left(\mu_D+E_D\right)}{2E_D\cdot\left(\lambda-E_q+E_D\right)\cdot\left(\lambda+E_q+E_D\right)}\\ +\dfrac{f^{BE}\left(\mu_D-E_D\right)}{2E_D\cdot\left(\lambda-E_q-E_D\right)\cdot\left(\lambda+E_q-E_D\right)}\end{array}\right)\cdot$$

The equation (69) is numerically more stable than (70). On the other hand, with (70), we notice that the baryons loop function is a "modified version" of the $B_0$ function evoked previously. This one is a complex function with real variables. This implies that $i\cdot\nu_m$, i.e. $k_0$, must be necessarily real. It prevents to work in the zone where the baryons are unstable. Anyway, we use the diquark mass (contained in $E_D$), thus we are limited to the zone where the diquark is stable. Indeed, when the diquark is unstable, $m_D$ becomes complex, which makes non-trivial the use of (69) or (70).

# 6. « Triangle » functions

The three-quark "triangle" function, figure 1, noted $\Gamma$ in our work and in the literature [4], intervenes in the calculation of cross sections. It is defined by:

$$\Gamma\left(i\cdot\nu_m,\vec{k}\ ;\ i\cdot\alpha_l,\vec{p}\right) \tag{71}$$
$$=-N_c\cdot\frac{i}{\beta}\cdot\sum_n\int\frac{\mathrm{d}^3q}{(2\pi)^3}\cdot Tr\left[iS^{f_1}\left(i\cdot\omega_n,\vec{q}\right)\cdot i\gamma_5\cdot iS^{f_2}\left(i\cdot\omega_n-i\cdot\alpha_l,\vec{q}-\vec{p}\right)\cdot i\gamma_5\cdot iS^{f_3}\left(i\cdot\omega_n-i\cdot\nu_m,\vec{q}-\vec{k}\right)\right].$$

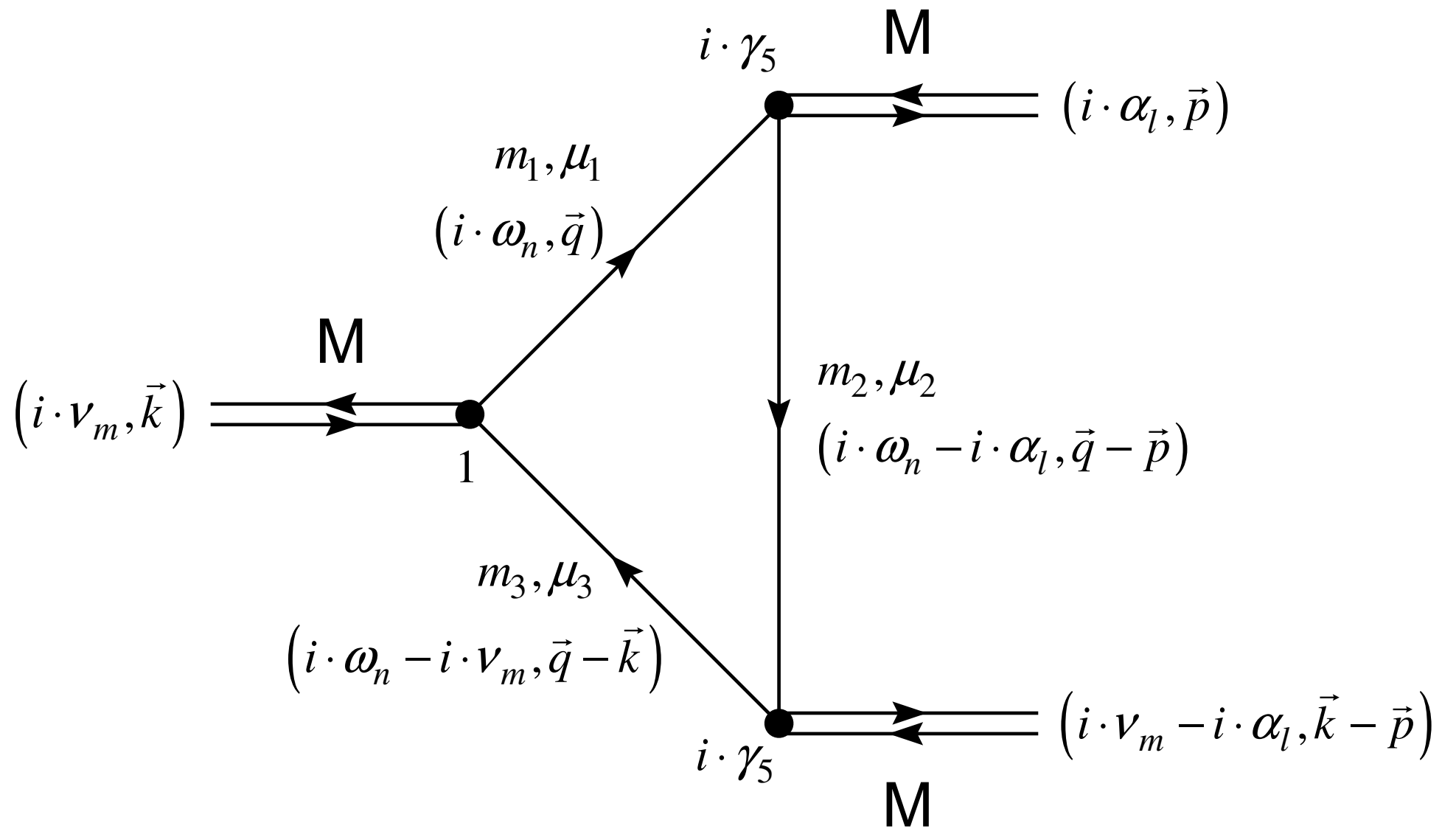

**Figure 1.** Schematization of the $\Gamma$ triangle function.

We check that $i\cdot\nu_m$, $i\cdot\alpha_l$ and $i\cdot\nu_m-i\cdot\alpha_l$ are bosonic frequencies. These three energies are those of the three mesons of figure 1. Also, $i\cdot\omega_n$, $i\cdot\omega_n-i\cdot\alpha_l$ et $i\cdot\omega_n-i\cdot\nu_m$ are fermionic: they are the energies of the three quarks in the triangle. The trace present in the equation (71) is a non-null term only if the number of $\gamma_5$ matrices is even. It imposes that if we want two pseudo-scalar mesons as outgoing particles, the incoming meson must inevitably be scalar (vertex 1), or maybe vectorial.

We obtain, as in [4]:

$$\Gamma\left(i\cdot\nu_m,\vec{k}\ ;\ i\cdot\alpha_l,\vec{p}\right) \tag{72}$$

$$= -\frac{N_c}{8\pi^2}\left[\begin{array}{l} \left(m_3-m_2\right)B_0\left(\vec{k}-\vec{p},m_2,\mu_2,m_3,\mu_3,i\cdot\nu_m-i\cdot\alpha_l\right) \\ +\left(m_1-m_2\right)B_0\left(\vec{p},m_1,\mu_1,m_2,\mu_2,i\cdot\alpha_l\right) \\ +\left(m_1+m_3\right)B_0\left(\vec{k},m_1,\mu_1,m_3,\mu_3,i\cdot\nu_m\right) \\ +\left(\begin{array}{l} m_1^{\ 2}\cdot\left(m_3-m_2\right) \\ +m_2^{\ 2}\cdot\left(m_1+m_3\right) \\ +m_3^{\ 2}\cdot\left(m_1-m_2\right)-2\cdot m_1\cdot m_2\cdot m_3 \\ m_3\left(\vec{p}^2-\left(i\cdot\alpha_l-\mu_2+\mu_1\right)^2\right) \\ -m_2\left(\vec{k}^2-\left(i\cdot\nu_m-\mu_3+\mu_1\right)^2\right) \\ +m_1\left(\left(\vec{p}-\vec{k}\right)^2-\left(i\cdot\alpha_l-i\cdot\nu_m-\mu_2+\mu_3\right)^2\right) \end{array}\right) C_0\left(\vec{p},\vec{k},m_1,\mu_1,m_2,\mu_2,i\cdot\alpha_l,m_3,\mu_3,i\cdot\nu_m\right) \end{array}\right].$$

# Appendix E

# Calculations with non-null chemical potentials

## 1. General techniques

When we work with mesons, diquarks or with baryons, we have similarities, which we are going to describe now. Firstly, we consider $i \cdot \nu_m$, which is the loop function argument corresponding to the "gross" energy of the particle loop. At non-null chemical potentials, $i \cdot \nu_m$ is modified by the addition or the subtraction of the particles chemical potentials that compose the loop.

The general rule for a composite particle/anti-particle built with two particles-antiparticles $p_1$ and $p_2$ is:

- we subtract $\mu_{p_1}$ if $p_1$ is a particle or a charge conjugate antiparticle, i.e. which moves right on a Feynman diagram.
- we add $\mu_{p_1}$ if $p_1$ is an anti-particle or a charge conjugate particle, which moves left on a Feynman diagram.

The technique is identical for $p_2$ by replacement of the notations. The "new" argument $i \cdot \nu_m$ is thus:

$$k_0 = i \cdot \nu_m \pm \mu_{p_1} \pm \mu_{p_2} . \tag{1}$$

We can propose a physical interpretation of this relation (1). Indeed, we recall that the definition of the particle's chemical potential corresponds to the energy that is necessary to add this particle in the system. We thus understand the sign difference between the particles and the anti-particles. Consequently, in (1), the composite particle energy $i \cdot \nu_m$ is modified by the quantity $\pm\mu_{p_1} \pm \mu_{p_2}$ to create the particles/anti-particles $p_1$ and $p_2$ starting from the vacuum; then they are linked together to finally form the wanted composite particle/anti-particle, with the energy $k_0$.

Then, if we want to work in a medium dominated by the antimatter, the chemical potentials become negative. For example in such cases, $k_0$ can become negative. This is incompatible with the previously studied functions, and the variable $k_0$ is expected to give the particle's mass... As a solution, we use the relation (19) from the previous appendix:

$$B_0\left(\vec{k}, m_1, \mu_1, m_2, \mu_2, -k_0, \beta, \Lambda\right) = B_0\left(\vec{k}, m_2, \mu_2, m_1, \mu_1, k_0, \beta, \Lambda\right) . \tag{2}$$

Therefore, we consider:

$$\begin{cases} B_0\left(\vec{k}, m_1, \mu_1, m_2, \mu_2, k_0, \beta, \Lambda\right) & \text{if } k_0 > 0 \\ B_0\left(\vec{k}, m_2, \mu_2, m_1, \mu_1, k_0' = -k_0, \beta, \Lambda\right) & \text{if } k_0 < 0 \end{cases}. \quad (3)$$

For each particle type, we will see hereafter how this relation modifies the loop functions and its physical interpretation.

# 2. Mesons

The study of the mesons is the simplest one, because generally the mesons loop functions are not invariant by two-flavor exchange between the quark and the anti-quark. Physically, in the general case, $q_1\overline{q}_2 \neq q_2\overline{q}_1$, except for some mesons ($\pi_0$ ...). The two possible cases are represented in table 1.

| $i\cdot\nu_m > 0$ | $i\cdot\nu_m < 0$ |
|---|---|
| $i\cdot\omega_n, \vec{p}$<br>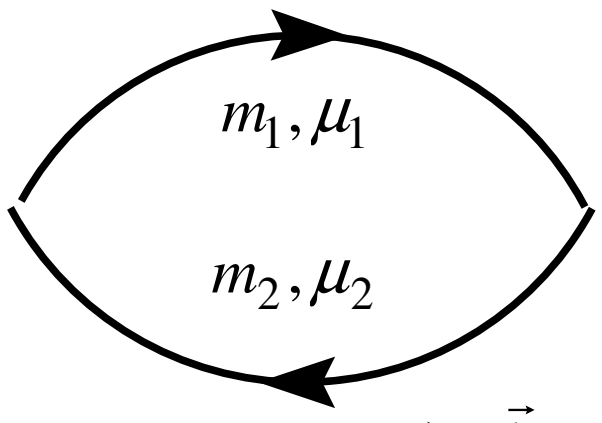<br><br>$i\cdot\omega_n - i\cdot\nu_m, \vec{p}-\vec{k}$<br>$= i\cdot\omega_n - \lvert i\cdot\nu_m\rvert, \vec{p}-\vec{k}$<br><br>$k_0 = \lvert i\cdot\nu_m\rvert - \mu_1 + \mu_2$<br>$q_1\overline{q}_2$ | $i\cdot\omega_n, \vec{p}$<br>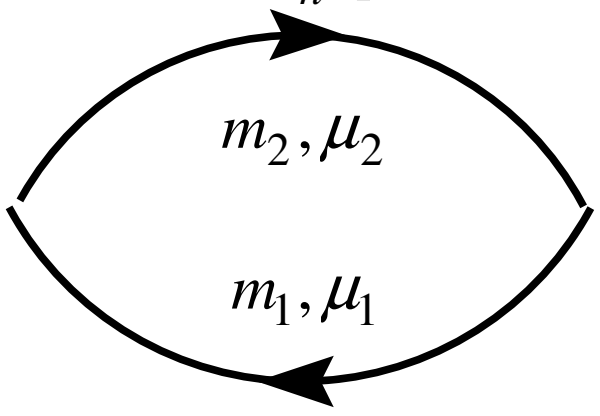<br><br>$i\cdot\omega_n - i\cdot\nu_m, \vec{p}-\vec{k}$<br>$= i\cdot\omega_n + \lvert i\cdot\nu_m\rvert, \vec{p}-\vec{k}$<br><br>$k_0 = -\lvert i\cdot\nu_m\rvert - \mu_2 + \mu_1$<br>$\overline{q_2\overline{q}_1} = q_1\overline{q}_2$ |

**Table 1.** Possible graphs for mesons.

The case $i\cdot\nu_m - \mu_1 + \mu_2 > 0$ corresponds to the left column of this table. It is associated with the ordinary situation: $i\cdot\nu_m > 0$. Therefore, we have the particle $q_1\overline{q}_2$, composed by a quark $q_1$ and by an anti-quark $\overline{q}_2$. According to the rules we previously saw:

$$k_0 = \lvert i\cdot\nu_m\rvert - \mu_1 + \mu_2 . \quad (4)$$

$i \cdot v_m - \mu_1 + \mu_2 < 0$ concerns the case exposed to the right-hand side of table 1. It corresponds to the second line of (3). To have equivalence with the first case, it is enough to say that $i \cdot v_m < 0$. We thus consider the antiparticle $\overline{q_2 \overline{q}_1}$ made by the quark $q_2$ and by the anti-quark $\overline{q}_1$, which is equivalent to the particle $q_1 \overline{q}_2$. We have:

$$k_0 = -|i \cdot v_m| - \mu_2 + \mu_1 . \tag{5}$$

# 3. Diquarks

In agreement with (42) of the previous appendix, the exchange of the two quarks leaves invariant the polarization function. Physically, it corresponds to $[q_1 q_2] = [q_2 q_1]$. We thus have two additional cases compared to mesons. The four possible cases are exposed in table 2. Another difference is the loop functions are built differently. In order to be able to employ the same functions $A$ and $B_0$ as the ones employed for mesons, we use a trick in our calculations: we invert the energies–momenta of the two particles in the loop function. Compared to the equations found for the mesons, the consequence is the addition of a minus sign in front of $i \cdot v_m$.

Another point to be underlined is the use of charge conjugate particles. Its effect is to change the sign of the chemical potential of the concerned particle. The loop function invariance by two quarks exchange indicates that the signs placed in front of the chemical potentials must be identical: $-\mu_1, -\mu_2$ for the diquarks and $+\mu_1, +\mu_2$ for the anti-diquarks.

| $i\cdot\nu_m > 0$ | $i\cdot\nu_m < 0$ |
|---|---|
| $i\cdot\omega_n - i\cdot\nu_m, \vec{p}-\vec{k}$<br>$= i\cdot\omega_n - \lvert i\cdot\nu_m\rvert, \vec{p}-\vec{k}$<br>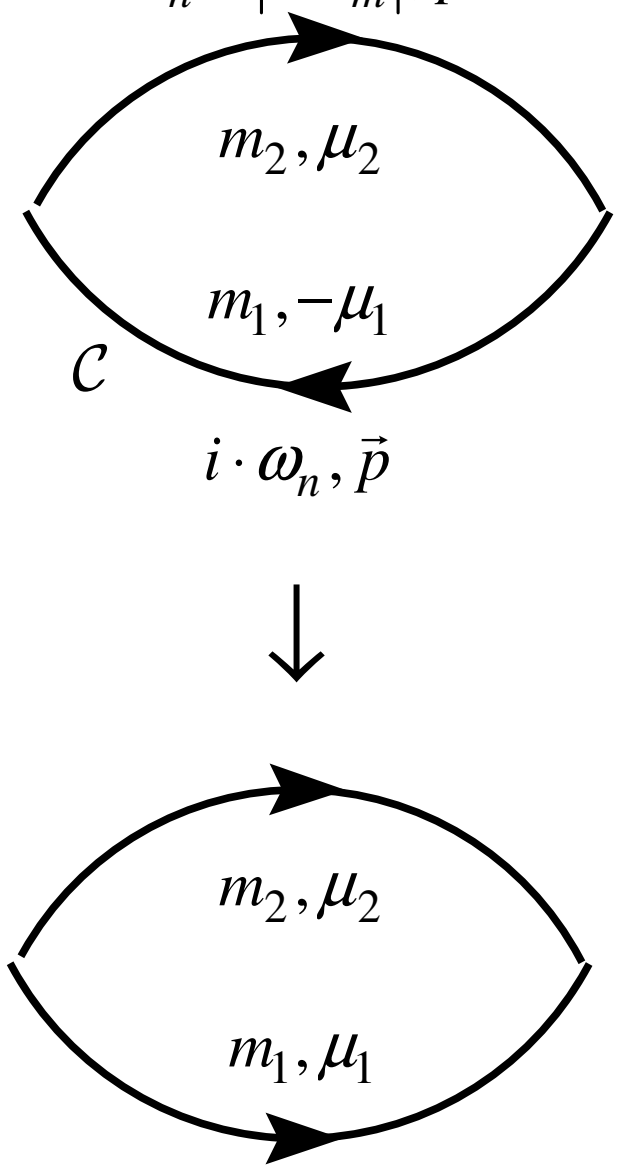<br><br>$k_0 = -\lvert i\cdot\nu_m\rvert - \mu_2 - \mu_1$<br>$q_2 q_1$ | $i\cdot\omega_n - i\cdot\nu_m, \vec{p}-\vec{k}$<br>$= i\cdot\omega_n + \lvert i\cdot\nu_m\rvert, \vec{p}-\vec{k}$<br>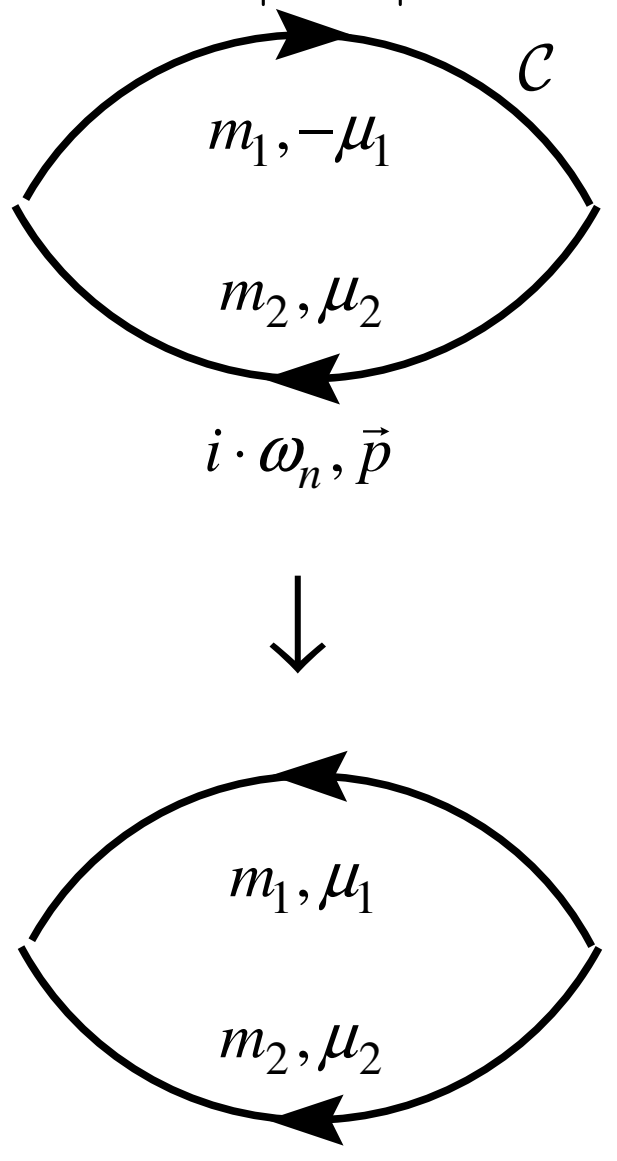<br><br>$k_0 = \lvert i\cdot\nu_m\rvert + \mu_1 + \mu_2$<br>$\overline{\bar{q}_1\bar{q}_2} = q_1 q_2$ |
| $i\cdot\omega_n - i\cdot\nu_m, \vec{p}-\vec{k}$<br>$= i\cdot\omega_n - \lvert i\cdot\nu_m\rvert, \vec{p}-\vec{k}$<br>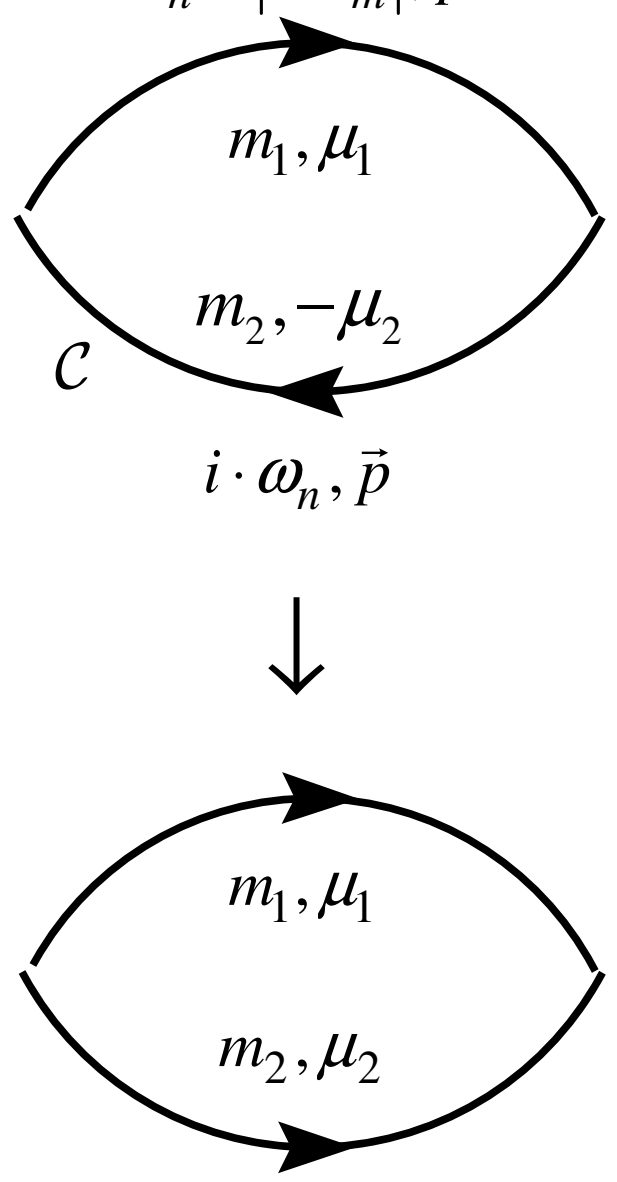<br><br>$k_0 = -\lvert i\cdot\nu_m\rvert - \mu_1 - \mu_2$<br>$q_1 q_2$ | $i\cdot\omega_n - i\cdot\nu_m, \vec{p}-\vec{k}$<br>$= i\cdot\omega_n + \lvert i\cdot\nu_m\rvert, \vec{p}-\vec{k}$<br>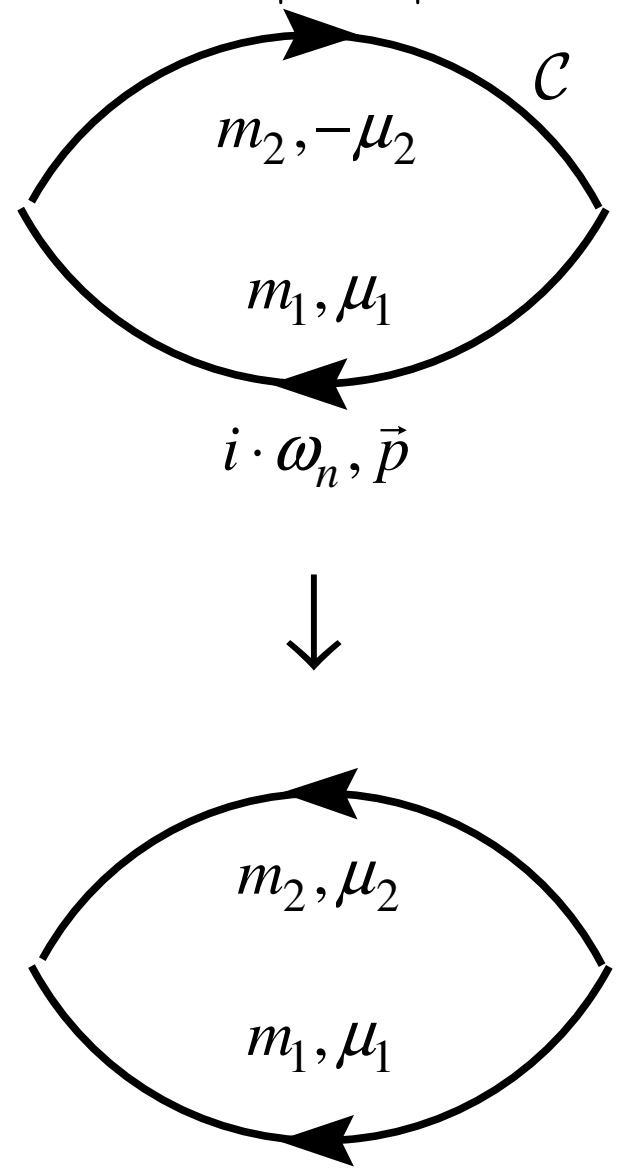<br><br>$k_0 = \lvert i\cdot\nu_m\rvert + \mu_2 + \mu_1$<br>$\overline{\bar{q}_2\bar{q}_1} = q_2 q_1$ |

**Table 2.** Possible graphs for diquarks.

The case $-i \cdot \nu_m - \mu_1 - \mu_2 > 0$, corresponds to the column on the left of table 2. In this case, $i \cdot \nu_m > 0$ , therefore it is associated with the diquark $q_1 q_2$. We have:

$$k_0 = -|i \cdot \nu_m| - \mu_1 - \mu_2 . \tag{6}$$

The case $-i \cdot \nu_m - \mu_1 - \mu_2 < 0$ is associated with the column on the right of table 2. We have there the two possible cases of the antiparticle $\overline{\bar{q}_2 \bar{q}_1}$ . Of course, it is equivalent to $q_2 q_1$. Here, $i \cdot \nu_m < 0$; we thus obtain:

$$k_0 = |i \cdot \nu_m| + \mu_2 + \mu_1 . \tag{7}$$

# 4. Baryons

About the baryons, the things are formally identical to what we said for the diquarks. The only difference is we replace one of the two quarks by a diquark. Moreover, it was seen equation (68) of the previous appendix that the two components $\Pi^{(1)}$ and $\Pi^{(2)}$ that model the baryon, although structurally different, are equivalent. It corresponds to the passage from the first to the second line, in the table 3.

If $-i \cdot \nu_m - \mu_q - \mu_D > 0$, we consider the two cases of the left-hand column of table 3. We have real particles, so $i \cdot \nu_m > 0$. As with the diquarks, we have:

$$k_0 = -|i \cdot \nu_m| - \mu_q - \mu_D . \tag{8}$$

If $-i \cdot \nu_m - \mu_q - \mu_D < 0$, i.e. the two cases of the right-hand column of table 3, we have the antiparticle $\overline{\bar{D}\bar{q}}$ , which is equivalent to $Dq$. Here, $i \cdot \nu_m < 0$, so:

$$k_0 = |i \cdot \nu_m| + \mu_q + \mu_D . \tag{9}$$

| $i\cdot\nu_m > 0$ | $i\cdot\nu_m < 0$ |
|---|---|
| $i\cdot\omega_n - i\cdot\nu_m, \vec{p}-\vec{k}$<br>$= i\cdot\omega_n - \lvert i\cdot\nu_m\rvert, \vec{p}-\vec{k}$<br>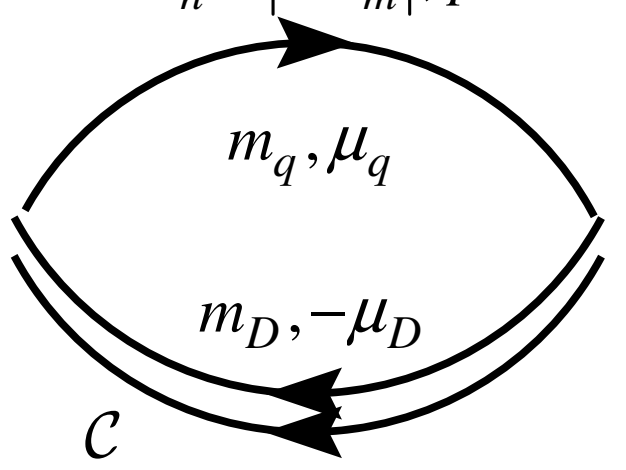<br>$i\cdot\omega_n, \vec{p}$<br>↓<br>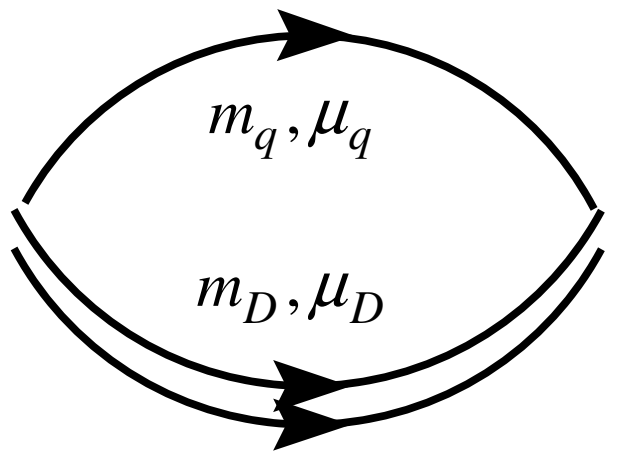<br>$k_0 = -\lvert i\cdot\nu_m\rvert - \mu_q - \mu_D$<br>$qD$ | $i\cdot\omega_n - i\cdot\nu_m, \vec{p}-\vec{k}$<br>$= i\cdot\omega_n + \lvert i\cdot\nu_m\rvert, \vec{p}-\vec{k}$<br>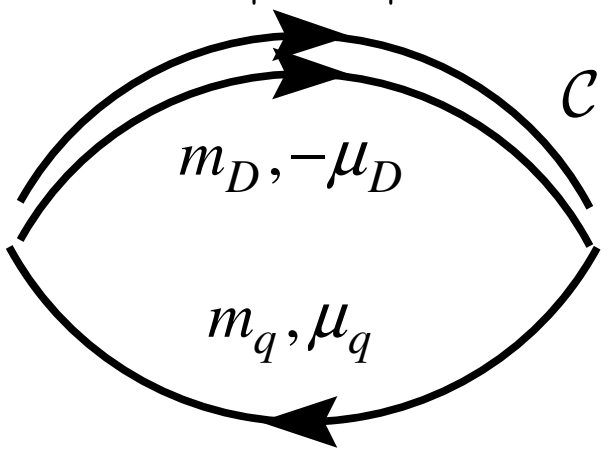<br>$i\cdot\omega_n, \vec{p}$<br>↓<br>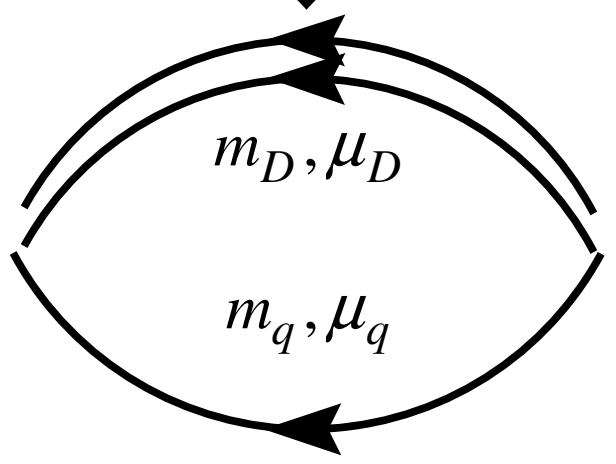<br>$k_0 = \lvert i\cdot\nu_m\rvert + \mu_D + \mu_q$<br>$\overline{\overline{D}\overline{q}} = Dq$ |
| $i\cdot\omega_n - i\cdot\nu_m, \vec{p}-\vec{k}$<br>$= i\cdot\omega_n - \lvert i\cdot\nu_m\rvert, \vec{p}-\vec{k}$<br>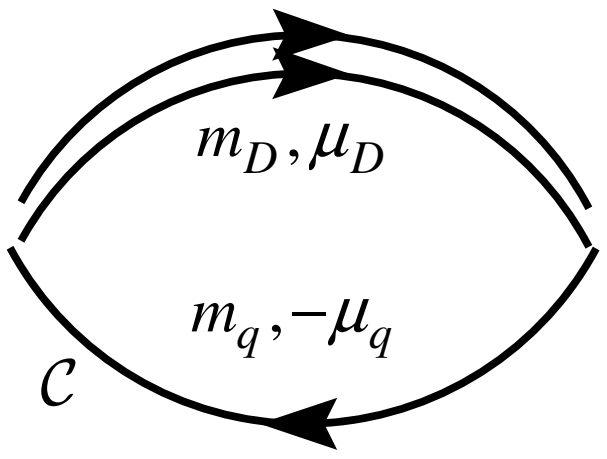<br>$i\cdot\omega_n, \vec{p}$<br>↓<br>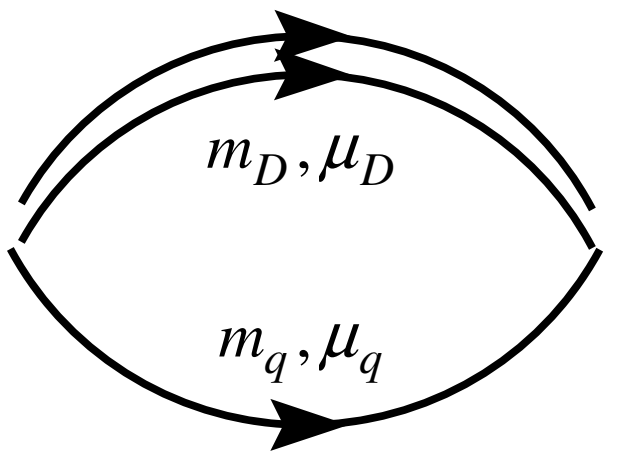<br>$k_0 = -\lvert i\cdot\nu_m\rvert - \mu_D - \mu_q$<br>$Dq$ | $i\cdot\omega_n - i\cdot\nu_m, \vec{p}-\vec{k}$<br>$= i\cdot\omega_n + \lvert i\cdot\nu_m\rvert, \vec{p}-\vec{k}$<br>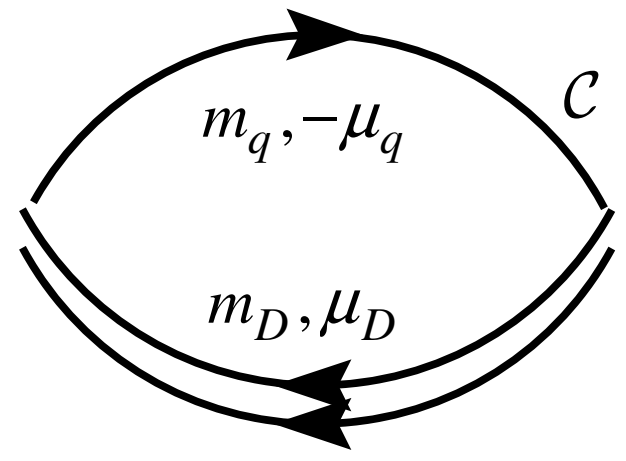<br>$i\cdot\omega_n, \vec{p}$<br>↓<br>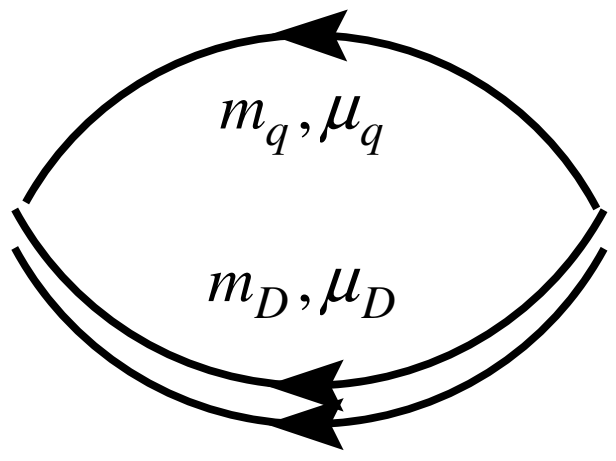<br>$k_0 = \lvert i\cdot\nu_m\rvert + \mu_q + \mu_D$<br>$\overline{\overline{q}\overline{D}} = qD$ |

**Table 3**. Possible graphs for baryons.

# Appendix F

# Elements of kinematics

The section 1 of this appendix was published in *J. Phys. G: Nucl. Part. Phys.* **39** 105003.
The sections 2, 3 intervene in *Phys. Rev.* C **89** 065204.

## 1. Mandelstam variables

### 1.1 General case

Let us consider an unspecified collision that includes two incoming particles and two outgoing ones. The collision is schematized by the figure 1, in which $p_i$ designates the four-momentum of the particle *i*. Thanks to the conservation of the energy and momenta, we write:

$$p_1 + p_2 = p_3 + p_4 \,. \tag{1}$$

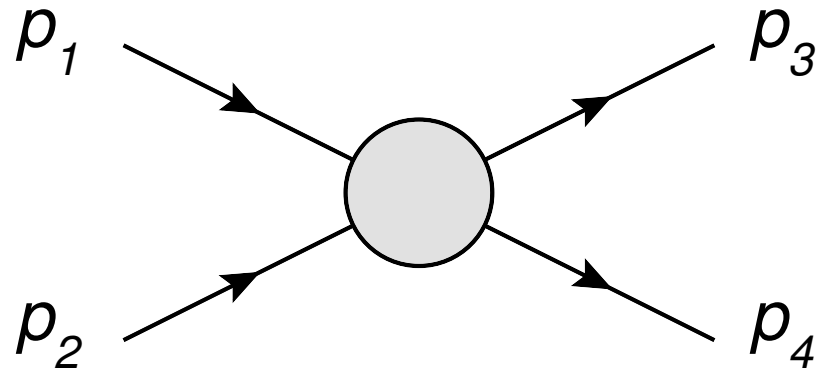

**Figure 1.** Schematization of the collisions treated in our work.

With the Mandelstam variables *s*, *t*, *u*, we have the relations:

$$\begin{cases} s = (p_1 + p_2)^2 = (p_3 + p_4)^2 \\ t = (p_3 - p_1)^2 = (p_4 - p_2)^2 \\ u = (p_4 - p_1)^2 = (p_3 - p_2)^2 \end{cases} \,. \tag{2}$$

We note $m_i$ the mass of the particle *i*. It comes:

$$s + t + u = m_1^2 + m_2^2 + m_3^2 + m_4^2 \;, \tag{3}$$

$$\begin{aligned} &2p_1 \cdot p_2 = s - m_1^2 - m_2^2 \,, && 2p_1 \cdot p_3 = m_1^2 + m_3^2 - t \,, && 2p_1 \cdot p_4 = m_1^2 + m_4^2 - u \,, \\ &2p_3 \cdot p_4 = s - m_3^2 - m_4^2 \,, && 2p_2 \cdot p_4 = m_2^2 + m_4^2 - t \,, && 2p_2 \cdot p_3 = m_2^2 + m_3^2 - u \,. \end{aligned} \tag{4}$$

## 1.2 Center of mass reference frame

The * used hereafter indicates that the concerned physical variable is expressed in the center of mass reference frame of the two incoming particles. This one is defined as the reference frame in which $\vec{p}_2^* + \vec{p}_1^* = \vec{0}$. We have the following relations between the momenta $\vec{p}_i^*$, the energies $E_i = \sqrt{(\vec{p}_i)^2 + m_i^2}$ and the Mandelstam variables [1]:

$$\begin{aligned} E_1^* &= \frac{s + m_1^2 - m_2^2}{2\sqrt{s}} & \|\vec{p}_1^*\| &= \frac{\sqrt{\left(s - (m_1 + m_2)^2\right)\cdot\left(s - (m_1 - m_2)^2\right)}}{2\sqrt{s}} \\ E_2^* &= \frac{s - m_1^2 + m_2^2}{2\sqrt{s}} & \vec{p}_2^* &= -\vec{p}_1^* \\ E_3^* &= \frac{s + m_3^2 - m_4^2}{2\sqrt{s}} & \|\vec{p}_3^*\| &= \frac{\sqrt{\left(s - (m_3 + m_4)^2\right)\cdot\left(s - (m_3 - m_4)^2\right)}}{2\sqrt{s}} \\ E_4^* &= \frac{s - m_3^2 + m_4^2}{2\sqrt{s}} & \vec{p}_4^* &= -\vec{p}_3^* \end{aligned} . \tag{5}$$

Furthermore, $t \in [t_-, t_+]$, with:

$$t_\pm = m_1^2 + m_3^2 - 2\cdot E_1^* \cdot E_3^* \pm 2\cdot\|\vec{p}_1^*\|\cdot\|\vec{p}_3^*\| . \tag{6}$$

# 2. Calculation of the impact parameter

The momenta of the incoming particles *1* and *2* are initially expressed in the laboratory frame. So, a first step to estimate the impact parameter $b^*$ is to apply a Lorentz boost on these momenta, in order to work in the center of mass reference frame of these two particles:

$$\begin{cases} \vec{p}_1^* = \vec{p}_1 + \vec{v}_{CM}\left((\Gamma - 1)\dfrac{\vec{p}_1 \cdot \vec{v}_{CM}}{\|\vec{v}_{CM}\|^2} - \Gamma E_1\right), \\ E_1^* = \Gamma\left(E_1 - \vec{p}_1 \cdot \vec{v}_{CM}\right) \end{cases} \tag{7}$$

where $\vec{v}_{CM}$ is the velocity of the center of mass reference frame. $\vec{v}_{CM}$ is written as:

$$\vec{v}_{CM} = \frac{\vec{p}_1 + \vec{p}_2}{E_1 + E_2}, \tag{8}$$

and the Lorentz factor is:

$$\Gamma = \frac{1}{\sqrt{1 - (v_{CM}/c)^2}}, \tag{9}$$

in which the speed of light $c$ is set equal to 1. Equation (7) is then simply adapted for the momentum of the incoming particle 2. The same procedure is applied to the positions of these particles 1 and 2, to express them in the center of mass reference frame. We obtain then, respectively, $\vec{r}_1^*$ and $\vec{r}_2^*$. We pose $\vec{r}^* = \vec{r}_1^* - \vec{r}_2^*$ and $\vec{p}^* = \vec{p}_2^* - \vec{p}_1^*$. We have $\vec{p}_1^* + \vec{p}_2^* = \vec{0}$.

Consequently, the choice $\vec{p}^* = \vec{p}_2^*$ gives the same results in the calculation performed hereafter.

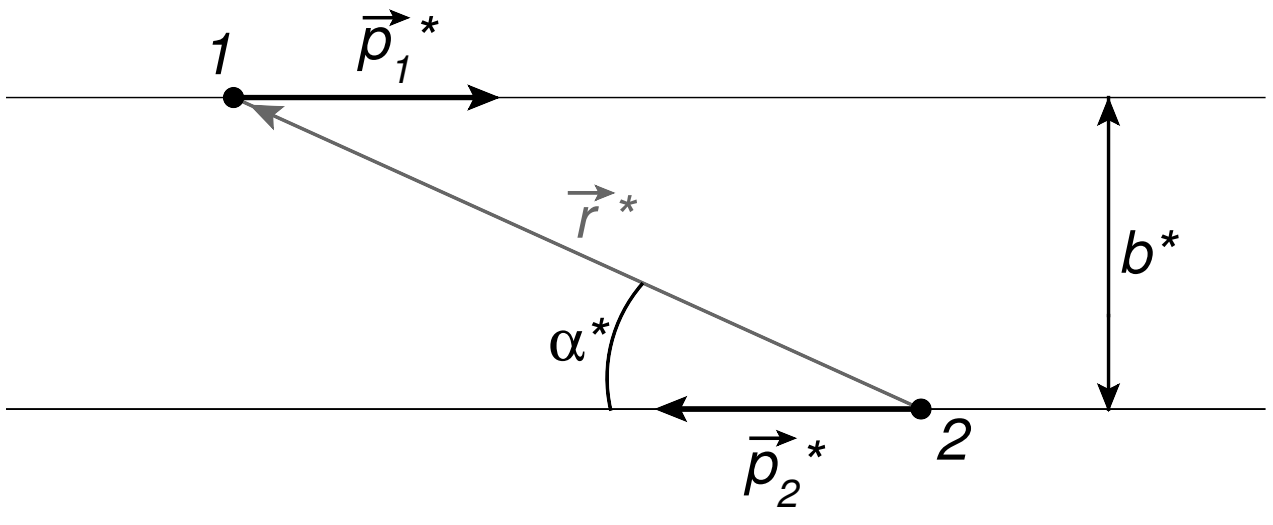

**Figure 2.** Determination of the impact parameter $b^*$.

With the figure 2, the equations (10) are found by geometrical considerations:

$$\cos\left(\alpha^*\right) = \frac{\vec{r}^* \cdot \vec{p}^*}{\left\|\vec{r}^*\right\| \left\|\vec{p}^*\right\|} \qquad \sin\left(\alpha^*\right) = \frac{b^*}{\left\|\vec{r}^*\right\|}. \tag{10}$$

Thanks to the relation $\sin\left(\cos^{-1}(x)\right) = \sqrt{1-x^2}$, the impact parameter $b^*$ established in the center of mass reference frame is expressed, as in [2]:

$$b^* = \sqrt{\left\|\vec{r}^*\right\|^2 - \left(\frac{\vec{r}^* \cdot \vec{p}^*}{\left\|\vec{p}^*\right\|}\right)^2}. \tag{11}$$

In fact, the first equation of (10) indicates if the particles are approaching ($\cos\left(\alpha^*\right) \geq 0$) or moving away ($\cos\left(\alpha^*\right) < 0$). This information cannot be supplied only by (11). In the case $\cos\left(\alpha^*\right) < 0$, the collision procedure is aborted for these two particles (see chapter 7).

# 3. Scattering angle and outgoing particles momenta

The figure 3 describes the used method to estimate the scattering angle $\theta^*$, in which $\sigma$ is the cross section associated with the considered reaction $1+2 \rightarrow 3+4$. Also, $\vec{r}^*$ is a vector that connects the two particles *1* and *2* at the precise instant of their interaction. This one is supposed to be punctual in time and space. In addition, $\vec{p}_3^*$ and $\vec{p}_4^*$ represent, respectively, the momenta of the outgoing particles *3* and *4*. Their modulus is found with (5).

Our approach takes the hard spheres model as a starting point, but the sum of the radius of the particles *1* and *2* is replaced here by $\sqrt{\sigma/\pi}$. Then, the particle *1* becomes the particle 3 after a rebound on the sphere, according to Snell-Descartes' law of reflection, whereas the particle *2* becomes the particle *4*. Thanks to geometrical considerations, the scattering angle $\theta^*$ is expressed as:

$$\theta^* = \pi - 2\sin^{-1}\left(b\sqrt{\pi/\sigma}\right). \tag{12}$$

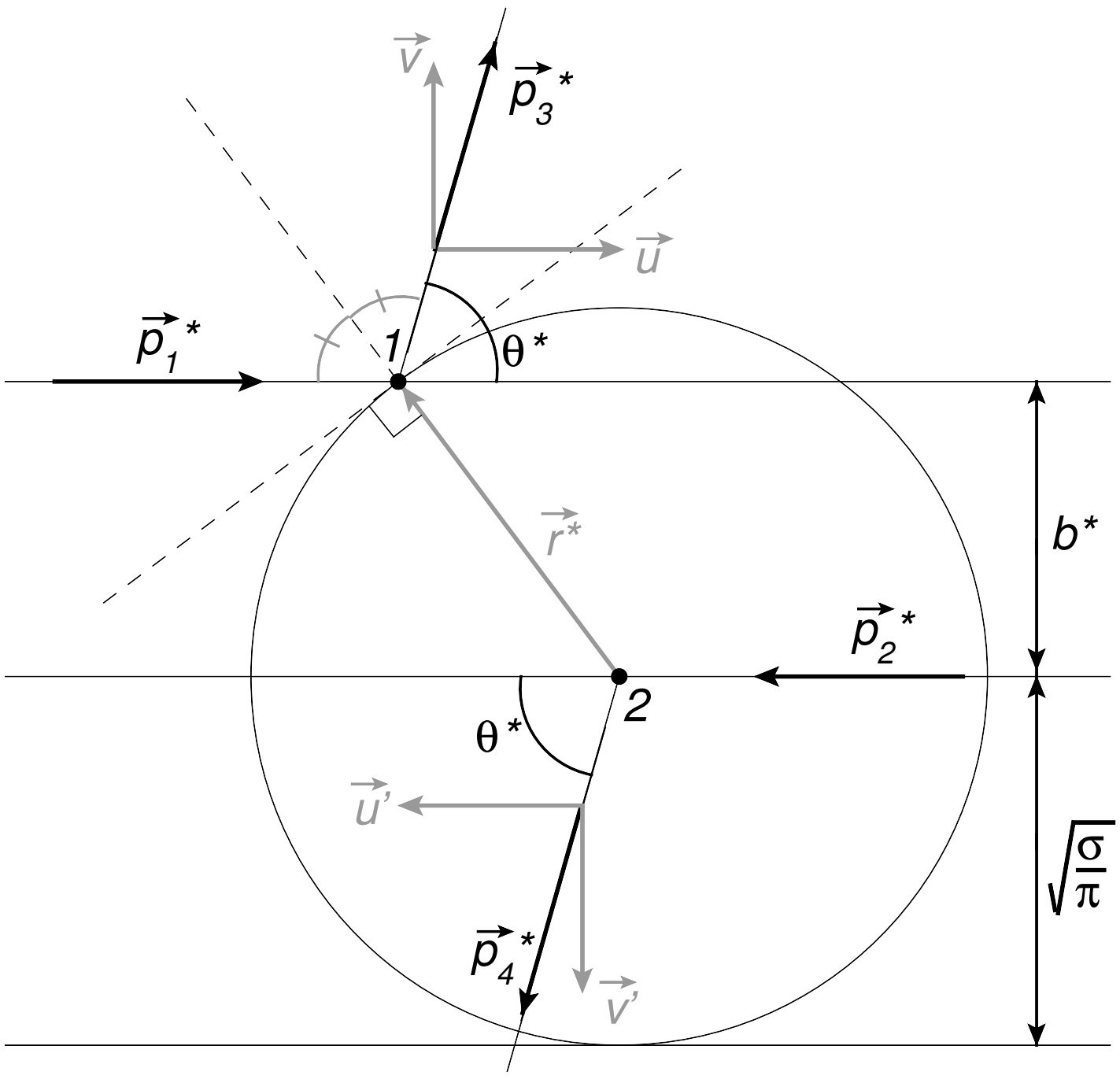

**Figure 3.** Estimation of the scattering angle $\theta^*$.

Thanks to this angle, the coordinates of the vectors $\vec{p}_3^*$ and $\vec{p}_4^*$ are then fully determinable. About the particle *3,* we create an orthonormal coordinate system $(\vec{u},\vec{v})$. We define $\vec{u}$ as an unitary vector built with $\vec{p}_1^*$:

$$\vec{u} = \frac{\vec{p}_1^*}{\left\|\vec{p}_1^*\right\|}. \tag{13}$$

To define the unitary vector $\vec{v}$, we note that $\vec{p}_3^*$ is in the plane containing at the same time $\vec{p}_1^*$ and $\vec{r}^*$. Thus, $\vec{u},\vec{v},\vec{p}_3^*$ must be coplanar. There are two possible vectors for $\vec{v}$. The both are collinear, but they have opposite directions. We choose the one that is oriented towards the opposite of the particle *2*. After calculations, it comes:

$$\vec{v} = \frac{\vec{r}^* - \left(\vec{r}^* \cdot \vec{u}\right)\vec{u}}{\left\|\vec{r}^* - \left(\vec{r}^* \cdot \vec{u}\right)\vec{u}\right\|}. \tag{14}$$

$\vec{p}_3^*$ is then projected in this coordinate system $(\vec{u},\vec{v})$: $\vec{p}_3^* = \left\|\vec{p}_3^*\right\| \cdot \cos\left(\theta^*\right) \cdot \vec{u} + \left\|\vec{p}_3^*\right\| \cdot \sin\left(\theta^*\right) \cdot \vec{v}$. The procedure could be similar for $\vec{p}_4^*$, or we can use the fact that $\vec{p}_4^* = -\vec{p}_3^*$, according to (5).

Finally, a Lorentz boost is applied to these momenta $\vec{p}_3^*$ and $\vec{p}_4^*$, in order to express them in the laboratory frame.

# Appendix G

# Final simulation

The vocation of this last appendix is to represent the evolution of one of our complete simulations, studied in the chapter 7. In this part of the work, the isospin symmetry was not taken into account. So, a great number of particles is considered. Indeed, we included in our simulation, the $u,d,s$ quarks, their anti-quarks, the nine pseudo-scalar mesons, the three scalar diquarks, their anti-diquarks, the octet baryons and the associated anti-baryons. Therefore, for the colored versions in this thesis, we propose hereafter color conventions, in order to represent the treated particles. This color code is based on the trichromatic synthesis known in optics or photography, figure 1 for the quarks/antiquarks and the mesons, and figure 2 for the diquarks/antidiquarks and the baryons/antibaryons. Also, the diameter of the disks used to represent each particle allows recognizing its type. More precisely, the quarks are represented by the most reduced disks. The mesons have got little more extended disks, then we have the diquarks, and the greater diameter is associated to baryons.

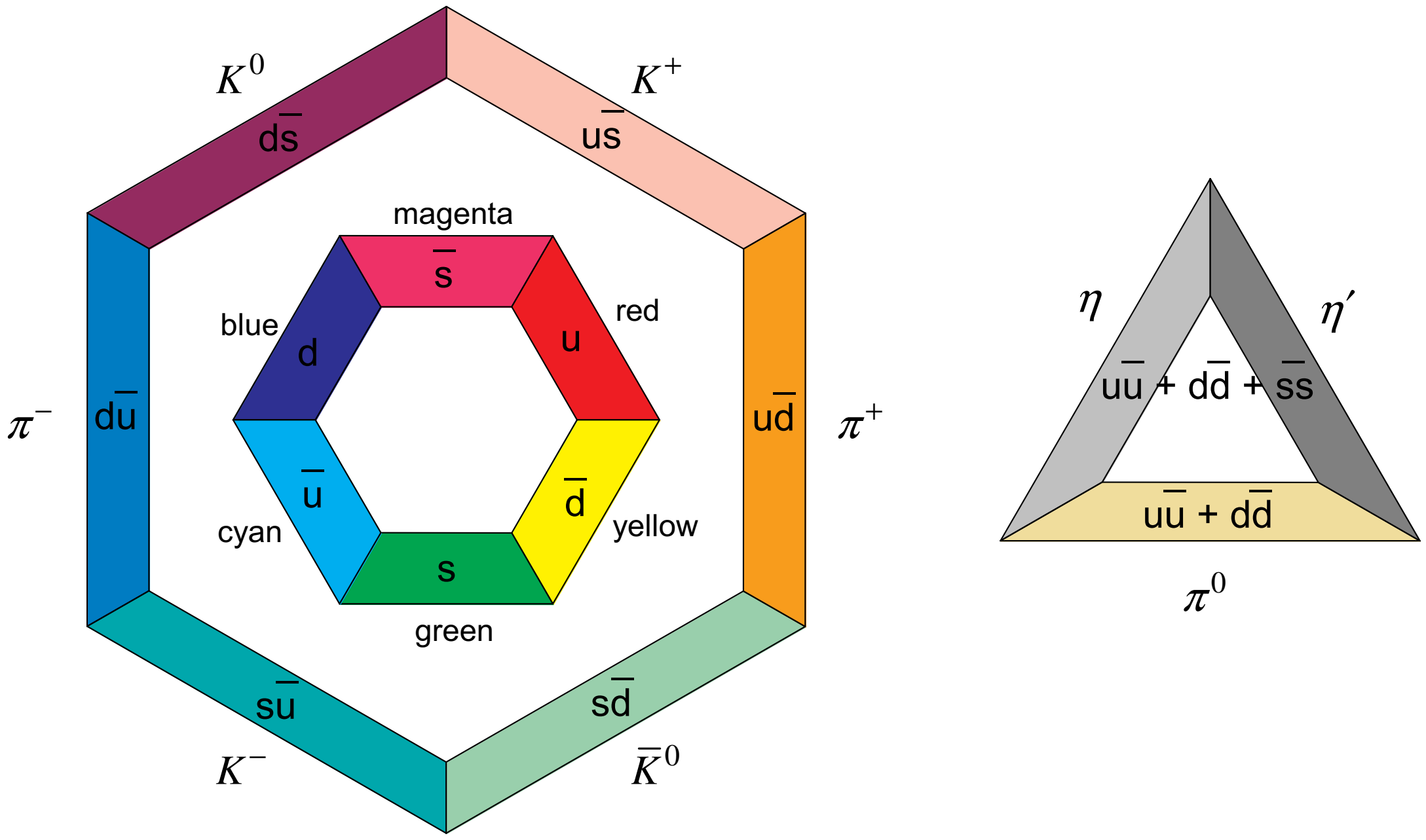

**Figure 1.** Color code for the quarks/antiquarks and mesons.

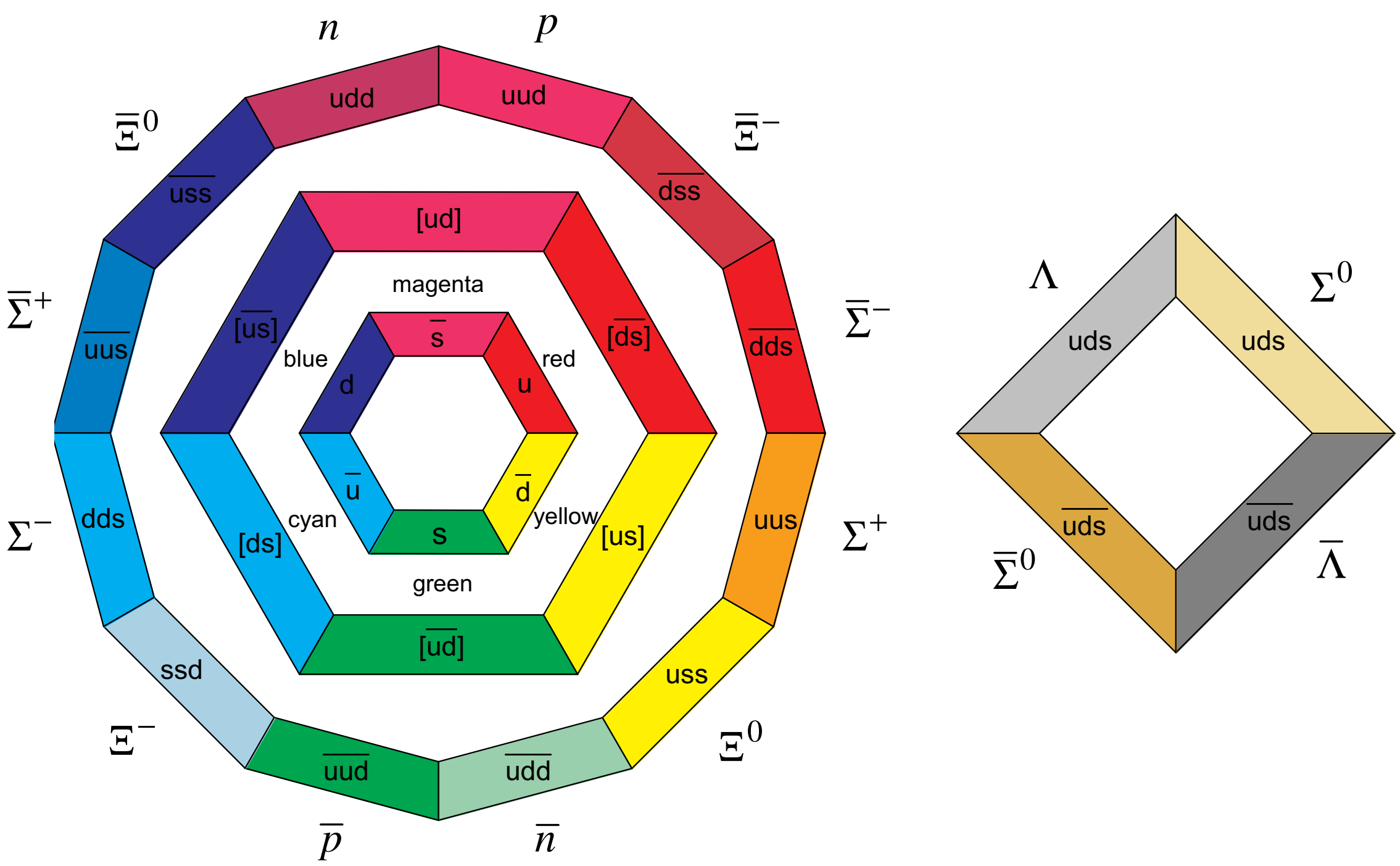

**Figure 2.** Color code for the diquarks/anti-diquarks and baryons/anti-baryons.

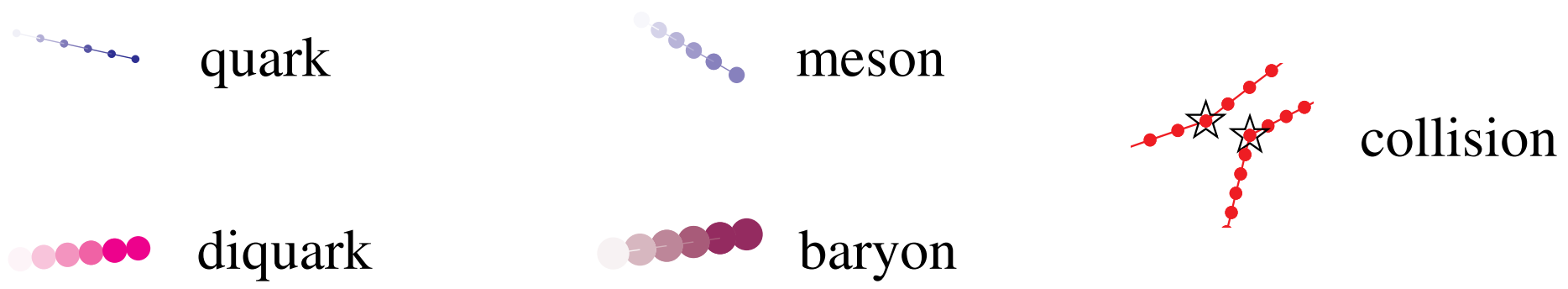

**Figure 3.** Representation of the included particles.

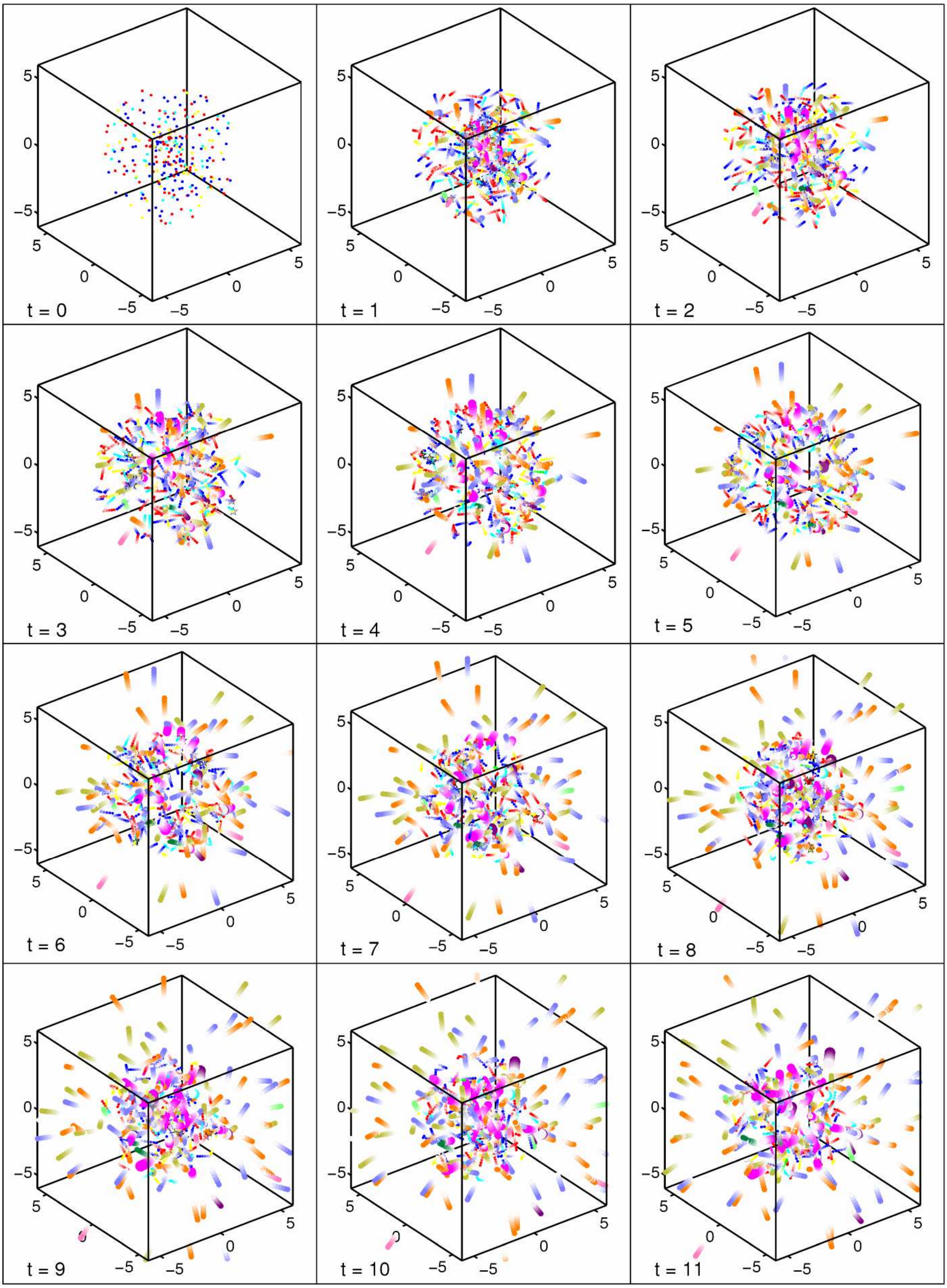

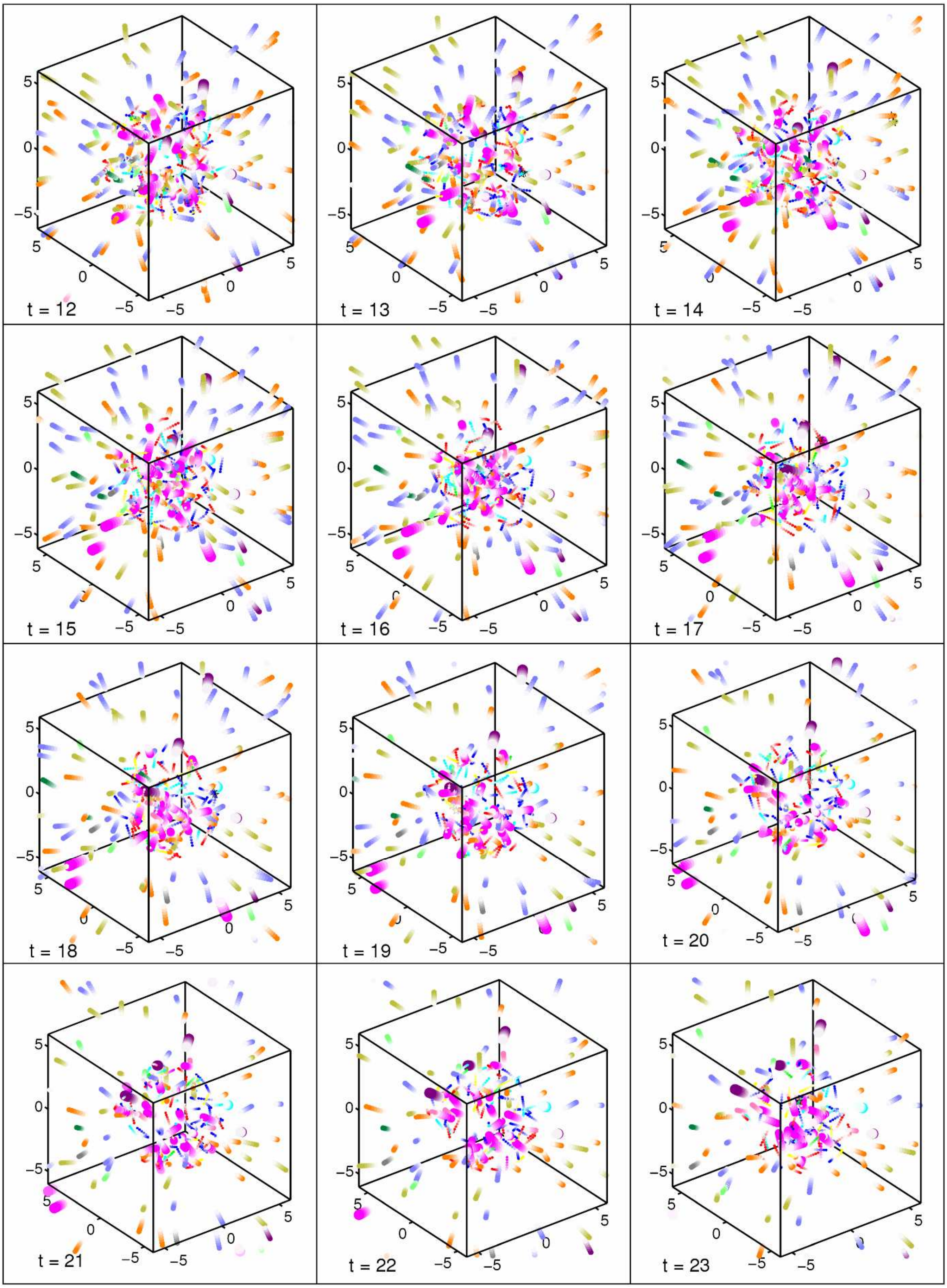

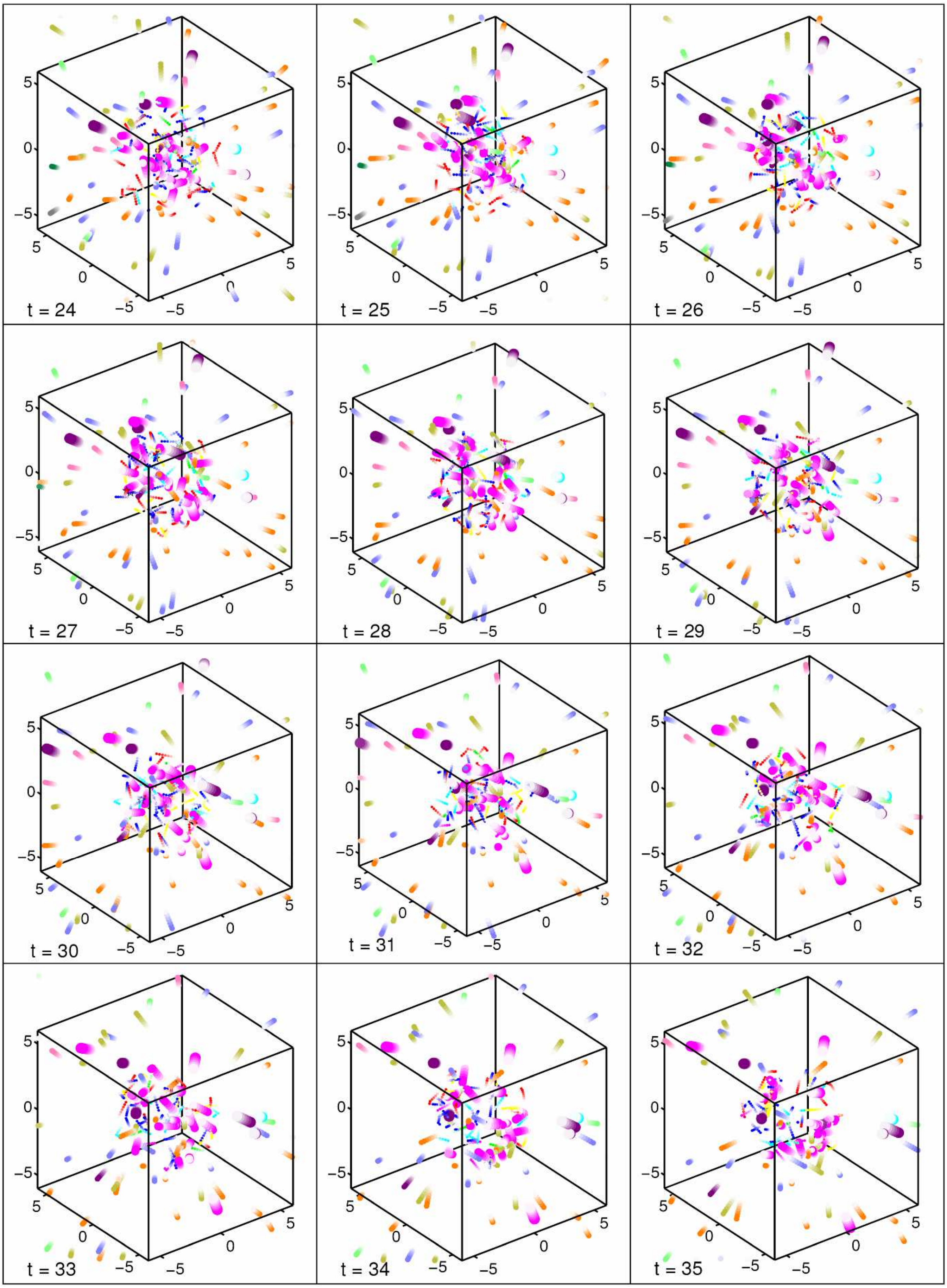

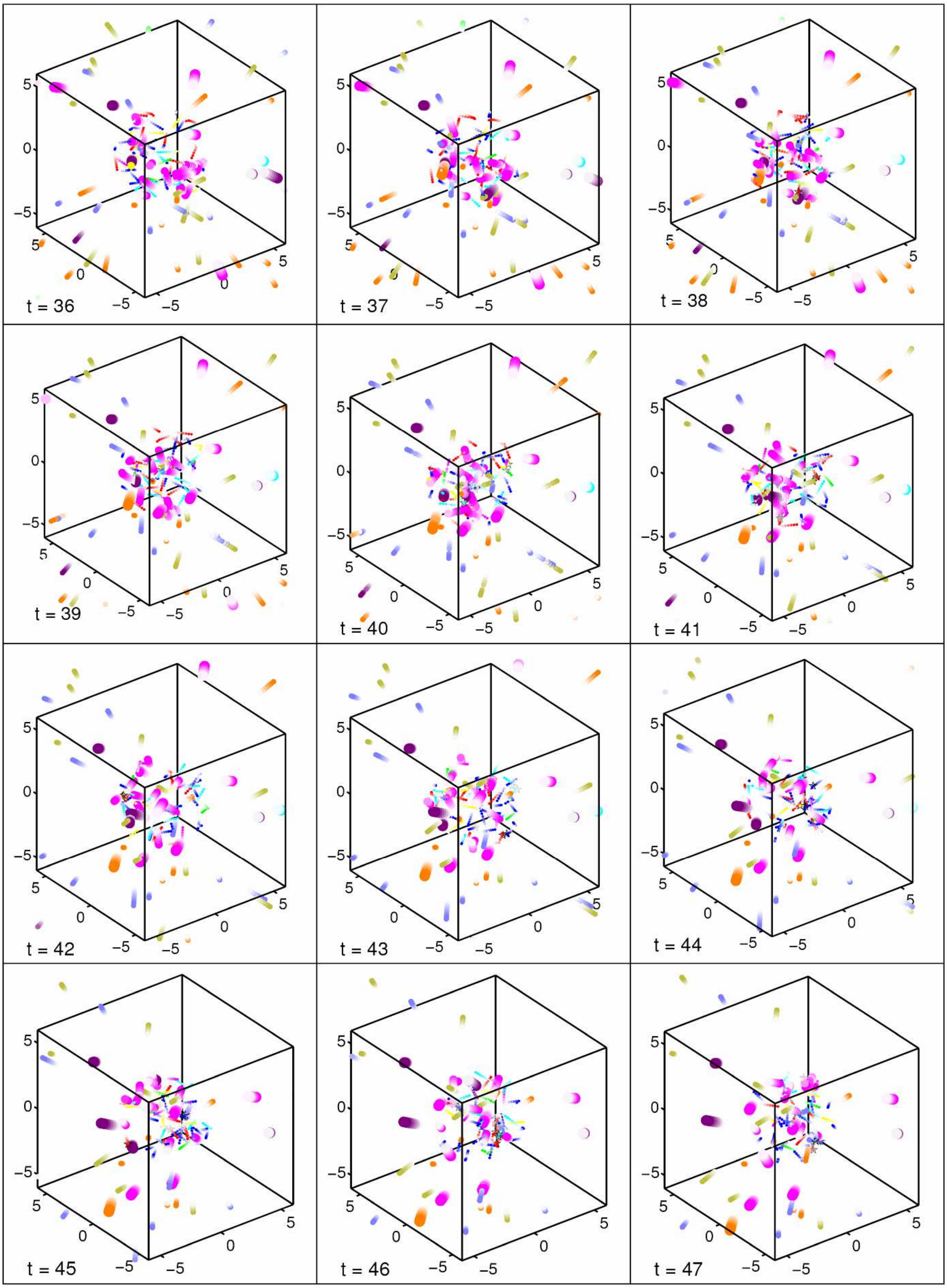

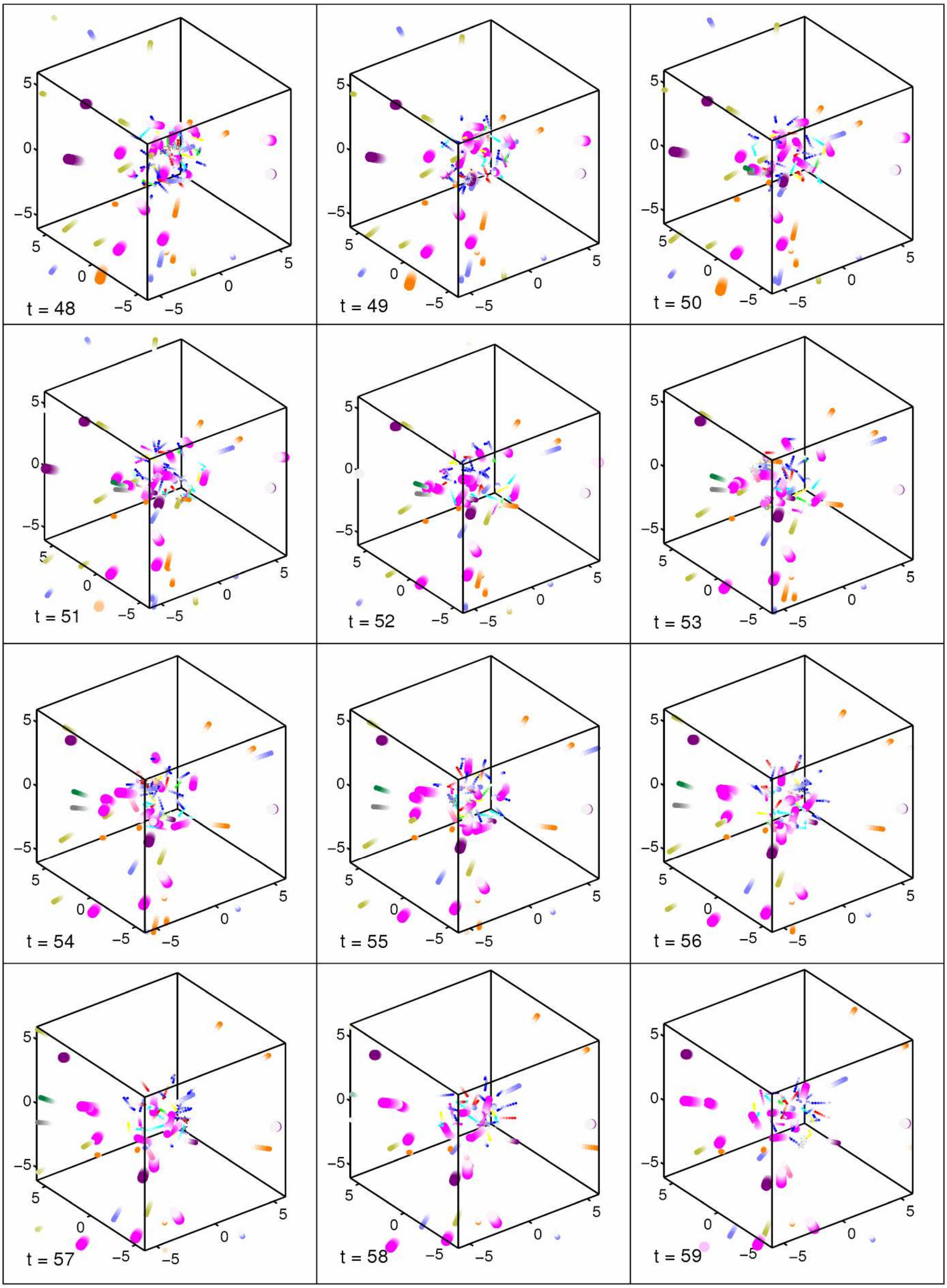

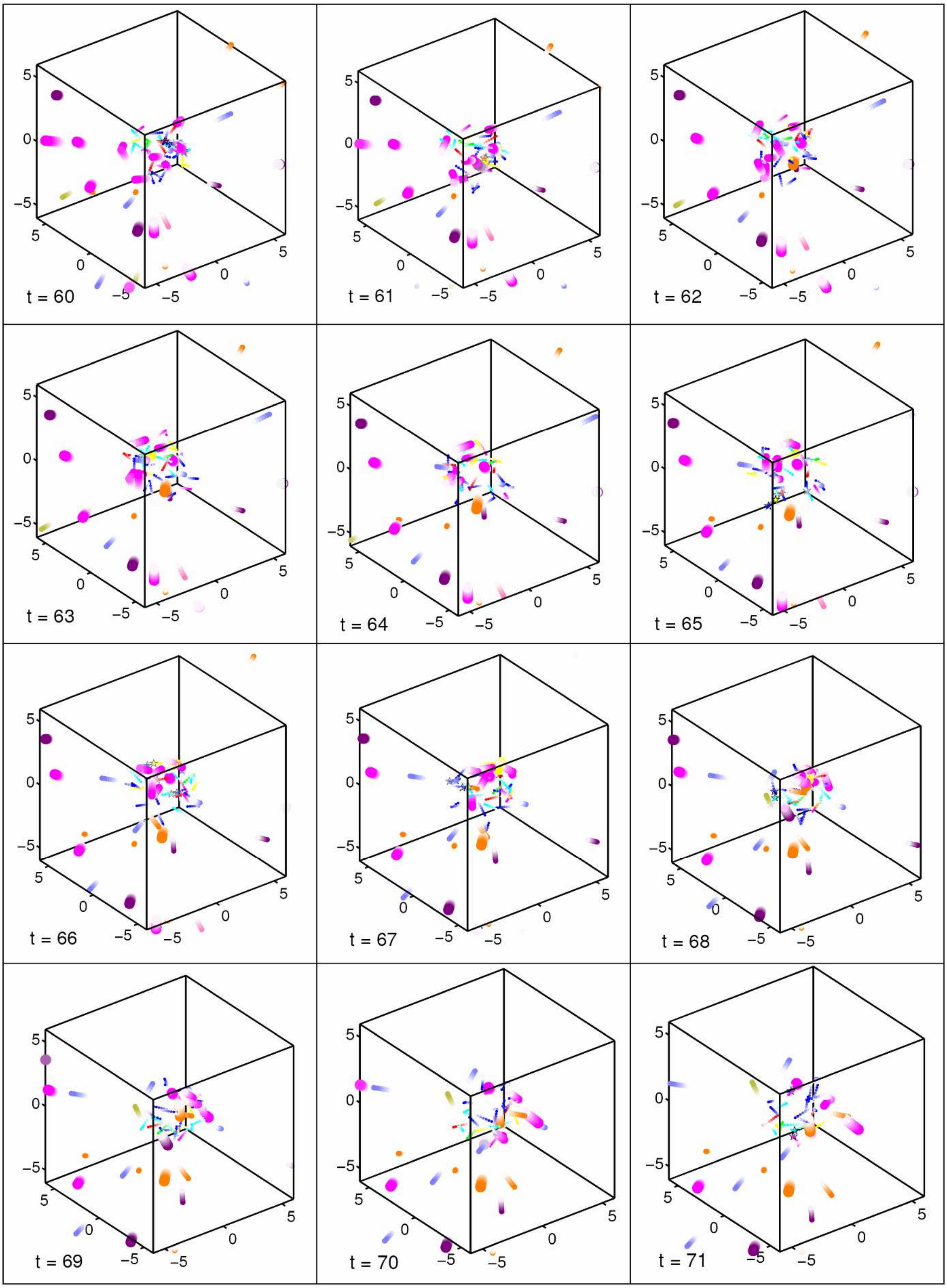

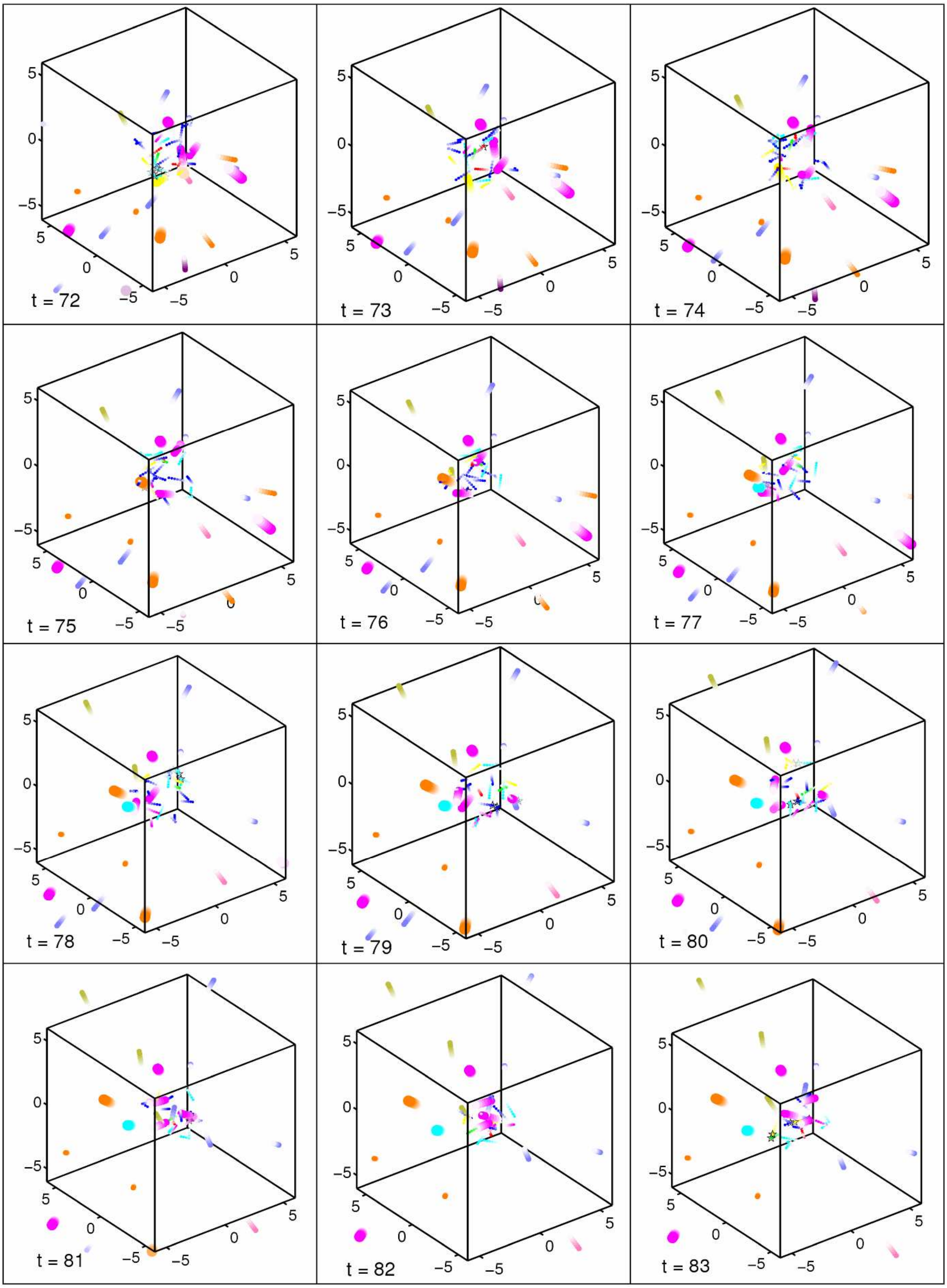

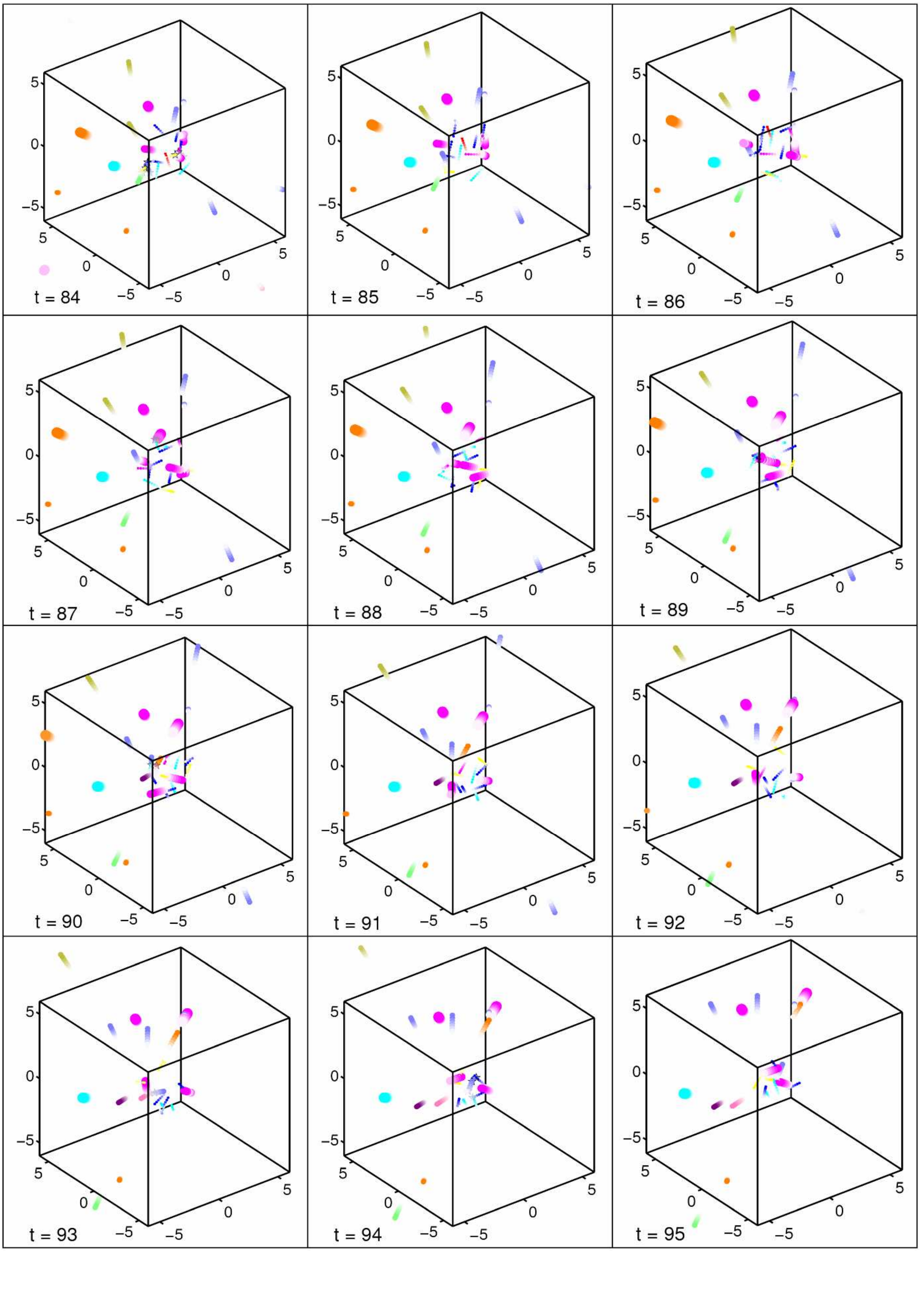

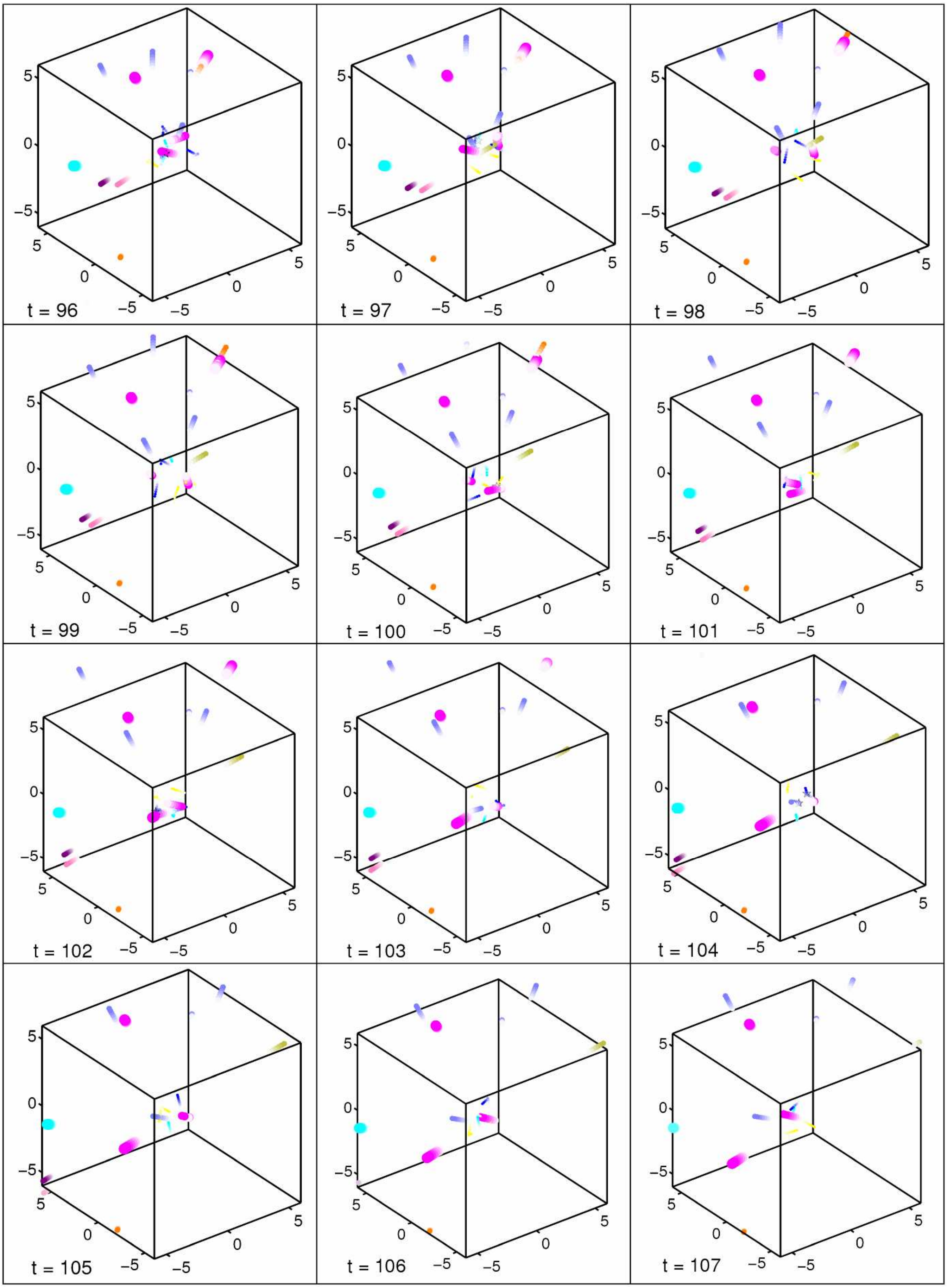

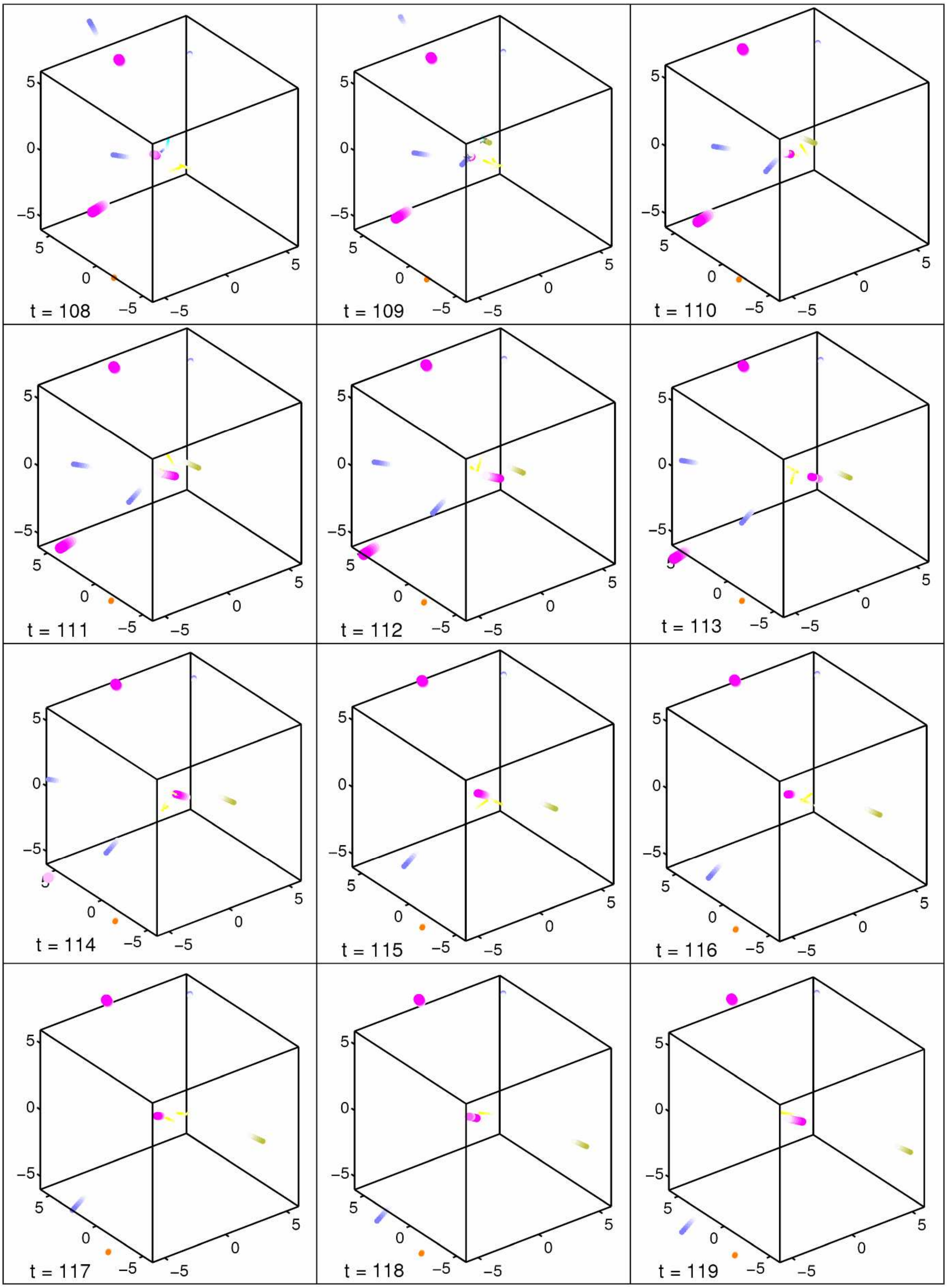

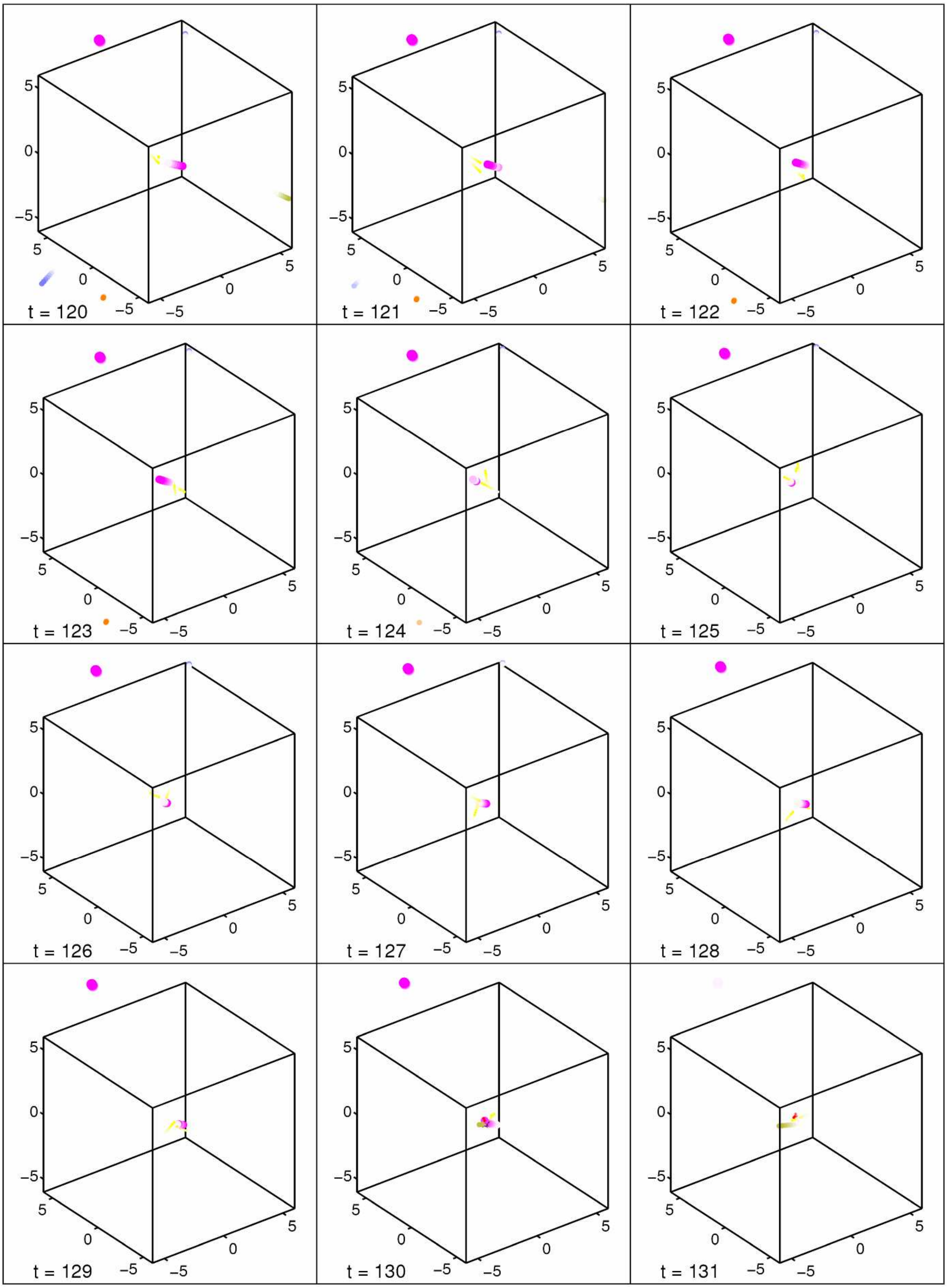

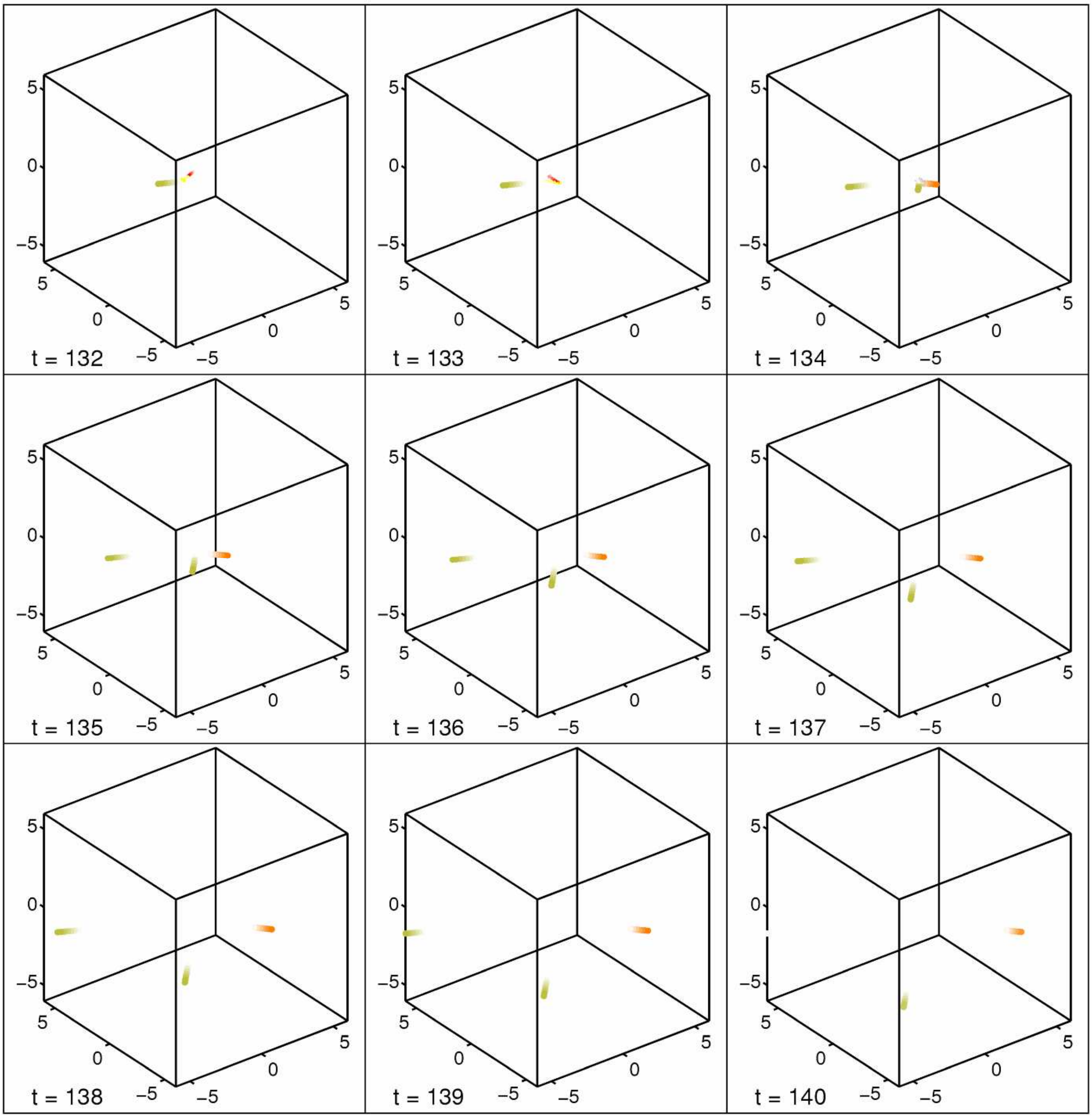

**Author**: Eric BLANQUIER

**Title**: The Polyakov, Nambu and Jona-Lasinio model and its applications to describe the sub-nuclear particles

**Thesis supervisor**: Patrice RAYNAUD

---

**Abstract:**

To study the high energy nuclear physics and the associated phenomenon, as the QGP/hadronic matter phase transition, the Nambu and Jona-Lasinio model (NJL) appears as an interesting alternative to the Quantum Chromodynamics, which is not solvable at the considered energies. Indeed, the NJL model allows describing the quarks physics, at finite temperatures and finite densities. Furthermore, in order to try to correct a limitation of the NJL model, i.e. the absence of confinement, it was proposed to couple the quarks/antiquarks to a Polyakov loop: it forms the PNJL model.

The objective of this thesis is to see the possibilities offered by the NJL and PNJL models, to describe the relevant sub-nuclear particles (quarks, mesons, diquarks and baryons), to study their interactions, and to proceed to a dynamical study involving these particles.

After a recall of the useful tools, we modeled the *u, d, s* effective quarks and the mesons. Then, we described the baryons as quarks–diquarks bound states. A part of the work concerned the calculations of the cross-sections associated with the possible reactions that include these particles. Then, we incorporated these works in a computer code, in order to study the cooling of a quarks/antiquarks plasma and its hadronization. In this study, each particle evolves in a system in which the temperature and the densities are local parameters. We have two types of interactions: one due to the collisions, and a remote interaction, notably between quarks. Finally, we studied the properties of our approach: qualities, limitations, and possible evolutions.

---

---

**Administrative topic**: Plasma engineering

---

**Laboratory**: Laboratoire Plasma et Conversion d'Energie (LAPLACE) – UMR 5213,
118, Route de Narbonne 31062 Toulouse cedex 9, France

**Auteur** : Eric BLANQUIER

**Titre** : Le modèle de Polyakov, Nambu et Jona-Lasinio et ses applications pour décrire les particules sub-nucléaires

**Directeur de thèse** : Patrice RAYNAUD

---

**Résumé :**

Pour étudier la physique nucléaire des hautes énergies et les phénomènes associés, comme la transition de phase quark-gluon-plasma/matière hadronique, le modèle de Nambu et Jona Lasinio (NJL) constitue une alternative intéressante à la Chromodynamique Quantique, non solvable aux énergies considérées. En effet, le modèle NJL permet de décrire la physique des quarks à températures et densités finies. D'autre part, afin de tenter de corriger une limitation de ce modèle, l'absence de confinement, il a été proposé un couplage des quarks/antiquarks à une boucle de Polyakov, formant le modèle PNJL.

L'objectif de cette thèse est de voir les possibilités offertes par les modèles NJL et PNJL, afin de décrire les particules sub-nucléaires pertinentes (quarks, mésons diquarks et baryons), d'étudier leurs interactions et de mener une étude dynamique avec ces particules.

Après un rappel des outils pertinents, nous avons modélisé les quarks effectifs *u*, *d*, *s*, et les mésons. Ensuite, nous avons décrit les baryons comme des états liés quarks–diquarks. Une part du travail a concerné le calcul des sections efficaces liées aux réactions possibles avec ces particules. Nous avons incorporé ces travaux dans un code de calcul pour étudier le refroidissement d'un plasma de quarks/antiquarks et son hadronisation. Dans cette étude, chaque particule évolue dans un système où la température et les densités sont des paramètres locaux. Les interactions entre particules sont de deux types : interactions par collisions et interactions à distance, notamment entre quarks. Finalement, nous avons étudié les propriétés de notre approche : qualités, limitations et évolutions possibles.

---

---

**Discipline administrative** : Ingénierie des PLASMAS

---

**Laboratoire** : Laboratoire Plasma et Conversion d'Energie (LAPLACE) – UMR 5213,
118, Route de Narbonne 31062 Toulouse cedex 9